\documentclass[titlepage, a4paper, oneside, doublespace]{book}
\usepackage[UKenglish]{babel}
\usepackage{color}
\usepackage{amsfonts}
\usepackage{amssymb}
\usepackage{amsmath}
\usepackage{graphicx}
\usepackage{psfrag}

\usepackage{amsmath}
\usepackage{amssymb}
\usepackage{dsfont}
\usepackage[all,arc,dvips,arrow,curve,cmactex,frame,tips]{xy}
\usepackage{mathrsfs}
\def\Nl{{\mathchoice
{\setbox0=\hbox{$\displaystyle\rm N$}\hbox{\hbox to0pt
{\kern0.4\wd0\vrule height0.9\ht0\hss}\box0}}
{\setbox0=\hbox{$\textstyle\rm N$}\hbox{\hbox to0pt
{\kern0.4\wd0\vrule height0.9\ht0\hss}\box0}}
{\setbox0=\hbox{$\scriptstyle\rm N$}\hbox{\hbox to0pt
{\kern0.4\wd0\vrule height0.9\ht0\hss}\box0}}
{\setbox0=\hbox{$\scriptscriptstyle\rm N$}\hbox{\hbox to0pt
{\kern0.4\wd0\vrule height0.9\ht0\hss}\box0}}}}

\def\Kl{{\mathchoice
{\setbox0=\hbox{$\displaystyle\rm K$}\hbox{\hbox to0pt
{\kern0.4\wd0\vrule height0.9\ht0\hss}\box0}}
{\setbox0=\hbox{$\textstyle\rm K$}\hbox{\hbox to0pt
{\kern0.4\wd0\vrule height0.9\ht0\hss}\box0}}
{\setbox0=\hbox{$\scriptstyle\rm K$}\hbox{\hbox to0pt
{\kern0.4\wd0\vrule height0.9\ht0\hss}\box0}}
{\setbox0=\hbox{$\scriptscriptstyle\rm K$}\hbox{\hbox to0pt
{\kern0.4\wd0\vrule height0.9\ht0\hss}\box0}}}}
\def\Zl{{\mathchoice
{\setbox0=\hbox{$\displaystyle\rm Z$}\hbox{\hbox to0pt
{\kern0.4\wd0\vrule height0.9\ht0\hss}\box0}}
{\setbox0=\hbox{$\textstyle\rm Z$}\hbox{\hbox to0pt
{\kern0.4\wd0\vrule height0.9\ht0\hss}\box0}}
{\setbox0=\hbox{$\scriptstyle\rm Z$}\hbox{\hbox to0pt
{\kern0.4\wd0\vrule height0.9\ht0\hss}\box0}}
{\setbox0=\hbox{$\scriptscriptstyle\rm Z$}\hbox{\hbox to0pt
{\kern0.4\wd0\vrule height0.9\ht0\hss}\box0}}}}
\def\Ql{{\mathchoice
{\setbox0=\hbox{$\displaystyle\rm Q$}\hbox{\hbox to0pt
{\kern0.4\wd0\vrule height0.9\ht0\hss}\box0}}
{\setbox0=\hbox{$\textstyle\rm Q$}\hbox{\hbox to0pt
{\kern0.4\wd0\vrule height0.9\ht0\hss}\box0}}
{\setbox0=\hbox{$\scriptstyle\rm Q$}\hbox{\hbox to0pt
{\kern0.4\wd0\vrule height0.9\ht0\hss}\box0}}
{\setbox0=\hbox{$\scriptscriptstyle\rm Q$}\hbox{\hbox to0pt
{\kern0.4\wd0\vrule height0.9\ht0\hss}\box0}}}}
\def\Rl{{\mathchoice
{\setbox0=\hbox{$\displaystyle\rm R$}\hbox{\hbox to0pt
{\kern0.4\wd0\vrule height0.9\ht0\hss}\box0}}
{\setbox0=\hbox{$\textstyle\rm R$}\hbox{\hbox to0pt
{\kern0.4\wd0\vrule height0.9\ht0\hss}\box0}}
{\setbox0=\hbox{$\scriptstyle\rm R$}\hbox{\hbox to0pt
{\kern0.4\wd0\vrule height0.9\ht0\hss}\box0}}
{\setbox0=\hbox{$\scriptscriptstyle\rm R$}\hbox{\hbox to0pt
{\kern0.4\wd0\vrule height0.9\ht0\hss}\box0}}}}
\def\Cl{{\mathchoice
{\setbox0=\hbox{$\displaystyle\rm C$}\hbox{\hbox to0pt
{\kern0.4\wd0\vrule height0.9\ht0\hss}\box0}}
{\setbox0=\hbox{$\textstyle\rm C$}\hbox{\hbox to0pt
{\kern0.4\wd0\vrule height0.9\ht0\hss}\box0}}
{\setbox0=\hbox{$\scriptstyle\rm C$}\hbox{\hbox to0pt
{\kern0.4\wd0\vrule height0.9\ht0\hss}\box0}}
{\setbox0=\hbox{$\scriptscriptstyle\rm C$}\hbox{\hbox to0pt
{\kern0.4\wd0\vrule height0.9\ht0\hss}\box0}}}}
\def\Hl{{\mathchoice
{\setbox0=\hbox{$\displaystyle\rm H$}\hbox{\hbox to0pt
{\kern0.4\wd0\vrule height0.9\ht0\hss}\box0}}
{\setbox0=\hbox{$\textstyle\rm H$}\hbox{\hbox to0pt
{\kern0.4\wd0\vrule height0.9\ht0\hss}\box0}}
{\setbox0=\hbox{$\scriptstyle\rm H$}\hbox{\hbox to0pt
{\kern0.4\wd0\vrule height0.9\ht0\hss}\box0}}
{\setbox0=\hbox{$\scriptscriptstyle\rm H$}\hbox{\hbox to0pt
{\kern0.4\wd0\vrule height0.9\ht0\hss}\box0}}}}
\def\Ol{{\mathchoice

{\setbox0=\hbox{$\displaystyle\rm O$}\hbox{\hbox to0pt
{\kern0.4\wd0\vrule height0.9\ht0\hss}\box0}}
{\setbox0=\hbox{$\textstyle\rm O$}\hbox{\hbox to0pt
{\kern0.4\wd0\vrule height0.9\ht0\hss}\box0}}
{\setbox0=\hbox{$\scriptstyle\rm O$}\hbox{\hbox to0pt
{\kern0.4\wd0\vrule height0.9\ht0\hss}\box0}}
{\setbox0=\hbox{$\scriptscriptstyle\rm O$}\hbox{\hbox to0pt
{\kern0.4\wd0\vrule height0.9\ht0\hss}\box0}}}}
\def\be{\begin{equation}}
\def\ee{\end{equation}}
\def\ba{\begin{eqnarray}}
\def\ea{\end{eqnarray}}
\def\mh{\mathcal{H}}
\def\mm{\mathcal{M}}
\def\mb{\mathcal{B}}
\def\mau{\mathcal{U}}
\def\ma{\mathcal{A}}
\def\mi{\mathcal{I}}
\def\mj{\mathcal{J}}
\def\md{\mathcal{D}}
\def\mc{\mathcal{C}}
\def\ml{\mathcal{L}}
\def\mo{\mathcal{O}}
\def\mq{\mathcal{Q}}
\def\mp{\mathcal{P}}
\def\me{\mathcal{E}}

\def\mx{\mathcal{\chi}}
\def\mg{\mathfrak{g}}

\def\g{{\cal G}}

\newcommand{\p}{\phi}
\newcommand{\te}{\theta}
\newcommand{\s}{\psi}
\newtheorem{theorem}{Theorem}[chapter]
\newtheorem{proof}{Proof}[chapter]
\newtheorem{axiom}{Axiom}[chapter]

\newtheorem{definition}{Definition}[chapter]
\newtheorem{example}{Example}[chapter]
\newtheorem{diagram}{Diagram}[chapter]
\newtheorem{corollary}{Corollary}[chapter]
\newtheorem{lemma}{Lemma}[chapter]
\newcommand\Sets{{\bf Sets}}
\newcommand\Sub{{\rm Sub}}
\newcommand\Om{\underline{\Omega}}
\newcommand\Sig{\underline{\Sigma}}
\newcommand\Si{\Sigma}
\newcommand\Ga{\Gamma}

\newcommand\op{{\rm op}}
\newcommand\cl{{\rm cl}}
\newcommand\Ob{{\rm Ob}}
\newcommand\Hi{{\cal H}}
\newcommand\Hione{\Hi_{t_1}}
\newcommand\Hitwo{\Hi_{t_2}}
\newcommand\V{\mathcal{V}}

\newcommand\Subone{\Sub(\Sig^{\Hi_1})}
\newcommand\Subtwo{\Sub(\Sig^{\Hi_2})}
\newcommand\intt{\Sets^{(\V(\Hi_1)\times\V(\Hi_2))^\op}}
\newcommand\inttonetwo{\Sets^{(\V(\Hi_{t_1})\times\V(\Hi_{t_2}))^\op}}
\newcommand\ps[1]{\underline{#1}}
\newcommand\bra[1]{\langle #1|\,}
\newcommand\ket[1]{\,|#1\rangle}

\newcommand\w{\mathfrak{w}}
\newcommand\la{\langle}
\newcommand\ra{\rangle}
\newcommand\ha{\hat\alpha}

\begin{document}
\begin{titlepage}
\begin{center}
{\Huge{\bf Approaches To Quantum Gravity}} \vspace{0.5cm}

    {\LARGE Dissertation\\\vspace{0.5cm}
    }
    {\normalsize{zur Erlangung des akademischen Grades}} \\\vspace{0.5cm}
    {\normalsize doctor rerum naturalium}\\ \vspace{.5cm}
    {\normalsize (dr. rer. nat.)}\\ \vspace{.5cm}
    {\normalsize im Fach Physik}\\ \vspace{1cm} 
    {\normalsize eingereicht an der \\\vspace{.5cm}
    Mathematisch-Naturwissenschaftlichen Fakult\"at I\\\vspace{.5cm}
    Humboldt-Universit\"at zu Berlin} \vspace{1.5cm}

    {\normalsize von\\\vspace{.5cm}
    Frau Master-Phys. Cecilia Flori\\\vspace{.5cm}
geboren am 18.11.1980 in Rom}\\ \vspace{1cm}  

\end{center}
\begin{flushleft} 
Pr\"asident der Humboldt-Universit\"at zu Berlin:\\\vspace{0.2cm}
Prof. Dr. Christoph Markschies\\\vspace{.5cm}
Dekan der Mathematisch-Naturwissenschaftlichen Fakult\"at I:\\\vspace{0.2cm}
Prof. Dr. Lutz-Helmut Sch\"on\\\vspace{1cm}
Gutachter:
\begin{enumerate}
\item[1.] Prof. Dr. Christopher J. Isham 
\item[2.] Prof. Dr. Jan Plefka
\item[3.] Prof. Dr. Thomas Thiemann 
\end{enumerate}
\mbox{ }\\[5pt]
Eingereicht am: 07-09-2009\\\vspace{0.5cm}
Tag der m\"undlichen Pr\"ufung: 
\end{flushleft}
\newpage

\begin{center}
\underline{\LARGE{\bf{Abstract}}}\\\vspace{1cm}
\end{center}

One of the main challenges in theoretical physics over the last five decades has been to reconcile quantum mechanics with general relativity into a theory of quantum gravity.  However, such a theory has been proved to be hard to attain due to i) conceptual difficulties present in both the component theories (General Relativity (GR) and Quantum Theory);
ii) lack of experimental evidence, since the regimes at which quantum gravity is expected to be applicable are far beyond the range of conceivable experiments. Despite these difficulties, various approaches for a theory of Quantum Gravity have been developed.

In this thesis we focus on two such approaches: Loop Quantum Gravity and the Topos theoretic approach. 
The choice fell on these approaches because, although they both reject the Copenhagen interpretation of quantum theory, their
underpinning philosophical approach to formulating a quantum theory of gravity are radically different.
In particular LQG is a rather conservative scheme, inheriting all
the formalism of both GR and Quantum Theory, as it tries to bring to its logical extreme consequences
the possibility of combining the two. On the other hand, the Topos approach involves the idea that a radical change of perspective is needed in
order to solve the problem of quantum gravity, especially in regard to the fundamental concepts of
`space' and `time'. Given the partial successes of both approaches, the hope is that it might be possible to find a
common ground in which each approach can enrich the other.

This thesis is divided in two parts: in the first part we analyse LQG, paying particular attention to the semiclassical properties of the volume operator. Such an operator plays a pivotal role in defining the dynamics of the theory, thus testing its semiclassical limit is of uttermost importance.\\
We then proceed to analyse spin foam models (SFM), which are an attempt at a covariant or
path integral formulation of canonical Loop Quantum Gravity (LQG). In particular, in this thesis we propose a new SFM, whose path integral is defined in terms of the Holst action rather than the Plebanski action (used in current SFM).
This departure from current SFM has enabled us to solve, explicitly, certain constraints which seem rather problematic in the current SFM.

In the second part of this thesis we introduce Topos theory and how it has been utilised to reformulate
quantum theory in a way that a consistent quantum logic can be defined.
Moreover, we also define a Topos formulation of history quantum theory. The striking difference of
this approach and the current consistent-history approach is that, in the former no fundamental role is
played by the notion of a consistent sets (set of histories which do not interfere with each other) while,
in the latter, such notions are central. This is an exciting departure since one of the main difficulty in
the consistent-history approach is how to choose the correct consistent set of history propositions, since
there are many sets, most of which incompatible. However, we have shown that in our Topos formulation of
history quantum theory truth values can be assigned to any history proposition, therefore the notion
of a consistent sets of propositions is unnecessary.
This implies that at the level of quantum gravity it could be possible to assign truth values to any
proposition about four-metrics (which can be considered as the GR analogue of a `history').

\newpage
\begin{center}
\underline{\LARGE{\bf{ Zusammenfassung}}}\\\vspace{1cm}
\end{center}

\noindent
In dieser Arbeit besch\"aftigen wir uns mit zwei Ansätzen zur Quantengravitation (QG), die einander kontr\"ar gegen\"uberstehen:
\begin{enumerate}
\item[-] Erstens mit der Loop Quantum Gravity (LQG), einem eher konservativen Ansatz zur QG, dessen Startpunkt eine Hamiltonsche Formulierung der klassischen Allgemeinen Relativit\"atstheorie (ART) ist, 

\item[-] zweitens mit der sogenannten Topos-Theorie, angewandt auf die Allgemeine Relativit\"atstheorie, die die mathematischen Konzepte der Quantentheorie (und m\"oglicherweise auch der ART) radikal umformuliert, was eine immense Redefinition von Konzepten wie Raum, Zeit und Raumzeit zur Folge h\"atte.
\end{enumerate}
Der Grund für die Wahl zweier so verschiedener Ansätze als Gegenstand dieser Arbeit liegt in der Hoffnung begr\"undet, dass sich diese beiden Ans\"atze auf einen gemeinsamen Ursprung zur\"uckf\"uhren lassen k\"onnen und somit gegenseitig erg\"anzen k\"onnen. 

Im ersten Teil dieser Arbeit f\"uhren wir den allgemeinen Formalismus der LQG ein und gehen dabei insbesondere auf den semiklassischen Sektor der Theorie ein; insbesondere untersuchen wir die semiklassischen Eigenschaften des Volumenoperators. Dieser Operator spielt in der Quantendynamik der LQG eine tragende Rolle, da alle bekannten dynamischen Operatoren auf den Volumenoperator zur\"uckgef\"uhrt werden k\"onnen. Aus diesem Grund ist es außerordentlich wichtig zu \"uberpr\"ufen, dass der klassische Limes des Volumenoperators wirklich mit dem  klassischen Volumen \"ubereinstimmt.

Anschließend besch\"aftigen wir uns mit sogenannten Spin Foam Modellen (SFM), welche als ein kovarianter oder Pfadintegralzugang zur kanonischen LQG angesehen werden k\"onnen. Diese Spin Foam Modelle beruhen auf einer Langrange-Formulierung der LQG mittels einer kovarianten sum-over-histories Beschreibung. Die Entwicklung eines Lagrange-Zuganges zur LQG wurde motiviert durch die Tatsache, dass es in der kanonischen Formulierung der LQG \"uberaus schwierig ist, \"Ubergangsamplituden auszurechnen.
Allerdings weichen die Spin Foam Modelle, die wir in dieser Arbeit behandeln in einem entscheidenden Punkt von den bisher in der Literatur diskutierten ab, da wir die Holst-Wirkung \cite{47} und nicht die Palatini-Wirkung als Ausgangspunkt nehmen. Dies erm\"oglicht es uns, explizit gewisse Zwangsbedingungen zu l\"osen, was in den gegenw\"artig diskutierten SFM problematisch scheint.

Im zweiten Teil dieser Arbeit f\"uhren wir in die Topos-Theorie ein und rekapitulieren, wie diese Theorie benutzt werden kann, um die Quantentheorie derart umzuformulieren, dass eine konsistente Quanten-Logik definiert werden kann.
Dar\"uber hinaus definieren wir auch eine Topos-Beschreibung der Quantentheorie in der sum-over-histories Formulierung. Unser Ansatz entscheidet sich vom gegenw\"artigen consistent-histories Ansatz vor allem dadurch, dass das Konzept der konsistenten Menge (eine Menge von Historien, die nicht mit sich selbst interferieren) keine zentrale Rolle spielt, w\"ahrend es in letzterem grundlegend ist. Diese Tatsache bietet einen interessanten Ausgangspunkt, da eine der Hauptschwierigkeiten im consistent-histories Ansatz darin besteht, die richtige konsistente Menge der Propositionen von Historien zu finden: Im allgemeinen gibt es viele solcher Mengen, und die meisten davon sind nicht miteinander kompatibel.
Wir zeigen, dass in unserer Topos-Beschreibung der sum-over-histories Quantentheorie jeder Proposition von Historien Wahrheitswerte zugeteilt werden können; daher ist das Konzept einer konsistenten Menge von Propositionen redundant.
Dies bedeutet, dass es im Rahmen einer Quantengravitationstheorie m\"oglich sein k\"onnte, jeder Proposition von vierdimensionalen Metriken (welche als allgemein relativistisches Analogon einer Historie angesehen werden k\"onnen) einen Wahrheitswert zuzuweisen.
\newpage

\tableofcontents
\listoffigures
\end{titlepage}
\chapter{Introduction}
One of the main challenges in theoretical physics in the past
fifty years is to define a theory of quantum gravity, i.e., a
theory which consistently combines general relativity and quantum
theory. However, not withstanding the great effort that has been
put into discovering such a theory, physicists cannot even all
agree on what such a theory should look like. The most that has
been agreed is that quantum theory and general relativity should
appear as limits of the theory in the appropriate regimes.

The reasons for this elusiveness in a quantum theory of gravity
are manifold. The difficulties which arise are of two types:
`factual' and conceptual. The factual reasons are the following:
\begin{enumerate}
\item[i)] The regimes at which quantum gravity is expected to be applicable
(Planck length $10^{-35}$m and Planck energy $10^{28}$ev) are far
beyond the range of conceivable experiments. This lack of
empirical results makes it difficult to test any proposal for a
quantum theory of gravity.

\item[ii)] Given the range of potential applications of a possible quantum theory of gravity
(just after the big-bang) there is not even any agreement on what
sort of data and predictions such a theory might have.
\end{enumerate}

On the conceptual side, the problems facing quantum gravity are of
two sorts:
\begin{enumerate}
\item[i)] Conceptual obstacles that arise from the individual component theories,
i.e., general relativity and quantum theory.

\item[ii)] Conceptual obstacles that arise from trying to combine such
theories.
\end{enumerate}

The presence of such obstacles might make one wonder what actually
guides researchers in developing possible theories of quantum
gravity: i.e., how can one define a conceptual framework in a
mathematical consistent language which could represent an unknown
quantum theory of gravity for which we have no tangible
experimental evidence (beyond the limiting situations in which GR or quantum theory, respectively, apply alone). Arguably, the main guiding principle is a
philosophical prejudice of what the theory should look like,
mainly based on the success of mathematical constructs for
theories that are believed to be closely connected.

However, we will not develop such a line of thought here. Instead,
we will simply analyse two of the current proposals for a theory
of quantum gravity. In this respect, it is interesting to note
that the different approaches to quantum gravity are based on
whether quantum theory and/or the current ideas of space and time,
i.e. GR (general relativity), are to be taken as fundamental or
not.

In those approaches in which both GR and quantum theory are
considered fundamental, the strategy to define a quantum theory of
gravity is to find an algorithm with which to quantise the metric
tensor, which is now regarded as a normal field.

If instead, only GR (respectively quantum theory) is regarded as
fundamental, a quantum theory of gravity is defined by adapting
quantum theory (respectively GR) to accommodate GR (respectively
quantum theory).

Alternatively, one can adopt the view that both GR and quantum
theory emerge from a deeper theory, which presupposes a drastic
change of our notions of space and time.

We will now briefly describe the two kinds of conceptual problem
mentioned above, although a detailed analysis of all the
conceptual problems of GR and quantum theory will not be given,
since this would take us beyond the scope of this thesis. What we
will do, instead, is to state those conceptual problems which are
most related to quantum gravity.

\paragraph*{Quantum Theory}

The central difficulty in quantum theory is that, to date, an agreed upon
interpretation of the theory does not exist. In fact there are several interpretations,
each of which rests on how fundamental the mathematical formalism is considered to be.
Each such interpretation will lead to different conceptual problems when applied to quantum gravity. \\
We will now analyse a few of them:
\begin{enumerate}
\item [1.] \emph{Copenhagen Interpretation}. \\The main postulates of this
interpretation are:
\begin{enumerate}
\item relative frequency interpretation of probability;

\item clear distinction between a classical realm and a quantum realm.
\end{enumerate}
Clearly, in this context, space and time are classical concepts
and thus belong to the classical realm. It follows that a quantum
theory of gravity which adopts the Copenhagen interpretation will
have to overcome the conceptual contradiction of applying quantum
concepts to `quantities' (space and time) which are essentially
classical.

Moreover, the Copenhagen interpretation of quantum theory leads to
the problem in quantum cosmology of how to define an observer with
respect to which measurements and, thus, probabilities are
defined. This problem is due to the fact that in a cosmological
context we are dealing with a closed system.

\item [2.] \emph{Many Worlds View}\\
The main feature of this interpretation of quantum mechanics is
the rejection of any concepts that cannot be described in purely
quantum-theoretical terms. Thus, the notion of external observer,
classical realm and the like, are rejected. In this context, the
process of state-vector reduction can be interpreted either in
terms of branching, where each branch represents a physical
reality\footnote{To be precise, the main postulates of the many
worlds view are the following: i) the state vector of a closed
system is a  superposition of eigenstates of a preferred quantity
(how to choose such a quantity is one of the problems in this
interpretation); ii) each of the components represents a real
definite value of this preferred quantity; iii) although there is
no collapse of the state vector we can only see one component of
the vector (explaining such a process in a mathematically rigorous
way is another of the problems facing this interpretation).}, or
in terms of decoherence. In this way there is no `real' collapse
of the state vector, although we end up seeing only one of the
various possibilities.

A quantum theory of gravity that adopts the many worlds
interpretation of quantum theory will have the following features:
\begin{enumerate}
\item it should accommodate the `branching' within the topological changes of space;

\item the notion of a quantum state for a closed system can be defined in the
many worlds view, since there is no postulated splitting between
the classical realm and the quantum realm;

\item the notion of decoherence and, subsequently, of consistent histories might help
to overcome the `problem of time'. In fact, as we will explain in
detail later, in the consistent histories approach time admits
different interpretations from those of standard classical physics
\cite{hartel1995, temporall}.
\end{enumerate}

\item[3.]\emph{Hidden Variables}\\
The main idea of the hidden-variables interpretation is that in
order to overcome the measurement problem the existence of some
extra variables is postulated. The rules of evolution of these
quantities is specified by the theory. Thus, the main features of
a hidden variable theory is that it is deterministic and realist
(in the sense that quantities exist independently of the
observer).

The most studied of these approaches is the deBroglie-Bohm
pilot-wave approach in which each point particle has a
well-defined position which is not known by the observer. The
evolution of such particles is guided by the Schr\"odinger 
equation. One conceptual problem with this approach is that the
guiding equation requires an absolute notion of time with respect
to which the positions of the particles evolve. This is in obvious
contradiction with Lorentz-invariance and relativity of
simultaneity central in GR.
\end{enumerate}

This ends the list of the major interpretations of quantum theory
and their conceptual problems. We will now discuss the conceptual
problems of general relativity that have particular importance in
the context of quantum gravity.

\paragraph*{General Relativity}
The conceptual difficulties in GR which are relevant for quantum
gravity are mainly two:
\begin{itemize}
\item[a)] \emph{Notion of spacetime points}\\
Are spacetime points real physical quantities or are they purely
mathematical constructs induced by the utilisation of
set-theoretic ideas in GR\footnote{It should be noted that both GR
and quantum theory agree on the definition of a spacetime as a
differentiable manifold.}?

Given Einstein's hole argument \cite{philosophy} and the
diffeomorphism invariance of GR, it would seem that spacetime
points should not be considered as physical entities but, solely,
as mathematical objects (points) in a model of spacetime based on
set theory. However, it could also be the case that spacetime
points are derived concepts of some more complex structure that is
taken to be the fundamental definition of spacetime.

The question which arises in quantum gravity is how much of these
spacetime concepts of GR does it inherit? As we will see, the
various  approaches to quantum gravity differ depending on how
fundamental the conception of spacetime implied by GR is
considered to be.

\item[b)] \emph{Role of diffeomorphism invariance}\\
Diffeomorphism invariance plays an important role in both 
classical GR and quantum gravity. We recall that a diffeomorphism
$\phi$ is a bijective $C^{\infty}$-map between manifolds whose
inverse is $C^{\infty}$. Thus the diffeomorphism group $\md$ is
given by the collections of invertible maps
$\phi:\mm\rightarrow\mm$ that preserve the differential structure
of $\mm$.

In GR there are two types of diffeomorphism invariance:
\emph{passive diffeomorphism invariance}, which represents an
invariance under change of coordinates and \emph{active
diffeomorphism invariance} which relates different objects in
$\mm$ under shifts of points of the manifold to other points.

The problems induced in quantum gravity by the diffeomorphism
invariance of GR is how to implement such an invariance. Various
approaches to quantum gravity differ according to how this is
done. For example, for canonical approaches to quantum gravity the
(spatial) diffeomorphism group is exactly the one given by
classical GR; for perturbative string theory the (spacetime)
diffeomorphism group is a subgroup of $\md$; while for
super-string theory, since GR appears only at the low-energy
limit, (target space) diffeomorphism transformations do not play such a prominent
role.
\end{itemize}

At the start of this Section we mentioned that the conceptual
problems affecting quantum gravity are of two types: those of the
individual ingredient theories---GR and quantum theory---and those
coming from attempts to combine them.

The former have been described above, while an important example
of  the latter is the so-called `problem of time'. This problem is
a consequence of the radically different conception of time
present in GR and in quantum theory. In particular, in normal
quantum theory 'time' is a labelling parameter related to the
fixed background structure.

The existence of a fixed causal structure is very important in
quantum theory: for example, the commutation relations of quantum
fields are heavily dependent on such causal structure. On the
other hand, in GR time is dynamical: indeed, the spacetime
manifold can be foliated into spacelike hypersurfaces in many
different ways, none of which is preferred. It follows that the
causal structure of spacetime is itself a dynamical quantity,
which, being influenced by matter, varies from one model to
another.

From this brief analysis it is easy to understand the difficulty
of trying to combine GR with quantum theory: namely, how to
formulate quantum theory with a fluctuating causal structure? As
we will see, each approach to quantum gravity tackles this issue
in different ways.

Now that we have briefly analysed the conceptual problems that a
possible theory of quantum gravity has to face, we will introduce
the two proposals for a theory of quantum gravity that are
analysed in this thesis. As stated earlier, the radical difference
between these proposals theories lies in the precise role assigned
to GR and quantum theory. Consequently, the conceptual problems
faced by each will differ accordingly. These candidates are:

\paragraph*{Loop Quantum Gravity}
Loop quantum gravity (LQG) is a canonical approach and, as such,
its starting point is classical GR which is to be quantised
through some quantisation algorithm. In this approach, both
quantum theory and GR are regarded as being fundamental and most
of their mathematical formalism and conceptual framework are
inherited by the ensuing quantum theory of gravity.

The technical details of LQG will be given in the next chapter. In
the present Section, we will focus on those conceptual aspects of
both GR and quantum theory that are of particular relevance to
this programme. These are, respectively:
\begin{enumerate}
\item[1.]
In LQG the Copenhagen interpretation of quantum mechanics seems
not to be applicable, since spacetime can be foliated in any way
and no preferred splitting of space and time is required (or even
possible).

More, a Copenhagen interpretation of quantum gravity would require
a fixed background metric: something which is not present in LQG,
unless some preferred foliation is chosen in some external way.\\
As an alternative interpretation of quantum mechanics LQG adopts the consistent histories interpretation.
Essentially in a consistent history interpretation the density matrix is not unitary, but instead follows a certain history (path) in the set of all possible histories
 which do not interfere among them. In this setting the probability of a given history to occur can be calculated (see Chapter \ref{cha:hist} for a detailed description). However, it is still debatable whether a) all histories are realised at once, but they don't communicate with eachother (Everretian interpretation); b) only  one history (the one we experience) is realised but the future is undetermined.\\
It should be noted that by adopting the consistent history interpretation of quantum theory the following problems are solved:
\\
i) No need of an external observer to give meaning to probabilities (closed system problem).
ii) No state vector collapse.
ii) No arrow of time problem: direction of time comes from the fact that there is an initial density matrix but no final one.\\
For a discussion of the above ideas the reader is referred to \cite{1a}, \cite{1b}

\item [2.]
LQG  adopts, more or less, the spacetime conception of classical
GR. In fact, the spacetime manifold $\mm$ is considered to be
diffeomorphic to $\Sigma\times \Rl$, where $\Sigma$ is the
3-dimensional spatial manifold.
\end{enumerate}
The problem of time seems particularly relevant in LQG, since it
is a background independent formulation of quantum gravity.
Attempts to solve this problem have been made by trying to
introduce time as being defined by a physical clock. What these
clocks might be is not unanimously agreed upon (see \cite{1b} and
references therein). \\[5pt]

\paragraph*{Topos Approach}
Here we are being very optimistic since, to date, there is no
topos\footnote{Roughly speaking a topos is a category which is similar to $Sets$: 
 fundamental mathematical properties (disjoint union, Cartesian product,
 etc) have a topos analogue.} formulation of quantum gravity as such. However, there is
well-developed idea on how topos theory can be used in general to
describe theories of physics including, potentially, a theory of
quantum gravity.

The key idea is that constructing a theory of physics involves
finding a representation, in a topos, of a certain formal
language\footnote{A formal language is a deductive system of
reasoning made of atomic variables, relations between such
variables, and rules of inference. In this context it is assumed
that each system has a formal language attached to it and which
provides a deductive system based on intuitionistic logic. }, that
is attached to the system under investigation (see \cite{andreas1}
for a detailed analysis). Thus the topos approach consists in
first understanding at a fundamental level what a theory of
physics and associated conceptual framework should look like and,
then, applying these insights  to quantum gravity. In this
context, a radically new way of thinking about space, time is
suggested: for example, the possibility that both GR and quantum
theory are `emergent' theories.

Since a topos formulation of quantum gravity has yet to be
developed, it is difficult to guess precisely which conceptual
difficulties and novelties could arise in such a theory. However,
a reformulation of quantum theory and its history formulation has
recently been carried out in \cite{andreas5, andreas, andreas2,
andreas3, andreas4, 46ah} and from these works it is clear how the
Copenhagen interpretation of quantum mechanics can be replaced
with a more realist interpretation.

The details of how such a more realist interpretation is achieved
are given in subsequent Sections. Here it suffices to say that the
scheme involves a synthesis of the many-worlds view and that of
extra variables. In particular, of the latter it retains the fact
that quantities have more values than those defined through the
eigenvalue-eigenstate link, while of the former it retains the
fact that these extra values are defined in terms of standard
quantum theory. This alternative interpretation of quantum theory
has been coined \emph{neo-realist}.\\

In the following we will analyse in detail the two programmes
mentioned above for developing a quantum theory of gravity: namely
LQG and the topos approach. As can be easily understood from what
has been said so far, these two approaches to define a quantum
theory of gravity are very different.


Obviously, there are many more approaches to quantum gravity and a
detailed analysis and comparison of each would be a very demanding
job, albeit a very useful one. However, in this thesis, as we have
said already, only two of these approaches will be analysed. The
choice fell on LQG and the topos approach because, although they
both reject the Copenhagen interpretation of quantum theory, their
underpinning philosophical approach to formulating a quantum
theory of gravity are radically different. In particular, the
topos approach involves the idea that a radical change of
perspective is needed in order to solve the problem of quantum
gravity, especially in regard to the fundamental concepts of
`space' and `time'. On the other hand, LQG is a rather
conservative scheme, inheriting as it does all the formalism of
both GR and quantum theory as it tries to bring to its logical
extreme consequences the possibility of combining the two.

Given the partial successes of both approaches, the hope is that
it might be possible to find a common ground in which, each
approach can enrich the other.

This thesis is divided into two parts: the first is concerned with
LQG and the second with the topos approach. The main topics
developed in part I and part II are, respectively, the following:

\vspace{0.6cm}\noindent \textbf{PART I}
\paragraph*{Mathematical formulation and derivation of LQG}
A promising proposal for a theory of quantum gravity is Loop
Quantum Gravity (LQG)---a non-perturbative, background-independent
quantum field theory \cite{1b},\cite{1a}.

The starting point of LQG is classical general relativity (GR),
reformulated as an Hamiltonian theory with constraints on the
phase space variables. In particular, we have the $SU(2)$ gauge
transformations, spatial diffeomorphisms, and the Hamiltonian
constraints.

This structure  can be canonically quantised, so that the constraint equations are promoted to
quantum constraint operators defined on a kinematical Hilbert
space, $\mathcal{H}_{kin}$. The strategy adopted for quantising a
system with constraints is that of Dirac. This consists in
quantising the unconstrained system, thus obtaining
$\mathcal{H}_{kin}$. The constraints are then implemented as
operators on $\mathcal{H}_{kin}$, such that the physical states
are annihilated by such operators. The physical Hilbert space
$\mathcal{H}_{phy}$ is then the space of solutions to all the
constraints. The dynamics of the theory is governed by the
Hamiltonian constraint $H$.

Two central problems in this approach are (i) \emph{constraint program}: to extract concrete
solutions for the Hamiltonian constraint; and (ii) to define an
inner product on $\mathcal{H}_{phy}$.

It has been shown that both $\mathcal{H}_{kin}$  and the
geometrical operators, such as the volume, area and length
operator can be rigorously defined. Moreover, the spectrum of
these operators is discrete \cite{v5}. However, it has still not been
possible to carry these results over to the physical Hilbert
space.

\paragraph*{Analysis of the semiclassical properties of the Volume operator}
An important part of the research programmes of LQG is to
understand the semiclassical properties of this theory. This is
vital in order to relate it to classical general relativity.

In the papers \cite{30b} \cite{46ah} an analysis of the
semiclassical properties of the volume operator  was performed
using coherent states on graphs. In particular, in \cite{30b} the
analysis was done with respect to dual-cell coherent states, while
in \cite{46ah} area-complexifier coherent states were used.\\
In both cases the inputs needed to construct such states were: \\
i) the choice of a complexiﬁer;\\
ii) the choice of a graph.\\
The definition of the complexifier for dual cell coherent states was given in terms of the flux operator and depended on a collection of surfaces defined by a polyhedronal partition of the spatial manifold. On the other hand the definition of the complexifier for area-complexifier coherent states was given in terms of area operators and depended on a collections of surfaces obtained through a parquette of foliations of the spatial manifold.\\
Regarding the choice of graph, for practical reasons, it is common to choose graphs that are topologically regular that is, have constant valence for each vertex.

These studies have shown that, as far as dual-cell coherent states are
concerned, the correct semiclassical properties of the volume
operator are obtained only if the graphs, representing the quantum
states of space, are 6-valent. On the other hand, if
area-complexifier coherent states are considered, the correct
semiclassical limit is attained only with 1) an artificial
rescaling of the complexified connection (see Section \ref{s2.5}); and 2) particular
embeddings of the 4-valent and 6-valent graphs within the set of
surfaces on which the complexifier depends.

However, the combinations of Euler angles for which such
embeddings are attained have measure zero in $SO(3)$ and are,
therefore, negligible. Thus the area-complexifier coherent states
are not the correct tools by which analyse the semiclassical
properties of the volume operator.

This result has interesting consequences in the field of spin foam
models, since the current spin foam models are all based on
boundary spin networks of valence {\sl four}. Motivated by this,
we have developed in  \cite{30a} an alternative spin foam model
constructed on a discretisation of the manifold in terms of
hypercubes rather than 4-simplices. Dual two-skeletons of such a
`cubulated' manifold would lead to 6-valent graphs. This
alternative spin foam model is called the `cubulated spin foam
model'.

\paragraph*{Spin-foam models}
Spin-foam theory is supposed to provide the dynamical aspects of
LQG and can be used as a tool for computing the quantum-gravity
`transition amplitudes'. More precisely, spin foam models are an
attempt to provide a path-integral formulation of LQG.

At each time step, in LQG, a quantum state of geometry is
represented by a graph labelled by spin quantum numbers which
carry information about the geometry of the space. Such a graph is
called a {\sl spin network}. A spin foam can be interpreted as a
history of such spin networks.

In 2+1 dimensions it has been shown \cite{perez1} that this
interpretation is indeed possible since, in this case, the
boundary states exactly match the states of LQG.

The 4-dimensional theory is much harder and few rigorous results
are known. The most successful spin foam model in four dimensions
is the Euclidean quantum-gravity model of  Barrett and Crane (the
`BC-model'). Although this model has some very interesting
properties, it is not physically correct in the sense that (i) it
does not always reproduce the correct low-energy limit; (ii) the
boundary states do not match; and (iii) the volume operator is ill
defined.

In \cite{415},\cite{26}, \cite{426a}, \cite{413} it was shown that
the problems of the BC-model can be traced back to the way in
which certain constraints are imposed.

The partition function for spin networks in the BC-model is
constructed using the well-known partition function of
$BF$-theory\footnote{$BF$-theory is a topological quantum field
theory \cite{48f}  whose action in D+1 dimensions is given by
$S_{BF}=\int_MTr(B\wedge F) $ where $F$ is the curvature of a
$SO(4)$ connection, and $B$ is a Lie-algebra valued two-form. A
detailed analysis is given later in this thesis.} but with the
addition of some extra constraints (the so-called `simplicity'
constraints). This procedure is adopted because it results in the
$BF$-action reducing to the Palatini action for GR. In the
$BC$-model these extra constraints are imposed as strong-operator
constraints of the form $\hat{C}_{n}\psi=0$, i.e., as if they were
first-class constraints.

However, it was argued in \cite{26}, \cite{{426a}}, \cite{413}
that, since the constraints in question are, in fact, second
class, they should be applied weakly in the form
$\langle\phi\hat{C}_n,\psi\rangle=0$, in order not to loose any
physical degrees of freedom. This strategy is very fruitful and
solves some of the problems in the BC-model.

Among the residual problems there is the fact that the solutions
of the simplicity constraints are not unique, and fall into two
sectors: the `topological sector' and the `gravitational sector'.

We are interested only in the gravitational sector. It was shown
in \cite{415}, that the model developed in \cite{26}, \cite{413}
is related to the topological sector, rather than to the
gravitational one.  A model for the gravitational sector was
developed in \cite{415}.

In all the above-mentioned models, the spin foams whose boundaries
are spin-network functions are constructed by discretising the
spacetime manifold  in terms of 4-simplices (triangulation). In
particular, spin foam and spin-network functions are defined in
terms of the dual 2-skeleton of such a triangulation. As a
consequence, the only allowed valence number of such a function is
four.

This poses some problems since it was shown in
\cite{30b},\cite{46ah} that the correct semiclassical properties
of the volume operator in LQG requires graphs whose valence is six.
This issue must be addressed before a spin foam model can be
interpreted as a path-integral formulation of LQG, because the
volume operator plays a prominent role in the implementation of
the Hamiltonian constraint.

Another problem that arises in all BC-type spin foam models is the
issue of ultra-locality. In fact, in these models the simplicity
constraints are applied to a single  4-simplex, thereby ignoring
any interaction between the various simplices. In \cite{415} this
issue was addressed and a solution proposed.

Interestingly, it was discovered in \cite{d133} that the BC-model
admits an interpretation  as a Feynman graph of a group field
theory. Moreover, it was shown in \cite{rovelli2} that {\sl any}
local spin foam model, whose transition amplitude is given in
terms of two complexes, can be interpreted as a Feynman graph of a
group field theory (GFT). This suggests that GFT may be a
structure that underlines  {\sl any} attempt to define a theory of
quantum gravity in a background-independent way
\cite{oriti1},\cite{oriti2}, \cite{tesid}, \cite{fridel2}.

\paragraph*{Cubulated Spin-Foam Model}
The novelties of this new spin foam model \cite{30a} are the
following:
\begin{enumerate}
\item The starting point is the Palatini action of GR rather than the BF-action.
As a result, it is no longer necessary to impose the simplicity constraints. This
avoids, from the outset, the problem of interpreting the extra solutions to
such constraints.

\item  In the canonical spin foam models, the variables $F$ and $B$ in the
action are discretised on the dual faces $f(t)$ of the triangulation, and
on the faces $t$ of the triangulation, respectively. In particular the action gets discretised as follows
\be
\int Tr(B\wedge F) =\sum_t Tr(B(t)A(\partial f(t)))
\ee
However, upon such a discretization it is not possible to define a disjoint action of the gauge group on both $B(t)$ and $A(\partial f(t)$ since there is no point of intersection between the loop $\partial f(t)$ and the triangle $t$

On the other hand, in the cubulated spin foam model, all the
variables are discretised in terms of geometric elements of the
original cubulation of the manifold. Thus they all transform in
terms of  elements of the same group. This  leads to the important
result that the cubulated spin foam model is manifestly gauge
invariant.

\item The absence of simplicity constraints solves the problem of ultra-locality
and leads to a transition amplitude that takes into account the interaction terms
coming from the boundary terms of every hypercube.

\item In the cubulated spin foam model, instead of performing the sum over
hypercubes we intend to take the continuum limit (as the
hypercubes get smaller and smaller ) as it is done in dynamical
triangulation \cite{loll1} and \cite{loll2}. This is done for
calculation simplicity.
\end{enumerate}
Since the path integral, as defined in the cubiculated spin foam
model is developed starting from the Palatini action, a different
type of measure, other than the one used in BF-theory is needed.
This is derived in \cite{cecilia3} where the Hamiltonian analysis
\cite{holst2}, \cite{48c} of the Holst action \cite{47} is carried
out, and the new measure is defined utilising the strategy
developed in \cite{47}.

\newpage
\vspace{0.6cm}\noindent \textbf{PART II}

\paragraph*{Introduction to Topos Formulation of Quantum Theory}
The topos reformulation of quantum theory aims at finding a more
realistic interpretation so as to avoid certain conceptual
problems that are inherent in the normal interpretation of theory (see
chapter \ref{che:topostheory} for a detailed analysis).

The strategy adopted to attain a more realist interpretation is to
make quantum theory `look like' classical theory. The reasons why
topos theory was chosen as the mathematical framework to achieve
this goal are:
\begin{enumerate}
\item[--] Classical physics uses $Sets$ as its mathematical structure. A topos is a category which `looks like' $Sets$, in particular any mathematical construct present in set theory has a topos theoretical analogue. This implies that the underpinning mathematical structures which renders classical theory a realist theory can be mimicked in terms of topos theory.

\item[--] In Classical physics, Boolean logic, which is a distributive logic, arises as the internal logic of subsets in $Sets$. In topos theory, it's internal logic arises in a similar manner, namely as the logic of subobjects of a given object. Similarly as in classical theory, such a logic is a distributive logic.
\end{enumerate}

Moreover, the Kochen-Specker theorem of quantum mechanics
suggested the need of introducing the notions of a \emph{context}
which would represent a classical snapshot. Specifically, such
contexts were identified with abelian subalgebras of the algebra
of bounded operators $\mb(\mh)$, since only within such
subalgebras can quantum theory `look like' classical theory.

All this motivated the choice of the topos of presheaves (see appendix for a detailed definition) over the
category of abelian subalgebras, as the correct topos to utilise
in the reformulation of quantum theory.

From a mathematical perspective, in order to make quantum theory
`look like` classical physics, the first step is to identify which
underpinning mathematical constructs render classical theory
realist and, then, define a topos analogue of such constructs in
the context of quantum theory.

This is precisely what was done in \cite{andreas1},
\cite{andreas2}, \cite{andreas3}, \cite{andreas4},
\cite{andreas5}, \cite{isham1}, \cite{isham2}, \cite{isham3},
\cite{isham4}, \cite{isham5}. What the mathematical structures are
and how the topos analogue is defined will be described in Section
\ref{s.topos}.

In the topos reformulation of quantum theory it is possible to
assign truth values to any single-time proposition. However, as
will be explained in detail in Section \ref{ssingletruth}, the set
of truth values is larger than the classical boolean set $\{\text{true, false}\}$.

\paragraph*{Histories Approach to Quantum Theory}
History theory originated in part as an attempt to describe closed
systems in quantum mechanics in the light of a possible theory of
quantum cosmology. Indeed, the familiar Copenhagen interpretation
of quantum theory is inadequate for considering closed systems,
since it employs probabilities defined in terms of a sequence of
repeated measurements by an external `observer'. This is one
aspect of the posited fundamental division between system and
observer which, of course, is inappropriate for a theory of
cosmology.

The most studied history theory is the so-called
`consistent-history' approach. In this approach, the
system-observer division is avoided via a formalism that makes it
possible to assign probabilities without making use of any
measurement-induced, state-vector reduction.
The key ingredient that allows such an assignment of probabilities
is the `decoherence functional', $d$, which is a map from the
space of (pairs of) all histories to the complex numbers.

Roughly speaking, the decoherence functional, $d(\alpha,\beta)$
measures  the interference of two histories ($\alpha,\beta$).
Furthermore,  when applied to a single history $\alpha$, the real
number $d(\alpha,\alpha)$ can be interpreted as the probability of
that history  being realised. A set of histories which do not
interfere with each other is called a {\sl consistent set}.

In \cite{gm2}, \cite{gm3} a path-integral approach to consistent
histories was developed. In this approach each history is seen as
a subset of paths in configuration space, and the decoherence
functional between any two histories is represented as an
appropriate path integral.

However, although this interpretation facilitates the computation
of the decoherence functional for inhomogeneous
histories,\footnote{In consistent-history theory a distinction is
made between {\sl homogeneous} and {\sl inhomogeneous} histories.
A {\sl homogeneous history} is any time-ordered sequence of
projection operators, while an {\sl inhomogeneous history} arises
when two disjoint homogeneous histories are joined using the
logical connective "or" ($\vee$).} it lacks a well-motivated
mathematical definition of that concept.

A solution to this problem was proposed in \cite{consistent3} as
part of a new approach to history theory known as the `History
Projection Operator' (HPO) scheme. The main idea is to represent
homogeneous history propositions with tensor products of the
projection operators that represent the single-time propositions.
Such tensor products are themselves projection operators and can
be used in the obvious way to define inhomogeneous histories. In
this way one obtains a temporal quantum logic.

However, in any approach to  consistent-histories theory, HPO or
otherwise, there remains the problem of how to deal with the
plethora of different, incompatible consistent sets.\footnote{Two
consistent sets are said to be {\sl incompatible} if they cannot
be joined together to form a bigger set}.

One possibility is to single out one specific set using some basic
physical principle.  An attempt in this direction was discussed in
\cite{gm}, which used a measure of the quasi-classicality of
consistent sets that is sharply peaked.

A more radical approach is to accept the plethora of
$d$-consistent sets and interpret it as some sort of  `many
worlds' view, as it was done in \cite{consistent2}. The
originality of this approach lies in the fact that, by using a
novel mathematical structure---namely topos
theory\footnote{Roughly speaking, a topos is a category with some
special extra structure that makes it behave, in certain critical
ways, like the category of sets. In particular, there is an
internal logic---a Heyting algebra---that is the analogue of the
Boolean algebra in set theory. Rather strikingly, each topos
provides an alternative to the category of sets in the foundations
of mathematics}---it is possible  to obtain a new logic by which
to interpret the probabilistic predictions of the theory, when all
$d$-consistent sets are taken into account simultaneously.
However, in this approach the notion of probability and,
therefore, the decoherence functional, is still central.

\paragraph*{Topos Formulation of Histories Theory}
Recently, a more general and fruitful way of implementing topos
theory in physics was put forward in \cite{andreas1},
\cite{andreas2}, \cite{andreas3}, \cite{andreas4},
\cite{andreas5}. There it is argued that, in order to define a
quantum theory of gravity, certain conceptual obstacles, present
in quantum theory itself, must first be overcome. The suggestion
is to do this by redefining the mathematical structure of quantum
theory using topos theory and, in such a way that, in the
appropriate topos, quantum theory is made to `look like' classical
physics.

Such a reformulation of quantum theory leads to the possibility of
constructing more general, {\sl neo-realist}\footnote{A
`neo-realist' theory is one in which truth values of propositions
have a meaning outside of the concepts of measurement, external
observer etc.} theories in which the ideas of continuum (in the
sense of real numbers) and probability  play no fundamental role.

The decentralisation of the concept of probability resembles the
motivation for the development of consistent-histories theory. In
that respect it would  be extremely interesting to see if it were
possible to define a new version of consistent-histories that
utilises this novel, topos-based, formulation of quantum theory.

This is indeed possible, as we have shown in \cite{cecilia}. In
particular we have investigated the possibility of constructing a
topos version of history theory using some of the ideas employed
in the topos formulation of normal quantum theory given in
\cite{andreas1},\cite{andreas2}, \cite{andreas3}, \cite{andreas4},
\cite{andreas5}. The ensuing theory is a new history version of
quantum theory.

In \cite{andreas1}, \cite{andreas2}, \cite{andreas3},
\cite{andreas4}, \cite{andreas5}  truth values are assigned to
single-time propositions which are represented by particular
objects in the topos. In \cite{cecilia} we have extended these
ideas to sequentially-connected propositions, i.e., time-ordered
sequences of propositions.  A key ingredient is a development of a
temporal logic of Heyting algebras which is a temporal structure
that exploits the existence of a well-defined concept of a tensor
product of two Heyting algebras.

The existence of these tensor products suggests a natural
candidate for a topos analogue of the HPO formalism of history
quantum theory. It is striking that, in this new theory no
fundamental role is played by the notions of decoherence
functional or consistent sets. This is an exciting departure from
the standard consistent-history formulation of quantum theory,
where the notion of the decoherence functional is central.

The main attraction of the topos formulation of history theory
comes from considerations of quantum gravity. In fact, to date,
the consistent-history approach is the only approach that allows
quantum statements about four-metrics (which can be considered as
the GR analogue of a `history').

The reason is that any other quantum gravity approach is mainly
concerned with the quantum effects in the three-geometry of space,
not the four-geometry of spacetime. However, a difficulty in the
consistent-history approach is how to choose the correct
consistent set of history propositions, since there are many sets,
most of which incompatible.

However, in \cite{cecilia} it was shown that Heyting-algebra
valued truth values can be assigned to {\it any} history proposition,
therefore the notion of a consistent sets of propositions is
unnecessary. This implies that at the level of quantum gravity, it
could be possible to assign truth values to {\it any} proposition about
spacetime, not just space.

\chapter*{}
\vspace*{\fill}
\begingroup
\centering
\begin{center}
{\Huge{\textbf{PART I}}}
\end{center}
\endgroup
\vspace*{\fill}
\newpage

\chapter{Hamiltonian Formalism of General Relativity}
\label{cha:hgr}
In this chapter we will describe General Relativity (GR) as an Hamiltonian system. This is a necessary step in order to apply the concept of Canonical quantisation. 

The first instance of describing GR in Hamiltonian language was done in 1962 and was called Arnowitt-Deser-Misner (ADM) formalism \cite{1}. However, in order to quantise such a formulation of GR, ulterior developments of the ADM formalism were undertaken, leading to a formulation of GR as a gauge field theory, whose elementary variables are SU(2) connections (Ashtekar connections) and electric fields. GR thus became an Hamiltonian system with constraints represented by additional conditions on the phase space variables. In particular: \emph{SU(2)-gauge}, \emph{diffeomorphism} and \emph{Hamiltonian constraint}.

If we consider an Hamiltonian system with constraints, then it is possible to quantise such a system through the well known Dirac quantisation procedure for an Hamiltonian system with constraints \cite{2}. Essentially, what this procedure amounts to is to first quantise the unconstrained phase space, so as to obtain a kinematical Hilbert space $\mh_{kin} $. The constraint equations are then promoted to quantum constraint
operators defined on $\mh_{kin} $. \\
Since at the classical level constraints are supposed to vanish on the constraint hypersurface of the phase space, at the quantum level we require that the physically relevant state be annihilated by the constraint operators. The space of solutions for all the constraints is then the Physical Hilbert space $\mh_{phy}$. 

The detailed analysis of such a quantisation will be described in chapter \ref{cha:qunt}. In this chapter we will only describe the derivation of GR as an Hamiltonian system with constraints.
\section{ADM Action}
In order to proceed with the derivation of GR as an Hamiltonian system with constraints, the first step is to split the spacetime manifold M into space and time. Such a split is called a 3+1 split. This split is necessary since it allows for a definition of velocity and, therefore, conjugate momenta in terms of the configuration variables. Moreover, diffeomorphism invariance is maintained since this split is kept arbitrary, i.e. is not fixed once and for all. In particular, two different splits of the manifold M can be related by a diffeomorphism.

In order to carry out a 3+1 splitting of $M$ we utilise the fact that, at the classical level, it is possible to assume that the topology of $M$ is such that $M\cong\Rl\times\sigma$ for a fixed three dimensional manifold $\sigma$ of arbitrary topology. \\
This assumption is justified by a theorem due to Geroch \cite{3} which states that:\\
if spacetime is globally hyperbolic\footnote{A spacetime $M$ is globally hyperbolic if it possesses a \emph{Cauchy surface}, that is, if there exist spacelike surfaces which are connected to all the other points in $M$ (but not on the surface) by causal curves. In detail, a Cauchy surface $S$ is a space like surface, such that no two points on that surface are related 
in a causal way 
and such that the domain of dependence $D(S)$ (the set of all points $p\in M$, such that every past and future inextendible curve through a point $p$ intersects $S$) is the initial space-time manifold $M$, i.e. $D(S)=M$. Pictorially a Cauchy surface can be seen as an instant of time throughout the universe. } then it is necessary of such a topology\footnote{It should be noted that the implementation of such a restriction on the topology of $M$ at the quantum level is non trivial, since, as expected, topological changes occur. In fact, typical states in LQG correspond to complete degenerate spatial topology. This is not the case of semiclassical states. For a detailed analysis see \cite{4}, \cite{5}, \cite{6} and references therein}. 

The fact that $M\cong\Rl\times\sigma$ induces a foliation of $M$ into hypersurfaces\footnote{A hypersurface is an embedded m-1 submanifold. Given an n-dim manifold $N$ and an m-dim manifold $M$, a $C^k$ map $\phi:N\rightarrow M$ is said to be an embedding iff $N\rightarrow \phi(N)$ is an injection and, for each open subset $V\in N$, the subset $\phi(V)$ is open in the induced subset topology, i.e. the topology derived from the open sets of the form $\phi(V)\cap U$ where $U$ is an open set in $M$ (i.e. $\phi$ is a regular embedding)} $\Sigma_t:=X_t(\sigma)$, where $X_t:\sigma\rightarrow M$ is the regular embedding, such that $X_t(x):=X(t,x)$.
That is to say, it is possible to define a diffeomorphism $M\rightarrow \Rl \times\sigma$ where $\sigma$ is a fixed 3 dimensional manifold of arbitrary topology. Any two such foliations can be related as follows: consider two diffeomorphism $X:M\rightarrow \Rl\times \sigma$ and  $X^{'}:M\rightarrow \Rl\times \sigma^{'}$, given any other diffeomorphism $\phi\in Diff(M)$, $\phi$ can be written as $\phi=X^{'}\circ X^{-1}$, thus $X^{'}=\phi\circ X$. This implies that any two foliations of $M$ are related by a diffeomorphism, i.e. the arbitrariness of the foliation of $M$ is equivalent to $Diff(M)$.

Given this foliation of M our aim is to perform a 3+1 decomposition of the Einstein-Hilbert action 
\begin{equation}\label{equ:act}
S=\frac{1}{k}\int_Md^4x\sqrt{|det(g_{\mu\nu})|}R^{(4)}
\end{equation}
where c=1 and $k:=16\pi G_N$, as defined in the spatial manifold $\sigma$ which gets embedded as the hypersurface $\Sigma_t$. 

For the time being we will define the quantities we need directly on the hypersurfaces $\Sigma_t$ and, then, pull them back through the embedding $X_t$ to define the respective quantities on the manifold $\sigma$. The reason for this is that in $\Sigma_t$ it is possible to compare spatial tensor fields (which are the ones we are interested in since $\sigma$ is a spatial manifold) with arbitrary tensor fields restricted to $\Sigma_t$, since both are defined on a subset of $M$\footnote{Specifically, since $X_t$ is an embedding, $\Sigma_t$ is a submanifold of M, therefore, any quantity defined on M can be restricted to $\Sigma_t$, in particular any tensor field defined in $M$ can be restricted to $\Sigma_t$. Therefore, any restricted tensor field in $\Sigma_t$ can be compared to a spatial tensor, which is only defined on the subset $\Sigma_t$ of M, since $\Sigma_t$ is the embedding of a spatial manifold.}.

In order to define the spatial tensors needed to write the analogue of action \ref{equ:act} as defined on $\sigma$, we first of all need to parametrise the hypersurfaces $\Sigma_t$ in terms of $N$ and $N^{\mu}$, called the shift function and the lapse vector, respectively. Together these two vectors form the deformation vector 
\begin{equation}
T^{\mu}(X):=(\frac{\delta X^{\mu}(t,x)}{\delta t})_{|X=X(x,t)}=:N(X)n^{\mu}(X)+N^{\mu}(X)
\end{equation}
where $n^{\mu}$ is the unit normal to the hypersurface $\Sigma_t$, i.e. $g_{\mu,\nu}n^{\mu}n^{\nu}=-1$. It follows that $N(X)n^{\mu}$ is orthogonal to the hypersurface, while $N^{\mu}$ is tangential to $\Sigma_t$; i.e.$g_{\mu,\nu}n^{\mu}X^{\nu},a=0$. This follows form Frobenious theorem. The unit normal $n$ is also required to be proportional to an exact one-form, i.e. $n=n_{\mu}dX^{\mu}=Fdf$. 
\begin{figure}[htb]
 \begin{center}
 \psfrag{a}{$\Sigma_{t_7}$}\psfrag{b}{$\Sigma_{t_6}$}\psfrag{c}{$\Sigma_{t_5}$}\psfrag{d}{$\Sigma_{t_4}$}\psfrag{e}{$\Sigma_{t_3}$}\psfrag{f}{$\Sigma_{t_2}$}\psfrag{g}{$\Sigma_{t_1}$}
 \psfrag{m}{$\mm$}\psfrag{h}{$\vec{N}\delta t$}\psfrag{x}{$x$}\psfrag{j}{$\dot{X}\delta t$}\psfrag{i}{$N\vec{n}\delta t$}
  \includegraphics[scale=0.7]{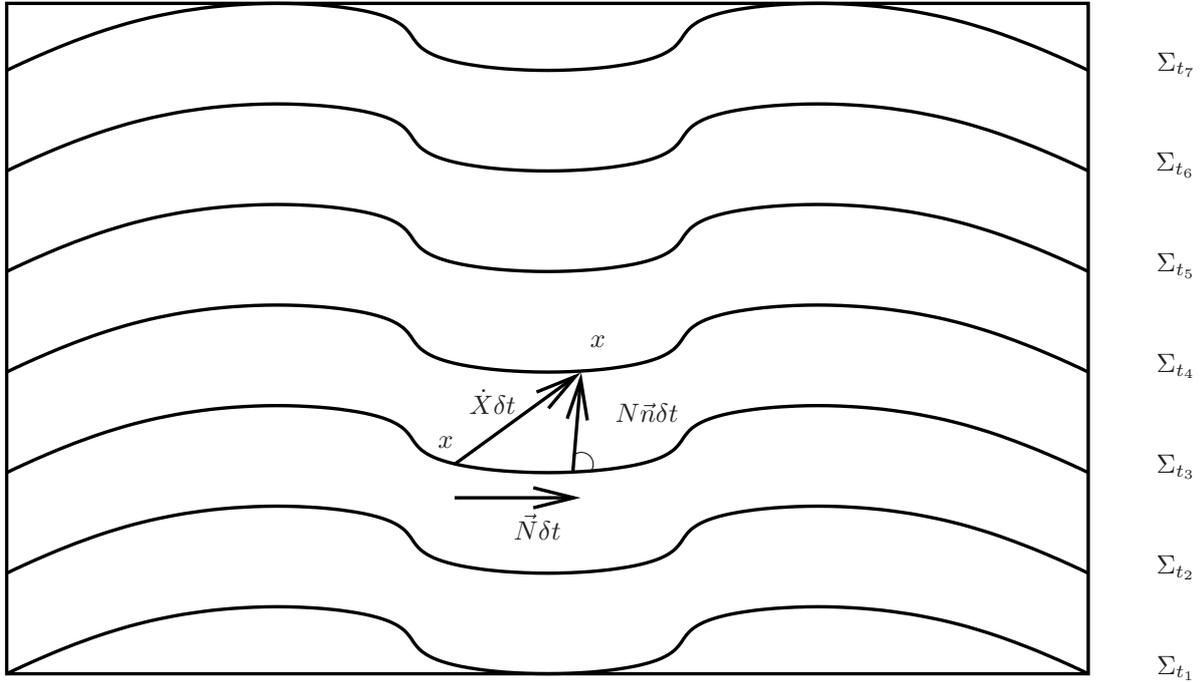}
\caption{Foliation of space-time into spacelike hypersurfaces $\Sigma_t:=X_t(\sigma)$. Here $\vec{N}$ represents the shift vector, while $N$ represents the lapse function.\label{fig:split} }
\end{center}
  \end{figure}
As can be deduced from the picture \ref{fig:split}, the conceptual significance of the quantities $T^{\mu}(X)$, $N^{\mu}$ and $N$ are as follows:
\begin{enumerate}
\item [1)]
\textbf{Deformation vector $T^{\mu}(X)$} : represents how hypersurfaces change with time, therefore it represents the differences between  hypersurfaces at different t`s. $T^{\mu}(X)$ is timelike $-N^2+g_{\mu\nu}N^{\mu}N^{\nu}\le 0$ and $N$ positive everywhere (since we want future directed foliation).
\item [2)] \textbf{Shift vector $N^{\mu}$} : represents the shift of position of a point as it``evolves" between different hypersurfaces.
\item [3)]
\textbf{Lapse function $N$}: is the function which indicates the shift in the orthogonal direction to the hypersurface $\Sigma_t$ and it indicates the time passed between the surface $\Sigma_{t_1}$ and $\Sigma_{t_2}$. We take $N$ to be positive everywhere in accordance with the requirement that $T^{\mu}(X)$ has to be positive everywhere, i.e. a future directed foliation.
\end{enumerate}
The requirements of positivity and future directedness of $T^{\mu}(X)$ reduce the possible embedding $X_t:\sigma\rightarrow M$ to a particular subset, dynamically constrained by the metric tensor $g_{\mu\nu}$.\\

So far we have described, in detail, how the foliation of M takes place in terms of the embedding $X_t$. The second step in constructing the analogue of the action \ref{equ:act} is to define the various quantities which appear in it, as referred to $\Sigma_t$. In particular, we define the \emph{first} and \emph{second fundamental forms} of $\Sigma_t$ 
\begin{equation}
q_{\mu\nu}:= g_{\mu\nu}-s n_{\mu}n_{\nu}\hspace{.5in} 
K_{\mu\nu}:=q^{\rho}_{\mu}q^{\sigma}_{\nu}\nabla_{\rho}n_{\sigma}\end{equation}
which are also called the ADM-metric and intrinsic curvature, respectively. It is easy to see that both the above tensors are spatial, since they vanish when contracted with $n^{\nu}$. We still need to define the Ricci scalar $R^{(4)}$ in terms of the Ricci scalar $R^{(3)}$ in the three dimensional submanifold $\Sigma_t$. This can be done through the construction of a covariant differential $D_{\mu}$ \footnote{ $\nabla$ is said to be a covariant differential with respect to a metric g if the following conditions hold i) $\nabla g=0$ (metric compatibility) ii) it is torsion free $[\nabla_{\mu},\nabla_{\nu}]f=0$ for all $f\in C^{\infty}(M)$.} with respect to the metric of Euclidean signature $q_{\mu\nu}$ on $\Sigma_t$. Thus we want $D_{\mu}$ to be a covariant differential on spatial tensors only, such that i) $D_{\mu}q_{\nu\rho}=0$ and ii) $D_{[\mu}D_{\nu]}f=0$ for scalars $f$. \\
It turns out that it is possible to define the covariant differential $D_{\mu}$ in terms of the covariant differential $\nabla_{\mu}$, compatible with $g_{\mu\nu}$ as follows: 
\begin{equation}\label{equ:cov}
D_{\mu}f:=q^{\nu}_{\mu}\nabla_{\nu}\tilde{f}\hspace{.5in}D_{\mu}u_{\nu}:=q^{\rho}_{\mu}q^{\sigma}_{\mu}\nabla_{\rho}\tilde{u}_{\sigma}
\end{equation}
where $\tilde{u}_{\sigma}$ is a spatial tensor field, i.e. $\tilde{u}_{\sigma}n^{\sigma}=0$.
Equation \ref{equ:cov} uncovers the fact that $D_{\mu}$ is nothing more than the spatial projection of the result of the application of $\nabla_{\mu}$. Here the quantities $\tilde{u}$ and $\tilde{f}$ are arbitrary smooth extensions of $f$ and $u$, respectively into a neighbourhood of $\Sigma_t$ in M. 

We can now define the Ricci curvature tensor $R^{(3)}$ on $\Sigma_t$ in terms of the above defined quantities $D_{\mu}$, $q_{\mu\nu}$ and $K_{\mu\nu}$. In particular we define the \emph{Codacci equation}
\begin{equation}\label{equ:codacci}
R^{(4)}=R^{(3)}-s[K_{\mu.\nu}K^{\mu.\nu}-(K^{\mu}_{\mu})^2]+2s\nabla_{\mu}(n^{\nu}\nabla_{\nu}n^{\mu}-n^{\mu}\nabla_{\mu}n^{\nu})
\end{equation}  
being really interested in the action defined with respect to $\sigma$, we now pull back all the quantities we have defined so far from $\Sigma_t$ to $\sigma$, through the pullback embedding $X^*:\Sigma_t\rightarrow\sigma$ and we obtain the following:
\begin{align}
q_{ab}(t,x)&:=(X^{\mu}_{,a}X^{\nu}_{,b}q_{\mu,\nu})(X(x,t))=g_{\mu\nu}(X(t,x))X^{\mu}_{,a}(t,x)X^{\nu}_{,b}(t,x)\\
K_{ab}(t,x)&:=(X^{\mu}_{,a}X^{\nu}_{,b}K_{\mu,\nu})(X(x,t))=(X^{\mu}_{,a}X^{\nu}_{,b}\nabla_{\mu}n_{\nu})(X(x,t))\\
R^{(3)}(t,x)&:=(R^{(3)}_{\mu\nu\rho\sigma} q^{\mu\rho}q^{\nu\sigma}(X(x,t)))(R^{(3)}_{\mu\nu\rho\sigma}X^{\mu}_{,a}X^{\nu}_{,b}X^{\rho}_{,c}X^{\sigma}_{,d})X(x,t)q^{ac}(x,t)q^{bd}(x,t)
\end{align}

It can be shown that the Ricci scalar $R^{(3)}$ is equal to the curvature scalar $R$ as defined in terms of the Christoffel symbols for $q_{ab}$.

We can now write the action \ref{equ:act} as defined with respect to the 3+1 split
\begin{equation}\label{equ:s}
S=\frac{1}{k}\int_{\Rl}dt\int_{\sigma}d^3x\sqrt{|det(q_{ab})|}|N|(R^{(3)}-s[K_{ab}K^{ab}-K^2])
\end{equation}
where we have dropped the total differential term $2s\nabla_{\mu}(n^{\nu}\nabla_{\nu}n^{\mu}-n^{\mu}\nabla_{\mu}n^{\nu})$ in the definition of $R^{(4)}$ ( see equation \ref{equ:codacci} ), since it can easily be obtained be applying the variational principle. Moreover, because of the covariance of the volume form $\Omega(X):=\sqrt{|det(g)|}d^{D+1}X$, its pull back 
$X^*\Omega(x,t):=\sqrt{|det(X^*g)|}dtd^{D}x$ is entirely determined by the identity $det(X^*g)=sN^2det(q_{ab})$.
\section{General Constraint Hamiltonian System }
The expression for the action given by equation \ref{equ:s} is not yet in canonical form ($\int dt(p\dot{q}-H)$), i.e. 
it does not dependent only on momenta, position and Hamiltonian.\\
In order to cast it into a canonical form, we need to perform a Legendre transformation from the Lagrangian density, defined as a function of configuration variables and velocity $q_{ab}$, $\dot{q}_{ab}$, $N$, $\dot{N}$, $N^a$, $\dot{N}^a$ to an Hamiltonian density, which is a function of the configuration variables and associated conjugate momentum , i.e. $q_{ab}$, $P_{ab}$, $\Pi$, $N$, $\Pi_a$. \\Transforming the conjugate momenta we obtain:
\begin{align}
P_{ab}(t,x)&:=\frac{\partial S}{\partial\dot{q}_{ab}(t,x)}=-s\frac{|N|}{N_k}\sqrt{det(q)}[K^{ab}-q^{ab}(K^c_c)]\\
\Pi(t,x)&:=\frac{\partial S}{\partial\dot{N}(t,x)}=0\\
\Pi_a(t,x)&:=\frac{\partial S}{\partial\dot{N}^a(t,x)}=0\\
\end{align}
The fact that the conjugate momenta $\Pi$ and $\Pi_a$ of N and $N^a$, respectively are zero, implies that the Lagrangian density in equation \ref{equ:s} is singular, i.e. the Legendre transformation 
\begin{align}\rho_L:T_*(\mathcal{C})&\rightarrow T^*(\mathcal{C})\\
(q,\dot{q})&\mapsto (q, p(q,\dot{q}))
\end{align} is a surjection only, therefore it is not invertible. The non invertibility of the Legendre transform implies that we are dealing with an Hamiltonian system with constraints. In order to quantise such a system one needs to follow the strategy developed by Dirac \cite{2}.
In this particular case under scrutiny, because of the singularity of the Lagrangian density, it is only possible to solve $\dot{q}_{ab}$ in terms of $\dot{q}_{ab}$, $N$, $N^a$ and $P_{ab}$, but for $\dot{N}$ and $\dot{N}^a$ we only obtain the \emph{primary constraints}
\begin{equation}\label{equ:constraints}
C(t,x):=\Pi(t,x)=0\hspace{.3in}C^a(t,x):=\Pi^a(t,x)=0
\end{equation} \indent Following Dirac constraint theory, the fact that the conjugate momenta $\Pi$ and $\Pi_a$ are zero, implies that $N$ and $N^a$ are not physically important variables. In fact, it turns out that they are chosen arbitrarily, therefore, we can multiply them by Lagrangian multipliers $\lambda(t,x)$ and $\lambda_a(t,x)$ and perform the Legendre transformation for the remaining variables. \\
Neglecting possible occurring boundary terms we obtain the following action 
\begin{equation}\label{equ:action}
S=\int_{\Rl}\int_{\sigma}d^Dx(\dot{q}_{ab}P^{ab}+\dot{N}\Pi+\dot{N}^a\Pi_a-[\lambda C+\lambda^aC_a+N^aH_a+|N|H])
\end{equation} 
where
\begin{align}\label{ali:hamdiff}
H_a&:=-2q_{ab}D_bP^{bc}\nonumber\\
H&:=_(\frac{s}{\sqrt{det(q)}}[q_{ac}q_{bd}-\frac{1}{D-1}q_{ab}q_{cd}]P^{ab}P^{cd}+\sqrt{det(q)}R)
\end{align} are the (\emph{spacial}) \emph{Diffeomorphism} and the \emph{Hamiltonian} constraints, respectively. It is straight forward to see that by varying \ref{equ:action} with respect to $\lambda(t,x)$ and $\lambda^a(t,x)$ one reproduces the \emph{primary constraints} \ref{equ:constraints}.

For a fixed $t\in \Rl$ the quantities $q_{ab}(t,x)$, $P_{ab}(t,x)$, $\Pi(t,x)$, $N(t,x)$, $\Pi_a(t,x)$ $N^a(t,x)$ are points in the infinite dimensional phase space $\mathcal{M}$, which carries the following symplectic structure $\Omega$ (Poisson brackets).
\begin{align}
\{P^{ab}(t,x), q_{cd}(t,y)\}&=\frac{k}{2}\delta^c_a\delta^d_b\delta^3(x,y)\nonumber\\
\{\Pi(t,x), N(t,y)\}&=\frac{k}{2}\delta^3(x,y)\nonumber\\
\{\Pi^a(t,x), N_b(t,y)\}&=\frac{k}{2}\delta^a_b\delta^3(x,y)
\end{align}
where all other possible Poisson brackets vanish identically.\\
Because of these primary constraints, the consistency of the dynamics of the system requires that we obtain secondary constraints
\begin{equation}\label{equ:con}
H(x,t)=0\hspace{.3in}H_a(x,t)=0
\end{equation}
Specifically one requires the primary constraints to be preserved under evolution of the system. Since the evolution of the system is defined in terms of Poisson brackets with respect to the Hamiltonian, we take the Poisson brackets of the constraints with the Hamiltonian and impose them to be equal to zero, thus obtaining
\begin{equation}
\{C(t,x),H\}=H(t,x)(\frac{N}{|N|}(t,x))=0\hspace{.3in}\{C(t,x),H\}=H_a(t,x)(\frac{N}{|N|}(t,x))=0
\end{equation} Since $N\neq0$ the equation \ref{equ:con} follows.

\begin{figure}[htb]
 \begin{center}
 \psfrag{a}{$\hat{\mm}$}\psfrag{b}{$[m]$}\psfrag{c}{$\bar{\mm}$}\psfrag{d}{$\mm$}
 \includegraphics[scale=0.5]{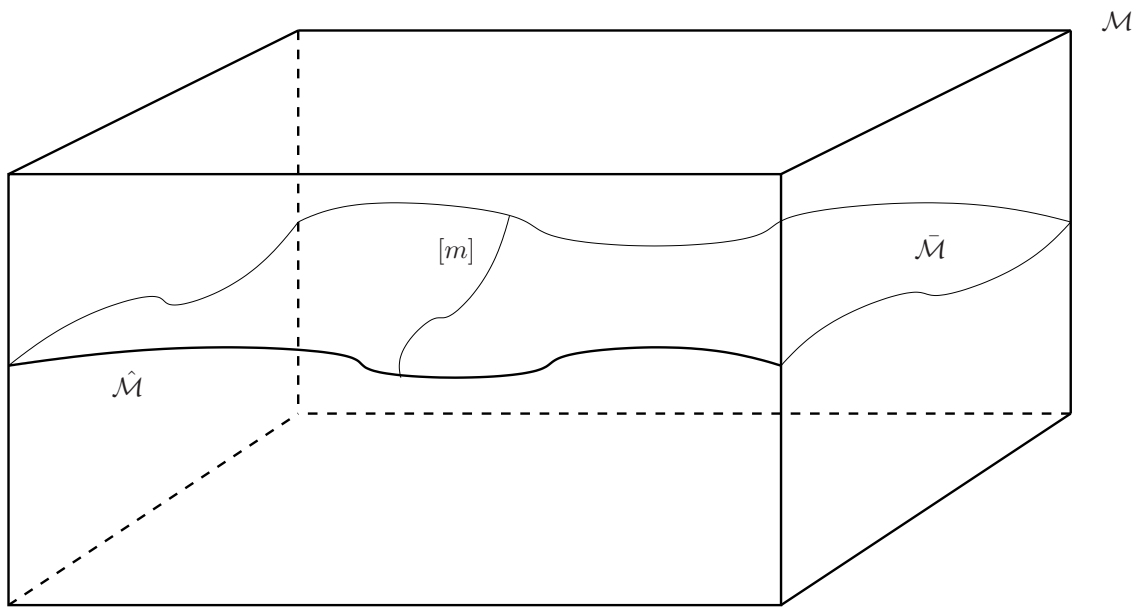}
 \caption{$\mathcal{M}$ represents the total phase space, $\mathcal{\bar{M}}$ represents the constraint hypersurface, the reduced phase space is given by $\hat{\mm}$ while the gauge orbits are denoted by $[m]$. \label{fig:constraint surface} }
  \end{center}
   \end{figure}

This implies that the Hamiltonian density
\be
{\bf H}:=\frac{1}{k}[\lambda C+\lambda^aC_a+N^aH_a+|N|H]
\ee
is constrained to vanish at each point in $\sigma$. It follows that General Relativity is a constrained Hamiltonian system with no true Hamiltonian. \\
Fortunately, the evolution of the \emph{secondary constraints} does not produce any other constraints. This implies that the constrained surface, which we denote by $\bar{\mm}$ and represents the submanifold of $\mm$ where the constraints hold, is preserved under the motions generated by such constraints (see figure \ref{fig:constraint surface}). It follows that all the constraints are \emph{first class constraints}, which determine co-isotropic constraint submanifolds, as opposed to \emph{second class constraints} which, instead, determine symplectic constraint submanifolds.\\
Since $C=\Pi$ and $C_a=\Pi_a$ are constrained to vanish on $\bar{\mm}$, the only terms which remain to be analysed in the Hamiltonian are $N$, $N_a$, $q_{ab}$ and $P^{ab}$. The equations of motion for the shift and lapse functions are  $\dot{N}^a=\lambda^a$ and $\dot{N}=\lambda$, respectively. Since the parameters $\lambda$ and $\lambda^a$ are completely arbitrary it follows that the trajectory of the lapse and shift vectors are completely arbitrary. Moreover, since the terms $\lambda C$ and $\lambda^a C_a$ are independent of the terms $q_{ab}$ and $P^{ab}$, the equations of motion of the latter will leave the former unaffected. This implies that, instead of utilising the full Hamiltonian ${\bf H}=\int_{\sigma}d^Dx[\lambda C+\lambda^a C_a+N^aH_a+|N|H]$, we can instead only consider the reduced form of the Hamiltonian constraint, since we are only interested in the variables $q_{ab}$ and $P^{ab}$, that is, we can only consider  ${\bf H}=\int d^Dx[N^aH_a+|N|H]$. Therefore the action becomes 
\begin{equation}\label{equ:actionadm}
S=\frac{1}{k}\int_{\Rl}\int_{\sigma}d^Dx(\dot{q}_{ab}P^{ab}-[N^aH_a+|N|H])
\end{equation}
This is the so called \emph{canonical Arnowitt-Deser-Misner (ADM) action}.
 
We are considering $q_{ab}$, then the constraints $H_a$
%
generates on all of $\mathcal{M}$ diffeomorphisms on $\mathcal{M}$ that preserve $\Sigma_t$, while $H$ generates diffeomorphisms on $\mathcal{M}$that are orthogonal to $\Sigma_t$. However, this is only true when the equations of motion $\dot{q}_{ab}=\{{\bf H},q_{ab}\}$ are satisfied. \\
This implies that the spatial diffeomorphisms on $\mathcal{M}$ induce diffeomorphisms on the phase space, which divide $\mathcal{\bar{\mathcal{M}}}$ into orbits of equivalence classes: $\hat{\mathcal{M}}=\{[m],m\in\mathcal{\bar{\mathcal{M}}}\}$. On the other hand, if we consider $P^{ab}$, then its evolution with respect to $H_a$ generates spatial diffeomorphism, while its variation with respect to $H$ generates diffeomorphism which are orthogonal to $\Sigma_t$ only on shell, i.e. only if the Vacuum Einstein equations $R^{(4)}_{\mu\nu}-\frac{1}{2}g_{\mu\nu}R^{(4)}=0$ are satisfied.

Summarising, what we have done so far is to first define constraints on the phase space $\mathcal{M}$, so as to select one particular hypersurface $\mathcal{\bar{M}}$  where the constraints $\Pi=0$, $\Pi_a=0$, $C=0$ and $C_a=0$ hold. Then, we have defined the gauge motions (Poisson brackets w.r.t. constraints) which are defined on all $\mathcal{M}$, but have the property that they leave $\mathcal{\bar{M}}$ invariant, therefore, each point $m$ on $\mathcal{\bar{M}}$ will not leave $\mathcal{\bar{M}}$ under the gauge transformations. What this implies is that $\mathcal{\bar{M}}$  gets divided into orbits of equivalent classes $[m]$. The set of all these orbits defines the reduced face space and Dirac observable depend only on these orbits (where the physics happens).

\section{New Variables}
In the previous section we have defined the canonical form of the ADM action. The ADM phase space is coordinatised by the variables $q_{ab}$ and $P^{ab}$, which satisfy the following Poisson algebra
\be\label{equ:admpoisson}
\{P^{ab}(x), q_{cd}(y)\}=\frac{k}{2}\delta^a_c\delta^b_d\delta^3(x,y)\hspace{.5in}\{P^{ab}(x),P^{cd}(y)\}=\{q_{ab}(x),q_{cd}(y)\}=0
\ee
\indent However, to date, it has not been possible to define a background-independent representation of such an algebra, which also accounts for the Hamiltonian constraint. The strategy adopted to overcome this problem is to extend the ADM-phase space and quantise the resulting Poisson algebra. This extended phase space is chosen such that its symplectic reduction, with respect to a 
certain extra constraint (Gauss constraint), will reproduce the ADM phase space with the original diffeomorphism and Hamiltonian constraints. \\
Moreover, since the constraint with respect to which we perform the symplectic reduction is the Gauss constraint of an SO(3) gauge theory, it follows that, as far as rotationally invariant observables are concerned, the only ones we are interested in, both the ADM system and the extended one are completely equivalent and we can as well work with the latter. After extending the ADM-phase space, an ulterior process is needed, namely, a canonical transformation on the extended phase space. 

Such a transformation resulted in the derivation of the Ashtekar variables. The advantage of such variables is that they render the constraints polynomial, thus easier to work with. 

Summarising, the process of constructing the new variables is actually two-fold: 
\begin{enumerate}
\item[i)] extension of the ADM phase space; 
\item[ii)] canonical transformation\footnote{A canonical transformation is a transformation which leaves the underlining Poisson algebra unchanged.} on the extended phase space. In particular such transformation will consist of two parts : a) A constant Wheyl rescaling b) an affine transformation.
\end{enumerate}
\indent We will now briefly describe the derivation of the Ashtekar variables. 
\\The first step is to introduce the co-3-bein fields $e$ such that the ADM metric can be written as
\be\label{equ:cobein}
q_{ab}:=e^j_ae^k_b\delta_{jk}
\ee
Equation \ref{equ:cobein} is invariant under local SO(3) rotation ($e_a^i\rightarrow O^i_je^j_a$), therefore $e^i_a$ contains three extra degrees of freedom which are not present in $q_{ab}$. It is precisely in this sense that we have `enlarged' the ADM phase space, since we have introduced extra gauge degrees of freedom. Such degrees of freedom will result in a Gauss constraints (see below). It follows that to reproduce the ADM metric we need to restrict such degrees of freedom. Next, we define the extrinsic curvature to be 
\be\label{equ:curvature}
K_{ab}:=-sK^j_{(a}e^k_{b)}\delta_{jk}
\ee 
where $K^i_a$ is an $su(2)$ valued one form. Since $K^i_a$ is a symmetric tensor field it has to satisfy the following constraint:
\be
G_{ab}:=K^i_{[a}e^k_{b]}\delta_{ik}=0   
\ee
which can be written as 
\be
G_{ik}:=K_{a[j}E^a_{k]}=0
\ee
where 
\be
E^a_j:=\frac{1}{2}sng(det(e))\epsilon_{jkl}\epsilon^{abc}e^k_be^l_c=\sqrt{det(q_{ab})}e_j^a
\ee
represents the densitised triad, which, because of equation \ref{equ:cobein} represents the dual of a Lie algebra valued pseudo 2-form.

The extended phase space is then coordinatised by the variables $(E^a_j(x), K^j_a(x))$, which undergo the following Poisson algebra
\be
\{E^a_j(x),K^k_b(y)\}=\frac{k}{2}\delta_b^a\delta^k_j\delta^3(x,y)\hspace{.5in} \{E^a_j(x),E^b_k(y)\}=\{K^j_a(x),K^k_b(y)\}=0
\ee 
It is then possible to define a new set of variable $ (\tilde{q}_{ab},\tilde{P}_{ab}) $ as functions of $E^a_j(x)$ and $K^j_a(x)$, such that they 
reproduce the usual ADM variables once the constraint $G_{jk}=0$ is applied:
\ba\label{ali:qpnew}
\tilde{q}_{ab}&:=&|det(E^c_j)|E^j_aE^k_b\delta_{jk}\nonumber\\
\tilde{P}^{ab}&:=&\frac{2}{|det(E^c_j)|}E^a_jE^d_kK^l_{[d}\delta^d_{c]}E^c_l\delta^{jk}
\ea
where $E^a_jE^k_a=\delta^k_j$. These new variables undergo the following Poisson algebra
\ba
\{\tilde{q}_{ab}(x), \tilde{q}_{cd}(y)\}&=&0\nonumber\\
\{\tilde{P}^{ab}(x), \tilde{P}^{cd}(y)\}&=&-k\Big(\frac{det(e)}{4}(\tilde{q}^{bc}G^{ad}+\tilde{q}^{bd}G^{ac}+\tilde{q}^{ac}G^{bd}+\tilde{q}^{ad}G^{bc})\Big)(x)\delta^3(x,y)\nonumber\\
\{\tilde{P}^{ab}(x),\tilde{q}_{cd}(y)\}&=&\frac{k}{2}\delta^a_c\delta^b_d\delta^3(x,y)
\ea
which is equivalent to \ref{equ:admpoisson} when $G_{jk}=0$

The Hamiltonian and the Diffeomorphism constraint can now be written in terms $(E^a_j(x), K^j_a(x))$ as follows:
\ba\label{ali:hamdiff2}
H_a &:=& 2sD_b[K^j_aE^b_j-\delta^a_bK^j_cE^c_j]\\
H &:=&\frac{-s}{\sqrt{det(\tilde{q}_{ab})}}(K^l_aK^j_b-K^j_aK^l_b)E^a_jE^b_l-\sqrt{det(\tilde{q}_{ab})}R
\ea
which again are equivalent to \ref{ali:hamdiff} up to terms proportional to $G_{jk}$.

By substituting \ref{ali:hamdiff2} and \ref{ali:qpnew} in \ref{equ:actionadm} we obtain the `extended action'
\be
S:=\frac{1}{k}\int_{\Rl}\int_{\sigma}d^3x\Big(2\dot{K}^j_aE^a_j-[\Delta^{jk}G_{jk}+N^aC_a+NC]\Big)
\ee
which is reduced to the ADM action in \ref{equ:actionadm} by a symplectic reduction with respect to the constraint $G_{jk}$. It follows that with respect to 
rotationally invariant observables, the ADM system and the extended one can be considered equivalent.

Given the above extended system, the remaining aim now is 
to write the constraint $G_{jk}$ in such a form that it becomes the \emph{Gauss constraint} of an SO(3) gauge theory, i.e. $G_{jk}$ should be of the form 
$G_{jk} = (\partial_a E^a + [A_a , E^a ])_{jk}$
for some $so(3)$ connection $A$. This will lead to the definition of the Ashteker variables.
The steps for such a derivation are
\begin{itemize}
\item[a)]\textbf{A constant Wheyl rescaling}: For any non-vanishing complex number $\beta\neq 0$ called the \emph{Immirzi Parameter}, the rescaling 
\be\label{equ:rescaling}
(K^j_a, E^a_j)\rightarrow (^{(\beta)}K^i_a:=\beta K^j_a;\hspace{.2in} ^{(\beta)}E^a_j:=\frac{E^a_j}{\beta})
\ee
is a canonical transformation which leaves invariant the rotational constraint $G_{jk}$. Moreover, the spin connection $\Gamma^i_a$, which can be considered as an 
extension of the spacial covariant derivative $D_a$ from tensors to generalised tensors having also an $so(3)$ index, turns out to be invariant under the rescaling in equation \ref{equ:rescaling}
\be
(^{(\beta)}\Gamma^i_a):=\Gamma^i_a(^{(\beta}E)=\Gamma^i_a(E)
\ee
This is a consequence of the fact that by writing $\Gamma^i_a$ as a function of $E^a_j$, it is possible to show, that, $\Gamma^i_a$ is a homogeneous rotational function of degree \emph{zero} in $E^a_j$ and its first derivatives.\\
\indent A similar result holds for the Christophel-symbols $\Gamma^c_{ab}$ with respect to $q_{ab}$, since they are homogeneous and rational functions of degree \emph{zero} in $q_{ab}$ and its first derivatives.
It follows that 
\be
D_aE^a_j=\partial_aE^a_j+\Gamma_{aj}^{\mbox{}k}E^a_k=\delta_aE^a_j+\epsilon_{jkl}\Gamma^k_aE^a_l=D_a(^{(\beta)}E^a_j)=0
\ee i.e. the total covariant differential $D_a$ transforms invariantly under the rescaling in equation \ref{equ:rescaling}.
\item[b)] \textbf{Affine transformation}\\
Given the results above, it is possible to write the rotational constraint as follows:
\be
G_j=0+\epsilon_{jkl}(^{(\beta)}K^k_a)(^{(\beta)}E^a_l)=\partial_a(^{(\beta)}E^a_j)+\epsilon_{jkl}[\Gamma^k_a+(^{(\beta)}K^k_a)](^{(\beta)}E^a_l)=:^{(\beta)}\mathcal{D}_a^{(\beta)}E^a_j
\ee
The above equation suggests the introduction of the new connection 
\be
(^{(\beta)}A^j_a):=\Gamma^j_a+(^{(\beta)}K^j_a)
\ee
also called the Asthekar-connection. The introduction of this new connection renders the constraint $G_j$ in the exact form of a Gauss law constraint used in SU(2) gauge theories.

The Pair $(^{(\beta)}A^j_a,^{(\beta)}E_j^a)$ forms a canonically conjugate pair, i.e.
\be
\{^{(\beta)}A^j_a(x), ^{(\beta)}A^k_b(y)\}=\{^{(\beta)}E_j^a(x), ^{(\beta)}E_k^b(y)\}=0\hspace{.4in}\{^{(\beta)}E_j^a,^{(\beta)}A^k_b(y)\}=\frac{k}{2}\delta^a_b\delta^k_j\delta^3(x,y)
\ee
As the last step we use such a conjugate pair to re-write both the Hamiltonian and the diffeomorphism constraints
\ba\label{ali:cong}
H_a&=&-2s\sigma_{ijk}^{(\beta)}F^{i}_{ab}\mbox{ }^{(\beta)}E^{a}_j\mbox{ }^{(\beta)}E^b_k+^{(\beta)}K_a^iG_i\nonumber\\
H&=&\frac{\beta^2}{\sqrt{|det(^{(\beta)}E_j^a\beta)|}}\big[\epsilon_{ijk}^{(\beta)}F^{i}_{ab}\mbox{ }^{(\beta)}E^{a}_j\mbox{ }^{(\beta)}E^b_k+2^{(\beta)}E^a_jD_aG_i\big]+\nonumber\\
& &+(\beta^2-s)\frac{(^{(\beta)}K_b^j\mbox{ }^{(\beta)}E_j^a)(^{(\beta)}K_a^J\mbox{ }^{(\beta)}E_j^b-(^{(\beta)}K_c^j\mbox{ }^{(\beta)}E_j^c)^2}{\sqrt{|det(^{(\beta)}E^a_j\beta)|}}
\ea
\end{itemize}
where $^{(\beta)}F_{ab}^{i}:=2\partial_{[a}^{(abeta)}A_{b]}^i+\epsilon_{ijk}^{(\beta)}A^{j(\beta)}_aA_b^k$.\\
We note that both the constraints in equation \ref{ali:cong} involve the Gauss constraint. In fact it is possible to symbolically write them as follows:  $H_a=H^{'}_a+f_a^jG_j$ and $H=H^{'}+f^jG_j$.

Since the rescaling transformation was a canonical one, it leaves the Poisson brackets of the first class constraints $G_j$ $H_a$ and $H$ unchanged\footnote{Recall that for first class constraints, the Poisson bracket of such constraints with any other constraint is given by a linear combination of the constraints.}. Since the Gauss constraint $G_j$ generates a subalgebra of the constraint algebras, then the modified system $H^{'}_a$, $H^{'}$ and $G_j$ is itself a first class system and generates the same constraint surfaces of the phase space, as defined for the original first class constraints  $H_a$ and $H$. Obviously, the algebra of the modified system will differ from the one defined by the original Hamiltonian and Diffeomorphism constraint. However, such an algebra will coincide on the constraint hypersurface $G_j=0$. It follows that the Einstein-Hilbert action can be written as 
\be
S=\frac{1}{k}\int_{\Rl}\int_{\sigma}d^3x\Big(\mbox{}^{(\beta)}\dot{A}^i_a\mbox{}^{(\beta)}E^a_y-[\Lambda^iG_i+N^aH^{'}_a+NH^{'}]\Big)
\ee


\chapter{Quantisation Program for Systems with Constraints}
\label{cha:qunt}
\section{Outline of quantisation strategy}\label{s outline}
In this section we briefly describe the steps involved in the process of quantising a system with constraints.

The main idea put forward by Dirac is to first quantise the unconstrained system, resulting in a kinematical Hilbert space and, only afterwards, apply the constraints as operator equations on the physical states. For example, given a symplectic manifold $\mathcal{M}$ with a Poisson structure $\Omega$ on it and a set of first class constraints $C_I$ ( where $I\in\mathcal{I}$ for some label set $\mathcal{I}$), then in order to apply Dirac's algorithm we do the following:
\begin{enumerate}
\item[i)] First of all quantise the unconstrained system obtaining, in such a way, the kinematical Hilbert space ($\mathcal{H}_{kin}$) in which, the
set of elementary real functions on the full phase space are represented by self-adjoint operator, such that $\{,\}\rightarrow \frac{-i}{\hbar}[,]$. 
\item [ii)] Since the constraints are real functions on the phase space, we should represent them as self-adjoint operators $\hat{C}_I$ in $\mathcal{H}_{kin}$. In other words, we require the representation of the Poisson algebra of $\mathcal{M}$ on $\mathcal{H}_{kin}$ to be such that the constraints $C_I$ can be represented as well defined self-adjoint operators $\hat{C}_I$ in $\mathcal{H}_{kin}$. The physical states will then be those states which are annihilated by the constraints i.e. $\hat{C}_I\Psi=0$ \footnote{It should be noted, however, that if the constraint algebra only closes with structure function, then this strategy should not be adopted. See \cite{117} for a detail analysis} 
\item[iii)] Define the notion of an inner product with respect to the physical states. This will define the physical Hilbert space $\mh_{phy}$.
\item[iv)] Find a complete set of gauge invariant observables \footnote{We can anticipate that observables will be represented by a densely defined (In a normed space $ X$, a linear operator $ A\colon\mathscr{D}(A)\subset X\to X$ is said to be densely defined if $ \mathscr{D}(A)$ is a dense vector subspace of $ X$) Hermitian 
(or self-adjoint) linear operator 
acting on the physical space.}
\end{enumerate}
The reason why it is more convenient to first quantise and then constrain is two-fold:
\begin{enumerate}
\item [1)] only gauge invariant quantities (i.e. quantities which Poisson commute with all the constraints) are physically relevant. These quantities are called the Dirac observables . There are two types of such observables, namely, the \emph{strong} Dirac observables which Poisson commute with the constraints everywhere on the manifold $\mathcal{M}$ and \emph{weak} Dirac observables,
which only commute on the constrained hypersurface $\bar{\mathcal{M}}$. Constraining before quantising would imply the full knowledge of all Dirac observables which, in principle, is extremely hard to obtain. 
\item [2)] Admitting it would be possible to obtain full knowledge of all Dirac observables, it would be very hard to find a representation of the corresponding Poisson algebra.
\end{enumerate}

In what follows we will describe, in detail, the steps needed to apply Dirac's algorithm for quantising a system with constraints.\\
Let us consider a constrained symplectic manifold $(\mathcal{M},\Omega)$, the steps in the quantisation algorithm are the following:\\[5pt]
I) \textbf{Classical Poisson *-algebra $\mathcal{B}$ }\\[5pt]
The first step in order to quantise a system is to find a suitable set $S$ of so called ``elementary" variables which coordinatise the phase space $\mathcal{M}$, such that any function on $\mathcal{M}$ can be written in terms of them, i.e. $S$ separates the points of $\mathcal{M}$. The requirements which these elementary variables need to satisfy are the following:\\
a) $S$ has to form a closed Poisson subalgebra of the full Poisson algebra $C^{\infty}(\mathcal{M})$. This is required since canonical quantisation implies replacing Poisson brackets by $i\hbar$ times the corresponding commutator relation.\\
b) $S$ has to be closed under complex conjugation. This is required since adjoints of operators are obtained by quantising complex conjugates.

The object which satisfies all the above requirements is a Poisson *-subalgebra $\mathcal{B}$ of $C^{\infty}(\mathcal{M})$.
%
%
%
This procedure of choosing $\mathcal{B}$ is sometimes called choice of polarization. There might be various choices of $\mb$, however, the guiding principles in this case would be i) simple behaviour under gauge transformations generated by the constraints such that, the Dirac observables will not be complicated functions of the elementary variables and, thus, easily quntisable; ii) $\mb$ should be minimal in the sense that removing any of its elements would not make $\mb$ separate the points in $\mm$; iii) the symplectic structure between the elements of $\mb$ should be as simple as possible. 

One way to proceed in the construction of $\mb$ is as follows:\\
suppose $\mm$ is a cotangent bundle $T^*(\mp)$ over some configuration space $\mp$, then $\mb$ can be identified with the Lie subalgebra $\mq$ of $Fun(\mp)\times V(\mp)$ , where $Fun(\mp)$ is the algebra of smeared functions over $\mp$ and $V(\mp)$ the space of vector fields over $\mp$.\\
Such a subalgebra is generated by certain chosen elements of $Fun(\mp)$ and corresponding Hamiltonian vector fields on $\mm$ of smeared momentum functions which preserve $Fun(\mp)$, i.e. they are elements of $V(\mp)$. The Lie structure of $\mq$ is given by 
\be
[(f,v),(f^{'},v^{'})]=(v[f^{'}]-v^{'}[f],[v,v^{'}])
\ee
where $[v,v^{'}]$ is the usual Lie bracket between two vector fields and $v[f^{'}] $ represents the action of the vector field on the function $f^{'}$.
\\[5pt]
II) \textbf{Quantum *-Algebra $\mau$}\\[5pt]
We now want to promote the classical *Poisson sub-algebra $\mb$ to a quantum *algebra such that the Poisson brackets are replaced by commutation relations and complex conjugation by involution\footnote{Given an algebra $\ma$, an involution is an anti linear automorphisms on $\ma$ such that i) it reverses the order $(z_1a+z_2b)^*:=\bar{z}_1a^*+\bar{z}_2b^*$, $(ab)^*:=b^*a^*$; ii) it squares to the identity $(a^*)^*:=a$ for $a,b\in\ma$ and $z_1,z_2\in\Cl$. }.\\
In order to construct the quantum *-algebra $\mau$ out of $\mb$ we first of all consider the tensor algebra $T(\mb)$ over $\mb$, defined as follows:\\
for any non-negative integer $k$ the $k$th power of $\mb$ is defined to be the tensor product of $\mb$, $k$ times with itself
\be
T^k(\mb) = \mb^{\otimes k} = \mb\otimes \mb \otimes \cdots \otimes \mb.
\ee
The tensor algebra $T(\mb)$ is then defined to be the direct sum of $T^k(\mb)$ for $k=0,1,2,\cdots$
\be
T(\mb):= \bigoplus_{k=0}^\infty T^k(\mb) = \Cl\oplus \mb \oplus (\mb\otimes \mb) \oplus (\mb\otimes \mb\otimes \mb) \oplus \cdots=\Cl\oplus\oplus_{k=1}^{\infty}\otimes_{n=1}^k\mb
\ee
where $T^0\mb=\Cl$ is the ground field $\Cl$. It follows that the elements in $T(\mb)$ are $a=(a_0,a_1,\cdots,a_n,\cdots)$ for $a_0\in\Cl$ and $a_n=a_{1n}\otimes a_{2n}\otimes\cdots\otimes a_{nn}$ for $a_{kn}\in\mb$.\\
Multiplication in $T(\mb)$ is defined through the canonical isomorphism
\be
T^k\mb \otimes T^l \mb \rightarrow T^{k + l}\mb
\ee
as follows:
\be
(a\otimes b)_n=\sum_{k+l=n}a_k\otimes b_l;\hspace{.5in}a_k\otimes b_l=a_{1k}\otimes a_{2k}\otimes\cdots\otimes a_{kk}\otimes b_{1l}\otimes b_{2l}\otimes\cdots\otimes b_{ll}
\ee
Addition, multiplication by a scalar and involution are, instead, defined in the following way:
\ba
(a+b)_n&:=&a_n+b_n\nonumber\\
(za)_n&:=&za_n= (za_{1n})\otimes a_{2n}\cdots \otimes a_{nn}=a_{1n}\otimes a_{2n}\cdots \otimes (za_{nn})\nonumber\\
a^*&=&\bar{a}_0\oplus\mbox{}\otimes^{\infty}_{n=1}a^*_n:\hspace{.5in}a^*_n=\bar{a}_{nn}\otimes\bar{a}_{(n-1)n}\cdots\otimes\bar{a}_{1n}
\ea
Then, in order to obtain the desired algebra $\mau$, we divide $T(\mb)$ by the two sided ideal\footnote{Given a subalgebra $\mi$ of an algebra $\ma$, we say $\mi$ is a right (left) ideal of $\ma$ iff $ba\in\mi$ ($ab\in\mi$) for all $a\in\ma$, $b\in\mi$. A two sided ideal is both a left and right ideal. } generated by elements of the form
\be
a_i\otimes b_i-b_i\otimes a_i-i\hbar\{a_i,b_i\}
\ee
for $a_i, b_i\in\mb$.

There are, however, certain domain issues arising when constructing $\mau$ as done above. In fact, not all elements of $\mb$ are bounded (most are not). As a consequence not all operators in $\mau$ will be bounded. This implies that such operators can only be defined on dense subsets of the Hilbert space\footnote{$\Phi$ is a dense subset of $\mathcal{H}_{aux}$ iff $\forall\Psi\in\mathcal{H}_{aux}$, $\epsilon > 0$, $\exists\Psi^1\in\Phi$ such that $||\Psi-\Psi^1||<\epsilon$ (normally $\Phi$ is a space of smooth functions of rapid decrease).}. Such a subset is called the domain of the operator. If two operators do not share the same domain, then questions concerning their commutation relations are ill defined. To avoid such issues, it is convenient to choose to map each element $a\in\mb$ to a \emph{bounded} function of it rather than $a$ itself. This is acceptable as long as we ensure that such functions still separate the points\footnote{A function $f$ on $\mm$ is said to separate the points in $\mm$ iff for all $x\neq y\in\mm$ $\exists f$ such that $f(y)\in f(x)$.} in $\mm$. To attain this, given any unbounded element $a\in\mb$ we define the one parameter family of unitary operators $t\mapsto W_t:=exp(ita)$ for $t\in\Rl$. Such operators both separate the points in $\mm$ and approximate $1_{\mau}+ita$ for $t\rightarrow 0$. In this situation, the two sided ideal needed to define the algebra $\mau$ is generated by the elements
\ba
W_s(a)W_t(b)W_{-s}(a)&:=&W_t\Bigg(\sum_{n=0}^{\infty}\frac{(is\hbar)^n}{n!}\{a,b\}_n\Bigg)\nonumber\\
(W_s(a))^*&:=&W_{-s}(a)=(W_s(a))^{-1}
\ea
where $\{a,b\}_n:=\{a,\{a,b\}_{n-1}\}$ and $\{a,b\}_0=b$ is the iterated Poisson bracket.
\\[5pt]
III) \textbf{Representation of $\mau$}\\[5pt]
We now want to find a representation of the quantum *Poisson-algebra $\mau$ in a Hilbert space (see definition \ref{def:repalgebra}), i.e. a function $\pi:\mau\rightarrow\mathcal{L}(\mathcal{H}_{kin})$ into the subalgebra of linear operators in $\mh_{kin}$. 
It follows that, for all operators  $\pi([w])$, the relations $\pi([w]^*)=\pi([w])^{\dagger}$, $\pi([w][w^1])=\pi([w])\pi([w^1])$, $\pi(z[w]+z^1[w^1])=z\pi([w])+z^1\pi([w^1])$ are required to hold. Moreover, such a representation should map constraints to self-adjoint operators.
However, there will be many inequivalent representations\footnote{Given two representations $\pi_1:\mau\rightarrow\ml(\mh_1)$ and $\pi_2:\mau\rightarrow\ml(\mh_2)$ we say that they are equivalent if there exists a unitary map $U:\mh_1\rightarrow\mh_2$ such that $\pi_2(a)=U\pi_1(a)U^{-1}$ for all $a\in\mau$.} which could be possible candidates for the representation of $\mau$. In order to choose from them the correct one, stronger physical assumptions have to be taken into account. For the algebra of LQG, such stronger physical assumptions exist and lead to a unique representation \cite{7}, \cite{8}, \cite {9}.  
\\[6pt]
IV)  \textbf{Solve the constraints}\\[5pt]
In order to find the physical Hilbert space $\mh_{phy}$ we need to find those states $\phi\in \mh_{kin}$ for which $\hat{C}_I\phi=0$ and such that $\mh_{phy}$ satisfies the reality condition. However, there is a problem since the operators $\hat{C}_I$ will have a continuous spectrum including the value \emph{zero}, therefore the eigenvectors $\phi$ will not belong to $\mh_{kin}$ since, in general, they will not be square integrable in $\mh$. Such states are called \emph{generalised eigenfunctions}\footnote{An elementary example is as follows: consider a function $f:x \mapsto e^{ix}$. This is an ``eigenvector" of the differential operator $-i\frac{d}{dx}$ on the real line $\Rl$. However $f$ is not square-integrable for the usual Borel measure on $\Rl$.}.

In order to overcome such a problem one can choose between two different strategies:
\begin{enumerate}
\item [1)] \textbf{Redefined Algebraic Quantisation (RAQ)}\\
The main idea behind RAQ is that instead of imposing the constraints on the physical states one modifies the inner product of the theory.\\

\noindent The essential steps in the process of RAQ are as follows:\\
as it is , $\mh_{kin}$ is to `small' to contain all the solutions to the constraints, therefore, what one does is to enlarge $\mh_{kin}$ by first defining a dense subspace\footnote{A subset $A$ of a topological space $X $ is called dense (in X) if any point in $X$ can be ``well-approximated" by points in $A$, i.e. $A$ is dense in $X$ if for any point $x \in X$, any neighbourhood of $x$ contains at least one point from $A$.
Alternatively, $A$ is dense in $X$ if the only closed subset of $X$ containing $A$ is $X$ itself. This can also be expressed by saying that the closure of $A$ is $X$, or that the interior of the complement of $A$ is empty. This definition implies that if we have a subset A in X which is dense in X, then the topology $\tau_A$ on A would have at least the same open sets as in the induced topology from X. In fact, the induced topology is $\tau_I:=\{S\cap A|S \text{ open in } X\}$, but from definition of dense subspace the intersection $S\cap A$ is never zero since for any point x in X, any neighbourhood of x contains at least one point from A. Moreover, if every set in a topology $\tau_I$ is also in a topology $\tau_A$, we say that $\tau_A$ is finer than $\tau_I$, i.e. bigger.} $\md_{kin}\subset\mh_{kin}$ on which the constraint operators can be defined. One then constructs the algebraic dual $\md^*_{kin}$ of $\md_{kin}$, i.e. the space of linear functionals on $\md_{kin}$, such that it is possible to define the following topological inclusion:
\be
\md_{kin}\hookrightarrow\mh_{kin}\hookrightarrow\md_{kin}^*
\ee
where the topology on $\md_{kin}^*$ is the weak *-topology of pointwise convergence\footnote{A net $\phi^{\alpha}$ in $\md^*_{kin}$ converges to $\phi$ iff for any $f\in\md_{kin}$ the net of complex numbers $\phi^{\alpha}(f)$ converges to $\phi(f)$.} which is coarser that the norm topology on $\mh_{kin}$. $\md_{kin}$ instead is equipped with the relative topology induced by $\mh_{kin}$.

The next step is to define the space of solutions to the constraints, i.e. define a subspace $\md_{phy}^*\subset\md^*_{kin}$ such that
\be\label{equ:const}
[(\hat{C}_I)^{'}l](f):=l(\hat{C}^{\dagger}_If)=0\mbox{ }\forall I\in\mi\, f\in\md_{kin}. l\in \md^*_{kin}
\ee
However, the physical Hilbert space $\mh_{phy}$ can only be defined on a subspace of $\md_{phy}^*$, since, otherwise, the algebra $\mo_{phy}$ of physical observables would be realised as an  algebra of bounded operators, since such operators would be defined everywhere in $\md_{phy}^*$. Instead, what we want is an algebra of unbounded operators, since these are the only physically relevant ones. For this reason we only turn a subset of $\md_{phy}^*$ into the physical Hilbert space such that, given a dense subspace $\md_{phy}\in \mh_{phy}$, $\mo_{phy}$ is densely defined on it. Then, similarly as to the kinematical case we obtain the following topological inclusion:
\be
\md_{phy}\hookrightarrow\mh_{phy}\hookrightarrow\md_{phy}^*
\ee

The last step is to define the physical inner product on $\mh_{phy}$, in such a way that the adjoint $\bullet$ in the physical inner product would coincide with the adjoint in the kinematical inner product, i.e.
\be
\langle\psi,\hat{O}^{'}\psi^{'}\rangle_{phy}=\langle(\hat{O}^{'})^{\bullet}\psi,\psi^{'}\rangle_{phy}=\langle(\hat{O}^{\dagger})^{'}\psi,\psi^{'}\rangle_{phy}
\ee
A definition of such an inner product can be carried out through the \emph{rigging map} construction which is an anti-linear map:
\be
\nu:\md_{kin}\rightarrow \md_{phy}^*
\ee
such that 
\begin{enumerate}
\item[i)]
\be
\langle\nu(f),\nu(f^{'})\rangle_{phy}:=[\nu(f^{'})](f)\mbox{}\forall\mbox{}f,f^{'}\in\md_{kin}
\ee
is a positive semidefinite sesquilinear form\footnote{Given a complex vector space $V$, a map $\phi:V\times V\rightarrow \Cl$ is said to be sesquilinear if it is linear in one argument and antilinear in the other, i.e. 
   $\phi(x + y, z + w) = \phi(x, z) + \phi(x, w) + \phi(y, z) + \phi(y, w)$ and $\phi(a x, b y) = \bar{a} b\,\phi(x,y)$ for all $x,y,z,w\in V$ and all $a, b \in\Cl$.}
\item[ii)]
\be
\hat{O}^{'}\nu(f)=\nu(\hat{O}f)\,\forall f\in\md_{kin}\mbox{ and for any } \hat{O}\in\mo_{phy} 
\ee
that is, the dual action of any operator $\hat{O}\in\mo_{phy} $ preserves the space of solutions.
\end{enumerate} 
The actual construction of the rigging map can be carried out through the process of \emph{group averaging}. However, in order to apply such construction we need to assume that the constraints operators are self-adjoint. The group averaging proposal is as follows: \\
given that $(\hat{C}_I)_{I\in\mi}$ are self-adjoint and form a Lie algebra, we can exponentiate them to obtain a group of unitary operators\footnote{Note that, as defined, $\hat{U}(g)$ is a unitary representation of the Lie group G generated by the constraint operators $(\hat{C}_I)_{I\in\mi}$. } $t^I\rightarrow exp\big(i\sum_{I}t^I\hat{C}_I\big)=:\hat{U}(g)$ where $(t^I)_{I\in\mi}\in T$ and $T$ is chosen such that the exponential map is a bijection on the component of identity. In terms of such unitary operators equations \ref{equ:const} becomes
\be\label{equ:uconst}
[(\hat{U}(g))^{'}l](f):=l(\hat{U}^{\dagger}(g)f)=l(f)\,\forall l\in\md^*_{kin}
\ee
i.e. $\hat{U}(g)$ acts trivially on the physical states.\\
\mbox{  } For the case in which G is a finite compact Lie group, then there exists a unique Haar measure $\mu_H$ which is invariant under both left- and right-translation and under inversion. This feature enables us to define the rigging map as follows:
\ba
\nu:\md_{kin}&\rightarrow& \mh_{phy}\nonumber\\
f&\mapsto&\nu(f):=\int_{G}d\mu_H(g)\langle\hat{U}(g)f,\cdot\rangle_{kin}=\int_Td\mu_H(t)\langle exp(t^I\hat{C}_If,\cdot\rangle_{kin}
\ea 
with physical inner product
\be
\langle\nu(f).\nu(f^{'})\rangle_{phy}:=[\nu(f^{'})](f)
\ee

The problem with the RAQ is that it only works if i) the constraint operators are self-adjoint ii) they form a Lie algebra iii) they are first class iv) the Lie group they generate is locally compact (with respect to the appropriate topology).\\
Such conditions imply that in the case of LQG, the RAQ could, in principle, only be used for the diffeomorphism constraint. However, since the uniqueness of the Haar measure is only guaranteed if the compact Lie group is finite dimensional, even for the case of the diffeomorphism constraint the inner product will not be unique, since we have infinitely many such constraints.

\item[2)]\textbf{Direct Integral Decomposition (DID)}\\
In contrast to the process of RAQ, the DID strategy for determining the physical Hilbert space is to directly solve the constraint. The main idea behind DID is that, for any separable\footnote{A topological space $X$ is called separable if it contains a countable dense subset, i.e. if there exists a sequence $\{ x_n \}_{n=1}^{\infty}$  of elements in $X$ such that every non-empty open subset of the space contains, at least, one element of the sequence.} Hilbert space $\mh$, there exist a self-adjoint operator $A$, such that we can represent $\mh$ as the direct integral of Hilbert spaces
\be
\mh\cong\int_{\Rl}d\mu(\lambda)\mh_{\lambda}
\ee 
where the operator $A$ acts on each $\mh_{\lambda}$ by multiplication by $\lambda$.

The physical Hilbert space is then associated with $\lambda=0$. However, the measure $\mu$ is unique up to equivalence. In fact, two measures are said to be equivalent if the set for which they are zero are the same. Therefore the Hilbert spaces $\mh_{\lambda}$ are unique only up to sets of measure zero.\\
\mbox{  } In the case of LQG, the Hilbert space, although non separable, can be decomposed into an uncountable sum of separable Hilbert spaces $\mh_{\lambda}$  which are left invariant from the action of the \emph{Master constraint} $\hat{\bf{M}}$ (see section \ref{ssolvingconst}).  Since $\hat{\bf{M}}$ is a self-adjoint operator and acts on each $\mh_{\lambda}$ by multiplication by $\lambda$, the DID can be applied to each separable Hilbert space separately and the physical Hilbert space we are interested in will be identified for $\lambda=0$. The physical inner product will be then given by $\langle\cdot, \cdot\rangle _{\mh_0}$.\\
\mbox{  } The essential steps of the process of DID, as applied to a general Hilbert space $\mh$, can be summarised as follows:
\begin{enumerate}
\item [1)] Express the Hilbert space $\mh$ as 
\be
\mh:=\bigoplus_{k\in K}\mh_k
\ee
where each of the individual $\mh_k$ are orthogonal to each other and are constructed through the completion of the sets
\be
s_i:=\{\sum_{l=1}^nz_l\hat{E}(B_l)\Omega_i|B_l\text{ measurable }, z_l\in\Cl\}
\ee
where $\Omega_i$ is a vector in $\mh$, such that $||\Omega_i||=1$ and $\hat{E}(B_l)$ is the projection operator on the measurable set $B\subset \Rl$.
\item[2)]
Define a unitary map 
\ba
U_1:\mh&\rightarrow& \bigoplus_k\mh_k\nonumber\\
\psi&\mapsto& U_i(\psi):=(f_k)_{k\in K}\mbox{ for } f_k\in L_2(\sigma(\hat{A}, d\mu_{\Omega_k})
\ea
where $\psi:=\sum_k f_k(\hat{A})\Omega_k=\sum_k \sum_l z_{k_l}\hat{E}(B_{k_l})\Omega_k=\sum_k \sum_l z_{k_l}\chi_{B_{k_l}}(\hat{A})\Omega_k$ and $d\mu_{\Omega_k}$ is the spectral measure.
\item[3)]
Introduce a new\footnote{It should be noted that the standard probability measure is define for every Borel set $B_l$, however, in this context we define a new measure which complies with the requierements of the Radon-Nikodym theorem} positive probability measure defined in terms of the spectral measure 
\be\label{equ:measure}
\mu(B_l)=\sum_ka_k\mu_{\Omega_k}(B_l)
\ee
GIven such a measure, it is possible to apply the Radon-Nikodym theorem\footnote{For any measurable space $X$, if there exists a $\sigma$-finite measure $\mu$ on it, such that $\mu $ is absolutely continuous with respect to a $\sigma$-finite measure $\mu^{'}$ on $X$, then there is a measurable function $f$ on $X$ taking values in $[0,\infty)$, such that
$\mu(A) = \int_A f \, d\mu^{'}$ for any measurable set $A$. (Note that any $\sigma$-finite measure $\mu$ on a space X is equivalent to a probability measure on $X$.}, obtaining 
\be\label{equ:measure}
\mu_{\Omega_k}(B_l)=\int_{B_l}\rho_k(\lambda)d\mu(\lambda)
\ee
This is needed since we want to introduce disjoint measurable sets $S_N:=\{\lambda\in\Rl; N_{\lambda}=N\}$, where $N_{\lambda}=N$ indicates the number of  $\rho_k(\lambda)>0$ . Given these sets it is possible to decompose $\Rl$ in terms of them, such that the inner product of two vectors in $\mh$ can be written in terms of sums over such sets ($S_N$), with respect to the newly defined measure, thus obtaining the equality 
\ba\label{ali:inpro}
\langle\psi,\tilde{\psi}\rangle_{\mh}&=&\sum_k\int_{\Rl}\bar{f}_k(\lambda)\tilde{f}_k(\lambda)\rho_k(\lambda)d\mu(\lambda)\nonumber\\
&=&\sum_{N=1}^{\infty}\int_{S_N}d\mu(\lambda)\sum_{k=1}^{\infty}\overline{[\sqrt{\rho_k(\lambda)}f_k(\lambda)]}[\sqrt{\rho_k(\lambda)}\tilde{f}_k(\lambda)]
\ea Since for each $\lambda$ only $N_{\lambda}$ of the terms $\rho_k$ will contribute in \ref{ali:inpro}, it is possible to interpret the sum over k as a scalar product in $\Cl^{N_{\lambda}}$. Therefore, for each $\lambda$ we obtain a Hilbert space $\mh_{\lambda}\cong \Cl^{N_{\lambda}}$.

It is now possible to define the map 
\ba
U_2:L_2(\sigma(\hat{A},d\mu_{\Omega_K})&\rightarrow& \prod_{\lambda\in\Rl}\mh(\lambda)\nonumber\\
(f_k)_{k\in K}&\mapsto&(g(\lambda))_{\lambda\in \Rl}:=\Big(\sum_{l=1}^{N_{\lambda}}\sqrt{\rho_{k_l}(\lambda)}f_{k_l}(\lambda)e_l\Big)_{\lambda\in \Rl}
\ea
which maps a discrete series to a continuous one.
\item [4)] Compose the two maps $U_2$ and $U_1$ so to obtain $U=U_2\circ U_1$ 
\ba
\mh&\rightarrow &\int_{\lambda\in\Rl}\mh_{\lambda}\nonumber\\
\psi&\mapsto&(g(\lambda))_{\lambda\in\Rl}
\ea
which give the desired integral decomposition of $\mh$ which can be written as 
\be
U\mh\cong\int_{\Rl}d\mu(\lambda)\mh_{\lambda}
\ee
It is easy to show that the operator $\hat{A}$ acts on each $\mh_{\lambda}$ by multiplication of $\lambda$.
\end{enumerate}
\end{enumerate}
\section{Loop Quantum Gravity}
In this section we will describe how the quantisation procedure described above is carried out in the context of LQG. The first step is to define the classical algebra $\mb$. However, in order to do that we first of all need to introduce various geometrical notations.
\subsection{Configuration Space and the Classical Algebra $\mb$}
\label{s configurationspace}
In the following, we will assume that the manifold $\sigma$ is a semianalytic, connected, locally compact and orientable 3-dimensional manifold.
\begin{definition}
Given a set $C$ of continuous, oriented, piecewise semianalytic, parametrised, compactly supported curves embedded in $\sigma$, an element $c\in C$ is defined to be a map:
\ba
c:[0,1]&\rightarrow& \sigma\nonumber\\
t&\mapsto& c(t)
\ea  such that :
\begin{enumerate}
\item [i)] $\exists$ a finite number $n$ and a partition $[0,1]=[t_0=0,t_i]\cup[t_1,t_2]\cup\cdots\cup[t_{n-1},t_n]$.
\item [ii)] $c$ is continuous at $t_k$, $k=1,\cdots, n-1$.
\item [iii)] $c$ is real semianalytic in $[t_{k-1}, t_k]$, $k=1,\cdots, n$.
\item [iv)] $c(t_{k-1}, t_k)$ $k=1,\cdots, n-1$ is an embedded one dimensional submanifold of $\sigma$. Moreover there is a compact subset of $\sigma$ containing $c$.
\end{enumerate}
\end{definition}
\mbox{  } From condition $iv)$ of the above definition it follows that, although a curve $c$ can be self-overlapping and self-intersecting, since it is only an immersion (need not be injective), however, for the intervals $(t_{k-1}, t_k)$ the curve $c$ is actually a regular embedding\footnote{Given an immersion $f:M_1\rightarrow M_2$, if $f$ is injective, then $f$ is called an \emph{embedding}. Moreover, if the differentiable structure on $f(M_1)$ induced by $M_2$ ( given by the atlas $\{V_J\cap f(M_1),\rho_J\}$ where $\{V_J,\rho_J\}$ is an atlas on $M_2$) coincides with the differentiable structure induced by $M_1$ (given by the atlas $\{f(U_I),\phi_I\circ f^{-1}\}$, where $\{U_I,\phi_I\}$ is an atlas of $M_1$), then $f$ is called a \emph{regular embedding}. }, therefore, it can not come arbitrarily close to itself.  

It is also possible to establish whether two curves are equivalent or not.
\begin{definition}\label{def:equ}
Two curves $c$ and $c^{'}$ are said to be equivalent $c\sim c^{'}$ iff
\begin{enumerate}
\item $b(c)=b(c^{'})$, $f(c)=f(c^{'})$.
\item $c^{'}$ is equivalent to $c$ up to a finite number of retracings\footnote{ A finite number of retracings of a curve $c^{'}$ means that $c^{'}=c_1^{'}\circ\tilde{c}^{'}_1\circ(\tilde{c}^{'}_1)^{-1}\circ\cdots\circ c_{n-1}^{'}\circ\tilde{c}^{'}_{n-1}\circ(\tilde{c}^{'}_{1-1})^{-1}\circ c^{'}_n$ for some finite number $n$ and curves $c^{'}_k$, $\tilde{c}^{'}_l$, $k=1,\cdots,n$, $l=1,\cdots, n-1$.} and a semianalitic reparametrization\footnote{A semianalytic parametrization of $c^{'}$ is defined through a diffeomorphisms $f:[0,1]\rightarrow [0,1]$ such that $c\circ f=c^{'}=c_1^{'}\circ\cdots \circ c^{'}_n$. }.
\end{enumerate}
\end{definition}
The definition of beginning and end point of a curve is defined below
\begin{definition}
Given a curve $c$ its beginning point, final point and range are defined to be, respectively
\be
b(c):=c(0), \hspace{.3in} f(c):=c(1), \hspace{.3in} r(c):=c([0,1])
\ee
If two curves $c_1$ and $c_2$ are such that $f(c_2)=b(c_1)$, it is possible to define the composition through the map $\circ:C\times C\rightarrow C$ as 
\be
(c_1\circ c_2)(t):=\begin{cases}c_1(2t)& t\in[0,\frac{1}{2}\\
c_2(2t-1)& t\in[\frac{1}{2},1]
\end{cases}
\ee
Inversion is instead defined through the map $^{-1}:C\rightarrow C$ as follows:
\be
c^{-1}(t):=c(1-t)
\ee
\end{definition}
It can be shown that the equivalence relation in definition \ref{def:equ} is both transitive, reflexive and symmetric.\\ The set of equivalence classes of curves is denoted by $\mp$, while an equivalence class of curves (or paths) is denoted by $p_c:=[c]_{\sim}$.

 We are now ready to introduce the concept of an edge $e$.
\begin{definition}
An edge $e\in \mp$ is an equivalence class of curves $c_e\in\mc$ which is semianalytic in all of $[0,1]$. The range of $e$ is defined as follows:  $r(e):=r(c_e)$, therefore the edges $e$ do not contain retracings.
\end{definition}
 It can be easily shown that $p_{c_1}\circ p_{c_2} =p_{c_1\circ c_2}$ and $p_c^{-1}=p_{c^{-1}}$ are well defined. This structure is reminiscent of a group structure, however, compositions of paths are not defined for all paths and there is no natural identity element on $\mp$, rather we have trivial paths $p_c\circ p_{c^{-1}}=b(p_c)$. Such a structure is called a groupoid.
\begin{definition}
A set $\ma$ is a \emph{groupoid} if there exists a unitary operation $i:\ma\rightarrow\ma$; $a\mapsto a^{-1}$ and a partial function $f:\ma\times\ma\rightarrow\ma$, which is not necessarily defined for all possible pairs of $\ma$-elements. 
\end{definition}
The categorical\footnote{See Appendix for the definition of a category and related concepts.} definition of a groupoid is as follows:
\begin{definition}
A \emph{groupoid} is a category in which each morphisms is an isomorphisms. 
\end{definition}  
In particular, the 3-dimensional manifold $\sigma$ can be turned into a groupoid category as follows:
\begin{definition}
The category  $\me$ of points and paths is defined such that: i) objects are the points $x\in\sigma$ ii) morphisms: $Hom(x,y):=\{p\in\mp; \, b(p)=x,\, f(p)=y\}$, i.e. paths between points. 
\end{definition}
Composition and identity in $\me$ are defined as above.  \\
A few more definitions regarding edges $e$ and what can be constructed through them, are necessary.
\begin{definition}
A graph $\gamma$ in $\sigma$ is a collection of edges, such that for any two pairs of edges they intersect at most in their end points, which are called vertices $(v)$.
\end{definition}
The collection of all vertices in a graph is denoted by $V(\gamma)$, while the set of all edges in $\gamma$ is denoted $E(\gamma)$
\begin{definition}
Given a graph $\gamma$, for any vertex $v\in V(\gamma)$ and edge $e\in E(\gamma)$ we have the following quantity:
\be
\sigma(e,v)=\begin{cases}+1 &\text{iff $b(e)=v$ then $e$ is \emph{outgoing} w.r.t. $v$} \\
-1 & \text{iff $f(e)=v$, then $e$ is \emph{ingoing} w.r.t. $v$}
\end{cases}
\ee
\end{definition}
Moreover, given a piecewise analytic surface $S$ (see definition \ref{def:picsurf}), the edges $e\in\gamma$ can have different relations with respect to $S$.
\begin{definition}\label{def:picsurf}
A surface $S$ is called piecewise analytic if it is a finite union of entire analytic, connected, embedded (D-1)-dimensional submanifolds (faces) $s_I$ of $\sigma$ (without boundary), whose closures intersect, at most, in their boundaries such that: \begin{enumerate}
\item[1)] The boundaries themselves are piecewise analytic (D-2)-submanifolds.
\item[2)] The union of all the analytic submanifolds is a connected $C^{(0)}$ (D-1)-dimensional submanifold (without boundary).
\item[3)] The closure of $S$ is contained in a compact (D-1) dimensional $C^{(0)}$ submanifold with boundary.
\item[4)] $S$ is contained in an open neighbourhood $U$ such that $U-S=U_+\cup U_-$ where $U_+$ and $U_-$ are disjoint non-empty open sets. We then say that $ S$ is orientable.
\end{enumerate}
\end{definition}
Given the above definition the edges of a graph can be divided into 4 classes: 
\begin{enumerate}
\item If $e\cap S=b(e)$ is an isolated intersection point and the edge lies in $U_-$, then the edge is called a \emph{down} edge.
\item If $e\cap S=b(e)$ is an isolated intersection point and the edge lies in $U_+$, then the edge is called an \emph{up} edge.
\item If $e\cap\bar{S}=e$, i.e. $e$ is contained in the closure of a face $S$, then the edge is called an \emph{inside} edge.
\item If $e\cap S=\emptyset$, then the edge is called an \emph{outside} edge\footnote{Not that this situation includes the case that $e$ intersects the boundary $\delta S:=\bar{S}-S$, since S has no boundary. }.
\end{enumerate} 
An ulterior relation between a graph $\gamma$ and a surface $S$ is given when all non-transversal points of intersection of  $\gamma$ with $S$ are vertices of $\gamma$. In this case $\gamma$ is said to be \emph{adapted} to the surface $S$.

Given the above definitions we are now ready to define the classical algabra $\mb$ for LQG. The conditions on such an algebra are i) $\mb$ has to be background independent ii) the Poisson bracket has to be non-distributional iii) we require the basic variables to have not so complicated transformation properties.\\ The Ashtekar connection and the densitised triads produce a Poisson algebras that is distributional ($\delta$ term appears), therefore one has to define an appropriate smearing of them. However, since the gauge transformation of $A^i_a$ and $E_j^a$ are $A^g\mapsto dgg^{-1}+gAg^{-1}$ and $E^g\mapsto gEg^{-1}$, respectively, any smearing in 3-dimensions of such functions would transform in a very complicated way. 

The solution to this was given by Wilson in \cite{wilson}, where he proposed to smear the connection $A^j_a$ along a one dimensional curves and, then, take the path ordered exponential obtaining, in such a way, the \emph{holonomy} of the connection $A^j_a$. The possibility of smearing $A^j_a$ along a one-dimensional curve is a direct consequence\footnote{The relation betwen p-forms and p-dimensional submanifolds is given by the Poincare' duality. } of the fact that $A$ is a one form and, as such, can be integrated along a differentiable curve resulting in an element of $SU(2)$.\\
The precise definition of the \emph{holonomy} of a connection is as follows:
\begin{definition}
Given a curve $c:[0,1]\rightarrow \sigma$ in $\sigma$, the holonomy $h_c(A)\in SU(2)=G$ of a connection $A$ along the curve $c$ is defined to be the unique solution to the differential equation in a local trivialisation
\be
\frac{d}{ds}h_{c_s}(A(c(s)))=h_{c_s}(A(c(s)))A^j_a(c(s))\frac{\tau_j}{2}\dot{c}^a(s),\mbox{ }h_{c_0}(A(c(0)))=1_G
\ee
where $c_s(t):=c(st)$ and $s\in[0,1]$, therefore
\be
h_c(a)=\mp exp(\int_c A)=1_G+\sum_{n=1}^{\infty}\int_0^1dt_1\int_{t_1}^1 dt_2\cdots\int_{t_{n-1}}^1 dt_n A(c(t_1))\cdots A(c(t_n))
\ee
\end{definition}

In the above definition, $\mp$ denotes the path ordering symbols and orders the smallest path to the left. Given the transformation of $A^j_a$, it follows that $h_c(A^g)=g(c(0))h_c(A)g(c(1))^{-1}$, i.e. the holonomy transforms locally under gauge transformations.\\
From the expression of the holonomy we note that it is invariant under reparametrization, therefore the holonomy depends only on equivalence class of curves, rather than single curves, i.e. $A(p_c):=h_c(A)=\mathcal{P}exp(\int_cA)$. This dependence implies the following relations:
\ba
A(p_c o p_c^{'})&=&h_{c o c^{'}} (A)=(\mathcal{P}exp\int_{coc^{'}})A=\mathcal{P}exp\int_c\int_{c^{'} }(A)=h_{c}(A) o h_{c^{'}} (A)=A(p_c)A( p_c^{'})\nonumber\\
A(p_c^{-1})&=&h_{c^{-1}}(A)=\mathcal{P}exp\int_{c^{-1}}(A)=\mathcal{P}exp\Big(-\int_{c}(A)\Big)=(h_c(A))^{-1}=A(p_c)^{-1}
\ea

However, the above mentioned properties are those required for a homeomorphisms, therefore we conclude that for each connection $A\in\ma$, its holonomy is a homeomorphisms from the set of all paths (i.e. all $p_c$) to the gauge group G, i.e. $h(A):\mathcal{P}\rightarrow G$. The fact that, for each element $A\in\ma$, $h$ maps $A$ to an element $h(A)\in G$, implies that there exists a map $H:\ma\rightarrow Hon(\mp,G)$. Such a map is an injection such that $\ma\subset Hom(\mp,G)$. This can be easily seen if we recall the bundle theoretic definition of connections, namely: given a bundle $(P,\pi,\sigma)$  ($P$ is a right $G$ space) a connection is a smooth assignment at each point $x$ in the base space of a vertical and horizontal subspaces of the tangent space of the bundle. Since the only trivial bundle occurs when $\sigma$ is 3 dim and G=SU(2), in general, we will obtain as many different spaces of connections  $\mathcal{A}_P$ ($P$ indicates the bundle it referees to) as there are possible bundles.

From above we see that for each  bundle $P$, the space of connections $A_P$ for that bundle gets mapped to $Hom(\mp, G)$. This implies that $Hom(\mathcal{P},G)$ must contain all possible $\mathcal{A}_P$ for all possible bundles P. Moreover, $Hom(\mp, G)$ depends only on $\sigma$, not on the bundle $P$ therefore, it will contain all possible spaces $\ma$ at once ($Hom(\mp, G)$ can also be shown to contain distributional elements). 
Therefore, given a bundle $P$, we can form the subset inclusion map $i:\mathcal{A}_P\rightarrow Hom(\mathcal{P},G)$,
i.e. $i$ is injective but not surjective.\\

We recall that our aim is to define the classical algebra $\mb$. To obtain a closed algebra the conjugate electric field $E_j^a$ should be smeared along 2 dimensional surfaces. Therefore, we obtain
\be
E_n(S)=\int_S n^j(*E)j
\ee
where $n^j$ is a Lie algebra valued scalar function.\\
Such a construction follows naturally from the fact that $E_j^a$ is dual to a Lie valued pseudo-2-form $(*E)^j_{ab}:=\epsilon_{abc}E^c_j-sgn(det(e))\epsilon_{jkl}e^k_je^l_b$, which can be integrated background independently over a surface.

The above can be formalised in the following definition:
\begin{definition}
The electric flux of the Lie algebra valued vector density $E_j^a$ through a piecewise analytic surface $S$ is defined as follows:
\be
E_n(S)=\sum_i\int_{s_i} n^j(*E)j=-1\frac{1}{2}\sum_i\int_{s_i} Tr(n* E)
\ee
where $s_i$ are the faces of $S$ such that $S=\bigcup_is_i$.
\end{definition}

The classical configuration space is then coordinatised by the holonomies of smooth connections $A\in\ma$ and the electrical fluxes (conjugate momentum). The Poisson brackets they satisfy are the following:
\be 
\{E_n(S), h_c(A)\}=k\int_{\sigma} d^3x\Big(\frac{\partial E_n(S)}{\partial E^a_j(x)}\Big)\Big(\frac{\partial h_c(A)}{\partial A_a^j(x)}\Big)
\ee 
However, if one computes the above Poisson brackets, it turns out that, in those situations for which the curves $c$ lie in the surface $S$, we get infinite contributions resulting in a non well defined Poisson bracket. The solution to this problem is to first perform a regularisation of both the holonomy and the electric flux by ulteriorly smearing them in 3 dimensions, then, perform the Poisson bracket between the regularised quantities and, finally, remove the regulator and, hopefully, end up with a non-distributional, simplectic structure of $E_n(S)$ and $h_c(A)$.

\begin{figure}[htb]
 \begin{center}
 \psfrag{a}{$s=-\epsilon$}\psfrag{b}{$s=0$}\psfrag{c}{$s=+\epsilon$}\psfrag{D}{$D^{\epsilon}_S$}\psfrag{S}{$S$} \includegraphics[scale=0.5]{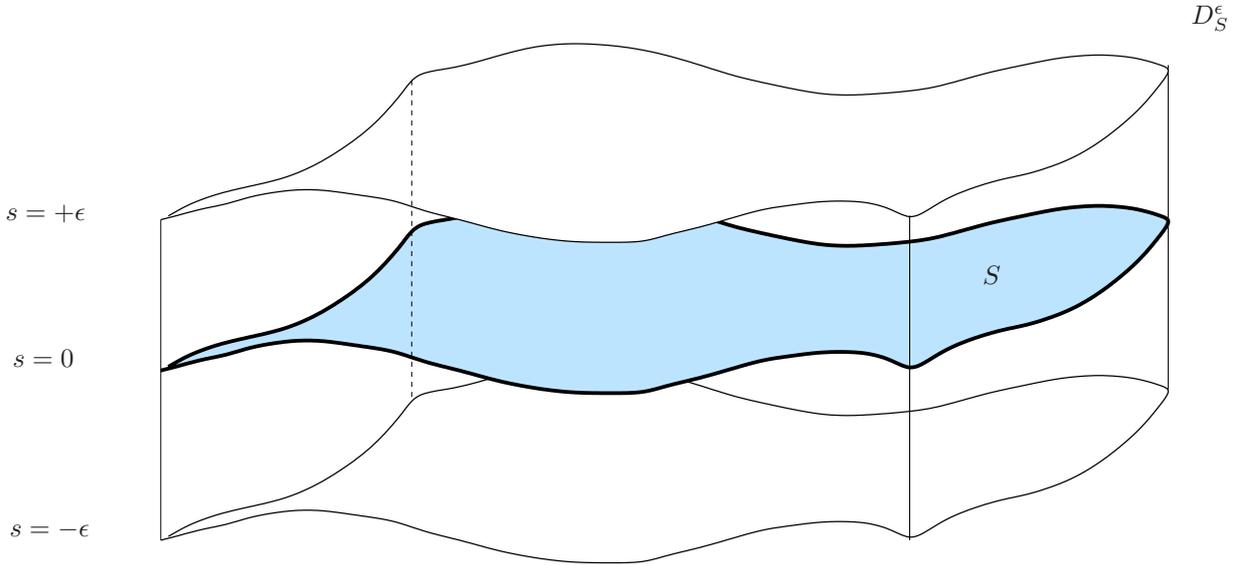}
\caption{Smearing in three dimensions of the surface $S$ on which the electric flux vector field is integrated over. This results in a disc $D^{\epsilon}_S$.\label{fig:renorm2} }
\end{center}
  \end{figure}

\begin{figure}[htb]
 \begin{center}
 \psfrag{a}{$P$}\psfrag{b}{$T^{\epsilon}_P$} \includegraphics[scale=0.5]{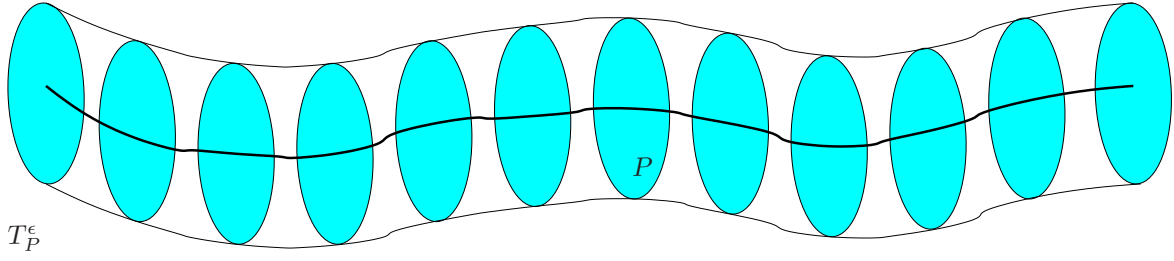}
\caption{Regularisation of the holonomy in three dimensions resulting in the tube $T^{\epsilon}_P$ . The centre of the tube is the path $p_c$. \label{fig:renorm1} }
\end{center}
  \end{figure}

This can be done \cite{1a}, \cite{1b} by smearing the path along a tube whose centre is the path itself (see figure \ref{fig:renorm1}) and, smearing the flux along a disc, whose centre is the surface $S_0$ where the flux was originally defined (see figure \ref{fig:renorm2}).

In the present work we will not go into the detail of how such a regularisation is carried out, the interested reader is referred to \cite{1b}, instead, what we will do is to point out the main results. In particular, utilising the above regularisation strategy, it is possible to define the regularised holonomy and flux, as elements of the subalgebra of the product algebra of the Lie *-algebra of smooth functions of the connection and smooth vector field thereon, i.e. $C^{\infty}(\ma)\times V^{\infty}(\ma)$.
\begin{definition}
The classical Poisson algebra $\mb$ is identified with the Lie *-algebra of $Cyl^{\infty}(\ma)\times V^{\infty}(\ma)$ generated by the smooth cylindrical functions\footnote{ A (smooth) cylindrical function on a graph $\gamma$ is a function which essentially identifies each connection in terms of its holonomies along edges. Specifically $f$ is a cylindrical function iff $f : \ma\rightarrow \Cl$, such that
there is a smooth function $F : G^E\rightarrow C$ with $f(A) = F(A(e_1),\cdots,A(e_E))$. See definition \ref{def:cylf}} $Cyl^{\infty}(\ma)$ and the flux vector fields $\tilde{E}_n(S)\in V^{\infty}(\ma)$ on $Cyl^{\infty}(\ma)$, such that \begin{enumerate}
\item [i)] $A(e)^*:=A(e^{-1})^T$
\item [ii)] $\tilde{E}_n(S)^*:=\tilde{E}_n(S)$
\end{enumerate}
It follows that the involution in $\mb$ is simply the complex conjugation.
\end{definition}
The flux vector filed $\tilde{E}_n(S)$ is such that, given a smooth cylindrical function $f:\ma\rightarrow \Cl$ we get
\be
(\tilde{E}_n(S)[f])(A)=\frac{1}{2}\sum_{k=1}^N\epsilon(e_k,S)\Big[\frac{n^j(b(e_k))\tau_j}{2}A(e)\Big]_{AB}\frac{\partial F_N}{\partial A(e_k)_{AB}}(\{A(e_k)\}_{k=1\cdots N})
\ee
where $F_N:G^N\rightarrow\Cl$, $AB$ indicate the SU(2) indices of the holonomy, i.e. $A(e_k)_{AB}$ and $\epsilon(e,S)$ takes the values $+1,-1,0,0$ depending whether the edges are of type up, down, inside or outside with respect to $S$, respectively. 
\subsubsection{Topology on the Space of Generalised Connections}
Having defined the classical algebra $\mb$, our aim, in this section, is to equip $\mathcal{A}\subset Hom(\mathcal{P},G)$ with a topology, so to develop a measure theory on $\mathcal{A}$. This can be achieved in two different ways:\\
the first method requires the notions of projective limit and direct product, while the second method is a C*-algebra approach. In the following we will briefly outline the important steps of each of the above mentioned methods. For a detailed description see \cite{1b} and references therein.\\

\textbf{Projective limit approach}\\[5pt]
The general outline of the first method is the following:\\
first of all we identify $Hom(\mathcal{P},G):=\bar{\ma}$ with the distributional extention of $\ma$. This was shown in 
\ref{s configurationspace}. The aim is then to equip $\bar{\ma}$ with a topology. The procedure for achieving this consists of various steps.
\begin{enumerate}
\item We first introduce the notion of tame subgroupoid $l(\gamma)$ of $\mathcal{P}$, however, in order to do this certain definitions are required:
\begin{definition}\mbox{ }\\
\begin{enumerate}
\item [i)] An oriented graph $\gamma$ is defined to be a graph generated by an independent set of edges\footnote{A finite set of edges  $\{e_i,\cdots, e_n\}$ is called independent if they intersect at most at their beginning point $b(e_k)$ and their final point $f(e_k)$. A path is defined to be a set of independent edges.} $\{e_i,\cdots, e_n\}$, i.e. $\gamma:=\bigcup_{k=1}^{n}r(e_k)$ where $r(e_k)\subset \gamma$ carries the orientation induced by $e_k$. $E(\gamma)$ is defined as the set of maximally semianalytic segments of $\gamma$ together with their orientations. The set of vertices of $\gamma$ is, instead, defined with respect to $E(\gamma)$ as follows: $V(\gamma)=\{b(e), f(e); e\in E(\gamma)\}$. The set of all oriented graphs is denoted by $\Gamma^w_0, where $w$ stands for semianalytic while $0$ indicates the compact support.$.
\item [ii)] Given a graph $\gamma$, $l(\gamma)\subset\mp$ is defined to be the subgroupoid of $\mp$ with objects $v(\gamma)$ and morphisms $E(\gamma)$. If $\gamma\in  \Gamma^w_0$, then the subgrupoid $l(\gamma)$ is called a tame subgroupoid.
\end{enumerate}
\end{definition}
\item We then define the set of all homomorphisms from the subgrupoid $l(\gamma)$ to $G$ as  $Hom(l(\gamma),G)=X_l$. It should be noted that the set $\mathcal{L}$ of all subgroupoids $l(\gamma)$ can be equipped with the structure of a partially ordered ($l\le l^{'}$ iff $l$ is a subgrupoid of $l^{'}$) directed set \cite{1b}. We will omit the proof of this here.
\item Equip $X_l$ with a compact Hausdorff topology through the pullback of the map 
\ba\label{ali:pl}
\rho_l:X_l&\rightarrow& G^{|E(\gamma)|}\nonumber\\
x_l&\rightarrow &\{x_l(e)\}_{e\in E(\gamma)}
\ea
Such a map is a bijection since any $x_l\in X_l$ is uniquely determined by the group elements $x_l(e)$ for $e\in E(\gamma)$. Moreover, since the group $G^n$ is a compact Hausdorff group for any finite $n$, the induced topology on $X_l$ through $\rho_l$, will be a compact Hausdorff.
\item We define the notion of a \emph{projective family} and a \emph{projective limit} of a projective family:
\begin{definition}\mbox{ }\\
\begin{enumerate}
\item [i)]
Given a partially ordered, directed index set $\ml$, then $(X_l, p_{l^{'}l})_{l\le l^{'}\in\ml}$ is a projective family which consists of sets $X_l$ labelled by $\ml$, together with surjective projections
\ba
p_{l^{'}l}:X_{l_1}&\rightarrow& X_l\hspace{.4in}\forall l\le l^{'}\nonumber\\
x_{l^{'}}&\mapsto& x_{l^{'}}|_l
\ea
such that 
\be\label{equ:conscond}
p_{l^{'}l}\circ p_{l^{''}l{'}}=p_{l^{''}l}\hspace{.4in}\forall l\le l^{'}\le l^{''}
\ee
\item [ii)] Given a projective family $(X_l, p_{l^{'}l})_{l\le l^{'}\in\ml}$ then, the projective limit $\bar{X}$ is defined to be the subset of the direct product $X_{\infty}:=\prod_{l\in\ml}X_l$ such that
\be
\bar{X}:=\{(x_l)_{l\in\ml}\in X_{\infty}|p_{l^{'}l}(x_{l^{'}})=x_l\mbox{ }\forall l\le l^{'}\}
\ee
\end{enumerate}
\end{definition}
It is, then, possible to show that the projections $p_{l^{'}l}$ are surjections and are continuous. This feature will be useful to carry out the next step.
\item Provide $X_{\infty}$ with the Tychonov topology as follows:
\begin{definition}
The Tychonov topology on the direct product $X_{\infty}:=\prod_{l\in\ml}X_l$  of topological spaces $X_l$ is defined to be the weakest topology, such that all the projections
\ba
p_l:X_{\infty}&\rightarrow X_l\nonumber\\
(x_{l^{'}})_{l^{'}\in\ml}&\mapsto& x_l
\ea
are continuous\footnote{The net $x^{\alpha}=(x^{\alpha}_l)_{l\in\ml} $ converges to $x=(x_l)_{l\in\ml}$ iff $x^{\alpha}_l\rightarrow x_l$ $\forall l\in\ml$.}.
\end{definition} 
Moreover, Tychonov theorem states that if the individual topological spaces $X_l$ are compact, then the product space $X_{\infty}:=\prod_{l\in\ml}X_l$ is compact in the Tychonov topology. This theorem will be essential in equipping $\bar{X}$ with a compact topology. 
\item Provide $\bar{X}$ with a compact topology identified with the subspace topology induced by $X_{\infty}$. In order to carry out such a requirement we first need to show that indeed $\bar{X}\subset X_{\infty}$ is a \emph{closed} subspace of $ X_{\infty}$, since closed subspaces of a compact space are compact in the subspace topology. The proof that $\bar{X}$ is a closed subspace of $ X_{\infty}$ consists in showing that for any convergent net in $\bar{X}$, the limiting point will lie in  $\bar{X}$. \\
Moreover, it turns out that both $\bar{X}$ and $X_{\infty}$ are Hausdorff spaces.\\
The above results converge in the following theorem:
\begin{theorem}
Given the set $\ml$ of all tame subgroupoids of $\mp$, the projective limit $\bar{X}$ of the spaces $X_l=Hom(l, G)$, $l\in \ml$ is a compact Hausdorff space in the induced Tychonov topology whenever G is a compact Hausdorff topological group .
\end{theorem}
\item The last step in our endevour of equipping $Hom(\mp,G)$ with a topology is to identify $\mathcal{\bar{A}}=Hom (\mathcal{P}, G)$ with $\bar{X}$ through a bijection map, which would endow $\mathcal{\bar{A}}$ with the Hausdorff topology of $\bar{X}$. Such a bijective map is given by  $\Phi:Hom(\mathcal{P},G)\rightarrow\bar{X}$; $H\rightarrow (H_{|l})_{l\in\mathcal{L}}$. We will omit the proof here, however the interested reader is referred to \cite{1b}. What, instead, we will do is to state the definition to which the above points ($1\rightarrow 7$) culminate to.
\begin{definition}
The space $\bar{\ma}:=Hom(\mp,G)$ of homomorphisms from the set $\mp$ of semianalytic paths to the compact Hausdorff group G, which was identified to the projective limit $\bar{X}$ of the space $X_l=Hom(l,G),$ where $\ml$ is the set of tame subgroupoids of $\mp$, is called the space of distributional connections over $\sigma$ and is equipped with a compact Hausdorff topology in the induced Tychonov topology of $X_{\infty}$ 
\end{definition} 
\end{enumerate}

\textbf{C*-Algebra Approach}\\[5pt]The second method of defining a topology on $\mathcal{\overline{A}}$ is called the C*-algebra approach. The main idea behind this method is that of identifying $\mathcal{\overline{A}}$ with the Gel'fand spectrum of a particular type of C*-algebra, which is a compact Hausdorff space in the Gel'fand topology. \\
The advantage of this method is that it is more general, since it does not make use of any underlying graph $\gamma$.

In what follows we will analyse the essential steps of this approach. The starting point will be a partially ordered, directed set $\mathcal{L}$ labelling \emph{any} compact Hausdorff spaces $X_l$ with surjective and continuous projections $p_{l^{'}l}:X_{l^{'}}\rightarrow X_l$ for $l\le l^{'}$, such that the consistency condition in \ref{equ:conscond} is satisfied. \\
We also consider the projective limit $\overline{X}$ and the direct product $X_{\infty}$ both with Tychonov topology, with respect to which they are Hoursdoff and compact. What we then do is to define the space of cylindrical functionals $Cyl(\overline{X})$ on $\overline{X}$ and show that its completion $\overline{Cyl(\overline{X})}$, with respect to some norm, is an Abelean $C^*$-algebra. As such, we can then apply Gel'fand's theorem to define an isometric isomorphism between $\overline{Cyl(\overline{X})}$ and the space of continuous functionals on its spectrum. \\
Such isometric isomorphism induces a homeomorphism between $\overline{X}$ and $Hom(\overline{Cyl(\overline{X})},\Cl)$ which, then, translates the Gel'fand isomorphism into an isomorphisms between the $C^*$-algebra $\overline{Cyl(\overline{X})}$ and the continuous functions on the projective limit. The homomorphisms between $\overline{X}$ and $Hom(\overline{Cyl(\overline{X})},\Cl)$ induces the desired compact Hausdorff topology on $\overline{X}$ purely in functional analytic terms, without references to underlying graphs.
\begin{enumerate}
\item As a first step we will define what cylindrical functions on the projective limit are.
\begin{definition} \label{def:cylf} Given the space $C(X_l)$ of continuous, complex valued functions on $X_l$ we define their union to be 
\be
Cyl^{'}(\overline{X}):=\bigcup_{l\in\ml}C(X_l)
\ee
such that, for any two functions $f, f^{'}\in Cyl^{'}(\overline{X})$, it is possible to find labels $l,l^{'}\in\ml$, so that $f_l\in C(X_l)$ and $f_{l^{'}}\in C(X_{l^{'}})$.
\end{definition}
 The space $Cyl^{'}(\overline{X})$ can be equipped with an equivalence relation as follows:
\begin{definition}\label{def:equivalencerel}
Given two functions $f, f^{'}\in Cyl^{'}(\overline{X})$, such that $f_l\in C(X_l)$ and $f_{l^{'}}\in C(X_{l^{'}})$ for some $l,l^{'}\in\ml$, we say that $f$ and $f^{'}$ are equivalent, i.e. $f\sim f^{'}$ if
\be\label{equ:pullb}
p^*_{l^{''}l} f=p^*_{l^{''}l^{'}}f^{'}\mbox{  }\forall l,l^{'}\le l^{''}
\ee
where $p^*_{l^{''}l}:X_l \rightarrow X_{l^{''}}$ is the pullback of $p_{l^{''}l}:X_{l^{''}}\rightarrow X_l $ (similarly $p^*_{l^{''}l^{'}}$)
\end{definition}
It can be shown that, once equation \ref{equ:pullb} holds for a particular $l\in \ml$, then it holds for any other $l^{'}$, such that $l\le l^{'}$. The proof of the above statement rests on the fact that $p_{l^{''}l}$ is a surgective map which satisfies the consistency condition in \ref{equ:conscond}. For a detailed proof and discussion the reader is referred to \cite{1b}.

Given the definition of equivalence on $Cyl^{'}(\overline{X})$ we can, then, define the space of cylindrical functionals on $\overline{X}$, as the space $Cyl^{'}(\overline{X})$ modulo the equivalence relation $\sim$ in definition \ref{def:equivalencerel}.
\begin{definition}
The space of cylindrical functionals on the projective limit $\overline{X}$ is defined to be the space of equivalence classes 
\be
Cyl(\overline{X}):=Cyl^{'}(\overline{X})/\sim
\ee
The equivalence class of a function $f\in Cyl^{'}(\overline{X})$ will be denoted as $[f]_{\sim}$ 
\end{definition}
\item The second step is to show that the space of cylindrical functions $Cyl(\overline{X})$ is a unital Abelian C*-algebra. In order to do so we will first show that it is a *-algebra. This requires the definition of operations between functions in $Cyl(\overline{X})$. However, two elements $f, f^{'}\in \overline{Cyl(\overline{X})}$ will generally belong to equivalence classes defined for different labels, i.e. $f=[f_{l}]_{\sim}$ and $f=[f_{l^{'}}]_{\sim}$ where $f_{l_i}\in C(X_{l_i})$. Therefore, we need a way of comparing any element in $\overline{Cyl(\overline{X})}$. It turns out that such a comparison is possible. In particular, it can be shown that for any two functions $f,f^{'}\in Cyl(\overline{X})$ there exists a common label $l\in \ml$ and $f_l, f^{'}_{l}$ such that $f=[f_l]_{\sim}$ and $f^{'}=[f^{'}_l]_{\sim}$. This property allows us to define all the operations in $Cyl(\overline{X})$, which turn $Cyl(\overline{X})$ into an Abelean *-algebra. 
\begin{lemma}
$Cyl(\overline{X})$ is an Abelean *-algebra defined by the following operations: 
\be
f+f^{'}:=[f_l+f_l^{'}]_{\sim}\mbox{,  }ff^{'}:=[f_lf^{'}_l]_{\sim}\mbox{,  }zf:=[zf_l]_{\sim}\mbox{,  }f^*:=\overline{f}:=[\overline{f}_l]_{\sim}\mbox{,  }\forall f,f^{'}\in Cyl(\overline{X})
\ee
where $z\in \Cl$ and $\overline{f}$ represents the complex conjugate.
\end{lemma}
It can also be shown that $Cyl(\overline{X})$ contains the unit element and can be equipped with the norm
\be
||f||:=sup_{x_l\in X_l}|f_l(x_l)|
\ee
which is well defined and independent of the chosen representative $f_l$.
\\
The completion $\overline{Cyl(\overline{X})}$ of $Cyl(\overline{X})$, with respect to such a norm, is a unital Abelean $C^*$-algebra. 

\item The last step is to show that $\overline{X}$ is a compact Hausdorff space with respect to the Gel'fand topology. This is done by defining a homomorphisms between $\overline{X}$ and the spectrum of cylindrical functions $\Delta(\overline{Cyl(\overline{X})})$ which, because of the Gel'fand theorem, is a compact Hausdorff space with respect to the Gel'fand topology. In detail, since $\overline{Cyl(\overline{X})}$ is a $C^*$-algebra, it is now possible to define, through the Gel'fand transform theorem, the following isometric isomorphism 
\ba
\bigvee:\overline{Cyl(\overline{X})}&\rightarrow& C(\Delta(\overline{Cyl(\overline{X})}))\nonumber\\
f&\mapsto&\tilde{f}\hspace{.8in}\text{such that}\hspace{.1in}\tilde{f}(\chi):=\chi(f)
\ea
This isomorphism turns the spectrum $\Delta(\overline{Cyl(\overline{X})})$ into a compact Hausdorff space in the Gel'fand topology, the weakest topology in which all the $\tilde{f}, f\in Cyl(\overline{X})$ are continuous.
We can now define the desired homomorphisms between $\overline{X}$ and $\Delta(\overline{Cyl(\overline{X})})$ as follows:

\ba
\mx:\overline{X}&\rightarrow& \Delta(\overline{Cyl(\overline{X})})\nonumber\\
x=(x_l)_{l\in\ml}&\mapsto&\mx(x)
\ea
where $[\mx(x)](f):=f_l(p_l(x))$ for $f=[f_l]_{\sim}$. The proof that $\mx$ is indeed an isomorphism can be found in \cite{1b}.\\
The above homomorphisms implies that the closure of the space of cylindrical functions $\overline{Cyl(\overline{X})}$ may be identified with the space of continuous functions $C(\overline{X})$ on the projective limit $\overline{X}$.
\end{enumerate} 
The importance of this second approach is that it was possible to define $\bar{X}$ as a compact Hausdorff space solely utilising $C^*$-algebra constructions, while leaving the index set $\ml$ and thus $X_l$ as general as possible. Therefore it has a wider scope than the first approach in which we had to restrict our analysis to subgrupoids $l(\gamma)$, which are graph dependent.

\subsection{Quantum Algebra $\mau$}
We now turn to the second step in the process of quantisation, namely the quantum representation $\mau$ of the classical algebra $\mb$. The first requirement is that the operators representing the holonomy and the flux have to be bounded operators, so as to avoid domain questions. For the case of operators representing holonomies, these will necessarily be bounded. In fact, as we previously stated, holonomies take values in a compact group, therefore, cylindrical functions, which are bounded functions of generalised connections\footnote{ Recall that a cylindrical function $f$, when $f$ is continuous, is defined on a finite number of independent edges, therefore it is a bounded function on some finite power of $G$.} will be promoted to bounded operators.

Problems arise when trying to define an operator associated to the flux vector fields $\tilde{E}_n(S)$. In fact such fields are analogous to momentum operators and, thus, are associated with differential operators which are unbounded. In order to overcome domain problems, which arise when dealing with unbounded operators, we will adopt the same strategy previously employed, namely use Weyl elements.
\begin{definition}
Given a flux vector field $\tilde{E}_n(S)\in \mb$, then for $t\in\Rl$ we can define the associated Weyl element as
\be\label{equ:W}
W^n_t(S):=exp\Big(-it[\frac{i\beta l^2_p}{2}\tilde{E}_n(S)]\Big)
\ee
where $\beta$ is the Immirzi parameter and $l_p=\hbar k$ is the Planck length.
\end{definition}
Given the above definition\footnote{ Note that for a general vector field  $X\in\mb$ generated by $\tilde{E}_n(S)$, the associated Weyl element is defined by replacing $\tilde{E}_n(S)$ by $X$ in \ref{equ:W}. }, the desired quantum algebra $\mau$ is defined as follows:
\begin{definition}
The algebra $\mau$ is generated by all the cyilindrical functions $f$ and all Weyl elements $W^n_t(S)$, such that the following relations are satisfied
\ba
[f,f^{'}]&=&0\nonumber\\
W^n_t(S)fW^n_t(S)^{-1}&=&(W^n_t(S))\cdot f\nonumber\\
W^n_t(S)W^{n^{'}}_{t^{'}}(S^{'})W^n_t(S)^{-1}&=&exp\Big(\frac{t^{'}\beta l^2_p}{2}\sum_{k=0}^{\infty}\frac{(t^{'}\beta l^2_p/2)^k}{k!}[\tilde{E}_n(S),\tilde{E}_{n^{'}}(S^{'})]_{(k)}\Big)
\ea
and the involution is 
\be
f^*:=\bar{f};\mbox{    }(W^n_t(S))^*:=W^n_{-t}(S)=(W^n_t(S))^{-1}
\ee
Similar relations hold for all vector fields in $\mau$.
\end{definition}
The commutation $[A,B]_{k}$ is inductively defined by $[A,B]_{0}=B$ and $[A,B]_{k}=[A,[A,B]]_{k-1}$.
\subsection{Representation of the Algebra $\mau$}
Now that we have defined the quantum algebra $\mau$ we need to define its representation in a Hilbert space, i.e. we want to find a *-morphisms between $\mau$ and a subset of linear operators on a Hilbert space $\mh$.

The strategy we will adopt to define a representation of $\mau$ is to first define a measure on the space $\overline{\ma}$, with respect to which a Hilbert space structure with associated inner product can be derived. The Hilbert space thus obtained is the kinematical Hilbert space $\mh_{kin}\cong L_2(\overline{\ma},d\mu_0)$. An orthonormal basis for $\mh_{kin}$ can be defined in terms of the spin network functions $T_{\gamma,\vec{\pi},\vec{m},\vec{n}}$. It is then possible, utilising Peter and Weyl theorem, to express $\mh_{kin}$ as a direct sum of orthogonal subspaces, each dependent, in some yet to be defined sense, on graphs $\gamma$.\\ 
Moroever, the representation $L_2(\overline{\ma},d\mu_0)$ of $\mau$ obtained above, can be derived as a unique GNS representation of a certain state. \\ 
\subsubsection{Measure on $\overline{\ma}$}
In order to define a measure on the configuration space $\overline{\ma}$, we will utilise Riesz-Markow theorem since it allows to define a family of consistent measures, which are compatible with the projective limit structure.
In particular, Riesz-Markow theorem is as follows:
\begin{theorem}
Given a compact Hausdorff space $X$ and a positive linear functional $\Lambda:C(X)\rightarrow \Cl$
on the space of continuous, complex-valued functions of compact support in $X$, then there exists a
$\sigma$−algebra $U$ on $X$, which contains the Borel $\sigma$−algebra and a unique positive measure $\mu$ on $U$, such
that $\Lambda$ is represented by $\mu$, i.e. 
\be
\Lambda_{\mu}(f)=\int_Xd\mu(x)f(x)\hspace{.5in}\forall f\in C(X)
\ee

$\mu$ has the following properties:
\begin{itemize}
\item[1)] $\mu(K) < \infty$ if $K \subset X $ is compact.
\item[2)] If $S^{'}\subset S\in U$ and $\mu(S) = 0$ then $S^{'}\in U$.
\item[3)] $\mu$ is regular.
\item [4)] For any $S \in U$ and any $\epsilon > 0$ there exist a closed set $C$ and an open set $O$ such that $C \subset S \subset O$ and $\mu(O-C) <\epsilon$.
\item [5)] For any $S \in U$ there exist sets $C^{'}$ and $O^{'}$ which are respectively countable unions and intersections
of closed and open sets, respectively, such that $C{'}\subset S \subset O^{'}$ and $\mu(O^{'}-C^{'}) = 0$.
\end{itemize}
\end{theorem}
Given the above definition, it is possible to obtain a unique Borel probability measure for each positive linear functional on a compact Hausdorff space if we normalise the measure, such that $\mu(X)=1$. We are now interested to apply this theorem to the space $\overline{\ma}=\overline{X}$, which is a compact Hausdorff space. However, we want the measure defined through the Riesz-Markov theorem to be compatible, in a yet to be defined sense, with the projective structure of $\overline{X}$. This is achieved by applying Riesz-Markov theorem to both $\overline{X}$ and $X_l$. Compatibility of the measures is then obtained by requiring that the functional $\Lambda_{\mu}$ on $\overline{X}$, restricted to $X_l$, is equivalent to the functional $\Lambda_{\mu_l}$ defined on $X_l$. The result of such a process results in the following definition:
\begin{definition}\label{def:consistent}
A family of measures $(\mu_l)_{l\in\ml}$ on the projections $X_l$ of a family $(X_l,p_{ll^{'}})_{l\le l^{'}\in\ml}$ is said to be consistent iff
\be
(p_{l^{'}l})_*\mu_{l^{'}}:=\mu_{l}\circ p_{l^{'}l}^{-1}=\mu_l\hspace{.5in}\forall l\le l^{'}
\ee
where $p_{l^{'}l}:X_{l^{'}}\rightarrow X_l$ are continuous-onto projections, and $(p_{l^{'}l})*\mu_{l^{'}}$ is the pushforward of $\mu_{l^{'}} $
\end{definition}

To understand the above definition, let us consider a probability measure $\mu $ on $\overline{X}$. We then define a positive linear functional on $X_l$ as follows:
\ba\label{ali:measure1}
\Lambda_{\mu|_{X_l}}:C(X_l)&\rightarrow&\Cl\nonumber\\
f_l&\mapsto&\Lambda_{\mu}(f_l)_{|_{X_l}}:=\int_{\overline{X}}d\mu(x)(p^*_lf_l)(x)
\ea  
The positivity requirement is satisfied by the fact that integrals over positive functions are always positive. However, since $X_l$ is a compact Hausdorff space, then, by Riesz-Markov theorem, there exists a unique Borel probability measure $\mu_l$, such that 
\ba\label{ali:measure2}
\Lambda_{\mu_l}:C(X_l)&\rightarrow&\Cl\nonumber\\
f_l&\mapsto&\Lambda_{\mu_l}(f_l)=\int_{X_l}d\mu_l(x_l)f_l(x_l)
\ea
For the two measures $\mu$ and $\mu_l$ to be consistent we require \ref{ali:measure1} and \ref{ali:measure2} to satisfy $ \Lambda_{\mu}(f_l)_{|_{X_l}}=\Lambda_{\mu_l}(f_l)$ or, equivalently, $\Lambda_{\mu_l}(p^*_lf_l)= \Lambda_{\mu_l}(f_l)$ for all $f_l\in C(X_l)$. By using the fact that measurable functions can be approximated by simple functions and that measurable simple functions can be approximated by continuous functions, we can write condition $\Lambda_{\mu_l}(p^*_lf_l)= \Lambda_{\mu_l}(f_l)$  as follows:
\ba
\Lambda_{\mu_l}(p^*_l\chi_{S_l}) &= &\Lambda_{\mu_l}(\chi_{l}) \nonumber\\
\int_{\overline{X}}d\mu(x)\chi_{p^{-1}_lS_l}(x) &=&\int_{X_l}d\mu_l(x_l)\chi_l(x_l)\nonumber\\
\mu(p^{-1}_{l}S_l)&=&\mu_l(S_l)
\ea
where $\chi_{S_i}$ is the characteristic function of $S_i$ and  $S_l\in X_l$ is any measurable set. The consistency condition for measures is thus $\mu\circ p^{-1}_l=\mu_l$, which actually represents the cylindrical projection of the measure $\mu$. It follows that, given any $l\le l^{'}$ then $\mu_l =\mu_{l^{'}}\circ p_{l^{'}l}^{-1}$.\\  We have shown that, given a regular Borel probability measure on $\overline{X}$, then $(\mu\circ p^{-1}_l=\mu_l)_{l\in\ml}$ defines a consistent family of Borel probability measures on $X_l$. \\
However, also the converse is true, namely: given a consistent family of Borel probability measures on $X_l$, it is possible to define a unique Borel probability  measure $\mu$ on $\overline{X}$, such that $\mu\circ p^{-1}_l=\mu_l$ is satisfied. \\To prove the above statement let us define a continuous linear functional on $Cyl(\overline{X})$
\ba
\Lambda^{'}_{\mu}Cyl(\overline{X})&\rightarrow& \Cl\\
f=[f_l]_{\sim}=p^*_lf_l&\mapsto&\Lambda^{'}_{\mu}(f)=\int_{X_l}d\mu_lf_l(x_l)
\ea 

The positivity of $\Lambda^{'}_{\mu}$ is given by the fact that each of the $\mu_l$ are positive. Since $Cyl(\overline{X})\subset \overline{Cyl(\overline{X})}$ and $\overline{Cyl(\overline{X})}$ is a unital $C^*$-algebra, it follows that i) $\Lambda^{'}_{\mu}$ is continuous ii) it can be uniquely and continuously extended to $\overline{Cyl(\overline{X})}$. Moreover, as it was previously shown, the Gel'fand theorem ensures that $C(\overline{X})$ is isomorphism to $\overline{Cyl(\overline{X})}$. We thus obtain 
\ba\label{ali:measure3}
\Lambda_{\mu}: C(\overline{X})&\rightarrow& \Cl\\
f=[f_l]_{\sim}=p^*_lf_l&\mapsto&\int_{X_l}d\mu_lf_l(x_l)
\ea
where $\Lambda_{\mu}$ is the extention of $\Lambda^{'}_{\mu}$.
The condition $\mu\circ p^{-1}_l=\mu_l$ means that \ref{ali:measure3} is independent of the chosen representative. By applying Riesz-Markov theorem we find a unique Borel probability measure $\mu$ such that 
\be
\Lambda_{\mu}(f)=\int_{\overline{X}}d\mu(x)f(x)
\ee

We now would like to apply the above results to the space $\overline{A}=Hom(\mp,G)$ which we identified with $\overline{X}$. To do so we actually have to specify the cylindrical functions in terms of tame subgroupoids of $\mp$, since $\overline{A}$ is identified with the space of homomorphisms from the  groupoid $\mp$ to $G$. In particular, for each tame subgroupoid $l(\gamma)=l$ of $\mp$,  $X_l=Hom(l(\gamma),G)$, therefore an element $x_l\in X_l$ is identified by the set of image points $\{x_l(e)\}_{e\in E(\gamma)}$ (being $x_l$ an homomorphisms).\\ Recalling equation \ref{ali:pl}, we can identify $\{x_l(e)\}_{e\in E(\gamma)}=\rho_{l}(x_l)$ by a collection of elements of G (=SU(2) for LQG). It follows that, given a continuous function $F_l:G^{|E(\gamma)|}\rightarrow \Cl$ we can write any $f_l\in C(X_l)$ as 
\be
f_l(x_l)=F_l(\rho_{l}(x_l))=F_l(\{x_l(e)\}_{e\in E(\gamma)})=(\rho^*_lF_l)(x_l)
\ee
Such a definition of continuous function allows us to work directly with finite powers of $G$. This is an advantage since we know that $G$ is equipped with a normalised Haar measure and, thus, we can define a positive linear functionals on $X_l$ in terms of such measure. In particular, defining $\rho_{l^{'}l}:G^{E(\gamma^{'})}\rightarrow G^{E(\gamma)}$ in terms of $p_{l^{'}l}:X_l^{'}\rightarrow X_l$ as $ \rho_{l^{'}l}:=\rho_l\circ p_{l^{'}l}\circ \rho_{l^{'}}^{-1}$ we obtain the following:
\begin{definition}
Given the set $\ml$ of all tame subgroupoids of $\mp$ and identifying $X_l=Hom(l,G)$ with $G^{|E(\gamma)|}$ through the map $\rho_l:X_l\rightarrow G^{|E(\gamma)|}$ if $l=l(\gamma)$, then for any $f\in C(X_l)$ we have
\ba\label{ali:functional}
\Lambda_{\mu_{O_l}}:C(X_l)&\rightarrow &\Cl\nonumber\\
f_l&\mapsto&\Lambda_{\mu_Ol}(f_l)=\int_{X_l}d\mu_{0_l}(\rho^*_lF_l)(x_l) :=\int_{G^{|E(\gamma)|}}\Big[\prod_{e\in E(\gamma)}d\mu_H(h_e)\Big]F_l(\{h_e\}_{e\in E(\gamma)})
\ea
where $\mu_H$ is the Haar measure which is invariant under left and right translations, since G is compact.
\end{definition}
It can be shown that the functional in equation \ref{ali:functional} is positive for all $l\in \ml$, it defines a consistent family $(\Lambda_{\mu_{0_l}} )_{l\in\ml}$ and $\Lambda_{\mu_{0_l}}(X_l)=1$. By Riesz-Markov theorem and utilising definition \ref{def:consistent} it follows that the family of measures $(\mu_{0_l})_{l\in\ml}$, that represents such functionals, is a consistent family. \\For a detailed proof see \cite{1b}.\\
From the discussion at the beginning of this section we know that, for a given family of consistent measures, there exists a unique measure on the projective limit. This is the desired measure $\mu_0$ on $\overline{\ma}$.

Summarising: we have shown that there is a one to one relation between probability measures, defined on projective limits, and a consistent family of probability measures on the corresponding projective family of sets. This correspondence was achieved through the Riesz-Markov theorem, which was applied to $\overline{X}$ and each $X_l$ being all conpact Hausdorff spaces.  Utilising this correspondence we were able to define a probability measure of the configuration space $\overline{\ma}$, which is identified with the projective limit of a projective family of sets. However, in this case, the index set is restricted to tame subgroupoids, therefore, we had to explicitly express the functions $f_l\in C(X_l)$ in terms of such subgroupoids.

Because of the existence of a pullback from each space $X_l$ to $G^{|E(\gamma)|}$ it was possible to define $f_l$ in terms of maps $F_l:G^{|E(\gamma)|}\rightarrow\Cl$. This has enabled us to define the positive linear functional $\Lambda_{\mu_{O_l}}$ required for the application of Riesz-Markov theorem in terms of the Haar measure on G, which insured that, for each subgroupoid $l$, $(\Lambda_{\mu_{0_l}} )_{l\in\ml}$ is a consistent family and $\Lambda_{\mu_{0_l}}(X_l)=1$. It follows that the family of measures $(\mu_{0_l})_{l\in\ml}$ that represents such functionals is a consistent family. Such a family induces the unique probability measure $\mu_0$ on $\overline{\ma}$.

It is now possible to equip the quantum configuration space with a Hilbert space as follows:
\begin{definition}\label{def:a-l}
The Hilbert space $\mh_{kin}$ is defined to be the space of square integral functions over $\overline{\ma}$ with respect to the measure $\mu_0$, i.e.
\be
\mh_{kin}:=L^2(\overline{\ma},d\mu_0)
\ee
$\mh_{kin}$ is called the kinematical Hilbert space, $\mu_0$ is called the Ashtekar-Lewandowski measure and $\overline{\ma}$ the Ashtekar-Isham configuration space.
\end{definition}
The Ashtekar-Lewandowski measure $\mu_0$ has some interesting properties namely:\\
i) the support of the measure $\mu_0$ is on the non-smooth (distributional) connections. This entails that the set of smooth connections $\ma$ is contained in a measurable subset of $\overline{\ma}$ which has measure zero.\\
ii) $\mu_0$ is faithful\footnote{We say that a measure $\mu$ on a space $X$ is faithful iff for all $f\neq 0\in C(X)$, $\Lambda_{\mu}(|f|)\ge 0$. For the case of a projective limit $\overline{X}$, it is possible to show that a measure $\mu$ on it is faithful iff $\mu_l$ is faithfull for all $l\in\ml$. }. This is a consequence of the fact that, for each $l\in\ml$, $\mu_{0_l}$ are finite powers of the Haar measure which is faithful and, thus, are themselves faithful.\\
iii) $\mu_0$ is both gauge and diffeomorphic invariant (see Section \ref{s gns}).\\

\subsubsection{Orthonormal Basis for $\mh_{kin}$}
\begin{figure}[htb]
 \begin{center}
\psfrag{A}{$\gamma$}\psfrag{B}{$\gamma^{'}$}\psfrag{a}{$v_1$}\psfrag{b}{$v_4$}\psfrag{c}{$v_5$}\psfrag{d}{$v_6$}\psfrag{e}{$j_{e_1}$}\psfrag{f}{$j_{e_2}$}\psfrag{g}{$j_{e_3}$}\psfrag{h}{$j_{e_4}$}\psfrag{i}{$j_{e_5}$}\psfrag{j}{$\tilde{v}$}\psfrag{k}{$j_{e^{'}_2}$}\psfrag{l}{$j_{e^{''}_2}$}\psfrag{m}{$j_{e^{''}_3}$}\psfrag{n}{$j_{e^{'}_3}$}\psfrag{o}{$j_{e^{''}_4}$}\psfrag{p}{$j_{e^{'}_4}$}\psfrag{q}{$j_{e^{''}_1}$}\psfrag{r}{$j_{e^{'}_1}$}\psfrag{t}{$j_{e^{''}_5}$}\psfrag{s}{$j_{e^{'}_5}$}
 \includegraphics[scale=0.5]{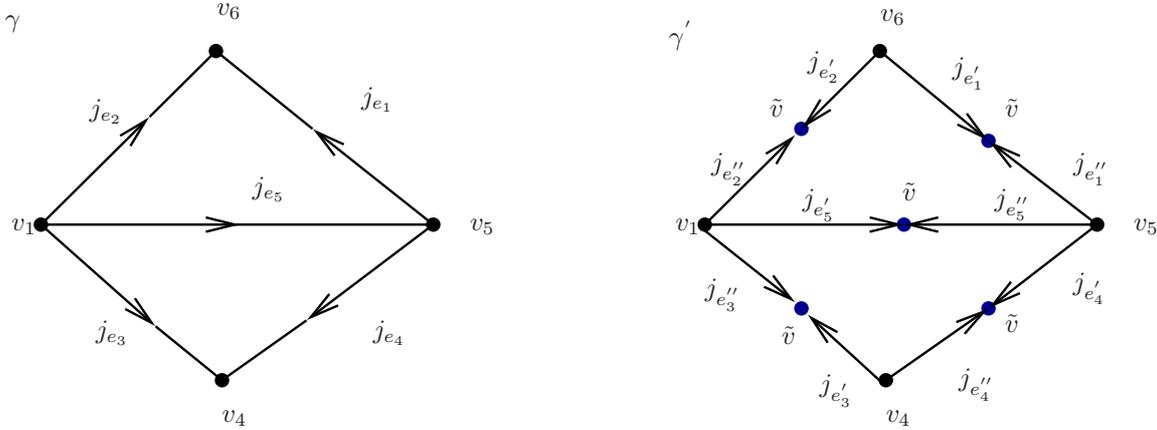}
 \caption{Introducing virtual vertices $\tilde{v}$ \label{fig:virtual} }
\end{center}
 \end{figure}

In this section we will introduce the notion of a \emph{spin network function} (SNF) which, as we will show, provides an orthonormal basis for $\mh_{kin}$. Since we are in the context of LQG, we will define such \emph{spin network function} over SU(2) although, in principle, they can be defined over any compact Lie group G.
\begin{definition}
Given a set of irreducible representations $\Pi=\{\pi_j|j=\frac{n}{2}\text{ with }n>0\in\Nl\}$ of SU(2) and a subgrupoid $l=l(\gamma)$, it is possible to associate to each edge $e\in E(\gamma)$ a non-trivial irreducible representations $\pi_e\in \Pi$. The set of all such assigned representations is denoted by $\vec{\pi}=(\pi_e)_{e\in E(\gamma)}$. A gauge variant \emph{spin network function} is then defined as follows
\ba\label{ali:snfvariant}
T_{\gamma, \vec{\pi},\vec{m},\vec{n}}:\overline{\ma}&\rightarrow &\Cl\nonumber\\
A&\mapsto&\prod_{e\in E(\gamma)}\sqrt{2j_e+1}[\pi_e(A(e))]_{m_en_e}
\ea
where $\vec{m} := \{m_e\}_{e\in E(\gamma)}$ , $\vec{n}: = \{n_e \}_{e\in E(\gamma)}$ with $m_e , n_e = 1, .., \sqrt{2j_e+1}$ label
the matrix elements of the representation.

\end{definition}
However, we would like to construct gauge \emph{invariant} spin networks functions. This can be done by introducing the concept of an intertwiner
\begin{definition}
Given two vector spaces $V_1$ and $V_2$ such that we have two linear representations on them, $R_1$ and $R_2$, respectively, then an \textbf{Intertwiner Operator I} is defined as a linear map $I:V_1\rightarrow V_2$ such that the following diagram commutes:
\[\xymatrix{
V_1\ar[rr]^{I}\ar[dd]_{R_1(g)}&&V_2\ar[dd]^{R_2(g)}\\
&&\\
V_1\ar[rr]_{I}&&V_2\\
}\]
for all $g\in G$
\end{definition}
The gauge \emph{invariant} spin networks functions are then defined to be:
\begin{definition}\label{def:invsnf}
Gauge-invariant SNF are obtained by restricting the gauge variant ones to intertwiners which project on the trivial representation, thus obtaining
\be\label{equ:snfinvariant}
T_{\gamma,\vec{\pi},\vec{I}|_{I_v\in I_{v(\vec{\pi},\pi_v^t)}}}:\overline{\ma}\rightarrow \Cl
\ee
\end{definition}
To understand why the introduction of intertwiners has enabled us to render \ref{ali:snfvariant} invariant we need to analyse, in detail, the result of applying 
a gauge-variant SNF to a generic graph $\gamma$. The first step is to render the graph in its \emph{standard form}, such that the edges, at each vertex, are outgoing. This can be achieved by splitting the edges in two and introducing a virtual vertex $\tilde{v}$ at the splitting point as depicted in figure \ref{fig:virtual}. 
Such virtual vertices will always have ingoing edges incident at them.\\ Generally, given a graph $\gamma$, with $N$ edges, its standard form $\gamma^{'}$ will have $2N$ edges. The introduction of virtual vertices allows to write each edge $e\in\gamma$ as $e=e_i\circ e^{-1}_j $ such that, denoting the virtual vertex common to $e_i$ and $e_j$ by $\tilde{v}$, we have $b(e_i)=b(e)$, $b(e_j)=f(e)$ and $f(e_i)=f(e_j)=\tilde{v}$. The reason why it is possible to work directly with the standard form of a graph rather than the graph itself, is because the gauge transformation of the holonomies coincides in both cases, i.e. $A^g(e_i)(A^g(e_j))^{-1}=A^g(e)$ for all $g\in G$. Moreover, the introduction of virtual vertices does not alter the representations associated to the original edges, i.e. $\pi_e=\pi_{e_i}=\pi_{e_j}$. From this discussion it follows that equation \ref{ali:snfvariant} can be written as follows:
\ba
T_{\gamma, \vec{\pi},\vec{m},\vec{n}}(A)&=&\prod_{e\in E(\gamma)}\sqrt{2j_e+1}[\pi_e(A(e))]_{m_en_e}\nonumber\\
&=&\prod_{e\in E(\gamma)}\sqrt{2j_e+1}[\pi_e(A(e_i))]_{m_el_e}[\pi_e(A(e_j)^{-1}]_{l_en_e}
\ea
By applying a gauge transformation, it can be shown that $T_{\gamma, \vec{\pi},\vec{m},\vec{n}}$ transforms trivially at the virtual vertices $\tilde{v}_i$. We are now interested in the behaviour of the original vertices $v\in V(\gamma)$ under a gauge transformation, i.e. we want to analyse how variant SNF transform at these vertices under gauge transformation. To answer this question it is convenient to re-express $T_{\gamma, \vec{\pi},\vec{m},\vec{n}}(A)$ as
\be
T_{\gamma, \vec{\pi},\vec{m},\vec{n}}(A)=\prod_{e^{'}\in E(\gamma^{'})}[\pi_{e^{'}}(A(e^{'}))]_{m_{e^{'}}n_{e^{'}}}=\prod_{v\in V(\gamma^{'})}\prod_{e^{'}\in E^b_v(\gamma^{'})}[\pi_{e^{'}}(A(e^{'}))]_{m_{e^{'}}n_{e^{'}}}
\ee
where $E^b_v(\gamma^{'})$ is the set of outgoing edges at vertex $v$\footnote{We recall at this point that for graphs in standard form the non-virtual vertices have only outgoing edges incident at them.}.\\ Note that we have omitted the trivial vertices and any other factor which, likewise, transforms trivially. Under a gauge transformation we then obtain
\be\label{equ:rep}
T_{\gamma, \vec{\pi},\vec{m},\vec{n}}(A^g)=\prod_{v\in V(\gamma^{'})}\prod_{e^{'}\in E^b_v(\gamma^{'})}\Bigg[\otimes_{v\in V(\gamma^{'})}\Bigg(\otimes_{e^{'}\in E^b_v(\gamma^{'})}\pi_{e^{'}}(g(v))\Bigg)\Bigg(\otimes_{e^{'}\in E^b_v(\gamma^{'})}\pi_{e^{'}}(A(e^{'}))\Bigg)\Bigg]_{m_{e^{'}}n_{e^{'}}}
\ee
However, since the group we are considering is SU(2) (compact group), it follows that any representation can be decomposed into a sum of irreducible representations, i.e. every representation is completely reducible.
\be\label{equ:decomposition}
\otimes_{e^{'}\in E^b_v(\gamma^{'})}\pi_{e^{'}}(g(v))=\oplus_{k}\pi^{'}_k(g(v))
\ee
In this context an intertwiner $I_v^{(\pi^{'})}$ is an element of the set $\mi(\vec{\pi}, \pi^{'}_v)$ of all representations occurring in \ref{equ:decomposition}, that are equivalent to the irreducible representation $\pi^{'}$, where $\pi^{'}$ is an element in the collection of fixed representatives for each equivalence class of irreducible representations of SU(2). By choosing a particular intertwiner at each vertex and collecting such chosen intertwiners, we can form a vector $\vec{I}:=\{I^{(\pi^{'})}_v\}_{v\in V(\gamma)}$. 

As explicitly shown from equation \ref{equ:rep}, for each vertex $v\in V(\gamma)$, $T_{\gamma, \vec{\pi},\vec{m},\vec{n}}(A)$ transforms in the tensor product representation, which can be projected into the representation associated to $I^{(\pi^{'})}_v$ by contracting it with the corresponding intertwiner. The resulting function $T_{\gamma,\pi,\vec{I}}$ is still a cylindrical function over $\overline{\ma}$ which, reintroducing the virtual vertices, can be written as 
\be
T_{\gamma,\pi,\vec{I}}(A):=\prod_{v\in V(\gamma)}\Bigg[I^{(\pi^{'}_{v})}_{v}\Bigg]_{\{m_{e^{'}}\}_{v=b(e^{'})}}\prod_{\tilde{v}\in V(\gamma^{'})}\Bigg[I^{(\pi^{'}_{\tilde{v}})}_{\tilde{v}}\Bigg]_{\{n_{e^{'}}\}_{\tilde{v}=f(e^{'})}}\Bigg[\pi_{e^{'}}(A(e^{'}))\Bigg]_{m_{e^{'}}n_{e^{'}}}
\ee
where $(I_{\tilde{v}}^{(\pi^{'}_{\tilde{v})}})_{\{n_{e^{'}}\}_{\tilde{v}=f(e^{'}}}$ are the intertwiners associated to the virtual vertices.

By varying both the intertwiners $I^{(\pi^{'})}_v$ and the representations $\pi^{'}_v$, the functions $T_{\gamma,\pi,\vec{I}}$ span exactly the same space as do $T_{\gamma, \vec{\pi},\vec{m},\vec{n}}$, thus nothing is lost when passing from one set of functions to the other. Definition \ref{def:invsnf} is then obtained by restricting the representations $\pi_v^{'}$ onto which the intertwiners $I^{(\pi^{'})}_v$   project to, to the trivial representation.
\\

The importance of both the gauge-variant spin network functions and the gauge-invariant spin network functions, lies in the following theorem:
\begin{theorem}\mbox{}\begin{enumerate}
\item [i)]
The gauge variant spin-network states provide an orthonormal basis for the Hilbert space $L_2 (\overline{\ma}, d\mu_0 )$ (provided we restrict to non-trivial representations).
\item [ii)]
The gauge invariant spin-network states provide an orthonormal basis for the Hilbert space $L_2 (\overline{\ma}/G, d\mu_0 )$, i.e. the Hilbert space in which the Gauss constraint has been solved.
\end{enumerate}

\end{theorem}
A proof of this theorem can be found in \cite{1b}.

\begin{figure}[htb]
 \begin{center}
\psfrag{A}{$J_1$}\psfrag{B}{$J_2$}\psfrag{C}{$J_3$}\psfrag{D}{$J_4$}\psfrag{E}{$J_5$}\psfrag{F}{$J_6$}\psfrag{G}{$J_7$}\psfrag{H}{$J_8$}\psfrag{I}{$J_9$}\psfrag{J}{$J_{10}$}\psfrag{K}{$J_{11}$}\psfrag{L}{$J_{12}$}\psfrag{M}{$J_{13}$}\psfrag{N}{$J_{14}$}
 \includegraphics[scale=0.7]{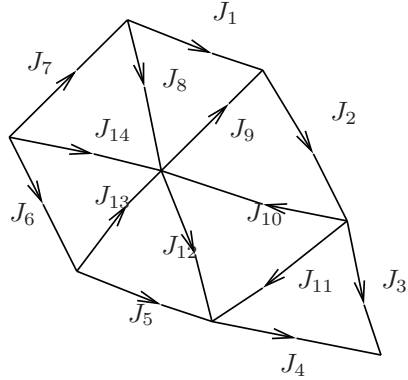}
 \caption{A SNF on a graph $\gamma$ defines a labelling of edges of the graph by spin $j_i$ and magnetic $m_i$, $n_i$ quantum numbers. For representational simplicity we have denoted $J_i=(j_i, m_i, n_i)$.\label{fig:snf} }
\end{center}
  \end{figure}

A useful way of conceptualising SNF is as a quantum state of space, i.e. it describes quantised three geometry (figure \ref{fig:snf}) 
\mbox{}\\[5pt]
\subsection{GNS Construction}
\label{s gns}
In this section we will first describe, in general, the process of Gel'fand-Nemark-Segal (GNS) construction and, then, show how the representation $L_2(\overline{A}, d\mu)$ can be obtained as a unique GNS representation.\\

In order to understand the main theorem regarding GNS construction we, fist of all, need to define the notion of a state and of a representation of an algebra.
\begin{definition}
Given a *-algebra $\mau$, a state $w$ of such an algebra is defined as a positive linear functional $w:\mau\rightarrow \Cl$, i.e. $w(a^* a)\geq 0$ for all $a\in\mau$. When $\mau$ is unital, then $w(1_{\mau})=1$.
\end{definition}
\begin{definition}\label{def:repalgebra}
\mbox{}\\
i) Given a *-algebra $\mau$, a representation consists of a pair $(\mh,\pi)$ where $\pi:\mau\rightarrow \ml(\mh)$ is a morphisms into the linear algebra of operators in $\mh$, with common and invariant dense domain.
\be
\pi(za+z^{'}a^{'})=z\pi(a)+z^{'}\pi(a^{'}),\hspace{.5in}\pi(ab)=\pi(a)\pi(b),\hspace{.5in}\pi(a^*)=(\pi(a))^{\dagger}
\ee
ii) If $ker(\pi)=\{0\}$, then the representation is said to be faithful.\\
iii) If $\pi(a)\psi=0$ for all $a\in\mau$ $\Rightarrow$ $\psi=0$, then $\pi$ is non-degenerate\\
iv) An element $\Omega\in \mh$ is called a cyclic vector if the set of states $\{\pi(a)\Omega:a\in \mau\}$ is a common dense and invariant domain in $\mh$. In this case $\pi$ is called a cyclic representation.\\
v) A representation is irreducible if every vector in a common dense and invariant domain is cyclic.
\end{definition}
The GNS construction is based on the following theorem:
\begin{theorem}\label{the:gns}
(GNS construction)\\
Given a state $w$ of a unital *-algebra $\mau$, the GNS data $(\mh_w,\pi_w,\Omega_w)$ consists of a Hilbert space $\mh_w$, a cyclic representation $(\mh_w,\pi_w)$ of $\mau$ and a normed cyclic vector $\Omega_w\in \mh_w$ (called the vacuum vector), such that 
\be\label{equ:gns}
w(a)=\langle\Omega_w,\pi(a)\Omega_w\rangle_{\mh_w}
\ee
The GNS data are uniquely (up to unitary equivalence) determined by equation \ref{equ:gns}
\end{theorem}
In order to fully understand this theorem and how it is derived, we first need to show that the Hilbert space $\mh_w$ is constructed as the Cauchy completion of an equivalence class of vectors. The equivalence relation is given in terms of a left ideal $\mi_w$.
\begin{lemma}
Given a positive linear functional $w:\mau\rightarrow \Cl$ on a *-algebra $\mau$, the set 
\be
\mi_w=\{a\in\mau| w(a^*,a)=0\}
\ee
is a left ideal on $\mau$, such that $a\in \mi_w$ iff $w(a^*,b)=0$ for all $b\in \mau$
\end{lemma}
A proof can be found in \cite{1b} and references therein.\\
We can now construct the quotient space $\tilde{\mh}_w:=\mau/\mi_w$ with elements $\psi_a\in\tilde{\mh}_w$ represented by the equivalence class $\psi_a:=[a]=\{a+b|b\in\mi_w\}$. The inner product in $\tilde{\mh}_w$ is defined as $\langle\psi_a,\psi_b\rangle_w:=w(a^*,b)$. This is well defined since:\\
i) it is \emph{Sesquilinear}: for $\alpha,\beta\in\Cl$
\ba
\langle\alpha(\psi_{a_1}+\psi_{a_2}),\beta(\psi_{b_1}+\psi_{b_2})\rangle&=&w((\alpha(a_1+a_2))^*,\beta(b_1+b_2))\\
&=&\overline{\alpha}\beta w(a^*_1b_1)+\overline{\alpha}\beta w(a_1^*b_2)+\overline{\alpha}\beta w(a_2^*b_1)+\overline{\alpha}\beta w(a_2^*b_2)\nonumber\\
&=&\overline{\alpha}\beta\langle\psi_{a_1},\psi_{b_1}\rangle_w+\overline{\alpha}\beta\langle\psi_{a_1},\psi_{b_2}\rangle_w+\overline{\alpha}\beta\langle\psi_{a_2},\psi_{b_1}\rangle_w+\overline{\alpha}\beta\langle\psi_{a_2},\psi_{b_2}\rangle_w\nonumber
\ea
ii) It is \emph{positive semi-definite} from the properties of $w$
\be
\langle\psi_a,\psi_a\rangle_w:=w(a^*,a)\geq 0
\ee
iii) 
\be
\langle\psi_a,\psi_a\rangle_w:=w(a^*,a)=0\hspace{.2in}\Leftrightarrow\hspace{.2in}a\in\mi_w\hspace{.2in}\Rightarrow \psi_a=0
\ee
The Cauchy completion $\mh_w$ of $\tilde{\mh}_w$ gives us the representation $(\mh_w,\pi_w)$ of $\mau$ given by 
\be
\pi_w(a)\psi_b:=\psi_{ab}
\ee
Since $\mi_w$ is a left ideal, for any representative of an equivalence class the following holds:
\be 
\psi_a+\psi_b:=\psi_{a+b}\hspace{.2in}\psi_a\psi_b:=\psi_{ab}\hspace{.2in}z\psi_a:=\psi_{za}
\ee
We know that if $\mau$ is unital then there exists a cyclic vector $\Omega_w\in\mh_w$ which we identify with $\Omega_w:=\psi_1$, such that any other element in $\mh_w$ can be derived from $\psi_1$ 
\be
\psi_a=\psi_{a1}=\pi_w(a)\psi_1=\pi_a\Omega_w
\ee
It follows that $(\mh_w,\pi_w, \Omega_w)$ is a cyclic representation of $\mau$. Thus
\be
w(a)=\langle\Omega_w,\pi_w(a)\Omega_w\rangle_w
\ee
This is precisely the GNS construction of the theorem \ref{the:gns}.\\

Now that we have given the general outline of what a GNS construction is, we can then show that the representation obtained in the previous section is a \emph{unique} GNS representation.\\
In particular the representations allowed from the general GNS construction of theorem \ref{the:gns} are many and nonequivalent.\\ 
In order to select representations compatible with the requirement of LQG, additional assumptions (mostly coming from physics reasoning) are required. These assumptions are:\\
i) irreducibility of the representation.\\
ii) The states derived from the representations have to be invariant under the algebraic analogue of the symmetries present in the classical theory, which are: semianalytic diffeomorphisms on the spacial manifold (see definition \ref{defsemdif}) and SU(2) gauge transformations.

We will first consider the first requirement. To this end we recall that in LQG the quantum algebra is a unital *-algebra, which contains invertible elements and $\pi(1)=id_{\mh}$. It follows that the representation is non degenerate, therefore we can apply theorem \ref{the:irrep}, together with definition \ref{def:repalgebra} to show that we have an irreducible representation.
\begin{theorem}\label{the:irrep}
Non degenerate representations of the generators of a *-algebra by bounded operators are a direct sum of cyclic representations. 
\end{theorem} 
We now turn to requirement ii) above. To this end we need the definition of a semianalytic diffeomorphism.
\begin{definition}\label{defsemdif}
Given the group $H(\sigma)$ of all homeomorphisms of a spatial manifold $\sigma$, the semianalytic diffeomorphisms $Diff^w_{sa}(\sigma)$ is a subgroup of $H(\sigma)$ which preserves the set of all semianalytic edges and semianalytic faces.
\end{definition}
Recalling that the configuration space in LQG is defined to be the space of connections $\ma$ defined on a principal G-bundle (P, $\pi$, $\sigma$) for a compact group G, the cotangent bundle $T^*\ma$ equipped with a simplectic structure becomes the phase space. \\
In this setting an automorphisms of the principal G-bundle is defined as follows:
\begin{definition}\label{def:semidif}
An automorphisms of a principal G-bundle (P, $\pi$, $\sigma$) is a pair of mas $F:P\rightarrow P$ and $f:\sigma\rightarrow \sigma$, such that 
\be
F\circ\pi=f\circ\pi
\ee 
i.e. F maps fibers to fibers and
\be
\forall g\in G,\hspace{.2in}\forall p\in P\hspace{.2in} f(p\cdot g) = f (p) \cdot g .
\ee
Aut(P) := $\mathfrak{Q}$ is called the automorphism group of P .
\end{definition}
In the case in which we restrict $f$ to semianalytic diffeomorphisms on $\sigma$, then we have a semianalytic bundle automorphisms.\\
Given a local trivialisation of the G-bundle (P, $\pi$, $\sigma$), $\mathfrak{Q}$ can be written as the semidirect product $\mathfrak{Q}=\g\ltimes Diff(\sigma)_{sa}^w $ of the gauge group and the diffeomorphism group, such that $\forall\phi^{'}\in \mathfrak{Q}$, $\phi^{'}=(\lambda_g,\phi)$, where $\phi\in Diff(\sigma)^w_{sa}$ and $\lambda_g\in \g$  \\
The group $\mathfrak{Q}$ has a natural action on the basic variables $A(e)$ and $E_n(S)$ of LQG. Since canonical transformations preserve the Poisson brackets between such variables, we obtain  a $\mathfrak{Q}$ action on the algebra $\mb$ through automorphisms of the algebra.
\begin{definition}
An automorphism of a  *-algebra $\mb$ is an isomorphism of $\mb$ which is compatible with the algebraic structure. \\
Given a group G, then G is said represented on the algebra $\mb$ by the following group automorphisms:
\be
\alpha:G\rightarrow Aut(\mb)\mbox{;  } g\mapsto \alpha_g\hspace{.25in}\text{iff}\hspace{.25in}\alpha_{g_1}\circ\alpha_{g_2}=\alpha_{g_1g_2}
\ee
\end{definition}
The action of the group $\mathfrak{Q}$ on $\mb$ is then defined through the automorphisms $\alpha_{\phi^{'}}=(\alpha_g,\alpha_{\phi})$ as follows:
\ba
\alpha_g((f,\tilde{E}_n(S))&:=&(p^*_lf_l(\{g(b(e))A(e)(g(f(e)))^{-1}\}_{e\in E(\gamma)}),\tilde{E}_{Ad_{g^{-1}(n)}}(S))\nonumber\\
\alpha_{\phi}((f,\tilde{E}_n(S))&:=&(p^*_lf_l(\{A(\phi(e))\}_{e\in E(\gamma)}), \tilde{E}_{\phi^{-1}(n)}(\phi(S)))
\ea
where $\tilde{E}_n(S)$ is the flux vector field.\\ 

The action of $\mathfrak{Q}$ can be lifted to the quantum algebra $\mau$ as follows:\\
a) the action can be extended to smooth cylindrical functions on $\overline{\ma}$, i.e  $\ma\rightarrow \overline{\ma}$;\\
b) the semianalytic gauge transformations can be generalised to arbitrary discontinuous ones $\g\rightarrow \overline{\g}:=Fun(\sigma, G)$. \\
By the above procedures $\mathfrak{Q}$ becomes a bundle automorphisms of $\mau$.\\  Requirement ii) (from the previous page), for a correct representation, implies that the states on $\mau$ be invariant with respect to the automorphisms group $\mathfrak{Q}$.
\begin{definition}
A state $w:\mau\rightarrow \Cl$ of an algebra $\mau$ is invariant with respect to an automorphisms $\alpha$ if $w\circ \alpha=w$. Given a group G, w is invariant for G if it is invariant for all $\alpha_g$, $g\in G$.
\end{definition}
Moroever, we also require that the automorphisms be unitary implemented, i.e. we would like a unitary representation of $\mathfrak{Q}$ on $\mh_w$. To this end we consider the following theorem and corollary
\begin{theorem}\label{thestateoperator}
Given a state $w$ of a unital *- algebra $\mau$ that is invariant under an automorphisms $\alpha\in Aut(\mau)$, then there exists a unique unitary operator $\hat{U}_w$ on the GNS Hilbert space $\mh_w$, such that
\be
\hat{U}_w\pi_w(a)\Omega_w=\pi(\alpha(a))\Omega_w
\ee
\end{theorem}
It follows that, if the states $w$ is G invariant then, for each $g\in G$, we would obtain a unitary operator acting on $\mh_w$, i.e. $\hat{U}_w(g)\pi_w(a)\Omega_w=\pi(\alpha_g(a))\Omega_w$ for all $g\in G$. In this way we would obtain a unitary representation of G on $\mh_w$. This is precisely the content of the next corollary.
\begin{corollary}
Given a unitary *- algebra $\mau$ and a G-invariant state $w$, then there exists a unitary representation $g\mapsto U_w(g)$ of $G$ on $\mh_w$, such that
\be
U_w(g)\pi_w(a)\Omega_w=\pi(\alpha_g(a))\Omega_w
\ee
\end{corollary}
In the case $G$ is the symmetry group $\mathfrak{Q}$, we obtain a unitary representation of the classical symmetries as required.\\

However, the requirements of irreducibility and of a unitary implementation of the classical symmetry group are not sufficient to single out a unique GNS representation. A third requirement is necessary, namely, we require that the representation of cylindrical functions be discontinuous, while the representation of the electric fluxes be smooth.\\
With the introduction of this third requirement, in \cite{7}, it was shown that it is possible to single out a unique representation of the quantum algebra $\mau$. In particular, the following Lewandowski-Okolow-Sahlmann-Thiemann (LOST)-theorem was proved:
\begin{theorem} (LOST Theorem)
There exists a unique semi-weakly smooth $\mathfrak{Q}$-invariant state $w$ on $\mau$. Moreover, the corresponding cyclic GNS construction is irreducible.
\end{theorem}
For an explicit proof and related discussion see \cite{7}.\\

Of particular importance is the fact that the state $\Lambda_{\mu_0}$ on $\mau$ is invariant under the symmetry group $\mathfrak{Q}$. To understand why this is the case, let us recall the action on $\overline{\ma}$ of the gauge group $\overline{\g}$, and the semianalitic-diffeomorphism group $Diff_{sa}^w(\sigma)$ are respectively:
\ba
\lambda_g&:&\overline{\g}\times\overline\ma\rightarrow\overline\ma ;\hspace{.25in} x\mapsto\lambda_g(x)\mbox{ where }\hspace{.15in}[\lambda_g(x)]p:=g(b(p))x(p)g(f(p))^{-1}\mbox{ }\forall p\in \mp\nonumber\\
\delta_g&:&Diff_{sa}^w(\sigma)\times \overline{\ma}\rightarrow \overline{\ma};\hspace{.25in}x\mapsto\delta_g(x):=(x_{\phi(l)})_{l\in\ml}\mbox{ where }\hspace{.15in}\phi(l):=l(\phi(\gamma))
\ea
It can be shown that both group actions are invariant with respect to the projective structure on $\overline{\ma}$, i.e.
\ba
p_{l^{'}l}(\lambda_g^{l^{'}}(x_{l^{'}}))&=&\lambda^l_g(p_{l^{'}l}x_{l^{'}})=\lambda_g^l(x_l)\nonumber\\
p_{\phi(l)\phi(l^{'})}(\delta^{l^{'}}_{\phi}(x_{l^{'}}))&=&\delta_{\phi}^l(p_{l^{'}l}x_{l^{'}})=\delta_{\phi}^l(x_l)
\ea
We now want to show that $(\lambda_g)_*\Lambda_{\mu_0}=\Lambda_{\mu_0}$ and $(\delta_{\phi})_*\Lambda_{\mu_0}=\Lambda_{\mu_0}$. Specifically, for any $f=p^*_l f_l\in C(\overline{\ma})$ and $F_l\in C(G^{|E(\gamma)|})$ we obtain
\ba
\Lambda_{\mu_0}(\lambda_g^* f)&=&\Lambda_{\mu_0}(\lambda^*_gp^*_lf_l)=(p_l)_*\Lambda_{\mu_0}((\lambda_g^l)^*f_l)=\Lambda_{\mu_{0_l}}((\lambda_g^l)^*f_l)\nonumber\\
&=&\int_{G^{|E(\gamma)|}}\Big[\prod_{e\in E(\gamma)}d\mu_H(h_e)\Big]F_l\big(\{g(b(e))h_eg(f(e))^{-1}\}_{e\in E(\gamma)}\big)\nonumber\\
&=&\int_{G^{|E(\gamma)|}}\Big[\prod_{e\in E(\gamma)}d\mu_H(g(b(e))^{-1}h^{'}_eg(f(e)))\Big]F_l\big(\{h^{'}_e\}_{e\in E(\gamma)}\big)\nonumber\\
&=&\int_{G^{|E(\gamma)|}}\Big[\prod_{e\in E(\gamma)}d\mu_H(h^{'}_e)\Big]F_l\big(\{h^{'}_e\}_{e\in E(\gamma)}\big)\nonumber\\
&=&\Lambda_{\mu_{0_l}}(f_l)=\Lambda_{\mu_0}(f)
\ea
where we have performed the change of variables $h^{'}_e\rightarrow g(b(e))h_eg(f(e))^{-1}$ and used the invariance of the Haar measure.\\
Similarly we obtain
\ba
\Lambda_{mu_0}(\delta^*_{\phi}f)&=&\Lambda_{\mu_0}(\delta^*_{\phi}p^*_lf_l)=\Lambda_{\mu_0}(p^*_{\phi(l)}(\delta^l_{\phi})^*f_l=\Lambda_{\mu_{0_{\phi^{-1}(l)}}}(\delta^l_{\phi})^*f_l\nonumber\\
&=&\int_{G^{|E(\gamma)|}}\Big[\prod_{e\in E(\phi(\gamma))}d\mu_H(h_e)\Big]F_l\big(\{h_e\}_{e\in E(\phi(\gamma))}\big)\nonumber\\
&=&\int_{G^{|E(\gamma)|}}\Big[\prod_{e\in E(\gamma)}d\mu_H(h_e)\Big]F_l\big(\{h_{e}\}_{e\in E(\gamma)}\big)\nonumber\\
&=&\Lambda_{\mu_{0_l}}(f_l)=\Lambda_{\mu_0}(f)
\ea
where we have relabelled the holonomies $h_{\phi(e)}$ by $h_e$.\\
The invariance of $\Lambda_{\mu_0}$ with respect to $\overline{\g}$ and $Diff_{sa}^w(\sigma)$ implies that the associated measure $\mu_0$ is invariant under these symmetries. 
\subsection{Solving the Constraints}
\label{ssolvingconst}
In this section we will discuss the solution of the constraints present in LQG, namely:\\
\textbf{Gauss constraint}, \textbf{Diffeomorphisms constraint} and \textbf{Hamiltonian constraint}.
\subsubsection{Gauss Constraint}
At the classical level the Gauss constraint is given by 
\be
G(\Lambda):=\int_{\sigma}d^3 x\Lambda^j(x)(D_aE^a_j)(x)=-\int_{\sigma}d^3x(D_a\Lambda^j)(x)E^a_j(x)
\ee
where $D_a$ is the covariant derivative and $\Lambda^i$ is an $su(2)$ valued function on $\sigma$ (smeared field). Similarly, as it was done for the electric flux, such an expression for the Gauss constraint needs to be regularised, such that we obtain a family of vector fields $G_l(\Lambda)\in V^{\infty}(X_l)$.\\
We then extend the action of the vector field to $\overline{\ma}$ and obtain a well defined self-adjoint operator with dense domain in $C^1(\overline{\ma})$, as follows:
\be
\hat{G}(\Lambda)f:=\hat{G}_l(\Lambda)f_l=\frac{i\beta l_p^2}{2}\sum_{v\in V(\gamma)}\Lambda_j(v)\Big[\sum_{e\in E(\gamma);v=b(e)}R^j_e-\sum_{e\in E(\gamma);v=f(e)}L^j_e\Big]f_l
\ee
where $R^j_e$ and $L^j_e$ are the right and left vector fields on $G$, respectively.
Each operator $\hat{G}_l(\Lambda)$ is an infinitesimal generator of $SU(2)$ gauge transformations. It follows that finite gauge transformations are generated by the one-parameter unitary group generated by $\hat{G}_l(\Lambda)$. 

Utilising the fact that gauge-invariant SNF form an orthonormal basis for $\mh^{\g}_{kin}$\footnote{ $\mh^{\g}_{kin}$ indicates the space of solution to the Gauss constraint but not the Diffeomorphic or Hamiltonian constraint.}, it is possible to express the space of solutions of $\hat{G}(\Lambda)$ as follows
\be
\mh^{\g}_{kin}=\oplus_{\gamma,\vec{\pi}}\mh^{\gamma}_{\gamma,\pi,\vec{I}\in\mi_v(\vec{\pi},\pi_v^t)}\subset\mh_{kin}
\ee
For a detailed derivation the reader is referred to \cite{1b} and reference therein.
\subsubsection{Diffeomorphic Constraint}
We now turn our attention to the Diffeomorphisms constraint which at the classical level is 
\be
H_a(\vec{N})=-2s\int_{\sigma}d^3x\epsilon_{ijk}N^aF^J_{ab}(x)E^b_j(x)
\ee
where $s=\pm 1$ depending if the signature is Euclidean or Lorentzian, respectively.\\
In order to promote such a constraint to an operator we recall that the state $\Lambda_{\mu_0}$ is invariant under the action of $Diff_{sa}^w(\sigma)$. It follows, from theorem \ref{thestateoperator} that, for each $\phi\in Diff_{sa}^w(\sigma)$ we can associate a unitary operator, whose action of the SNF is as follows:
\be
\hat{U}(\phi)T_s:=T_{\phi(s)}\hspace{.3in}\forall \phi\in Diff_{sa}^w(\sigma)
\ee 
where $\phi(s):=(\phi(e), \pi_e, m_e, n_e)$.

It turns out that the action of $\hat{U}(\phi)$ is not weakly continuous\footnote{It should be noted, at this point, that the discontinuity of the diffeomorphic action is deeply rooted in the distributional character of $\overline{\ma}=Hom(\mp, G)$. In fact, if two paths in $\mp$ differ slightly, a distributional connection can assign completely independent values to them. Notice that the distance of any two points is a gauge-variant quantity. In fact, since we are dealing with a diffeomorphism-invariant theory, any two points can be taken as far apart or as close together as one finds fit. This can  be done by applying any diffeomorphisms and measuring the distance with respect to any metric.}, therefore, from Stones's theorem the Lie algebra of $Diff_{sa}^w(\sigma)$ can not be defined on $\mh_w$. However, the constraint equation $\hat{C}(\vec{N})\psi=0$ is equivalent to $\hat{U}(\phi)\psi=\psi$ therefore, although the representation of $\mau$ seems not to support the constraints as operators on $\mh_{kin}$, it is nonetheless still a well suited representation, since it supports the equivalent constraint equation $\hat{U}(\phi)\psi=\psi$.\\
We can now safely try to find solutions to the diffeomorphism constraint. This can be done through the process of RAQ described in section \ref{s outline}. The first step is to find an algebraic distribution  $l\in D^*$ (where $D=C^{\infty}(\overline{\ma})$), such that the following equation holds (analogue of equation \ref{equ:uconst})
\be
l(\hat{U}^{\dagger}(\phi) f)=l((\hat{U}(\phi))^{-1}f)=l(\hat{U}(\phi^{-1}f)=l(f)\hspace{.25in}\forall \phi\in Diff_{sa}^w(\sigma)\mbox{ }f\in D
\ee
However, since the SNF are dense in $D$, it is possible to write the above equation in terms of them as follows:
\be\label{equ:tcondition}
l(\hat{U}(\phi^{-1})T_s)=l(T_s)\hspace{.25in}\forall \phi\in Diff_{sa}^w(\sigma)\mbox{, }s\in S
\ee
where $S$ is the set of SNF labels $s$. \\
Any algebraic distribution ($l\in D^*$) is completely specified if it is defined pointwise in $D$, i.e. by the set of all its values $l(T_s)$, therefore we define all $l\in D^*$ as follows:
\be
 l=\sum_{s\in S}l_s\langle T_s,\cdot \rangle_{kin}
\ee
where $l(T_s)=:l_s$.\\
By inserting the above definition of $l$ in \ref{equ:tcondition}, the new condition on the algebraic distribution $l$ now becomes
\be
l_s=l_{\phi(s)}\hspace{.25in}\forall \phi\in Diff_{sa}^w(\sigma)\mbox{, }s\in S
\ee
Such a condition can be interpreted as an equivalence requirement, thus it is useful to introduce the following orbits
\be
[s]:=\{\phi(s)|\phi\in Diff_{sa}^w(\sigma)\}
\ee
The general solution for the Diffeomorphisms constraint is then given by 
\be\label{equ:nconst}
l=\sum_{[s]}c_{[s]}l_{[s]}=\sum_{[s]}c_{[s]}\sum_{s^{'}\in [s]}\langle T_{s^{'}},\cdot\rangle
\ee
where $c_{[s]}$ are complex coefficients which only depend on the orbits. The term
$l_{[s]}:=\sum_{s^{'}\in [s]}\langle T_{s^{'}},\cdot\rangle$ is the algebraic functional associated with an orbit, and it is such that
\be
l_{[s]}(T_{s}):=\sum_{s^{'}\in [s]}\langle T_{s^{'}},T_{s}\rangle=\sum_{s^{'}\in [s]}\delta_{s^{'}s}=\chi_{[s^{'}]}(s)
\ee
This implies that the sum in equation \ref{equ:nconst} is finite when $l$ is acting on a SNF $T_s$.\\

The last step in the process of RAQ is to find a rigging map $\eta:D\rightarrow D^*\subset \mh_{diff}$, in terms of which it is possible to define the inner product in $\mh_{diff}$ (the space of solutions of the diffeomorphisms constraint). Such a map is given by 
\be
\eta(T_s):=\eta_{[s]}l_{[s]}
\ee
where $\eta_{[s]}$ are some positive coefficients ($\eta_{[s]}>0$).\\ The rigging map $\eta$ allows us to map any element $f=\sum_s f_sT_s\in D$ to the element $\eta(f)=\sum_s f_s\eta(T_s)\in D^*$.\\
The inner product then becomes
\be\label{equ:innerprod}
\langle\eta(T_s),\eta(T_{s^{'}})\rangle_{Diff}:=\eta(T_{s^{'}})[T_s]=\eta_{[s^{'}]}\chi_{[s^{'}]}(s)
\ee
It can be shown that \ref{equ:innerprod} has the following properties:\\
i) \emph{linearity}: this follows from the linearity of $l_{[s]}$.\\
ii) \emph{Positive semi-definiteness}: this is a consequence of the fact that the coefficients $\eta_{[s]}$ are assumed to be real and positive.\\
iii) \emph{Hermicity}:
\be
\eta_{[s^{'}]}\chi_{[s^{'}]}(s)=\langle\eta(T_s), \eta(T_{s^{'}})\rangle_{diff}=\overline{\langle\eta(T_{s^{'}}), \eta(T_{s})\rangle}_{diff}=\overline{\eta(T_s)[T_{s^{'}}]}=\eta_{[s]}\chi_{[s]}(s^{'})
\ee
We now turn our attention to the Hamiltonian constraint.

\subsubsection{Hamiltonian Constraint}
The Hamiltonian constraint is central in the development of LQG as a quantum theory of gravity since it governs the dynamics of the theory and thus, if solved, would allow the possibility of making predictions which are central in testing the validity of a theory.\\
However, the Hamiltonian constraint is much more difficult to solve than the Diffeomorphic constraint and the Gauss constraint, for two reasons:
\begin{enumerate}
\item[1)] The Hamiltonian constraint is non-linear. This causes UV problems.
\item[2)] Due to the presence of structure functions, the algebra between spatial diffeomorphic constraints and Hamiltonian constraints is not a Lie algebra.
\end{enumerate}
There are two different attempts in implementing the Hamiltonian constraint which overcome the UV problem, however, only one of them will overcome the second problem. In the following, we will briefly describe both attempts\\[5pt]
\textbf{Regularised Hamiltonian Constraint}\\[5pt]
We recall that the classical Hamiltonian constraint is 
\ba
H(N)&=&\frac{1}{k}\int_{\sigma}d^3x N(x)\frac{1}{\sqrt{|det(E_j^a(x))|}}\Big[\epsilon_{ijk}F_{ab}^i(x)E_j^a(x)E_k^b(x)\Big]+\nonumber\\
& &+(1-s)\frac{(K^j_b(x)E^a_j(x))(K^j_a(x)E^b_j(x))-(K^j_c(x)E^c_j(x))^2}{\sqrt{|det(E_j^a(x))|}}
\ea
The term $\frac{1}{\sqrt{det(q)}}$ is problematic since it is non polynomial. However, it was shown in \cite{46d} that such a prefactor can be absorbed into a commutation relation of well defined operators. \\
This is done as follows: \\
the first step is to express the Hamiltonian constraint in terms of the \emph{Euclidean Hamiltonian constraint} $H_e$ and a non-Euclidean part $\tilde{H}$: 
\be
H(N)=H_e(N)+(1-s)\tilde{H}
\ee
By introducing the following classical identity 
\be
sign(det(e))\frac{E^a_jE^b_l\epsilon_{jkl}}{\sqrt{det(q)}}(x)=\epsilon^{abc}e^j_a(x)=\frac{4}{k}\epsilon^{abc}\{V(R),A^j_a(x)\}
\ee
where $V(R)$ is the classical volume of a region $R$, i.e. 
\be
V(R)=\int_Rd^3x\sqrt{det(q)}=\int_Rd^3x\sqrt{|det(E)|}
\ee 
we can express $H_e$ as
\be\label{equ:euclideanpart}
H_e(N)=-\frac{8}{k}\int_{\sigma}d^3x N(x)\epsilon^{abc}Tr(F_{ab}(x)\{A^j_a(x), V(R)\})
\ee
On the other hand, by introducing the integrated densitised trace of extrinsic curvature
\be
K:=\int_{\sigma}K^i_aE^a_i=\{H_e(1), V(\sigma)\}
\ee
which satisfies the following Poisson bracket
\be
K_a^j(x)=\frac{k}{2}\{K,A^j_a(x)\}
\ee
it is possible to express the remaining part of the Hamiltonian constraint as
\be\label{equ:otherterm}
\tilde{H}=-\frac{64}{k^4}\int_{\sigma}d^3x N(x)\epsilon^{abc}Tr\Big(\{A_a(x), K\}\{A_b(x), K\}\{A_c(x), V(R)\}\Big)
\ee
The aim of doing this is to be able to express the Hamiltonian constraint in terms of holonomies. This is desirable, since the Hilbert space $\mh_{kin}$ is defined in terms of generalised holonomy functions and $\hat{H}$ acts on $\mh_{kin}$.

\begin{figure}[htb]
\begin{center}
\psfrag{a}{$e_I$}\psfrag{b}{$e_J$}\psfrag{c}{$e_K$}\psfrag{d}{$\tilde{e}$}\psfrag{v}{$v$} \psfrag{e}{$\gamma$}    
\includegraphics[scale=0.5]{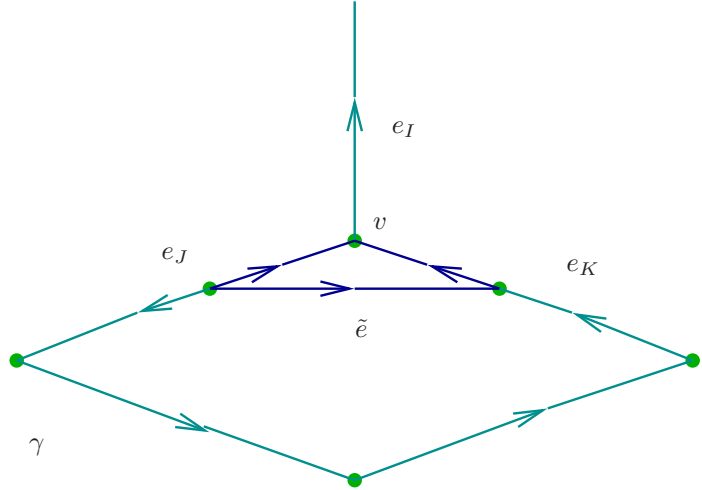}
\caption{Action of the Hamiltonian constraint $\hat{H}^{\epsilon}(N)$, where $\tilde{e}$ represents the added edge. \label{fig:loophamiltonian}}
\end{center}
 \end{figure}

In order to express both $H_e$ and $\tilde{H}$ in terms of holonomies we have to introduce a triangulation $T(\epsilon)$ of the manifold $\sigma$ in terms of tetrahedrons $\Delta$, whose volume is given by $\frac{\epsilon^3\sqrt{2}}{12}$.\\
In the following we will denote the three edges singling out a given tetrahedron as $e_I(\Delta)$, $e_J(\Delta)$, $e_K(\Delta)$ and $v(\Delta)$ denotes the common vertex. The orientation of $\Delta$ is given by the determinant of the tangents of the three edges defining $\Delta$. Moreover, we denote by $\alpha_{IJ}(\Delta)$ the loop joining the two edges $e_I(\Delta)$ and $e_J(\Delta)$, i.e. $\alpha_{IJ}(\Delta)=e_I(\Delta)\circ a_{IJ}\circ e_{J}(\Delta)^{-1}$ where $ a_{IJ}$ is the arch connecting the end point of $e_I(\Delta)$ and $e_J(\Delta)$.\\
In terms of the above triangulation, equations \ref{equ:euclideanpart} and \ref{equ:otherterm} become
\ba
H_e(N)&=&lim_{\epsilon\rightarrow 0}\frac{16}{2k^2}\sum_{\Delta\in T(\epsilon)}N(v(\Delta))\epsilon^{IJK}Tr\Big(A(\alpha_{IJ}(\Delta))A(e_K(\Delta))\{A^{-1}(e_K(\Delta)), V(R_{v(\Delta)})\}\Big)\nonumber\\
\tilde{H}(N)&=&im_{\epsilon\rightarrow 0}\frac{64}{3k}\sum_{\Delta\in T(\epsilon)}\epsilon^{IJK}N(v(\Delta))Tr\Big(A(e_I(\Delta))\{A^{-1}(e_I(\Delta)), K(\Delta)\}A(e_J(\Delta))\{A^{-1}(e_J(\Delta)), K(\Delta)\}\nonumber\\
& & A(e_K(\Delta))\{A^{-1}(e_K(\Delta)), V(R_{v(\Delta)})\}\Big)
\ea
Moreover, by writing $K=\{H_e(1), V(\sigma)\}$, both terms in the Hamiltonian are written solely in terms of the volume operator and the holonomy for which, well defined operators on $\mh_{kin}$ exist. We thus obtain the following operators corresponding to the  Hamiltonian constraint: 
\be
\hat{H}(N)=\hat{H}_e+(N)\hat{\tilde{H}}(N) 
\ee
where
\ba
\hat{H}^e(N)&=&lim_{\epsilon\rightarrow 0}\frac{16}{2i\hbar k^2}\sum_{\Delta\in T(\epsilon)}N(v(\Delta))\epsilon^{IJK}Tr\Big(\hat{A}(\alpha_{IJ}(\Delta))\hat{A}(e_K(\Delta))[\hat{A}^{-1}(e_K(\Delta)),\hat{V}(R_{v(\Delta)})]\Big)\nonumber\\
&=&lim_{\epsilon\rightarrow 0}\hat{H}_e^{\epsilon}(N)
\ea
and 
\ba
\hat{\tilde{H}}(N)&=&im_{\epsilon\rightarrow 0}\frac{64}{3i\hbar k}\sum_{\Delta\in T(\epsilon)}\epsilon^{IJK}N(v(\Delta))Tr\Big(\hat{A}(e_I(\Delta))[\hat{A}^{-1}(e_I(\Delta)),\hat{K}(\Delta)]\hat{A}(e_J(\Delta))[\hat{A}^{-1}(e_J(\Delta)),\hat{K}]\nonumber\\
& &\hat{A}(e_K(\Delta))[\hat{A}^{-1}(e_K(\Delta)),\hat{V}(R_{v(\Delta)})]\Big)
\nonumber\\
&=&lim_{\epsilon\rightarrow 0}\hat{\tilde{H}}^{\epsilon}(N)
\ea  
A detailed analysis and proof of the existence of the limit $\epsilon\rightarrow 0$ can be found in \cite{46d}.\\
With respect to a SNF $T_{\gamma,\vec{\pi},\vec{m},\vec{n}}$, the action of the regularised Hamiltonian is
\be
\hat{H}^{\epsilon}(N)T_{\gamma,\vec{\pi},\vec{m},\vec{n}}=\sum_{v\in V(\gamma)}N(v)\hat{H}^{\epsilon}_vT_{\gamma,\vec{\pi},\vec{m},\vec{n}}
\ee
It follows that regularised Hamiltonian only acts on the vertices of the graph $\gamma$. In particular, given any non-planar\footnote{The Hamiltonian constraint acts trivially on vertices with only planar edges incident on them, since the volume operator does.} triplets of edges intersecting a common vertex $v$, the Hamiltonian constraint acts on that vertex $v$ by adding a closed loop at the vertex, which contains only one extra edge (see figure \ref{fig:loophamiltonian} ). \\
\\
The new vertices formed by the action of the Hamiltonian only have planar edges incident on them, so no further action of the Hamiltonian constraint is possible on them. The repeated action of the Hamiltonian constraint will create a self-similar structure around each vertex, as it was shown in \cite{46d}.

In these papers an algorithm for finding solutions of the above constructed Hamiltonian constraint was put forward, but it is still not clear whether such solutions have zero or infinite norm with respect to the physical inner product, since such product is not yet defined. This is a consequence of the fact that group averaging techniques used to define inner products can not be applied to situations in which the  constraint algebras has structure functions.

The way delineated above of defining the Hamiltonian constraint does not solve the second issue mentioned at the start of this section, namely that the Dirac algebra formed by the Hamiltonian constraints and the Diffeomorphic constraints is not a Lie algebra. This problem, however, can be solved by adopting a different method of formulating the Hamiltonian constraint. This is the so called \textbf{Mater constraint program} \cite{v4} which we now turn to.\\[5pt]
\textbf{Master Constraint} \\[5pt]
We will now introduce the \emph{Master constraint program} carried out in \cite{v4}. For pedagogical reasons we will reiterate the issues that this program was set out to solve.
\begin{itemize}
\item[1)] \emph{Introduction of a non standard topology}. In \cite{46d} it was shown that the limit $\epsilon\rightarrow 0$ for the Hamiltonian constraints exists as a well defined operator. The proof rested on diffeomorphic invariance. However, it is not possible to define the Hamiltonian constraint directly on $\mh_{diff}$, since it is spatially diffeomorphism invariant, i.e. the Dirac algebra $\md$ of the constraints does not preserve $\mh_{diff}$ ($\{H_a, H\}\propto H$ ). \\
In other words, the spatial diffeomorphism constraints form a subalgebra but not an ideal of $\md$.\\
This implies that it is not possible to work directly with $\mh_{diff}$ and, consequently, it is not possible to use the standard strong or weak topology defined on it. What has to be done, instead, is to introduce a different, unconventional topology.
\item[2)] \emph{No generators for Diffeomphism operators}. The one-parameter subgroups of spatial diffeomorphisms are not weakly continuous, therefore, from Stone's theorem, it is not possible to define a self-adjoint operator corresponding to the diffeomorphism constraint. This, in turn, implies that it is not possible to implement, at the quantum level, the Poisson bracket between two Hamiltonian constraints, since it is proportional to the diffeomorphic constraint, i.e. $\{H(N), H(N^{'})\}\propto H_a(q^{-1}(dNN^{'}-NdN^{'}))$.
\item[3)]\emph{No true Lie algebra}. The fact that we get a structure function ($q^{-1}(dNN^{'}-NdN^{'})$) in the Poisson brackets, rather than a structure constant, implies that $\md$ is not a true Lie algebra, thus it is not possible to use group averaging techniques and RAQ to solve the constraints and to define observables.
\end{itemize}
To overcome such problems the \emph{Master constraint program} \cite{v4} has been introduced. The main idea, in this program, is to replace the infinite Hamiltonian constraints with a single \emph{Master} constraint. \\
Let us first analyse how this is done for a general classical theory with constraint $C_J$ and associated simplectic structure $\{M, \{,\}\}$ . Here $\mj$ represents some index set, such that $\mj = D\times X$; $J = (j, x)$, where $D$ is a discrete label set and $X$ is a topological space. We then write $C_J = C_j(x)$. 

In this setting the Master constraint is defined to be the weighted sum of the single constraints as follows:
\be
\mathbf{M}:=\frac{1}{2}\int_Xd\mu(x)\sum_{j,k\in D}q^{jk}(x)C_j(X)C_k(x)
\ee
where $\mu$ is a measure on $X$ and $q^{jk}\in C^{\infty}(M)$ is a metric function.\\
The following lemma shows that the constraint surface induced by $\mathbf{M}$ is the same as that induced by all the constraints $C_J$
\begin{lemma}\label{lem:constraintsurface}
The constraint hypersurface C of M defined by
\be
C := \{m\in M| C_J(m) = 0  \text{ for } \mu; J\in\mj  \} 
\ee
is equivalent to the hypersurface defined by 
\be
C = \{m\in M| M(m) = 0\} 
\ee
\end{lemma}
In order to complete the definition of $\mathbf{M}$ as an alternative to the Hamiltonian constraint, we need to show that it is possible to define Dirac observables in terms of $\mathbf{M}$. \\
To this end, let us first recall the notion of a Dirac observable.
\begin{definition}\mbox{ }
\begin{itemize}
\item[i)]
A function $O\in C^{\infty}(M)$ is called a \textbf{weak Dirac observable} iff
\be
\{O,C(N)\}_{|C} = 0
\ee
\item[ii)]
A function $O\in C^{\infty}(M)$ is called a \textbf{strong Dirac observable} iff
\be
\{O,C(N)\} = 0 
\ee
\end{itemize}
\end{definition}
It follows that, every strong Dirac observable is a weak Dirac observable. \\
In terms of the Master constraint Dirac observables are defined as follows:
\begin{theorem}\label{thediracobservables}
A function $O\in C^{\infty}(M)$ is a weak Dirac Observable if and only if
\be
\{O, \{O,\mathbf{M}\}\}_{\mathbf{M}=0} = 0
\ee
\end{theorem}
The proof of this theorem can be found in \cite{v4}. \\The double brackets were necessary, since any single time function on $M$ has vanishing Poisson bracket with the Master constrain, thus making week Dirac observables undetectable. \\
If we now apply the above definitions to the case of GR we would obtain the following extended master constraint:
\be
\mathbf{M} :=\int_{\sigma}d^3x\frac{H^2+q^{ab}H_aH_b+\delta^{jk}G_jG_k}{\sqrt{det(q)}}
\ee
where $G_j$, $H_a$ and $H$ are the Gauss, diffeomorphism and Hamiltonian constraint, respectively. The weighted sum is chosen such that $\mathbf{M}$ is diffeomorphism invariant.\\
By applying lemma \ref{lem:constraintsurface} it can be shown that the master constraint reproduces the same constraint surface as do the single constraints. In fact, the requirement that $\mathbf{M}=0$ is equivalent to the requirement that all three constraints are zero, i.e.
\be
\mathbf{M}=0\cong \Big(G_j(\lambda^j)=0\Big)\wedge \Big(H_a(N^a)=0\Big)\wedge \Big(H(N)=0\Big)
\ee
for any smearing function $\lambda^j$ , $N^a$ and $N$.\\
Now that we have grouped all the constraints into a single one, the resulting algebra is trivial
\be
\{\mathbf{M},\mathbf{M}\}=0
\ee
Moreover, utilising theorem \ref{thediracobservables} it is possible to define weak Dirac observables in terms of double Poisson brackets with the Master constraint.

For the time being we are interested in solving the Hamiltonian constraint, therefore we will restrict the definition of the Master constraint only for the Hamiltonian. 
This is the so called {\it non-extended Master constraint}. Thus, at the classical level, we get
\be
\mathbf{M} :=\int_{\sigma}d^3x\frac{H^2(x)}{\sqrt{det(q)(x)}}=\int_{\sigma}d^3x\big(\frac{H}{\sqrt[4]{det(q)}}\big)(x)\int_{\sigma}d^3y\delta(x,y)\big(\frac{H}{\sqrt[4]{det(q)}}\big)(y)
\ee
The constraint algebra $\md$ is now replaced by the \emph{Master Constraint algebra} $\mathfrak{M}$  
\ba
\{H_a(\vec{N}), H_a(\vec{N}^{'})\}&=&-k\ml_{\vec{N}}(\vec{N})\nonumber\\
\{H_a(\vec{N}), \mathbf{M}\}&=&0\nonumber\\
\{\mathbf{M}, \mathbf{M}\}&=&0
\ea 
In order to find a suitable operator corresponding to $\mathbf{M}$ we procede in a similar manner, as it was done in the previous section. Specifically, we discretize the manifold $M$ through a triangulation $T(\epsilon)$ and the Hamiltonian constraint $H(N)$ can then be written as the Riemannian sum
\be
H(N) = lim_{\epsilon\rightarrow 0}\sum_{\Delta\in T(\epsilon)}N(v(\Delta))H(\Delta)
\ee
where $H(\Delta)=H(\chi_{\Delta})$. The discretised version of the Master constraint then becomes
\be
\mathbf{M} = lim_{\epsilon\rightarrow 0}\sum_{\Delta\in T(\epsilon)}\frac{H(\Delta)^2}{V(\Delta)}
\ee
where $V (\Delta)=\int_{\Delta} d^3x\sqrt{| det(E)|}$ represents the volume. \\

By introducing the quantity
\be
C(\Delta) :=\frac{H(\Delta)}{\sqrt{V (\Delta)}}=\int_{\Delta} Tr\Big(F\wedge\frac{ \{A, V (\Delta)\}}{\sqrt{V (\Delta)}}\Big)= 
2\int_{\Delta}Tr\Big(F \wedge \{A,\sqrt{V (\Delta)}\}\Big)
\ee
where in the last equality we used the identity $\{., V (\Delta)\}\sqrt{V (\Delta)}\} = 2\{.,\sqrt{V (\Delta)}\}$, it is now possible to write the Master constraint as follows:
\be
\mathbf{M} = lim_{\epsilon\rightarrow 0}\sum_{\Delta\in T(\epsilon)}C^2(\Delta)=lim_{\epsilon\rightarrow 0}\sum_{\Delta\in T(\epsilon)}\overline{C(\Delta)}C(\Delta)
\ee
This expression for the Master constraint is very convenient, since $C(\Delta)$ can be quantised in exactly the same way as it was done for $H(\Delta)$ in the previous section. The resulting operator corresponding to the Master constraint is \cite{v4}
\be
\hat{\mathbf{M}}T_{[s_2]}:=\sum_{[s_1]}Q_{\mathbf{M}}(T_{[s_1]},T_{[s_2]})T_{[s_1]}
\ee
where $T_{[s]}:=l_{[s]}/\sqrt{\eta_{[s]}}$, as defined in the previous section, and $Q_{\mathbf{M}}$ is defined to be
\be\label{equ:quadraticform}
Q_{\mathbf{M}}(l, l^{'})=\sum_{[s]}\eta_{[s]}\sum_{v\in V(\gamma(s_0([s])))}\overline{l(\hat{C}^{\dagger}_v T_{s_0([s])})}l^{'}(\hat{C}^{\dagger}T_{s_0([s])})
\ee
where $s_0([s])$ indicates a representative of the equivalence class $[s]$.

In \cite{v4} it was shown that the quadratic form in \ref{equ:quadraticform} is closable and induces a unique positive, self-adjoint operator $\hat{\mathbf{M}}$ on $\mh_{diff}$ which contains the point zero spectrum, i.e the kernel of the Master constraint is equivalent to the kernel of the Hamiltonian constraint.

\begin{figure}[htb]
\begin{center}
\psfrag{a}{$e(v)$}\psfrag{d}{$(e^{'})^{-1}(v_1)$}\psfrag{f}{$v_1$}\psfrag{c}{$e(v_1)$}\psfrag{b}{$(e^{'})^{-1}(v)$} \psfrag{e}{$v$} 
\includegraphics[scale=0.5]{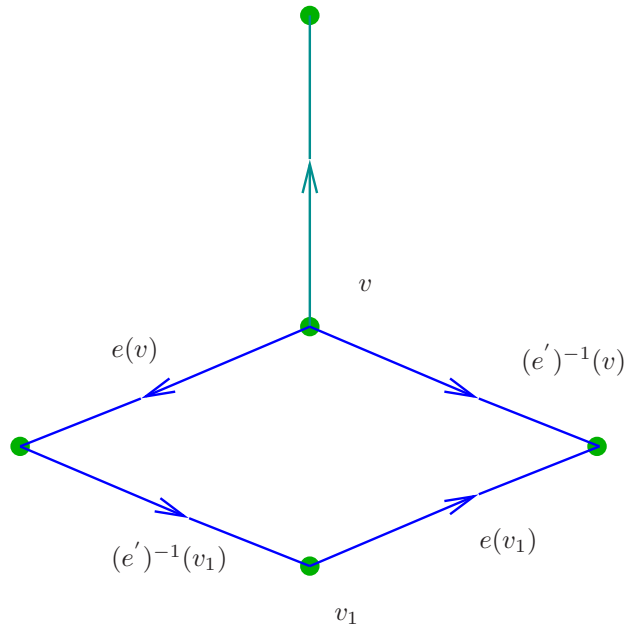}
\caption{Action of the non-graph changing Master constraint $\mathbf{M}$ \label{fig:minimalloop}}
\end{center}
 \end{figure}

The reason why we had to define the Master constraint on $\mh_{diff}$, instead of defining it directly on $\mh_{kin}$, is because it is a graph-changing diffeomorphism invariant operator. In fact it was shown in \cite{v4} that the only diffeomorphism invariant operators, which can be defined on $\mh_{kin}$, are those not involving the connection $A$ but only $E$, for example the volume operator ( see Section \ref{s4.0} ). \\
However, it is possible to define a non-graph-changing version of the Master constraint operator on $\mh_{kin}$. The way this is done is by defining $\mathbf{M}$ in the spin network basis and, then, for each $T_{s}$ and each $v\in \gamma(s)$ one must define a unique diffeopmorphic rule that singles out the loop produced by the action of $\mathbf{M}$ as an already existing loop. Such a rule is called the minimal loop rule
\begin{definition}\label{def:minimalloop}
For a graph $\gamma$, a vertex $v\in V(\gamma)$ and two different edges $e, e^{'}\in E(\gamma)$, both starting at $v$, a loop
$\alpha_{\gamma,v,e,e^{'}}$ in $\gamma$ which starts along $e$ and ends along $(e^{'})^{-1}$ is said to be minimal iff
there is no other loop with the same properties, which contains fewer edges of $\gamma$. (see figure \ref{fig:minimalloop})
\end{definition}
In the eventuality that there is more than one minimal loop then one averages over them.\\
Since we can quantise the non-graph changing Master Constraint Operator as a positive operator on $\mh_{kin}$, then it is possible, by using the semiclassical techniques developed in \cite{21, 21, 46f}, to check its semi-classical properties.

However, the non-graph-changing Master constraint is anomalous, i.e. it does not contain the zero value on its spectrum. A possible way to deal with this problem is to subtract from the Master constraint operator the minimum value of the spectrum: $\hat{\mathbf{M}}-\lambda_{min}$, as long as $\lambda_{min}$ is finite and of order $\hbar$. \\
The modified master constraint has the same classical limit as the original one, thus the master constraint program delineated above is still valid.

So far we have only considered the non-extended Master constraint, however, it is also possible to quantise the extended Master constraint, both in a graph-changing and graph-non changing fashion. Similarly to the Master constraint operator, also for the extended Master constraint operator only the non-graph-changing version can be utilised,  when evaluating semiclassical properties. \\
It turns out that, when utilising the extended master constraints, due to diffeomorphism invariance, the information regarding how the graphs are embedded in the spatial manifold are lost and, so, it is also the information on how the edges of graphs are knotted and braided.

The only information that is left is which vertices are connected with other vertices and for how many times. This implies that the only information retained in the extended Master constraint is of an algebraic nature (the topological information is lost). This feature has motivated the development of Algebraic Quantum Gravity (AQG) \cite{51tina} \cite{v9}. In the context of AQG, the semiclassical limit of the extended Master constraint was analysed and it turned out that such a limit reproduces the correct infinitesimal generators of General Relativity (see \cite{18} and \cite{20} for a detailed analysis). 
\chapter{Semiclassical Analysis}
In this chapter we will discuss the tools and techniques utilised to investigate the semiclassical limit of a theory. This is particularly important when testing the validity of a quantum theory as the correct quantisation of a classical theory. For example in the case of LQG, in order to verify if this theory is indeed a quantisation of GR, one has to check whether in the classical regime LQG reduces to GR. 

Such a semiclassical analysis is carried out in terms of the so called \emph{semiclassical state} which, roughly speaking, are states close (in some yet to be specified way) to some given classical geometry. A particular class of \emph{Semiclassical states}, namely classical coherent states and their application for analysing certain semiclassical properties of LQG, will be the topics of the following sections.  

\section{Review of Semiclassical Coherent States}

One of the major unsolved problems in Loop Quantum Gravity is the
verification if, in the classical limit, this theory reduces to
General Relativity, i.e. if there exist certain semiclassical
states $\psi$ in the Hilbert space $\mathcal{H}$, such that
expectation value $\langle\psi,\hat{O}_i\psi\rangle$ of the
operators $\hat{O}_i\in\mathcal{B}(\mathcal{H})$ on $\mathcal{H}$,
with respect to these states, coincides with the classical values
O(m) of the respective observables, where $m\in\mathcal{M}$ is the
point in the manifold at which we evaluate $O$. 

So, given a quantum
theory X, which we identify with the triplet
$(\mathcal{H},\langle\cdot,\cdot\rangle,\hat{\mathcal{O}})$ consisting
of a Hilbert space $\mathcal{H}$ with scalar product
$\langle\cdot,\cdot\rangle$ and a *-subalgebra $\hat{\mathcal{O}}$ of
the algebra $\mathcal{L}(\mathcal{H})$ of linear operators on
$\mathcal{H}$, what we are looking for are those states in
$\mathcal{H}$ that enable us to define a quantisation, i.e. a map:
\begin{equation*}
d:(\mathcal{M},\{\cdot,\cdot\},\mathcal{O})\rightarrow (\mathcal{H},\langle\cdot,\cdot\rangle,\hat{\mathcal{O}})
\end{equation*}
where the triplet $(\mathcal{M},\{\cdot,\cdot\},\mathcal{O})$, consisting of a phase space $\mathcal{M}$ with Poisson brackets $\{\cdot,\cdot\}$ and a *-Poisson subalgebra $\mathcal{O}$ of the Poisson algebra $\mathcal{B}^{\infty}(\mathcal{M})$ of smooth functions on $\mathcal{M}$, which separate the points in $\mathcal{M}$, represents the classical limit of $(\mathcal{H},\langle\cdot,\cdot\rangle,\hat{O})$ \\
The definition of the classical limit of a theory can be
formalised as follows
\begin{definition}
A triple $(\mathcal{M},\{\cdot,\cdot\},\mathcal{O})$ is said to be the classical limit of $(\mathcal{H},\langle\cdot,\cdot\rangle,\hat{\mathcal{O}})$ iff there exists a quantisation map
\begin{equation}\label{equ:map}
\psi:\mathcal{M}\rightarrow\mathcal{H}\hspace{.3in}m\mapsto\psi_m\hspace{.3in}\text{with}\hspace{.3in}||\psi_m||^2=1
\end{equation}
such that, for all self-adjoint operators $\hat{O}\in \hat{\mathcal{O}}$ and for generic points $m\in\mathcal{M}$ at which $O(m)\neq 0$ and $|O(m)|$ is bigger than the fluctuations, the following conditions hold:
\begin{enumerate}
\item \emph{Expectation value property}\\
\begin{equation*}
|\frac{\langle\psi_m,\hat{O}\psi_m\rangle}{O(m)}-1|\ll 1
\end{equation*}
\item \emph{Infinitesimal Ehrenfest property}
\begin{equation*}
|\frac{\langle\psi_m,\frac{[\hat{O},\hat{O}^{\prime}]}{i\hbar}\psi_m\rangle}{\{O,O^{\prime}\}(m)}-1|\ll 1
\end{equation*}
\item \emph{Fluctuation property}
\begin{equation*}
|\frac{\langle\psi_m,\hat{O}^2\psi_m\rangle}{\langle\psi_m,\hat{O}\psi_m\rangle^2}-1|\ll 1
\end{equation*}
\end{enumerate}
\end{definition}
The reason why the dequantisation map $d$ reduces to the map \ref{equ:map} is because the value of an observable $O$ depends on the point $m\in \mathcal{M}$, on which we evaluate $O$ (i.e. $O(m)$), therefore we need to associate a state $\psi\in \mathcal{H}$ to each point $m\in \mathcal{M}$ in order for the dequantisation to be possible.\\
The inverse of the above process, i.e. the process of canonical quantisation is instead defined as follows:
\begin{definition}
A triple $(\mathcal{H},\langle\cdot,\cdot\rangle,\hat{O})$ is a quantisation of a triple $(\mathcal{M},\{\cdot,\cdot\},\mathcal{O})$ iff there exists a *-Lie algebra homomorphism called a representation of $\mathcal{O}$
\begin{equation} \label{equ:lie}
\Lambda:\mathcal{O}\rightarrow\hat{\mathcal{O}},\hspace{.3in}O\mapsto\hat{O}
\end{equation}
with the following properties
\begin{equation*}
\widehat{zO+z^{\prime}O^{\prime}}=z\hat{O}+z^{\prime}\hat{O}^{\prime},\hspace{.1in}\hat{O}\dagger=\hat{\bar{O}}\hspace{.1in}\text{and} \hspace{.1in}[\hat{O},{O}^{\prime}]=i\hbar\widehat{\{O,O^{\prime}\}}
\end{equation*}
for all $O,O^{\prime}\in\mathcal{O}$ and $z,z^{\prime}\in\Cl$. The algebra $\mathcal{O}$ is called the algebra of elementary observables.
\end{definition}\textbf{Consequences of the above properties}\\
\begin{enumerate}
\item
If we define an operator $\hat{z}:=\hat{O}+ix\hat{O}^{\prime}$ where $x\in\Rl$, then
property 1) is satisfied iff $\hat{z}\psi_m=z(m)\psi_m$, i.e. $\psi_m$ is an eigenstate of $\hat{z}$
with eigenvalues $z(m)\in \Cl$. $x$ is called the quenching parameter since the fluctuations of $\hat{z}$ and $\hat{O}$ agree when $x = 1$.\\
What the above means is that, once we define the operator $\hat{z}:=\hat{O}+ix\hat{O}^{\prime}$ then, for the expectation value property to be satisfied all we need to do is to define a relation between z and a point $m\in\mathcal{M}$.
\item Infinitesimal Ehrenfest property
is satisfied iff  $\hat{O}$ and $\hat{O}^{\prime}$ belong to a set $\hat{\mathcal{O}}$, such that the representation theory is satisfied, i.e. $\Lambda:\mathcal{O}\rightarrow\hat{\mathcal{O}}$ is defined for both $\hat{O}$ and $\hat{O}^{\prime}$.
\item Fluctuation property
is satisfied iff the commutation relation $[\hat{O},\hat{O}^{\prime}]$ is of order unity. In fact, by defining the states $\psi_m$ as eigenstates of $\hat{z}:=\hat{O}+ix\hat{O}^{\prime}$ with eigenvalue $z$, it can be shown that the Heisenberg uncertainty inequality is saturated
\begin{equation}
\langle_{\psi}(\Delta\hat{O})^2\rangle_{\psi}\langle(\Delta\hat{O}^{\prime})^2\rangle_{\psi}=\frac{1}{4}|\langle[\hat{O},\hat{O}^{\prime}]\rangle_{\psi}|^2
\end{equation}
Therefore, if the commutation relation $[\hat{O},\hat{O}^{\prime}]$ is of order unity, the fluctuation $ \Delta\hat{O}$ of the operators is small and the Fluctuation property holds.
\end{enumerate}
Following the above discussion we are now ready to define \textbf{Semiclassical Coherent states}.\\
\begin{definition}
A system of semiclassical states $\{\Psi_m\}_{m\in\mathcal{M}}$ is called coherent iff the following properties hold:
\begin{enumerate}
\item Overcompletness property:\\
$\exists$ a measure $\nu$ on $\mathcal{M}$ such that
\begin{equation*}
id_{\mathcal{H}}=\int_{\mathcal{M}}d\nu(m)\psi_{m}\langle\psi_m,\cdot\rangle
\end{equation*}
holds
\item Minimal uncertainty property:\\
$\hat{\mathcal{O}}$ can be generated by a set of annihilator operators $\hat{z}$ and their adjoint creation operator $\hat{z}^{\dagger}$ such that
\begin{equation*}
\hat{z}\psi_m=z(m)\psi_m
\end{equation*}
and the map
\begin{equation}
z:\mathcal{M}\rightarrow\mathcal{Z}\hspace{.3in}z\mapsto z(m)
\end{equation}
is a bijection. Here $\mathcal{Z}$ represents a complex manifold.
\item Peakedness Property:\\
The overlap function
\begin{equation}
|\langle\psi_z\psi_{z^{\prime}}\rangle|^2
\end{equation}
is sharply peaked at $z=z^{\prime}$
\end{enumerate}
\end{definition}
The overcompletness property enables one to expand generic states in terms of coherent states.
Overcompleatness rather than completeness is required since, a basis for a separable Hilbert space $\mathcal{H}$ is a countable number of states but the states $\psi_z$, which depend on the continuous parameter $z$, are not countable.

The set of states $\{\psi_z\}$ which satisfies the conditions for semiclassical states is not unique. This implies that a given quantum theory can have distinct classical limits. So the aim in the context of LQG is to find at least one set of semiclassical states, which well approximate the elementary observables of General Relativity.
 A type of semi-classical states are the {\it complexifier coherent states}, i.e. coherent states which are generated by a complexifier (see below for definition).
This method was first introduced in \cite{22}, \cite{31}, \cite{417} and has been subsequently used in other contexts \cite{34}, \cite{46ah}, \cite{30b}, \cite{18,20}, \cite{10}, and \cite{46g, 15}. 

The advantage of this approach is that it can be applied to any system, whose phase space can have a cotangent bundle structure (i.e. $\mathcal{M}=T^*\mathcal{P}$) and its application does not require any conditions on the Hamiltonian of the system. This is of particular relevance in GR where no true Hamiltonian is a priori available. The semiclassical limit of the Hamiltonian constraint, in turn, can not be tested by semiclassical states which it annihilates, thus we need kinematical coherent states.

In what follows, we will briefly review how the complexifier method is used to construct coherent
states for cotangent bundles over a compact group, then we will apply the complexifier machinery to
construct coherent states on a graph.
\section{Complexifier Coherent States}
\label{s2}
In this section we will review the complexifier method to construct 
coherent states. We will not go into all the detail, for a complete exposition and analysis see \cite{21} 
and references therein. 

\subsection{General Complexifier Method}
\label{s2.1}

Let us consider a symplectic manifold ${\cal M}=T^\ast({\cal C})$ that is a 
cotangent bundle over a configuration manifold $\cal C$, which may be 
infinite dimensional (we will suppress any indices in what follows). 
\begin{definition} \label{def2.1} ~\\
A complexifier $C:\;{\cal M} \to \mathbb{R}^+$ is a sufficiently smooth, 
positive function on 
$\cal M$ with dimension of an action which has the following scaling 
behaviour 
\be \label{2.1}
\lim_{\lambda \to \infty} \frac{C(q,\lambda p)}{\lambda}=\infty
\ee
where $(q,p)$ are the canonically conjugate, real configuration and 
momentum 
coordinates on $\cal M$.
\end{definition} 
The reasons for these restrictions will become evident in a moment. With the 
aid of $C$ we define 
\be \label{2.2}
z:=\exp(-i\{C,.\})\cdot q=\sum_{n=0}^\infty\;\frac{(-i)^n}{n!}\;
\{C,q\}_{(n)}
\ee
where $\{C,f\}_{(0)}:=f,\;\{C,f\}_{(n+1)}:=\{C,\{C,f\}_{(n)}\}$. The meaning 
of ``sufficiently smooth'' is that all coefficients in the Taylor 
expansion (\ref{2.2}) exist. 

Notice that (\ref{2.2}) defines a (complex 
valued) canonical transformation, hence $\{z,z\}=\{\bar{z},\bar{z}\}=0$ 
(this is non trivial
when $\dim({\cal C})\ge 2$). The scaling behaviour implies that 
$z,\;\bar{z}$ can be used as coordinates for $\cal M$, in fact, $z\in 
{\cal C}^{\mathbb{C}}$ defines a complex polarisation of $\cal 
M$.  

We now assume that $\cal M$ can be quantised, that is, given the classical 
Poisson$^\ast$ algebra defined by 
$\{q,q\}=\{p,p\}=0,\;\{p,q\}=1_{{\cal M}};\;\;
\bar{q}=q,\;\bar{p}=p$, there exist a representation $(q,p)\mapsto (\hat{q},\hat{p})$
on a Hilbert space of the form ${\cal 
H}=L_2(\overline{{\cal C}},d\mu)$\footnote{It should be noted that in the finite dimensional case $\overline{{\cal C}}={\cal C}$ 
while in the the infinite dimensional case ${\cal C}\subset 
\overline{{\cal C}}$, i.e. $\overline{{\cal C}}$ is a suitable 
distributional extension.}, such that the operators satisfy (assuming 
careful domain definitions)
$[\hat{q},\hat{q}]=[\hat{p},\hat{p}]=0,\;[\hat{p},\hat{q}]=i\hbar\;1_{{\cal 
H}};\;\;
\hat{q}^\dagger=\hat{q},\;\hat{p}^\dagger=\hat{p}$. Here 
$\overline{{\cal C}}$ comes with some topology and $\mu$ is a Borel 
measure on it. 


Assuming that also $C$ has a quantisation $\hat{C}$ as a positive, self 
adjoint operator (in field theories this is non trivial due to 
operator ordering and operator product expansion questions), it is possible to construct 
the operator representation of (\ref{2.2}) 
by substituting Poisson brackets by commutators divided by $i\hbar$
\be \label{2.3}
\hat{z}:=\sum_{n=0}^\infty\;\frac{(-i)^n}{n!\;(i\hbar)^n}\;
\{\hat{C},\hat{q}\}_{(n)}=e^{-\hat{C}/\hbar}\;\hat{q}\;e^{\hat{C}/\hbar}
\ee
This formula explains the dimension restriction on $C$. The operator 
$e^{\pm \hat{C}/\hbar}$ is well defined via the spectral theorem.
We will refer to $e^{-\hat{C}/\hbar}$ as the {\it heat kernel} and to 
$\hat{z}$ as the {\it annihilation operator}. 

The $\delta$ distribution $\delta_{q_0}$ on the subset of ${\cal H}$ consisting of the 
continuous functions and with support at $q_0\in 
\overline{{\cal C}}$ is defined by
\be \label{2.4}
\delta_{q_0}[\psi]:=\psi(q_0):=<\delta_{q_0},\psi>:=\int_{\overline{{\cal 
C}}}\;
d\mu(q)\;\delta_{q_0}(q) \psi(q)
\ee
where $\delta_{q_0}(q)$ is the integral kernel of the unit operator.
The {\it coherent state} for $z\in \overline{{\cal C}}^{\mathbb{C}}$ is defined as  ``heat kernel evolution'' followed by analytic 
continuation:

\be \label{2.5}
\psi_z:=[e^{-\hat{C}/\hbar} \delta_{q_0}]_{q_0\to z}
\ee
In order for this expression to be well defined, the function 
$e^{-\hat{C}/\hbar} \delta_{q_0}$ must not only be in $\cal H$ but also 
analytic in $q_0$. This explains the positivity and scaling requirement 
on $C$, which 
makes sure that the heat kernel is a 
damping operator such that, at least for separable $\cal H$, the function
(\ref{2.5}) is not only normalisable but also analytic\footnote{Here we have 
assumed that the map ${\cal C}\to {\cal C}^{\mathbb{C}};\;q\mapsto z$
in (\ref{2.2}) has an extension to some $\overline{{\cal 
C}}^{\mathbb{C}}$.}.

It is now possible to verify the following:
\be \label{2.6}
\hat{z} \; \psi_z=z\; \psi_z
\ee
Thus, $\psi_z$ is an eigenfunction of the annihilation operators 
$\hat{z}$ which explains the notion ``coherent state''.
As it is well known, property (\ref{2.6}) implies that the uncertainty 
relation for the self-adjoint operators 
\be \label{2.7}
\hat{x}:=[\hat{z}+\hat{z}^\dagger]/2,\;\;
\hat{y}:=-i[\hat{z}-\hat{z}^\dagger]/2
\ee
is saturated, that is
\be \label{2.8}
[<\hat{x}^2>_z-(<\hat{x}>_z)^2]\;
[<\hat{y}^2>_z-(<\hat{y}>_z)^2]=\frac{1}{4}\;|<[\hat{x},\hat{y}]>_z|
\ee
where $<.>_z:=<\psi_z,.\;\psi_z>/||\psi_z||^2$ denotes the expectation 
value with respect to $\psi_z$ (notice that $\psi_z$ is in general not 
automatically normalised). This is a second property commonly attributed 
to coherent states \cite{31}. 

Finally, under certain technical assumptions spelled out in \cite{32},
the completeness relation
\be \label{2.9}
1_{{\cal H}}=\int_{\overline{{\cal C}}} \; d\mu(q_0)\; \delta_{q_0}\;  
\delta_{q_0}[.]
\ee
implies that there exists a measure $\nu$ on $\overline{{\cal 
C}}^{\mathbb{C}}$ such that
\be \label{2.10}
1_{{\cal H}}=\int_{\overline{{\cal C}}^{\mathbb{C}}} \; d\nu(z)\; 
\psi_z\; <\psi_z,.>
\ee
This concludes the general discussion. The interested reader may verify
\cite{21,23} 
that the coherent states for Maxwell Theory on Minkowski space result 
from the complexifier
\be \label{2.11}
C=\frac{1}{2\kappa^2}\int_{\mathbb{R}^3}\;d^3x\; \delta_{ab} E^a 
\sqrt{-\Delta}^{-1} E^b
\ee
where $E^a$ is the Maxwell electric field, $\Delta$ is the 
Laplacian on $\mathbb{R}^3$, $\kappa$ is the electric 
charge and 
$\alpha=\hbar\kappa^2$ is the Feinstruktur constant.

\subsection{Complexifiers for Background Independent Gauge Theories}
\label{s2.2}

As explained in detail in \cite{33,34}, gauge theories with 
compact gauge group $G$ provide 
an almost perfect arena for the general theory summarised in the 
previous 
section. Let us explain this in detail. First of all we need to identify the following elements: i) {\it Classical Phase Space}, ii) {\it Configuration Space} iii) {\it Hilbert space} iv) {\it Complexifer}. 

In the case of LQG, we recall from Chapter \ref{cha:qunt}, that the first three ingredients are the following: 
\begin{itemize}
\item[1.] {\it Classical Phase Space}\\
The role of $\cal C$ is 
played by some space of smooth 
connections $\cal A$ over some $D-$dimensional spatial manifold 
$\sigma$. The role of 
$\cal M$ is then $T^\ast {\cal A}$. The configuration and the
momentum coordinates, on this phase space, are real 
valued connection one 
forms $A_a^j(x)$ (potentials) and Lie algebra valued vector densities
$E^a_j(x)$ (electric fields) respectively, which enjoy the following 
Poisson brackets:
\be \label{2.12}
\{A_a^j(x),A_b^k(y)\}=
\{E^a_j(x),E^b_k(y)\}=0,\;\;
\{E^a_j(x),A_b^k(y)\}=\kappa\;\delta^a_b\;\delta_j^k\;
\delta(x,y)
\ee
Here $\kappa$ denotes the square of the coupling constant of the gauge 
theory, $a,b,c,..=1,..,D$ denote spatial tensor indices and 
$j,k,l,..=1,..,\dim(G)$ denote Lie algebra indices. We will assume 
that $G$ is connected, semisimple and we take the convention that the 
internal metric is 
just $\delta_{jk}$. 
\item[2.] {\it Distributional Configuration Space}\\
Now consider
arbitrary, finite, piecewise analytic (more precisely semianalytic 
\cite{14}) graphs embedded in $\sigma$. This can be thought of as collections 
of edges $e$, that is, piecewise analytic one dimensional paths which 
intersect, at most, in their endpoints. The collection of such end points is called the set $V(\gamma)$ of 
vertices of $\gamma$. The set of edges of $\gamma$, is instead denoted by $E(\gamma)$.\\
Given $\gamma$, let us
consider functions {\it cylindrical over $\gamma$} (see definition \ref{def:cylf}) of the form
\be \label{2.11a}
f:\;{\cal A}\to \mathbb{C};\;
A\mapsto f(A)=f_\gamma(\{A(e)\}_{e\in E(\gamma)})
\ee
where $f_\gamma:\;G^{|E(\gamma)|} \to \mathbb{C}$ is a complex valued 
function on $|E(\gamma)|$ copies of $G$ and $A(e)$ denotes the holonomy
of $A$ along $e$. 

Functions of the form (\ref{2.11a}) form an Abelian 
$^\ast$ algebra under pointwise operations with the involution given by 
complex conjugation. We can turn it into an Abelian 
$C^\ast-$algebra,
usually called Cyl (cylinder functions) with respect to the sup-norm on 
$\cal A$ that is
\be \label{2.12a}
||f||:=\sup_{A\in {\cal A}} \;|f(A)|
\ee
As shown in \cite{35} and briefly explained in section \ref{s configurationspace}, Abelian $C^\ast-$algebras $\mathfrak{A}$ are 
isometric isomorphic to the Abelian $C^\ast-$algebra which consists 
of continuous functions on a compact Hausdorff space 
$\Delta(\mathfrak{A})$, called the spectrum of $\mathfrak{A}$. \\
Denote 
the spectrum of Cyl by $\overline{{\cal A}}$. Its geometrical 
interpretation is as a space of generalised connections in the sense that
the holonomy of 
$A\in \overline{{\cal A}}$ satisfies all the usual algebraic relations, 
satisfied by smooth holonomies: 
$A(e\circ e')=A(e) A(e')$ if the end point of $e$ is the beginning point 
of $e'$ and $A(e^{-1})=(A(e))^{-1}$. However, neither smoothness or continuity are required. The 
topology on $\overline{{\cal A}}$ is the Gel'fand topology which, in 
this case, is equivalent to the requirement that a net of generalised connections 
converges when the corresponding net of holonomies, for all 
possible paths, converges. For more detail see 
\cite{13,36,1b}.
\item[3.] {\it Hilbert Space}\\
Being a compact Hausdorff space, a natural set of representations
of the Poisson$^\ast-$algebra generated by all the holonomies and all 
the electric fluxes through codimension 1 (piecewise analytic) surfaces 
$S$ 
should be of the form ${\cal H}=L_2(\overline{{\cal A}},d\mu)$, where 
$\mu$ is a Borel probability measure. It turns out that all cyclic
representations that carry a unitary representation of the 
diffeomorphism group Diff$(\sigma)$ are of this form \cite{14} and the 
corresponding measure, first discovered in \cite{13}, is unique.
See e.g. \cite{1b} for detail. For our purposes it is enough to  
know that $\cal H$ admits a natural orthonormal basis, called spin 
network functions (SNWF) defined in equation \ref{ali:snfvariant}. 
%
\end{itemize}
The next step is to identify a possible complexifier. The complexifier for Maxwell theory,
displayed in (\ref{2.11}), is motivated by the fact that the associated 
annihilation operators are precisely those that enter the Maxwell 
Hamiltonian. In General Relativity there is no a priori Hamiltonian
but there is the Hamiltonian constraint. Hence one might be tempted 
to choose a complexifier whose associated annihilation operator is 
related to the Hamiltonian constraint. 

Unfortunately, the Hamiltonian 
constraint is, in contrast to Maxwell theory, neither polynomial nor 
does it have a quadratic piece with respect to which a perturbation 
scheme can be defined. Hence, the notion of an annihilation 
operator, defined by the Hamiltonian constraint, is ill 
defined\footnote{The situation slightly improves when a physical 
Hamiltonian is available, see \cite{5}.}. On the other hand, in this context we are only interested in 
constructing coherent states which well approximate our
elementary holonomy and flux operators, defined on the 
kinematical Hilbert space (on which the Hamiltonian constraint is not 
satisfied). 

Such states are then utilised to verify whether other kinematical operators (i.e. 
not 
invariant under the gauge motions generated by the spatial 
diffeomorphism and Hamiltonian constraints), such as the volume operator,    
have been correctly quantised. Therefore, the motivation to use the Hamiltonian 
constraint, as a selection criterion for the complexifier, is less strong.

In lack of a better selection criterion, we take here a practical 
attitude:we would like to consider a complexifier which comes close 
to the Maxwell one (\ref{2.11}), which obviously satisfies all 
the requirements of definition \ref{def2.1}. Since we have 
applications in General Relativity in mind, we must preserve background 
independence and, therefore, the Minkowski background Laplacian entering
(\ref{2.11}) must be replaced by something background metric independent.

One possibility is to use a background independent Laplacian, which 
depends on the dynamical 3-metric of $q_{ab}$ with triad 
$E^a_j/\sqrt{|\det(E)|}$. However, this would lead to a very complicated object
with which no practical calculations are possible. In fact, the 
practical use of coherent states in Maxwell theory rests on the fact 
that (\ref{2.11}) is quadratic in the momenta (electric fields) which 
leads to states that are basically Gaussians in both the position and 
the momentum representation.

This motivates to keep our complexifier quadratic in the momenta as 
well. Furthermore, we must preserve $G$ invariance. For Abelian 
gauge theories the electric fields are already gauge invariant but 
not for non Abelian gauge theories.

Thus a first attempt would be to define as complexifier
\be \label{2.15}
C\propto \int_\sigma \; d^3x\; q_{ab} \; E^a_j E^b_k 
\delta^{jk}/\sqrt{\det(q)}
\ee
where the background metric $\delta_{ab}$ in (\ref{2.11}) has been replaced 
by the dynamical metric and, in order to make (\ref{2.15}) spatially 
diffeomorphism invariant, we have included a density factor 
$1/\sqrt{\det(q)}$. However, it is easy to see that (\ref{2.15}) becomes 
\be \label{2.16} 
C\propto V=\int_\sigma \; d^3x\; \sqrt{|\det(E)|}
\ee
the volume functional. While it satisfies the requirements of a 
complexifier and admits a quantisation as a positive self-adjoint 
operator, its spectral decomposition is not analytically available,
so that $C=V$ is not practically useful.

Hence, what we need is a gauge invariant, background independent 
expression, quadratic in the
electric fields 
which preferably is non vanishing everywhere on $\sigma$ and 
which can be expressed in terms of (limits of) electric 
fluxes, since 
only those are well defined in the quantum theory. In \cite{21} it is 
shown that in non Abelian gauge theories no quadratic 
complexifier, based strictly on fluxes exists, that meets all these 
requirements. The way out is to give up the the requirement that the 
complexifier is composed out of the fluxes but to allow more general 
objects than fluxes. There are basically two proposals in the literature.
The first \cite{10} replaces fluxes by gauge covariant fluxes. The second 
\cite{21} replaces the fluxes by areas. We will review these two proposals 
separately.

In what follows we assume that, as in General Relativity, the canonical 
dimension of $E^a_j$ is cm$^0$ and that of $A_a^j$ is cm$^{-1}$ so that 
(\ref{2.17}) has dimension 
cm$^{D-1}$. Since the kinetic term in the canonical action is
$\int_{\mathbb{R}} dt \int_\sigma d^D E^a_j \dot{A}_a^j/\kappa$ 
it follows that $\hbar\kappa$ has dimension cm$^{D-1}$.

\subsubsection{Gauge Covariant Flux Complexifiers}
\label{s2.2.1}

Given a surface $S$ we select a point $p(S)\in S$.
Furthermore, for each point 
$x\in S$ we choose a path $\rho_S(x)\subset S$ within $S$ with beginning 
point $p(S)$ and ending point $x$. We denote the path system by ${\cal 
P}_S$. 
We recall that the gauge covariant flux of $E$ through 
$S$ subordinate to the path system ${\cal P}_S$ and the edge $e_S$ is 
defined by 
\be \label{2.16a}
E_j(S)\tau_j:=\int_S\; 
{\rm Ad}_{A(\rho_S(x))}((\ast E)(x))
\ee
Here $i\tau_j$ are the Pauli matrices, $\ast E=\frac{1}{2}\epsilon_{abc}
dx^a\wedge dx^b \; E^c_j \tau_j$ and Ad denotes the adjoint action of $G$ 
on its Lie algebra. Obviously, (\ref{2.16a}) transforms in the adjoint 
representation under gauge transformations at $p(S)$.

Let $\cal S$ be a collection of surfaces with associated path systems 
${\cal P}_S$ for each $S\in{\cal S}$ and 
$\mu$ a measure on $\cal S$. Let $K$ be any positive definite, measurable  
function on 
$\cal S$.
A gauge covariant flux complexifier (GCFC) is defined by 
\be \label{2.16b}
C:=\frac{1}{2 L^{D-1} \kappa} \int_{{\cal S}}\; d\mu(S) \;K(S)\;
[-\frac{1}{2}{\rm Tr}(E(S)^2)]
\ee
Here $L$ is a parameter of dimension of length and we assume both 
$\mu,\;K$ to be dimensionless.

The mostly studied case is when $D=3$ and $\cal S=C_2(P)$ is a discrete 
set of 
oriented surfaces 
which coincide with the faces (its sub 2-complex) of a polyhedronal cell 
partition $P$ of 
$\sigma$. In this case $\mu$ is just the counting measure and, for 
convenience, one chooses $K=1$. 
We will denote the associated complexifier by $C_P$. In this case the 
complexified connection is given by
\be \label{2.16c}
Z_a^j(x)=A_a^j(x)-\frac{i}{L^2}\sum_{S\in C_2(P)} 
\int_S\; \frac{1}{2}\epsilon_{abc} dy^b\wedge dy^c \;\
[{\rm Tr}(E(S)\;{\rm Ad}_{\rho_S(x)}(\tau_j))]
\; \delta(x,y)
\ee
Notice that the series involved in $Z_a^j$ terminates at the first term.
This is because when computing the second iterated Poisson bracket there
is a double sum over surfaces involved, however, since the paths 
$\rho_S(x)$ are disjoint from $S'$ for $S'\not= S$ 
there is no contribution from $S'\not=S$ to 
$\{C_P,A_a^j(x)\}_{(2)}$. For $S'=S$ there is, in principle, a contribution 
but by the regularisation \cite{1b} the classical flux does not 
Poisson act on paths lying in its associated surface.

This connection is distributional but, fortunately, we are only 
interested 
in the integral of (\ref{2.16c}) over one dimensional paths $e$ given by
\be \label{2.16e}
i L^2\int_e dx^a[Z_a^j(x)-A_a^j(x)]=\sum_{S\in C_2(P)}\;\sum_{x\in S\cap 
e} \; \sigma_x(S,e)
[{\rm Tr}(E(S)\;{\rm Ad}_{\rho_S(x)}(\tau_j))]
\ee
where 
\be \label{2.16f}
\sigma_z(S,e)=\frac{1}{2} \int_e\; dx^a\; \epsilon_{abc}\; \int_S \; 
dy^b\wedge dy^c \delta(x,y) \delta_{x,z}
\ee
is the signed intersection number at $z\in e\cap S$, which here we have 
assumed to be an interior point (otherwise there is an additional factor 
of $1/2$, see \cite{1b}).

\subsubsection{Area Compexifier}
\label{s2.2.2}

Let $\cal S$ be a collection of surfaces, $\mu$ a measure on $\cal S$ and 
$K(S,S')$ a positive definite integral kernel. An area complexifier 
is given by the expression 
\be \label{2.16g}
C=\frac{1}{a^{D-1} \kappa} 
\int_{{\cal S}}\; d\mu(S) \;
\int_{{\cal S}}\; d\mu(S) \; K(S,S')\; {\rm Ar}(S)\; {\rm Ar}(S')
\ee
where $a$ is a parameter of dimension of length. Here Ar$(S)$ is 
the gauge invariant ``modulus of the electric flux''
\be \label{2.17}
{\rm Ar}(S):=\int_S \; \sqrt{{\rm Tr}([\ast E]^2)}
\ee
which, in General Relativity, has the meaning of the {\it area of S}.

The most studied case arises from a diagonal and constant integral kernel
and suitable choices of $\cal S$ and $\mu$, respectively.
\begin{definition} \label{def2.2} ~\\
i) A stack $s$ in $\sigma$ is a D-dimensional submanifold with the topology
of $\mathbb{R}\times (0,1]^{D-1}$.\\
ii)
A stack family $S=\{s_\alpha\}$ is a partition of $\sigma$ into stacks 
which are mutually disjoint.\\
iii) 
D families of foliations $F_I,\;I=1,..,D$ of $\sigma$ generated by 
vector fields
$\partial/\partial t^I,\;I=1,..,D$ are said to be linearly independent 
if the vector fields $\partial/\partial t^I$ are everywhere linearly
independent.\\
iv) 
D stack families $S^I$ are said to be linearly independent, provided that 
there exist D linearly independent foliations $F_I$,
such that the leaves of the foliation 
$F_I$ is transversal to every 
stack in $S^I$. That is, the intersection $s^I_{\alpha t}$ of any leaf 
$L_{It},\;t\in 
\mathbb{R}$ of $F_I$ 
with any stack $s^I_\alpha$ in $S^I$, called a plaquette, has topology 
$(0,1]^{D-1}$.\\ 
v) 
The collection of the plaquettes $s^I{\alpha t}$ is called a 
parquette at time t within $L_{It}$.
\end{definition}
In general $\sigma$ will have to be partitioned into pieces each of which admits 
$D$ linearly independent foliations. Below we will construct the 
complexifier for one such piece, the complete complexifier is 
then the sum over the individual pieces. 

The complexifier defined by $D$ linearly independent stack families 
is now defined by
\be \label{2.18}
C:=\frac{1}{2\kappa a^{D-1}} \sum_{I=1}^D \sum_\alpha 
\int_{\mathbb{R}}\; 
dt\;
[{\rm Ar}(p^I_{\alpha t})]^2  
\ee
Here we take the foliation parameter $t$ to be dimensionless, $a$ is 
a parameter with dimension cm$^1$ so that $C/\hbar$ is dimension free
and $p^I_{\alpha t}=s^I_\alpha \cap L_{It}$ denotes the plaquette at 
time $t$ within the stack $s^I_\alpha$ in direction $I$.
For Abelian gauge theories also the following simpler expression is 
available      
\be \label{2.18a}
C:=\frac{1}{2\kappa a^{D-1}} \sum_{I=1}^D \sum_\alpha 
\int_{\mathbb{R}}\; 
dt\; [E_j(p^I_{\alpha t})]^2  
\ee
which uses the gauge invariant flux rather than the areas.

Let us now compute the complexified connections. Notice that, due to the 
fact that each stack is foliated by squares with half open and half 
closed boundaries, for each $x\in \sigma$ and each direction $I$ there 
exists a unique stack $s^I_\alpha(x)$ corresponding to a label 
$\alpha_I(x)$, such that $x\in s^I_\alpha$. 
Likewise, for each direction $I$ there exists a unique leaf $L_It(x)$ 
corresponding to a time $t_I(x)$, such that $x\in L_{It}$. \\
Consider 
the one parameter family of embeddings $X^I_{\alpha t}:\;[0,1)^{D-1}\to 
p^I_{\alpha t}$, then there exists a unique $u_I(x)$ such that 
$x=X^I_{\alpha_I(x) t_I(x)}(u_I(x))$. We now set
\ba \label{2.19}
J_I(x) &:=& |\det(\frac{\partial X^I_{\alpha t}(u)}{\partial (t, 
u)})|_{\alpha=\alpha_I(x),t=t_I(x),u=u_I(x)}
\nonumber\\
n_a^I(x) &:=& \frac{1}{(D-1)!} \epsilon_{ab_1 .. b_{D-1}}\; 
\epsilon_{l_1 
.. 
l_{D-1}} 
\frac{\partial X^{I b_1}_{\alpha t}(u)}{\partial u^{l_1}} ..
\frac{\partial X^{I b_{D-1}}_{\alpha t}(u)}{\partial u^{l_{D-1}}}
\ea
For the non Abelian
complexifier we find 
\be \label{2.20}
Z_a^j(x)=A_a^j(x)-\frac{i}{a^{D-1}} E^b_j(x)\sum_I \frac{n_b^I(x) 
n_a^I(x)}{J_I(x)}\;
\frac{{\rm Ar}(p^I_{\alpha_I(x) t_I(x)})}{\sqrt{[E^c_k(x) n_c^I(x)]^2}} 
\ee
while, for the Abelian one, we obtain 
\be \label{2.21}
Z_a^j(x)=A_a^j(x)-\frac{i}{a^{D-1}} \sum_I \frac{n_a^I(x)}{J_I(x)}\;
E_j(p^I_{\alpha_I(x) t_I(x)})
\ee
Notice that in both cases the imaginary part of $Z_a^j$ is only quasi 
local in $E^a_j$, that is, we can recover $E^a_j$ from $Z_a^j$ only up 
to the resolution provided by the parquettes.

\subsection{Coherent States for Background Independent Gauge Theories}
\label{s2.3}

We are now ready to compute the coherent states. The first step is to write 
the $\delta$ distribution as 
\be \label{2.22}
\delta_{A_0}=\sum_s \; T_s(A_0)\;  <T_s,.>
\ee
where the sum is over all spin network labels $s=(\gamma,\pi,m,n)$,
hence the coherent state is given by
\be \label{2.23}
\psi_Z=\sum_s \; T_s(Z) \; <e^{-\hat{C}/\hbar} T_s,.>
\ee
Here $\hat{C}$ is obtained by replacing in (\ref{2.16a}), (\ref{2.18}) 
or 
(\ref{2.18a}), the gauge covariant flux, area or flux 
functionals by the gauge covariant flux, area or flux operator 
\cite{10,7,37} respectively, which are 
positive, self-adjoint operators with pure point spectrum only. 

It remains to compute the action of the heat kernel and, for this purpose, 
we restrict to the case $D=3$. Again, we do this 
separately for the two types of complexifiers.

\subsubsection{Gauge Covariant Flux Coherent States}
\label{s2.3.1}

There is, in principle, an operator ordering problem involved in the 
quantisation of (\ref{2.16a}), however, the regularisation described in section \ref{s configurationspace}, 
shows that there is no action of the operator valued distribution 
$\ast E(x)$ on a holonomy $A(p)$ if $\ast E(x)$ is smeared over an 
infinitesimal surface element of a surface in which the path $p$ lies.
Let us introduce the matrices 
\be \label{2.23a}
O_{jk}(g):=-\frac{1}{2}\;{\rm Tr}(\tau_k {\rm Ad}_g(\tau_j))
\ee
where we have assumed the normalisation ${\rm Tr}(\tau_j 
\tau_k)=-2\delta_{jk}$. Since $G$ is compact, we can always embed into 
a subgroup of some $U(N)$ so that 
$\overline{\tau_j^T}=-\tau_j,\;\overline{g}^T=g^{-1}$, hence $O_{jk}(g)$ 
is real valued. Moreover, the identity 
$O_{jk}(g)=O_{kj}(g^{-1})$, as well as the fact that Ad acts on Lie$(G)$,
i.e. $Ad_g(\tau_j)=O_{jk}(g)\tau _k$, reveals that 
\be \label{2.23b}
O_{jk}(g) O_{jl}(g)=\delta_{kl}  
\ee
so that $g\mapsto O_{jk}(g)$ is a subgroup of $O(\dim(G))$. 

The known quantisation of the non gauge covariant flux \cite{1b,37}, 
together with the above mentioned trivial action on 
$O_{jk}(A(\rho_S(x))$, reveal that 
\be \label{2.23c}
\widehat{E_j(S)} T_{\gamma,j,m,n} = i\ell_P^2 \sum_{e\in E(\gamma)} \;
\sum_{x\in S\cap e}\; \sigma_x(S,e) \; O_{jk}(A(\rho_S(x)))\; 
\frac{1}{4} X^k_e \; T_{\gamma,j,m,n}
\ee
where $X^k_e$ is the right invariant vector field of $G$ acting on 
$g=A(e)$, specifically $X^k_e={\rm Tr}(\tau_j g \partial/\partial g^T)$.
Here we have assumed that the graph has been adapted to $S$ by suitable 
subdivisions of edges, such that each edge of $\gamma$ is either 
outgoing from an 
isolated 
intersection point or completely lies within $S$ or lies completely 
outside $S$. 

Formula (\ref{2.23c}) can now be plugged into (\ref{2.16b}). Since, again, 
there is no action of $\widehat{E(S)}$ on $\rho_S(x)$ we find 
\be \label{2.23d}
\widehat{E_j(S)}^2 T_{\gamma,j,m,n} = -\ell_P^4 \sum_{e,e'\in E(\gamma)} 
\;
\sum_{x\in S\cap e}\; \sigma_x(S,e) \;
\sum_{y\in S\cap e'}\; \sigma_y(S,e') 
\; O_{kl}(A(\rho_S(x)^{-1}\circ \rho_S(y)))\; 
\frac{1}{16} X^k_e X^l_{e'} \; T_{\gamma,j,m,n}
\ee

The appearance of the matrix $O_{kl}(A(\rho_S(x)^{-1}\circ \rho_S(y)))$
makes the computation of the spectrum of (\ref{2.23d}) rather difficult 
for a general graph. However, it becomes simple in case that the 
graph is such that the surface $S$ has only a single isolated 
intersection point $x$ with the graph. In that case (\ref{2.23d}) 
becomes
\be \label{2.23e}
\widehat{E_j(S)}^2 T_{\gamma,j,m,n} = -\ell_P^4 
[\sum_{e\in E(\gamma)} \;\sum_{x\in S\cap e}\; \sigma_x(S,e) \;
\frac{1}{4} X^j_e]^2 \; T_{\gamma,j,m,n}
\ee
One can now introduce, similar as done in \cite{37}, the vector fields 
\be \label{2.23f}
Y_S^{j \pm}=-i\sum_{\sigma_x(e,S)=\pm 1} X^j_e/2,\;   
Y_S^j=Y^{j +}_S+Y^{j -}_S
\ee
so that we obtain the linear combinations of Casimir operators   
\be \label{2.23g}
\widehat{E_j(S)}^2 T_{\gamma,j,m,n} = \frac{\ell_P^4}{4}
[2 (Y^{j +}_S)^2+2 (Y^{j -}_S)^2- (Y^j_S)^2] \; T_{\gamma,j,m,n}
\ee
$T_{\gamma, j, 
m,n}$ is gauge invariant at $x$ when $x$ is an interior point of a single edge
$e=(e_1)^{-1}\circ e_2$ intersected transversally by the surface $S$, so that 
$\sigma_x(S,e_1)=-\sigma_x(S,e_2)=\pm 1$. In this case equation (\ref{2.23g}) further simplifies to
\be \label{2.23h}
\widehat{E_j(S)}^2 T_{\gamma,j,m,n} = \ell_P^4\;
[-i X^j_e/2]^2 \; T_{\gamma,j,m,n}
\ee
For $G=U(1)^3$ or $G=SU(2)$ the eigenvalues of 
$(-iX^j_e)^2$ are given by $(n^j_e)^2$ and $j_e(j_e+1)$, respectively.
This special situation arises when $\gamma$ is a graph dual to the polyhedronal cell complex 
complexifier, i.e. there is 
precisely one edge $e$ of $\gamma$ which intersects a given face $S$ 
and it does so transversally. 

\subsubsection{Area Coherent States}
\label{s2.3.2}

For each direction $I$, each graph $\gamma$ and each stack $\alpha$, 
the Lebesgue measure 
of the set of times $t$, such that $p^I_{\alpha t}$ contains a vertex of 
$\gamma$ or that $p^I_{\alpha t}$ contains entire segments of edges of 
$\gamma$, vanishes. From the properties of the area operator and flux 
operator, it follows that these time points do not contribute to the heat 
kernel evolution and, therefore, we may assume, without loss of generality, 
that each $p^I_{\alpha t}$ intersects the edges of $\gamma$, at most, 
transversally in an interior point. Now consider in the non Abelian 
case for natural numbers 
$N_e\in \mathbb{N}_0$ the set 
\be \label{2.24}
S^{I \alpha \gamma}_N:=\{t\in \mathbb{R};\;
|p^I_{\alpha t}\cap e|=N_e\}
\ee
where we have used the following shorthand notation: $N:=\{N_e\}_{e\in E(\gamma)}$. This is the set 
of parquettes within stack $s^I_\alpha$, which intersect edge $e$ 
precisely $N_e$ times transversally. 
Likewise, consider in the Abelian case for integers $N_e \in 
\mathbb{Z}$ the set
\be \label{2.25}
S^{I \alpha \gamma}_N:=\{t\in \mathbb{R};\;
\sum_{x\in p^I_{\alpha t}\cap e} \sigma(p^I_{\alpha,t},e)_p=N_e\}
\ee
where, for any surface $S$ intersecting $e$ transversally, the number 
$\sigma(S,e)_p$ for $p\in S\cap e$ takes the value $+1$ or $-1$ 
if the 
orientations of $S$ and $e$ at $p$ agree or disagree, respectively.\\
This set represents the set of parquettes within stack $p^I_\alpha$ whose signed 
intersection number with edge $e$ is precisely $N_e$.

In both cases let
\be \label{2.26}
l^{I\alpha \gamma}_N:=\int_{S^{I \alpha}_N}\; dt
\ee
be the Lebesgue measure or {\it length} of those sets. These length 
functions are needed in order to define a cylindrically consistent 
family of heat kernels, as it was first observed in \cite{34}. Then 
the action of the complexifier on SNWF is diagonal, i.e.
\be \label{2.27}
\frac{\hat{C}}{\hbar} T_s =\lambda_s T_s
\ee
The corresponding eigenvalues, for $G=SU(2)$, are given by
\be \label{2.28}
\lambda_s=\frac{\ell_P^2}{2a^2}\sum_{I,\alpha}\; \sum_N\; 
l^{I \alpha \gamma}_N 
\; [\sum_{e\in E(\gamma)}\; N_e\; \sqrt{j_e(j_e+1)}]^2
\ee
while for $G=U(1)^3$ they are given by 
\be \label{2.29}
\lambda_s=\frac{\ell_P^2}{2a^2}\sum_{I,\alpha}\; \sum_N\; 
l^{I \alpha \gamma}_N 
\; [\sum_{e\in E(\gamma),j}\; N_e\; n^j_e]^2
\ee
Here we have used the fact that the irreducible, non trivial representations of 
$SU(2)$ 
are given by positive, half integral spin quantum numbers $j_e\not=0$, 
while for 
$U(1)^3$
they are given by triples of integers $n_e^j\not=0,\;j=1,2,3$. 
Furthermore, with $\kappa=8\pi G_{{\rm Newton}}$, $\ell_P^2=\hbar 
\kappa$ is the Planck area. The ratio $t:=\ell_P^2/a^2$ is known as the 
classicality parameter. Without dynamical input this is a free 
parameter for our coherent states, that decides up to which scale 
the fluctuations of operators are negligible.

\subsection{Gauge Covariant Flux versus Area Coherent States}
\label{s2.5}

Consider the case that the plaquttes are much smaller than 
the edges with respect to the three metric to be approximated by the 
coherent states and, that, the edges do not wiggle much on the scale of 
the plaquettes. Then, for each direction $I$ the number of 
stacks that do not 
contain a vertex of $\gamma$, but still intersect $\gamma$ drastically, 
outnumbers the number of stacks 
that do contain a vertex. 

Moreover, among the vertex free stacks, the 
number of stacks 
that intersect only one 
edge completely outnumbers the ones that intersect more than one edge. 
Finally,
among those with single edge intersections, the number of stacks that 
intersect the respective edge 
once, completely outnumbers the ones that do so more than once. Therefore, 
in these cases the expressions (\ref{2.28}) and (\ref{2.29})
can be replaced with good approximation by simpler expressions of the 
form 
\be \label{2.33}
\lambda_s=\frac{\ell_P^2}{2a^2}
\sum_{e\in E(\gamma)}\; l^\gamma_e \; j_e(j_e+1)
\ee
and 
\be \label{2.34}
\lambda_s=\frac{\ell_P^2}{2a^2}
\sum_{e\in E(\gamma)}\; l^\gamma_e \; [n_e^j]^2
\ee
respectively, where the length function $l^\gamma_e=g^\gamma_{ee}$ solves 
$l^\gamma_{e\circ 
e'}=l_e+l_{e'},\;l_{e^{-1}}=l_e$ in order that the complexifier 
has cylindrically consistent projections. 
This is the form of the heat kernel eigenvalue considered for the 
states in \cite{34}. As shown in \cite{21}, these eigenvalues  
cannot come from a known classical
complexifier, so that the complexification map $A\mapsto Z$, without 
which the $Z$ label of the coherent state has no relation to the phase 
space point to be approximated, is unknown. When using 
the 
complexifier coming from a polyhedronal cell complex, a concrete 
relation between $Z$ and the phase space can be given for 
specific 
graphs, then the above eigenvalues arise as we saw in section \ref{s2.3.1},
(\cite{21}).

Let us also check that the area complexification map $Z$ in 
(\ref{2.20}) and (\ref{2.21}) comes close to the gauge covariant flux 
one (\ref{2.16e}), at least on certain graphs. 
Let $\gamma$ be a graph dual to the cell complex $P$. Thus, for each 
edge $e$ there is a unique face $S_e$ which intersects $e$ in an 
interior point transversally, such that $\sigma_{S_e\cap e}(S_e,e)=+1$
and no other face intersects $e$.
Then the gauge covariant flux complexification map, at the 
level 
of the holonomies is given by \cite{10}
\be \label{2.35}
A(e)\mapsto g_e(Z):=Z_\gamma(e)=\exp(-i\tau_j E_j(S_e)/L^2)\; A(e) 
\ee
For $SU(2)$, $i\tau_j$ are the Pauli matrices while for 
$U(1)^3$ $i\tau_j=1,\;j=1,2,3$.  In contrast, the area complexification 
map
is given, at the level of the holonomies, by
\be \label{2.36}
A(e)\mapsto Z(e)={\cal P} \exp(\int_e Z^j \tau_j)
\ee
where $\cal P$ denotes path ordering and $Z_a^j$ is given in 
(\ref{2.20}) and (\ref{2.21}) for non Abelian and Abelian cases, respectively. Now for sufficiently 
``short'' edges we have $Z(e)\approx \exp(\int_e [Z-A]^j\tau_j) A(e)$
to leading order in the edge parameter length. 
If we assume that $E^a_j$ is slowly varying at the scale of the 
plaquettes, then we have ${\rm Ar}(p^I_{\alpha_I(x) t_I(x)}) \approx
\sqrt{[E^a_j(x) n_a^I(x)]^2}$ so that 
(\ref{2.20}) is approximated by  
\be \label{2.37}
\int_e (Z^j-A^j)\approx -\frac{i}{a^2}\sum_I 
\int_0^1 \; \frac{\dot{e}^a(t) n_a^I(e(t))}{J_I(e(t))} \; 
n_b^I(e(t)) E^b_j(e(t))
\ee
where we have assumed that $e$ is the embedded interval $[0,1]$.
Now consider the case that the graph is cubic and that the stack 
family and the graph are aligned as follows:\\
suppose that we have an embedding
$X:\; \mathbb{R}^3 \to \sigma; s \mapsto X(s)$. For $\epsilon_{IJK}=1$ 
we define $X^I_t(u^1,u^2):=X(s^I=t,s^J=u^1,s^K=u^2)$. 

This determines 
linearly independent 
foliations $F^I$ with leaves $L_{It}=X^I_t(\mathbb{R}^2)$. 
The corresponding stack families are labelled by 
$\alpha:=(\alpha^1,\alpha^2)\in \mathbb{Z}^2$ and defined by 
$X^I_{\alpha t}:\; [0,1)^2 \to \sigma;\;X^I_{\alpha t}(u):=
X^I_t([\alpha^1+u^1]l,[\alpha^2+u^2]l)$ where $l>0$ is a 
certain parameter. The edges of the 
cubic graph are labelled by vertices $v=(v^1,v^2,v^3)\in \mathbb{Z}^3$ 
and directions $I$ and are defined for $\epsilon_{IJK}=1$ by 
$e_{v,I}:\;[0,1]\to \sigma; e_{v I}(t):=X(s^I=[v^I+t]\delta, 
s^J=v^J\delta, s^K=v^K\delta)$ where $\delta>0$ is another 
parameter. 

In this situation, (\ref{2.37}) can be further simplified to
\be \label{2.38}
\int_{e_{v I}} (Z^j-A^j)\approx 
-\frac{i}{a^2} \delta\;
\int_0^1 \;  n_b^I(e_{v I}(t)) E^b_j(e_{I v}(t))
\approx -\frac{i}{a^2} \delta E_j(p^I_v)
\ee
where $p^I_v$ is any plaquette in the stack in the $I$ direction intersected 
by $e_{I v}$. 

Thus, for cubic 
graphs, which are the only ones considered so far in semiclassical 
calculations, we get a close match 
between (\ref{2.35}) and (\ref{2.38}), whenever the cubic graph and the 
stack families are aligned. However, there is still an 
important difference, namely:\\ 
the parameter area $l^2$ of the plaquette $p^I_v$ in (\ref{2.38}) 
has no a priori relation to 
the parameter length $\delta$ of the edge $e_{v I}$, while the 
parameter area of the dual face $S_{e_{v I}}$  
in (\ref{2.35}) is of the order $\delta^2$. 
These considerations reveal that the individual plaquettes 
of the stacks cannot be interpreted as the faces of a dual graph
although, roughly, $[\delta/l]^2$ of them combine to a face. Hence the 
states considered in \cite{21} are genuinely different from those in 
\cite{10}.\\
This will 
turn out to be important.

In fact, as will be discussed later, in order to be able to perform practical calculations
for $SU(2)$ with off diagonal edge metrics, we need $l\ll \delta$ so that the edge metric is close 
to diagonal for generic graphs. 
It turns out that if we use the same 
parameter $a$ both in the label $Z$ of the state and for the classicality 
parameter $t=\ell_P^2/a^2$, then the expectation value of the volume 
turns out to be 
of the order of $(l/\delta)^3$ too small. Hence, there is a tension 
between the possibility to perform practical calculations and the 
correctness of the classical limit. 

The only analytical calculation 
possible with $l=\delta$ uses a graph which is aligned with the stacks 
and thus is necessarily cubic. While the result of that calculation 
results 
in the correct classical limit, this calculation is of limited interest 
because we have already seen above that for this case the coherent states 
of \cite{21} reduce to those of \cite{10} for which we knew already that 
the classical limit is correct. 

However, if one wants to test the semiclassical limit 
for graphs of non cubic topology, this can be done with the states of 
\cite{10} without limitation. On the other hand, with the states of \cite{21} this is 
possible if we 
redefine $Z_a^j \to A_a^j+\frac{a^2}{b^2}(Z_a^j-A_a^j)$ where $b\ll a$.
This rescaling is actually not in the spirit of the complexifier 
programme, but it repairs the semiclassical limit of {\it all operators} 
built from the fluxes.   
It will then turn out that for graphs that satisfy $l/\delta=b/a$ the 
correct classical limit results for $n=6$ only. 

As already mentioned 
in the introduction, one could rescale the label of the coherent state 
by a different amount, in order to reach the correct semiclassical limit 
of the volume operator for one and only one $n\not=6$. However, that 
would destroy the correct semiclassical limit of other operators such as 
areas. Hence the rescaling by $(b/a)^2$ is harmless in the sense that 
it reproduces the semiclassical limit of all operators, while 
$n-$dependent rescalings do not. 

Also with respect to the states of \cite{10} the value $n=6$ is 
singled out. The fact that the cut off states of \cite{21} have 
acceptable semiclassical behaviour, only when both the corresponding cut off 
graph and the label of the coherent state satisfy certain restrictions 
imposed by the structure that defines the complexifier, in this case, the 
size of the parquettes is similar to the restrictions imposed 
by the polyhedronal cell complex complexifier \cite{10}, namely that the 
graph be dual to it.

\subsection{Cut -- Off Coherent States}
\label{s2.4}

Formulae (\ref{2.23}), (\ref{2.23d}) (\ref{2.28}) and (\ref{2.29}) 
display the 
coherent states in closed form. Unfortunately, although the eigenvalues 
of the heat kernel grow quadratically with the representation weight,
these states are still not normalisable because the Hilbert space is not 
separable, or in other words, the SNWF are labelled by the continuous 
parameter $\gamma$. In view of the uniqueness result when 
insisting on background independence, the non separability is 
not avoidable and one must accept it. The observation is 
that (\ref{2.23}) defines a 
well defined 
distribution on the dense subset of $\cal H$, consisting of the finite 
linear span of SNWF. To 
extract normalisable information from $\psi_Z$ we introduce the notion 
of a cut -- off state labelled by a graph $\gamma$. These are defined by
\be \label{2.30}
\psi_{Z,\gamma}:=\sum_{s; \gamma(s)\subset \gamma}\; T_s(Z)\; 
<e^{-\hat{C}/\hbar} T_s,.>
\ee
That is, the sum over all spin networks $s=(\gamma(s),\pi(s),m(s),n(s))$ 
is truncated or {\it cut off} to those whose
graph entry $\gamma(s)$ is a subgraph of the given $\gamma$. The Ansatz 
is then to use $\psi_{Z,\gamma}$ for suitable $\gamma$ as a 
semiclassical 
state.    

Notice that both (\ref{2.28}) and (\ref{2.29}), respectively can be 
rewritten in the form
\be \label{2.31}
\lambda_s=\frac{t}{2}\sum_{e,e'}\; l^\gamma_{e,e'}\; 
\sqrt{j_e(j_e+1)}\; \sqrt{j_{e'}(j_{e'}+1)}
\ee
and 
\be \label{2.32}
\lambda_s=\frac{t}{2}\sum_{e,e'}\; l^\gamma_{e,e'}\; 
n^j_e\; n^j_{e'}
\ee
where the {\it edge metric} 
\be \label{2.32a}
l^\gamma_{e,e'}=
\sum_{I,\alpha}\; \sum_N\; 
l^{I \alpha \gamma}_N\; N_e N_{e'}
\ee
has entered the stage. Such non diagonal edge metrics have already 
appeared in other background dependent contexts \cite{23,24}. 
The edge metric decays quickly off 
the diagonal because, for most edge pairs $e\not=e'$, there is no 
direction 
and no stack in that direction intersecting both $e,e'$, which means 
that $l^{I\alpha \gamma}_N=0$ for $N_e,\;N_{e'}\not=0$ for such edge 
pairs. It is for this 
reason that we will be able to actually carry out our calculations.   

Using the edge metric, formulas (\ref{2.20}), (\ref{2.21}) and 
(\ref{2.28}), (\ref{2.29}) admit an interesting reformulation: \\
the {\it signed intersection number} between a path $e$ and a surfaces 
$S$ is defined by (adopting convenient parametrisations)
\ba \label{2.32b}
\sigma(S,e) &:=& \int_e dx^a\;\int_S \; dy^b \; dy^c\; 
\frac{1}{2}\;\epsilon_{abc}
 \;
\delta(x,y)
=\int_0^1\; dt\int_{[0,1]^2}\; d^2u\;
[\epsilon_{abc} \dot{e}^a(t)
\frac{\partial S^b(u)}{\partial u^1}
\frac{\partial S^c(u)}{\partial u^2}]\; \delta(e(t),S(u))
\nonumber\\
&=& \sum_{x\in S\cap e}\; \sigma_x(S,e)
\ea
while the {\it intersection number} is given by
\be \label{2.32c}   
|\sigma|(S,e):=
=\int_0^1\; dt\int_{[0,1]^2}\; d^2u\;
|\epsilon_{abc} \dot{e}^a(t)
\frac{\partial S^b(u)}{\partial u^1}
\frac{\partial S^c(u)}{\partial u^2}|\; \delta(e(t),S(u))
\ee
Both expressions can be regularised in such a way that entire segments 
of $e$, that lie inside $S$, do not contribute to the integral \cite{37}.
Notice that $|\sigma|(e,S)\not=|\sigma(e,S)|$, then it is not difficult 
to see that for $SU(2)$
\be \label{2.32d}
l^\gamma_{e,e'}=\sum_{\alpha,I} \; \int \; dt\; 
|\sigma|(e,p^{\alpha I}_t)\; |\sigma|(e',p^{\alpha I}_t)
\ee
while for $U(1)^3$ 
\be \label{2.32e}
l^\gamma_{e,e'}=\sum_{\alpha,I} \; \int \; dt\; 
\sigma(e,p^{\alpha I}_t)\; \sigma(e',p^{\alpha I}_t)
\ee
To verify (\ref{2.32d}), (\ref{2.32e}) it is easier to use directly
the action of non Abelian area and Abelian flux operators, respectively 
on the corresponding SNWF \cite{37} (with only transversal 
intersections)
\ba \label{2.32f}
{\rm Ar}(S) T_{\gamma,j,m,n} &=& \ell_P^2 \;[\sum_{e\in E(\gamma)}\; 
|\sigma|(e,S)\; \sqrt{j_e(j_e+1)}]\; T_{\gamma,j,m,n}
\nonumber\\
E_j(S) T_{\gamma,n} &=& \ell_P^2 \;[\sum_{e\in E(\gamma)}\; 
\sigma(e,S)\; n_e^j]\; T_{\gamma,n}
\ea 
and to plug this formula into the expression for $C$. An alternative 
proof is by realising that in the non Abelian or Abelian case, 
respectively 
\be \label{2.32g}
\chi_{S^{\alpha I}_{N}}(t)=\prod_{e\in E(\gamma)} 
\delta_{|\sigma|(p^{\alpha I}_t,e),N_e},\;\;
\chi_{S^{\alpha I}_{N}}(t)=\prod_{e\in E(\gamma)} 
\delta_{\sigma(p^{\alpha I}_t,e),N_e}
\ee
where $\chi_S$ denotes the characteristic function of a set. When 
plugging (\ref{2.32g}) into (\ref{2.32a}) and solving the Kronecker 
$\delta$'s when carrying out the sum over the integers $N$, one obtains
(\ref{2.32d}) and (\ref{2.32e}) respectively. 

From the easily verifiable properties of the (signed) intersection 
numbers
\ba \label{2.32h}
\sigma(e\circ e',S)&=&\sigma(e,S)+\sigma(e',S),\;
\sigma(e^{-1},S)=-\sigma(e,S);\nonumber\\
|\sigma|(e\circ e',S)&=&|\sigma|(e,S)+|\sigma|(e',S),\;
|\sigma|(e^{-1},S)=|\sigma|(e,S)
\ea
it follows immediately that 
\be \label{2.32i}
l^\gamma(e\circ e',e\circ 
e')=l^\gamma(e,e)+l^\gamma(e',e')+2l^\gamma(e,e'),\;\;
l^\gamma(e^{-1},e^{-1})=l^\gamma(e,e)
\ee
This is precisely the generalisation to non diagonal edge metrics
of the cylindrical consistency conditions of the complexifier 
\cite{21,34}. Notice that for the general area complexifier 
(\ref{2.16g}) 
we arrive instead at the edge metrics
\ba \label{2.32l}
l^\gamma_{e,e'} &=&\int_{{\cal S}}\;d\mu(S)\; \int_{{\cal S}}\;d\mu(S') \;
|\sigma|(S,e)\;K(S,S')\; |\sigma|(S',e'),\nonumber\\
l^\gamma_{e,e'} &=&\int_{{\cal S}}\;d\mu(S) \;\int_{{\cal S}}\;d\mu(S) \; 
\sigma(S,e)\;K(S,S')\; \sigma(S',e')
\ea

Finally we have for any edge $e$
\be \label{2.32j}
\int_e\;  dx^a\; ia^2[Z_a^j-A_a^j](x) = \sum_{I,\alpha} \int\; dt\; 
{\rm Ar}(p^{I \alpha}_t)
\int_0^1 \; ds\;
\frac{(n_c^I E^c_j)(e(s))}{\sqrt{[(n_b^I E^b_j)(e(s))]^2}}\;
\; \int d^2u \;[\dot{e}^a(s) n_a^{\alpha I t}(u) \delta(p^{\alpha 
I}_t(u),e(s))]
\ee
in the non Abelian case while for the Abelian case 
\be \label{2.32k}
\int_e\;  dx^a\; ia^2[Z_a^j-A_a^j](x) = \sum_{I,\alpha} \int\; dt\; 
E_j(p^{\alpha I}_t) \;\sigma(p^{\alpha I}_t,e)
\ee
Interestingly, if $E$ does not vary too much on the scale of a 
plaquette, then (\ref{2.32j}) actually reduces to (\ref{2.32k}), which is 
written directly in terms of the signed intersection number and 
plaquette fluxes. This will be useful later on when we compute 
expectation values.

\subsection{Replacing $SU(2)$ by $U(1)^3$}
\label{s2.6}

The considerations of previous sections have revealed that practically useful cut -- 
off states will be based on graphs, which are much coarser than the 
parquets so that the edge metric is diagonal in very good approximation.
We will restrict to such graphs in the calculations that follow and find 
independent confirmation for that restriction, as well in the form of the 
quality of the semiclassical approximation. Assuming exact diagonality
and thus suppressing the corrections coming from off -- diagonality, 
which we will show to be small under the made coarseness assumptions,
the cut -- off states in fact factorise
\be \label{2.39}
\psi_{Z, \gamma}=\prod_{e\in E(\gamma)} \; \psi_{Z,\gamma,e}
\ee
where for $SU(2)$
\be \label{2.40}
\psi_{Z,\gamma,e}(A)=\sum_{2j=0}^\infty\; (2j+1)\; 
e^{-\frac{t}{2} l^\gamma_e j(j+1)} \;
\chi_j(g_e A(e)^{-1})
\ee
while for $U(1)^3$
\be \label{2.41}
\psi_{Z,\gamma,e}(A)=\sum_{n\in \mathbb{Z}^3}\;  
e^{-\frac{t}{2} l^\gamma_e \sum_j (n^j)^2} \;
\chi_n(g_e A(e)^{-1})
\ee
Here $\chi_j$ and $\chi_n$ denote the character of the 
$j-$th and $n-$th irreducible 
representation of $SU(2)$ and $U(1)^3$, respectively. 

Under the 
assumptions made above, the edge metrics 
$l^\gamma_e$ are identical for both groups because, while 
$l^{I \alpha \gamma}_N$ is defined for non negative integers $N$ only in 
the 
case of $SU(2)$ while for $U(1)^3$ all integers are allowed, for the 
graphs under consideration for each edge $e$ only either $N_e=+1$ or 
$N_e=-1$ leads to non vanishing $l^{I \alpha \gamma}_N$ so that these 
numbers, in fact, coincide and since we take the diagonal elements of the 
edge metric (\ref{2.32a}) both signs lead to the same $l^\gamma_e$.

Finally, if $(A_0, E_0)$ is the phase space point to be approximated 
and from which we calculate $Z=Z(A_0,A_0)$ via 
(\ref{2.20}) and (\ref{2.21}), then for $SU(2)$ we have 
\be \label{2.42}
g_e\approx \exp(-i\tau_j P_0^j(e)) \; \exp(\tau_j \int_e A_0),\;\;
P_0^j(e)=\frac{1}{b^2} \sum_I \int_0^1\; dt \; \frac{\dot{e}^a(t) 
n_a^I(e(t))}{J_I(e(t))}\; [E^b_{0j}(e(t)) n_b^I(e(t))]\;    
\ee
while for $U(1)^3$ we have 
\be \label{2.43}
g_e=(g_e^j)_{j=1}^3,\;\; g_e^j=\exp(-P_0^j(e)+i\int_e A^j_0) 
\ee
where, as before, we have made the approximation 
\be \label{2.44}
\frac{{\rm Ar}(p^I_{\alpha_I(x) t_I(x)})}{\sqrt{[E^c_k(x) n_c^I(x)]^2}} 
\approx 1
\ee
which is valid if $E_0$ is slowly varying at the scale of the 
plaquettes.\\
\\
Thus, given $Z=Z(A_0,E_0)$, we have the following abstract situation 
under the 
made assumptions:\\
1) For each edge $e$ there exist vectors $P_0^j(e), A_0^j(e)$ such that for 
$SU(2)$ we have \\
$g_e\approx \exp(-i\tau_j P_0^j\tau_j) \exp(\tau_j 
A^j_0(e))\in SL(2,\mathbb{C})=SU(2)^{\mathbb{C}}$ while for $U(1)^3$ we 
have \\
$g_e=(e^{-P^j_0(e)+iA_0^j(e)})_{j=1}^3\in 
(\mathbb{C}-\{0\})^3=(U(1)^3)^{\mathbb{C}}$.\\
2) The coherent states adopt, approximately, the product form 
$\psi_{Z,\gamma}\approx \prod_{e\in E(\gamma)} \; \psi_{g_e}$ where 
\be \label{2.45}
\psi_g(h)=\sum_{2j=0}^\infty\; (2j+1)\; e^{-t l^\gamma_e j(j+1)/2}\;
\chi_j(g h^{-1})
\ee
for $h\in SU(2)$ while
\be \label{2.46}
\psi_g(h)=\sum_{n\in \mathbb{Z}^3}\; e^{-t l^\gamma_e 
\sum_{j=1}^3 n_j^2}\;
\chi_n(g h^{-1})
\ee
for $h\in U(1)^3$.\\
\\
Now, as anticipated in the introduction, using the tools of 
semiclassical perturbation theory \cite{18} we are able to calculate 
the expectation value of the volume operator $V$ of LQG, with respect to 
the 
correct $SU(2)$ coherent states, in terms of the 
expectation value of a certain operator $Q$. Here $V=\root 4 
\of{Q}$,  
which we display 
explicitly in the next section and which is a sixth order polynomial
in the right invariant vector fields $X^j_e$ on $SU(2)$, where $X^j_e$ 
acts on $h_e$ in (\ref{2.45}).

The crucial
observation, made in \cite{10}, is that if we simply replace the 
$SU(2)$ right invariant vector fields in $Q$, by $U(1)^3$ right 
invariant vector fields $X^j_e$ acting on $h_e$ in (\ref{2.46}) and, if 
we replace the $SU(2)$ coherent states (\ref{2.45}) by the related 
$U(1)^3$ 
coherent states in (\ref{2.46}), then the remarkable fact is that {\it 
the expectation values of polynomials of right invariant vector 
fields actually 
coincide to zeroth order in $\hbar$}. By the same argument, this will 
be also true if we perform the right invariant vector field replacement 
already at the level of $V$ rather than $Q$. This observation was also key 
in the semiclassical analysis of \cite{46g, 15, 20}. 

This feature is maybe not 
as surprising as it looks at first sight because, after all, the coherent 
states for both groups have to approximate the same phase space points. 
The underlying reason is  
that the classical phase space of the $SU(2)$ theory (i.e. the range of 
fields and the 
symplectic 
structure) and of the fictive $U(1)^3$ theory actually coincide. It is 
only when 
we add the dynamics of the theory, as for instance the Gauss constraint,
that we see a difference. The Gauss law is taken into account in two 
ways, first by using 
the appropriate group coherent states, here $SU(2)$ or $U(1)^3$ 
respectively, which is dictated by the fact that the underlying 
holonomies take values in the appropriate group. Secondly, one can 
construct quantum Gauss constraint invariant coherent states 
\cite{10,38} by averaging over the gauge group action at the vertices.
Denote this group averaging map by $\eta$. Then, as shown in 
\cite{10,38}, we have that $<\eta(\psi_{Z,\gamma}, A 
\eta(\psi_{Z,\gamma})>$
and $\psi_{Z,\gamma}, A\psi_{Z,\gamma}>$ agree to zeroth order in 
$\hbar$ (notice that the Gauss invariant Hilbert space is an honest 
subspace of the kinematical Hilbert space so that the same inner 
product can be used) for any Gauss invariant
operator $A$ such as the volume operator, because the overlap function 
between coherent states, peaked at different phase space points, is 
sharply peaked\footnote{In more detail we have 
\be \label{2.47}
\eta(\psi_{Z,\gamma})=\int_{G^{|V(\gamma)|}}\; \prod_{v\in 
V(\gamma)} 
\; d\mu_H(g_v)\; \alpha_g(\psi_{Z,\gamma})
\ee
where 
$\alpha_g(\psi_{Z,\gamma})(A)=\psi_{Z,\gamma}(\alpha_g(A))$ and 
$[\alpha_g(A)](e)=g(b(e)) A(e) g(f(e))^{-1}$ where $b(e)$ and $f(e)$ 
respectively denote beginning and final point of $e$, respectively.
Now, due to gauge covariance of the coherent states we have 
$\alpha_g(\psi_{Z,\gamma})=\psi_{\alpha_{g^{-1}}(Z),\gamma}$ so that
the gauge invariant coherent state expectation value of a gauge 
invariant operator becomes (using the invariance properties of the 
Haar measure)
\be \label{2.48}
\frac{<\eta(\psi_{Z,\gamma}), A\eta(\psi_{Z,\gamma})>}
{||\eta(\psi_{Z,\gamma})||^2}
=\frac{  \int_{G^{|V(\gamma)|}}\; \prod_{v\in V(\gamma)} \; d\mu_H(g_v)\;
<\psi_{\alpha_g(Z),\gamma},A\psi_{Z,\gamma}>}
{\int_{G^{|V(\gamma)|}}\; \prod_{v\in V(\gamma)} \; d\mu_H(g_v)\;
<\psi_{\alpha_g(Z),\gamma},\psi_{Z,\gamma}>}
\ee
From \cite{10} we know, for gauge invariant polynomials $A$ in 
right invariant vector fields, that the  
peakedness property 
\be \label{2.49}
<\psi_{Z',\gamma},A \psi_{Z,\gamma}>=
\frac{<\psi_{Z,\gamma}, A \psi_{Z,\gamma}>}{||\psi_{Z,\gamma}||^2}
\;<\psi_{Z',\gamma},\psi_{Z,\gamma}>\;[1+O(\hbar)]
\ee
holds. Now the claim is immediate.}. This justifies the use 
of the kinematical states when analysing semicalssical properties.\\
\\
\\
So far we have showed that using kinematical $U(1)^3$ coherent states is a convenient approximation 
for actual $SU(2)$ coherent state expectation value calculations for 
Gauss invariant operators if one 
is only interested in the zeroth order in $\hbar$. At non vanishing 
orders in $\hbar$ there will be differences but we are not interested in 
them in this context. One may wonder whether the argument 
made above, namely using kinematical rather than Gauss invariant 
coherent states also survives when considering the spatial 
diffeomorphism constraint. This issue, currently 
under investigation, is more complicated in part because it is not 
completely obvious which distributional extension of the classical 
diffeomorphism group one should use \cite{39}. However, since we are looking at the {\it 
local} volume operator which is not spatially diffeomorphism 
invariant, expectation value calculations with respect to 
spatially diffeomorphism invariant coherent states are meaningless. It 
is the local volume which enters the Hamiltonian and Master constraint 
and verifying the semiclassical limit of those only makes sense at the
kinematical Hilbert space level (one cannot check the correct classical 
limit of a constraint on its kernel). Once this limit is verified, one 
has 
confidence that the physical Hilbert space defined by the Hamiltonian 
constraint is correct.
\section{Regular Simplicial, Cubical and Octahedronal Cell Complexes}
\label{s3}
In this section we will describe a general method of how to embed 
graphs of valence $n=4,6,8$ with respect to the stack
families. This can be done by starting from regular dual 
simplicial 
(tetrahedronal), cubical and octahedronal partitions of the three 
manifold $\sigma$. For the definition of coherent states of \cite{10} this embedding 
is not needed except that it shows the existence of (regular) polyhedral
cell complexes dual to $n=4,6,8$ valent graphs such that all cells of 
that complex are platonic solid bodies, i.e. tetrahedra, cubes and 
octahedra respectively.

\begin{figure}[hbt]
\begin{center}
 \includegraphics[scale=0.3]{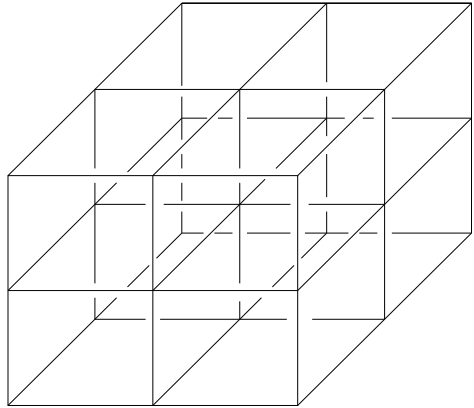}
\caption{Cubic cell decomposition.}
 \label{fig2}
\end{center}
\end{figure}

\begin{figure}[hbt] 
\begin{center}
 \includegraphics[scale=0.3]{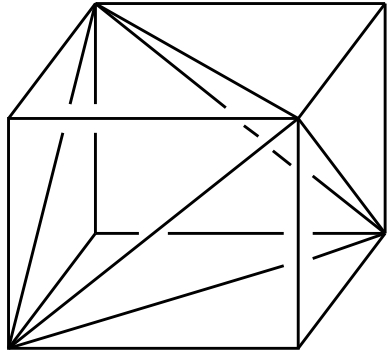}
\caption{Type A triangulation of a cube.}
\label{fig3}
\end{center}
 \end{figure}

\begin{figure}[hbt] 
\begin{center}
 \includegraphics[scale=0.3]{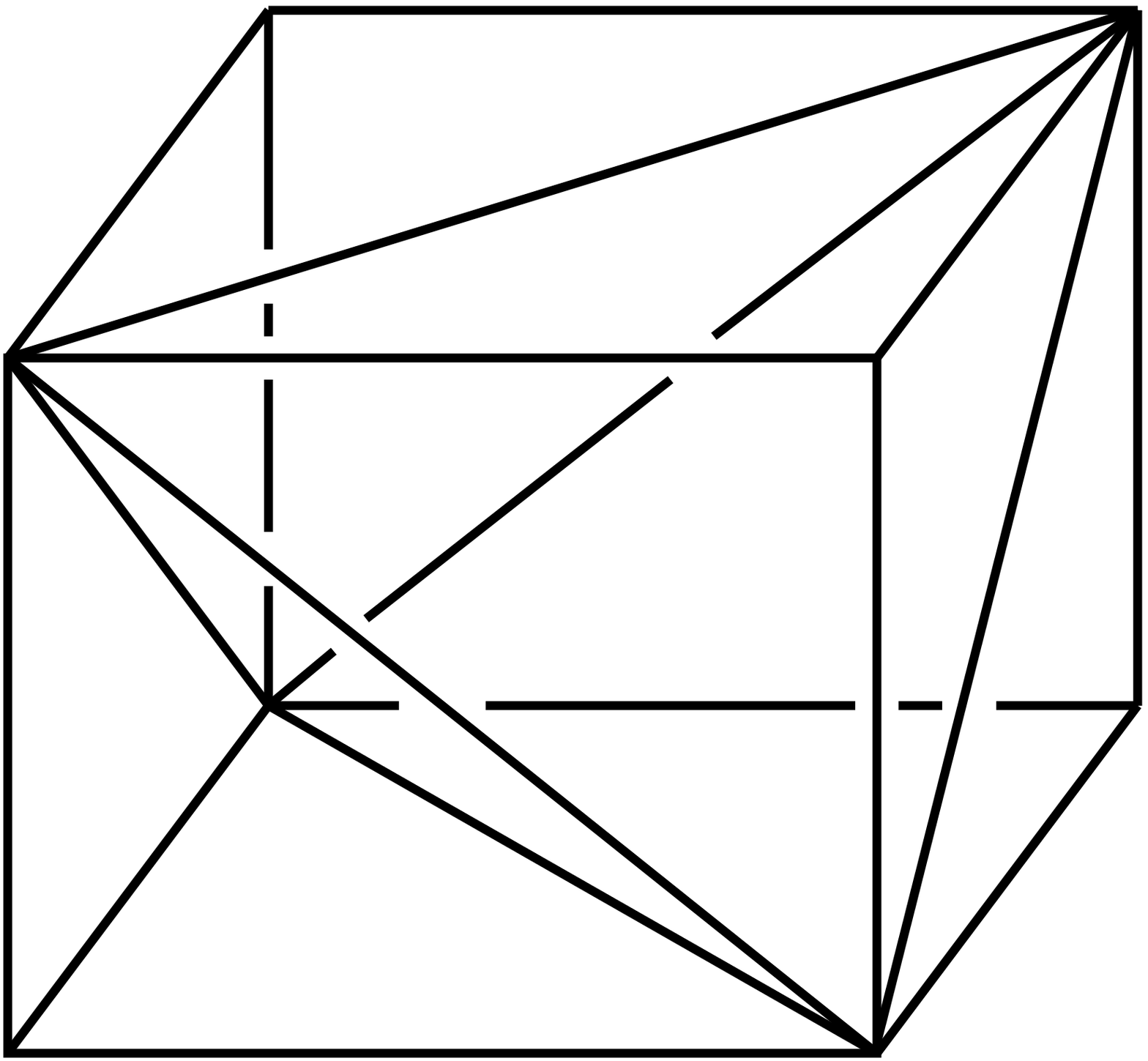}
\caption{Type B triangulation of a cube.}
\label{fig4}
\end{center}
 \end{figure}

\begin{figure}[hbt] 
\begin{center}
 \includegraphics[scale=0.3]{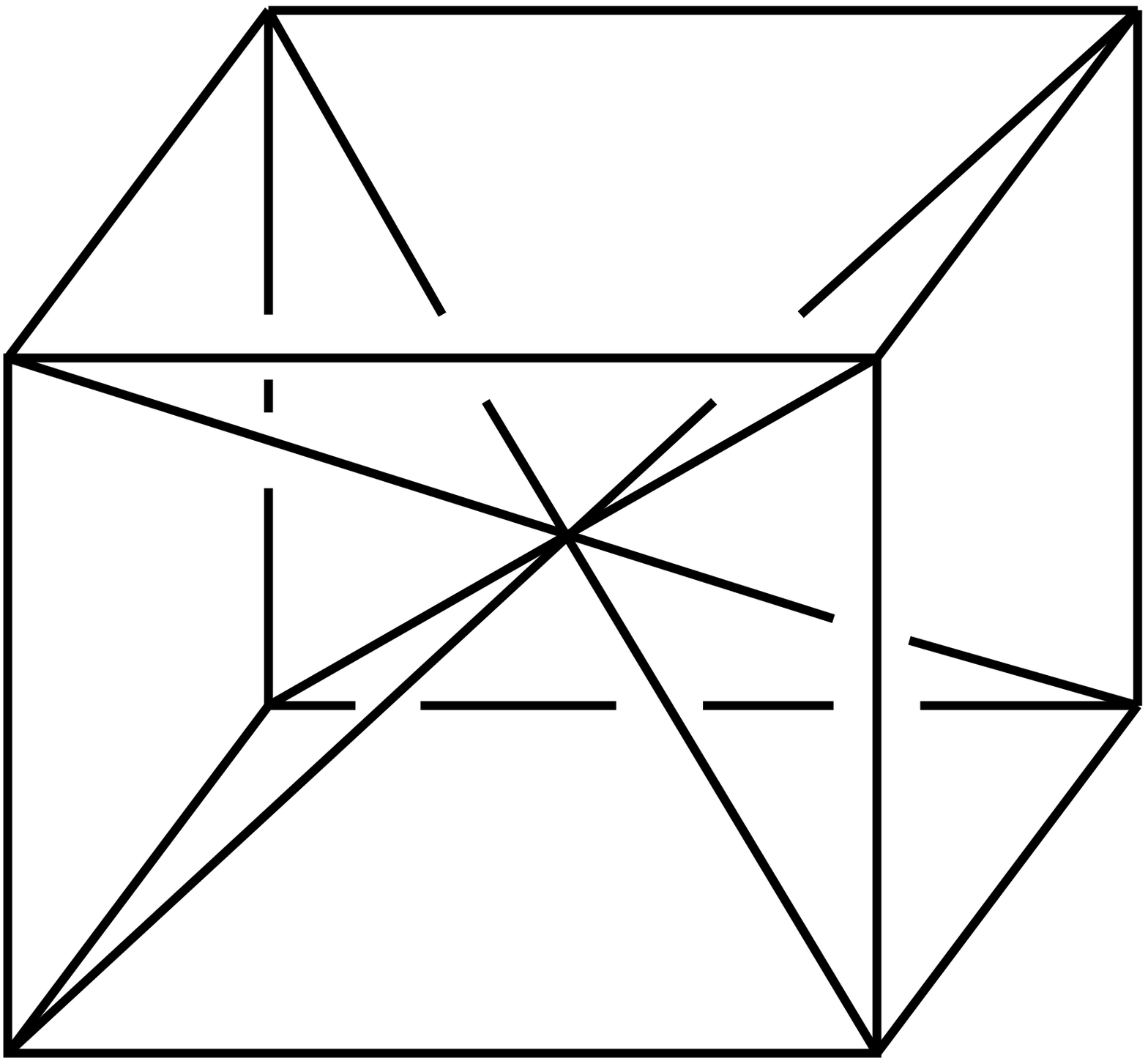}
\caption{Decomposition of a cube into six pyramids.}
\label{fig5}
\end{center} \end{figure}

In fact, it is possible to define such partitions all from refinements 
of cubical decompositions such as sketched in figure \ref{fig2}.
We perform the analysis 
for each chart 
$X:\;\mathbb{R}^3\to \sigma$ separately and use the Euclidean metric on 
$\mathbb{R}^3$ in the following definitions. 
\begin{definition} \label{def3.1} ~\\
i) A cubical partition of $\mathbb{R}^3$ is defined by the cubes 
$c_n,\;n\in \mathbb{Z}^3$ where 
\be \label{3.1}
c_n=\{s\;\in \mathbb{R}^3;\; s^I=n^I+t^I,\;I=1,2,3\}
\ee
The boundary faces (squares) of $c_n$ are taken with outward 
orientation.\\
ii) A simplicial partition of $\mathbb{R}^3$ subordinate to a cubical one 
is defined as follows:\\
first draw in $c_{(0,0,0)}$ diagonals on the boundary squares, such that 
the 
diagonals on opposite squares are orthogonal. Specifically, in the face 
defined by $s^I=0;\; s^J,s^K\in [0,1]^2;\; \epsilon_{IJK}=1$ the 
diagonal is the line $t\mapsto (s^I=0,s^J=t,s^K=t),\;t\in [0,1]$, while    
in the face 
defined by $s^I=1;\; s^J,s^K\in [0,1]^2;\; \epsilon_{IJK}=1$ the 
diagonal is the line $t\mapsto (s^I=1,s^J=t,s^K=1-t),\;t\in [0,1]$.\\
Now continue this pattern of orthogonal diagonals in opposite faces to 
the six cubes adjacent to $c_0$ where common faces have the same 
diagonal. This also defines the remaining four diagonals in those six 
cubes by connecting the endpoints of the already present two diagonals. \\
Finally continue this process for all cubes.\\
The face diagonals define altogether five tetrahedra that partition each 
cube. We will take their boundary triangles with outgoing orientation.\\
iii) An octahedronal partition of $\mathbb{R}^3$ subordinate to a cubical
one is defined as follows:\\
For each cube draw the unique four space diagonals. Specifically in 
$c_{(0,0,0)}$ these are the 
lines $t\mapsto (t,t,t),\;(t,t,1-t),\;(t,1-t,t),\;(1-t,t,t);\;\;
t\in [0,1]$. These partition each cube into six pyramids with common tip
in the barycentre of the cube and with the six faces of the cube as 
their bases. Now glue two pyramids in adjacent cubes 
along their common base. Obviously, two glued pyramids define an 
octahedron which we take with outgoing orientation. 
\end{definition}  
\indent The basic building blocks of the tetrahedronal and octahedronal 
decompositions are displayed in figures \ref{fig3}, \ref{fig4} and 
\ref{fig5}, respectively.
When gluing the bases of the pyramids along the faces of the original 
cubes one obtains an octahedronal decomposition as displayed in figure 
\ref{fig6}.
\begin{figure}[hbt] 
\begin{center}
 \includegraphics[scale=0.3]{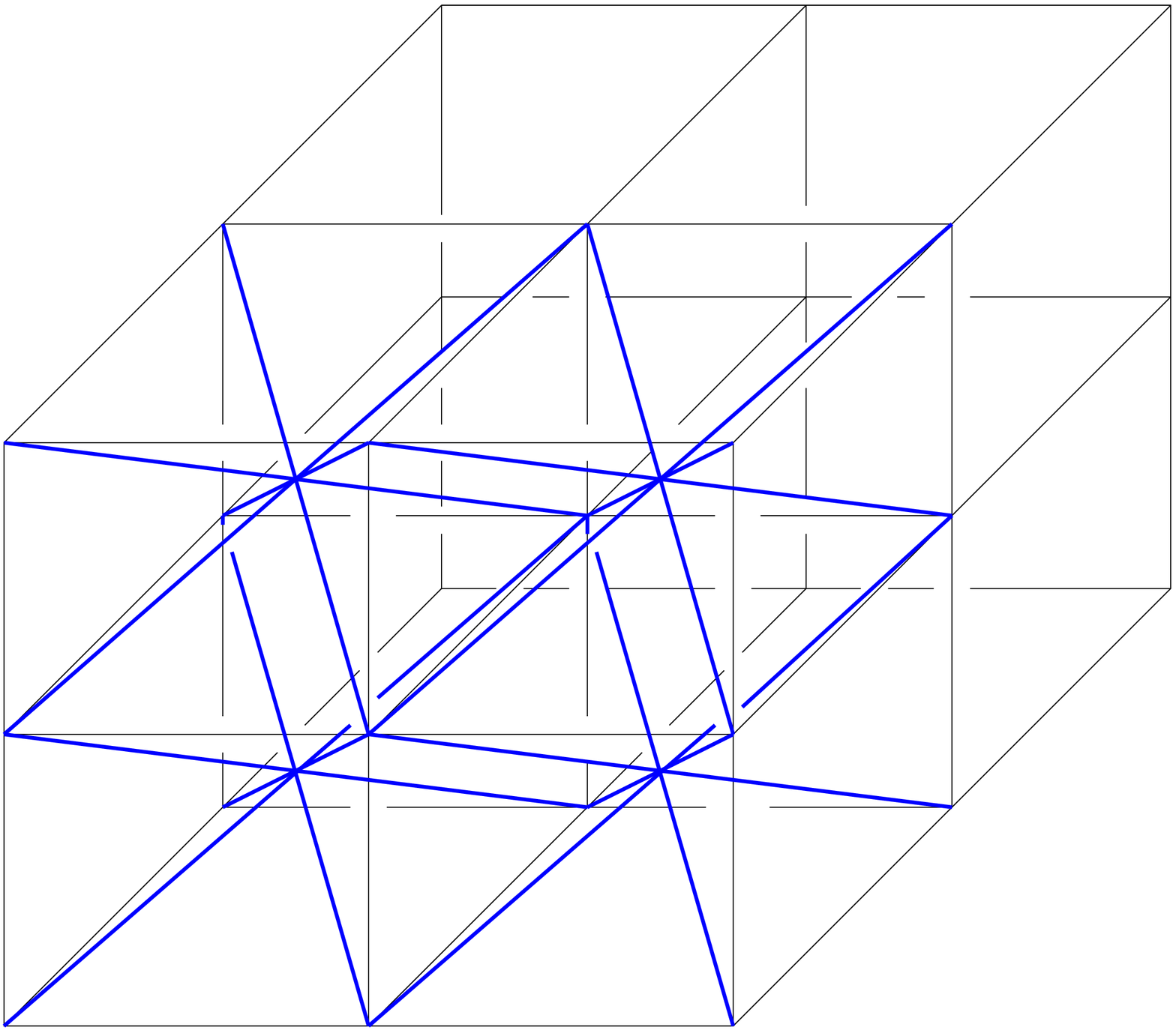}
\caption{Octahedronal decomposition.}
\label{fig6} 
\end{center} 
\end{figure}
It is maybe not completely obvious that the 
drawing of the diagonals that define the tetrahedra is a 
consistent and unique prescription. To 
see this, we use the checkerboard visualisation displayed in 
figure {fig7}: %
\begin{figure}[hbt] 
\begin{center}
 \includegraphics[scale=0.3]{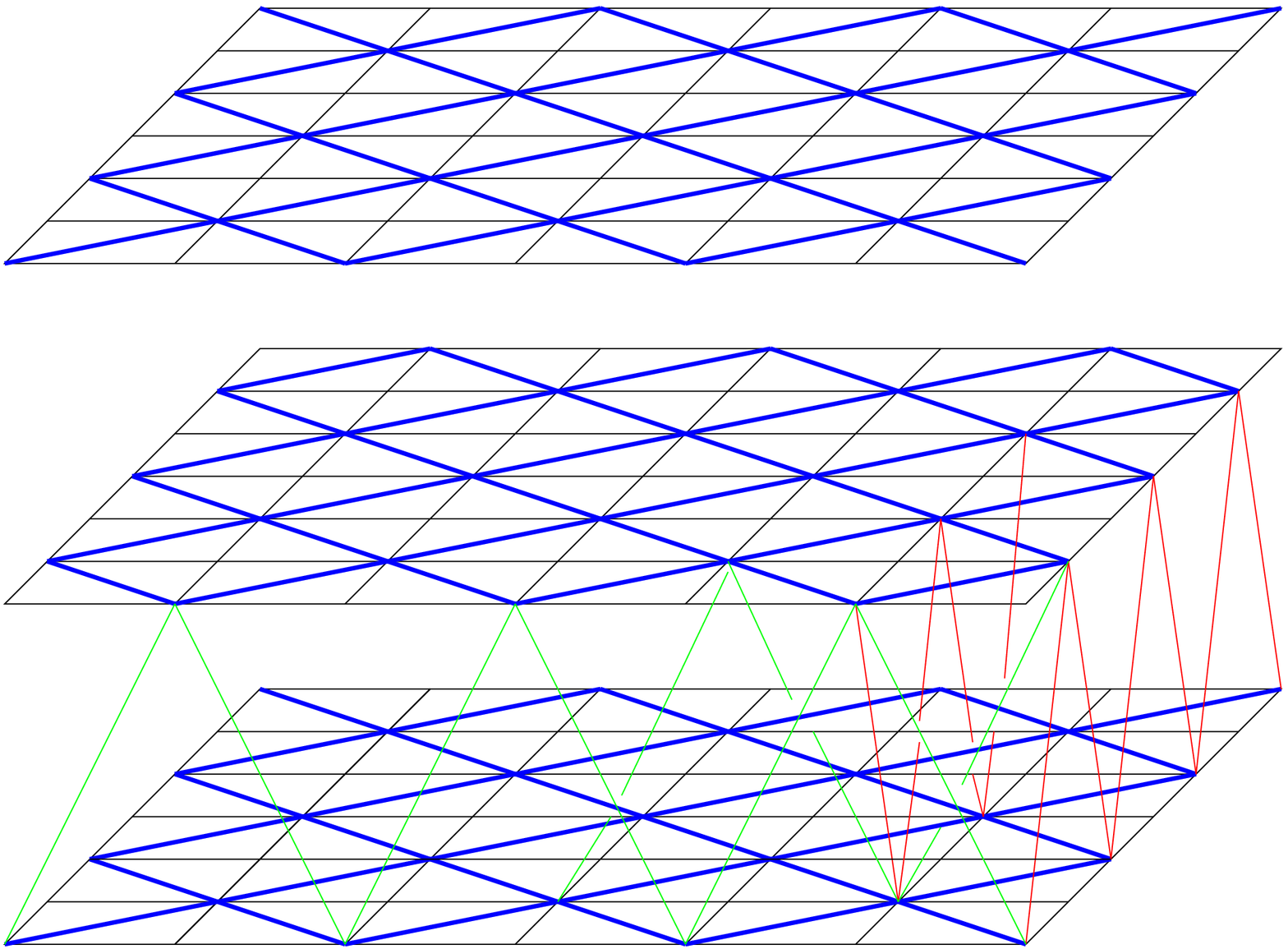}
\caption{Checkerboard visualisation of the triangulation.}
\label{fig7}
\end{center} 
\end{figure}
first draw all 
plaquettes in the $s^3=n\in\mathbb{Z}$ layers. 
Now take the $n=0$ layer and draw the diagonal for the plaquette in that 
layer that belongs to $c_{(0,0,0)}$, as prescribed in the definition.
Define that plaquette as ``black''. Now turn the $n=0$ layer into a 
checkerboard in the unique way consisting of black and white plaquettes.
The other layers $n\not=0$ are also turned uniquely into  
checkerboards by asking that checkerboards in adjacent layers are  
complementary, i.e. if the plaquette $(n^1,n^2,n^3)$ is white (black)
then the plaquette $(n^1,n^2,n^3\pm 1)$ is black (white).\\ 
Draw diagonals in plaquettes of opposite colour orthogonally to each 
other. This defines face diagonals in the $s^3=n$const. layers. 
These have the property that they form squares in each layer, which 
lie at an angle of $\pi/4$ relative to the plaquettes and which are such 
that only every second plaquette corner is a vertex of these squares.
We will refer to such corners that are vertices as ``used''. 
It is easy to see that in adjacent layers, used plaquette corners lie 
above unused ones.  Now draw the 
remaining face diagonals in the $s^1,s^2=n=$const. layers by connecting 
the used corners in adjacent layers using the appropriate diagonals of 
the cubes. This results in the triangulation depicted in figure 
\ref{fig8}.
\begin{figure}[hbt] 
\begin{center} 
\includegraphics[scale=0.3]{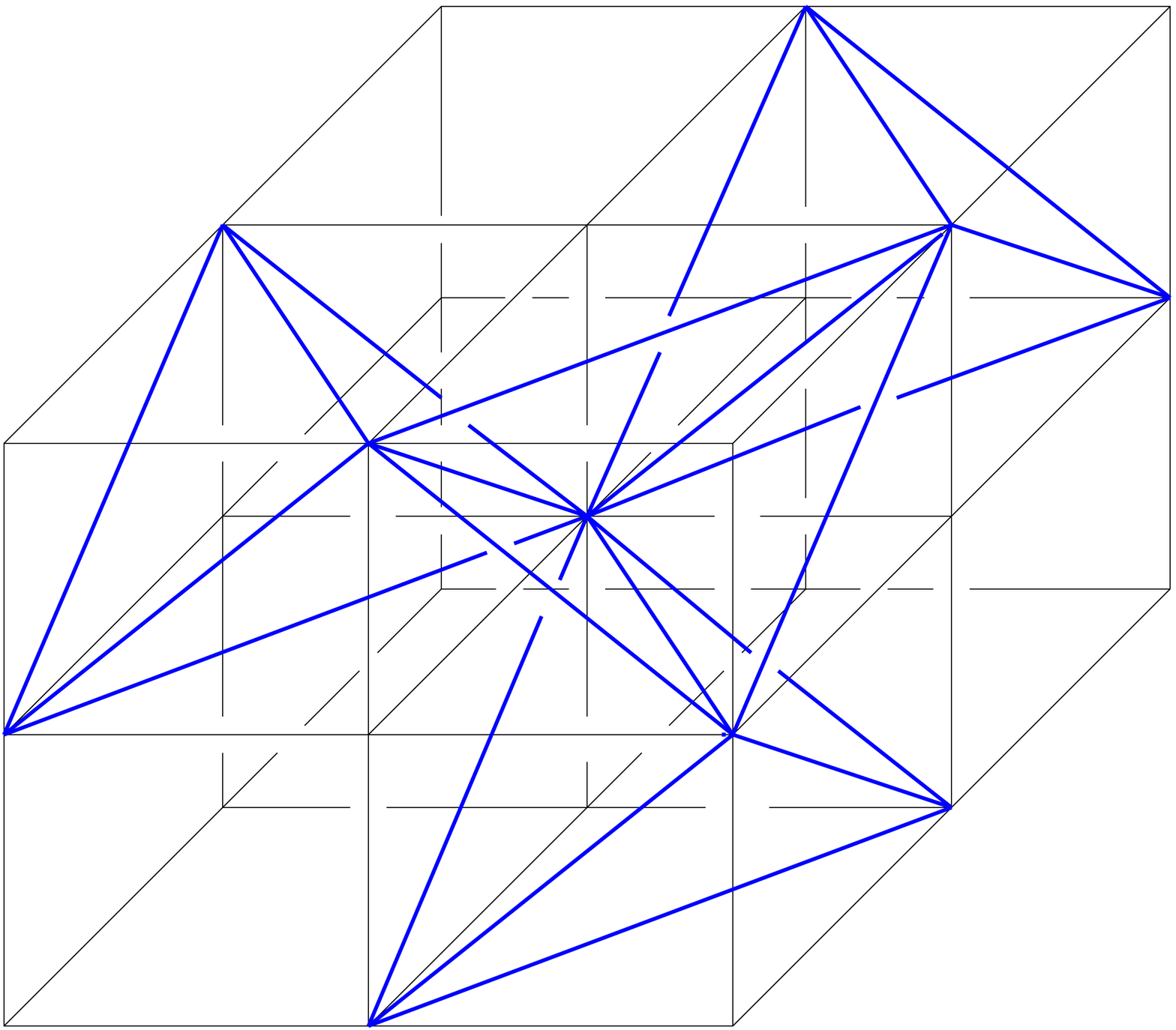}
\caption{Triangulation.}
\label{fig8}
\end{center} \end{figure}
We now define the graphs dual to these particular polyhedronal 
decompositions.
\begin{definition} \label{def3.2} ~\\
The graph in $\mathbb{R}^3$ dual to the above simplicial, cubical and 
octahedronal cell complexes is obtained by connecting the barycentres of 
adjacent tetrahedra, cubes and octahedra respectively by straight lines 
through their common triangles, squares and triangles respectively. 
Here the barycentre of a region $R\subset \mathbb{R}$ is defined as 
usual by 
\be \label{3.3}
B(R)=\frac{\int_R\; d^3s\; (s^1,s^2,s^3)}{\int_R\; d^3s}
\ee
\end{definition}
The advantage of the explicit definition of the cell complex is that we 
can explicitly label the edges and vertices of the dual graph. This is 
of course only feasible for sufficiently regular graphs, otherwise we 
run into difficult bookkeeping problems. 
\begin{itemize}
\item[1.] {\it Cubical Graph}\\
The barycentres of the cubes 
$c_n$ are evidently the points 
$v_n:=(n^1+\frac{1}{2},n^2+\frac{1}{2},n^3+\frac{1}{2})$ which form 
the
vertices of the dual graph. The edges $e_{n,I},\;I=1,2,3$, which connect
the vertices with labels $n$ and $n+b_I$ respectively, where $b_I$ is 
the standard unit vector $(b_I)^J=\delta_I^J$, have the explicit 
parametrisation 
$e_{n,I}(t)=v_n+tb_I,\; t\in [0,1]$. The other three edges adjacent to 
$v_n$ are ingoing and are given by $e_{n-b_I,I}$.
These 
edges form the 1 skeleton of 
another cubical cell complex, which is just shifted by the vector 
$(\frac{1}{2},\frac{1}{2},\frac{1}{2})$ from the original one.
\item[2.] {\it Tetrahedronal graph}\\
The tetrahedronal graph is the most complicated one because there are two 
different types of simplicial decompositions of a cube into five 
tetrahedra. Type A corresponds to the case that the vertices of the 
internal tetrahedron within a standard unit cube are given by \\
$(0,0,0),\;(1,1,0),\;(1,0,1),\;(0,1,1)$ while type B has vertices at
$(1,0,0),\;(0,1,0),\;(0,0,1),\;(1,1,1)$. These types alternate in adjacent 
cubes, as we move in any of the three coordinate directions. 
Hence, by defining the cube $c_0$ to be of type A., 
the triangulation is completely specified. Indeed, the type of 
$c_n$ is A if $n^1+n^2+n^3$ is even and of type B otherwise.  

\mbox{  }To determine the dual graph, we first discuss the barycentres of the 
tetrahedra for the two 
types separately for 
a standard unit cube, as well as the edges of the dual graph that lie 
within it. The vertices of and the edges 
in $c_n$ follow then by translation 
by $n=n^I b_I$. Notice that 
a tetrahedron $T$ based at $v$ and spanned by vectors $e_I$, that is
$T=\{v+t^I e_I;\; 0\le t^I\le 1;\; t^1+t^2+t^3\le 1\}$, has 
barycentre at $B(T)=v+\frac{1}{4}(e_1+e_2+3_3)$.
\begin{itemize}
\item[A.] {\it Type A}\\
The barycentre of the interior tetrahedron coincides with the barycentre 
$v_0:=\frac{1}{2}(1,1,1)$ of the cube. The barycentres of the remaining 
four 
exterior tetrahedra 
based at vertices\\  $(1,0,0),\;(0,1,0),\;(0,0,1),\;(1,1,1)$
respectively, are at
$v^A_1:=\frac{1}{4}(3,1,1),\;v^A_2=\frac{1}{4}(1,3,1),\;
v^A_3=\frac{1}{4}(1,1,3),\;v^A_4=\frac{1}{4}(3,3,3)$, respectively. 
Accordingly, the dual edges within the cube are 
$e^A_\alpha=v^A_\alpha-v_0,\;\alpha=1,2,3,4$.
\item[B.] {\it Type B}\\
The barycentre of the interior tetrahedron coincides with the barycentre 
$v_0:=\frac{1}{2}(1,1,1)$ of the cube. The barycentres of the remaining 
four 
exterior tetrahedra 
based at vertices  
$(0,0,0),\;(1,1,0),\;(1,0,1),\;(0,1,1)$\\ 
respectively are at
$v^B_4:=\frac{1}{4}(1,1,1),\;v^B_3=\frac{1}{4}(3,3,1),\;
v^B_2=\frac{1}{4}(3,1,3),\;v^B_1=\frac{1}{4}(1,3,3)$, respectively. 
Accordingly, the dual edges within the cube are 
$e^B_\alpha=v^B_\alpha-v_0,\;\alpha=1,2,3,4$.
\end{itemize}
It remains to describe the dual edges that result from gluing the faces of 
the exterior tetrahedra of adjacent cubes. But this is simple because each 
of the exterior tetrahedra 
within a cube has three triangles as faces, which lie in the three 
coordinate planes, hence the gluing is between those triangles which 
result from drawing the respective face diagonal within a boundary square 
of a cube. 

Hence, each cube has twelve edges perpendicular to the twelve 
boundary triangles of the exterior tetrahedra, which are adjacent to the 
four barycentres of those exterior tetrahedra. Altogether we can identify 
six possible gluings:\\
1) either going from type A to type B when 
moving along the positive $I$ direction and gluing along the  
$s^I=$const. plane;\\
2) or going from type B to type A when 
moving along the positive $I$ direction and gluing along the  
$s^I=$const. plane.\\
As one may check, the type A to type B gluing
in $I$ direction corresponds to two dual edges running from vertices $v_I^A$
to a $v_4^B$ and from vertices $v_4^A$ to $v_I^B$, respectively. 
Likewise,
the type B to type A gluing
in $I$ direction corresponds to two dual edges running from vertices $v_J^A$
to $v_K^B$ and from vertices $v_K^A$ to $v_J^B$, respectively 
where $\epsilon_{IJK}$. In all cases, these $I$ direction edges have 
coordinate length $\frac{1}{2}$ as one may easily calculate.\\
\mbox{  } Altogether, we can now easily describe the dual lattice as follows:\\
the vertices are labelled $v_{n,\alpha},\;\alpha=0,1,2,3,4$ with 
$v_{n,0}=n+v_0$ and $v_{n,\alpha}=n+v^A_\alpha,\;\alpha=1,2,3,4$ if 
$n^1+n^2+n^3$ is even while $v_{n,\alpha}=n+B^A_\alpha,\;\alpha=1,2,3,4$ 
if $n^1+n^2+n^3$ is odd. The edges are labelled by 
$e_{n,\alpha},\;\alpha=1,2,3,4$ and $e_{n,I,j},\;I=1,2,3,\;j=1,2$
where $e_{n,\alpha}(t)=n+v_0+t(v^A_\alpha-v_0$ if $n^1+n^2+n^3$ is even,  
$e_{n,\alpha}(t)=n+v_0+t(v^B_\alpha-v_0)$ if $n^1+n^2+n^3$ is odd,  
$e_{n,I,1}(t)=n+v_I^A+\frac{t}{2}b_I$ and    
$e_{n,I,2}(t)=n+v_4^A+\frac{t}{2}b_I$ if $n^1+n^2+n^3$ is even
and finally    
$e_{n,I,1}(t)=n+v_J^A+\frac{t}{2}b_I$ and    
$e_{n,I,2}(t)=n+v_K^A+\frac{t}{2}b_I$ if $n^1+n^2+n^3$ is odd
where $\epsilon_{IJK}=1$. 
\item[3.] {\it Octahedronal Graph}\\
Each cube contains six pyramids or halves of 
the octahedra. Therefore, the barycentre of an octahedron coincides with 
the barycentre of the common boundary face of the two cubes that contain 
it. It follows that the octahedra may be labelled by $o_{n,I}$ 
corresponding to the vertices $v_{n,I}=n+\frac{1}{2} b_J+\frac{1}{2} 
b_K;\; \epsilon_{IJK}=1$ which define its barycentre. Such an octahedron
has the property that it has a common base of two pyramid halves which 
lies in the $s^I=$const. plane.   
For the vertex $v_{n,I}$ we define four edges 
$e_{n,I,J,\sigma},\;J\not=I;\;\sigma=\pm$ outgoing from it through the 
explicit 
parametrisation
$e_{n,I,j}(t):=v_{n,I}+\frac{t}{2}(b_I+\sigma b_J)$, which connects 
the vertices $v_{n,I}$ and $v_{n+\frac{1}{2}(1+\sigma) b_J,J}$. 
Notice that these edges lie in the $(I,J)$ or $(I,K)$ plane but there 
are no edges in the $(J,K)$ plane adjacent to $v_{n,I}$.
The other four edges adjacent to $v_{n,I}$ have ingoing orientation.\\
As an aside, notice that the 1-skeleton of an octahedral cell complex, as 
defined 
above, is an eight valent graph after removing the edges of the original 
cubes. 
\end{itemize}
The basic building blocks of the dual graphs are displayed in figures
\ref{fig9}, \ref{fig10}, \ref{fig11} and \ref{fig12}, respectively.
\begin{figure}[hbt] 
\begin{center} 
\includegraphics[scale=0.3]{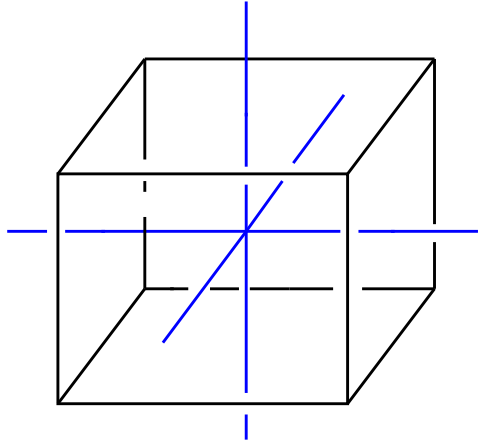}
\caption{Cube and dual six vallent graph.}
\label{fig9}
\end{center} \end{figure}
\begin{figure}[hbt] 
\begin{center}
 \includegraphics[scale=0.3]{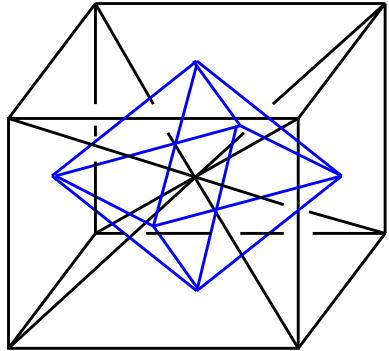}
\caption{Octahedron and dual eight valent graph.}
\label{fig10}
\end{center} \end{figure}
\begin{figure}[hbt] 
\begin{center} \includegraphics[scale=0.3]{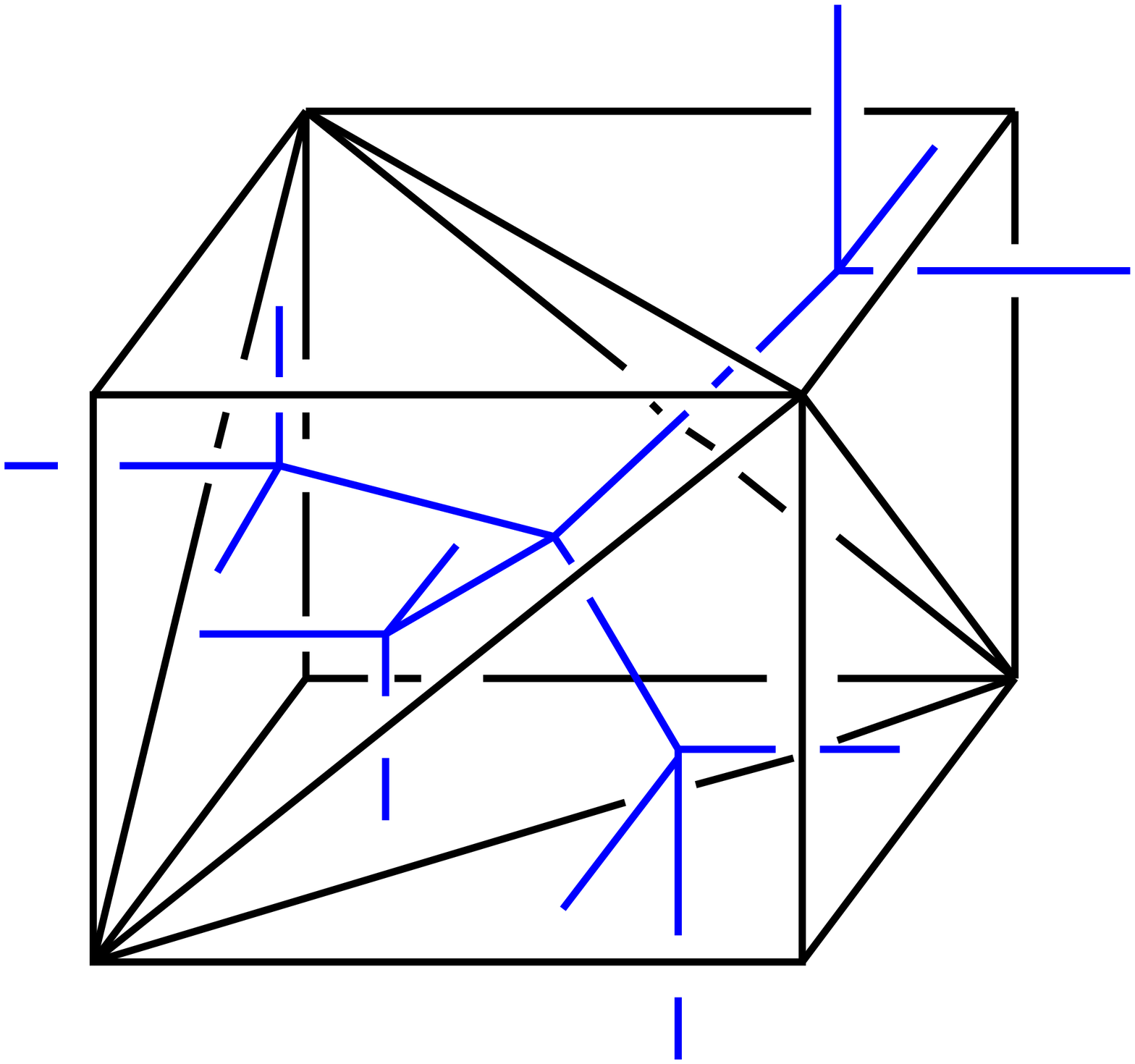}
\caption{Type A triangulation of a cube and dual four valent graph.}
\label{fig11}
\end{center} \end{figure}
\begin{figure}[hbt] 
\begin{center} 
\includegraphics[scale=0.3]{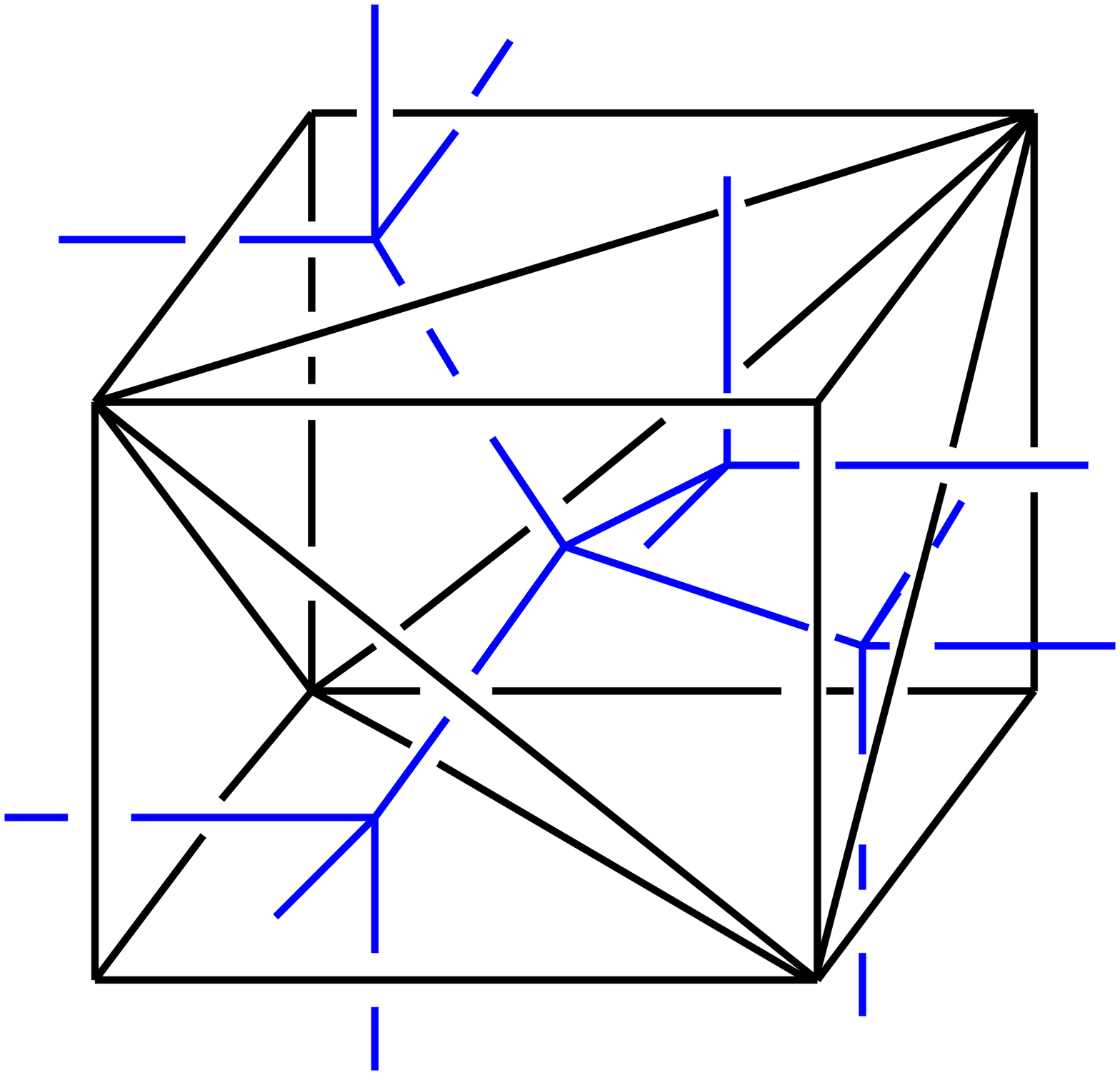}
\caption{Type B triangulation of a cube and dual four valent graph.}
\label{fig12}
\end{center} \end{figure}
The connection of the tetrahedronal lattice with the diamond lattice is as 
follows:\\
For each cube of type A or B respectively, keep the interior 
tetrahedron. Now move the barycentres of the remaining exterior tetrahedra 
into that corner of the cube which is also a corner of the tetrahedron 
under consideration. In this process, the edges dual to the faces of the 
interior tetrahedron become halves of the spatial diagonals of the cube.
Finally, drop all the other edges which were running between the 
barycentres of the exterior tetrahedra. The result is a diamond lattice. 
Its basic building blocks are depicted in figures \ref{fig13} and 
\ref{fig14}, respectively.
\begin{figure}[hbt] 
\begin{center}
 \includegraphics[scale=0.3]{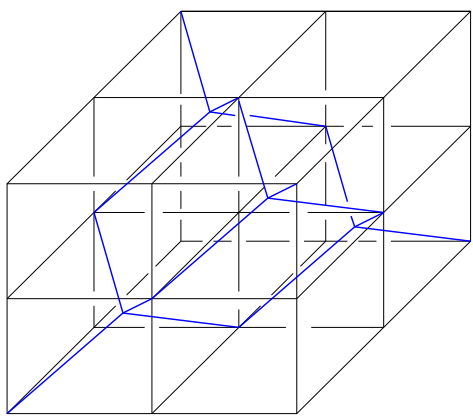}
\caption{Type A diamond cell with occupied lower, left front cube.}
\label{fig13}
\end{center} \end{figure}
\begin{figure}[hbt] 
\begin{center} 
\includegraphics[scale=0.3]{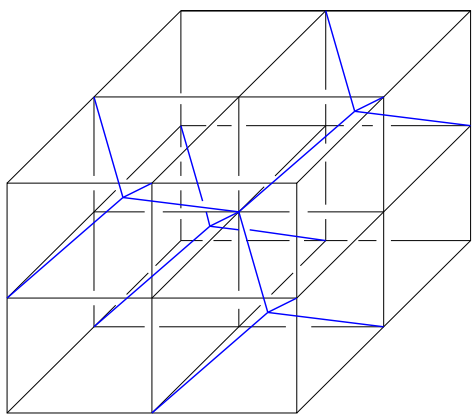}
\caption{Type B diamond cell with unoccupied lower, left front cube.}
\label{fig14}
\end{center} \end{figure}
It is also four valent, however, it does not have a piecewise 
linear polyhedronal complex 
dual to it (i.e. whose faces (which are subsets of linear planes) are in 
one to one correspondence with the 
edges). It does have a cell complex dual to it, if one gives up 
piecewise linearity by suitably rounding off corners but that is 
inconvenient to describe analytically. On the other hand, the natural 
polyhedronal complex consisting of the 
interior 
tetrahedra of the original cubes with the cubes deleted consists 
of those tetrahedra, as well octahedra which surround half of the corners 
of the original cubes. Only half of the triangle faces of those 
octahedra are penetrated by the edges of the diamond lattice.

\begin{figure}[hbt] 
\begin{center}
\includegraphics[scale=0.3]{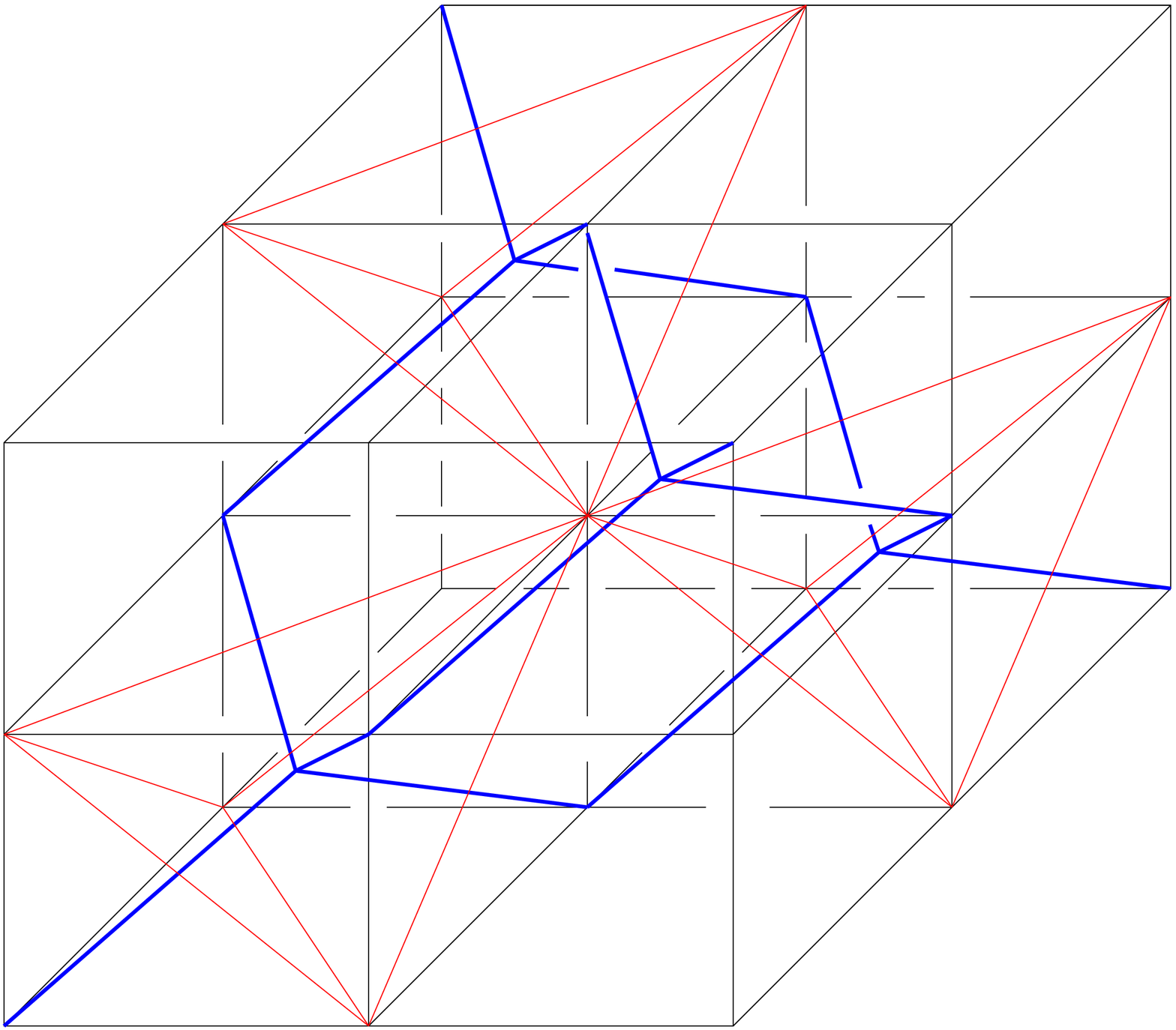}
\caption{Dual diamond cell of type A}
\label{fig15}
\end{center} \end{figure}

\begin{figure}[hbt] 
\begin{center} \includegraphics[scale=0.3]{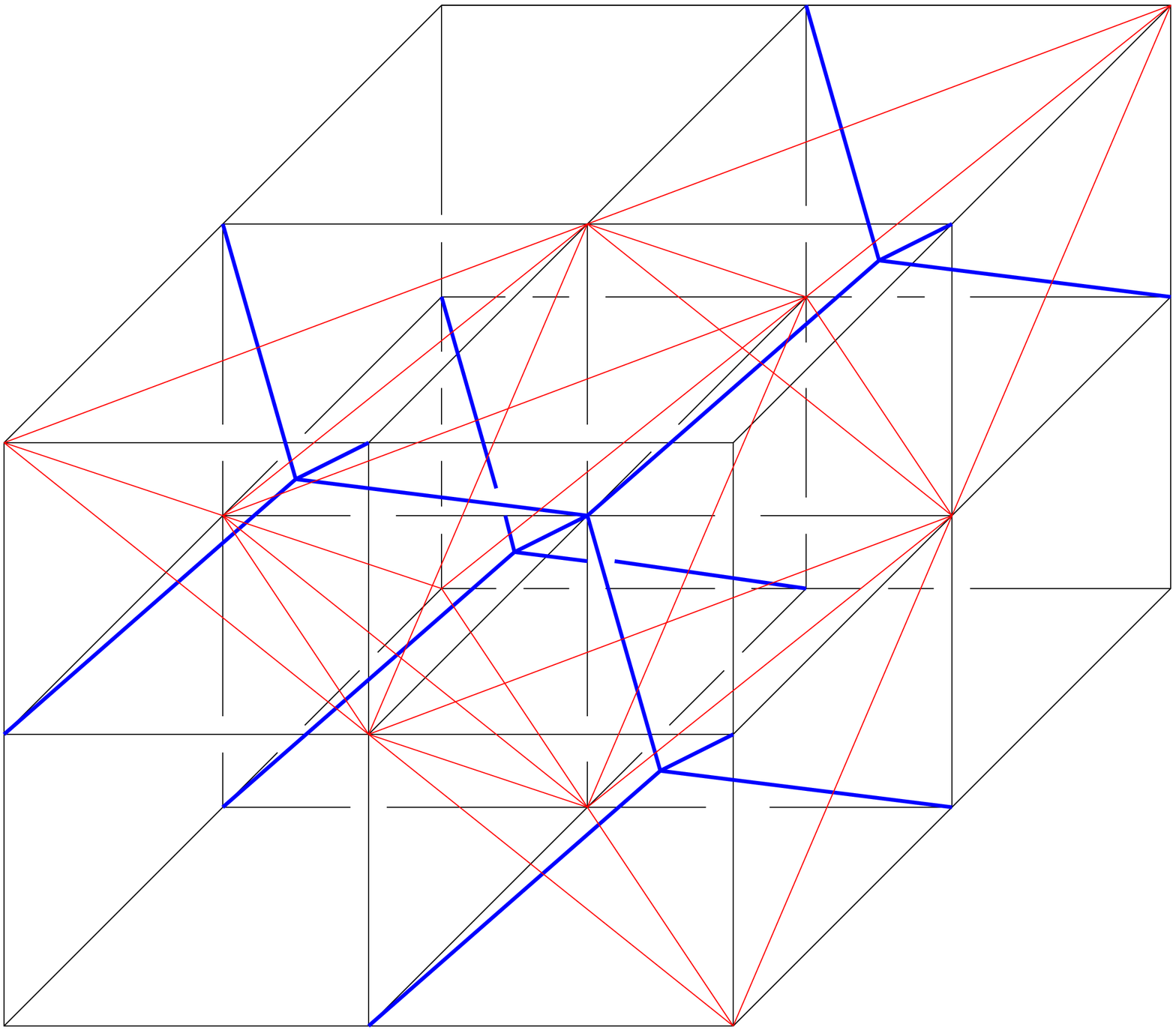}
\caption{Dual diamond cell of type B}
\label{fig16}
\end{center} \end{figure}

\begin{figure}[hbt] 
\begin{center} \includegraphics[scale=0.3]{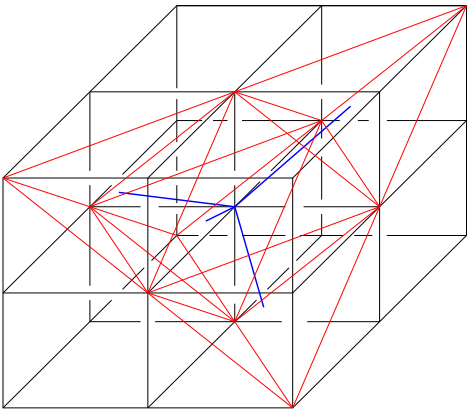}
\caption{Dual diamond cell of type B with only the central four valent 
vertex left.}
\label{fig17}
\end{center} \end{figure}

\begin{figure}[hbt] 
\begin{center} \includegraphics[scale=0.3]{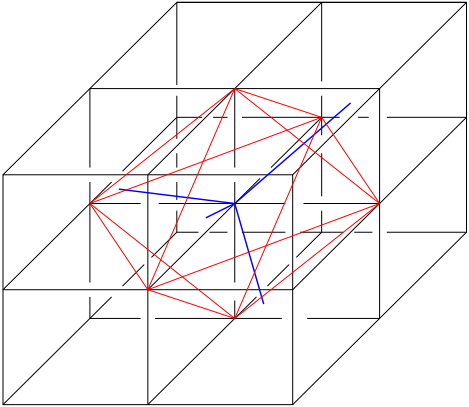}
\caption{Dual diamond cell of type B with only the central four valent
vertex left and keeping only the faces adjacent to the vertex.}
\label{fig18}
\end{center} \end{figure}

\begin{figure}[hbt] 
\begin{center} \includegraphics[scale=0.3]{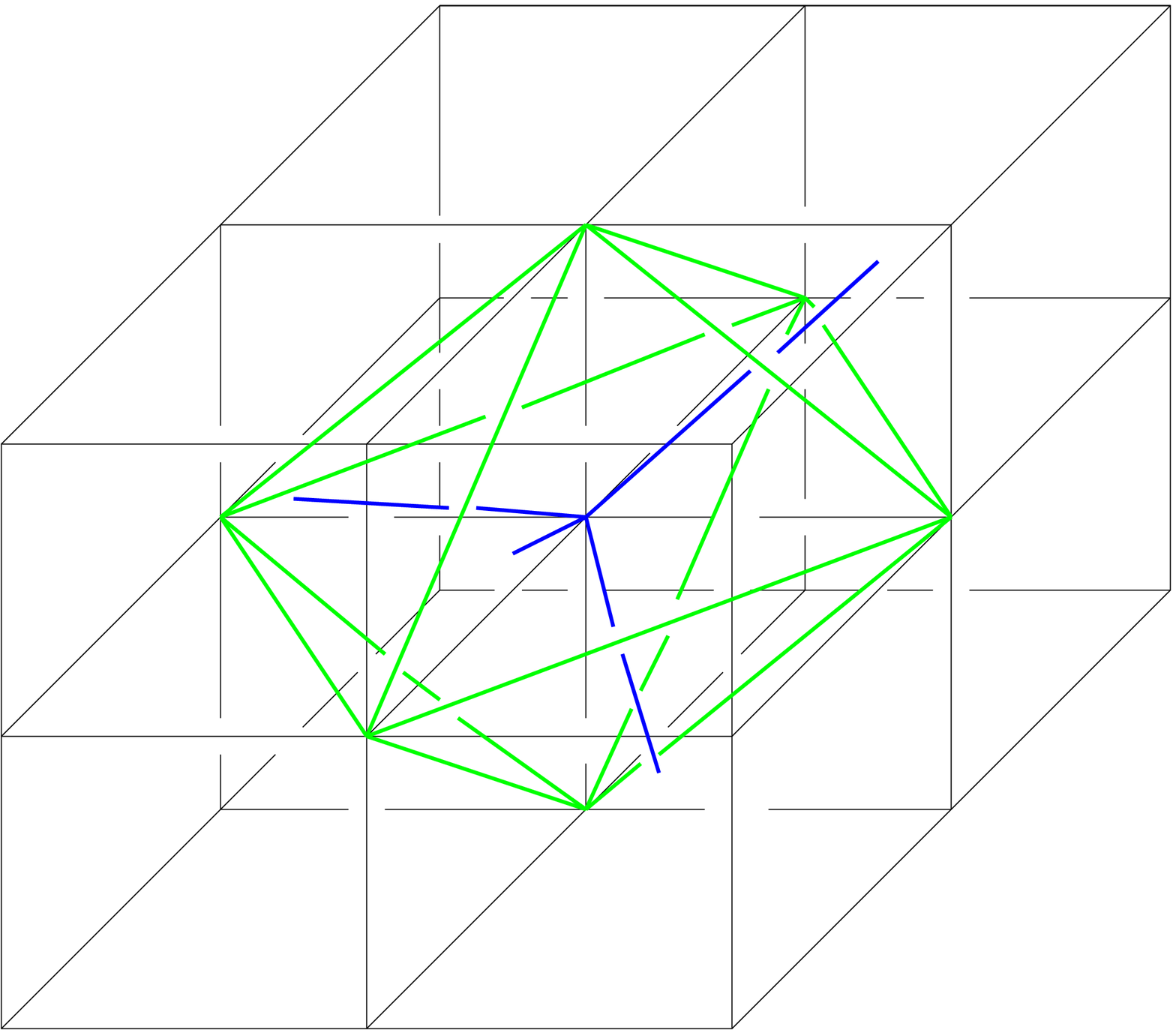}
\caption{Dual diamond cell of type B with only the central four valent
vertex left and keeping only the faces adjacent to the vertex, 
highlighting its octahedronal cell structure.}
\label{fig19}
\end{center} \end{figure}

The building of this semi dual polyhedronal cell complex consisting of 
tetrahedra and octahedra is visualised in figures 
\ref{fig15}, \ref{fig16}, \ref{fig17}, \ref{fig18} and \ref{fig19}, 
respectively. 
In order to achieve the desired duality, one has to 
fill in the original cubes again which then triangulate those 
octahedra into eight tetrahedra. This then results in the additional 
vertices and edges that we have described and depicted in figures
\ref{fig11} and \ref{fig12}.

\chapter{Expectation Value of the Volume Operator}
In this chapter we will analyse the
semiclassical properties of the volume operator with respect to
both classes of coherent state : \emph{dual cell coherent states} and \emph{area complexifier coherente states}.
The
result of our analysis is that if we use the former states\cite{10},
{\it the correct semiclassical limit is attained with these states
for $n=6$ only}. If instead we use the latter states \cite{21}, the
correct semiclassical limit is attained \emph{only} for:\\
1) artificial rescaling of the coherent state label;\\
2) particular embeddings of the 4-valent and 6-valent graphs, with
respect to the set of surfaces on which the complexifier depends.

However, the combinations of Euler angles, for which such
embeddings are attained, have measure zero in SO(3), and are,
therefore, negligible. Thus the {\it area complexifier coherent
states are not the correct tools with which to analyse the
semiclassical properties of the volume operator}.

If one wants to obtain embedding independence, a possible strategy
is to sample over graphs (Dirichlet-Voronoi sampling \cite{40}), as
outlined in \cite{10}. What this strategy amounts to is that,
instead of singling out one particular coherent state
$\psi_{\gamma, m}$---as defined in terms of a single graph
$\gamma$---one considers an ensemble of coherent states
constructed by averaging the one-dimensional projections
$\hat{P}_{\gamma, m}$ onto the states $\psi_{\gamma, m}$, over a
subset $\Gamma_m$ of the set of all allowed graphs. In other
words, one considers a mixed state (with an associated density
matrix) rather than a single coherent state. In such a way, if the
subset $\Gamma_m$ is big enough, it can be shown (\cite{10}) that
it is possible to eliminate the embedding dependence (the
`staircase problem'\footnote{Roughly the staircase problem can be
stated as follows: consider an area operator $\hat{A}_S$ for a
surface $S$. If we compute the expectation value for $\hat{A}_S$,
with respect to a coherent state $\psi_{\gamma, m}$, such that the
surface $S$ intersects transversely one and only one edge $e$ of
$\gamma$, then the expectation value of the area operator
coincides with the classical value $A(m)$. However, if the surface
$S$ lies transversally to the edges, then we do not obtain the
correct classical limit.}).

It is straightforward to deduce that the area complexifier
coherent states cannot be used to construct embedding-independent,
mixed coherent states because of condition 2) above. We thus claim
that \emph{area complexifier coherent states should be ruled out
as semiclassical states altogether, if one wants to attain
embedding independence}. Instead, one should use the flux coherent
states, as it was done in \cite{10}. For such states we will show
that {\it the correct semiclassical limit is attained only for
$n=6$ }. In other words, {\it up to now, there are no
semiclassical states known other than those with cubic-graph
topology!}

Thus the implication of our result for LQG is that the
semiclassical sector of the theory is spanned by SNWF that are
based on cubic graphs. This has some bearing for spin foam models
\cite{25}, which are supposed to be---but, so far, have not been
proved to be---the path-integral formulation of LQG. Spinfoams
are certain state-sum models that are based on simplicial
triangulations of four manifolds whose dual graphs are therefore
5-valent. The intersection of this graph with a boundary
three-manifold is 4-valent and, therefore, we see that spin foam
models, based on simplicial triangulations, correspond to boundary
Hilbert spaces spanned by spin-network states based on 4-valent
graphs only\footnote{ As an aside, whether this boundary Hilbert
space of spin foams really can be interpreted as the 4-valent
sector of LQG is a subject of current debate, even with the recent
improvements \cite{26}, \cite{415}, \cite{416}, \cite{413} in the Barrett--Crane model \cite{d111}.
There are two problems: first, the boundary connection predicted
by spin foams does not coincide with the LQG connection \cite{28},
secondly, the 4-valent sector of the LQG Hilbert space is not a
superselection sector for the holonomy flux algebra of LQG. In
fact, the LQG representation is known not only to be cyclic but
even irreducible \cite{28a}. Therefore the 4-valent sector is not
invariant under the LQG algebra.}. However, we have proved that
the correct semiclassical states, for analysing the semiclassical
properties of the volume operator, are the {\it gauge covariant
flux} states. For such states, only those of cubic topology give
the correct semiclassical value of the volume operator.

Even if the mismatches between the 4-valent sector of LQG and the
boundary Hilbert space of spin foams could be surmounted, the
result of our analysis seems to be that {\it the boundary Hilbert
space of current spin foam models does not contain any
semiclassical states!} This, apparently, contradicts recent findings
that the graviton propagator, derived from spin foam models, is
correct \cite{29a}, \cite{29}. However, it is notable that these latter
results only show that the propagator has the correct fall-off
behaviour: the correct tensorial structure has not yet been
verified.

One straightforward way of possibly repairing this situation is to
generalise spin foam models to allow for arbitrary---in
particular, cubic---triangulations, as suggested in \cite{48e,30a}.
\section{Volume Operator}
\label{s4.0}

The classical expression for the volume of a region $R$ of a 
semianalytical three
dimensional manifold $\sigma$ is:
\be \label{4.0.1}
V_R:=\int_R \; d^3x \; \sqrt{\det(q)}=\int_R \; d^3x\; \sqrt{|\det E|}
\ee
where $q_{ab}$ is the three metric. The version of the volume operator 
\cite{v8} consistent with the triad quantisation \cite{v9} that enters the 
quantum dynamics \cite{v3} has cylindrically consistent projections 
$\widehat{V}_{R,\gamma}$ given by 
\be \label{4.0.2}
\hat{V}_R=
\sum_{v\in V(\gamma)\cap R}\hat{V}_{\gamma,v}
\ee 
where
\be \label{4.0.3}
\hat{V}_{\gamma,v}=\ell_P^3\; \sqrt{|\frac{1}{8}
\sum_{e_I, e_J, e_K,\\ I\leq J\leq K\leq N|v\in e_I\cap e_J\cap e_K }
\epsilon^{ijk}\epsilon(e_I,e_J,e_K)X^{e_I(v)}_iX^{e_J(v)}_jX^{e_I(v)}_k|}
\ee
Here $N$ denotes the valence of the vertex, $\ell_P^2=\hbar\kappa$ is 
the Planck area, 
$X^{e_I(v)}_i={\rm Tr}([\tau_i h_I]^T \partial/\partial h_I)$ are 
right 
invariant vectors on $SU(2)$ acting on the holonomy $h_I:=A(e_I)$
($i \tau_j=\sigma_j$ are the Pauli matrices)  
and $\epsilon(e_I,e_J,e_K)$ is called the orientation function, which is 
defined as follows:
\be \label{4.0.4}
\epsilon(e_I,e_J,e_K)=\left\{\begin{array}{lll}
1,\text{iff}\hspace{.05in}\dot{e}_I,\dot{e}_J,\dot{e}_K
\text{are linearly independent at v and positively oriented}\\
-1,\text{iff}\hspace{.05in}\dot{e}_I,\dot{e}_J,\dot{e}_K
\text{are linearly independent at v and negatively oriented}\\
0,\text{iff}\hspace{.05in}\dot{e}_I,\dot{e}_J,\dot{e}_K
\text{are linearly dependent at v}
\end{array}
\right.
\ee
Here we take the convention that the edges at $v$ have been taken with 
outgoing orientation, hence if in $\gamma$ the orientation of an edge 
$e$ adjacent to 
$v$ is actually ingoing, just apply the above expression to 
$\psi'(..,h_e^{-1},...):=\psi(..,h_e,..)$.
 
From (\ref{4.0.2}), we deduce that the volume operator is a sum  
of contributions, one for each vertex. Therefore, in the expectation 
value calculations that follow, it will be sufficient to calculate 
the expectation values for each $\hat{V}_{\gamma, v}$ separately and, 
then, to add the contributions. Notice that each of these contributions 
is of the form $V_{\gamma, v}=\root 4 \of{Q_{\gamma,v}}$, where 
$Q_{\gamma,v}$ is minus the square of the expression appearing between 
the modulus labels $|..|$ in (\ref{3.3}) and, therefore, it is a sixth order 
polynomial in the $SU(2)$ right invariant vector fields.

We will now proceed to calculate the general expression for the 
expectation value of the volume operator for an $n=4,6,8$ valent graph.
\section{Expectation Values of the Volume Operator for Dual Cell Complex 
Coherent States}
\label{s5} 

In this section we compute the expectation value of the volume operator 
with respect to the dual cell complex coherent states of \cite{10}. In 
order to carry it out explicitly, we have to specify the 
graph and the dual cell complex. Here we focus our attention on 
arbitrary graphs with the following properties:\\
1. All vertices have 
constant valence $n=4,6,8$ \\
2. The dual cell complex consists only of 
tetrahedra, cubes and octahedra, respectively.

Such graphs and dual cell 
complexes exist as we have explicitly shown in section \ref{s3}. This is all 
we need for the purpose of this section, more specifics about the graph 
and the complex are not needed. 

We can actually perform a full $SU(2)$ calculation as follows: \\
The coherent states are explicitly given by \cite{10}
\be \label{5.1}
\psi_{Z,\gamma}=\prod_{e\in E(\gamma)}\; \psi_{Z,e},\;\;
\psi_{Z,e}(A)=\sum_{2j=0}^\infty\; e^{-t j(j+1)/2}\; \chi_j(g_e(Z) 
A(e)^{-1})
\ee
where $t=\ell_P^2/L^2$ and $g_e(Z)$ is given by (\ref{2.35}). 
The volume operator expectation value is given by 
\be \label{5.2}
<V(R)>_{Z,\gamma}=\sum_{v\in V(\gamma)\cap R} 
\;<V_{\gamma,v}>_{Z,\gamma}
\ee
Notice that due to the product form of (\ref{5.1}), the expectation value 
$<V_{\gamma,v}>_{Z,\gamma}$ only involves the edges adjacent to $v$. 
As we have seen in the previous section we have $V_{\gamma,v}=\root 4 
\of{Q_{\gamma, v}}$. By the arguments presented in the introduction, 
the zeroth order in $\hbar$ of $<V_{\gamma,v}>_{Z,\gamma}$ is given by
$\root 4 \of{<Q_{\gamma, v}>_{Z,\gamma}}$. Since $Q_{\gamma,v}$ is a 
polynomial in right invariant vector fields, the results of \cite{10}
reveal that, to zeroth order in $\hbar$, the expectation value of 
any polynomial in the right invariant vector fields $i\ell_P^2 X^j_e$ is 
simply
obtained by replacing it by $E_j(S_e)$ which is given in 
(\ref{2.16a}).

It follows that to zeroth order in 
$\hbar$ we have $<Q_{\gamma,v}>_{Z,\gamma}=[P_{\gamma,v}(E)]^2$ where  
\be \label{5.3}
P_{\gamma,v}(E)=\frac{1}{48}\sum_{e\cap e'\cap e^{\prime\prime}} \;
\epsilon_{e,e',e^{\prime\prime}})\; \epsilon^{jkl}\;
E_j(S_e)\; E_k(S_{e'}) \; E_l(S_{e^{\prime \prime}})
\ee
Notice that, for 
sufficiently fine graphs, we can drop the holonomies along the paths 
$\rho_e(x)$ involved in the definition of $E_j(S_e)$ as we approach 
the continuum. It is then clear that the correct
expectation value of the volume operator is reached, provided that 
(\ref{5.3}) approximates the volume, as specified by $E^a_j$, of the 
cell of the polyhedronal complex, which is bounded by the faces $S_e$
involved in (\ref{5.3}).

To do this, we 
use the fact that for sufficiently fine graphs a polyhedron $P$ in 
$\sigma$ dual to a vertex of the graph lies in the domain of a chart
$Y$, so that $P$ is the image under $Y$ of a standard polyhedron $P_0$
in $\mathbb{R}^3$. Introducing 
\be \label{5.4}
n_a^I(s)=\frac{1}{2}\; \epsilon_{abc} \; \epsilon^{IJK}\;
\frac{\partial Y^b(s)}{\partial s^J}\;
\frac{\partial Y^c(s)}{\partial s^K}\;
\ee
and setting $P=Y(P_0)$ we immediately find that 
\be \label{5.5}
{\rm Vol}(P)=\int_P \; d^3x\; \sqrt{|\det(E)(x)|}
=\int_{P_0} \; d^3s\; \sqrt{|\det(\tilde{E}(s)|}
\ee
where
\be \label{5.6}
\tilde{E}^I_j(s)=E^a_j(Y(s))\; n_a^I(s)
\ee
Now for sufficiently fine graphs (\ref{5.6}) is approximately constant 
over $P_0$, so that 
\be \label{5.7}
{\rm Vol}(P)
\approx  \sqrt{|\det(\tilde{E}(s)|}_{Y(s)=v}\; 
{\rm Vol}_0(P_0)
\ee
where 
\be \label{5.8}
{\rm Vol}_0(P_0)=\int_{P_0} \; d^3s 
\ee
is the volume of the standard polyhedron with respect to the Euclidean
metric on $\mathbb{R}^3$.  

The idea behind this rewriting is that the fluxes $E_j(S_e)$
can be approximated by specific linear combinations of the 
$[\tilde{E}^I_j(s)]_{Y(s)=v}$, 
so that 
a direct comparison between (\ref{5.3}) and (\ref{5.8}) is possible.  
This is because a boundary face $S$ is also the image under $Y$ of a 
standard face $S^0$ in $\mathbb{R}^3$, so that (dropping the holonomies 
along the $\rho_e(x)$ as explained)
\be \label{5.9}
E_j(S)
=\int_S \; \frac{1}{2}\; \epsilon_{abc}\; dx^b\wedge dx^c \;E^a_j(x)
=\int_{S^0} \; \frac{1}{2}\; \epsilon_{IJK}\; ds^J\wedge 
ds^K \;
\tilde{E}^I_j(s)
\approx [\tilde{E}^I_j(s)]_{Y(s)=v} \; {\rm F}^I(S^0)
\ee
where 
\be \label{5.10} 
{\rm F}_I(S^0)
=\int_{S^0} \; \frac{1}{2}\; \epsilon_{IJK}\; ds^J\wedge 
ds^K \;
\ee 
is the $I$ component of the Euclidean flux through $S_0$. Thus, plugging 
(\ref{5.10}) into (\ref{5.3}) we find 
\be \label{5.11}
|P_{\gamma,v}(E)|^{1/2} \approx \sqrt{|\det(\tilde{E}(s))}_{Y(s)=v}\;
{\rm Vol}_0(v)
\ee
where 
\be \label{5.12}
{\rm Vol}_0(v)
=\sqrt{|\frac{1}{48}\sum_{e\cap e'\cap e^{\prime\prime}} 
\;
\epsilon_{e,e',e^{\prime\prime}})\; \epsilon^{IJK}\;
F_I(S^0_e)\; F_J(S^0_{e'}) \; F_K(S^0_{e^{\prime \prime}})|}
\ee
It remains to compare (\ref{5.8}) and (\ref{5.12}).
All of this still holds for general graphs. In order to 
test the correctness of the expectation value for specific, simple 
situations, we restrict our attention to graphs with the above specified properties but of valence 
$n=4,6,8$. Thus we 
know that for each vertex $v$ the faces $S_e$ dual to the edges $e$ 
adjacent to $v$ form the surface of a tetrahedron, cube and octahedron, 
respectively. Thus we just have to compare (\ref{5.3}) with the volume 
of such platonic bodies as measured by $E^a_j$. 
We will discuss the three cases separately.

\subsubsection{Tetrahedron}
\label{5.2.1}

A standard tetrahedron is the subset
\be \label{5.13}
T_0=\{s\in \mathbb{R}^3:\;0\le s^I\le 1;\;I=1,2,3,\;s^1+s^2+s^3\le 1\}
\ee
It has four boundary triangles given by 
\ba \label{5.14}
t^0_I &=& \{s\in \mathbb{R}^3:\; s^I=0,\; 0\le s^J, s^K\le 
1,\;s^J+s^K\le 1;\;\;\epsilon_{IJK}=1\}
\nonumber\\
t^0_4 &=& \{s\in \mathbb{R}^3:\;0\le s^I\le 1;\;I=1,2,3,\;
s^1+s^2+s^3=1\}
\ea
We easily compute 
\be \label{5.15}
{\rm Vol}_0(T_0)=\frac{1}{6}
\ee
while (remembering that the surfaces carry outward orientation if the 
edges are outgoing from $v$)  
\be \label{5.16}
F_I(t^0_J)=\frac{1}{2} \delta_{IJ},\; F_I(t^0_4)=-\frac{1}{2}
\ee
Let us label the edges adjacent to $v$ by $e_1,..,e_4$ where $e_\alpha$ 
is dual to $Y(t^0_j),\;j=1,2,3,4$, then
\ba \label{5.17}
{\rm Vol}_0(v) &=&
=\sqrt{|\frac{1}{8}\sum_{1\le j <k<l\le 4} 
\;
\epsilon_{e_j,e_k,e_l})\; \epsilon^{IJK}\;
F_I(t^0_j)\; F_J(t^0_k) \; F_K(t^0_l)
|}
\nonumber\\
&=& \frac{1}{8} \sqrt{|
\epsilon(e_1,e_2,e_3)
-\epsilon(e_1,e_2,e_4)
-\epsilon(e_1,e_3,e_4)
-\epsilon(e_2,e_3,e_3)
|}
\ea
which still depends on the sign factors. Hence, the expectation value 
takes values in the range $0,\frac{1}{8},\;
\frac{\sqrt{2}}{8},\;
\frac{\sqrt{3}}{8},\;\frac{1}{4}$,
none of which coincides with $\frac{1}{6}$. For the explicit four valent 
graph that we constructed in Section (\ref{s3}), each triple among the 
four edges has linearly independent tangents at $v$ and the expectation 
value is given by $\frac{\sqrt{2}}{8} > \frac{1}{6}$, which is too large.

\subsubsection{Cube}
\label{s5.2.2}

A standard cube is the subset
\be \label{5.18}
C_0=\{s\in \mathbb{R}^3:\;0\le s^I\le 1;\;I=1,2,3\}
\ee
It has six boundary squares given by 
\ba \label{5.19}
s^0_{I +} &=& \{s\in \mathbb{R}^3:\; s^I=1,\; 0\le s^J, s^K\le 
1;\;\;\epsilon_{IJK}=1\}
\nonumber\\
s^0_{I -} &=& \{s\in \mathbb{R}^3:\; s^I=0,\; 0\le s^J, s^K\le 
1;\;\;\epsilon_{IJK}=1\}
\ea
We easily compute 
\be \label{5.20}
{\rm Vol}_0(T_0)=1
\ee
while (remembering that the surfaces carry outward orientation if the 
edges are outgoing from $v$)  
\be \label{5.21}
F_I(s^0_{J \sigma})=\sigma \delta_{IJ}
\ee
with $\sigma=\pm$. 

Let us label the edge dual to $Y(s^0_{I \sigma})$ by $e_{I\sigma}$, then 
the expectation value becomes 
\be \label{5.22}
{\rm Vol}(v)=
\sqrt{|
\frac{1}{48} \sum_{I,J,K;\sigma_1,\sigma_2,\sigma_3} \;
\epsilon(
e_{I\sigma 1},e_{J\sigma 2},   
e_{K\sigma 3})\; \sigma_1\sigma_2\sigma_3 \; \epsilon_{IJK}
|} \ee
which again depends on the precise embedding of the graph. For an actual 
cubical graph constructed in Section \ref{s3}, the edges $e_{I +}, e_{I 
-}$ are analytic continuations of 
each other, so that the orientation factor vanishes if two or more edges 
carry the same direction label $I$, otherwise, there are more 
contributions. Which orientation factors are allowed has been analysed in 
detail in \cite{17}. In the case of the actual cubical graph we have 
$\epsilon(
e_{I\sigma 1},e_{J\sigma 2},   
e_{K\sigma 3})=\sigma_1 \sigma_2 \sigma_3 \epsilon_{IJK}$, so that 
(\ref{5.22}) becomes
\be \label{5.23}    
{\rm Vol}(v)=
\sqrt{|
\frac{1}{6} \sum_{I,J,K} \; \epsilon_{IJK}^2|}=1
\ee
which coincides with (\ref{5.20}).

\subsubsection{Octahedron}
\label{s5.2.3}

A standard octahedron is the subset
\be \label{5.24}
O_0=\{s\in \mathbb{R}^3:\; |s^3|\le \frac{1}{2},\; |s^1|,|s^2| \le
\frac{1}{2} -|s^3|\}
\ee
It has eight boundary triangles given by 
\be \label{5.25}
t^0_{I \sigma \sigma'} = \{s\in \mathbb{R}^2:\; 
0\le \sigma' s^3 \le \frac{1}{2},\;s^I=\sigma(\frac{1}{2}-|s^3|),\;
|s^J|\le \frac{1}{2} -|s^3|\}
\ee
where $I,J=1,2;\;I\not=J;\;\sigma,\sigma_3=\pm$. 

We easily compute 
\be \label{5.26}
{\rm Vol}_0(O_0)=\frac{1}{3}
\ee
while (remembering that the surfaces carry outward orientation if the 
edges are outgoing from $v$)  
\be \label{5.27}
F_I(t^0_{J \sigma \sigma'})=\frac{1}{4}[\sigma \delta_{IJ}+\sigma' 
\delta_{I3}]
\ee

Labelling the edge dual to $Y(t^0_{I \sigma \sigma_3})$ by $e_{I \sigma 
\sigma_3}$ we find for the expectation value 
\be \label{5.28}
{\rm Vol}_0(v) = \sqrt{|
\frac{1}{48\cdot 64} \sum_{{I_1,I_2,I_3=1,2} \atop
{\sigma_1,\sigma_2,\sigma_3,
\sigma'_1,\sigma'_2,\sigma'_3=\pm}}\;
\epsilon(
e_{I_1 \sigma_1 \sigma_1'},
e_{I_2 \sigma_2 \sigma_2'},
e_{I_3 \sigma_3 \sigma_3'})\;
[\sigma_1 \sigma_2 \sigma_3' \epsilon^{I_1 I_2}
+\sigma_1 \sigma_2' \sigma_3' \epsilon^{I_3 I_1}
+\sigma_1' \sigma_2 \sigma_3 \epsilon^{I_2 I_3}]
|}
\ee
where $\epsilon^{IJ}$ is the alternating 
symbol for $I,J=1,2$ with $\epsilon^{12}=1$. Expression (\ref{5.28})
is already very complicated to analyse for the most general 
edge configuration and, again, we refer to \cite{17} for a comprehensive 
discussion. However, for the case of the 
graphs constructed in section \ref{s3} the situation becomes simple 
enough. Namely, in this case the eight edges $e_{I \sigma \sigma'}$ have 
the property that $e_{I, \sigma, \sigma'}$ and $e_{I, -\sigma, -\sigma'}$ 
are analytic 
continuations of each other. This implies that $\dot{e}_{I, \sigma, 
\sigma'}(0)=\sigma' \dot{e}_{I, \sigma\sigma',+}(0)$ where $e_{I \sigma 
\sigma'}(0)=v$ is the common starting point of all edges. Since 
$\epsilon(e,e',e^{\prime\prime})={\rm sgn}(\det(\dot{e}(0),\dot{e}'(0),
\dot{e}^{\prime\prime}(0)))$ is completely skew in 
$e,e',e^{\prime\prime}$, in this case we can simplify (\ref{5.28}) 
to
\ba \label{5.29}
{\rm Vol}_0(v) &=& \sqrt{|
\frac{1}{48\cdot 64} \sum_{{I_1,I_2,I_3=1,2}\atop
{\sigma_1,\sigma_2,\sigma_3,\sigma'_1,\sigma'_2,\sigma'_3=\pm}}\;
\epsilon(
e_{I_1, \sigma_1 \sigma_1',+},
e_{I_2, \sigma_2 \sigma_2',+},
e_{I_3,\sigma_3 \sigma_3',+})
}
\nonumber\\
&& \overline{\times\;
[\sigma_1\sigma_1' \sigma_2 \sigma_2' \epsilon^{I_1 I_2}
+\sigma_1\sigma_1' \sigma_3 \sigma_3' \epsilon^{I_3 I_1}
+\sigma_2\sigma_2' \sigma_3 \sigma_3' \epsilon^{I_2 I_3}]
|}
\ea
Since (\ref{5.29}) only depends on $\tilde{\sigma}_I=\sigma_I \sigma_I'$,
after proper change of summation variables, (\ref{5.29}) turns into 
\be \label{5.29a}
{\rm Vol}_0(v) = \sqrt{|
\frac{1}{48\cdot 8} \sum_{{I_1,I_2,I_3=1,2}\\
{\sigma_1,\sigma_2,\sigma_3=\pm}}\;
\epsilon(
e_{I_1, \sigma_1,+},
e_{I_2, \sigma_2,+},
e_{I_3, \sigma_3,+})\;
[\sigma_1 \sigma_2 \epsilon^{I_1 I_2}
+\sigma_1 \sigma_3 \epsilon^{I_3 I_1}
+\sigma_2 \sigma_3 \epsilon^{I_2 I_3}]
|}
\ee
Being $\epsilon(
e_{I_1, \sigma_1,+},
e_{I_2, \sigma_2,+},
e_{I_3, \sigma_3,+})$ and $\sigma_1 \sigma_2 \epsilon^{I_1 I_2}$ are 
both antisymmetric under the simultaneous exchange $(\sigma_1 
I_1) \leftrightarrow (\sigma_2 I_2)$ etc. we may further simplify 
(\ref{5.29a}) to 
\ba \label{5.30}
{\rm Vol}_0(v) &=& \sqrt{|
\frac{1}{48\cdot 4} \sum_{\sigma_1,\sigma_2,\sigma_3=\pm}\;
[\sum_{I_3} \;\sigma_1 \sigma_2\;
\epsilon(e_{1, \sigma_1,+},e_{2, \sigma_2,+},e_{I_3, \sigma_3,+})
+\sum_{I_1} \;\sigma_2 \sigma_3\;
\epsilon(e_{I_1, \sigma_1,+},e_{1, \sigma_2,+},e_{2, \sigma_3,+})
}
\nonumber\\
&& \overline{+\sum_{I_2} \;\sigma_3 \sigma_1\;
\epsilon(e_{1, \sigma_1,+},e_{I_2, \sigma_2,+},e_{2, \sigma_3,+})
]
|}
\ea
Carrying out the respective sums over $I_1,I_2,I_3$ and using the fact that 
$\epsilon(e,e',e^{\prime\prime})$ is completely skew we can bring all
orientation factors into one of the two  
standard forms 
$\epsilon(e_{1,\sigma_1,+},e_{1,\sigma_2,+},e_{2,\sigma_3,+})$
and
$\epsilon(e_{2,\sigma_1,+},e_{2,\sigma_2,+},e_{1,\sigma_3,+})$,
respectively. After proper relabelling of the $\sigma_I$ we find that
\be \label{5.31}
{\rm Vol}_0(v) = \sqrt{|
\frac{1}{16\cdot 4} \sum_{\sigma_1,\sigma_2,\sigma_3=\pm}\; \sigma_3\;
[\sigma_2\;
\epsilon(e_{1, \sigma_1,+},e_{1, \sigma_2,+},e_{2, \sigma_3,+})
+\sigma_1\;
\epsilon(e_{2, \sigma_1,+},e_{2, \sigma_2,+},e_{1, \sigma_3,+})
]
|}
\ee
Since $\epsilon(e_{I, \sigma_1,+},e_{I, \sigma_2,+},e_{J, \sigma_3,+})$
is skew in $\sigma_1,\sigma_2$ the sum over $\sigma_2$ 
collapses to the term $\sigma_2=-\sigma_1$ and (\ref{5.31}) becomes
\ba \label{5.32}
{\rm Vol}_0(v) 
&=& \sqrt{|
\frac{1}{16\cdot 4} \sum_{\sigma_1,\sigma_3=\pm}\; 
\sigma_3\;\sigma_1\;
[-
\epsilon(e_{1, \sigma_1,+},e_{1, -\sigma_1,+},e_{2, \sigma_3,+})
+
\epsilon(e_{2, \sigma_1,+},e_{2,-\sigma_1,+},e_{1, \sigma_3,+})
]
|}
\nonumber\\
&=& \sqrt{|
\frac{1}{16\cdot 2} \sum_{\sigma_3=\pm}\; 
\sigma_3\;
[-\epsilon(e_{1,+,+},e_{1,-,+},e_{2, \sigma_3,+})
+
\epsilon(e_{2,+,+},e_{2,+,+},e_{1, \sigma_3,+})
]
|}
\ea
Finally, using 
$\epsilon(e_{I,+,+},e_{I,-,+},e_{J, \sigma_3,+})=
\sigma_3\; \epsilon(e_{I,+,+},e_{I,-,+},e_{J,+,+})$ 
and \\
$\epsilon(e_{1,+,+},e_{1,-,+},e_{2,-,+})=
\epsilon(e_{2,+,+},e_{2,-,+},e_{2,+,+})=1$ 
we find 
\be \label{5.33}
{\rm Vol}_0(v)=\frac{1}{2\sqrt{2}}
\ee
which does not agree with (\ref{5.26}). \\

Interestingly, for both valence 
$n=4$ or $n=8$ the expectation value is larger than the expected value 
with the same ratio $3/(2\sqrt{2})$. 
In general, for generic edge 
configurations and for higher and higher valence, the 
expectation value will probably also be larger in ratio than the 
expected 
volume.   
This is because for a vertex of valence $n$ the number of ordered 
triples of edges, contributing to the expectation value is given by 
${n\choose 3}$ and, for appropriate choice of the orientation factors,
these terms all contribute with the same sign. Such a choice is always 
possible up to topological 
obstructions discussed to some extent
in \cite{17}. For large $n$ the polyhedron dual to the vertex will 
approach more and more a sphere triangulated into $n$ polygonal faces of 
typical 
unit area $4\pi/n$. Hence we expect the leading $n$ 
behaviour of the expectation value to be 
given by $\sqrt{\frac{1}{8}\; n^3/6\; 
(4\pi/n)^3}=\sqrt{8\pi^3/6}=4\pi/3\sqrt{3\pi/4}$, while 
the expected volume should approach $4\pi/3$.    

Surely, we have not shown that, for graph topologies different from a 
cubical one, the expectation value of the volume operator, with 
respect to the dual cell complex coherent states, cannot be matched with 
the classical volume value. This is because one can allow degenerate 
triples which decrease the volume expectation value. However, the 
discussion reveals that the question for which graphs the expectation 
value comes out correctly, is far from trivial and even for natural 
choices the only admissible graph topology is the cubical one.

Notice that the expectation value is insensitive to the embedding of the 
graph relative to the dual cell complex, as long as the graph is dual to 
it. For non dual embeddings or graph topologies, which do not match the 
cell complex topology at all, the expectation value will be completely 
off the correct value. This demonstrates that the cut -- off graph must 
lie within a certain class, which is adapted to the cell complex. 

Summarising, we have shown that the only 
known states of LQG, which are semiclassical 
for the volume operator, must be based on cubic cut -- off graphs.
This looks surprising at first but can, perhaps, be understood intuitively 
by the following reasoning.
 
The volume operator is a derived operator and arises from the 
known representation of the flux operator on the Hilbert space. 
The derivation involves a regularisation step which involves cubes 
surrounding the vertices of the graph in question, on whose faces 
the fluxes are located. In order to take the limit in which the cubes 
shrink to the vertices and, in order to make the result independent of 
the relative orientation between cubes and graphs, an averaging 
procedure must be applied. Hence one might be tempted to say that the 
fact that cubical topology is singled out rests on the cubical 
regularisation. 

However, this is not the case. Namely, cylindrical 
consistency and background 
independence alone already fix the cylindrical projections of the 
volume operator up to a global constant, as proved explicitly in 
\cite{7,8}. The constant depends on the averaging procedure chosen 
and on whether one uses tetrahedra rather than cubes in the 
regularisation. However, 
consistency between volume and flux quantisation fixes 
that factor \cite{9} and rules out the operator \cite{7}. That is to 
say, there is no freedom left in defining the volume operator and, 
therefore, the detail of the regularisation do not matter; it is a 
regularisation independent result. Hence, the preference for cubic 
graphs in the semiclassical analysis must have a different origin.

To see what it is, notice that the volume operator at a vertex involves 
a sum over ordered triples of edges adjacent to the vertex of which  
only those with 
linearly independent tangents contribute. If the vertex has valence $n$
then typically there are ${n \choose 3}$ contributions \cite{17}. 
They all contribute with equal weight (up to sign) which is the unique
factor determined in \cite{9}. That constant is such that each triple 
contributes as if (the tangents of) a triple of edges spans a 
corresponding 
parallelepiped. However, it is clear that generally far less than 
${n \choose 3}$ parallelepipeds are sufficient to triangulate a 
(dual) neighbourhood of the vertex and, thus, it is not surprising that 
large
valence cut -- off graphs will not give rise to good semiclassical 
states. On the other hand, unless the graph is cubic, even at low 
$n=4$ the parallelepiped volume contribution per triple is too high
for the triangulation of a tetrahedron. We have seen both effects at 
work in the previous section.  

This result has two implications: either one is able to find new types 
of states, which are not constructed by the complexifier method or by 
different complexifiers than the ones employed so far, such that the 
correct semiclassical behaviour is recovered also for graphs of 
different than cubic topology. Or, if that turns out to be impossible,
one should accept this result and conclude that, in order that the 
boundary Hilbert space of spin foam models has a semiclassical sector,
one should generalise them to more general than simplicial 
triangulations of the four manifold, as advocated in \cite{48e,30a}.

\section{ Expectation Values of the Volume Operator for Area Coherent States}
\label{s4}

In this Section we compute the expectation value of the operator
$\hat{V}_{\gamma,v}$ for an arbitrary $n$-valent vertex, $v$, for
the stack family coherent states using the replacement of $SU(2)$
by $U(1)^3$. This uses the calculational tools developed in previous sections. 
We may, therefore,
replace the $SU(2)$ right-invariant vector fields by $U(1)^3$
right-invariant vector fields $X^j_{e_I(v)}=i
h^j_I\partial/\partial h^j_I$ acting on $h^j_I:=A^j(e_I(v))$. The
crucial simplification is that these vector fields mutually
commute. Their common eigenfunctions are the spin-network
functions which, for $U(1)^3$, take the explicit form \be
\label{4.0.5} T_{\gamma,n}(A)=\prod_{e\in E(\gamma)}\;
\prod_{j=1}^3\; [A^j(e)]^{n_j^e} \ee We will refer to them as
`charge network states' because the $n_j^e\in \mathbb{Z}$ are
integer valued. Using the spectral theorem we may immediately
write down the eigenvalues of $\hat{V}_{\gamma,v}$ on
$T_{\gamma,n}$ as \be \label{4.0.6} \lambda_{\gamma,n,v}=\ell_P^3
\sqrt{\Big|\frac{1}{8} \sum_{{v\in e_I\cap e_J\cap e_K } \atop
{1\le I\leq J\leq K\leq N}} \epsilon^{ijk}\epsilon(e_I,e_J,e_K)
\big[n^{e_I}_i\; n^{e_J}_j \; n^{e_K}_k\big]\Big|} \ee

What follows is subdivided into four parts. We begin by performing
the calculation for a general graph. This leads to the inverse of
the edge metric which, for large graphs, is beyond analytical
control. In the second part we restrict the class of graphs, which
let us perform perturbative computations of the inverse of the
edge metric. This gives a good approximation of the actual
expectation value. In the third and fourth parts we consider the
dependence of our results on the relative orientation of the graph,
with respect to the family of stacks.

\subsection{Expectation Value of the Volume Operator for a General n-Valent Graph}
\label{s4.1}

In this Section we drop the graph label and set
$t^{jk}_{ee'}:=\delta^{jk}\; t l^\gamma_{e e'},\;t:=\ell_P^2/a^2$.
This gives a positive, symmetric bilinear form on vectors
$n:=(n^e_j)_{e\in E(\gamma),\;j=1,2,3}$ that is defined by
$t(n,n'):=\sum_{e,e',j,k} n^e_j n^{\prime e'}_k t^{jk}_{e e'}=n^T
\cdot t\cdot n'$. We also set $Z^T \cdot n:=\sum_{e, j} Z^j(e)
n^e_j$.

The coherent state associated with an $n$-valent graph, in which
more than one edge intersects a given plaquette $S$, is as
follows: \be \label{4.1} \psi_{Z,\gamma} =\sum_n\;e^{-\frac{1}{2}
n^T\cdot t\cdot n}\; \; e^{Z^T \cdot n}\; T_{\gamma, n} \ee The
norm of the coherent states is given by \be \label{4.2}
||\psi_{Z,\gamma}||^2 =\sum_n \; e^{-n^T \cdot t \cdot n}\; e^{2
P\cdot n} \ee where \be \label{4.2a} P^j(e)=i\int_e
(Z-A)=:\frac{1}{b^2} E^e_j \ee The length parameter, $b$, that
appears here is generally different from the parameter, $a$, that
enters the classicality parameter $t=\ell_P^2/a^2$, as explained
in \cite{30b}.

The expectation value for the volume operator is \be \label{4.3}
<\hat{V_v}>_{Z,\gamma}=
\frac{\langle\psi_{Z,\gamma}\hat{V_{\gamma,v}}
\psi_{Z,\gamma}\rangle}{||\psi_{Z,\gamma}||^2} =\frac{\sum_n \;
e^{-n^T \cdot t \cdot n}\; e^{2 P\cdot n}
\lambda_{\gamma,v}(n)}{||\psi_{Z,\gamma}||^2} \ee where \be
\label{4.4} \lambda_{\gamma,v}(n)=\ell_P^3
\sqrt{\Big|\frac{1}{48}\sum_{e\cap e^{\prime}\cap
e^{\prime\prime}=v}\;\epsilon(e,e^{\prime},e^{\prime\prime})\;
\det_{e e' e^{\prime\prime}}(n)\Big| } \ee are the eigenvalues of
the volume operator $\hat{V}_{\gamma,v}$. We have introduced the
notation \be \label{4.5} \det_{e e' e^{\prime\prime}}(n):=
\epsilon^{jkl}\;n^e_j n^{e'}_k n^{e^{\prime\prime}}_l \ee

The semiclassical limit of the volume operator is obtained from
(\ref{4.3}) in the limit of vanishing $t$. That is, it is the
zeroth-order in $t$ of the expansion of (\ref{4.3}) in powers of
$t$. Since (\ref{4.3}) converges slowly for small values of $t$,
we will perform a Poisson transform which replaces $t$ by
$\frac{1}{t}$, which converges quickly. To this end, analogous to
\cite{46g, 15}, we introduce the following variables \ba \label{4.6}
t_e&:=&t_{ee},\; T_e:=\sqrt{t_e},\;x^e_j:= T_e n^e_j,\;
y^e_j(x):=x^e_j/T_e,\nonumber\\
 C^e_j&:=&P^e_j/T_e,\;w^e_j(n):=n^e_j/T_e,\;
A_{e e'}^{jk}:= A_{e e'} \delta^{jk}:=\frac{t_{ee'}}{\sqrt{t_e
t_{e'}}} \ea Notice that the diagonal entries of $A$ equal unity.
The off-diagonal ones, however, are bounded from above by unity, by
the Schwarz inequality applied to the scalar product defined by
$A$ and are restricted to the six-dimensional subspace restricted
to vectors, with non-zero entries for the $e,e'$ components only.

Then (\ref{4.3}) turns into \be \label{4.7} <\hat{V_v}>_{Z,\gamma}
=\frac{\sum_n \; e^{-x^T \cdot A \cdot x}\; e^{2 C\cdot x}
\lambda_{\gamma,v}(y(x))} {\sum_n \; e^{-x^T \cdot A \cdot x}\;
e^{2 C\cdot x}} \ee Applying the Poisson transform to (\ref{4.7})
we obtain (recall $N:=|E(\gamma)|$) \be \label{4.8}
<\hat{V_v}>_{Z,\gamma} = \frac{\sum_n \; \int_{\mathbb{R}^{3 N}}\;
d^{3N}x\; e^{-2\pi i w(n)^T \cdot x} e^{-x^T \cdot A \cdot x}\;
e^{2 C\cdot x} \lambda_{\gamma,v}(y(x))} {\sum_n \;
\int_{\mathbb{R}^{3 N}}\; d^{3N}x\; e^{-2\pi i w(n)^T \cdot x}
e^{-x^T \cdot A \cdot x}\; e^{2 C\cdot x}} \ee

In order to perform the Gaussian integrals in (\ref{4.8}) we
notice that, by construction, $A$ is a positive-definite,
finite-dimensional matrix, so that its square root, $\sqrt{A}$,
and its inverse are well defined via the spectral theorem. Hence,
we introduce, as new integration variables \be \label{4.9}
z^e_j:=\sum_{e',k} (\sqrt{A})^{ee'}_{jk} x^{e'}_k,\;
y^e_j(z)=\frac{1}{T_e} x^e_j = \frac{1}{T_e}\sum_{e' k}
\big([\sqrt{A}]^{-1}]\big)^{ee'}_{jk} z^{e'}_k \ee The Jacobian
from the change of variables drops out in the fraction, and
(\ref{4.8}) becomes \be \label{4.10} <\hat{V_v}>_{Z,\gamma} =
\frac{\sum_n \; \int_{\mathbb{R}^{3 N}}\; d^{3N}z\; e^{2(C-2\pi i
w(n))^T \cdot \sqrt{A}^{-1} \cdot z} e^{-z^T \cdot z}\;
\lambda_{\gamma,v}(y(z))} {\sum_n \; \int_{\mathbb{R}^{3 N}}\;
d^{3N}z\; e^{2(C-2\pi i w(n))^T \cdot \sqrt{A}^{-1} \cdot z}
e^{-z^T \cdot z} } \ee

Now one would like to shift $z$ into the complex domain by
$\sqrt{A}^{-1}(C-i \pi w)$ and then perform the ensuing Gaussian
integral. This is unproblematic for the denominator of (\ref{4.8}),
which is analytic in $z$, however, the numerator is not. The
careful analysis  in \cite{46g, 15} shows the existence of branch cuts
in $\mathbb{C}^{3N}$ of the fourth-root function involved. In
turn, this shows  that, in the semiclassical limit, both numerator
and denominator are dominated by the $n=0$ term, while the
remaining terms in the series are of order $\hbar^\infty$ (i.e
they decay as $\exp(-k_n/t),\;t=\ell_P^2/a^2$ for some $k_n>0, \;
\lim_{n\to \infty} k_n=\infty$). See \cite{46g, 15} for the detail.

The upshot is that to any polynomial order in $\hbar$ we may
replace (\ref{4.10}) by \be \label{4.11} <\hat{V_v}>_{Z,\gamma} =
\frac{\int_{\mathbb{R}^{3 N}}\; d^{3N}z\; e^{-z^T \cdot z}\;
\lambda_{\gamma,v}(y(z+\sqrt{A}^{-1} C))} {\int_{\mathbb{R}^{3
N}}\; d^{3N}z\; e^{-z^T \cdot z} } \ee which is now defined
unambiguously because the argument of the fourth root is the
square of a real number.

The denominator of (\ref{4.11}) simply equals $\sqrt{\pi}^{3N}$.
Therefore, the only $\hbar$-dependence of (\ref{4.11}) lies in the
numerator in the function $\lambda_{\gamma,v}$. We now note that
the eigenvalues of the volume operator come with a factor of
$\ell_P^3$, as displayed in (\ref{4.4}). Pulling it under the
square root into the modulus, and noticing that the modulus is a
third-order polynomial in $y$, we see that \be \label{4.12}
\lambda_{\gamma,v}\big(y(z+\sqrt{A}^{-1} C)\big) = \sqrt{
\Big|\frac{1}{48}\sum_{e\cap e'\cap e^{\prime\prime}=v}
\epsilon(e, e',e^{\prime\prime}) \det_{e,
e',e^{\prime\prime}}\big(\ell_P^2 y(z+\sqrt{A}^{-1} C)\big)\Big| }
\ee where we have \ba \label{4.13} \big[\ell_P^2
y(z+\sqrt{A}^{-1})C\big]^e_j &=& \frac{\ell_P^2}{T_e}\sum_{e',k}
\big[\sqrt{A}^{-1}\big]^{e e'}_{jk} \big[z+\sqrt{A}^{-1}
C\big]^{e'}_k
\nonumber\\
&=& \frac{\sqrt{t}}{T_e}\sum_{e',k}\;\Big( a^2 \sqrt{t}
\big[\sqrt{A}^{-1}\big]^{e e'}_{jk} z^{e'}_k
+\frac{\sqrt{t}}{T_{e'}} \big[A^{-1}\big]^{e e'}_{jk}
\Big(\frac{a}{b}\Big)^2 E^{e'}_k\Big) \ea where (\ref{4.2a}) has
been used in the last line.

To extract the leading order in $t$ of (\ref{4.11}) is now easy.
First note that the matrix elements of  both $A$ and
$T_e/\sqrt{t}$ are of order unity. Since $a$ is some macroscopic
length scale, the first term that is proportional to $z$ in
(\ref{4.13}) is therefore of order $\sqrt{t}$, while the second is
of zeroth-order in $t$. Therefore,
$F:=\lambda_{\gamma,v}\big(y(z+\sqrt{A}^{-1} C)\big)$ is of the
form \be \label{4.14} F(z\sqrt{t})=\root 4 \of{Q+P(z\sqrt{t})} \ee
where $P$ is a certain sixth-order polynomial in $z \sqrt{t}$ with
no zeroth-order term, while $Q$ is independent of $z$. Moreover,
$Q+P(z\sqrt{t})$ is non-negative for all $z$ because it is the
square of a third-order polynomial in $z$. In particular, this
holds at $z=0$, and therefore $Q$ is also a non-negative number.
Then,  provided $Q>0$, we can define \be \label{4.15}
f(z\sqrt{t}):=\frac{F(z\sqrt{t})}{\root 4\of{Q}}=:\root 4
\of{1+R(z\sqrt{t})} \ee where $R$ is a sixth-order polynomial with
no zeroth-order term which is bounded from below by $-1$. Now, as
in \cite{18}, we exploit the existence of  $r>0$ such that \be
\label{4.16} 1+\frac{1}{4} R-r R^2\le f\le 1+\frac{1}{4} R \ee for
all $R\ge -1$. Inserting this estimate into (\ref{4.11}) we can
bound the integral from above and below because the Gaussian is
positive. The integrals over $R$ and $R^2$ are finite and are at
least of order $t$ because odd powers of $z$ do not contribute to
the Gaussian integral.

It follows that to zeroth-order in $t$ we have \be \label{4.17}
<\hat{V_v}>_{Z,\gamma} =\sqrt{\Big|\frac{1}{48} \sum_{e\cap e'\cap
e^{\prime\prime}=v} \; \epsilon(e,e',e^{\prime\prime}) \;
\det_{e,e',e^{\prime\prime}}(Y)\Big|} \ee where \be \label{4.18}
Y^e_j:=\Big(\frac{a}{b}\Big)^2\sum_{e',k}\; \frac{t}{T_e T_{e'}}
\; [A^{-1}]^{e e'}_{jk} E^{e'}_k \ee

This is as far as we can go with the calculation for a general
graph. Notice that the inverse of the edge metric appears in this
expression and, for a general graph, this is beyond analytical
control. Therefore we will now make restrictions on the graph so
as to analyse (\ref{4.18}) further.\\

\begin{figure}[hbt]
\begin{center}
 \includegraphics[scale=0.3]{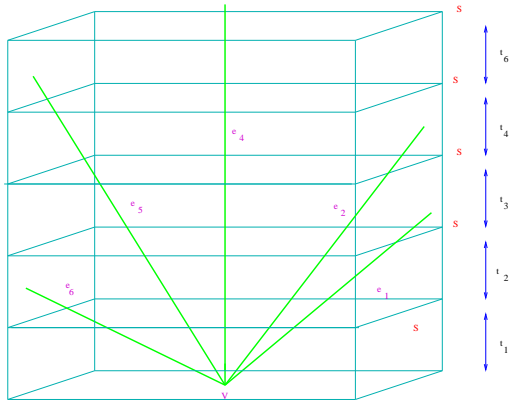}
\caption{Example of a tame, coarse and non-aligned
graph\label{fig0}}
\end{center} \end{figure}

\noindent The assumptions about the class of graphs to be considered are as
follows:
\begin{itemize}
\item[1.] {\it Coordinate Chart}\\
The graph, the region $R$ and the families of stacks lie in a
common coordinate chart $X:\;\mathbb{R}^3\to \sigma$. This is not
a serious restriction, because the general situation may be reduced
to this one by appropriately restricting attention to the various
charts of an atlas that covers $\sigma$.
\item[2.] {\it Tame Graphs}\\
We assume that the graph is {\it tame} with respect to the stacks.
By this we mean that for each direction $I$, and each stack
$\alpha$, a given edge, $e$, of the graph enters and leaves that
stack at most once. This means that the graph does not `wiggle'
too much on the scale of the plaquettes. Analytically, it means
that $l^{I \alpha \gamma}_N$ vanishes whenever $|N_e|\ge 2$ for
any $e$, and that, for given $e$, the number $l^{I \alpha
\gamma}_N$ is non-vanishing at most for either $N_e=+1$ or
$N_e=-1$ but, not both, and independently of $\alpha$. Finally, it
means that the sets $S^{I \alpha \gamma}_N$ are connected.
\item[3.] {\it Coarse Graphs}\\
We assume that the graph is much coarser than the plaquettation, in
the sense that any edge intersects many different stacks in at
least one direction $I$.
\item[4.] {\it Non-Aligned Graphs}\\
We exclude the possibility that distinct edges are `too aligned'
with each other, in the sense that the number of stacks that they
commonly traverse is much smaller than the number of stacks that
they individually traverse .
\end{itemize}
Pictorially, the situation therefore typically looks as in figure
\ref{fig0}.
A consequence of the tameness, coarseness and `alignedness'
assumption is that $|t^\gamma_{e,e'}|\ll t^\gamma_e=t^\gamma_{e
e}, \; t^\gamma_{e'}=t^\gamma_{e' e'}$ for all $e,e'$, as it is
immediately obvious from the formulae displayed in (\ref{2.32a}),
because the number of stacks with $|N_e|=|N_{e'}|=1$ will be very
much smaller than the number of stacks with $|N_e|=1, N_{e'}=0$ or
$|N_{e'}|=1, N_e=0$. Hence the edge metric will be almost
diagonal. This is important because we need its inverse, which can
only be calculated with good approximation (that is, for large,
 semiclassically relevant graphs) if it is almost
diagonal. The graphs that we will eventually consider are
embeddings of subgraphs dual to tetrahedronal, cubical or
octahedronal triangulations of $\mathbb{R}^3$. These correspond to
embeddings of regular $4$-,$6$-,$8$-valent lattices, which ensure
the non-alignedness property.

Thus, without loss of generality, we may choose the stacks and
plaquettes as follows:\\
using the availability of the chart
$X:\;\mathbb{R}^3\to \sigma;\;s\to X(s)$ we consider the
foliations $F^I$ defined by the leaves
$L_{It}:=X^I_t(\mathbb{R}^2)$ where for $\epsilon_{IJK}=1$ we set
$X^I_t(u^1,u^2):=X(s^I:=t,\;s^J:=u^1,s^K:=u^2)$. The stacks are
labelled by $\alpha=(\alpha^1,\alpha^2)\in \mathbb{Z}^2$, the
corresponding plaquettes are given by $p^I_{\alpha
t}=\{X^I_t([\alpha+u]l);\;u\in [0,1)^2\}$ where $l>0$ is a
positive number.

Likewise, using the availability of the chart, we take the edges
of the graph to be embeddings of straight lines in $\mathbb{R}^3$
(with respect to the Euclidean background metric available there),
that is, $e(t)=X(s_e+v_e \delta t)$ where $v_e$ is a vector in
$\mathbb{R}^3$ and $e(0)=X(s_e)$ defines the beginning point of
the edge.

After these preparations, we can now analyse (\ref{4.17}) and
(\ref{4.18}) further. Recall that \be \label{4.19}
E^e_j=\sum_{\alpha,I}\; \int\; dt\; E_j(p^{\alpha I}_t)
\sigma(p^{\alpha I}_t,e) \ee and \be \label{4.20}
t^\gamma_{ee'}=\sum_{\alpha I} \int\; dt\;
 \sigma(p^{\alpha I}_t,e)
\; \sigma(p^{\alpha I}_t,e') \ee By the assumption about the
graphs made above, the signed intersection number takes at most
the numbers $\pm 1$ and independently of $\alpha$, so that
$\sigma(p^{\alpha I}_t,e)^2=\sigma^I_e \sigma(p^{\alpha I}_t,e)$
for certain $\sigma^I_e=\pm 1$ which takes the value $+1$ if the
orientation of $e$ agrees with that of the leaves of the
foliation, $-1$ if it disagrees, and $0$ if it lies inside a leaf.
If we assume that the electric field $E^a_j$ is slowly varying at
the scale of the graph (and hence at the scale of the plaquettes
as well), then we may write \be \label{4.21} E^e_j \approx \sum_I
t^\gamma_{e e} \sigma^I_e E_j(p^I_v) \ee where
$p^I_v=p^{\alpha_I(v) I}_{t_I(v)}$ and $v$ is the vertex at which
$e$ is adjacent and which is under consideration in
$V_{\gamma,v}$. It follows that (\ref{4.18}) can be written as \be
\label{4.22} Y^e_j=\Big(\frac{a}{b}\Big)^2\sum_{e',k}\;
\frac{t}{T_e T_{e'}} \; \big[A^{-1}\big]^{e e'}_{jk} \sum_I
t^\gamma_{e' e'} \sigma^I_e E_k(p^I_v)
=\Big(\frac{a}{b}\Big)^2\sum_{e'}\; \sqrt{\frac{t^\gamma_{e'
e'}}{t^\gamma_{e e}}} \; \big[A^{-1}\big]^{e e'}\sum_I \sigma^I_e
E_j(p^I_v) \ee where we have used $T_e^2=t t^\gamma_{e e}$.

Now, by construction, $A=1+B$ with $B$ off-diagonal and with small
entries \be \label{4.23} B_{ee'}=\frac{t^\gamma_{e
e'}}{\sqrt{t^\gamma_{ee} t^\gamma_{e' e'}}} \ee which are of the
order of $l/\delta$, since two distinct edges will typically only
remain in the same stack for a parameter length $l$, while the
parameter length of an edge is $\delta$. Now notice that under the
assumptions we have made, we have $l^\gamma_{e e'}=0$ if $e,e'$
are not adjacent. Define $S_e$ to be the subset of edges which are
adjacent to $e$, then \ba \label{4.24} ||B x||^2 &=&
\sum_e\sum_{e', e^{\prime\prime}\in S_e} x_{e'} B_{e' e} B_{e
e^{\prime\prime}} x_{e^{\prime \prime}}
\nonumber\\
& \le & [\sup_{e,e'} B_{e e'}^2]\; \sum_e \; \big[\sum_{e'\in S_e}
\;x_{e'}\big]^2
\nonumber\\
&\le& \Big(\frac{l}{\delta}\Big)^2 \sum_e\; \Big[ \big(\sum_{e'\in
S_e} 1^2\big)^{1/2} \; \big(\sum_{e'\in S_e}
x_{e'}^2\big)^{1/2}\Big]^2
\nonumber\\
&\le & \Big(\frac{l}{\delta}\Big)^2\; M\; \sum_e\; \sum_{e'\in
S_e} x_{e'}^2
\nonumber\\
&=& \Big(\frac{l}{\delta}\Big)^2\; M\; \sum_{e'} x_{e'}^2 \;
\sum_{e} \chi_{S_e}(e')
\nonumber\\
&\le & \Big(\frac{l}{\delta}\Big)^2\; M^2\; ||x||^2 \ea Here, in
the second step we have estimated the matrix elements of $B$ from
above; in the third step we have applied the Schwarz inequality; in the
fourth step we have estimated $|S_e|\le M$, where $M$ is the maximal
valence of a vertex in $\gamma$; and in the sixth step we have
exploited the symmetry \be \label{4.25} \chi_{S_e}(e') = \left\{
\begin{array}{cc}
1 & :\;\;e'\cap e\not=\emptyset \\
0 & :\;\;e'\cap e=\emptyset
\end{array}
\right. = \chi_{S_e'}(e) \ee as well as the definition of the norm
of $x$.

It follows that  for $l/\delta< M$, $B$ is bounded from above by
unity.  Therefore, the geometric series
$A^{-1}=1+\sum_{n=1}^\infty (-B)^n$ converges in norm. Hence we
are able to consider the effects of a non-diagonal edge metric up
to arbitrary order, $n$, in $l/\delta$. Here we will consider
$n=1$ only and write $A^{-1}=1-(A-1)=2\cdot 1-A$. However, before
considering corrections from the off-diagonal nature of $A$ notice
that, to zeroth-order in $l/\delta$, equation (\ref{4.22}) becomes
simply \be \label{4.25a} Y^e_j =\Big(\frac{a}{b}\Big)^2 \sum_I
\sigma^I_e E_j(p^I_v) \ee Inserting (\ref{4.25a}) into
(\ref{4.17}) we find \be \label{4.26} <\hat{V_v}>_{Z,\gamma}\,
\approx\, \Big(\frac{a}{b}\Big)^3 \sqrt{\big|\det(E)(v)\big|}\;
\sqrt{\Big|\frac{1}{48} \sum_{e\cap e'\cap e^{\prime\prime}=v} \;
\epsilon(e,e',e^{\prime\prime}) \;
\det_{e,e',e^{\prime\prime}}(\sigma)\; \det(p_v)\Big|} \ee where
\be \label{4.27}
\det(p_v)=\frac{1}{3!}\epsilon_{IJK}\epsilon^{abc} \;
\int_{[0,1)^2}\; d^2u\; n_{a t_I(v)}^{\alpha_I(v) I}(u) \;
\int_{[0,1)^2}\; d^2u\; n_{b t_J(v)}^{\alpha_J(v) I}(u) \;
\int_{[0,1)^2}\; d^2u\; n_{c t_K(v)}^{\alpha_K(v) I}(u) \ee On
recalling that $n^{\alpha I}_{a t}=\epsilon_{abc} X^{I b}_{\alpha
t, u^1} X^{I c}_{\alpha t, u^2}$ with $X^{Ia}_{\alpha
t}(u)=X^a(s^I=t,s^J=\alpha^1+l u^1, s^K=\alpha^2+l u^2)$ for
$\epsilon_{IJK}=1$, we find \be \label{4.28} \det(p_v)\approx l^6
\big[\det(\partial X(s)/\partial s)\big]_{X(s)=v} \ee Hence
(\ref{4.26}) becomes \be \label{4.29} <\hat{V_v}>_{Z,\gamma}
\approx \Big(\frac{a l}{b}\Big)^3 \sqrt{\big|\det(E)(v)\big|}\;
\Big|\big[\det(\partial X(s)/\partial s)\big]_{X(s)=v}\Big|\;
\sqrt{\Big|\frac{1}{48} \sum_{e\cap e'\cap e^{\prime\prime}=v} \;
\epsilon(e,e',e^{\prime\prime}) \;
\det_{e,e',e^{\prime\prime}}(s)\Big|} \ee

We can draw an important conclusion from expression (\ref{4.29}).
Namely, the first \emph{three} factors approximate the classical
volume $V_v(E)$ as determined by $E$ of an embedded cube with
parameter volume $(al/b)^3$. When we sum (\ref{4.29}) over the
vertices of $\gamma$, which have a parameter distance, $\delta$,
from each other where $l\ll \delta$ by assumption, then the volume
expectation value only has a chance to approximate the classical
volume, when the graph is such that $\delta=a l/b$, or
$\delta/l=a/b$. This could never have been achieved for $b=a$ and it
explains why we had to rescale the labels of the coherent states
by $(a/b)^2$, while keeping the classicality parameter at
$t=\ell_P^2/a^2$. See our \cite{46ah} for a detailed
discussion. There we have also explained why one must have $\delta/l$
actually equal to $a/b$ and not just of the same order. In fact, while one
could use this in order to favour other valences of the volume
operator, the expectation value of other geometrical operators,
such as area and flux, would be incorrect.

Assuming $\delta/l=a/b$ we write (\ref{4.29}) as \be \label{4.30}
<\hat{V_v}>_{Z,\gamma}=:V_v(E)\; G_{\gamma,v} \ee thereby
introducing the graph geometry factor $G_{\gamma,v}$. It does not
carry any information about the phase space, only about the
embedding of the graph relative to the leaves of the
three-foliations. From the fact that (\ref{4.30}) reproduces the
volume of a cube up to a factor, we may already anticipate that
the geometry factor will be close to unity for, at most, a cubic
graph. Whether this holds for an arbitrary orientation of the
graph, with respect to the stack family, it will occupy a large part of
the analysis which follows.

\subsection{Analysis of the Graph Geometry Factor}
\label{s4.3} We start by investigating the behaviour of the graph
geometry factor $G_{\gamma,v}$ under diffeomorphisms, $\varphi$,
of $\sigma$, that is, under $G_{\gamma,v}\mapsto
G_{\varphi(\gamma),\varphi(v)}$, while the linearly-independent
families of stacks are left untouched. This will answer the
question of how much the geometry factor depends on the relative
orientation of the graph with respect to the stacks.

In fact, the orientation factor $\epsilon(e,e',e^{\prime\prime})$
is invariant under diffeomorphisms of the spatial manifold
$\sigma$. The signature factor \be \label{4.31}
\det_{e,e'e^{\prime\prime}}(\sigma)=\epsilon_{IJK} \sigma^I_e
\sigma^J_{e'} \sigma^K_{e^{\prime\prime}} \ee is obviously
invariant under any diffeomorphism that preserves the foliations
$F^I$, i.e. which map leaves onto leaves, because \be \label{4.32}
\sigma^I_e=\frac{1}{2}\int_e dx^a\, \epsilon_{abc} \int_{L_{It}}
dy^b\wedge dy^c \delta(x,y) \ee where $L_{It}$ is any leaf in $t$
which intersects $e$ transversely. Since we consider graphs, whose
edges are embedded lines in $\mathbb{R}^3$ with the same embedding
that defines the stacks, it follows that the geometry factor is
invariant under any embedded global translations in
$\mathbb{R}^3$.

Next, since global rescaling in $\mathbb{R}^3$ preserves the
foliations and the topological invariant (\ref{4.32}), the
geometry factor is also invariant under embedded global rescalings
of $\mathbb{R}^3$. Finally, any embedded global rotations of
$\mathbb{R}^3$, that preserves all the orientation factors
$\sigma^I_e$, will leave the geometry factors invariant. 

Since the
orientation factors only take the values $+1,-1, 0$ (depending on
whether an edge agrees, disagrees with the orientation of the
leaves, or lies within a leaf), there will be a vast range of
Euler angles for which this condition is satisfied, if the graph
is an embedded, regular lattice of constant valence\footnote{In
fact, for a random graph we may also have rotational invariance on
large scales.}. Hence, in order to check whether the geometry
factor is rotationally invariant under any rotation we only need to
worry about those rotations which lead to changes in the
$\sigma^I_e$. Likewise, if we rotate a graph which is dual to a
polyhedronal complex, we expect that the expectation value remains
invariant as long as the graph remains dual to the complex.

Fortunately, using the explicit formulae derived for the edges and
vertices for $n=4$-, $6$-, $8$-valent graphs displayed in
\cite{30b} we can  calculate the $\sigma^I_e$ for each edge $e$.
Intuitively, it is clear, that whenever many of the $\sigma^I_e$
change from $+1$ to $-1$,  we can expect a drastic change of the
expectation value. However, one has to take into account the
combined effect of these changes,  and this is what makes
rotational invariance possible. As a first step we determine the
action of a rotation on the sign factors.

\subsection{Calculation of the $\sigma^I_e$ Terms}
\label{s4.4}

 In what follows we will discuss the cases that show a drastic change in
 the value of
 $\det_{e,e^{'},e^{''}}(\sigma)=\epsilon_{IJK}\sigma_e^I\sigma_{e^{'}}^J
 \sigma_{e^{''}}^k$ caused by a change in the values of $\sigma^I_e$.
To carry out this calculation we will perform a rotation of each
of the three different types of lattice analysed so far: namely,
the $4$-, $6$- and $8$-valent lattices. 

These rotations will be
parametrised by Euler angles and will be centred at a particular
vertex of the lattice, for example $V_0$. The effects of a
rotation will depend on the distance of the vertices from the
centre of the rotation. In fact, the position of each vertex in
the lattice after rotation, will depend on both the distance from
the centre of the rotation and the Euler angles used in
 the rotation. Fortunately, the values of the terms $\sigma^I_e$ will
 not depend on the former
 but only on the latter.

This is easy to see since the value of $\sigma^I_e$ can be either
1, -1 or 0 depending on whether the edge is outgoing, ingoing, or
lies on the plaquette in the direction $I$. Thus it will only
depend on the angle the edge makes with the perpendicular to the
plaquette in any given direction, i.e. it will depend on the
angles the edge makes with respect to a coordinate system centred
at the vertex at which the edge is incident. Clearly, only the
values of the Euler angles of the rotation  will affect the angles
each edge has, with respect to the vertex at which it is incident.
In particular, since the graph we are using is regular, following
the rotation, all edges which were parallel to each other will
remain such and, thus, will have the same angles with respect to
the vertex at which they are incident. This implies that in order
to compute the values of the terms $\sigma^I$, we can consider
each vertex separately and apply the same rotation to each vertex
individually.

On the other hand, the distance from the centre of the rotation
affects the position of each vertex with respect to the plaquette
structure, and thereby affects both the values of the terms
$t_{ee^{'}}$ and the number of them that are different from zero.
These effects  can be easily understood with the aid of the
two-dimensional diagram (Figure \ref{fig1}).

\begin{figure}[htb]
\begin{center}
\psfrag{b}{$e$}\psfrag{c}{$e^{'}$}\psfrag{a}{$t_{e,e^{'}}$} 
\includegraphics[scale=0.6]{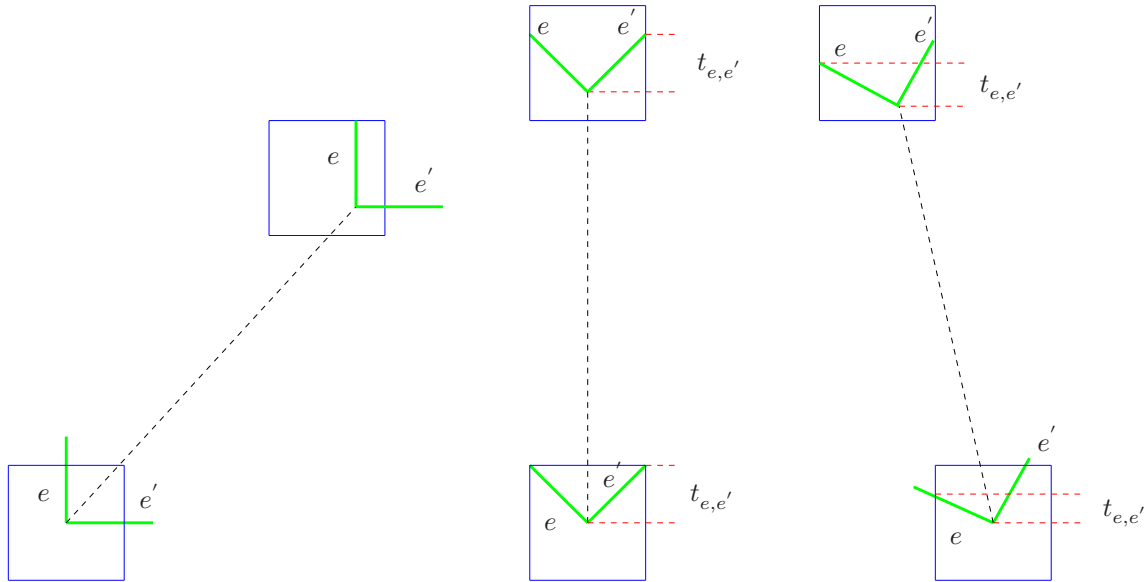}
\caption{Example of translation and rotation \label{fig1}}
\end{center}
 \end{figure}

It is clear that, for any two parallel edges, the angle each of
them has with respect to the vertex at which they are incident, is
independent of the distance of the edge from the centre of
rotation. On the other hand, the values of the $t_{ee^{'}}$ will
depend on both the rotation and the distance of the centre of
rotation, since the position of the rotated vertex, with respect to
the plaquette, depends on both these parameters. Therefore, we can
tentatively assume that two different geometric factors will be
involved in the computation of the volume operator:
\begin{enumerate}
\item[i)] $G_{\gamma,V}$, which indicates how the terms $\sigma^I$ are
affected by rotation. This geometric factor affects all orders of
approximation of the expectation value of the volume operator.
\item[ii)] $C_{\gamma, V}$, which indicates the  effect of rotation on the
terms $t_{ee^{'}}$. This term affects only the first- and
higher-order approximations of the expectation value of the volume
operator, not the zeroth-order.
\end{enumerate}

In what follows we will analyse the geometric term $G_{\gamma,
V}$, i.e. we will analyse the changes in the values of the
$\sigma_e^{I}$ due to a rotation applied at each vertex
independently. We will do this for the 4-, 6- and 8-valent graphs
separately. The geometric factor $C_{\gamma, V}$ will be analysed
in subsequent Sections.

As we will see, our calculations show that for all 4-, 6- and
8-valent graphs, the rotations that produce drastic change in the
values of $\det_{e,e^{'},e^{''}}(\sigma)
 =\epsilon_{IJK}\sigma_e^I\sigma_{e^{'}}^J\sigma_{e^{''}}^k$ have
 measure zero in $SO(3)$, since they occur for specific Euler angles rather
 than for a range of them.

Let us start with the 6-valent graph (Figure \ref{fig6}).
\begin{figure}[htbp]
\begin{center} \includegraphics[scale=0.7]{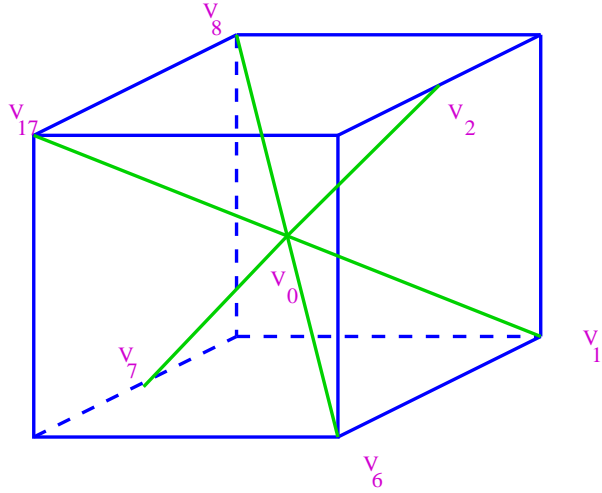}
\caption{6-valent vertex \label{fig6}}
\end{center}
 \end{figure}
From the discussion above, we need only consider the effects of
the rotation on one vertex, $V_0$.

In order to compute the change in the values of the individual
$\sigma_e^{I}$, we will divide the cube,
formed by the intersection of the plaquettes in the three
directions and containing the vertex we are analysing ($V_0$), into eight small sub-cubes.
\begin{figure}[htbp]
\begin{center}
 \includegraphics[scale=0.7]{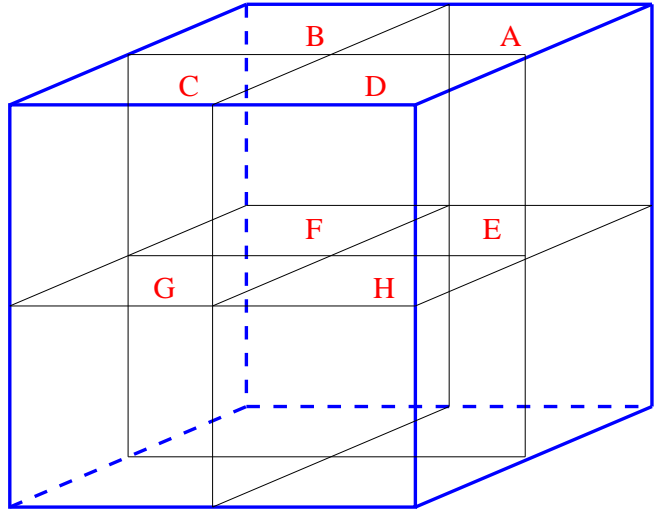}
\caption{Division of the cube in 8 sub-cubes \label{figdivision}}
\end{center}
 \end{figure}
It is then easy to see that, for each edge $e$, the corresponding
value of $\sigma_e^I$  depends on  the
 sub-cube in which it lies. In particular, we have the
following table for the values of $\sigma^I_e$.
\begin{center}
\begin{tabular}{l|l|l|l|l|l|l|l|l}
& A & B & C & D & E & F & G & H\\ \hline $\sigma_e^x$& + & - &- &
+& + & - & - & +\\ \hline $\sigma_e^y$& + & + &- & - & + & + & - &
-      \\ \hline $\sigma_e^z$& + & + &+& +&- & -& -& - \\ \hline
\end{tabular}
\end{center}

From the above table it is clear that when an edge moves from one
of the eight cubes to another, the values of each of the
$\sigma_e$ changes accordingly. Given any 6-valent vertex, each of
the six edges incident at a vertex will be in one distinct cube.
Moreover, since any two edges incident at a vertex can be either
co-planar or perpendicular (in the abstract pull-back space with
Euclidean metric), there are only certain combinations of allowed
positions. For instance, for the edges $e_1,e_2,e_3,e_4,e_5,e_6$
only the combinations \be C_1B_2D_3E_4F_5H_6\mbox{ , }
C_1D_2A_3F_4G_5E_6 \mbox{ , }A_1B_2C_3H_4E_5G_6 \mbox{ ,
}B_1A_2D_3H_4G_5F_6 \ee are allowed (here the notation $A_i$ means
that the edge $i$ lies in the cube $A$);  the combination
$A_1B_2C_3D_4G_5F_6$ is not allowed.

Because of the highly symmetric structure of the 6-valent graph we
do not have to analyse all possible combinations of all the six
edges incident at a vertex, since different combinations are
related by symmetry arguments. For example, the combination in
which edges $e_1,e_2,e_3,e_4,e_5,e_6$ lie in the cubes
$C_1D_2A_3F_4G_5E_6$, and the combination in which they lie in the
cubes $F_1G_2H_3C_4D_5A_6$ lead to the same value of
$|\det_{e,e^{'},e^{''}}(\sigma)|=
|\epsilon_{IJK}\sigma_e^I\sigma_{e^{'}}^J\sigma_{e^{''}}^k|$, and
an equal number of $\det_{e,e^{'},e^{''}}(\sigma)>0$ and
$\det_{e,e^{'},e^{''}}(\sigma)<0$, but obtained from different
triplets $e,e^{'},e^{''}$. In particular, any consistent
relabelling of the edges will produce the same overall result for
the determinants of the triplets. These symmetries reduce,
considerably, the number of cases that need to be analysed.

\begin{figure}[htbp]
\begin{center}
 \includegraphics[scale=0.7]{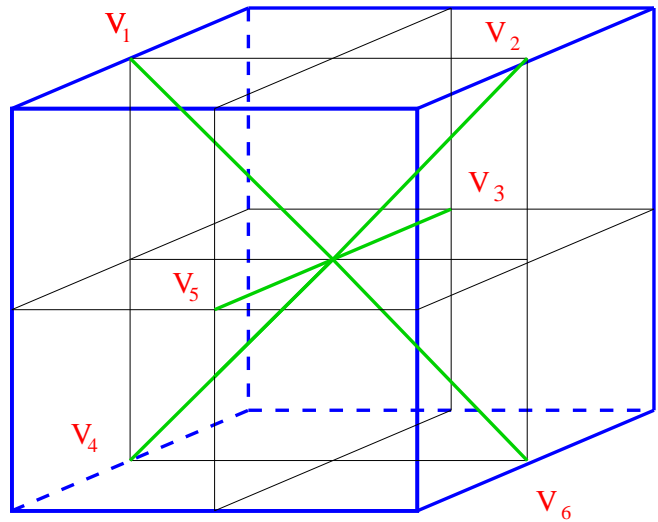}
\caption{Example of configuration with zero measure in
$SO(3)$\label{fig:9} }
\end{center}
\end{figure}

\begin{figure}[htbp]
\begin{center}
 \includegraphics[scale=0.7]{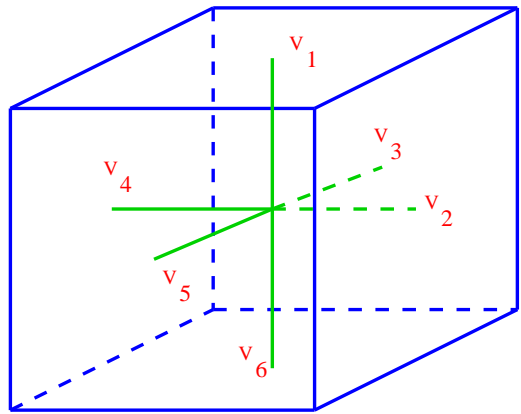}
\caption{Regular 6-valent graph \label{fig:orig}}
\end{center}
\end{figure}

In what follows, we consider the cases for which the edges
$e_1,e_2,e_3,e_4,e_5,e_6$ lie in the following combinations of
cubes: \be C_1B_2D_3E_4F_5H_6\mbox{ , }C_1D_2A_3F_4G_5E_6\mbox{ ,
}A_1B_2C_3H_4E_5G_6\mbox{ , }B_1A_2D_3H_4G_5F_6\ee For each of
these cases there will be sub-cases according to whether one edge
or more lie in a particular plaquette, or are parallel to a given
direction $I,J,K$. These sub-cases are the following:
\begin{enumerate}
\item No edge lies in any plaquette, or is parallel to any of the
directions.

In this case we obtain $|\det_{e,e^{'},e^{''}}(\sigma)|=4$ for all
triplets, but four of these triplets will have
$\det_{e,e^{'},e^{''}}(\sigma)=-4$ while the remaining four will
have $\det_{e,e^{'},e^{''}}(\sigma)=4$.

\item Only one edge lies in a particular plaquette (say the $J$ direction)

(see Figure \ref{fig6}). This edge and its co-linear edge will
have $\sigma^J_e$ equal to zero (J being the direction of the
plaquette in which the edge lies.)

In this case we obtain $\det_{e,e^{'},e^{''}}(\sigma)=-4$ for four
triplets, and $\det_{e,e^{'},e^{''}}(\sigma)=4$ for the remaining
four triplets.\footnote{Note that the geometric factor associated
to this edge orientation will coincide with the geometric factor
as derived from case 1). In this sense, case 2) can be seen as a
limiting case of 1)}

\item Two edges lie in two different plaquettes such that each of
these two edges and their respective co-linear edges will have
$\sigma^I_e$ equal to zero in the direction of the plaquette in
which they lie. In this case, because of the geometry of the
6-valent lattice, the remaining edges will each be parallel to a
given direction $J$, such that all but the $\sigma^J_e$ are zero.

In this case we obtain $\det_{e,e^{'},e^{''}}(\sigma)=2$ for four
triplets while the remaining four will have
$\det_{e,e^{'},e^{''}}(\sigma)=-2$. (See Figure \ref{fig:9}).

\item Each edge is parallel to a given direction such that all the $\sigma^I_e$ (for any $I,J,K$) are equal to zero,
except for the one in the direction to which the edge is parallel.
In this case we obtain $\det_{e,e^{'},e^{''}}(\sigma)=1$ for four
triplets and $\det_{e,e^{'},e^{''}}(\sigma)=-1$ for the remaining
four. (See Figure \ref{fig:orig})
\end{enumerate}
Only sub-cases 3 and 4 might lead to a change of value for the
geometric factor $G_{\gamma,V}$. However, cases 2, 3 and 4 have
measure zero in $SO(3)$.

As a demonstrative calculation on how this is derived we will
choose case 3. In particular, we select the configuration depicted
in Figure \ref{fig:9},
 which can be obtained by a rotation of the original configuration
in Figure \ref{fig:orig}.
Let us consider the linearly-independent triples comprised of the
edges that connect the barycentre of the cube to the vertices
$V_1$, $V_2$ and $V_3$. In the original configuration the
coordinates of these vertices  are (in what follows we will denote
the length of an edge, $e$, by $\delta_e=\delta$) \be
V_1=(0,0,\delta)\mbox{ , } V_2=(\delta,0,0)\mbox{ , }
V_3=(0,\delta,0)\ee By applying a general Euler rotation, whose
matrix representation is given in (\ref{a1}), the coordinates of
the rotated vertices become: \be V_1=(R_{13}\delta, R_{23}\delta,
R_{33}\delta)\mbox{ , } V_2=(R_{11}\delta, R_{21}\delta,
R_{31}\delta)\mbox{ , } V_3=(R_{12}\delta, R_{22}\delta,
R_{32}\delta)\ee

 Our task now is to determine which Euler angles
would give rise to the configuration in Figure \ref{fig:9}. Since
in such a configuration the edges $e_{01}$ (the edge joining the
barycentre of the cube to vertex $V_1$) and $e_{03}$ lie in the
plane $x$--$y$, while the edge $e_{01}$ is parallel to the
$z$-direction,
 the coordinates of the rotated vertices are
constrained by the following set of equations:
\begin{align}
&x_{V_1}=R_{13}= \sin\s \sin\te<0\hspace{.2in} y_{V_1}=R_{23}=
\cos\s \sin\te=0\hspace{.2in} z_{V_1}=R_{33}=\cos\te >0\\
& x_{V_2}=R_{11}=\cos \s \cos \p -\cos \te \sin\p \sin\s<0
\hspace{.2in} y_{V_2}=R_{21}-\sin\s \cos\p- \cos\te \sin\p \cos\te=0\\
&z_{V_2}=R_{31}=\sin\te \sin\p>0\\
&x_{V_3}=R_{12}\cos\s \sin\p +\cos\te \cos\p
\sin\s=0\hspace{.2in}y_{V_3}=R_{22}=-\sin\s \sin\te+\cos\te \cos\p
\cos\s>0\\ &z_{V_3}=R_{32}=-\sin\te \cos\p=0
\end{align}
By solving this set of equations we find that the Euler angles
 $\s$, $\p$ and $\te$, that give rise to the configuration in Figure
 \ref{fig:9} are
\begin{enumerate}
\item[i)] $\te=(n+1)\frac{\pi}{2}$ and $\s=(p+1)\frac{\pi}{2}$ for $n$=odd, $p$=even and $0<\te<\frac{\pi}{2}$
\item[ii)] $\te=(n+1)\frac{\pi}{2}$ and $\s=(p+1)\frac{\pi}{2}$ for $n$=even, $p$=odd and $\frac{3\pi}{2}<\te<2\pi$
\end{enumerate}
It follows that the arrangement of edges under scrutiny has
measure zero in $SO(3)$.

By a similar method it can be shown that whenever an edge lies in
a plaquette, or it is parallel to a plaquette, one of the Euler
angles will have to be equal to $\frac{n}{\pi}$ for $n$ odd or
even. Therefore, that arrangement will have measure zero. This is
not so for the general arrangement (number 1) delineated above.
However, for any such arrangement, the values for the orientation
factor and, subsequently, the geometric factor $G_{\gamma, V}$ will
always be the same and, in zeroth-order, it will not lead to any
changes of the expectation value of the volume operator.

Hence, the only cases of interest---i.e. the cases with measure
different from zero---will not lead to a rotational dependence of
the expectation value of the volume operator in zeroth-order. This
should not come as a surprise, since the geometry of a regular
6-valent graph is such that to each edge there corresponds a
co-linear one. Thus, whenever the term $\sigma^I_e$ for edge $e$
changes from -1 to 1, the term $\sigma^I_{e^{'}}$ of the co-linear
edge undergoes the inverse transformation. As a consequence there
will always be the same number of
$\det_{e,e^{'},e^{''}}(\sigma)=-4$ and
$\det_{e,e^{'},e^{''}}(\sigma)=4$, although the triplets involved
will be different in each case. It follows that  the overall value
of the geometric factor $G_{\gamma,V}$ remains constant.

A similar reasoning holds for the 8-valent graph, since here too
each edge has a corresponding co-linear edge. Therefore, there
will always be an equal number of $\sigma^I_e=1$ and
$\sigma^I_e=-1$. This implies that, as in the case for 6-valent
graph, when no edge lies on a plaquette, the value of the
expectation value of the volume operator for each 8-valent vertex
will be rotationally invariant. On the other hand, the orientation
of edges in an 8-valent graph, in which one or more edges lie in a
plaquette, or an edge is parallel to a given direction, have
measure zero in $SO(3)$, as it was the case for the 6-valent graph.
However, as previously stated, it is precisely such cases that
lead to a change in the value of the geometric factor $G_{\gamma,
V}$.

For the 4-valent case the situation is somewhat different since
there are no co-planar edges. Those arrangements of edges, with
respect to the stacks of plaquettes that cause drastic changes in
the values of the orientation factor, are the following:
\begin{enumerate}
\item No edge lies in any plaquette. In this case we obtain $|\det_{e,e^{'},e^{''}}(\sigma)|=4$ for all linearly-independent triplets.
\item Each edge lies in a given plaquette. This gives
$|\det_{e,e^{'},e^{''}}(\sigma)|=2$ for all linearly-independent
triplets.
\item One edge is aligned with a given plaquette, one edge lies in a given plaquette, and  the remaining edges do not lie in---and are not aligned to---any plaquette. In this case we obtain $|\det_{e,e^{'},e^{''}}(\sigma)|=1$ for two triplets , $|\det_{e,e^{'},e^{''}}(\sigma)|=2$ for one triplet, and
$|\det_{e,e^{'},e^{''}}(\sigma)|=4$ for the remaining triplet.
\end{enumerate}

Similar calculations to those for the 6-valent graph will then show  
that the cases 1 and 2  above have measure zero in $SO(3)$.

Summarising,  the discussion above shows that for all 4-, 6- and
8-valent graphs, those orientations of the edges with respect to
the stacks that cause a drastic change in the orientation factor,
have measure zero in $SO(3)$. Therefore, up to measure zero in
$SO(3)$, the geometric factor $G_{\gamma, V}$ for these graphs is
rotationally invariant.

\subsubsection{Computation of the geometric factor for 4-, 6- and 8-valent graphs }
\label{s 2.4.1} In this Section we will compute the geometric
factor $G_{\gamma,v}$ for the 4-, 6- and 8-valent graphs. We
recall from equation (\ref{4.29}) that the expression for the
geometric factor is \ba\label{ali:g}
G_{\gamma,v}:&=&\sqrt{\Big|\frac{1}{48}\sum_{e\cap e^{'}\cap
e^{''}=v}\epsilon(e, e^{'}, e^{''})\det_{e, e^{'},
e^{''}}(\sigma)\Big|}\\\nonumber
&=&\sqrt{\Big|\frac{1}{8}\sum_{1\leq i\le j\le k\leq
N}\epsilon(e_i, e_{j}, e_{k})\det_{e_i, e_{j},
e_{k}}(\sigma)\Big|} \ea where $N$ is the valence of the vertex
and $\det_{e_i, e_{j}, e_{k}}(\sigma)=\epsilon_{IJK}
\sigma^I_{e_i} \sigma^J_{e_j} \sigma^K_{e_k}$. In what follows we
will calculate $G_{\gamma, v}$ for the 4-, 6- and 8-valent graphs,
respectively. In particular (for each valence) we will analyse
each of the cases discussed in the previous Section which lead to
different values of orientation factor. Any sub-case of these
cases will lead to the same geometric factor.

\paragraph*{4-valent graph:}

We now compute the geometric factor for the 4-valent vertex for
different embeddings of the graph in the stack of surfaces.
\begin{enumerate}
\item
The most general situation is one in which none of the edges is
aligned to, or lies in, a given plaquette. Thus, for example,
consider the situation in which the edges $e_1$, $e_2$ $e_3$ and
$e_4$ are in the octants A, C, H and F, respectively (see Figure
\ref{fig:4general}). Such a combination has a non-zero measure in
$SO(3)$.

The values for $\det_{e_i, e_{j}, e_{k}}(\sigma)$ relative to this
case are given by
 \begin{align}\label{ali:g4}
 &\det_{e_1,e_{2},e_{3}}(\sigma)=\det_{e_1,e_{3},e_{4}}(\sigma)=-4\\\nonumber
 &\det_{e_1,e_{2},e_{4}}(\sigma)=\det_{e_2,e_{3},e_{4}}(\sigma)=4
 \end{align}
Inserting these values in (\ref{ali:g}) it gives \ba
G_{\gamma,v}&=&\sqrt{\Big|\frac{1}{8}\sum_{1\leq i\le j\le k\leq
4}\epsilon(e_i, e_{j}, e_{k})\det_{e_i, e_{j},
e_{k}}(\sigma)\Big|}\\\nonumber
&=&\sqrt{\Big|\frac{1}{8}\big(-4\epsilon(e_1, e_{2},
e_{3})-4\epsilon(e_1, e_{3}, e_{4})+4\epsilon(e_1, e_{2},
e_{4})+4\epsilon(e_2, e_{3}, e_{4})\big)\Big|}\\\nonumber
&=&\sqrt{\frac{1}{8}16} \ea
\begin{figure}[hbtp]
 \begin{center}
  \includegraphics[scale=0.7]{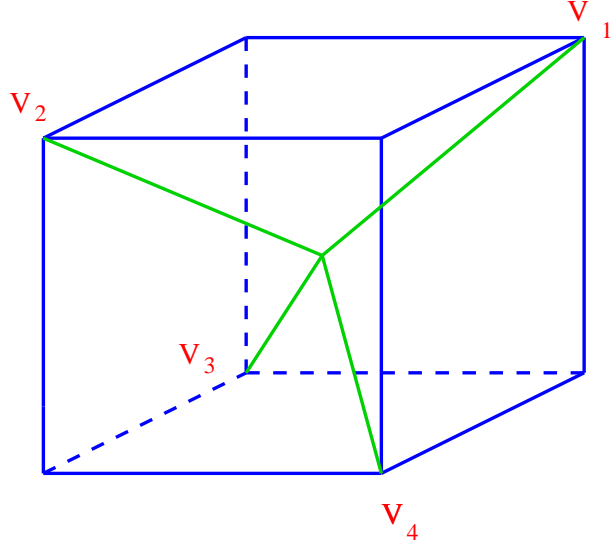}
 \caption{General  4-valent vertex }\label{fig:4general}
 \end{center}
 \end{figure}
\item If, instead, we consider the case in which each of the edges lies in a plaquette as, for example, it is depicted in Figure \ref{fig:4aligned}, then the value for the geometric factor is
\ba G_{\gamma,v}&=&\sqrt{\Big|\frac{1}{8}\sum_{1\leq i\le j\le
k\leq 4}\epsilon(e_i, e_{j}, e_{k})\det_{e_i, e_{j},
e_{k}}(\sigma)\Big|}\\\nonumber
&=&\sqrt{\Big|\frac{1}{8}\big(-2\epsilon(e_1, e_{2},
e_{3})-2\epsilon(e_1, e_{3}, e_{4})+2\epsilon(e_1, e_{2},
e_{4})+2\epsilon(e_2, e_{3}, e_{4})\big)\Big|}\\\nonumber
&=&\sqrt{\frac{1}{8}8} \ea
\begin{figure}[hbtp]
 \begin{center}
  \includegraphics[scale=0.7]{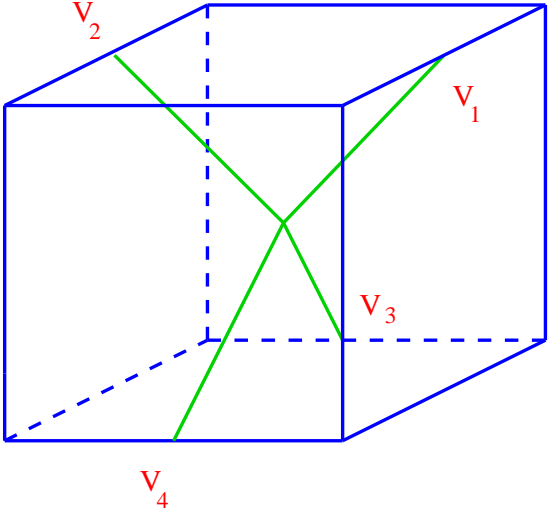}
 \caption{Aligned 4-valent vertex }\label{fig:4aligned}
 \end{center}
 \end{figure}
\item For the situation in which one edge lies in a plaquette and another edge is aligned with a plaquette in another direction (Figure \ref{fig:43}), we obtain\ba
G_{\gamma,v}&=&\sqrt{\Big|\frac{1}{8}\sum_{1\leq i\le j\le k\leq
4}\epsilon(e_i, e_{j}, e_{k})\det_{e_i, e_{j},
e_{k}}(\sigma)\Big|}\\\nonumber
&=&\sqrt{\Big|\frac{1}{8}\big(-1\epsilon(e_1, e_{2},
e_{3})-4\epsilon(e_1, e_{3}, e_{4})+1\epsilon(e_1, e_{2},
e_{4})+2\epsilon(e_2, e_{3}, e_{4})\big)\Big|}\\\nonumber
&=&\sqrt{\frac{1}{8}8} \ea
 \begin{figure}[hbtp]
 \begin{center}
  \includegraphics[scale=0.7]{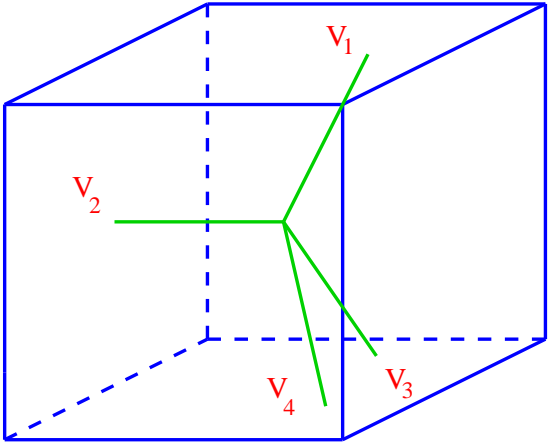}
 \caption{Semi-aligned 4-valent vertex }\label{fig:43}
 \end{center}
 \end{figure}
\end{enumerate}
However, we have proved above that the embeddings of the vertex, with
respect to the stack depicted in cases $2)$ and $ 3) $ have
measure zero in $SO(3)$.


\paragraph*{6-valent graph:}
We now compute the geometric factor for the 6-valent vertex in the
cases from 1 to 4, described in the previous Section and which lead to
different values of the signature factor.
\begin{enumerate}
\item We start with the most general embedding of a 6-valent vertex with respect to the stacks. For example, consider the case in which the edges $e_1$, $e_2$ $e_3$, $e_4$ $e_5$ and $e_6$ are in the octants A H E B C and G, respectively. We then obtain the following value for the geometric factor:
 \ba
G_{\gamma,v}&=&\sqrt{\Big|\frac{1}{8}\sum_{1\leq i\le j\le k\leq
4}\epsilon(e_i, e_{j}, e_{k})\det_{e_i, e_{j},
e_{k}}(\sigma)\Big|}\\\nonumber
&=&\sqrt{\Big|\frac{1}{8}\big(4\epsilon(e_1, e_{2},
e_{3})+4\epsilon(e_1, e_{3}, e_{4})+4\epsilon(e_1, e_{4},
e_{5})-4\epsilon(e_1, e_{2}, e_{5})-4\epsilon(e_2, e_{3},
e_{4})+4\epsilon(e_2, e_{5}, e_{6}})\\\nonumber &
&\overline{-4\epsilon(e_3, e_{4}, e_{6})-4\epsilon(e_4, e_{5},
e_{6})\big)\Big|}\\\nonumber &=&\sqrt{\frac{1}{8}4\times 8} \ea

\item For the geometric factor, when only \emph{one} edge and its co-planar edge lie in a plaquette (Figure \ref{fig6}), we obtain: $G_{\gamma,v}=2$.

\item For the case in which \emph{two} edges and their co-planar edge lie in two different plaquettes in two different directions, while the remaining edge and its co--planar edge are aligned with the plaquette in the third direction (Figure \ref{fig:9}), we obtain $G_{\gamma,v}=\sqrt{2}$.

\item For the case in which all the edges are aligned with the stacks
(Figure \ref{fig:orig}) we obtain  $G_{\gamma,v}=1$, since in that
case $|\det_{e_i,e_j,e_k}|=1$ for all linearly independent
triplets $e_i$, $e_j$, $e_k$. \end{enumerate} However, we have
proved above that cases 2), 3) and 4) have measure zero in
$SO(3)$.

\paragraph*{8-valent graph:}We now compute the geometric factor for the 8-valent vertex for
different embeddings of the graph, with respect to the stack of
surfaces.
\begin{enumerate}
\item
In the most general case, none of the edges lie in, or are aligned
to, a given plaquette: for example, when the edges $e_1$, $e_2$
$e_3$, $e_4$ $e_5$, $e_6$ $e_7$ and $e_8$ are in the octants $B$,
$C$, $A$, $D$, $H$, $E$, $G$ and $F$, respectively. This leads to
the following result \ba
G_{\gamma,v}&=&\sqrt{\Big|\frac{1}{8}\sum_{1\leq i\le j\le k\leq
4}\epsilon(e_i, e_{j}, e_{k})\det_{e_i, e_{j},
e_{k}}(\sigma)\Big|}\\\nonumber
&=&\sqrt{\Big|\frac{1}{8}\big(4\epsilon(e_1, e_{2},
e_{3})+4\epsilon(e_1, e_{2}, e_{4})-4\epsilon(e_1, e_{2},
e_{8})-4\epsilon(e_1, e_{2}, e_{7})-4\epsilon(e_1, e_{3},
e_{4})+4\epsilon(e_1, e_{3}, e_{6})}\\\nonumber &
&\overline{+4\epsilon(e_1, e_{3}, e_{8})+4\epsilon(e_1, e_{4},
e_{6})-4\epsilon(e_1, e_{4}, e_{7})+4\epsilon(e_1, e_{6},
e_{7})+4\epsilon(e_1, e_{6}, e_{8})-4\epsilon(e_1, e_{7},
e_{8})}\\\nonumber
& &\overline{-4\epsilon(e_2, e_{3}, e_{4})-4\epsilon(e_2, e_{3}, e_{5})\cdots\big)\Big|}\\
&=&\sqrt{\frac{1}{8}4\times 32} \ea

\item A more restricted case is when one edge and its co-planar edge are aligned with a plaquette in a given, different direction, while the remaining three edges and their co-planar edge lie in a given plaquette. Here  we obtain $G_{\gamma, V}=\sqrt{5}$.

\item A special case is when each edge lies in a given plaquette, this gives $G_{\gamma, V}=2\sqrt{2}$.
\end{enumerate}
Similarly to the 4- and 6-valent vertex above, arrangement $2)$
and $3)$ have measure zero in $SO(3)$.

From the discussion above of the geometric factor we can already
deduce that, ignoring off-diagonal entries of the edge metric $A$,
the expectation value of the volume operator gives the correct
semiclassical value only for combinations of edges that have
measure zero in $SO(3)$.

In fact, in zeroth-order in $\frac{l}{\delta}$ the expectation
value of the volume operator is given by \be
<\hat{V_v}>_{Z,\gamma} \approx \Big(\frac{a l}{b}\Big)^3
\sqrt{\big|\det(E)(v)\big|}\; \Big|[\det(\partial X(s)/\partial
s)]_{X(s)=v}\Big|\; \sqrt{\Big|\frac{1}{48} \sum_{e\cap e'\cap
e^{\prime\prime}=v} \; \epsilon(e,e',e^{\prime\prime}) \;
\det_{e,e',e^{\prime\prime}}(\sigma)\Big|} \ee where $(\frac{a
l}{b})^3 \sqrt{\big|\det(E)(v)\big|}\big|[\det(\partial
X(s)/\partial s)]_{X(s)=v}\big|$ approximates the classical volume
$V_v(E)$, as determined by $E$ of an embedded cube with parameter
volume $(al/b)^3$. It is  straightforward to see that the correct
semiclassical behaviour is attained for $G_{\gamma, V}=1$.

The fact that the correct semiclassical behaviour of the volume
operator is attained \emph{only} for cases in which the graph is
aligned to the plaquettation (6-valent case), or each edge lies in
a given plaquette (the 4-valent case), seems rather puzzling
since, both cases, have measure zero in $SO(3)$. This makes one
question the\textit{ prima facie} validity of utilising the area
coherent states, to compute the expectation value of the volume
operator. However, it is interesting to note that case $4)$ of the
6-valent graph is precisely what one gets when constructing such a
graph as the dual of a cubical cell complex.
We will now proceed to compute the  higher,
$\frac{l}{\delta}$-order dependence of the expectation value of
the volume operator for 4-, 6- and 8-valent graphs, respectively.

\section {The Higher, $\frac{l}{\delta}$-Order Dependence of the Expectation
Value of the Volume Operator} \label{s7}

In this Section we analyse the higher order contributions to the expectation value of the volume
operator for the 4-, 6-, and 8-valent graphs. 

The following Section is subdivided into four parts. In the first
we explain the general method to be applied in the subsequent
Sections. In the second, third and fourth parts we apply this
method to our 4-, 6- and 8-valent graphs, respectively. Each of
these subsections is itself subdivided into three parts: in the
first, the stack family and the cubulation that defines the
platonic-body cell complex dual to the graph are aligned (see
\cite{30b}); in the second we study the effect of a rotation; and
in the third we study the effect of a translation.

\subsection{Initial Preparations}
\label{s7.1}

As a first step towards computing the expectation value of the
volume operator, we must calculate the values of the quantities
$t^{\gamma}_e$ and $t^{\gamma}_{ee^{'}}$ defined in \cite{30b},
which indicate the number of surfaces, $s^I_{\alpha t}$, that the
edge $e$ intersects, and the number of surfaces, $s^I_{\alpha t}$,
which are intersected by both edges $e$ and $e^{'}$. Both these
quantities depend, explicitly, on how the graph is embedded in the
stack family $S^I$ (see \cite{30b}). In fact, the conditions for
two or more edges to intersect a common surface are the following:
\begin{enumerate}
\item[1)] Two edges $e_i$ and $e_j$ intersect the same plaquette,
$s^z_{\alpha
t}$, iff $0< \phi_i,\phi_j<\frac{\pi}{2}$ or $\frac{\pi}{2}< \phi_i,\phi_j<\pi$.\\

\item[2)] Two edges $e_i$ and $e_j$ intersect the same plaquette,
$s^x_{\alpha t}$, iff
$-\frac{\pi}{2}<\theta_i,\theta_j<\frac{\pi}{2}$ or
$\frac{\pi}{2}<\theta_i,\theta_j<\frac{3\pi}{2}$.\\

\item[3)] Two edges $e_i$ and $e_j$ intersect the same plaquette,
$s^y_{\alpha
t}$, iff $0< \theta_i,\theta_j<\pi$ or $\pi< \theta_i,\theta_j<2\pi$.\\

\item[4)] If we have equalities in any of the above conditions, such that the angles of each of the two edges correspond
to a different limiting case, we obtain
$t^I_{e_ie_j}=\{t_i\in\Rl|S^{I}_{t}\cap
e_k\neq\emptyset, k=i,j\}=\emptyset$.\\
\end{enumerate}
We can also have situations in which two or more edges intersect a
common plaquette in more than one stack. The
conditions for such occurrences are the following:\\[-5pt]
\begin{enumerate}
\item[a)] Given condition (1), two edges $e_i$ and $e_j$ will
intersect more than one $z$-stack iff $|\theta_i|+|\theta_j|<
\pi/2$ and such that
 $n\frac{\pi}{4}<\theta_i,\theta_j<(n+1)\frac{\pi}{4}$,
 where $n=\{1,2,3,4,5,6,7,8\}$.

\item[b)] Given condition (2), two edges  $e_i$ and $e_j$ will
intersect more than one $x$-stack iff condition (1) above is
satisfied and
$n\frac{\pi}{4}<\theta_i,\theta_j<(n+1)\frac{\pi}{4}$, where
$n=\{1,2,3,4,5,6,7,8\}$.

\item[c)] Given condition (3), two edges  $e_i$ and $e_j$ will
intersect more than one $y$-stack iff condition (1) above is
satisfied and
$n\frac{\pi}{4}<\theta_i,\theta_j<(n+1)\frac{\pi}{4}$, where
$n=\{1,2,3,4,5,6,7,8\}$.
\end{enumerate}

The conditions above imply that rotating the graph will change
the values of the $t^I_{e_ie_j}$ and also the number of the
$t^I_{e_ie_j}$ that are non-zero.

We will now briefly explain, with the aid of an easy example, the
strategy we will use to compute the terms
$\frac{t^I_{e_ie_j}}{\sqrt{t_{e_i}^It_{e_j}^I}}$, that are used in
the calculations of the expectation value of the volume operator
for the 4-,6- and 8-valent graphs. To this end, consider an edge,
$e\in\gamma$, of a generic graph, whose length is given by
$\delta$. This edge will intersect the stacks of plaquettes in
each direction a certain number of times. In particular, given a
length $l$ of a plaquette, each edge will have $n$ intersections
with the stacks of any given direction, where $n$ is identified
with the Gauss bracket $[\frac{c_i}{l}]$ and $c_i$ is proportional
to $\delta$, where $c_i$ for $i\in\{x,y,z\}$ are the coordinates
of the edge.

\begin{figure}[htbp]
\begin{center}
\psfrag{V}{$V_1$}
\includegraphics[scale=0.9]{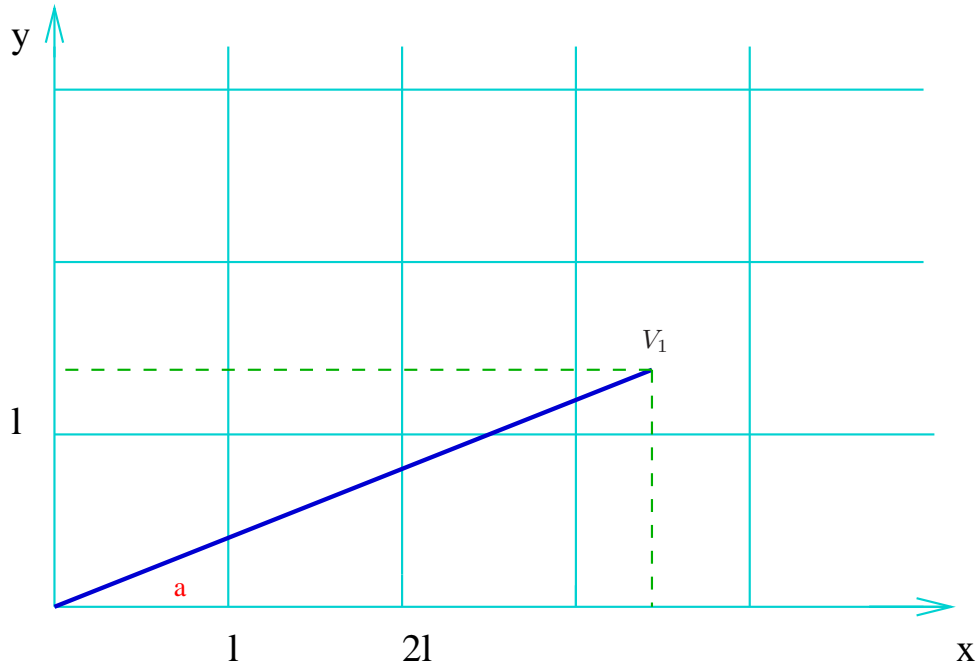}
\caption{Example in 2-dimensions}\label{fig:7}
\end{center}
\end{figure}

For example, in the two-dimensional case of Figure \ref{fig:7},
the values of $n$, in any given direction for vertex $V_1$ (or
equivalently the edge $e_{0,1}$ of length $\delta$), whose
coordinates are
$V_1=(\delta cos(a),\delta sin(a))$, would be $n_x=[\frac{\delta
cos(a)}{l}]$ and $n_y =[\frac{\delta sin(a)}{l}]$.

The values $n_i$, $i\in\{x,y,z\}$, depend on both the angle $a$
and the ratio $\frac{l}{\delta}$. Concomitantly, the expectation
value of the volume operator will also depend on such parameters.

The rotational dependence will be dealt with later. In the present
Section we will focus on the $\frac{l}{\delta}$ dependence. We
need to consider three different sub-cases:
\begin{enumerate}
\item $\frac{\delta}{l}>1$
\item $\frac{\delta}{l}=1$
\item $\frac{\delta}{l}<1$
\end{enumerate}
and determine which of the them leads to consistent solutions.

However, to obtain an expansion of $\sqrt{A}^{-1}$ we need to
perform a Taylor series. The condition for applying such an
expansion is that $\Vert A-1\Vert<1$. From the expression for the (square) matrix $A$
(see (\ref{4.23})) it is clear that the condition above is
satisfied iff
$m(m-1)\frac{t_{ee^{\prime}}}{\sqrt{t_et{_e^{\prime}}}}<1$ where
$m$ is the dimension of the matrix. As we will show
$\frac{t_{ee^{\prime}}}{\sqrt{t_et{_e^{\prime}}}}=C\times
\frac{l^{\prime}}{\delta_e}$ where C=constant and $l^{\prime}<l$,
thus the condition
$m(m-1)\frac{t_{ee^{\prime}}}{\sqrt{t_et{_e^{\prime}}}}<1$ becomes
$\l<<\delta_e$, i.e. we need to choose the parquette to be much
finer than the edge length (see Section \ref{s4.1}). If this
requirement is satisfied, then we can perform a Taylor expansion
of $\sqrt{A}$ obtaining
$\sqrt{A}=1+\frac{1}{2}(A-1)+\frac{1}{8}(A-1)^2+\mathcal{O}(A-1)^3$.
Actually, we are only interested in first-order terms, and so we
shall only consider the approximation $\sqrt{A}\simeq 1+\frac{1}{2}(A-1)$ whose inverse,
in first-order, is simply $(\sqrt{A})^{-1}\simeq 1-\frac{1}{2}(A-1)$.
Since the parquette length must be much finer than the edge length,
in the following we will consider only case (1) and
analyse whether it gives the correct semiclassical limit.

The first step in the calculation is to determine the range of
allowed positions for each vertex, $V_i$, of the graph with
respect to the plaquette. Since the graphs we consider are
regular, determining the position of one vertex suffices to derive
the positions of the remaining vertices in the graph.

\begin{figure}[htbp]
\begin{center}
\psfrag{a}{$e_i$}\psfrag{b}{$e_j$}
\includegraphics[scale=0.7]{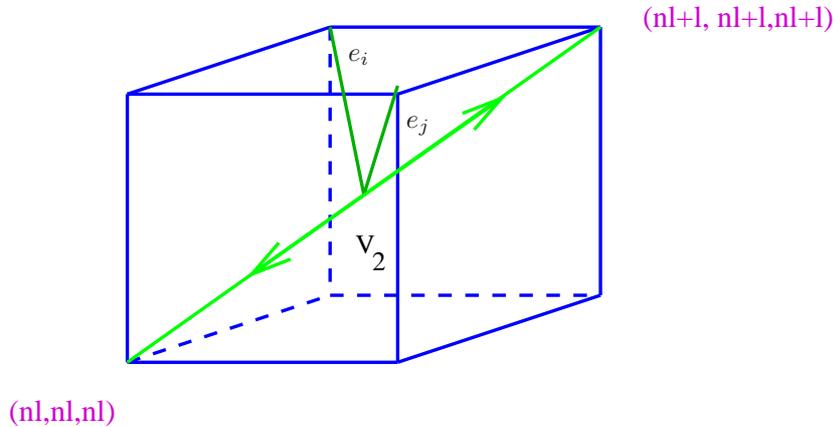}
\caption{Allowed positions of a vertex \label{fig:allowedpos}}
\end{center}
\end{figure}

As an explanatory example let us consider a regular 4-valent graph
$\gamma$, whose vertex $V_0$ coincides with the point $(0,0,0)$ of
the plaquettation and whose vertex $V_2$ (equivalently the edge
$e_{0,2}$) has coordinates
$(\frac{-\delta}{\sqrt{3}},\frac{\delta}{\sqrt{3}},\frac{\delta}{\sqrt{3}})$.
It follows that the range of allowed positions of $V_2$ is from
(nl, nl, nl) to (nl+l, nl+l, nl+l) where
$n_x=n_y=n_z=n=[\frac{\delta}{\sqrt{3}l}]$, as depicted in Figure
\ref{fig:allowedpos}.
\\It is straightforward to understand that different positions of $V_2$ will determine different values of $t_{e_ie_j}$ for any two edges $e_i$ and $e_j$ incident at $V_2$. A detailed analysis shows that the terms $t_{e_ie_j}$ differ according to which of the following conditions is satisfied:
\begin{enumerate}
\item[I)]   $|x_{V_2}|>|nl+\frac{l}{2}|$
\item[II)]  $|x_{V_2}|<|nl+\frac{l}{2}|$
\item[III)] $|x_{V_2}|=|nl+\frac{l}{2}|$
\end{enumerate}
Similar conditions apply for all vertices in $\gamma$.\\
Since the position of $V_2$ will determine the positions of all
other
 vertices, it is possible to establish which positions of $V_2$ will lead to different values of the terms $\frac{t_{e_ie_j}}{\sqrt{t_{e_i}t_{e_j}}}$ for all edges of all vertices of the graph $\gamma$. Such positions of $V_2$, for a regular 4-valent graph are:
\begin{enumerate}
\item[a)]  $|nl|\leq|x_{V_2}|\leq|nl+\frac{l}{6}|$
\item[b)]  $|nl+\frac{l}{6}|\leq|x_{V_2}|\leq|nl+\frac{l}{4}|$
\item[c)]  $|nl+\frac{l}{4}|\leq|x_{V_2}|\leq|nl+\frac{l}{2}|$
\item[d)]  $|nl+\frac{l}{2}|\leq|x_{V_2}|\leq|nl+\frac{l}{3}|$
\item[e)]  $|nl+\frac{l}{3}|\leq|x_{V_2}|\leq|nl+\frac{3l}{4}|$
\item[f)]  $|nl+\frac{3l}{4}|\leq|x_{V_2}|\leq|nl+2l|$
\end{enumerate}
For each such condition it is possible to derive the respective
conditions for both the $y$- and the $z$-coordinates in the
three-dimensional case. It turns out that similar relations hold
for the 6- and 8-valent graphs as well.

To explicitly compute the terms $t_{e_ie_j}$ we must choose one of
the above conditions ($a\rightarrow f$), each of which will lead
to different values for each $t_{e_ie_j}$. However, the
computation procedures are the same. In the calculations of
Sections \ref{s7.2} we will choose case (a).

To describe the method for computing the values of $t_{e_ie_j}$,
we go  back to a very simple example in two dimensions. We will
then give the general outline of how this calculation can be
generalised to the 3-dimensional case.

\begin{figure}[htbp]
\begin{center}
\psfrag{t}{$t^y_{e_i}$}\psfrag{s}{$t^y_{e_j}$}
\includegraphics[scale=0.7]{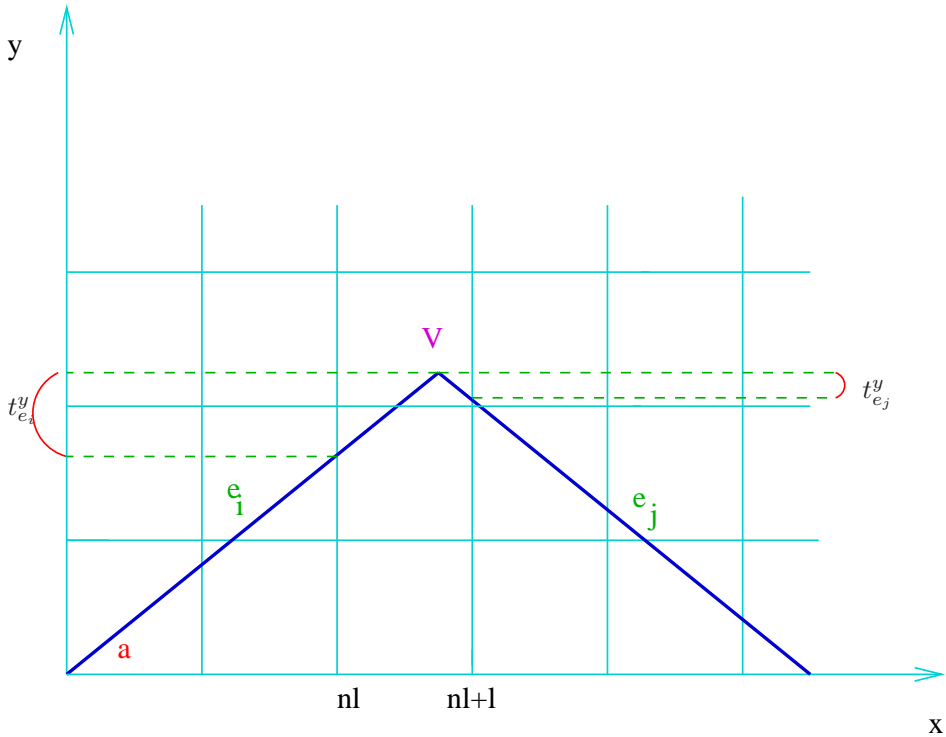}
\caption{Example in two dimensions \label{fig:2d}}
\end{center}
\end{figure}

Let us consider Figure \ref{fig:2d},
where we chose $x_{v_1}> nl+\frac{l}{2}$. For simplicity we assume that the vertex is symmetric with respect to the axis,
i.e. the angles, $\phi$, made by the two edges with respect to the
$y$-axis, are the same.

We now want to compute the values of
$t_{e_ie_j}=t^y_{e_ie_j}+t^x_{e_ie_j}+t^y_{e_ie_j}$ where for each
$t^k_{e_ie_j}$, $k=\{x,y,z\}$ we have $t^k_{e_ie_j}=t^k_{e_i}\cap
t^k_{e_j}$.

As a first step we compute for each edge, $e_i$, the value of
$t^y_{e_i}$ in the $y$-direction, obtaining \be
t_{e_j}^y=(nl+l-x_{v_1})\cot a\hspace{.5in}\text{ and
}\hspace{.5in} t_{e_i}^y=(x_{v_1}-nl)\cot a\ee Since the two edges
commonly intersect only one $y$ stack, in order to define the
value of $t^y_{e_ie_j}=t^y_{e_i}\cap t^y_{e_j}$ we need to
establish which of the two terms $t^y_{e_i}$ or $t^y_{e_j}$ is the
smallest. Thus, for example, \be
x_{v_1}-nl>nl+l-x_{v_1}\hspace{.5in}\text{ iff }\hspace{.5in}
x>nl+\frac{l}{2}\ee Since we have chosen $x_{v_1}> nl+\frac{l}{2}$
it follows that $t_{e_i}^y<t_{e_j}^y$ which implies that
$t^y_{e_ie_j}=t_{e_j}^y=(nl+l-x_{v_1})\cot a$. 
As it can be seen from Figure \ref{fig:2d}, there are no
intersections in the $x$ stacks, therefore we obtain \be
t_{e_ie_j}=t^y_{e_ie_j}+t^x_{e_ie_j}=t^y_{e_ie_j}\ee

 We now want to determine the values for $\frac{t_{e_{i}e_{j}}} {\sqrt{t_{e_{i}}t_{e_{j}}}}$ where, in this situation, $t_{e_{i}}=\tau_{e_{i}}^x+\tau_{e_{i}}^y=\delta\sin a+\delta\cos a=t_{e_{j}}$; therefore, $\frac{t_{e_{i}e_{j}}} {\sqrt{t_{e_{i}}t_{e_{j}}}}=\frac{(nl+l-x_{v_1})\cot a}{\sqrt{(\delta\sin a+\delta\cos a)^2}}$.

\begin{figure}[htbp]
\begin{center}
\psfrag{a}{$t^{y_1}_{e_i,e_j}$}\psfrag{b}{$t^{y_2}_{e_i,e_j}$}
\psfrag{f}{$e_j$}\psfrag{e}{$e_i$}
\includegraphics[scale=0.7]{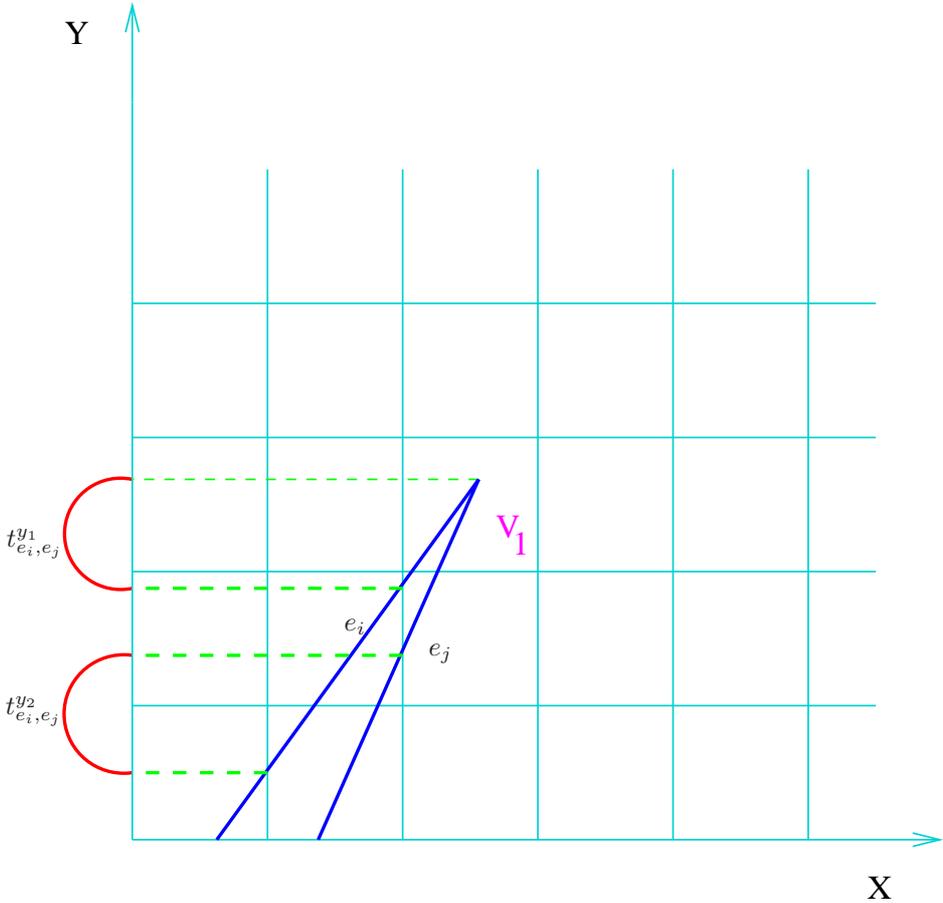}
\caption{The second example in two dimensions \label{pic:2examp}}
\end{center}
\end{figure}

This calculation is very simple since the intersection of the two
edges occurs only in one $y$ stack. But it could well be the case
that the angle between two edges is such that they intersect more
than one stack in a given direction. For example, consider Figure
\ref{pic:2examp}, always in two dimensions.
In this case we would have
$t^y_{e_ie_j}=t^{y_1}_{e_ie_j}+t^{y_2}_{e_ie_j}$.

Since in analysing the expectation value for the volume operator
we will be considering graphs formed by regular 4-, 6- and
8-valent lattice, it turns out that the angles $\theta_i$---the
angle formed by the projection on the edge on the $x$--$y$-plane
and the $x$-axis---and the angle, $\phi_i$, with respect to the
$z$-axis for any edge, are such that two or more edges can only
commonly intersect at most one plaquette in a given direction.

\begin{figure}[htbp]
\begin{center}
\psfrag{t}{$t^z_{e_i,e_j}$}\psfrag{a}{$e_i$}\psfrag{b}{$e_j$}
\includegraphics[scale=0.7]{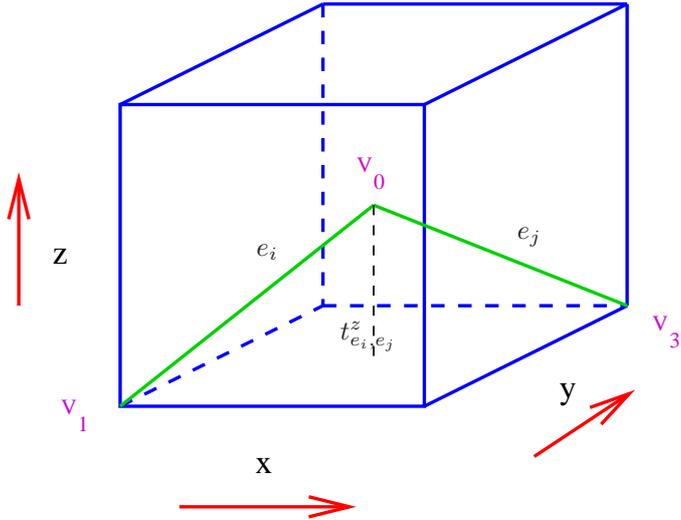}
\caption{Example in three dimensions \label{fig:3d}}
\end{center}
\end{figure}

When generalising the procedure described above for calculating
the values of $\frac{t_{e_{i}e_{j}}} {\sqrt{t_{e_{i}}t_{e_{j}}}}$
to the 3-dimensional case, some extra care is needed. In fact,
consider Figure \ref{fig:3d}.
It is clear that the values for $t^z_{e_i}$ can be computed with
respect to both the $x$- and the $y$-coordinates as follows:
\begin{equation}\label{equ:te}
^xt^z_{e_i}=a\times\cot\phi\frac{1}{\cos\theta}=a\times
Z^x_{e_i}\hspace{.5in}
^yt^z_{e_i}=c\times\cot\phi\frac{1}{\sin\theta}=c\times Z^y_{e_i}
\end{equation}
where
\begin{equation*}
a=\begin{cases}
x_{V_j}-n_{x_{V_j}}l & \text{iff the edge points in the negative x direction} \\
n_{x_{V_j}}l+l-x_{V_j} & \text{iff the edge points in the positive
x direction}
\end{cases}
\end{equation*}
and
\begin{equation*}
c=\begin{cases}
y_{V_j}-n_{y_{V_j}}l & \text{iff the edge points in the negative y direction} \\
n_{y_{V_j}}l+l-y_{V_j} & \text{iff the edge points in the positive
y direction}
\end{cases}
\end{equation*}

The term $^xt^z_{e_i}$ in these equations represents the value of
$t^z_{e_i}$ as computed with respect to the $x$-coordinate, while
$^yt^z_{e_i}$ is the value of $t^z_{e_i}$ as computed with respect
to the $y$-coordinate. The non-uniqueness of the computation of
the values $t^z_{e_i}$ implies that there is an extra difficulty
in the three-dimensional case. We will illustrate this
with the aid of an example.
\begin{figure}[htbp]
\begin{center}
\psfrag{a}{$^yt^{z}_{e_i}$}\psfrag{b}{$^xt^{z}_{e_i}$}
\psfrag{e}{$e_i$}\psfrag{c}{$V_j$}
\includegraphics[scale=0.8]{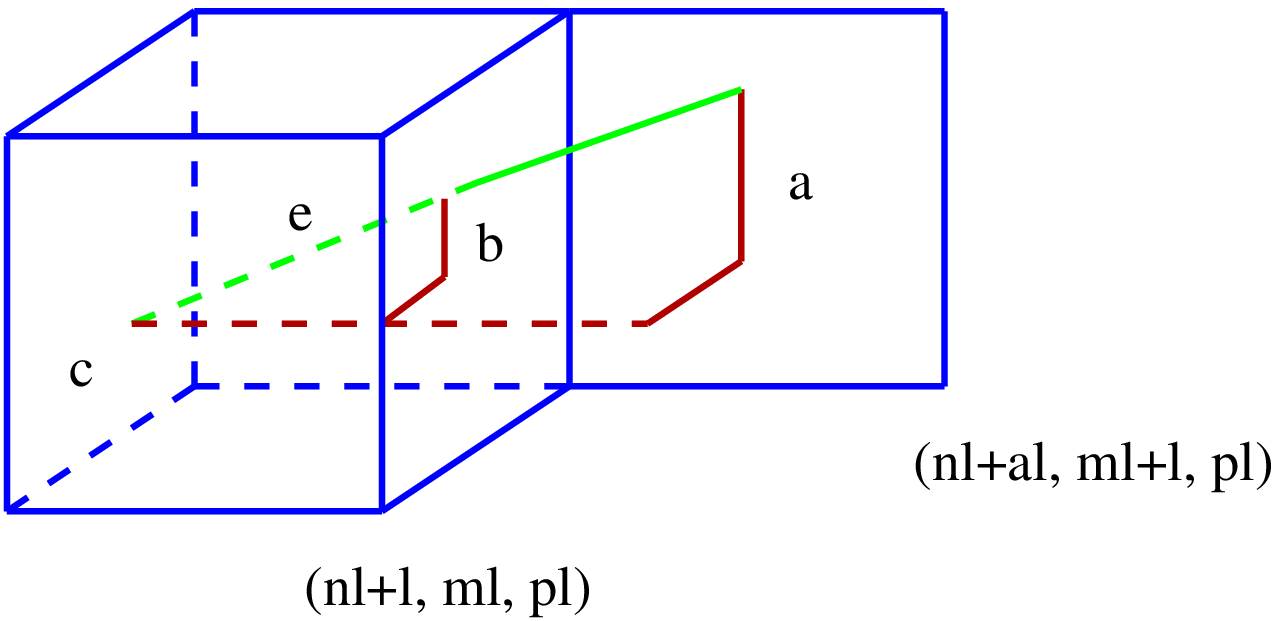}
\caption{Computation for the terms $t^z_{e_i}$ in 3-dimensions
\label{fig:difficulty}}
\end{center}
\end{figure}

Consider the edge $e_i$ in figure \ref{fig:difficulty}. The value
of $t^z_{e_i}$ can be computed with respect to both the $x$ and
the $y$ coordinate, thus obtaining $^xt^z_{e_i}$ or $^yt^z_{e_i}$,
respectively. However, it is clear from the diagram that the
intersection of the edge $e_i$, with the stack of plaquettes in the
$z$ direction containing the vertex $V_j$, is given by
$^xt^z_{e_i}$. On the other hand $^yt^z_{e_i}$
defines the intersection of the edge $e_i$ with the stacks of plaquettes in the $z$ direction containing the vertex \emph{plus} the stack in the $z$ direction delimited, in the $x$ direction, by the values $nl+l$ and $nl+al$.

This example shows that, given the values
$^xt^z_{e_i}$ and $^yt^z_{e_i}$, the intersection of the edge
$e_i$ with the stacks of plaquettes in the $z$ direction, which
contain the vertex $V_j$, is given by the smallest term, i.e.,
$t^z_{e_i}=^xt^z_{e_i}\cap ^yt^z_{e_i}$. It follows that, given
two edges $e_i$ and $e_j$, in order to find
$t^z_{e_i,e_j}:=t^z_{e_i}\cap t^z_{e_j}$ we first need to
establish whether $t^z_{e_i}=^xt^z_{e_i}$ or
$t^z_{e_i}=^yt^z_{e_i}$ and, similarly, for the edge $e_j$.
Once the value of the terms $t^z_{e_{i}}$ and $t^z_{e_{j}}$ is
determined, we can proceed as for the two-dimensional case and
identify $t^z_{e_{i}e_{j}}$ with the smallest $t^z$, i.e.
$t^z_{e_i,e_j}:=t^z_{e_i}\cap t^z_{e_j}$.

For intersections in the $x$ and $y$ stacks the procedure for
computing the values of $t_{e_{i}e_{j}}$ is essentially the same.
However, the formulae for the values of the individual terms,
$^jt^k_{e_i}$, are different. Specifically, for the $x$-direction
we have:
\begin{equation}
\label{equ:te2} ^zt^x_{e_i}=d\times\tan\phi\cos\theta=d\times
F^x_{e_i}\hspace{.5in}^yt^x_{e_i}=c\times\cot\theta=c\times
T^x_{e_i}
\end{equation}
 where
\begin{equation*}
d=\begin{cases}
z_{V_j}-n_{z_{V_j}}l & \text{iff the edge points upwards} \\
n_{z_{V_j}}l+l-z_{V_j} & \text{iff the edge points downwards}
\end{cases}
\end{equation*}
and $c$ is defined as above. For the $y$-direction we have
\begin{equation}
\label{equ:te3} ^zt^y_{e_i}=d\times\tan\phi\sin\theta=d\times
F^y_{e_i}\hspace{.5in}^xt^y_{e_i}=a\times\tan\theta=a\times
T^y_{e_i}
\end{equation}
 where $a$ and $d$ are defined as above.

When computing the values of $\frac{t_{e_{i}e_{j}}}
{\sqrt{t_{e_{i}}t_{e_{j}}}}$ in three dimensions, as for the
two-dimensional case, we need to compute the values for
$t_{e_{i}}$, which in this case are simply
$t_{e_{i}}=t^x_{e_{i}}+t^y_{e_{i}}+t^z_{e_{i}}=
\delta_{e_i}\cos(90-\phi_{e_i})\cos\theta_{e_i}+\delta_{e_i}
\cos(90-\phi_{e_i})\sin\theta_{e_i}+\delta_{e_i}\cos\phi_i$. The
explicit values of the terms $\frac{t_{e_{i}e_{j}}}
{\sqrt{t_{e_{i}}t_{e_{j}}}}$ obtained for the 4-, 6-, and 8-valent
graph which satisfies condition (a) above, namely
$|nl|\leq|x_{V_2}|\leq|nl+\frac{l}{6}|$, for the 4-valent graph
and an equivalent condition for the 6- and 8-valent graphs, are
given in the Appendix.

Since for all 4-, 6- and 8-valent graphs we are dealing with
symmetric lattices, after a certain number of vertices the values
for the terms $\frac{t_{e_{i}e_{j}}} {\sqrt{t_{e_{i}}t_{e_{j}}}}$
will  repeat, i.e., there will be a periodicity in the values of
the terms $\frac{t_{e_{i}e_{j}}} {\sqrt{t_{e_{i}}t_{e_{j}}}}$.
Therefore, in computing these values we need only consider those
vertices which comprise the periodicity cell, i.e., those vertices
for which the values of the term $\frac{t_{e_{i}e_{j}}}
{\sqrt{t_{e_{i}}t_{e_{j}}}}$ cannot be obtained through symmetry
arguments. As we will see later, this periodicity is different for
graphs of different valency.

We now proceed to compute the expectation value of the volume
operator for the 4-, 6- and 8-valent cases, utilising the values
of the terms $\frac{t_{e_{i}e_{j}}} {\sqrt{t_{e_{i}}t_{e_{j}}}}$
given in the Appendix.

\subsection{Analysis of the Expectation Value of the Volume Operator
for a 4-Valent Graph} \label{s7.2}

In this Section we will compute the expectation value of the
volume operator as applied to a 4-valent graph. We will first take into
 consideration the non-rotated graph. In establishing rotational
 and translational dependence of the expectation value, we will perform
 both a rotation by arbitrary
 Euler angles and a translation and, then, recalculate the
 expectation value. We will see
 that the contributions that come from the terms $\frac{t_{e_{i}e_{j}}}
 {\sqrt{t_{e_{i}}t_{e_{j}}}}$, which comprise the off-diagonal elements
 of the matrix $\sqrt{A}^{-1}$, are not trivial, thereby producing a strong
 rotational and translational dependence in the expectation value of the
volume operator in higher order in $\frac{l}{\delta}$.

\subsubsection{Expectation value of the volume operator for a
4-valent graph} \label{s7.2.1} 
\begin{figure}[htbp]
\begin{center}
 \includegraphics[scale=0.7]{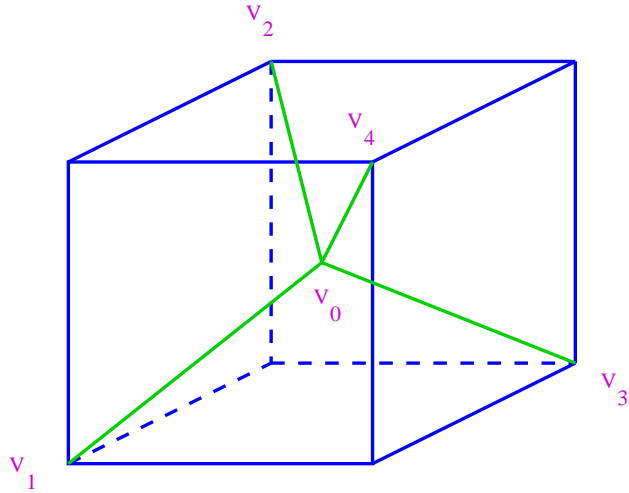}
\caption{General vertex of a 4-valent graph \label{fig:46a}}
\end{center}
\end{figure}

To calculate the expectation value
of the volume operator we will consider a 4-valent graph
constructed from the simplicial cell complex, as discussed in
\cite{30b}. We choose the vertex $V_0$ to be $V_0=(0,0,0)$, and
the angles $\phi_e=\cos^{-1}(\frac{1}{\sqrt{3}})$ and
$\theta_e=45^{\circ}$ for all $e\in\gamma$, such that we obtain the
configuration depicted in picture \ref{fig:46a}.
The periodicity cell for a 4-valent graph contains four vertices,
including $V_0$. The coordinates of the remaining three vertices
are $V_2 = \big(\frac{\delta}{\sqrt3},
\frac{\delta}{\sqrt3},\frac{\delta}{\sqrt3}\big)$; $V_8 =\big
(\frac{\delta}{\sqrt3}, 3\frac{\delta}{\sqrt3}
,\frac{\delta}{\sqrt3}\big)$; $V_{13} = \big(0,2
\frac{\delta}{\sqrt3} ,2\frac{\delta}{\sqrt3}\big)$.

It follows that the edges $e_{0,1}$, $e_{0,2}$, $e_{0,3}$ and
$e_{0,4}$ lie in the octants $G$, $B$, $E$ and $D$ respectively.
This implies that the geometric factor for the vertex $V_0$ will
be $G_{\gamma, V}=\sqrt{2}$. Because of the geometry of a regular
4-valent graph, it turns out that all the vertices comprising the
periodicity cell, $V_2$, $V_8$ and $V_{13}$ will have $G_{\gamma,
V_i}=\sqrt{2}$.

The following table gives the values obtained for the terms
$\frac{t_{e_{i}e_{j}}} {\sqrt{t_{e_{i}}t_{e_{j}}}}$ for the
4-valent graph that satisfies condition (a), as defined in the
previous Section, namely $|nl|\leq|x_{V_2}|\leq|nl+\frac{l}{6}|$.

It should be noted that, because of the geometry of the 4-valent
graph, the terms
$T^{\{x,y\}}_{e_i}$,$F^{\{x,y\}}_{e_i}$,$Z^{\{x,y\}}_{e_i}$ in
equations \ref{equ:te}, \ref{equ:te2} and \ref{equ:te3} are all
equal to $1$ for each edge $e_i$:
\begin{center}
\begin{tabular}{l|l}
$V_0$& $\text{six terms } \frac{t_{e_{i}e_{j}}}
{\sqrt{t_{e_{i}}t_{e_{j}}}}=0$\\ \hline $V_2$& $\text{six terms
}\frac{t_{e_{i}e_{j}}}
{\sqrt{t_{e_{i}}t_{e_{j}}}}=\big(\frac{\delta}{\sqrt{3}}-nl\big)\frac{1}{\delta\sqrt{3}}$\\
\hline $V_8$& $\text{six terms }\frac{t_{e_{i}e_{j}}}
{\sqrt{t_{e_{i}}t_{e_{j}}}}=\big(\frac{\delta}{\sqrt{3}}-nl\big)\frac{1}{\delta\sqrt{3}}$\\
\hline $V_{13}$& $\text{two terms }\frac{t_{e_{i}e_{j}}}
{\sqrt{t_{e_{i}}t_{e_{j}}}}=\big(\frac{2\delta}{\sqrt{3}}-2nl\big)\frac{1}{\delta\sqrt{3}}$\\
\hline
\end{tabular}
\end{center}
Here, $\delta$ is the length of the edge $e$. In order to apply
equation (\ref{4.17}), we first need to determine the values of
the term $\det_{e,e^{'}e^{''}}(\sqrt{A}^{-1})$ for each triplet of
linearly-independent edges $e,e^{'},e^{''}$. Using the fact that,
in first-order approximation,
$(\sqrt{A})^{-1}=1-\frac{1}{2}(A-1)$, the explicit expression for
$(\sqrt{A})^{-1}$ for the 4-valent graph under consideration is
\[ \left( \begin{array}{c|ccccccccccccccc}
&e_{0,1}&e_{0,2}&e_{0,11}&e_{0,12}&e_{2,7}&e_{2,5}&e_{2,6}&e_{13,9}&e_{13,8}&e_{13,10}&e_{13,11}&e_{8,13}&e_{8,14}&e_{8,15}&e_{8,5}\\\hline
e_{0,1}&1 & 0 & 0 & 0 & 0 & 0 & 0 & 0 & 0 & 0 & 0 & 0 & 0 & 0 & 0  \\
e_{0,2}&0 & 1 & 0 & 0 & \frac{1}{2}\alpha & -\frac{1}{2}\alpha & -\frac{1}{2}\alpha & 0 & 0 & 0 & 0 & 0 & 0 & 0 & 0 \\
e_{0,11}&0 & 0 & 1 & 0 & 0 & 0 & 0 & 0 & 0 & 0 & 0 & 0 & 0 & 0 & 0  \\
e_{0,12}&0 & 0 & 0 & 1 & 0 & 0 & 0 & 0 & 0 & 0 & 0 & 0 & 0 & 0 & 0  \\
e_{2,7}&0 & -\frac{1}{2}\alpha & 0 & 0 & 1 & -\frac{1}{2}\alpha & -\frac{1}{2}\alpha & 0 & 0 & 0 & 0 & 0 & 0 & 0 & 0  \\
e_{2,5}&0 &-\frac{1}{2} \alpha & 0 & 0 & -\frac{1}{2}\alpha & 1 & -\frac{1}{2}\alpha & 0 & 0 & 0 & 0 & 0 & 0 & 0 & 0  \\
e_{2,6}&0 & -\frac{1}{2}\alpha & 0 & 0 & -\frac{1}{2}\alpha & -\frac{1}{2}\alpha & 1 & 0 & 0 & 0 & 0 & 0 & 0 & 0 & 0 \\
e_{13,9}&0 & 0 & 0 & 0 & 0 & 0 & 0 & 1 & -\frac{1}{2}\alpha & -\frac{1}{2}\alpha & -\frac{1}{2}\alpha & 0 & 0 & 0 & 0 \\
e_{13,8}&0 & 0 & 0 & 0 & 0 & 0 & 0 & -\frac{1}{2}\alpha & 1 & -\frac{1}{2}\alpha & -\frac{1}{2}\alpha & 0 & 0 & 0 & 0 \\
e_{13,10}&0 & 0 & 0 & 0 & 0 & 0 & 0 & -\frac{1}{2}\alpha & -\frac{1}{2}\alpha & 1 & -\frac{1}{2}\alpha & 0 & 0 & 0 & 0  \\
e_{13,11}&0 & 0 & 0 & 0 & 0 & 0 & 0 & -\frac{1}{2}\alpha & -\frac{1}{2}\alpha & -\frac{1}{2}\alpha & 1 & 0& 0 & 0 & 0 \\
e_{8,13}&0 & 0 & 0 & 0 & 0 & 0 & 0 & 0 & 0 & 0 & 0 & 1 & 0 &  -\alpha & 0  \\
e_{8,14}&0 & 0 & 0 & 0 & 0 & 0 & 0 & 0 & 0 & 0 & 0 & 0 & 1 & 0 &  -\alpha  \\
e_{8,15}&0 & 0 & 0 & 0 & 0 & 0 & 0 & 0 & 0 & 0 & 0 &  -\alpha & 0 & 1 & 0 \\
e_{8,5}&0 & 0 & 0 & 0 & 0 & 0 & 0 & 0 & 0 & 0 & 0 & 0 & -\alpha & 0 & 1  \\
 \end{array} \right)\]
where $\alpha=\frac{t_{e_{i}e_{j}}}
{\sqrt{t_{e_{i}}t_{e_{j}}}}=\big(\frac{\delta}{\sqrt{3}}-nl\big)\frac{1}{\delta\sqrt{3}}$
and the terms $t_{e_{i}e_{j}}$, $t_{e_{i}}$ and $t_{e_{j}}$ are
computed using the techniques defined in the previous Section.

Now that we have an expression for the inverse of the matrix
$\sqrt{A}$ we can compute the expectation value of the volume
operator for each of the four vertices in the periodicity cell
and, then, sum their contributions.

We start with the vertex $V_0$. First consider the sub-matrix of
the matrix $(\sqrt{A})^{-1}$ formed by all the edges incident at
$V_0$. This is
\[\left( \begin{array}{c|ccccc}
&e_{0,1}&e_{0,2}&e_{0,11}&e_{0,12}\\\hline
e_{0,1}&1 & 0 & 0 & 0  \\
e_{0,2}&0 & 1 & 0 & 0  \\
e_{0,11}&0 & 0 & 1 & 0  \\
e_{0,12}&0 & 0 & 0 & 1 \\
 \end{array} \right)\]
Because of the geometry of the 4-valent graph, at each vertex
there are four triplets of linearly-independent edges.  Keeping
this in mind and computing the determinant of the matrices formed
by each such set of triplets, we obtain the following expression
for the expectation value of the volume operator at $V_0$:
\begin{equation}
\delta^3\sqrt{\frac{1}{8}}\sqrt{\det\big(E^a_j(u)\big)}\;\Big|16\big|\det\Big(\frac{\partial
X_S^a}{\partial(s,u^1,u^2)}\Big)\big|^2\Big|^{\frac{1}{2}}
\end{equation}
By a similar procedure for vertices $V_2$ and $V_3$ we obtain
\begin{equation}
\delta^3\sqrt{\frac{1}{8}}\sqrt{\det\big(E^a_j(u)\big)}\;\Big|16\Big(1-\frac{3\alpha^2}{4}-
\frac{\alpha^3}{4}\Big)\,\big|\det\Big(\frac{\partial
X_S^a}{\partial(s,u^1,u^2)}\Big)\big|^2\Big|^{\frac{1}{2}}
\end{equation}

In both cases, the sub-matrix of $\sqrt{A}^{-1}$ we consider is
\[\left( \begin{array}{cccc}
1 &  -\frac{1}{2}\alpha &  -\frac{1}{2}\alpha &  -\frac{1}{2}\alpha  \\
 -\frac{1}{2}\alpha & 1 &  -\frac{1}{2}\alpha &  -\frac{1}{2}\alpha  \\
 -\frac{1}{2}\alpha &  -\frac{1}{2}\alpha & 1 &  -\frac{1}{2}\alpha  \\
 -\frac{1}{2}\alpha &  -\frac{1}{2}\alpha &  -\frac{1}{2}\alpha & 1 \\
 \end{array} \right)\]
and then we compute the determinant of all the sub-matrices formed
by linearly-independent triplets of edges.

For the vertex $V_{4}$ we obtain
\begin{equation}
\delta^3\sqrt{\frac{1}{8}}\sqrt{\det(E^a_j(u)\big)}\;
\Big|16(1-\alpha^2)\Big|\det\Big(\frac{\partial
X_S^a}{\partial(s,u^1,u^2)}\Big)\Big|^2\Big|^{\frac{1}{2}}
\end{equation}
where we have used the sub-matrix
\[\left( \begin{array}{cccc}
1 & 0 & -\alpha & 0  \\
0 & 1 & 0 & -\alpha  \\
-\alpha & 0 & 1 & 0  \\
0 & -\alpha & 0 & 1 \\
 \end{array} \right)\]
Summing up these contributions we obtain
\begin{align}\label{ali:v4}
V_R&=\delta^3\sqrt{\frac{1}{8}}\;4\;\Big|\det\Big(\frac{\partial
X_S^a}{\partial(s,u^1,u^2)}\Big)\Big|
\;\;\Big|\Big(1\sqrt{|\det\big(E^a_j(u)\big)|}+\sqrt{|\det\big(E^a_j(u)\big)|}\Big(\big|1-\frac{3\alpha^2}{4}-
\frac{\alpha^3}{4}\,\big|\Big)^{\frac{1}{2}}\nonumber\\
&+\sqrt{|\det\big(E^a_j(u)\big)|}\Big(\big||1-\frac{3\alpha^2}{4}-
\frac{\alpha^3}{4}\,\big|\Big)^{\frac{1}{2}}+\sqrt{|\det\big(E^a_j(u)\big)|}\Big(\big|1-\alpha^2\,\big|\Big)^{\frac{1}{2}}
\end{align}
To first-order approximation we obtain
\begin{align}
V_R&=\delta^3
2\sqrt{2}\sqrt{|\det\big(E^a_j(u)\big)|}\;\Big|\det\Big(\frac{\partial
X_S^a}{\partial(s,u^1,u^2)}\Big)\;\;\Big|1 -\frac{1}{2}\Big(
\frac{3\alpha^2}{16} -\frac{\alpha^2}{4}\Big)+\mathcal{O}(3)
\end{align}

It should be noted that, although the term $\det(E^a_j(u))$ is
vertex dependent, we can safely assume that, to first-order in
$l/\delta$, the values will be the same for each vertex within
each periodicity cell that involves only an order of four
vertices. Thus this term can be factored out from the equation.
This first-order approximation will be used throughout. As mentioned previously this is justified since we
choose $\frac{l}{\delta}<<1$. It follows that the terms
$\alpha^2\propto \frac{l-x}{\delta}$, which are much smaller than
\emph{one} (see Section \ref{s4.1}).

The term proportional to $\alpha^2$ in the equation above
represents the $l/\delta$-correction for the expectation value of
the volume operator for a given region $R$. As in \cite{30b}, for
a general 4-valent graph, even in the zeroth-order approximation,
the expectation value for the volume of a given region $R$ does
not coincide with the classical value for the volume of that
region. Notably, there is no linear correction in $l/\delta$!

\subsubsection{Expectation value of the volume operator for a rotated
4-valent graph} \label{s7.2.2} We will now analyse how the results of
the calculations above depend on how the graph is embedded in
$\Rl^3$. Here we will consider rotational invariance;
translational invariance is discussed in the following subsection.

To analyse the rotational dependence of the expectation value of
the volume operator, we will perform a Euler rotation of the
graph with respect to some arbitrary Euler angles $\beta$, $\psi$,
$\alpha$ and, then, repeat the calculation. The transformation
matrix is
\[R= \left( \begin{array}{ccc}\label{a1}
\cos\p \cos\s-\sin\p \cos\te \sin\psi  & \cos\psi \sin\p+\cos\te \cos\p \sin\psi  & \sin\psi \sin\te   \\
-\sin\psi \cos\p-\cos\te \sin\p \cos\psi & -\sin\psi \sin\p+ \cos\te \cos\p \sin\psi & \cos\psi \sin\te \\
\sin\te \sin\p & -\sin\te \cos\p & \cos\te\end{array} \right)\]
The coordinates of the rotated vertices are then given by
$V^{\prime}_i=R\cdot\vec{V_i}$
\[\left(\begin{array}{c}
x_{V^{\prime}_i}\\
y_{V^{\prime}_i}\\
z_{V^{\prime}_i}
\end{array} \right)=
\left( \begin{array}{c}
R_{11}x_{v_i} + R_{12}y_{v_i} + R_{13}z_{v_i}  \\
R_{21}x_{v_i} + R_{22}y_{v_i} + R_{23}z_{v_i} \\
R_{31}x_{v_i} + R_{32}y_{v_i} + R_{33}z_{v_i}\end{array} \right)\]

Applying this transformation matrix to the 4-valent graph we
obtain the following new coordinates for the vertices:
\begin{eqnarray}
V^{\prime}_2&=&\Big((-R_{11}+R_{12}+R_{13})\frac{\delta}{\sqrt{3}},
(-R_{21}+R_{22}+R_{23})\frac{\delta}{\sqrt{3}},
(-R_{31}+R_{32}+R_{33})\frac{\delta}{\sqrt{3}}\Big)\\
V^{\prime}_8&=&\Big((R_{11}+3R_{12}+R_{13})\frac{\delta}{\sqrt{3}},
(R_{21}+3R_{22}+R_{23})\frac{\delta}{\sqrt{3}},(R_{31}+3R_{32}+R_{33})
\frac{\delta}{\sqrt{3}}\Big)\\
V^{\prime}_{13}&=&\Big(2(R_{12}-R_{13})\frac{\delta}{\sqrt{3}},
2(R_{22}-R_{23})\frac{\delta}{\sqrt{3}},2(R_{32}-R_{33})\frac{\delta}{\sqrt{3}}\Big)
\end{eqnarray}
The new angles between the rotated edges and the $x$,$y$,$z$-axes
can now easily be computed using elementary trigonometry.

As an explanatory example let us consider the edge $e_{0,2}$. To
find the angles this edge has with respect to the axes we need
first to compute the coordinates of the vector $\vec{e_{0,2}}$
starting at vertex $V_0$ and ending at vertex $V_2$. In this case,
the coordinates of $\vec{e_{0,2}}$ coincide with the coordinates
of the vertex $V_2$:  \be\vec{e_{2,0}} = \Big((−R_{11} + R_{12}
+ R_{13})\frac{\delta}{\sqrt{3}}, (−R_{21} + R_{22} + R_{23}
)\frac{\delta}{\sqrt{3}} , (−R_{31} + R_{32} + R_{33} )
\frac{\delta}{\sqrt{3}}\Big)\ee If instead we considered the edge
$e_{25}$ we would get \be \vec{e_{25}}=e_{02}-e_{05}=
\Big(-(R_{11}+R_{12}+R_{13})\frac{\delta}{\sqrt{3}},-(R_{21}+R_{22}+
R_{23})\frac{\delta}{\sqrt{3}},-(R_{31}+R_{32}+R_{33})
\frac{\delta}{\sqrt{3}}\Big)\ee

Once we have the coordinates for $\vec{e_{2,0}}$, the angle
$\phi_{e_{2,0}}$ it forms with respect to the $z$-coordinate, is
\be \label{equ:p}
\phi_{e_{2,0}}=\tan^{-1}\Big(\frac{\sqrt{x^2+y^2}}{z}\Big)=
\tan^{-1}\Big(\frac{\sqrt{(-R_{11}+R_{12}+R_{13})^2+(-R_{21}+
R_{22}+R_{23})^2}}{(-R_{31}+R_{32}+R_{33})}\Big)\ee The angle,
$\theta_{e_{2,0}}$, between the projection of the vector on the
$x$--$y$ plane and the $x$-axis is given by \be\label{equ:t}
\theta_{e_{2,0}}=\tan^{-1}\Big(\frac{y}{x}\Big)=
\tan^{-1}\left(\frac{-R_{21}+R_{22}+R_{23}}{-R_{11}+R_{12}+R_{13}}\right)\ee

In the same way we can obtain the angles for all the edges in our
graph in terms of the  elements of the transformation matrix. Thus
the orientation of each of the edges of the graph will depend on
the matrix elements of the transformation matrix, i.e., on the
Euler angles that parametrise the rotation.

In order to determine the rotational dependence of the expectation
value of the volume operator, we have performed a case study in
which the expectation values were computed for all possible
orientations of the graphs, that have non-zero measure in $SO(3)$.
Such possible orientations were described in Section \ref{s4.4}.
To aid calculational simplicity, these sub-cases are  defined in
terms of possible ranges of values for the angles $\theta$ and
$\phi$ for each edge in the graph, rather than on possible values
for the Euler angles.

In order to keep our results as general as possible, we performed
our subdivisions so as to cover all possible situations. This is
less tedious than it might seem since we are dealing with regular
lattices and, therefore, once  the angles for the edges of one
vertex are fixed, we immediately know the orientation of the edges
of all other vertices.

Let us choose $V^{\prime}_0$ as our reference vertex, with respect
to which the possible orientations of the edges are defined. The
edges incident at $V^{\prime}_0$ are $e_{0,1}$ , $e_{0,2}$ ,
$e_{0,3}$ , $e_{0,4}$. In defining the orientation we  use the
convention that both $0<\phi<2\pi$ and $0<\theta<2\pi$ increase
anti-clockwise.

Once  the rotational matrix has been applied, whatever the values
of the Euler angles might be, we will end up in a situation in
which two edges $e_i$, $e_j$ point upwards, i.e.,
$-\frac{\pi}{2}<\phi_{e_i}, \phi_{e_j}<\frac{\pi}{2}$, and the
remaining two edges point downwards, i.e.
$\frac{\pi}{2}<\phi_{e_k}, \phi_{e_l}<\frac{3\pi}{2}$. This is a
consequence of the geometry of the 4-valent graph. We will call
two edges pointing in the same up, or down, direction an `up' or
`down' couple, respectively. Since we are considering only those
edge orientations with measure \emph{non-zero} in $SO(3)$, the angles
of the edges $e_i$, $e_j$ of each up/down couple will satisfy the
following conditions: $|\theta_{e_i}|=|90+\theta_{e_j}|$ and
$|\phi_{e_i}|=|\phi_{e_j}-54.75|$

Given a particular choice of up and down couple we have to
specify in which octant (see Figure \ref{figdivision})
 each edge lies. This is required since different octants
induce different values for the geometric factor $G_{\gamma,V}$.
The angles $\theta_{e_i}$ and $\phi_{e_i}$ required for an edge
$e_i$ to lie in each of the octants are listed in Table
\ref{tab:4g} where, again, we use the convention that
$0<\phi_{e_i}<2\pi$ and $0<\theta_{e_i}<2\pi$, with both angles increasing in
an anticlockwise direction.
\begin{table}[h] \footnotesize
\begin{center}
\begin{tabular}{l|l|l|l|l}
& A & B & C & D \\ \hline $\phi_{e_i}$&
$\frac{3\pi}{2}<\phi_{e_i}<2\pi$ & $0<\phi_{e_i}<\frac{\pi}{2}$
&$0<\phi_{e_i}<\frac{\pi}{2}$  &
$\frac{3\pi}{2}<\phi_{e_i}<2\pi$\\ \hline
$\theta_{e_i}$&$0<\theta_{e_i}<\frac{\pi}{2}$&
$\frac{\pi}{2}<\theta_{e_i}<\pi$
&$\pi<\theta_{e_i}<\frac{3\pi}{2}$ &
$\frac{3\pi}{2}<\theta_{e_i}<2\pi$\\ \hline
\end{tabular}
 \end{center}
 \begin{center}
\begin{tabular}{l|l|l|l|l}
&  E & F & G & H\\ \hline $\phi_{e_i}$&
$\pi<\phi_{e_i}<\frac{3\pi}{2}$ &$\frac{\pi}{2}<\phi_{e_i}<\pi$&
$\frac{\pi}{2}<\phi_{e_i}<\pi$ &$\pi<\phi_{e_i}<\frac{3\pi}{2}$\\
\hline $\theta_{e_i}$& $0<\theta_{e_i}<\frac{\pi}{2}$&
$\frac{\pi}{2}<\theta_{e_i}<\pi$
&$\pi<\theta_{e_i}<\frac{3\pi}{2}$ &
$\frac{3\pi}{2}<\theta_{e_i}<2\pi$\\ \hline
\end{tabular}
\caption{Angle-ranges for each octant} \label{tab:4g}
\end{center}
\end{table}
However, because of the geometry of a 4-valent graph, the allowed
angle-ranges have to be restricted to those listed in Table
\ref{tab:4}.
\begin{table}[h] \footnotesize
\begin{center}
\begin{tabular}{l|l|l|l|l}
& A & B & C & D \\ \hline $\phi_{e_i}$&
$\frac{3\pi}{2}<\phi_{e_i}<2\pi-\sin^{-1}(\frac{1}{3}$ &
$\sin^{-1}(\frac{1}{3}<\phi_{e_i}<\frac{\pi}{2}$
&$\sin^{-1}(\frac{1}{3}<\phi_{e_i}<\frac{\pi}{2}$  &
$\frac{3\pi}{2}<\phi_{e_i}<2\pi-\sin^{-1}(\frac{1}{3}$\\ \hline
$\theta_{e_i}$&$0<\theta_{e_i}<\frac{\pi}{2}$&
$\frac{\pi}{2}<\theta_{e_i}<\pi$
&$\pi<\theta_{e_i}<\frac{3\pi}{2}$ &
$\frac{3\pi}{2}<\theta_{e_i}<2\pi$\\ \hline
\end{tabular}
 \end{center}
 \begin{center}
\begin{tabular}{l|l|l|l|l}
&  E & F & G & H\\ \hline $\phi_{e_i}$&
$\pi+\sin^{-1}(\frac{1}{3})<\phi_{e_i}<\frac{3\pi}{2}$
&$\frac{\pi}{2}<\phi_{e_i}<\pi-\sin^{-1}(\frac{1}{3})$&
$\frac{\pi}{2}<\phi_{e_i}<\pi-\sin^{-1}(\frac{1}{3})$
&$\pi+\sin^{-1}(\frac{1}{3}<\phi_{e_i}<\frac{3\pi}{2}$\\ \hline
$\theta_{e_i}$& $0<\theta_{e_i}<\frac{\pi}{2}$&
$\frac{\pi}{2}<\theta_{e_i}<\pi$
&$\pi<\theta_{e_i}<\frac{3\pi}{2}$ &
$\frac{3\pi}{2}<\theta_{e_i}<2\pi$\\ \hline
\end{tabular}
\caption{4-valent graph angle-ranges for each octant \label{tab:4}
}
\end{center}
\end{table}

Our calculations show that for all possible sub-cases of angle
arrangements defined in Table \ref{tab:4}, the expectation value
for the volume operator is rotational invariant \emph{only}
at the zeroth-order\footnote{This is a consequence of the fact
that the geometric factors $G_{\gamma, V_i}$, for each of the
sub-cases in Table \ref{tab:4}, will be the same (see Section
\ref{s 2.4.1})}, while higher-order terms \emph{are}
rotationally dependent. Therefore, in what follows, we will not
compute the expectation value for the volume operator as computed
for each possible orientation of the graph. Instead, we will choose
a particular sub-case of Table \ref{tab:4} and compute the
expectation value for such a sub-case. Specifically, we will choose
the case in which the arrangement of edges, incident at the
vertex $V_0$ after a rotation, is given by the following ranges:
\begin{align}
\label{ali:angle}
&0<\theta_{e_{0,2}}<\frac{\pi}{2}\hspace{.5in}\frac{3\pi}{2}<\phi_{e_{0,2}}<2\pi-\sin^{-1}\frac{1}{3}\nonumber\\ &\pi<\theta_{e_{0,4}}<\frac{3\pi}{2}\hspace{.5in}\sin^{-1}(\frac{1}{3})<\phi_{e_{0,4}}<\frac{\pi}{2}\nonumber\\
&\frac{\pi}{2}<\theta_{e_{0,1}}<\pi\hspace{.5in}\frac{3\pi}{2}<\phi_{e_{0,1}}<\pi-\sin^{-1}(\frac{1}{3})\nonumber\\
&\frac{3\pi}{2}<\theta_{e_{0,3}}<2\pi\hspace{.5in}\pi+\sin^{-1}(\frac{1}{3})<\phi_{e_{0,3}}<\frac{3\pi}{2}
\end{align}
This implies that the edges $e_{0,2}$, $e_{0,3}$, $e_{0,4}$ and
$e_{0,1}$ lie in the octants $A$, $H$, $C$ and $F$. From the
geometry of the 4-valent lattice, the angles of the edges incident
at all the other vertices follow.

There is a vast range of Euler angles for which the case above is
obtained but, for the sake of brevity, we will not list them
here. What is important, though, is that such case has a non-zero
measure in $SO(3)$.

It should be noted that different combinations of angles within
the angle ranges in (\ref{ali:angle}) lead to different outcomes for
the expectation value of the volume operator, since they lead to
different values of the terms
$\frac{t_{e_ie_j}}{\sqrt{t_{e_i}t_{e_j}}}$. However, in
zeroth-order, the expectation value of the volume operator will be
the same irrespectively of which angles satisfying (\ref{ali:angle})
we decide to utilise. In fact, the rotational dependence of the
expectation value of the volume operator, in the zeroth-order in
$\frac{l}{\delta}$, is determined \emph{solely} by the geometric
factors $G_{\gamma, V_i}$. For the case which we are analysing
(\ref{ali:angle}), the values of $G_{\gamma, V_i}$ will be the same
irrespectively of which sub-case of (\ref{ali:angle}) we analyse. On
the other hand, the dependence of the expectation value of the
volume operator on higher orders of $\frac{l}{\delta}$ \emph{is}
determined by the terms $\frac{t_{e_ie_j}}{\sqrt{t_{e_i}t_{e_j}}}$
and, therefore, will depend on the sub-cases we analyse.

This discussion shows that for higher orders in $\frac{l}{\delta}$ the expectation value of the volume operator \emph{is} rotational dependent since, as stated above, for differing angle-ranges that lead to the same geometric factors, the values of the terms $\frac{t_{e_ie_j}}{\sqrt{t_{e_i}t_{e_j}}}$ will differ.

We will now compute the expectation value of the volume operator
for the periodicity cell in the 4-valent graph, for the case in
which the angles of the edges incident at vertex $V_0$ satisfy
condition (\ref{ali:angle}).
The first step in order to compute the expectation value of the
volume operator is to define the matrix $\sqrt{A}^{-1}$, whose
off-diagonal entries are the terms
$\frac{t_{e_ie_j}}{\sqrt{t_{e_i}t_{e_j}}}$. This matrix
is given in the Appendix of \cite{46ah}. Although different combinations of angles satisfying condition (\ref{ali:angle}) will lead to different values of the terms $\frac{t_{e_ie_j}}{\sqrt{t_{e_i}t_{e_j}}}$, however, any such combination will lead to the same non-zero entries of the matrix $\sqrt{A}^{-1}$. This means that the pairs of edges commonly intersecting a plaquette in a given direction will coincide for \emph{any} combination of angles satisfying conditions (\ref{ali:angle}), even though the number $t_{e_i,e_j}$ of plaquettes they commonly intersect will differ in each case. Therefore, in computing the matrix $\sqrt{A}^{-1}$ we will not determine the precise value of the individual entries, but we will leave them as general as possible. Their precise values can be computed once a specific combination of angles satisfying (\ref{ali:angle}) is chosen.

Given the matrix $\sqrt{A}^{-1}$ we are then able to apply formula
(\ref{4.29}) for computing the expectation value of the volume
operator. As in the aligned case, we first compute the expectation
value of the volume operator for each of the four vertices and,
then, sum their contributions. In what follows, the term
$\frac{t_{e_i,e_j}}{\sqrt{t_{e_i}t_{e_j}}}$ is denoted by
$\alpha_{e_i,e_j}=\frac{t_{e_i,e_j}}{\sqrt{t_{e_i}t_{e_j}}}$. An
explicit form for these terms can be found in Section 11.1 of the
Appendix.

The expectation value for the volume operator for the entire
periodicity cell, up to first-order in $\frac{l}{\delta}$ is:
\begin{align}
&V_R=\delta^3\sqrt{\frac{1}{8}}\Big|\det\Big(\frac{\partial
X_S^a}{\partial(s,u^1,u^2)}\Big)\Big|\Big(
4\sqrt{|det\big(E^a_j(u)\big)|} \\
&+ 4\sqrt{|1 +\frac{1}{2}(- \alpha_{e_{8,10},e_{8,11}}^2 -
\alpha_{e_{8,10},e_{8,12}}^2 - \alpha_{e_{8,10},e_{8,5}}^2 - \alpha_{e_{8,11},e_{8,12}}^2 - \alpha_{e_{8,11},e_{8,5}}^2 - \alpha_{e_{8,12},e_{8,5}}^2)|}\sqrt{|det\big(E^a_j(u)\big)|}\nonumber\\
&+ 4\sqrt{|1 +\frac{1}{2}(- \alpha_{e_{13,18},e_{13,15}}^2 - \alpha_{e_{13,19},e_{13,15}}^2 - \alpha_{e_{13,19},e_{13,18}}^2 -\alpha_{e_{13,3},e_{13,15}}^2 - \alpha_{e_{13,3},e_{13,18}}^2 - \alpha_{e_{13,3},e_{13,19}}^2)|}\nonumber\\
&\times \sqrt{|det\big(E^a_j(u)\big)|}\nonumber\\
&+ 4\sqrt{|1 +\frac{1}{2}(- \alpha_{e_{2,0},e_{2,5}}^2 -
\alpha_{e_{2,0},e_{2,6}}^2 -  \alpha_{e_{2,0},e_{2,7}}^2 -
\alpha_{e_{2,5},e_{2,6}}^2 - \alpha_{e_{2,5},e_{2,7}}^2 -
\alpha_{e_{2,6},e_{2,7}}^2)|}
\sqrt{|det\big(E^a_j(u)\big)|}\;\Big)+\mathcal{O}(3)\nonumber
 \end{align}
By performing a Taylor expansion for each of the roots present in
the above formula, we can factor out the term
$\sqrt{|\det\big(E^a_j(u)\big)|}$ since, in the first-order
approximation that we are considering,
 they turn out to be the same for each
 vertex. Such an approximation is justified by the analysis performed in Section (\ref{s4.1}).
We then obtain
 \begin{align}
V_R=&\delta^32\sqrt{2}\sqrt{|\det\big(E^a_j(u)\big)|}
\Big(1+\frac{1}{4}( - \alpha_{e_{8,10},e_{8,11}}^2 -
\alpha_{e_{8,10},e_{8,12}}^2 - \alpha_{e_{8,10},e_{8,5}}^2 - \alpha_{e_{8,11},e_{8,12}}^2 - \alpha_{e_{8,11},e_{8,5}}^2 \nonumber\\
& -\alpha_{e_{8,12},e_{8,5}}^2- \alpha_{e_{13,18},e_{13,15}}^2 - \alpha_{e_{13,19},e_{13,15}}^2 - \alpha_{e_{13,19},e_{13,18}}^2 -\alpha_{e_{13,3},e_{13,15}}^2 - \alpha_{e_{13,3},e_{13,18}}^2  \nonumber\\
& - \alpha_{e_{13,3},e_{13,19}}^2-\alpha_{e_{2,0},e_{2,5}}^2 - \alpha_{e_{2,0},e_{2,6}}^2 - \alpha_{e_{2,0},e_{2,7}}^2 - \alpha_{e_{2,5},e_{2,6}}^2 - \alpha_{e_{2,5},e_{2,7}}^2 - \alpha_{e_{2,6},e_{2,7}}^2)\Big)\\
&=\delta^3\sqrt{|\det\big(E^a_j(u)\big)|}2\sqrt{2}\bigg(1
+\frac{1}{4}\big(-\sum^{\frac{n}{2}(n-1)}_{i,j=1;j\neq
}\alpha_{ji}^2\big)\;\Big|\det\Big(\frac{\delta
X_S^a}{\delta(s,u^1,u^2)}\Big)\Big|\bigg)
 \end{align}
The term $\frac{1}{2}(-\sum^{\frac{n}{2}(n-1)}_{i,j=1;j\neq
}\alpha^2_{ji})$ represents $l/\delta$-corrections. Each term
$\alpha_{i,j}$ is proportional to $C\frac{l^{\prime}}{\delta}$ for
$l^{\prime}<l$; $C$ is a constant that depends on the Euler angles
we chose. On the other hand the geometric factors
$G_{\gamma,V_{i}}$ for cases (\ref{ali:angle}) coincide with the
geometric factors as computed for any of the sub-cases in Table
\ref{tab:4}, i.e., $G_{\gamma, V_i}=2\sqrt{2}$. This implies that
although for such cases the expectation value of the volume
operator \emph{is} rotational invariant in zeroth-order,
nonetheless, it does \emph{not} reproduce the correct
semiclassical limit.

For those embeddings whose measure is \emph{zero} in $SO(3)$, the
geometric factor turns out to be different and, in zeroth-order in
$l/\delta$, leads to a different value of the expectation value of
the volume operator as computed for 4-valent graphs.

\subsubsection{Expectation value of the volume operator for a translated
4-valent graph} \label{s7.2.3}

In this Section we will analyse whether the expectation value of
the volume operator for the 4-valent graph is translational
invariant with respect to the plaquette.

To perform this analysis we consider our original aligned graph
and  translate it by an arbitrary vector $\vec{p}=(\epsilon_x,
\epsilon_y, \epsilon_z)$. The new coordinates for the vertices
are:
\begin{eqnarray}
V^{''}_0&=&(\epsilon_x,\epsilon_y,\epsilon_z)\mbox{ with }
\epsilon_x>\epsilon_y>\epsilon_z\\
V^{''}_2&=&\Big(-\frac{\delta}{\sqrt{3}}+\epsilon_x,\epsilon_y+\frac{\delta}{\sqrt{3}},\epsilon_z+\frac{\delta}{\sqrt{3}}\Big)\\
V^{''}_8&=&\Big(\frac{\delta}{\sqrt{3}}+\epsilon_x,\epsilon_y+3\frac{\delta}{\sqrt{3}},\epsilon_z+\frac{\delta}{\sqrt{3}}\Big)\\
V^{''}_{13}&=&\Big(\epsilon_x,\epsilon_y+2\frac{\delta}{\sqrt{3}},\epsilon_z-2\frac{\delta}{\sqrt{3}}\Big)
\end{eqnarray}

Similarly to the analysis for rotational invariance,  the
computation of the expectation value of the volume operator can be
divided into different sub-cases, each of which would lead to
different outcomes.

The first division is given by the choice of the signs and the
relations between $\epsilon_x$, $\epsilon_y$ and $\epsilon_z$,
i.e., whether they are positive or negative and whether one
coordinate is bigger or equal to another. Each of these
cases can be ultimately subdivided into sub-cases depending on the
relation between the ratio $b=\frac{\delta}{\sqrt{3}}$ and the
coordinates of the translational vector.

To carry out our calculations we choose the following:
\begin{enumerate}
\item[1)]$b>\epsilon_x>\epsilon_y>\epsilon_z>0$
\item[2)]$|V^i_k|-|V_k^j|>n^i_kl-n^j_kl\mbox{ for all } |V_k^i|>|V_k^j|$
\end{enumerate}
Altogether, such conditions will allow to determine both the
sign of the coordinates for each of the vertices of the translated
graph and, also, the magnitude relation between the coordinates of
each vertex.

However, it will transpire that our result is independent of which
case we decide to use to perform the calculations. In fact, as we
will see, in zeroth-order the expectation value of the volume
operator for a 4-valent graph is translation invariant up to
combinations of measure \emph{zero} in $SO(3)$. However, for
higher orders of approximation this will no longer be true.

As a first step in our calculations we need to specify the allowed
positions for each of the translated vertices. Due to the highly
symmetrical structure of the 4-valent graph, in order to determine
the allowed positions of each vertex, it suffices to find the
allowed positions of one reference vertex. We choose such a
reference vertex to be $V_0$, whose new coordinates are
$V_0=(\epsilon_x,\epsilon_y,\epsilon_z)$.

The number of stacks intersected by the vector that represents
vertex $V_0$ in the $x$, $y$ and $z$-directions are, respectively,
$n=[\frac{\epsilon_x}{l}]$, $m=[\frac{\epsilon_y}{l}]$ and
$p=[\frac{\epsilon_z}{l}]$ (where $[]$ indicates the Gauss
bracket). It follows that the allowed positions for vertex $V_0$
are given by the following ranges of each coordinate:
$nl<\epsilon_x<nl+l$, $ml<\epsilon_y<ml+l$ and
$pl<\epsilon_x<pl+l$.

It turns out that to carry out the calculations for the
expectation value of the volume operator we have to
restrict the value-range of the coordinates $\epsilon_x$,
$\epsilon_y$ and $\epsilon_z$. We choose $
nl<\epsilon_x<nl+\frac{l}{6}$, $ml<\epsilon_z<ml+\frac{l}{6}$ and
$pl<\epsilon_z<pl+\frac{l}{6}$

We will now compute the expectation value of the volume operator
of the periodicity lattice of the 4-valent graph. We will not give
the detail of all the calculations involved since they are quite
lengthy. However, the method utilised is the same as for the non-translated case, namely, for each of the four vertices comprising the
periodicity cell we consider the sub-matrix of $\sqrt{A}^{-1}$,
labelled by the four edges intersecting at the vertex. For each of
these sub-matrices, call them $M$, we compute the determinant of
the four $3\times 3$ sub-matrices of $M$ defined by the triplets
of linearly-independent edges. We then sum up the contributions
coming from each of the vertices. Similarly as for the aligned
4-valent graph we have
\begin{equation}
T^{\{x,y\}}_{e_i}=F^{\{x,y\}}_{e_i}=Z^{\{x,y\}}_{e_i}=1\hspace{.5in}
\forall  e_i\in \gamma
\end{equation}

The expression for the matrix $\sqrt{A}^{-1} $ is
\[ \left( \begin{array}{ccccccccccccccc}
1 & \frac{-C_0}{2} & \frac{-B_0}{2} & \frac{-C_0}{2} & 0 & 0 & 0 & 0 & 0 & 0 & 0 & 0 & 0 & 0 & 0  \\
\frac{-C_0}{2} & 1 & \frac{-C_0}{2} & \frac{-B_0}{2} & \frac{-A_2}{2} & \frac{-C_2}{2} & \frac{-A_2}{2} & 0 & 0 & 0 & 0 & 0 & 0 & 0 & 0 \\
\frac{-B_0}{2} & \frac{-C_0}{2} & 1 & \frac{-C_0}{2} & 0 & 0 & 0 & 0 & 0 & 0 & 0 & 0 & 0 & 0 & 0  \\
\frac{-C_0}{2} &\frac{-B_ 0}{2} & \frac{-C_0}{2} & 1 & 0 & 0 & 0 & 0 & 0 & 0 & 0 & 0 & 0 & 0 & 0  \\
0 & \frac{-A_2}{2}& 0 & 0 & 1 & \frac{-A_2}{2} & \frac{-C_2}{2} & 0 & 0 & 0 & 0 & 0 & 0 & 0 & 0  \\
0 &\frac{-C_2}{2}  & 0 & 0 & \frac{-A_2}{2} & 1 & \frac{-A_2}{2} & 0 & 0 & 0 & 0 & 0 & 0 & 0 & 0  \\
0 & \frac{-A_2}{2} & 0 & 0 & \frac{-C_2}{2} &\frac{-A_2}{2} & 1 &0 & 0 &0 & 0 & 0 & 0 & 0 & 0 \\
0 & 0 & 0 & 0 & 0 & 0 & 0 & 1 & \frac{-A_{13}}{2} & \frac{-C_{13}}{2} &\frac{-A_{13}}{2} & 0 & 0 & 0 & 0 \\
0 & 0 & 0 & 0 & 0 & 0 & 0 & \frac{-A_{13}}{2} & 1 & \frac{-A_{13}}{2} & \frac{-C_{13}}{2} & 0 & 0 & 0 & 0 \\
0 & 0 & 0 & 0 & 0 & 0 & 0 & \frac{-C_{13}}{2} & \frac{-A_{13}}{2} & 1 & \frac{-A_{13}}{2} & 0 & 0 & 0 & 0  \\
0 & 0 & 0 & 0 & 0 & 0 & 0 & \frac{-A_{13}}{2} & \frac{-C_{13}}{2} & \frac{-A_{13}}{2} & 1 &0&0& 0 & 0 \\
0 & 0 & 0 & 0 & 0 & 0 & 0 & 0 & 0 & 0 & 0 & 1 & \frac{-A_8}{2} & \frac{-C_8}{2} &\frac{-C_8}{2}  \\
0 & 0 & 0 & 0 & 0 & 0 & 0 & 0 & 0 & 0 & 0 &\frac{-A_8}{2} & 1 & \frac{-C_8}{2} & \frac{-C_8}{2}  \\
0 & 0 & 0 & 0 & 0 & 0 & 0 & 0 & 0 & 0 & 0 & \frac{-C_8}{2} & \frac{-C_8}{2} & 1 & \frac{-A_8}{2} \\
0 & 0 & 0 & 0 & 0 & 0 & 0 & 0 & 0 & 0 & 0 &\frac{-C_8}{2}& \frac{-C_8}{2} & \frac{-A_8}{2} & 1  \\
\end{array} \right)\]
where $A_i=(x_{V_i}-n_il)\frac{1}{\delta\sqrt{3}}$,
$B_i=(y_{V_i}-m_il)\frac{1}{\delta\sqrt{3}}$,
$C_i=(z_{V_i}-p_il)\frac{1}{\delta\sqrt{3}}$.

Applying the method described above we compute the expectation
value for the volume operator for one periodicity cell to be
\begin{align}
V_R=&\sqrt{\frac{1}{8}}\delta^3|\det\Big(\frac{\delta X_S^a}{\delta(s,u^1,u^2)}\Big)\Big|2\times\\
&\Bigg(\sqrt{|\det\Big(E^a_j(u)\Big)|}\sqrt{|4 - B_0^2 - 2 C_0^2 - B_0 C_0^2|}+\sqrt{|\det\Big(E^a_j(u)\Big)|}\sqrt{|4 - 2 A_2^2 - A_2^2 C_2 - C_2^2| }+\nonumber\\
&\sqrt{|\det\Big(E^a_j(u)\Big)|}\sqrt{|4 - 2 A_{13}^2 - A_{13}^2
C_{13} - C_{13}^2| }+\sqrt{|\det\Big(E^a_j(u)\Big)|}\sqrt{|4 -
A_8^2 - 2 C_8^2 - A_8 C_8^2|}\Bigg)
\end{align} In first-order approximation we then obtain
\begin{align}
&V_R=2\sqrt{\frac{1}{8}}\delta^3\sqrt{|\det\Big(E^a_j(u)\Big)|}
\Bigg(4 - \frac{1}{8}\Big(B_0^2 + 2 C_0^2 + B_0 C_0^2 + 2 A_2^2 + A_2^2 C_2 + C_2^2 +2 A_{13}^2+ A_{13}^2 C_{13} + C_{13}^2 +\nonumber\\
& A_8^2 + 2 C_8^2 + A_8 C_8^2\Big)\Bigg)+\mathcal{O}(4)\Bigg)\Big|\det\Big(\frac{\delta X_S^a}{\delta(s,u^1,u^2)}\Big)\Big|\nonumber\\
&=2\sqrt{2}\delta^3\sqrt{|\det\Big(E^a_j(u)\Big)|} \Bigg(1 -
\frac{1}{32}\Big(B_0^2+ 2 C_0^2 +  2 A_2^2 + C_2^2 +2A_{13}^2 +
  C_{13}^2  + A_8^2 + 2 C_8^2
\Big)+\mathcal{O}(3)\Bigg)\nonumber\\
&\times \Big|\det\Big(\frac{\delta
X_S^a}{\delta(s,u^1,u^2)}\Big)\Big|
\end{align}
where in the last equation we have only considered first-order
contributions obtained by the usual Taylor series of the square
root (see Section \ref{s4.1}). Thus, we were able to factor out
the term $\sqrt{\det(E^a_j(u))}$. As it is evident, the
corrections of second order in $l/\delta$  are not translationally
invariant.

\subsection{Analysis of the expectation value of the volume operator
for 6-valent graphs} \label{s7.3}

 In this Section we will calculate the expectation value of
 the volume operator for a 6-valent graph. First we  consider the
 non-rotated graph, then we will analyse the rotational and
 translational dependence of the expectation value by performing a
 rotation of the graph, followed by  a translation of the graph. We will
 then recalculate the expectation value.

\subsubsection{Expectation value of the volume operator for a general
6-valent graph} \label{s7.3.1}

Similarly as  for the 4-valent graph, we will analyse
the case for which $\frac{\delta}{l}>0$; the motivation for such a
choice was given in Section \ref{s4.1}. For computational
simplicity we will position the graph so that the $(0,0,0)$
coordinates of the graph coincide with the $(0,0,0)$ coordinates
of the plaquette. We also need to align the graph in such a way
that each vertex is symmetrical with
respect to the axis. Therefore we will choose, for each vertex $V_k$, the value $\phi_{e_{k,i}}=45$ for all edges $e_{k,i}$ incident at $V_k$ and $\theta_{e_{k,i}}=45$ for four edges, while the remaining two will have $\theta_{e_{k,i}}=0$. This edge orientation corresponds to the limiting case (2) described in Section \ref{s 2.4.1}.

As for the diamond lattice, we  choose the vertex $V_2$ as our
reference vertex with respect to which we determine the allowed
positions of all the remaining vertices of the graph. We also
choose the allowed values of the $x$-coordinate of $V_2$ to be
$nl<x_{V_2}<nl+\frac{l}{6}$, where, in this case,
$n=[\frac{\delta\sqrt{2}}{2l}]$. Using the same method used
in Section 3.1 we can compute all the terms
$\frac{t_{e_{i}e_{j}}} {\sqrt{t_{e_{i}}t_{e_{j}}}}$ for the
periodicity cell of the 6-valent graph that contains nine
vertices. Given the geometry of the 6-valent graph we have the
following values for the term in equations (\ref{equ:te})--(\ref{equ:te3})
\begin{align}
Z^{\{x,y\}}_{e_i}&=\begin{cases}\frac{2}{\sqrt{2}}& \text{ iff } \theta_{e_i}\neq 0\\
1&\text{ iff } \theta_{e_i}= 0
\end{cases}
\hspace{.5in}
T^{\{x,y\}}_{e_i}=\begin{cases}\frac{\sqrt{2}}{2}& \text{ iff } \theta_{e_i}\neq 0\\
1&\text{ iff } \theta_{e_i}= 0
\end{cases}\nonumber\\
F^x_{e_i}&=\begin{cases}1 & \text{ iff } \theta_{e_i}\neq 0\\
\infty& \text{ iff } \theta_{e_i}= 0
\end{cases}\hspace{.5in}
F^x_{e_i}=1\hspace{.5in} \forall e_i\in \gamma
\end{align}

The coordinates of the vertices of the periodicity cell are
\begin{eqnarray*}
V_0&=&(0,0,0)\\
V_2&=&\big(\delta\frac{\sqrt{2}}{2},0,\delta\frac{\sqrt{2}}{2}\big)\\
V_3&=&\big((\delta\frac{\sqrt{2}}{2}-(\delta\frac{1}{2},-\delta\frac{1}{2},2\delta\frac{\sqrt{2}}{2}\big)\\
V_4&=&\big(2\delta\frac{\sqrt{2}}{2},-2\delta\frac{1}{2},2\delta\frac{\sqrt{2}}{2}\big)\\
V_5&=&\big(0,-2\delta\frac{1}{2},0\big)\\
V_{13}&=&\big(\delta\frac{\sqrt{2}}{2},-2\delta\frac{1}{2},\delta\frac{\sqrt{2}}{2}\big)\\
V_{18}&=&\big(2\delta\frac{\sqrt{2}}{2}+\delta\frac{1}{2},-1\delta\frac{1}{2},\delta\frac{\sqrt{2}}{2}\big)\\
V_{15}&=&\big(\delta\frac{\sqrt{2}}{2}-\delta,-2\delta\frac{1}{2},3\delta\frac{\sqrt{2}}{2}\big)\\
V_{14}&=&\big(\delta\frac{\sqrt{2}}{2}+\delta\frac{1}{2},-3\delta\frac{1}{2},0\big)
\end{eqnarray*}

The values for the terms $\frac{t_{e_{i}e_{j}}}
{\sqrt{t_{e_{i}}t_{e_{j}}}}$ are given in the Section 2.1
of the Appendix in \cite{46ah}

Expanding the square root (see the analysis in Section \ref{s4.1})
and considering first-order terms we obtain the value for the expectation value of the volume operator for one
periodicity cell, consisting of \emph{nine} vertices

\begin{align}
V_R&=2\delta^3\sqrt{|\det\big(E^a_j(u)\big)|}\,\Big|\det\Big(\frac{\delta
X_S^a}{\delta(s,u^1,u^2)}\Big)
\Big|\Bigg(9+\frac{1}{2}\times\frac{1}{8}\big(- 2 (A^{'}_{14})^2 - 3 (A^{'}_{15})^2 - \frac{3 (A^{'}_{18})^2}{2} - \frac{5 (A_{3}^{'})^2}{2}\nonumber\\
& - A_{14}^2- 2 A_{15}^2   - \frac{5 A_{3}^2}{2} -
 \sqrt{2} A_{3}^2 - \frac{19 \alpha^2}{4}   -
 29 (\alpha^{'})^2  -
 3 \sqrt{2} A_{14} \beta - 3 \sqrt{2} A_{15} \beta - 2 \alpha \beta -
 2 \sqrt{2} \alpha \beta \nonumber\\
 &- \frac{125 \beta^2}{2} -
 32 \sqrt{2} \beta^2  - \frac{67 (\beta^{'})^2}{2})\;\Bigg)
\end{align}
Contrary to the dual cell complex coherent states (\cite{30b}) we find that, to zeroth-order in
$l/\delta$, the expectation value of the volume operator for a
6-valent graph does \textit{not} have the correct semiclassical
limit. On the other hand, if the graph is aligned to the
orientation of the plaquette we \emph{do} obtain the correct
semiclassical value. However, this embedding has measure
\emph{zero} in $SO(3)$.

\subsubsection{Expectation value of the volume operator for a rotated
6-valent graph} \label{s7.3.2}

We will now analyse the expectation value of the volume operator
for  a rotated 6-valent graph. As for the 4-valent graph,
different choices of Euler angles in the rotation give different
values of $t_{e_ie_j}$. Therefore, we will  once again have to
define sub-cases which are defined according to the possible
ranges of values for the angles $\phi_{e_i}$ and $\theta_{e_i}$
for each edge $e_i$. Because of the geometry of the 6-valent
lattice, we know that for any edge, $e$, of a given vertex there
exists a co-linear edge, $e^{\prime}$, which intersects the same
vertex. This implies that, given two co-linear edges $e$ and
$e^{\prime}$,  we can define the angles of $e$ (respectively
$e^{\prime}$) in terms of $e^{\prime}$ (respectively $e$) as
follows: $\phi_e=180^{\circ}-\phi_{e^{\prime}}$ and
$\theta_e=180^{\circ}-\theta_{e^{\prime}}$. Such relations reduce
the number of cases that need to be analysed.

We choose the vertex $V^{\prime}_0$ as our reference vertex. The
relations for the angles of the edges incident at $V^{\prime}_0$
are:
\begin{eqnarray}\label{eqn:cond}
\phi_{e_{0,7}}&=&180^{\circ}-\phi_{e_{0,2}}\nonumber\\
\phi_{e_{0,1}}&=&180^{\circ}-\phi_{e_{0,17}}\nonumber\\
\phi_{e_{0,6}}&=&180^{\circ}-\phi_{e_{0,8}} \nonumber\\
\theta_{e_{0,8}}&=&180^{\circ}-\theta_{e_{0,6}}\nonumber\\
\theta_{e_{0,17}}&=&180^{\circ}-\theta_{e_{0,1}}\nonumber\\
\theta_{e_{0,2}}&=&180^{\circ}-\theta_{e_{0,7}}
\end{eqnarray}
These relations imply that the allowed values of the angles of the
edges at a given vertex  fall into one of the following
groups:\begin{enumerate}
\item A given triplet of edges points upwards, i.e., their angle $\phi$ lies between $-\frac{\pi}{2}$ and $\frac{\pi}{2}$, and the triplet formed by their co-linear edges points downwards, i.e., their angle $\phi$ lies between $\frac{\pi}{2}$ and $\frac{3\pi}{2}$. This situation arises when none of the edges is aligned with one of the $x$, $y$, $z$-coordinates. However we have two distinct sub-cases which satisfy this arrangement of edges
\begin{itemize}
\item[a.] No edge lies in any plaquette.
\item[b.] Only one edge and its co-planar lie in a given plaquette (Figure \ref{fig6}).
\end{itemize}
\item A given couple of edges points upwards i.e., their angle $\phi$ lies between $-\frac{\pi}{2}$ and $\frac{\pi}{2}$, and their co-linear edges point downwards i.e., their angle $\phi$ lies between $\frac{\pi}{2}$ and $\frac{3\pi}{2}$. This situation arises when one edge (and subsequently its collinear edge) is aligned with one of the coordinates axis and, subsequently, the remaining two edges and their co-linear lie in two different plaquettes in the same direction (Figure \ref{fig:9})
\item Only one edge points upwards, i.e., its angle $\phi$ lies between $-\frac{\pi}{2}$ and $\frac{\pi}{2}$, and the co-linear edge points downwards i.e., its angle $\phi$ lies between $\frac{\pi}{2}$ and $\frac{3\pi}{2}$. This situation arises when all of the edges are aligned with the coordinate axis (Figure \ref{fig:orig}).
\end{enumerate}
A discussion of each of these cases and the respective value for
the geometric factor was carried out in Section \ref{s 2.4.1}. There it
was shown that only case 1a above has \emph{non-zero} measure in
SO(3), therefore we will restrict our analysis to such a case.

It is straightforward to see that case 1a can be divided into
further sub-cases according to the values of the $\theta$-angles
and the relations between the $\phi$-angles of each of the up/down
couples. In what follows, we will not give the results for all
possible choices. Instead, we will  choose a particular sub-case
and perform the calculations for the expectation value of the
volume operator with respect to this sub-case. As we will see,
these calculations show that, up to embeddings of measure zero in
$SO(3)$, the semiclassical behaviour of the volume operator, in
zeroth-order, does not depend on how the graph is rotated: \emph{a
fortiori}, it is independent of the particular case we have analysed.

\begin{figure}[htbp]
\begin{center}
\includegraphics[scale=0.5]{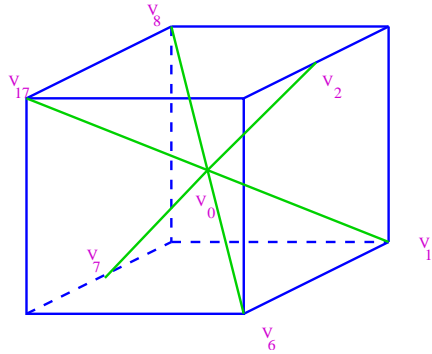}
 \caption{Regular 6-valent vertex.\label{fig:64}}
\end{center}
\end{figure}

In order to carry out a proper comparison between the
semiclassical behaviour of the volume operator, as applied to graphs
of different valence, we will apply the same Euler transformations
(i.e., with the same Euler angles) to each of the graphs we
consider. Since we have not specified the values of the Euler
angles, the only way to do this is to assume that after a
rotation, those edges which had the same angles used in both the
aligned 4-valent and 6-valent case will end up in the same octant.
For example, consider Figures \ref{fig:64} and \ref{fig:46} which
depict both 6-valent and 4-valent vertices, respectively. From such
pictures it is easy to see that the edges $e_{0,2}$ and $e_{0,3}$
of the 4-valent graph have the same $\theta$- and $\phi$-angles as
the edges, $e_{0,8}$, and, $e_{0,1}$, of the 6-valent graph.
Therefore, in the rotated case we will assume that  $e_{0,8}$,
and, $e_{0,1}$ lie in the same octants as $e_{0,2}$ and $e_{0,3}$,
respectively. It follows that the angle-ranges for the edges
incident at vertex $V_0$ for a 6-valent graph are:
\begin{align}\label{ali:ranges6}
&\frac{3\pi}{2}<\theta_{e_{0,2}}<\frac{7\pi}{4}\hspace{.5in}\frac{3\pi}{2}<\phi_{e_{0,2}}<2\pi\nonumber\\
&\frac{\pi}{4}<\theta_{e_{0,8}}<\frac{\pi}{2}\hspace{.5in}0<\phi_{e_{0,8}}<\frac{\pi}{2}\nonumber\\
&\frac{3\pi}{4}<\theta_{e_{0,17}}<\pi\hspace{.5in}0< \phi_{e_{0,17}}<\frac{\pi}{2}\nonumber\\
&\frac{\pi}{2}<\theta_{e_{0,7}}<\frac{3\pi}{4}\hspace{.5in}\frac{\pi}{2}<\phi_{e_{0,7}}<\pi\nonumber\\
&\frac{\pi}{2}<\theta_{e_{0,6}}<\frac{3\pi}{4}\hspace{.5in}\pi<\phi_{e_{0,6}}<\frac{3\pi}{2}\nonumber\\
&\frac{7\pi}{4}<\theta_{e_{0,1}}<2\pi\hspace{.5in}\pi<\phi_{e_{0,1}}<\frac{3\pi}{2}
\end{align}
Such conditions of the angles implies that the edges $e_{0,1}$,
$e_{0,2}$, $e_{0,6}$, $e_{0,7}$, $e_{0,7}$ and $e_{0,17}$ lie in
the octants $H$, $D$, $G$, $F$, $A$ and $B$, respectively.

Since we are considering a regular 6-valent lattice, the above
ranges of angles induce a relation on all the other angle-ranges
of the edges  for each vertex in the graph.
\begin{figure}[htbp]
\begin{center}
 \includegraphics[scale=0.5]{4valent.eps}
\caption{Regular 4-valent vertex\label{fig:46}}
\end{center}
\end{figure}
The coordinates of the rotated vertices are:
\begin{eqnarray}
V^{\prime}_0&=&\Big(0,0,0\Big)\nonumber\\
V^{\prime}_2&=&\Big((R_{11}+R_{13})\frac{\delta
\sqrt{2}}{2},(R_{21}+R_{23})\frac{\delta
\sqrt{2}}{2},(R_{31}+2R_{33})
        \frac{\delta \sqrt{2}}{2}\Big)\nonumber\\
V^{\prime}_3&=&\Big((R_{11}(\frac{\sqrt{2}-1}{2})-\frac{R_{12}}{2}+R_{13} \sqrt{2})\delta,(R_{21}(\frac{\sqrt{2}-1}{2})-\frac{R_{22}}{2}+R_{23} \sqrt{2})\delta,(R_{31}(\frac{\sqrt{2}-1}{2})-\frac{R_{32}}{2}+R_{33} \sqrt{2})\delta\Big)\nonumber\\
V_{4}^{\prime}&=&\Big((\sqrt{2}R_{11}-R_{12}+\sqrt{2}R_{13})\delta
, (\sqrt{2}R_{21}-R_{22}+\sqrt{2}R_{23})\delta ,
(\sqrt{2}R_{31}-R_{32}+\sqrt{2}R_{33})\delta \Big)\nonumber\\
V_{5}^{\prime}&=&\Big(-R_{12}\delta ,
R_{22}\delta, R_{32}\delta\Big)\nonumber\\
V_{13}^{\prime}&=&\Big((-\frac{\sqrt{2}}{2}R_{11}-R_{12}+\frac{\sqrt{2}}{2}R_{13})\delta
,
(-\frac{\sqrt{2}}{2}R_{21}-R_{22}+\frac{\sqrt{2}}{2}R_{23})\delta,
(-\frac{\sqrt{2}}{2}R_{31}-R_{32}+\frac{\sqrt{2}}{2}R_{33})\delta\Big)\nonumber\\
V_{18}^{\prime}&=&\Big((\frac{2\sqrt{2}+1}{2}R_{11}-\frac{1}{2}R_{12}+\frac{\sqrt{2}}{2}R_{13})\delta
,(\frac{2\sqrt{2}+1}{2}R_{21}-\frac{1}{2}R_{22}+\frac{\sqrt{2}}{2}R_{23})\delta,
(\frac{2\sqrt{2}+1}{2}R_{31}-\frac{1}{2}R_{32}+\frac{\sqrt{2}}{2}R_{33})\delta\Big)\nonumber\\
V^{\prime}_{15}&=&\Big(((\frac{\sqrt{2}}{2}-1)R_{11}-R_{12}+\frac{3\sqrt{2}}{2}R_{13})\delta,
((\frac{\sqrt{2}}{2}-1)R_{21}-R_{22}+\frac{3\sqrt{2}}{2}R_{23})\delta,((\frac{\sqrt{2}}{2}-1)R_{31}-R_{32}+\frac{3\sqrt{2}}{2}R_{33})\delta\Big)\nonumber\\
V_{14}^{\prime}&=&\Big((\frac{\sqrt{2}+1}{2}R_{11}-\frac{3}{2}R_{12})\delta
, (\frac{\sqrt{2}+1}{2}R_{21}-\frac{3}{2}R_{22})\delta,
(\frac{\sqrt{2}+1}{2}R_{31}-\frac{3}{2}R_{32})\delta\Big)
\end{eqnarray}
Similarly, as it was done for the 4-valent case, different choices of combination of angles satisfying conditions (\ref{ali:ranges6}) above will lead to different values for the terms $\frac{t_{e_i,e_j}}{\sqrt{t_{e_i}t_{e_j}}}$. However, the couples of edges commonly intersecting a given stack will coincide for any such combination. This implies that the matrix $\sqrt{A}^{-1}$ will have the same entries for any sub-case of (\ref{ali:ranges6}) but their specific values will be different.

Moreover, the geometric factor $G_{\gamma,V}$ of any sub-case of
(\ref{ali:ranges6}) will coincide. It follows that any combination
of angles that satisfies conditions (\ref{ali:ranges6}) will lead to the
same value in zeroth-order in $\frac{l}{\delta}$ of the
expectation value for the volume operator. Therefore, as it was done for
the 4-valent case, in order to compute the expectation value for
the volume operator, we will not specify a particular sub-case of
(\ref{ali:ranges6}), but leave the result as general as possible.

Given conditions  \ref{ali:ranges6} the expectation value of the
volume for the periodicity cell is
\be
2\delta^3\sqrt{|\det\big(E^a_j(u)\big)|}\bigg(9
+\frac{1}{2}\Big(-\frac{1}{16}\sum^{\frac{n}{2}(n-1)}_{i,j=1;j\neq
i }\rho_{j_k,i_k}^2 \Big)\bigg)\;\Big|\det\Big(\frac{\delta
X_S^a}{\delta(s,u^1,u^2)}\Big)\Big|
\ee
where we have expanded the square roots (see Section A.2.2 in \cite{30b}) and have considered just first-order terms. $\rho_{l_i,k_i}$ represent the off-diagonal entries of the
matrix $\sqrt{A}^{-1}$ (see Section A.2.2 in \cite{30b}), which denote the
value of the term $\frac{\sqrt{t_{e_{i,l},
e_{i,k}}}}{t_{e_{i,l}}t_{e_{i,k}}}$ for the edges $e_{i,l}$ and
$e_{i,k}$ incident at the vertex $V_i$.

The term proportional to $\rho_{j_k,i_k}^2$ represents the
higher-order corrections to the expectation value of the volume
operator. Each term $\rho_{i_k,j_k}$ is proportional to $C\times
\frac{l^{\prime}}{\delta}$ for $l^{\prime}<l$ where $C$ is a
constant that  depends on the Euler angles we choose. It follows
that only the zeroth-order of the expectation value of the volume
operator for the 6-valent graph is rotationally invariant, up to
embeddings with measure zero in $SO(3)$. However, only embeddings
which have measure \emph{zero} in $SO(3)$ (when the edges are
aligned to the plaquettes) give the correct semiclassical limit.

\subsubsection{Expectation value of the volume operator for a
translated 6-valent graph} \label{s7.3.3}

In this Section we will calculate the expectation value of the
volume operator for a translated 6-valent graph. To make the
comparison as accurate as possible, we translate the 6-valent
graph by a vector with more or less the same properties as the
vector with respect to which we translated the 4-valent graph,
namely:
\begin{enumerate}
\item[1)] $\frac{\delta }{2}\ge\epsilon_x\ge\epsilon_y\ge\epsilon_z\ge 0$
\item[2)] $\frac{\delta\sqrt{2}}{2}<\epsilon_i+\epsilon_j$
\item[3)]$\frac{\delta\sqrt{2}}{2}-\frac{\delta 1}{2}<\epsilon_z$
\item[4)]$|V^i_k|-|V_k^j|>n^i_kl-n^j_kl\mbox{ for all } |V_k^i|>|V_k^j|$.
\end{enumerate}

The coordinates of the translated vertices are
\begin{eqnarray}
V^{''}_0&=&(\epsilon_x,\epsilon_y,\epsilon_z)\\
V^{''}_2&=&\Big(\frac{\delta
\sqrt{2}}{2}+\epsilon_x,\epsilon_y,\frac{\delta
\sqrt{2}}{2}+\epsilon_z\Big)\\
V^{''}_3&=&\Big(\epsilon_x+\frac{\delta \sqrt{2}}{2}-\frac{\delta
}{2},\epsilon_y-\frac{\delta
}{2},\delta \sqrt{2}+\epsilon_z\Big)\\
V^{''}_4&=&\Big(\delta \sqrt{2}+\epsilon_x,\epsilon_y-\delta
,\delta \sqrt{2}+\epsilon_z\Big)\\
V^{''}_5&=&\Big(\epsilon_x,\epsilon_y-\delta ,
\epsilon_z\Big)\\
V^{''}_{13}&=&\Big(\frac{\delta
\sqrt{2}}{2}+\epsilon_x,\epsilon_y-\delta
,\frac{\delta \sqrt{2}}{2}+\epsilon_z\Big)\\
V^{''}_{18}&=&\Big(\delta
\sqrt{2}+\frac{\delta}{2}+\epsilon_x,\epsilon_y-\frac{\delta}{2},\epsilon_z+\frac{\delta
\sqrt{2}}{2}\Big)\\
V^{''}_{15}&=&\Big(\frac{\delta \sqrt{2}}{2}-\frac{\delta
2}{2}+\epsilon_x,\epsilon_y-\delta
,3\frac{\delta \sqrt{2}}{2}+\epsilon_z\Big)\\
V^{''}_{14}&=&\Big(\frac{\delta
\sqrt{2}}{2}+\frac{\delta}{2}+\epsilon_x,\epsilon_y-3\frac{\delta}{2},\epsilon_z\Big)
\end{eqnarray}
Similarly, as for the aligned case, we have
\begin{align}
Z^{\{x,y\}}_{e_i}&=\begin{cases}\frac{2}{\sqrt{2}}& \text{ iff } \theta_{e_i}\neq 0\\
1&\text{ iff } \theta_{e_i}= 0
\end{cases}
\hspace{.5in}
T^{\{x,y\}}_{e_i}=\begin{cases}\frac{\sqrt{2}}{2}& \text{ iff } \theta_{e_i}\neq 0\\
1&\text{ iff } \theta_{e_i}= 0
\end{cases}\nonumber\\
F^x_{e_i}&=\begin{cases}1 & \text{ iff } \theta_{e_i}\neq 0\\
\infty& \text{ iff } \theta_{e_i}= 0
\end{cases}\hspace{.5in}
F^x_{e_i}=1\hspace{.5in} \forall e_i\in\gamma
\end{align}


Given the conditions above the geometric facto will be $G_{\gamma}=2$ thus we obtain the following value for the
expectation value of the volume operator, as applied to the
periodicity cell:
\begin{align}\label{ali:6t}
V_R=&\delta^3\sqrt{|det\Big(E^a_j(u)\Big)|}\Big|det\Big(\frac{\delta
X_S^a}{\delta(s,u^1,u^2)}\Big)\Big|2\Bigg(9+\frac{1}{16}\Big( -
\frac{3 (A^{'}_0)^2}{2} - \frac{5 B_0^2}{2}- \sqrt{2} B_0^2   - 2
(B^{'}_0)^2 - B_0 C_0  - \frac{
 3 C_0^2}{4}\nonumber\\
 & - \frac{3 (C^{'}_0)^2}{2} - \frac{3 (A^{'}_2)^2}{2} - \frac{7 (B^{'}_2)^2}{2}  - 3 B_2^2 - 2 \sqrt{2} B_2^2  - (C^{'}_2)^2 - \frac{C_2^2}{2} - \frac{3 (A^{'}_3)^2}{2}- A_3^2 - \frac{5 (B^{'}_3)^2}{2}  - \frac{
 B^{'}_3 B_3}{2}  \nonumber\\
&- 3 B_3^2 -
 2 \sqrt{2} B_3^2  - \frac{
 3 (C^{'}_3)^2}{4} - \frac{3 (A^{'}_4)^2}{2} - \frac{A_4^2}{2} - 3 (B^{'}_4)^2  - 3 B_4^2 - 2 \sqrt{2} B_4^2- \frac{3 (C^{'}_4)^2}{4}  - \frac{C_4^2}{4}- 3 (A^{'}_5)^2\nonumber\\
 &   - A_5^2  - \sqrt{2} A_5 B_5 - \frac{3 B_5^2}{2} - \frac{13 (C^{'}_5)^2}{8} - B_5 C_5 - \frac{5 C_5^2}{8}  - \frac{3 (A^{'}_{13})^2}{2} - \frac{11 (B^{'}_{13})^2}{4} - 3 B_{13}^2 - 2 \sqrt{2} B_{13}^2  - \frac{3 (C^{'}_{13})^2}{4}\nonumber\\
 & - \frac{C_{13}^2}{2} - \frac{5 (A^{'}_{18})^2}{4} - \frac{A_{18}^2}{4} - 3 (B^{'}_{18})^2  - 3 B_{18}^2 - 2 \sqrt{2} B_{18}^2- \frac{3 (C^{'}_{18})^2}{4}  - \frac{C_{18}^2}{2} - 3 (A^{'}_{15})^2  - 2 A_{15}^2 - \frac{B^{'}_{15})^2}{2}\nonumber\\
 &-
  \sqrt{2} A_{15} B_{15} - 2 B_{15}^2 - \sqrt{2} B_{15}^2  - C_{15}^2 - \frac{B_{15} C_{15}^2}{8}- \frac{3 (A^{'}_{14})^2}{2} - 2 (B^{'}_{14})^2  - \frac{5 B_{14}^2}{2}- \sqrt{2} B_{14}^2 - \frac{
 3 (C^{'}_{14})^2}{2} \nonumber\\
 &- B_{14} C_{14} - \frac{3 C_{14}^2}{4}\Big)\;\Bigg)
\end{align}
 where the terms
 \begin{align}\label{ali:terms}
&A^{\prime}_i=\frac{|x_{V_i}|-n_i}{\delta(\sqrt{\sqrt{2}+1)}}\hspace{.5in}A_i=\frac{(|x_{V_i}|-n_il)}{\delta_e(1+\frac{\sqrt{2}}{2}}
\nonumber\\
&B^{\prime}_i=\frac{(|y_{V_i}|-m_il)}{\delta(\sqrt{\sqrt{2}+1)}}\hspace{.5in}B_i=\frac{(|y_{V_i}|-m_il)}{\delta_e(1+\frac{\sqrt{2}}{2}}
\nonumber\\
&C^{\prime}_i=\frac{(|z_{V_i}|-p_il)}{\delta(\sqrt{\sqrt{2}+1)}}\hspace{.5in}C_i=\frac{(|z_{V_i}|-p_il)}{\delta_e(1+\frac{\sqrt{2}}{2}}
\end{align}
are the off-diagonal matrix elements of $\sqrt{A}^{-1}$. 
The quantities $x_{V_i}$, $y_{V_i}$,
$z_{V_i}$ represent the $x$, $y$, $z$-coordinates of the vertex
$V_i$, respectively. As for the previous cases, we have expanded the square root in the expression for the expectation value of the volume operator and we have considered only first-order contributions (see Section \ref{s4.1}). Therefore, we were able to factor out the term
$\sqrt{\det\big(E^a_j(u)\big)}$, since we can assume that,
although it is vertex dependent, the values of this term to
first-order will be the same for each vertex. Due to the
appearance of the terms (\ref{ali:terms}), which are proportional to
the Euler angles, equation (\ref{ali:6t}) is translational
invariant (up to embeddings of measure zero in $SO(3)$), only at
zeroth-order.

\subsection{Analysis of the expectation value of the volume operator
for 8-valent graphs} \label{s7.4}

In this Section we will calculate the expectation value of the
volume operator for an 8-valent graph. As in the case of 4- and
6-valent graphs, we will first consider the non-rotated graph. We will
 then analyse the rotational and translational dependence of
the expectation value by performing a rotation and, then, a
translation of the graph; we then repeat the calculation. It
transpires that, even for the 8-valent graph, the off-diagonal
elements of the matrix have non-trivial contributions, that cause
the expectation value of the volume operator to be translationally
and rotationally dependent for higher orders than the zeroth-one.

\subsubsection{Expectation value of the volume operator for a general
8-valent graph} \label{s7.4.1}

As in the previous cases, we take the $(0,0,0)$ point of the
lattice to coincide with the $(0,0,0)$ point of the plaquette, and
each vertex to be symmetric with respect to the axis. The
coordinates of the vertices, comprising the periodicity cell, are
the following:
\begin{eqnarray}
V_1&=&\Big(\frac{-\delta}{\sqrt{3}},\frac{-\delta}{\sqrt{3}},\frac{\delta}{\sqrt{3}}\Big)\\
V_{9}&=&\Big(0,\frac{-2\delta}{\sqrt{3}},0\Big)\\
V_{12}&=&\Big(0,\frac{-2\delta}{\sqrt{3}},\frac{2\delta}{\sqrt{3}}\Big)\\
V_{4}&=&\Big(\frac{-\delta}{\sqrt{3}},\frac{-\delta}{\sqrt{3}},\frac{3\delta}{\sqrt{3}}\Big)
\end{eqnarray}
THe geometric factor is $G_{\gamma}=4$. For the 8-valent lattice we choose $V_1$ as our reference vertex.
The allowed value for its $x$-coordinate is
$nl<|x_{V_2}|<nl+\frac{l}{4}$, where
$n=[\frac{\delta}{\sqrt{3}l}]$. Similarly to the cases of  4- and
6-valent graphs, the allowed positions of the remaining vertices
in the periodicity cell can be computed from the allowed positions
of $V_1$. Because of the geometry of the 8-valent graph we obtain
\begin{equation}
T^{\{x,y\}}_{e_i}=F^{\{x,y\}}_{e_i}=Z^{\{x,y\}}_{e_i}=1\hspace{.5in}
\forall  e_i\in \gamma
\end{equation}
This results in the following values for the terms
$\frac{t_{e_{i}e_{j}}} {\sqrt{t_{e_{i}}t_{e_{j}}}}$ as computed
for the above five vertices.
\begin{center}
\begin{tabular}{l|l}
$V_0$& $\text{all} \frac{t_{e_{i}e_{j}}}
{\sqrt{t_{e_{i}}t_{e_{j}}}}=0$\\ \hline $V_1$& $\text{14 terms
equal to }\beta; \text{ 12 terms equal to } 2\beta$\\ \hline
$V_9$& $\text{8 terms equal to }2\beta$\\ \hline $V_{12}$&
$\text{8 terms equal to }4\beta; \text{ 4 terms equal to }
2\beta$\\ \hline $V_{e}$& $\text{1 term equal to }6\beta; \text{ 9
terms equal to } 2\beta; \text{ 12 terms equal to } \beta; \text{
2 terms equal to } 4\beta$\\ \hline
\end{tabular}
\end{center}
Here $\beta:=(\frac{\delta}{\sqrt{3}}-nl)\frac{1}{\sqrt{3}\delta}$
and it is proportional to the off-diagonal entries of the matrix
$\sqrt{A}^{-1}$.

The expectation value of the volume operator in first order approximation is:
\begin{align}
V_R=&4\delta^3
\sqrt{|det\Big(E^a_j(u)\Big)|}\;\Big|det(\frac{\delta
X_S^a}{\delta(s,u^1,u^2)})\Big|\;\Bigg(5+\frac{1}{2}\times\frac{1}{32}\Big(
324 \beta^2\Big)\Bigg)
\end{align}
In this case, the deviation from the classical value of the volume
of a region, $R$, is of the order four, even to zeroth-order in
$l/\delta$.

 \subsubsection{Expectation value of the volume operator for a rotated
 8-valent graph}
\label{s7.4.2}

We will now analyse the expectation value of the volume operator
for a rotated 8-valent graph. In order to make the comparison with
the 4- and 6-valent graphs as accurate as possible, we will rotate
the 8-valent graph by the same amount the other valence
graphs were rotated. It follows that the angles of the edges incident at $V_0$
will satisfy the following conditions:
\begin{enumerate}
\item[1)] $0<\theta_{0,2},\theta_{e_{0,8}}<\frac{\pi}{2}$, $\frac{3\pi}{2}<\theta_{0,3},\theta_{0,7}<2\pi$ $\frac{\pi}{2}<\theta_{0,1},\theta_{0,6}<\pi$ and $\pi<\theta_{0,4},\theta_{0,5}<\frac{5\pi}{4}$.
\item[2)]$\frac{3\pi}{2}<\phi_{e_{0,2}},\phi_{e_{0,3}}<2\pi-\sin^{-1}\frac{1}{3}$, $\pi+\sin^{-1}(\frac{1}{3})<\phi_{e_{0,7}},\phi_{e_{0,8}}<\frac{3\pi}{2}$,$\frac{3\pi}{2}<\phi_{e_{0,6}},\phi_{e_{0,5}}<\pi-\sin^{-1}(\frac{1}{3})$ and $\sin^{-1}(\frac{1}{3})<\phi_{e_{0,4}}, \phi_{e_{0,1}}<\frac{\pi}{2}$
\end{enumerate}

The angles for the co-linear edges are defined through the formula
$\theta_{e}=180^{\circ}-\theta_{e_{\text{collinear}}}$ and
$\phi_{e}=180^{\circ}-\phi_{e_{\text{collinear}}}$, respectively.
It follows that the edges $e_{0,1}$, $e_{0,2}$, $e_{0,3}$,
$e_{0,4}$, $e_{0,5}$, $e_{0,6}$, $e_{0,7}$ and $e_{0,8}$ lie in
the octants $B$, $A$, $D$, $C$, $G$, $F$, $H$ and $E$,
respectively. The coordinates of the rotated vertices are
\begin{eqnarray}
V^{\prime}_1&=&\Big((-R_{11}-R_{12}+R_{13})\frac{\delta}
{\sqrt{3}} , (-R_{21}-R_{22}+R_{23})\frac{\delta} {\sqrt{3}},
(-R_{31}-R_{32}+R_{33})\frac{\delta} {\sqrt{3}}\Big)\\
V^{\prime}_9&=&\Big(-2R_{12}\frac{\delta} {\sqrt{3}},
-2R_{22}\frac{\delta} {\sqrt{3}}, -2R_{32}\frac{\delta} {\sqrt{3}}\Big)\\
V^{\prime}_{12}&=&\Big((-R_{12}+2R_{13})\frac{\delta} {\sqrt{3}},
(-R_{22}+2R_{23})\frac{\delta} {\sqrt{3}},
(-2R_{32}+2R_{33})\frac{\delta} {\sqrt{3}}\Big)\\
V^{\prime}_{e}&=&\Big((-R_{11}-R_{12}+3R_{13})\frac{\delta}
{\sqrt{3}}, (-R_{21}-R_{22}+3R_{23})\frac{\delta} {\sqrt{3}},
(-R_{31}-R_{32}+3R_{33})\frac{\delta} {\sqrt{3}}\Big)
\end{eqnarray}

As it was done for the 4- and 6-valent graphs, in order to carry out
the calculations for the expectation value of the volume operator,
we would have to specify a particular combinations of angles
satisfying conditions 1) and 2) above. However, all combinations
satisfying 1) and 2) above lead to the same value, in zeroth-order
in $\frac{l}{\delta}$ of the expectation value of the volume
operator. Rotational dependence will only appear for higher
orders in $\frac{l}{\delta}$. Moreover any sub-case of 1) and 2)
will lead to the same couples of edges commonly intersecting
surfaces $s_{\alpha,t}^I$ in a given stack. Therefore, to leave
the result as general as possible, we will not specify a
particular sub-case of 1) and 2), but simply derive a general
expression for the expectation value of the volume operator given
conditions 1)and 2).

The expectation value of the volume operator for a periodicity
region, $R$, is then computed as
\begin{equation}
\delta^3\sqrt{|\det\big(E^a_j(u)\big)|}\;4\bigg(5+\frac{1}{2}\Big(-\frac{1}{32}\sum^{\frac{n}{2}(n-1)}_{i,j=1;
j\neq i }\alpha_{ji}^2\Big)\,\Big|\det\Big(\frac{\delta
X_S^a}{\delta(s,u^1,u^2)}\Big)\Big|\bigg)
\end{equation}
where the terms $\alpha_{ij}$ are the off-diagonal entries of the
matrix $\sqrt{A}^{-1}$ and the geometric factor is $G_{\gamma}=4$. 
Evidently, the higher-order corrections are angle dependent, while
the zeroth-ones are not. Therefore, as for the 4- and 6-valent
case, the expectation value of the volume operator for the
8-valent graph is rotational invariant, in zeroth-order up to
measure \emph{zero} in $SO(3)$. However, it does \emph{not} reproduce the correct semiclassical limit.

\subsubsection{Expectation value of the volume operator for a
translated 8-valent graph} \label{s7.4.3}

We now consider the translated 8-valent graph. As for the  4- and
6-valent graphs we  choose the following conditions on the
components of the translation vector:
\begin{enumerate}
\item[1)]$b>\epsilon_x>\epsilon_y>\epsilon_z>0$
\item[2)]$|V^i_k|-|V_k^j|>n^i_kl-n^j_kl\mbox{ for all } |V_k^i|>|V_k^j|$
\end{enumerate}
Similarly, as for the aligned 8-valent graph we have
\begin{equation}
T^{\{x,y\}}_{e_i}=F^{\{x,y\}}_{e_i}=Z^{\{x,y\}}_{e_i}=1\hspace{.5in}
\forall  e_i\in\gamma
\end{equation}

The coordinates of the translated vertices are
\begin{eqnarray}
V^{'}_0&=&(\epsilon_x,\epsilon_y,\epsilon_z)\\
V^{'}_1&=&\Big(-\frac{\delta} {\sqrt{3}}+\epsilon_x,-\frac{\delta} {\sqrt{3}}+\epsilon_y,\frac{\delta} {\sqrt{3}}+\epsilon_z\Big)\\
V^{'}_9&=&\Big(\epsilon_x,-2\frac{\delta} {\sqrt{3}}+\epsilon_y,\frac{\delta} {\sqrt{3}}+\epsilon_z\Big)\\
V^{'}_{12}&=&\Big(\epsilon_x,-\frac{\delta} {\sqrt{3}}+\epsilon_y,2\frac{\delta} {\sqrt{3}}+\epsilon_z\Big)\\
V^{'}_e&=&\Big(-\frac{\delta} {\sqrt{3}}+\epsilon_x,-\frac{\delta}
{\sqrt{3}}+\epsilon_y,3\frac{\delta} {\sqrt{3}}+\epsilon_z\Big)
\end{eqnarray}
\\

The value obtained for the volume of a region $R$ is, to first-order in $\frac{l}{\delta}$:
\begin{align}
&V_R=4\delta^3\sqrt{|det\Big(E^a_j(u)\Big)|} \Big|det\Big(\frac{\delta X_S^a}{\delta(s,u^1,u^2)}\Big)\Big|\times \Bigg\{ \sqrt{1+\frac{1}{32}\Big( - 4 A_0^2 - 12 B_0^2  - \frac{45 C_0^2}{2}  - 3 D_0^2 - E_0^2 -  F_0^2\Big)}\nonumber\\
&+ \sqrt{1+\frac{1}{32}\Big(-16 A_1^2 -  24 B_1^2 - D_1^2 -   -
E_1^2   - 3 F_1^2 \Big)}+\sqrt{1+\frac{1}{32}\Big( - 24 B_9^2
- 8 C_9^2  - \frac{9 D_9^2}{2}  - \frac{3 E_9^2}{2} - F_9^2 \Big)} \nonumber\\
&+ \sqrt{1\frac{1}{32}\Big(-24 A_{12}^2  - 12 B_{12}^2  - 4
C_{12}^2   - D_{12}^2  - E_{12}^2
 - 3 F_{12}^2 \Big)}\nonumber\\
&+ \sqrt{1+\frac{1}{32}\Big( - \frac{29 A_e^2}{2} - \frac{43 B_e^2}{2} - 2 C_e^2  - D_e^2  - \frac{3 E_e^2}{2}
- 3 F_e^2\Big)}\Bigg\}\nonumber\\
&=4\delta^3
\sqrt{|det\Big(E^a_j(u)\Big)|} \Big|det\Big(\frac{\delta X_S^a}{\delta(s,u^1,u^2)}\Big)\Big|\times\Bigg\{ 5+\frac{1}{2}\times\frac{1}{32}\Big( - 4 A_0^2 - 12 B_0^2  - \frac{45 C_0^2}{2}  - 3 D_0^2 - E_0^2 -  F_0^2\nonumber\\
&-16 A_1^2 -  24 B_1^2 - D_1^2 -   - E_1^2   - 3 F_1^2  - 24 B_9^2
- 8 C_9^2  - \frac{9 D_9^2}{2}  - \frac{3 E_9^2}{2} - F_9^2 -24 A_{12}^2  - 12 B_{12}^2 \nonumber\\
&  - 4 C_{12}^2   - D_{12}^2  -
E_{12}^2
 - 3 F_{12}^2- \frac{29 A_e^2}{2} - \frac{43 B_e^2}{2} - 2 C_e^2  - D_e^2  - \frac{3 E_e^2}{2}
- 3 F_e^2 \Big)\Bigg\}
\end{align}

where $A_i$, $B_i$, $C_i$ $D_i$, $E_i$ and $F_i$ are the matrix
elements of $\sqrt{A}^{-1}$. 
Again, translational invariance holds only at
zeroth-order up to measure \emph{zero} in $SO(3)$. However, it does \emph{not} reproduce the correct semiclassical limit.
\subsection{Discussion}
We have shown that if we use semiclassical states derived from the
area complexifier, then we do \emph{not} obtain the correct semiclassical
value of the volume operator, unless we perform an artificial
re-scaling of the coherent state label and we restrict our
calculation to the following special cases:
\begin{enumerate}
\item[1)] The edges of the graph are aligned with the orientation of the plaquettes (6-valent graph).
\item[2)] Two or more edges lie in a given plaquette (4-valent graph).
\item[3)] One edge is aligned with a given plaquette while a second edge lies in a given plaque (4-valent graph).
\end{enumerate}
However, such combination of edges have measure zero in $SO(3$).
For embeddings, whose measure in $SO(3)$ is non-trivial, we do not
obtain the correct semiclassical behaviour for the volume operator
for any valence of the graph.

This result suggests, strongly, that the area complexifier coherent
states are not the correct states with which to analyse
semiclassical properties in LQG. Moreover, as previously
mentioned, if embedding independence (staircase problem) is to be
eliminated, area complexifier coherent states should be ruled out
as semiclassical states altogether.
\chapter{Spin Foam}
In this Chapter we will introduce spin foam models. Essentially, a spin foam model represents a Lagrangian formulation of LQG given in terms of a covariant sum-over-histories formulation. The development of a Lagrangian formulation of LQG was motivated by the fact that in the Hamiltonian formulation of LQG given in Chapter \ref{cha:qunt}, it is very complicate to compute transition amplitudes. In fact spin foam models were born as a way of defining transition amplitudes in the context of LQG, but from a different prospective, namely as a sum-over-histories. \\
In particular, in Chapter \ref{cha:qunt}, we have shown that space is represented by spin networks. A spin foam is a time evolution of such spin networks, thus representing spacetime. Another way of defining a spin foam is as a world sheet of a spin network. However, one should keep in mind that a spin foam is purely a combinatorial object and does not `live' in a background, representing itself spacetime.\\

In the following we will give a precise definition of what a spin foam is and how it is constructed. We will then give concrete examples for 3- and 4-dimensions.
\section{Spin Foams}
As mentioned in previous sections, the starting point of LQG is classical general relativity (GR)
reformulated as an Hamiltonian theory with constraints.  This
structure  can be canonically quantised systematically,
so that the constraint equations are promoted to quantum
constraint operators, defined on a kinematical Hilbert space,
$\mathcal{H}_{kin}$. The dynamics of the theory is governed by the
Hamiltonian constraint $H$, whose solutions define the physical
Hilbert space, $\mathcal{H}_{phy}$.

There are, however, two central problems in this approach: (i) extracting concrete
solutions for the Hamiltonian constraint; and (ii) defining an
inner product on $\mathcal{H}_{phy}$ (see however, DID (MCP) in Section \ref{s outline} and \cite{v5}) .\\
An important approach to
both these problems is given by the theory of spin foam models
\cite{perez1},\cite{48f}.

In particular, spin foam theory is supposed to provide the dynamical aspects of
LQG and can be used as a tool for computing ``transition amplitudes" in a possible theory of quantum-gravity, more precisely, spin foam models are an
attempt to provide a path-integral formulation of LQG.

At each time step, in LQG, a quantum state of geometry is
represented  by a graph labelled by spin quantum numbers which
carry information about the geometry of the space. Such a graph is
called a {\sl spin network}. A spin foam can be interpreted as a
history of such spin networks.\\ Therefore, generally, a spin foam represents a possible history of the gravitational field and can be seen as a set of possible different transitions through different quantum states of space (states of 3-geometry as defined by LQG). However, care is needed when interpreting such transition amplitude, since LQG is a covariant theory in which there is no notion of time, thus transition amplitudes can only be interpreted as defining physical inner products.

Specifically we recall that LQG is a canonical quantisation of GR written in the Hamiltonian formalism. In such a formalism, the presence of gauge symmetries  (which for GR is diffeomorphism invariance) give rise to constraints on the phase space variables, such that the allowed states of the theory are constrained to lie in the constrained hypersurface. Moreover, the Poisson brackets, with respect to such constraints, give rise to gauge transformations on the constrained hypersurface. As a consequence, the reduced phase space, which represents the set of all physical states, is isomorphic to the space of orbits, such that, any two points on the same gauge orbit represent the same state.\\
In chapter \ref{cha:hgr} it was shown that the Hamiltonian in GR is nothing but a linear combination of constraints, thus time evolution is a pure gauge transformation. Therefore, given a constrained hypersurface, a spacetime can be formed by considering, as the time component, a one parameter family of gauge transformations (In ADM those would correspond to a choice of shift vector and the lapse function). It is  precisely such a notion of time that forces us to interpret path integrals as physical inner products.

\begin{figure}[htb]
 \begin{center}
\psfrag{H}{$\hat{H}$}\psfrag{i}{$\rho_{e_i}$}\psfrag{j}{$\rho_{e_j}$}\psfrag{k}{$\rho_{e_k}$}\psfrag{n}{$\rho_{e_n}$}\psfrag{m}{$\rho_{e_m}$}\psfrag{l}{$\rho_{e_l}$}\psfrag{x}{$i_{e_i,e_j,e_k}$}\psfrag{z}{$i_{e_i,e_l,e_n}$}\psfrag{o}{$\rho^{'}_{f_i}$}\psfrag{p}{$\rho^{'}_{f_j}$}\psfrag{q}{$\rho^{'}_{f_k}$}\psfrag{r}{$\rho^{'}_{f_l}$}\psfrag{s}{$\rho^{'}_{f_m}$}\psfrag{t}{$\rho^{'}_{f_n}$}\psfrag{y}{$i^{'}_{f_i,f_l,f_n}$}\psfrag{u}{$i^{'}_{f_i,f_l,f_j}$}
 \includegraphics[scale=0.8]{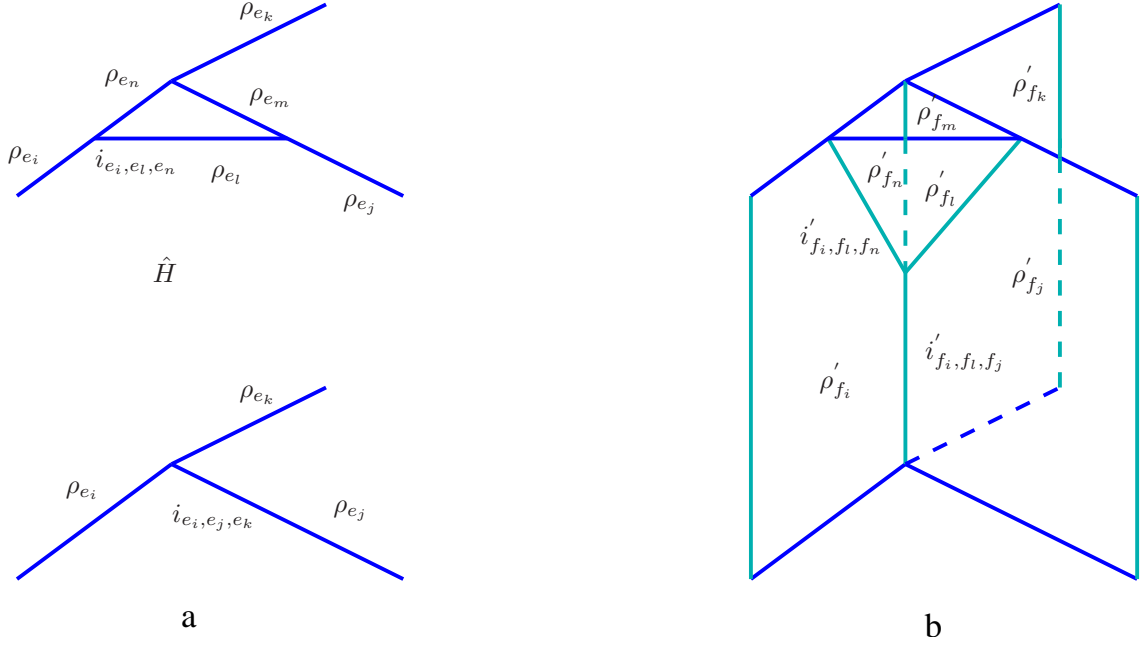}
\caption{Figure a) represents the action of the Hamiltonian constraint on a spin network, while figure b) shows the corresponding spin foam.  \label{fig:spin foam} }
\end{center}
  \end{figure}

Let us try to understand the conceptual motivation behind defining such an inner product in terms of path integrals. We know that in LQG the dynamics is governed by the Hamiltonian constraint which, however, is very difficult to solve since it changes the graphs/states to which you apply it. This implies that the physical Hilbert space is not known explicitly. This corresponds to the situation in classical GR where only few exact solutions are known. However, generally speaking, the physical Hilbert space is associated with the kernel of the constraints, therefore it can be defined through the projection 
\begin{equation}
P:\mathcal{H}_{kin}\rightarrow\mathcal{H}_{phys}
\end{equation} 
 defined as  \begin{equation}
P:=\sum_{x\in\sigma}\partial(\hat{H}^{\dagger}(x))=\int \mathcal{D}(N) e^{i\int_{\sigma}d^3x N(x)\hat{H}(x)}
\end{equation}
where $N(x)$ is the laps function. Therefore, the physical inner product can be heuristically defined as 
\be
\langle T_{[s]}, T_{[s^{'}]}\rangle_{phy}:=\langle T_{[s]}, \hat{P}T_{[s^{'}]}\rangle_{kin}
\ee
The idea is then to somehow construct a path integral for the amplitudes defined with respect to the operator $\hat{P}$, i.e. we want to give meaning, in the context of LQG, to the following heuristic expression: 
\begin{equation}\label{equ:tra}
\langle T_{[s]},\hat{P} T_{[s^{\prime}]}\rangle_{kin}=\int \mathcal{D}(N)e^{i\mathcal{S}(N)}:=\sum_n \frac{(i)^n}{n!}T_{[s^{\prime}]}([\hat{H}]^n T_{[s]})
\end{equation}
where $\mathcal{S}(N)=\int_{\sigma}d^3x N(x)\hat{H}(x)$.\\
The term $[\hat{H}]^n$ corresponds to a discrete $n$-step evolution from the initial spin network $T_{[s]}$ to the final $T_{[s^{\prime}]}$.
Figure \ref{fig:spin foam} describes a spin network evolution for $n=1$.\\
Such a history of spin network is precisely what a \emph{spin foam} is. 
\\ In order to give a precise definition of a spin foam, we first recall the definition of a  spin network:
\begin{definition}
 A spin network $\Psi$ is defined to be a triple $(\gamma, \rho,i)$ where:
\begin{enumerate}
\item [1)]  $\gamma$ is a 1-dimensional oriented complex (a graph).
\item [2)] $\rho$ is a labelling of each edge $e$ of $\gamma$ by an irreducible representation $\rho_e$ of G.
\item[3)] $i$ is a labelling of each vertex $v$ of $\gamma$  by an intertwiner such that, given a set of incoming edges $(e_1,e_2\cdots e_n)$ and outgoing edges $(e^{'}_1,e^{'}_2\cdots e^{'}_n)$ at $v$ we have
\be
i_v:\rho_{e_1}\otimes \rho_{e_2}\cdots \otimes \rho_{e_n}\rightarrow \rho_{e^{'}_1}\otimes \rho_{e^{'}_2}\cdots \otimes \rho_{e^{'}_n}
\ee
\end{enumerate}
\end{definition}
A spin foam of the form $F:\emptyset\rightarrow\Psi$ is then defined to be:
\begin{definition}
Given a spin network $\Psi= (\gamma,\rho, i)$, a spin foam $F:\emptyset\rightarrow\Psi$ is defined to be a 
triple $(k, \rho^{'},i^{'})$ where:
\begin{enumerate}
\item $k$, is a 2-dimensional oriented complex whose border is $\gamma$.
\item $\rho^{'}$, is a labelling of each face $f \in k$ by an irreducible representation $\rho^{'}_f$ of $ G$, such that for any edge $e\in k$, $\rho^{'}_f = \rho_e$ if $f$ is incoming\footnote{Given a face $f\in k$ and an edge $e_\in k$ then we have two possible relations: i) $b(e_i(f))=v_i(f)$ and $f(e_i(f))=v_{i+1}(f)$ in which case we say that the face $f$ is incoming with respect to the edge $e_i$, ii) $f(e_i(f))=v_i(f)$ and $b(e_i(f))=v_{i+1}(f)$ in which case we say that the face $f$ is incoming with respect to the edge $e_i$. Here $v_k(f)$ represents the kth vertex of the face $f$. } to $e$, while $\rho^{'}_f = (\rho_e)^*$ if $f$ is
outgoing to $e$.
\item $i^{'}$, is a labelling of each edge $e\in k$ not lying in $\gamma$ by an intertwiner
\be
i^{'}_e:\rho^{'}_{f_1}\otimes \rho^{'}_{f_2}\cdots \otimes \rho^{'}_{f_n}\rightarrow \rho^{'}_{f^{'}_1}\otimes \rho^{'}_{f^{'}_2}\cdots \otimes \rho^{'}_{f^{'}_n}
\ee
where $f_1,f_2\cdots f_n$ are the faces incoming to $e$, while $f^{'}_1,f^{'}_2\cdots f^{'}_n$
are the faces outgoing
from $e$. Each intertwiner $i^{'}$ is such that, for any vertex $v \in k$, $i^{'}_e=i_e$ after appropriate dualizations.
\end{enumerate}
\end{definition}
The relation between the underlying 1-dimensional oriented complex $\gamma$ of spin networks and the underlying 2-dimensional complex $k$ underlying the respective spin foam model can be better understood thourough the notion of affine maps. Specifically, given any 1-dimensional oriented complex $\gamma$ and a 2-dimensional oriented complex $k$, it is possible to construct a 2-dimensional complex from $\gamma$ via the product $\gamma\times [0,1]$. We then say that $\gamma$ borders $k$ iff there exists a 1:2:1 affine map $c:|\gamma|\times [0,1]\rightarrow k$ mapping each cell in $|\gamma|\times [0,1]$ to a unique cell in $k$, in such a way that the orientation is preserved. In this way each $n$-cell of $\gamma$ is seen as a face of a unique $(n+1)$-cell in $k$, therefore, each vertex $v\in \gamma$ is the source or target of a unique edge in $k$; each edge in $\gamma$ is the edge of a unique face in $k$ and so on.

\begin{figure}[htb]
 \begin{center}
\psfrag{F}{$F$}\psfrag{G}{$F^{'}$}
  \includegraphics[scale=0.7]{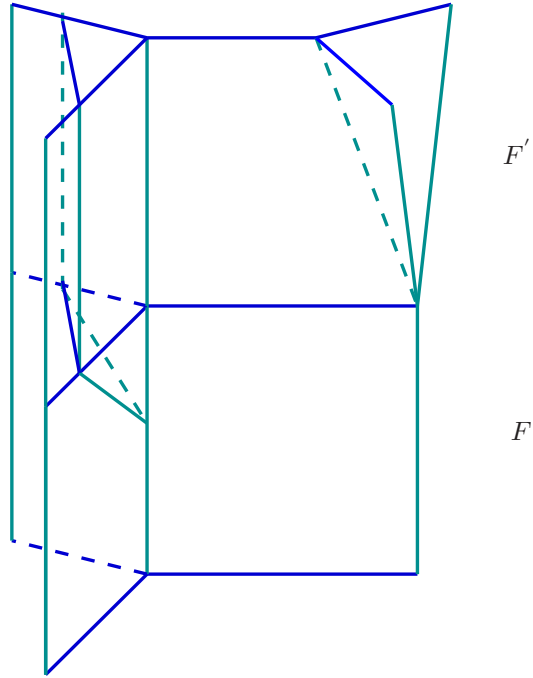}
\caption{Composition of two spin foams $F$ and $F^{'}$.\label{fig:sftransition} }
\end{center}
  \end{figure}

Alternatively, it is possible to define a spin foam as follows:
\begin{definition}
Given two spin networks $\Psi=(\gamma,\rho,i)$ and $\Psi^{\prime}=(\gamma^{\prime},\rho^{\prime},i^{\prime})$, the spin foam $F:\Psi\rightarrow\Psi^{\prime}$ is identified with the spin foam $F:\emptyset\rightarrow\Psi^*\otimes\Psi^{\prime}$ (where $\Psi^*\otimes\Psi^{\prime}$ has, as underlying spin network, the disjoint union of $\gamma\cup\gamma^{\prime}$ with the respective labellings $\rho^{\prime},\rho,i,i^{\prime}$). \\
$F:\emptyset\rightarrow\Psi^*\otimes\Psi^{\prime}$ is defined as the triple $(k,\tilde{\rho},\tilde{i})$ where:
\begin{enumerate}
\item $k$, is a 2 dimensional oriented complex which is bounded by the disjoint union of $\gamma\cup\gamma^{\prime}$.
\item $\rho^{'}$, is a labelling of each face $f\in k$ of irreducible representations $\rho^{'}_f$ of G.
\item $i^{'}$, is a labelling of each of the edges $e^{'}\in k$ not lying in the disjoint union of $\gamma\cup\gamma^{\prime}$ with intertwiners of the form
\begin{equation}
i^{'}_{e^{'}}:\rho^{'}_{f_1}\otimes\cdots \rho^{'}_{f_n}\rightarrow\rho^{'}_{f^{\prime}_1}\otimes\cdots\rho^{'}_{f^{\prime}_n}
\end{equation}
where the f and the $f^{\prime}$ represent, respectively, ingoing and outgoing faces to the edge $e^{'}$.
\end{enumerate}
Both the representation and intertwiner labelling have to satisfy certain compatibility conditions with the 1-complex $\gamma\cup\gamma^{\prime}=\beta$, namely:
\begin{enumerate}
\item The representations $\rho^{'}_f$, associated to faces f, which have as an edge $e$ of the 1-complex $\beta$ must be such that $\rho^{'}_f={\rho}_e$ if f is incoming to e and $\rho^{'}_f={\rho}_e^*$ (dual representation) if $f$ is outgoing to $e$.
\item For any vertex $v\in\beta$ : $i^{'}_e=i_e$ after appropriate dualization.
\end{enumerate}
\end{definition}
It is also possible to define equivalence classes of non-degenerate\footnote{A spin foam is said to be non degenerate if every vertex is the end of at least one edge, every edge of at least one face and every face is labelled by an irreducible representation of $G$. } spin foams, where two spin foams $F$ and $F^{'}$ are considered equivalent if one can be obtained from the other by a sequence of the following moves and their inverses: 
\begin{enumerate}
\item [i)] \emph{Affine transformation}:  $F^{'}$ is obtained from $F$ by affine transformation iff:
\textbf{a)} there is a one-to-one affine map $\phi$ which maps cells in $k$ to cells in $k^{'}$, in such a way that orientation is preserved, \textbf{b)} for each face $f\in k$ then $\rho_f=\rho^{'}_{\phi(f)}$,  \textbf{c)} for all $e\in k$ $i_e=i^{'}_{\phi(e)}$. 
\item [ii)]\emph{Subdivision}: $F^{'}$ is obtained from $F$ by subdivision iff: \textbf{a)} the oriented 2 complex $k^{'}$ is obtained by a subdivision of the oriented two complex $k$, \textbf{b)} if a face $f^{'}\in k^{'}$ is contained in a face $f\in k$, then $\rho^{'}_{f^{'}}=\rho_f$ \textbf{c)} if $e^{'}\in k^{'}$ is contained in an edge $e\in k$, then $i_e=i^{'}_{e^{'}}$ \textbf{d)} if $e^{'}\in k^{'}$ is shared by two faces in $ k^{'}$, both contained in the same face $f$ of $k$, then $i^{'}_{e^{'}}=1_{\rho(f)}$.
\item [iii)] \emph{Orientation reversal}. $F^{'}$ is obtained from $F$ by orientation reversal iff : \textbf{a)} $k$ and $k^{'}$ have the same cells but with (possibly) different orientations; \textbf{b)} if $f\in k$, then 
\be
\rho^{'}_f=\begin{cases} \rho_f&\text{ if $k$ and $k^{'}$ have the same orientation} \\
(\rho_f)^*&\text{ if $k$ and $k^{'}$ have opposite orientation}
\end{cases}
\ee
\textbf{c)} for all $e\in k$, $i_e=i^{'}$ after appropriate dualization.
\end{enumerate}
It is also possible to compose (equivalence classes of ) spin foams as follows:\\
given two spin foams $F:\Psi\rightarrow \Psi^{'}$ and $F^{'}:\Psi^{'}\rightarrow \Psi^{''}$, if we choose a representative of both $F$ and $F^{'}$ living in some space $\Rl^n$, such that the copy of the spin net $\Psi^{'}=(\gamma^{'}, \rho^{'}, i^{'})$ is the same for both, then the affine maps $c, c^{'}:\gamma^{'}\times [0,1]\rightarrow \Rl^n$ can be composed to a single map $f:\gamma^{'}\times [1,1]\rightarrow \Rl^n$. The composite spin foam $FF^{'}$ is defined to be such that, the underlying complex is the union of the underlying complexes of $F$ and $F^{'}$. All subcomplexes inherit the labellings from $F$ and $F^{'}$, except for the edges in $\gamma^{'}$, which get labelled by the (dualized) identity intertwiner.\\
Such a composition of spin foams is shown in picture \ref{fig:sftransition}.\\
Given the above definition of spin foams it is straightforward to interpret a spin foam as a dual 2-skeleton\footnote{ A dual skeleton of a (n)-dimensional manifold associates an (n-m)-simplex to each (m)-simplex.} of a triangulation of a manifold. Specifically, let us consider a triangulated $n$-dimensional manifold $M$ representing spacetime and a foliation of it given by $(n-1)$-oriented submanifolds $S_i$ representing space. Such submanifolds inherit the triangulation defined on $M$.

A dual 1-skeleton of such submanifolds $S_i$ defines a spin network. In this context a spin foam, which represents a history of a spin network, can be seen as a dual 2-skeleton of the triangulation of $M$, whose boundary is given by the dual 1-skeleton representing the spin network.

We know from LQG that all possible spin networks (all possible triangulations of $S$) represent gauge invariant states in $\mh_{kin}$. It follows that time evolution between states in $\mh_{kin}$ is identified with an operator $Z:\mh_{kin}\rightarrow \mh_{kin}^{'}$. To define such an operator it suffices to define the transition amplitude for one spin network only, since spin networks form a basis for $\mh_{kin}$. The idea is then to write the transition amplitude between two spin networks in terms of the sums of all possible spin foams (all possible triangulations of the manifold) joining the spin networks in question, i.e.
\be\label{equ:z}
Z(\Psi, \Psi^{'})=\sum_{F}Z(F)
\ee
where $F:\Psi\rightarrow \Psi^{'}$ defines a spin foam from $\Psi$ to $\Psi^{'}$. It is precisely in this sense that equation \ref{equ:tra} gets interpreted in the context of spin foams. In particular, the 
\emph{sum-over-paths} formulation of transition amplitudes in QFT gets translated into the \emph{sum-over-spin foams} formulation of transition amplitudes in LQG where, in this case, there is only a fictitious time parameter represented by a foliation of $M$ into space hypersurfaces $S$. However, it should be noted that a spin foam represents a \emph{gauge history} of a spin network, such that the sum in equation \ref{equ:z} is really a sum of gauge histories of the kinematical states. It is precisely such an averaging of gauge orbits, generated by the constraints that allows for a definition of $P$ as an operator which extracts the true degrees of freedom, thus projecting on $\mh_{phy}$. 

The motivation of this interpretation is given by the heuristically definition of $P$  
\be
P:=\int[dN]U(N)
\ee
where $U(N)=e^{i\int_{\sigma}d^3x N(x)\hat{H}(x)}$ is the operator generated by the constraints and, as such, it gets represented in terms of sums over gauge histories. However, this heuristic motivation is not mathematically correct since, due to the presence of the structure function, the set of constraints does not form a group, $U(N)$ is not self adjoint and $\mathcal{D}N$ is not a Haar measure for this group.\\

Interestingly it is possible to give a categorical definition of spin foams, namely we define the category $\mathcal{F}$ of spin foams to have as objects non-degenerate spin networks, and as morphisms spin foams between them. For the associativity and unit laws to hold the following equivalence relations (in addition to the one previously defined) have to be imposed:\\
1) $F(GH)\sim (FG)H $ for any spin foam $F,G,H$; 2) $1_{\Psi}F\sim F\sim F1_{\Psi}$, where $1_{\Psi}:\Psi\rightarrow \Psi$ is the left and right unit for any spin network $\Psi$.

In what follows we will describe a spin foam model in 3 and 4 dimensions. We will then proceed in defining a way to obtain, concretely, a spin foam model through BF-theory.
\section{Spin Foam Model in 3-Dimensions}
\label{ssfm3}
In this Section we will briefly describe a spin foam model for 3-dimensional quantum gravity. The reason for introducing such a model is that, despite its simplicity, nonetheless it sheads light on certain issues present in the case of 4-dimensional quantum gravity. The simplicity of the 3-dimensional case is due to the fact that, in 3-dimensions, GR becomes a topological theory, thus it does not have any local degrees of freedom, only global. As a consequence, such a theory can be easily quantised and a partition function of such a quantised theory can be defined.\\
As we will see, the partition function obtained in 3-dimensional spin foam models turns out to be an invariant of the manifold. This is a consequence of the fact that such models are invariant under changes of the triangulation of the manifold, which preserve the topology.

We will now describe, in detail, how a spin foam model is defined in 3-dimensional quantum gravity. We will only consider the Riemannian case (SU(2)). The Lorentzian case has been carried out in \cite{25}, and it adopts, essentially, the same procedure as the Riemannian one but, in addition, because of the non-compactness of the group (SO(2,1)), a gauge fixing is required to avoid divergences.\\

Our starting point will be the classical action of GR in 3-dimensions
\begin{equation}\label{equ:3d}
S[e,w]=\int_{\mathcal{M}}Tr(e\wedge F(w)) 
\end{equation}
where the tetrad fields $e^i$ and the connections $w$ are $su(2)$ Lie valued 1-forms, $F(w)=dw+w\wedge w=d_w w$ is the curvature, $d$ is the exterior derivative of 1-forms and $d_ww$ is the covariant derivative with respect to the connection $w$. The relation between the tetrad and the metric is as follows:
\be\label{equ:3dq}
g_{\mu,\nu}=\eta_{ij}e^i_{\mu}e^j_{\nu}
\ee 
where $\eta=(+, + ,+)$ since we are considering the Reimannin case. The equations of motion are
\be
d_we=0\hspace{.5in}F(w)=0
\ee
which indicate, respectively, the compatibility between the triad $e$ and the connection $w$ and that the connection should be flat everywhere, i.e. no local excitations are possible.\\
The symmetries of the action \ref{equ:3d} are:\\
i) local Lorenz gauge symmetries
\be
\delta_X^L w=d_wX\hspace{.5in}\delta_X^Le=[e,X]
\ee
for an arbitrary Lie algebra element $X\in\mg$.\\
ii) Translational symmetries
\be
\delta^T_\phi w=0\hspace{.5in}\delta_{\phi}^Te=d_w\phi
\ee
for any $\phi\in\mg$.\\
iii) Diffeomorphisms
\be
\delta_{\varphi}^Dw=d(i_{\varphi}w)+i_{\varphi}(d_w)\hspace{.5in}\delta^D_{\varphi}e=d(i_{\varphi}e)+i_{\varphi}(d_e)
\ee
Being a topological theory, the action \ref{equ:3d} can be interpreted as the action of a 3-dimensional topological BF-theory, where we would replace the tetrad $e^i$ by the Lie algebra valued 1-form $B$ ($B$ field) and, $w$, by the Lie algebra valued connection $A$. It is precisely such a similarity which allows us to apply all the tools for quantisation and definition of the partition function developed for BF-theory to the case at hand. In this respect, the first step in quantising the action \ref{equ:3d} is to perform a discretization of the manifold $M$ through an oriented triangulation $T$. Each of the variables present in the action are, then, associated with an element in the discretization. Moreover, since both $e$ and $w$ are 1-forms we want to associate them to 1-dimensional elements of the triangulation. The tetrad is integrated over the edges of the triangulation, thus we obtain a collection of Lie algebra elements, each associated to an edge $E^i_e=\int_e e^i(x)$.

The connections $w$ are, instead, associated to the edges (dual edges) of the simplicial complex $K^*$ dual\footnote{Given a simplicial complex $S$, its dual simplicial complex $S^*$ is defined by associating to any d-simplex in $S$ a (n-d)-simplex in $S^*$, where $n$ is the dimension of the manifold.} to the triangulation $K$. In particular, the connection gets integrated over the dual edges $e^*$ in $K^*$, thus obtaining holonomies with associated group elements $g_{e^*}$. The curvature $F(w)$ is then associated to the product of all such holonomies around a dual face $f^*$, i.e. $\prod_{e^*\subset\partial f}g_{e^*}=g_{f^*}$. Since each dual face is associated to an edge in the triangulation $T$ we, automatically, associate to each such edge its simplicial curvature. Moreover, by taking the logarithm of $g_{f^*}$ we obtain a Lie algebra element $U_e^i$. Given such a discretization, the action \ref{equ:3d} becomes
\be\label{equ:discretised3d}
S[E^i_e,U^i_e]=\sum_{e\in T}Tr(E^i_e, U^i_e)
\ee 
It can be shown that such an action, similarly as its continuum counterpart, is invariant under both i) Lorentz transformation ii)  discrete translation. However, full diffeomorphic invariance is lost due to the choice of a triangulation \cite{diffandsf}. \\

Now that we have discretized the action we want to quantise the resulting theory. Since we are working with simplicial complexes, in order to obtain a quantisation of such a theory we need to define the quantum analogues of each of the simplices involved. In particular, we need to find a quantum analogue of each of the variables contained in \ref{equ:discretised3d}, in such a way that a quantum state can be associated to each 2-dimensional surfaces (obtained by gluing together a collection of triangles along their common edges) and an amplitude, for each 3-dimensional manifold, is given by a collection of 3-simplices glued along common triangles. Precisely these amplitudes will be utilised to define transition amplitudes between quantum states in terms of path integrals. 

For simplicity, we first consider a single tetrahedron $\tau$. We know from standard geometry that a tetrahedron is uniquely defined in terms of the length (squared) of its 6 edges. Now, the variables to quantise are $E^i_e$ which, being associated to each edge $e^i$ of the tetrahedron, uniquely defines it\footnote{It should be noted that \emph{only} the geometrical information about the tetrahedron is obtained in this way, any other information is lost. However, we are trying to quantise the spacetime geometry, thus for our purpose such information suffices.}. The quantisation of an SU(2) Lie algebra element (such as $E^i_e$) is done by choosing a representation $j$ of SU(2) and associating the (lie algebra) element to an operator in the representation space $V^j$. \\
In particular, if we were to choose to associate to each edge an element $J^j_{e_i}$ of the basis of SU(2) in a given representation $j$, then, the operator associated to the edge length (squared) is the Casimir operator $C=J_{e_i}^j\cdot J_{e_k}^j$, which is diagonal on the representation space with eigenvalues given by $j_{e_i}(j_{e_i}+1)$. 

In such a way, for each representation $j$ we assign to an edge, we obtain the corresponding length $j_{e_i}(j_{e_i}+1)$ of that edge and the corresponding Hilbert space $V^j$. As a consequence we can identify the Hilbert space associated to an edge as the sum of the Hilbert space obtained by assigning different representations to that edge, i.e. $\oplus _{j_e}V^j_e=\mh_e$. Therefore, the quantisation procedure allows us to associate to each edge a Hilbert space $\mh_e$ with associated Casimir operators.

The next step it to construct the quantum state associated to a triangle. Each triangle can be uniquely specified by its three vectors provided that i) the \emph{closure constraint} holds, i.e. $E^i_{e_l}+E^i_{e_j}+E^i_{e_k}=0$ and ii) the Riemannian triangle inequalities hold.

For simplicity let us choose a specific assignment of representations to each edge $e$, then, from the discussion above, to each triangle we assign the Hilbert space comprised of the three Hilbert spaces associated to the edges of the triangle, namely, $V^{j_{e_1}, j_{e_2}, j_{e_3}}= V^{j_{e_1}}\otimes V^{j_{e_2}}\otimes V^{j_{e_3}}$. However, because of the closure constraint, the correct Hilbert space should be the space of invariant tensors $inv(V^{j_{e_1}}\otimes V^{j_{e_2}}\otimes V^{j_{e_3}})$, such that $\psi\in inv(V^{j_{e_1}}\otimes V^{j_{e_2}}\otimes V^{j_{e_3}})$ is $\psi:inv(V^{j_{e_1}}\otimes V^{j_{e_2}}\otimes V^{j_{e_3}})\rightarrow \Cl$.\\
Moreover, it can be shown that, by taking in consideration the quantum analogues of the Riemannian triangle inequalities, then the  quantum states associated to triangle is, up to a constant factor, uniquely determined by the edges of the triangle.
Specifically, for a given assignment of representations to edges, such a state is identified with the 3j-symbol, i.e.
\be
\psi=C_{m_1m_2m_3}^{j_1,j_2j_3}=\left( \begin{array}{ccc}
j_1 & j_2 & j_3 \\
m_1 & m_2 & m_3 \end{array} \right)
\ee
 If we now consider any possible assignment of representations to edges, then the total Hilbert space associated to each triangle is
\be
\mh_{e_i,e_l,e_k}=\oplus_{j_{e_i}j_{e_l}j_{e_k}}inv(V^{j_{e_i}}\otimes V^{j_{e_l}}\otimes V^{j_{e_k}})
\ee
To obtain the state associated to a general 2-dimensional face, we need to glue, along common edges,  each state $\psi$ coming from the individual triangles comprising the surface. Since the states are tensor products, the joining is done through the contraction of common indices (common edges).

At the beginning we have said that the building blocks for transition amplitudes between two states $\psi$ and $\psi^{'}$, as constructed above, are given from the amplitudes associated to single tetrahedrons. In topological field theories, for each representation $j$, such amplitudes are generally given by a map
\be
\otimes_i\psi_i=\otimes_i inv(V^{j_{e_i}}\otimes V^{j_{e_l}}\otimes V^{j_{e_k}})\rightarrow \Cl
\ee
The simplest map compatible with all the requirements so far encountered is given by the 6j-symbol, which is obtained by fully contracting the four 3j symbols associated to each of the triangles comprising the tetrahedron, i.e.
\be
\{6j\}=\left[ \begin{array}{ccc}
j_1 & j_2 & j_3 \\
j_4 & j_5 & j_6 \end{array} \right]
\ee
By allowing the edge length to vary, i.e. by considering all possible assignments of representations to edges, the amplitude associated to a tetrahedron becomes
\be
\Big(\prod_i\sum_{j_i}\Delta_{j_i}\Big)\{6j\}
\ee
where $\Delta_{j_i}=2j_i+1$ is the dimension of the representation $j$. For a general transition amplitude one has, then, to compute the product of each 6j-symbol coming from each tetrahedron and sum over representations for all the edges involved.

Up to now we have defined the quantised version of simplicial 3-geometry in such a way that, quantum states are associated to collection of triangles and amplitudes to collections of tetrahedrons.\\
We now want to apply this discretization method for defining the partition function for the action in \ref{equ:3d}. This will lead to the so called Ponzano-Regge model for 3d gravity.\\
At the continuum level the partition function is
\be
Z=\int\md e\md w e^{i\int_{M}tr(e\wedge f(w))}
\ee
which, upon the discretization outlined above becomes
\be
Z(K)=\int_{g_{e}}\prod_{e\in E}dE_e\int_{G^{E^*}}\prod_{e^*\in E^*}dg_{e^*}e^{i\sum_{e\in K}tr(E_e^iU_e^i)}
\ee
By performing the integral over the $E_e^i $ variable, we obtain        
\be
Z(K)=\int_{G^{E^*}}\prod_{e^*\in E^*}dg_{e^*}\prod_{f^*}\delta(g_{f^*})
\ee
This is merely an imposition of the flatness constraint on the connection. \\
Applying Plancherel formula $\delta(g_{f*})=\sum_{j_{f*}}\Delta_{j_{f*}}\chi^{j_{f*}}(g_{f*})$ and utilising the 1:2:1 correspondence between dual edges and faces described above, we obtain
\ba
Z(K)&=&\int_{G^{E_*}}\prod_{e^*\in E^*}dg_{e^*}\prod_{f^*}\sum_{j_{F^*}}\Delta_{j_{f^*}}\chi^{j_{f^*}}(g_{f^*})\nonumber\\
&=&\Big(\prod_{f^*}\sum_{j_{F^*}}\Big)\Big(\prod_{e^*}\sum_{SU(2)}dg_{e^*}\Big)\prod_{f^*}\chi^{j_{f^*}}(\prod_{e^*\in\partial f^*}g_{e^*})
\ea
We then expand the character functions as $\chi^{j}(\prod g)=\sum_m\prod D^j_{mm^{'}}(g)$ (Wigner formula).\\
Moreover, we notice that each dual edge is shared by three dual faces, therefore
\be\prod_{e^*}\int_{SU(2)}dg_{e*}\prod_{f*}\sum_m\prod_{e^*\in\partial f*} D^j_{mm^{'}}(g_{e^*})\xrightarrow{\text{becomes}}\prod_{e^*}\int_{SU(2)}dg_{e*}D^{j_{f_i^*e^*}}_{m_im^{'}_i}(g_{e^*}))D^{j_{f_k^*e^*}}_{m_km^{'}_k}(g_{e^*}))D^{j_{f_l^*e^*}}_{m_lm^{'}_l}(g_{e^*}))\ee
i.e. the integral of three representation functions with the same argument for each dual edge. Utilising the formula
\be
\int_{SU(2)}dg_{e*}D^{j_{f_i^*e^*}}_{m_im^{'}_i}(g_{e^*}))D^{j_{f_k^*e^*}}_{m_km^{'}_k}(g_{e^*}))D^{j_{f_l^*e^*}}_{m_lm^{'}_l}(g_{e^*}))=C_{m_im_km_l}^{j_ij_kj_l}C_{m^{'}_im^{'}_km^{'}_l}^{j_ij_kj_l}
\ee
we see that for each dual edge we associate two 3j-symbols or, alternatively, for each triangle we associate two 3j-symbols. The indexes $m $ and $m^{'}$ represent the two dual vertices incident at each dual edge or, equivalently, the two tetrahedron which share the common face. By contracting the indices that refer to the same tetrahedrons (or dual vertices) we can write the partition function as follows:
\be\label{equ:transamp}
Z(K)=Z(K^*)=\Big(\prod_{f^*}\sum_{j_{f^*}}\Big)\prod_{f^*}\Delta_{j_{f^*}}\prod_{v^*}(-1)^{c(j)}\left[ \begin{array}{ccc}
j_1 & j_2 & j_3 \\
j_4 & j_5 & j_6 \end{array} \right]_{v^*}
\ee
where $c(j)$ is a linear combination of the representations in the 6j-symbol for each vertex.

After defining an appropriate regularisation \cite{diffandsf} the resulting expression is the Ponzano-Regge spin foam model for 3d-gravity. In \cite{65d}, \cite{122d} it was shown that the asymptotic behaviour of the 6j symbol reproduces the discretized Regge
action for 3d gravity, i.e. the classical limit of the model is correct.\\

The fact that such a model is indeed a spin foam model as described in the previous Sections, comes from the fact that the dual 2-complex of the triangulation $K$, which has edges labelled by intertwiners and faces labelled by representation, can be seen as the underlying 2-complex of a spin foam. \\
In fact, let us now consider the boundary of the triangulation of the manifold which consists of triangles, labelled by intertwiners and edges labelled by representations. The dual of such a boundary is a 2-complex, whose (dual) edges are labelled by representations and (dual) vertices are labelled by intertwiners. This is precisely what a spin network is, whose underlying graph is the graph dual to the boundary of the triangulation.
Such spin networks are the kinematical states in 3-dimensional LQG but with the restriction of the valence being only three. Therefore, the kinematical states in the Ponzano-Regge model correspond to the kinematical states in 3d LQG. For any pair of such spin network, the dual 2 complex joining them represents the history of those spin networks, i.e. a spin foam.
It is in this context that transition amplitudes between spin networks are identified with the partition function in \ref{equ:transamp}, implemented with a sum over all spin foams ((dual) 2-complex), whose boundaries are the spin networks in question. This sum over spin foams was achieved in terms of group field theory (see Chapter \ref{chap:gft}).

We recall that the strategy of defining the projection operator from the Kinematical Hilbert space to the physical Hilbert space was via the definition of transition amplitudes between kinematical states. The discussion above uncovers the fact that the Ponzano-Regge model is a realisation of such a projection operator in the context of LQG, as it was proved in \cite{119d}.
However, such a proof does not hold in 4-dimensions, since it rests on the triangulation invariance of the model. This is not the case in 4-dimensions where, as we will see, a sum over triangulations is necessary to overcome triangulation dependence.

Since the Ponzano-Regge model represents the link between Regge calculus and quantum gravity in the following subsection we will give a brief description of Regge calculus.
\subsection{Regge Calculus}
In this section we are going to give a brief overview of what Regge calculus is. For a detailed analysis and recent progress see 
\cite{rwilliams}, \cite{reggeprogress}, \cite{reggereview}.

Regge calculus was born as an attempt to reformulate GR without the need of introducing any coordinate system. The aim of such a reformulation was to overcome certain problems present in GR when a continuum formulation of the theory is considered. For example, the problem of how to represent complicated topologies or the problem of finding numerical solution to Einstein's equations for generalised systems.\\ The starting point behind Regge calculus is to consider space (or spacetime) as a collection of n-dimensional flat simplices which are glued together by an identification of their (flat) (n-1)-dimensional simplices. In such a discretised manifold, the curvature resides in the (n-2)-dimensional simplices which get the name of \emph{hinges}. Thus the notion of a space (or spacetime) in which the curvature varies smoothly is rejected.

\begin{figure}[htb]
 \begin{center}
 \psfrag{a}{$\epsilon_{i}$}
  \includegraphics[scale=0.5]{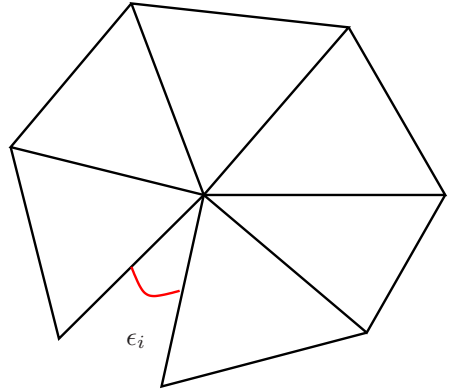}
\caption{Projection onto a plane of a set of triangles meeting at a vertex $p$. No curvature is present on the triangles or on the edges, but only in the vertex.\label{fig:2d-regge}}
\end{center}
  \end{figure}

In 2-dimensions it is very easy to give a visual example of how the curvature is defined in Regge calculus. Consider a dome which is tessellated by triangles. If we flatten the dome, then two triangles joint along an edge can be flattened without distortion, however, when a group of triangles meeting at a vertex is flattened, then there will be a gap. This gap represents the curvature present at the vertex and it is proportional to the size of the gap, which is called the \emph{deficit angle} $\epsilon$ and is given by 
\be
\epsilon=2\pi-\sum_{\text{ vertex angles }}
\ee
A graphical representation is given in figure \ref{fig:2d-regge}.\\
In 3-dimensions consider a tessellation of a 3D dome by flat tetrahedral glued together along flat triangles. If we consider a set of tetrahedra meeting at an edge, they will not fit together, but there will be a deficit angle, i.e. a dihedral angle. This angle represents the curvature concentrated on the edges and it is given by
\be   
\epsilon=2\pi-\sum_{\text{ dihedral angles at the edge }}
\ee
In 4-dimensions we consider 4-simplices joined along common tetrahedrons. In this case the \emph{hinges} are the flat triangles between the tetrahedrons where the 4-simplices meet.

In order to make a connection with GR, we need to decide which particular piecewise linear is an Einstein space, such that Einstein action can be evaluated. As a starting point we define our variables to be the edge lengths, which can be considered an equivalent of the continuum metric. To this end we first construct an analogue of Einstein's action in terms of the edge lengths, then apply the principle of stationary action to define Einstein equations for these edge lengths.\\
We recall that the Einstein action in $n$ dimensions is 
\be\label{equ:einsteinaction}
I=\frac{1}{16\pi G}\int R\sqrt{-g}d^nx
\ee
where $R$ is the scalar curvature. Since in a simplicial space the curvature is restricted along the hinges, equation \ref{equ:einsteinaction} was shown to be equivalent to the discretized action
\be\label{equ:reggeaction1}
I_R=\sum_{\text{hinges i}}F_i
\ee
where $F_i$ is the curvature associated to the $i$th hinge. Since the hinges are homogeneous the curvature is proportional to the volume of the hinge, i.e.
\be
F_i:=\overbrace{|\sigma_i|}^{\text{volume}}f(\overbrace{\epsilon}^{\text{deficit angle}})
\ee
where $f$ is a linear function of the deficit angle, i.e. $f(\epsilon_1+\epsilon_2)=f(\epsilon_1)+f(\epsilon_2)$. \\
By inserting the formula for the curvature in \ref{equ:reggeaction1} and considering the fact that any hinge can be seen as the superposition of two identical hinges such that $\epsilon_i=\epsilon_i^1+\epsilon_i^2$, we obtain
\be
I_R=\frac{1}{8\pi G}\sum_{\text{hinges i}}|\sigma_i|\epsilon_i
\ee
We now vary the above action with respect to the edge lengths $l_j$
\be\label{equ:variation}
\frac{\partial I_R}{\partial l_j}=\sum_{\text{hinges i}}\Big(\frac{\partial|\sigma_i|}{\partial l_j}\epsilon_i+|\sigma_i|\frac{\partial\epsilon_i}{\partial l_j}\Big)
\ee
Denoting by $\theta^q_i$ the dihedral angle of the two faces of the simplex q meeting at the hinge $i$, the expression for the deficit angle becomes 
\be
\epsilon_i=2\pi-\sum_{\text{q simplices meeting at hinge i}}\theta^q_i
\ee
Inserting this in the second term of the equation \ref{equ:variation}, we obtain 
\be
\sum_{\text{hinges }i }|\sigma_i|\frac{\partial \epsilon_i}{\partial l_j}=-\sum_{\text{hinges } i}\sum_{\text{simplices }j}|\sigma_i|\frac{\partial\theta_i^q}{\partial l_j}=-\sum_{\text{hinges } i}\Big(\sum_{\text{simplices }j}|\sigma_i|\frac{\partial\theta_i^q}{\partial l_j}\Big)
\ee
This expression turns out to be zero, the motivation being that a flux of a constant vector through a closed surface is zero \cite{regge}. \\Thus, the field equations are
\be
\sum_{\text{hinges i}}\frac{\partial|\sigma_i|}{\partial l_j}\epsilon_i=0
\ee
It would thus seem that there are as many equations as there are unknowns, providing a possibility for a complete solution for the edge lengths. However this is not the case, in fact there are the Regge analogues of the Bianchi identities (\cite{regge, rwilliams} and references there in). This implies that the equations are not all independent of each other.

\begin{figure}[htb]
 \begin{center}
   \includegraphics[scale=0.5]{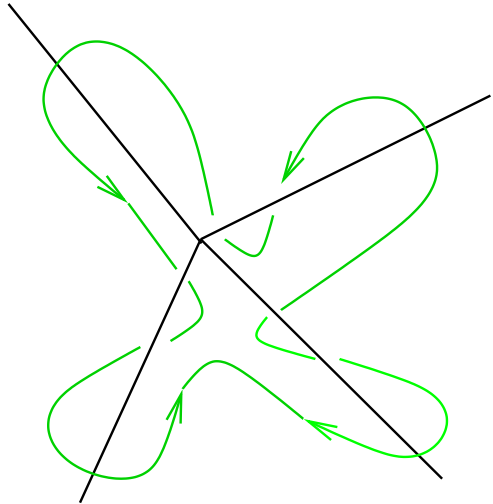}
\caption{Topologically trivial path.\label{fig:bianchidentity}}
\end{center}
  \end{figure}

In order to describe the Bianchi identities in Regge calculus we will consider an example in 3-dimensions. The generalisation in 4-dimensions is straightforward.\\ Let us consider a 4-valent vertex and a path which encircles each of the edges as shown in figure \ref{fig:bianchidentity}. 
Since in 3-dimensions the curvature is concentrated along the edges, if we parallel transport a vector along the path it will rotate. Thus, in our case, we would obtain a product of four rotation matrices one for each edge that the path encloses. However, if the path was such that it could be deformed, so as not to enclose any edges, i.e. it would be topologically trivial, then, the product of the four rotation matrices would equal the identity matrix. Therefore we obtain the following relation between the deficit angles for the edges meeting at a vertex:
\be 
\prod_{\text{hinges } i}exp(\epsilon_iU^i_{\alpha,\beta})=1
\ee 
where $U^i_{\alpha,\beta}$ are the rotation matrices associated to the edges.\\
For low order expansions of the above equation it is possible to recover the continuum version of the Bianchi identity.\\
This situation is analogous to the gauge freedom in the continuum, here we can freely specify an appropriate set of edge lengths.

The quantisation of the Regge action is through Euclidean path integral methods, thus one defines the heuristic partition function
\be\label{equ:pathregge}
Z=\int \md l_j e^{-I_R(l_j))}
\ee
The challenge for this quantisation strategy is to define the integration measure such that the discrete analogues of the diffeomorphisms invariance of the continuum limit\footnote{It should be noted that the Regge diffeomorphism invariance is still a problematic issue. In fact, there are two strategies to define such invariance, namely: i) Diffeomorphisms are transformations of the edge lengths which leave the geometry invariant. ii) Diffeomorphisms are transformations of the edge lengths which leave the action invariant.} is satisfied.
\\If one imposes a quantisation condition of the edge lengths, the integral in equation \ref{equ:pathregge} can be reduced to a summation.\\
\section{Spin Foam Model in 4-Dimensions}
\label{ssfm4}

\subsection{Palatini Formalism}
In this Section we will briefly describe the Palatini formalism and its properties. The main feature of this formalism is that it subordinates the role of the metric $g$ to that of the coframe field $e$ (or co-tetrad). 
The precise definition of a tetrad can be given with the aid of the following diagram:
\[\xymatrix{
TM\ar[rr]^{e}\ar[rdd]_p&&\mathcal{T}\ar[ldd]_{\pi}\\
&&\\
&M& \\
}\]
where $\mathcal{T}$ is a vector bundle over the spacetime M which is isomorphic to the tangent bundle and it is equipped with a metric $\eta$

Since M is an n-dimensional orientable manifold diffeomorphic to $\Rl^n$, then it follows that the tangent space TM is trivializable. 
Therefore the coframe $e$ can be identified with a choice of trivialization as follows:
\begin{align}
e:&TM\rightarrow\mathcal{T}\cong M\times\Rl^n\\
&T_xM\rightarrow\mathcal{T}\cong \{x\}\times\Rl^n\cong\Rl
\end{align}
Thus, what the coframe does is to define a coordinate basis for $T_xM$. In fact the above map can be factorised as follows:
\begin{align}
&TM\rightarrow B(M)\times\Rl^n\rightarrow M\times\Rl^n\\
&T_xM\xrightarrow{f}B(M)\times\Rl^n\xrightarrow{i_e} \{x\}\times\Rl^n\\
&V^{n}\partial_n\rightarrow [e,\vec{v}]\rightarrow \sum_{\mu=1}^n v^{\mu}e_{\mu}
\end{align}
where $B(M)$ is the frame field over M and $e_{\mu}$ is a basis. It is easy to see now that a tetrad field assigns a basis set to $T_xM$\\

The key idea of the Palatini action is to use the bundle $M\times \Rl^n$ to define ``objects" and, then, use the frame field to pullback this ``objects" on the bundle we are interested in, namely TM. This trick is needed since the bundle $M\times \Rl^n$ has a canonical inner product defined on it, which is lacking on TM. Specifically, given two sections $s$ and $s^{\prime}$ of $M\times \Rl^n$ the inner product is $\eta(s,s^{\prime})=\eta_{ij}s^is^j$ where $\eta$ is the internal metric of $\Rl^n$.\\ 
The bundle $TM$ can then be equipped with a metric by pulling back the metric on $M\times \Rl^n$, thus obtaining
\be
g(v,w)=\eta(e^{-1}v,e^{-1}w)
\ee
which in index notation becomes 
\begin{equation}
g_{\alpha\beta}=e^a_{\alpha}e^b_{\beta}\eta_{ab}
\end{equation}

Moroever, if $g$ corresponds to a classical solution of general relativity, then the coframe is actually an isomorphism and $g$ is non-degenerate. It is then possible to pull back a connection $w$ of the bundle $\mathcal{T}\cong M\times \Rl$ to a connection\footnote{Given a vector bundle $T\rightarrow M$ over a smooth manifold M and the space of smooth sections $S(E)$, a connection on $T$ is an $\Rl$-linear map
$\nabla :S(E) \rightarrow S(T\otimes T^*M)$ such that $ \nabla(\sigma f) = (\nabla\sigma)f + \sigma\otimes df$ holds for all smooth functions $f$ on M and all smooth sections $\sigma\in S(T)$.} on $TM$.
This is done as follows: suppose we have a section $s$ of $\mathcal{T}$, the differential of such section is given by $(D_{\mu}s)^a=\partial_{\mu}s^a+w^a_{\mu b}s^b$. The corresponding connection on $TM$ is then defined as $\nabla_v w=e^{-1}D_ve w$, which for $v=\partial_{\mu}$ becomes $(\nabla_{\mu}w)^{\alpha}=\partial_{\mu}w^{\alpha}+\Gamma_{\mu\beta}^{\alpha}w^{\beta}$ where $\Gamma_{\mu\beta}^{\alpha}:=e^{\alpha}_a(\delta^a_b\partial_{\mu}+w^a_{\mu b})e^b_{\beta}$. Now that we have pulled back both the metric and the connection to $TM$ we can now write the Palatini action as follows:
\be
S=\frac{1}{2k}\int_M*F^{IJ}\wedge e^I\wedge e^J+\frac{1}{\gamma} F^{IJ}\wedge e^I\wedge e^J
\ee
where $F^{IJ}$ is the curvature of $w$, $k$ is Newton's constant and $\gamma$ is the Immirzi parameter. Variation with respect to $w$ and $e$ gives back Einstein's equations.\\

In spin foam models the importance of the Palatini action is that it represents a subsector of the so-called Plebanski action which describes gravity as a constrained topological action. Plebanski action is a BF-type action and, therefore, there are known methods of how to quantise it and define a path integral. However, in order to obtain such a quantisation for the Palatini action, certain constraints have to be implemented. As we will see such an implementation of the constraints turns out to be non-trivial. Before going into the detail of how a spin foam model can be derived for the Palatini formulation of GR through the Plebanski action, we will first describe the precise tools needed to rigorously apply the discretization procedure mentioned for the 3-dimensional case. 
\section{Precise Definition of Tools of Discretization}
In this Section we will describe, in detail, the tools that are used in spin foams to discretize the manifold M. Such a discretization of the manifold is needed in order to regularize the theory. Moreover, the action utilised in standard spin foam models is the BF-theory action, with some constrains on the B field. 

Such a BF-theory is a topological theory, therefore the discretization of the manifold one needs to perform has to be compatible with the topological invariance of the theory, i.e. once discretized the BF-action, it still has to be topological invariant. The variables of the BF-action are p-forms (p depends on which dimensions we are working with), therefore, in order for the BF-theory to be topologically invariant, one needs to find the discrete version of those operations, which can be performed on such p-forms, while retaining the theory topological invariant. Specifically, we will define the discrete analogue of the wedge product, Hodge dual and the exterior derivative. 
\begin{definition}
A p-simplex, denoted $\sigma^p$ is identified to be the convex hull of P+1 vectors
which span a p-dimensional vector space, i.e.\begin{equation}
\sigma^p:=\{x\in\Rl^m|x=\sum_{i=0}^{p}t_iv_i;t_i\geq0;\sum_{i=0}^{p}t_i=1\}
\end{equation}
$\sigma^p$
is denoted as follows $\sigma^p=[v_0,v_1\cdots, v_p]$\end{definition}
Each p-simplex has an orientation depending on the order in which the vertices appear in the list $\sigma^p=[v_0,v_1,\cdots, v_p]$. It is possible to permute such an order so as to obtain an equal or opposite orientation of the simplex. Specifically, we say that given, a permutation $\pi\in S_{p+1}$, then the simplices $[v_0,v_1,\cdots, v_p]$ and $[\pi v_0,\pi v_1,\cdots, \pi v_p]$ are equally oriented if $\pi$ is even, otherwise they are opposite oriented.
\begin{definition}\label{def:bar}
Given a p-simplex, $\sigma^p=[v_0,v_1\cdots, v_p]$, the barycentric point $\hat{\sigma}^p$ is defined as follows:
\begin{equation}
\hat{\sigma}^p:=\frac{\sum_{k=0}^p v_k}{p+1}
\end{equation}
\end{definition}
By joining together certain complexes of different dimensions in a coherent manner it is possible to form the so called simplicial complexes.
\begin{definition}
A simplicial complex K is a collection of simplices $\sigma^p_i$ for $p=0,1,\cdots ,N$ and $i=1,\cdots ,N_p$ with the following properties:
\begin{enumerate}
\item all subsimplices for each simplex $\sigma^p_i$ belong to K.
\item Given two simplices $\sigma^p_i$ and $\sigma^q_j$ they can, at most, intersect in a common subsimplex which has opposite orientation when considered as being part of the two original simplices.
\end{enumerate}
\end{definition}
The interesting fact is that any differentiable manifold admits a discretizatoin in terms of the above defined simplicial complexes, i.e. admits a triangulation. However, there exists an isomorphic partition of the manifold in terms of dual complexes to the original simplicial complex.
 \begin{definition}
In $D$ dimensions, given any simplex $\sigma^p_{i_0}$ of a simplicial complex K, and considering all possible (D-p) tuples of simplices  $\sigma^{(p+k)}_{i_k}$ ($k=1,\cdots D-p$ and $1\leq i_k\leq N_{p+k}$) also belonging to K, such that \begin{enumerate}
\item for all $l=0,\cdots, D-p-1$ the simplex  $\sigma^{(p+l)}_{i_l}$is a face of $\sigma^{(p+l+1)}_{i_{l+1}}$ with induced orientation.
\item For each (D-p) tuple of simplices construct a (D-p)-simplex $[\hat{\sigma}^p_{i_0},\hat{\sigma}^{p+1}_{i_1}\cdots ,\hat{\sigma}^D_{i_{D-p}} ]$ in terms of the barycentric subdivision of each simplex.
\end{enumerate} The dual cell to the simplex $\hat{\sigma}^p_{i_0}$ is defined as follows
\begin{equation}
*_K[\hat{\sigma}^p_{i_0}]:=\cup_{\sigma^{p+l}_{i_l}\subset\delta\sigma^{p+l+1}_{i_{p+l+1}}; l+0,\cdots,D-p-1}[\hat{\sigma}^p_{i_0},\hat{\sigma}^{p+1}_{i_1}\cdots ,\hat{\sigma}^D_{i_{D-p}} ]
\end{equation}
By gluing together all such defined dual simplices among the common subsimplices we obtain the dual cell $K^*$ of K \footnote{An alternative definition of a k cell and its elements would be as follows:\\
1) \textbf{k-cell}: given a polyhedron (A polyhedron is defined to be a subset of $\Rl^n$ such that every point $x\in X$ has a neighbourhood of the form $\{ax+by:a,b\geq 0\mbox{ } a+b=1,y\in Y\}$ for $Y\subseteq X$ is compact) $X$, we say that $X$ is a \textbf{k-cell} iff the smallest affine space (vector space which has forgotten its origins) which contains $X$ is of dimension $k$. For example, in $\Rl^n$, 0-cells are identified with the points, 1-cells with compact intervals affinely embedded in $\Rl^n$, and 2-cells with convex compact polygons affinely embedded in $\Rl^n$.
\\
2) The \textbf{elements of k-cells} are:
\\
i) \emph{Vertex}: Given a point $x\in X$, define the union of all lines $L$ passing through x with X as $\langle x,X\rangle$, such that for each line $L$, $L\cap X$ is an interval with x as its interior. If $\langle x,X\rangle$ does not exist, then x is a vertex.\\
ii) \emph{Faces}: $\langle x,X\rangle \cap X$ is a face of X.\\
A \textbf{piecewise linear cell complex} is defined to be a collection $h$ of cells in $\Rl^n$ such that
\begin{itemize} 
\item  If $X\in h$ and Y is a face of X i.e. $Y\geq X$, then $Y\in h$.
\item If $X,Y\in h$ then $Y\cap X\in h$.
\end{itemize}
}.
It follows that the operation $*_K$ is a map as follows: \begin{equation}
^*K:C_p(K)\rightarrow C_{D-p}(K^*)
\end{equation}
where $C_p(K)$ indicate the p-chains\footnote{A p-chain is a formal linear combination of p-simplicies} of K.
\end{definition}
The term $\delta\sigma^{p+l+1}_{i_{p+l+1}}$ indicates the boundary of the simplex $\sigma^{p+l+1}_{i_{p+l+1}}$. In particular, given  a simplex $\sigma^{p}$, then $\delta\sigma^{p}$ is defined to be the set of point $t_k=0$, $k=0,\cdots p$ which form p+1 different p-1 simplices $\sigma^{p-1}_k:=[v_o,\cdots \hat{v}_k,\cdots v_p]$, such that $\delta\sigma^{p}=\cup_k\sigma^{p-1}_k $. The notation $\hat{v}_k$ indicates that the vertex $v_k$ is being omitted. The orientation of such boundaries will be equal to the orientation of the whole simplex if k is even, otherwise it will have an opposite orientation. \\
As we will see, both the operations $*_K$ and $\delta$, when applied to p-chains of a simplex K, actually represent the discretized analogue of the Hodge dual operation and the exterior derivative, respectively. However, the cell complex $K^*$ is not a simplicial complex. We will say more about it later.
\begin{definition}
Given a simplicial complex $K=\{\sigma^p_i;p=0\cdots D; i=0\cdots N_p\}$ then we define the following:
\begin{enumerate}
\item
A ``formal real" linear combination of the simplices $\sigma^p_{i}$ defines a vector space of p-chains $C_p(K)$.
\item It is possible to transform $C_p(K)$ into a Hilbert space by defining an inner product as follows:
\begin{equation}\label{equ:scalar}
\langle \sigma^p_{i},\sigma^p_{j}\rangle_K:=\delta_{ij} 
\end{equation}
for all $i,j=1\cdots N_p$. This implies that all the p-simplices provide an orthonormal basis for K. Given \ref{equ:scalar} it is possible to identify the dual space $C^p(K)$ of linear forms on $C_p(K)$ (of co-chains) with $C_p(K)$ itself.  
\item The boundary operation between p-chains is defined as follows:
\begin{align}
\delta_K:C_p(K)&\rightarrow C_{p-1}(K)\\\nonumber
\sigma^p_{i}&\mapsto\delta\sigma^p_{i}:=\sum_{k=0}^{k=p}(-1)^k[v_0\cdots\hat{v}_k\cdots v_p]
\end{align}
such that $\delta^2=0$. The adjoint (under the scalar product \ref{equ:scalar}) of $\delta_K$ is the coboundary operator $d_K:C_p(K)\rightarrow C_{p+1}(K)$. It is precisely this co-boundary operator that is the discrete analogue of the $*d*$ operation on p-forms, i.e. the dual of the exterior derivative for p-forms.
\end{enumerate}
\end{definition}
We will now define the analogue of the wedge product for p-forms. To this end, consider only those p-forms which form a p-chain and denote them by $\Lambda^p(K) $. We then can define the following:
\begin{definition}\mbox{}
\begin{enumerate}
\item
For a simplicial complex K the Whitney map is given by 
\begin{align}
W_K:C_p(K)&\rightarrow\Lambda^p(K)\\\nonumber
\sigma^p=[v_0\cdots v_p]&\mapsto p!\sum_{k=0}^p(-1)^kt_k dt_0\wedge\cdots\wedge\hat{dt_k}\wedge\cdots dt_p
\end{align}
($t_k$ are local coordinates of $\sigma^p$).
\item The de Rham map is given by 
\begin{align}
R_K:\Lambda^p(K)&\rightarrow C_p(K)\\\nonumber
w&\mapsto\langle R_K(w),\sigma^p\rangle_k:=\int_{\sigma^p}w
\end{align}
\item The wedge product on p-chains is defined as follows:
\begin{align}
\bigwedge_K:C_p(K)\times C_q(K)&\rightarrow C_{p+q}(K)\\\nonumber
\big(\sigma^p,\sigma^q\big) &\mapsto\sigma^p\wedge_K \sigma^q:=R_K(W_K(\sigma^p)\wedge W_K(\sigma^q))
\end{align}
\end{enumerate}

\end{definition} 
For the operations defined above, it is possible to define the following relations
\begin{theorem}
The operations defined by the Whitney map and the de Rham map obey the following relations:
\ba
\sigma^{(p)}\wedge_K \sigma^{(q)}&=&(-1)^{pq}\sigma^{(q)}\wedge_K\sigma^{(p)}\nonumber\\
d_K(\sigma^{(p)}\wedge_K\sigma^{(q)})&=&(d_K\sigma^{(p)})\wedge_K\sigma^{(q)}+(-1)^p\sigma^{(p)}\wedge_K(d_K\sigma^{(q)})\nonumber\\
R_K\circ W_K&=&id\nonumber\\
d\circ W_K&=&W_k\circ d_K\nonumber\\
d_K\circ R_K&=& R_K\circ d\nonumber\\
\int_{\sigma^{(p)}}W_K(\sigma^{(p)'})&=&\langle\sigma^{(p)},\sigma^{(p)'}\rangle_K
\ea
\end{theorem}
We mentioned above that the cell complex $K^*$ dual to K is not really a simplicial complex, therefore it is not possible to define $K^{**}$. This implies that we can not yet define the operation $^*k$ as the discretized analogue of the Hodge star. To be able to do so we need to introduce another simplex $B(K)$ of which $K^*$ is a subsimplex
\begin{definition}
Given a p-simplex $\sigma^p=[v_1\cdots v_p]$ its barycentric subdivision (defined in \ref{def:bar}) comprises $(p+1)!$ different p-simplices $\sigma_{\pi}^p$, one for each permutation $\pi\in S_{p+1}$ ($S_{p+1}$ is the symmetric group) as follows :
for all k-simplex $\sigma(k)_{\pi}=[v_{\pi(1)}\cdots v_{\pi(k)}]$ $k=0\cdots p$ its barycentric subdivision is
\begin{equation}
\hat{\sigma}(k)_{\pi}:=\frac{\sum_{l=0}^k v_{\pi(l)}}{k+1}
\end{equation} 
Define a simplex in terms of the barycentric points as follows:  $\sigma^p_{\pi}:=[\hat{\sigma}^0_{\pi},\cdots \hat{\sigma}^p_{\pi}]$.\\
The \textbf{Barycentric refinement} $B(K)$ is then defined as the collection of all $(p+1)!$ subdivisions of each p-simplex in K for all $p=0\cdots D$.
\end{definition}
Given the barycentric subdivision B(K) then the dual $K^*$ is defined as the union of p-simplices in B(K), hence $K, K^*\subseteq B(K)$. This implies that it is possible to extend all operations regarding $K$ to operations on $B(K)$. Moreover, since $C_p(K^*)\subseteq C_p(B(K))$ all operations can be extended to $K^*$, thus obtaining the following:
\begin{theorem}
For all $x\in C_p(K)$ and $y\in C_{D-p}(K^*)$ we have
\begin{align}
\langle *_{K}(x),y\rangle_{K^*}&=\frac{(D+1)!}{p!(D-p)!}\int_MW_{B(K)}(E(x))\wedge W_{B(K)}(E(y))\\\nonumber
\langle *_{K^*}(y),x\rangle_K&=\frac{(D+1)!}{p!(D-p)!}\int_MW_{B(K)}(E(y))\wedge W_{B(K)}(E(x))
\end{align}
where $E(x)$ are linear combinations of elements $x\in C_p(K)$ in terms of elements $C_p(B(K))$. The inner product in $K^*$ is defined in the same way as for $K$ by defining dual cells as orthonormal.\\ We can now define the operation of exterior derivative in terms of the operations $*_{K^*}$, $d_{K^*}$ and $*_K$ as follows:
\begin{align}
\partial_K&=(-1)^{p(D-p)}*_{K^*}\circ d_K\circ *_{K}\\\nonumber
\partial_{K^*}&=(-1)^{p(D-p)}*_{K}\circ d_K\circ *_{K^*}
\end{align}
\end{theorem}

We will now apply the discretization tools defined in this section to a general BF-theory action in order to derive a spin foam model.
\subsection{Spin Foam Models through BF Theory}
\label{sbft}
In this Section we will describe how spin foam models are obtained through BF-theory, in particular through the Plebanski action.
The general form of BF-action in 4-dimensions is 
\begin{equation}
S_{BF}=\int_M Tr(B\wedge F)
\end{equation}
where $B$ is a Lie algebra valued 2-form on the principal G-bundle $P$ under the adjoint representation and $F$ is the curvature of the connection $A$.\\
The discretization of such an action can be defined utilising the Whitney and the de Rham map, defined above as follows:
\ba
S_{BF}&=&\int_M Tr(W_{B(K)}(R_{B(K)}(B))\wedge W_{B(K)}(R_{B(K)}(F)))\nonumber\\
&=&Tr(\langle*_{K^*}(R_{B(K)}(F)),R_{B(K)}(B)\rangle_K)
\ea
Given an orthonormal basis $\sigma^{(2)}$ of $C_2(K)$ we obtain
\ba\label{ali:exact}
S_{BF}&=&\sum_{\sigma^{(2)}\in C_2(K)}Tr(\langle *_{K^*}(R_{B(K)}(F)),\sigma^{(2)}\rangle_K\langle\sigma^{(2)},(R_{B(K)}(B))\rangle_K)\nonumber\\
&=&\sum_{\sigma^{(2)}\in C_2(K)}Tr(\langle *_{K^*}(R_{B(K)}(F)),\sigma^{(2)}\rangle_K\int_{\sigma^{(2)}}B\nonumber\\
&=&\sum_{\sigma^{(2)}\in C_2(K)}Tr(\int_MW_{B(K)}(R_{B(K)}(F))\wedge W_{B(K)}(E(\sigma^{(2)})))\int_{\sigma^{(2)}}B\nonumber\\
&=&\sum_{\sigma^{(2)}\in C_2(K)}Tr\Big([\int_{*_{K(\sigma^{(2)})}} F][\int_{\sigma^{(2)}}B]\Big)
\ea
which is an exact result and independent of the triangulation $K$.

In order to define a path integral an ulterior discretization step is required, which is not exact. Specifically, we know that to each triangle in the original triangulation ($t(f)\in C_2(K)$) there corresponds a unique dual face $f^*\in C_2(K^*)$, therefore we can perform a sum over dual faces in \ref{ali:exact}.\\
Moreover, by approximating $\int_{f^*\in C_2(K^*)}F=1_4+F(f^*)+\cdots=U(\partial f^*)$ where $U(\partial f^*)$ is the holonomy of the SO(4) connection along the loop, $\partial f^*$, we can approximate \ref{ali:exact} by
\be
\sum_{f^*\in C_2(K^*)}Tr([\int_{t(f^*)\in C_2(K)}B] U(\partial f^*))
\ee
The term $1_4$ drops out of the trace, thus the approximation is correct.\\
The partition function then becomes
\ba\label{ali:pathint}
Z_{BF}&:=&\int\prod_{e^*\in C_1(K^*)}d\mu_H(g_e)\prod_{f^*\in C_2(K^*)}d^6[\int_{\sigma^{(2)}}B]exp(iS_{BF}(K^*)) \nonumber\\
&=&\int\prod_{e^*\in C_1(K^*)}d\mu_H(g_{e^*})\prod_{f^*\in C_2(K^*)}\sum_{I<J}\delta_{\Rl}(Tr(P_{IJ}U(\partial f^*)))
\ea
where in the last line we have performed the integration over the B field, resulting in a  $\delta$-distribution.\\
The elements $P_{IJ}$ are the generators of the algebra $su(4)$.

At the classical level we know that the solutions of the equation of motion of BF-theory are flat connections. We would like these solutions to be translated at the quantum level. However, the integrand in \ref{ali:pathint} has support on those elements $g\in SO(4)$, such that $g=g^T$, therefore even on elements $g\neq 1$. These ``extra" solutions get discharged by hand, thus obtaining
\be
Z_{BF}=\int\prod_{e^*\in C_1(K^*)}d\mu_H(g_{e^*})\prod_{f^*\in C_2(K^*)}\delta_{SO(4)}(U(\partial f^*))
\ee
Since there are as many dual faces as there are triangles, the choice of discretising the $B$ field on triangles and the curvature on dual faces allows us to get as many flatness conditions as there are holonomies. However, as explained in Section \ref{s5.1}, such a choice of discretisation will lead to issues related to gauge invariance.

\begin{figure}[htb]
 \begin{center}
 \psfrag{j}{$\rho_1$}\psfrag{h}{$\rho_2$}\psfrag{f}{$\rho_3$}\psfrag{d}{$\rho_4$}\psfrag{b}{$\rho_0$}\psfrag{i}{$\rho_{12}$}\psfrag{g}{$\rho_{23}$}\psfrag{e}{$\rho_{34}$}
\psfrag{c}{$\rho_{40}$}\psfrag{a}{$\rho_{01}$}\psfrag{k}{$\rho_{13}$}\psfrag{m}{$\rho_{41}$}\psfrag{l}{$\rho_{02}$}\psfrag{p}{$\rho_{24}$}\psfrag{o}{$\rho_{03}$}\includegraphics[scale=0.5]{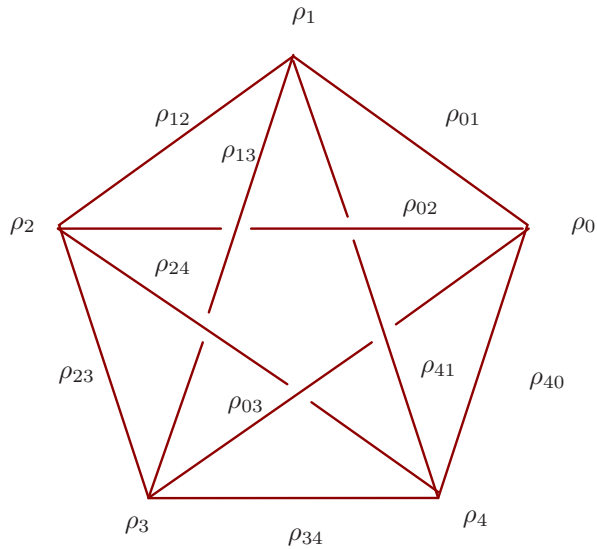}
\caption{The dual vertex associated to a tetrahedron in 4-dimensions. The links are labelled by representations.\label{fig:10j} }
\end{center}
  \end{figure}

By expanding the $\delta$-distribution using Peter-Weyl theorem, and performing the various integrals, the resulting expression for the partition function can be written in the following form \cite{1b}:
\be\label{equ:partfunc}
Z_{BF}=\sum_{\{\rho_{f^*}\}}\sum_{\{\rho_{e^*}\}}[\prod_{f^*\in C_2(K^*)}A_{f^*}(\{\rho_{f^*}\})][\prod_{e^*\in C_1(K^*)}A_{e^*}(\{\rho_{e^*}\})][\prod_{v^*\in C_0(K^*)}A_{v^*}(\{\rho_{f^*}\},\{\rho_{e^*}\})]
\ee
where the terms $A_{f^*}$, $A_{e^*}$ and $A_{v^*}$ are the amplitudes associated to the (dual) faces, (dual) edges and (dual) vertices, respectively. The terms $\rho_{f^*}$ are the representations assigned to each dual face, while $\rho_{e^*}$ are the intertwiners associated to each dual edge. The vertex amplitude is actually given by the 10j-symbol and it is diagrammatically depicted in figure \ref{fig:10j}.\\
Interestingly enough, \ref{equ:partfunc} is invariant under change of triangulation $K$, even after regularising it by cutting off the sum over representation (quantum groups), i.e. the model is a topological model.\\

So far we have described the method for obtaining a partition function for a general BF-theory. However, we are interested in deriving a partition function for a yet to be defined quantum theory of gravity. Therefore, the correct BF-action to utilise is the Plebanski action, since it reduces to the Palatini action of GR under certain constraints of the $B$ field. \\
The Plebanski action is given by
\be
S[B,A,\lambda,\mu]=\int[B^{IJ}\wedge F_{IJ}(A)+\lambda_{IJKL}B^{IJ}\wedge B^{KL}]
\ee      
where 
$\lambda_{IJKL}$ is a Lagrangian multiplier satisfying $\lambda_{IJKL}=-\lambda_{JIKL}=-\lambda_{IJLK}=\lambda_{KLIJ}$ and
the constraint $\epsilon_{IJKL}\lambda_{IJKL}=0$. Variation of the action with respect to $\lambda_{IJKL}$ results in the following \emph{simplicity constraint} on the $B$ field
\be
B^{IJ}\wedge B^{KL}=\epsilon^{IJKL}\frac{1}{4!}\epsilon_{MNPQ} B^{MN}\wedge B^{PQ}
\ee
which is equivalent to the existence of a co-tetrad $e^I$ such that 
\be
B^{IJ}=\pm e^I\wedge e^J\mbox{  or  }B^{IJ}=\pm \frac{1}{2}\epsilon_{IJKL}e^I\wedge e^J
\ee
The simplicity constraint allows for five different solutions, namely:
\be
 B^{IJ}=\pm e^I\wedge e^J;\hspace{.5in} B^{IJ}=\pm\frac{1}{2}\epsilon^{IJ}_{KL}e^K\wedge e^L;\hspace{.5in}\text{  degenerate solutions  }
\ee
The non degenerate part of the constraint can be written as follows:
\be\label{equ:nondegenerate}
\epsilon_{IJKL}B^{IJ}_{\mu\nu}B^{KL}_{\rho\sigma}=e\epsilon_{\mu\nu\rho\sigma}
\ee
which implies that 
\be
\epsilon_{IJKL}B^{IJ}_{\mu\nu}(B^{KL})^{\rho\sigma}=0
\ee
However, only $B^{IJ}=\frac{1}{2}\epsilon^{IJ}_{KL}e^K\wedge e^L$ reduces the Plebanski action to the Palatini action\footnote{It should be noted that $B^{IJ}=-\frac{1}{2}\epsilon^{IJ}_{KL}e^K\wedge e^L$ would imply only a global change of sign, thus at the classical level would still reproduce the Palatini action. }.

Moreover, the bivectors $B^{IJ}$ also satisfy the so called closure constraint: $\int_{\diamond}dB^{IJ}(x)=\int_{\partial(\diamond)}B^{IJ}(x)=\sum_t\int_tB^{IJ}(x)=\sum_tB^{IJ}(t)=0$, i.e. the bivectors associated to the triangles $t$ of a tetrahedron $\diamond$ sum to zero.\\
The discretization of the non constraint part of the action is carried out in an analogous way as for the general BF-action, while the constraint part is discretised as follows:
\be
\sum_{\Delta\in C_4(K)}\sum_{v\in V(\Delta)}\frac{\lambda_{IJKL}}{5}\epsilon^{ijkl}B^{IJ}(t^v_{ij}(\Delta))B^{KL}(t^v_{kl}(\Delta))
\ee
where the factor of $5$ is necessary since each 4-simplex contains $5$ vertices, and $\Delta$ represents a 4-simplex, $V(\Delta)$ is the set of all vertices for a given simplex and $t^v_{ij}(\Delta)$, $1\leq i<j\leq4$ are the 6 triangles incident at the vertex $v$ and whose boundary loop starts from $v$ along $e^v_i(\Delta)$ and ends at $v$ along $e^v_j(\Delta)^{-1}$. \\
Because of the definition of $t^v_{ij}(\Delta)$ it follows that: 
$t^v_{ij}(\Delta)=-t^v_{ji}(\Delta)$.\\
The complete discretised action then becomes
\be
S_P=\sum_{\sigma^{(2)}\in C_2(K)}Tr\Big([\int_{*_{K(\sigma^{(2)})}} F][\int_{\sigma^{(2)}}B]\Big)\sum_{\Delta\in C_4(K)}\sum_{v\in V(\Delta)}\frac{\lambda_{IJKL}}{5}\epsilon^{ijkl}B^{IJ}(t^v_{ij}(\Delta))B^{KL}(t^v_{kl}(\Delta))
\ee
We now analyse the discretised version of the simplicity constraint \ref{equ:nondegenerate}, which can be written as 
\be
\epsilon_{IJKL}B^{IJ}(t^v_{ij}(\Delta))B^{KL}(t^v_{kl}(\Delta))=\epsilon_{ijkl}\frac{1}{4!}\epsilon_{IJKL}\epsilon^{pqrs}B^{IJ}(t^v_{pq}(\Delta))B^{KL}(t^v_{rs}(\Delta))
\ee
It is easy to verify that the above constraint translates into the two following conditions:
\begin{itemize}
\item [i)] $\epsilon_{IJKL}B^{IJ}(t)B^{KL}(t^{'})=0$ iff  $t=t^{'}$ or $t\cap t^{'}=e$. This constraint implies that the fields $B$ associated to neighboring triangles or to the same triangle, are simple bivectors.
\item [ii)]  $\epsilon_{IJKL}B^{IJ}(t_{12})B^{KL}(t_{34})=\epsilon_{IJKL}B^{IJ}(t_{13})B^{KL}(t_{12})=\epsilon_{IJKL}B^{IJ}(t_{14})B^{KL}(t_{23})$ iff the six triangles $t_{ij}$ only share a common vertex of the 4-simplex.
\end{itemize}
It is straightforward to deduce that, if the triangle on which the bivectors are defined changes orientation, the bivectors will change sign.\\

Similarly, as for the continuum case, there are, excluding degenerate solutions, four solutions to the above constraint. In particular, the bivectors associated to each triangle can be: \\
i) $B^{IJ} $ ii) $-B^{IJ} $ iii) $*B^{IJ}=\epsilon^{IJ}_{KL}B^{KL} $ iv) $-*B^{IJ}=-\epsilon^{IJ}_{KL}B^{KL} $. \\
The first two cases correspond to well defined simplicial geometries, differing only by a global change of orientation, while the remaining have no geometric meaning at all.\\
The set of all constraints can be identified with the set $C_{\alpha}(\{\int_{t(f^*)\in C_2(K)}B\}_{f^*\in C_2(K^*)})$ for some set $\alpha$. These constraints are then implemented at the level of the action by inserting the following term in the action:
\be\prod_{\alpha}\int \frac{dt}{2\pi}e^{iC_{\alpha}(\{\int_{t(f^*)\in C_2(K)}B\}_{f^*\in C_2(K^*)})t}=\prod_{\alpha}\delta(C_{\alpha}(\{\int_{t(f^*)\in C_2(K)}B\}_{f^*\in C_2(K^*)}))\ee 
We then obtain, as a possible partition function for Plebanski action the following:
\ba
Z_P(K^*)&=&\int\Big[\prod_{e\in C_1(K^*)}d\mu_H(g_e)\Big]\Big[\prod_{f\in C_2(K^*)}d^6[\int_{t(f^*)\in C_2(K)}B]\Big]\Big[\prod_{\alpha}\delta(C_{\alpha}(\{\int_{t(f^*)\in C_2(K)}B\})\Big]\nonumber\\
& &\times  exp\Big(i\sum_{f^*}Tr(\int_{t(f^*)\in C_2(K)}B U(\partial f^*))\Big)
\ea
In order to derive the analogue of \ref{equ:partfunc} for the Plebanski action some approximations are needed. In particular, one has to impose the flatness conditions, i.e. $U(\partial f^*)=1$ \emph{before} performing the $B$ integral. As it was previously mentioned this is justified \emph{a posteriori} since, at the classical level, only flat connections are allowed.  However, it is an approximation which is put in by hand and it is not rigorously derived. Nonetheless one assumes the flatness constraint $U(\partial f^*)=1$. As a consequence, the commuting set of constraints can be replaced by a non-commuting set\footnote{Roughly this is a consequence of the following fact: given $S=\sum_{f^*}Tr(B_f U(\partial f^*))$ where $B_f=\int_{t(f^*)}B$, then $[X^{IJ}_f, S]=[X^{IJ}_f,Tr(B_f U(\partial f^*))]=Tr([P^{IJ}U_f(\partial f^*)]^TB_f$. By setting $U(\partial f^*)=1_{SO(4)}$ and writing $B_f=\sum_{I<J}P^{IJ}B^{IJ}_f$ ($Tr(P^{IJ}P^{KL})=-\delta^I_{[K}\delta^J_{L]}$) it follows that $[X^{IJ}_f, S]\sim B_f$. Therefore since $\Big[\prod_{\alpha}\delta(C_{\alpha}(\{X_f\}))\Big]e^{iS}=e^{iS}
\prod_{\alpha}\Big[e^{-iS}\delta(C_{\alpha}(\{X_f\}))e^{iS}\Big]=e^{iS}
\prod_{\alpha}\Big[\delta(e^{-iS}C_{\alpha}(\{X_f\})e^{iS})\Big]$, where the constraints are now defined using the $X$ as
$e^{-iS}\epsilon_{IJKL}X_f^{IJ}X_{f^{'}}^{KL}e^{iS}=\epsilon_{IJKL}(X_{f}^{IJ}+i[X_f^{IJ},S])(X_{f^{'}}^{KL}+i[X_{f^{'}}^{KL}S])$, equation \ref{equ:replacement} follows. }, i.e.
\be\label{equ:replacement}
\prod_{\alpha}\delta(C_{\alpha}(\{\int_{t(f^*)\in C_2(K)}B\}_{f^*\in C_2(K^*)}) \simeq \prod_{\alpha}\delta(C_{\alpha}(\{\int_{t(f^*)\in C_2(K)}X^{IJ}\}_{f^*\in C_2(K^*)})
\ee
where $\int_{t(f^*)\in C_2(K)}X^{IJ}:=Tr([P^{IJ}U(\partial f^*)]^T\frac{\partial}{\partial U(\partial f^*)}$ is a right invariant vector field on the copy of SO(4). \\
This replacement allows to perform the integral with respect to the $B$ field, thus obtaining
\ba\label{ali:first}
Z_{P}&=&\int\Big[\prod_{e\in C_1(K^*)}d\mu_H(g_e)\Big]\Big[\prod_{f\in C_2(K^*)}d^6[\int_{t(f^*)\in C_2(K)}B]\Big]\Big[\prod_{\alpha}\delta(C_{\alpha}(\{\int_{t(f^*)\in C_2(K)}X\})\Big]\nonumber\\
& & \times exp\Big(i\sum_{f^*}Tr(\int_{t(f^*)\in C_2(K)}B U(\partial f^*))\Big)\nonumber\\
&=&\int\Big[\prod_{e\in C_1(K^*)}d\mu_H(g_e)\Big]\Big[\prod_{\alpha}\delta(C_{\alpha}(\{\int_{t(f^*)\in C_2(K)}X\})\Big]\Big[\prod_{f\in C_2(K^*)}\delta(U(\partial(f^*))\Big]
\ea
However, if one considers all triangles at once, then it will not be possible to write \ref{ali:first} in the form of \ref{equ:partfunc}, where amplitudes related to each simplex in the simplicial complex $K^*$ are taken into consideration.

To solve this problem one simply considers each individual 4-simplex separately, thus ignoring interaction terms. The geometrical motivation for such a solution is given by analysing the quantum analogue of 4-dimensional simplicial geometry. \\
As it was done for the 3-dimensional case, a quantum 4-dimensional simplicial geometry can be derived by first defining the quantum analogues of the discretised $B$ fields and, then, constructing a ``quantum triangle" in terms of them. In this way a quantum state is associated to a collection of ``quantum tetrahedrons" glued together along common ``quantum triangles". Individual `` quantum 4-simplices" are, then, the building block to define transition amplitudes between quantum states.

Let us analyse how this is done in detail.
In order to quantise the $B$ fields we need to associate them to some operators acting on a certain Hilbert space. In order to achieve this we utilise the isomorphism that exists between the space of bivectors $\wedge^2\Rl^4$ ($\wedge^2\Rl^{3,1}$ for Lorentzian case) and the Lie algebra $so(4)$ ($so(3,1)$), such that each bivector $B^{IJ}(t)$ of a given triangle $t$ is associated with the generator of a Lie algebra, i.e. $B^{IJ}(t)\rightarrow *J^{IJ}(t):=\epsilon^{IJ}_{KL}J^{KL}(t)$.\\
However, it turns out that such a procedure leads to the wrong sector of solutions of the simplicity constraint. In order to get the desired solution of the simplicity constraints, i.e. the solutions that lead to the Palatini action, one has to associate each bivector to an element of the dual of the Lie algebra
\ba
\beta:\Lambda^2\Rl^n&\rightarrow& so(n)^*\nonumber\\
 (e \wedge f)(l)&\mapsto&\beta(e \wedge f)l = \eta(le, f)\hspace{.5in}\forall (e \wedge f)\in\Lambda^2\Rl^n;l\in so(n)
\ea
where $\eta$ represents the Riemannian or Lorenzian metric.\\
The dual Lie
algebra $so(n)^*$ has a natural Poisson structure called the flipped Poisson bracket, which was shown in \cite{d118} to be the correct structure to use.

If we then associate to each triangle $t$ a representation $\rho_t$, with associated representation space $V_t$, then the generators of the Lie algebra act on such a space as derivative operators. In this way it is possible to associate to each $B^{IJ}(t)$ an operator acting on $V_t$. \\
Normally, the representation one chooses for each triangle is the irreducible unitary representation. The reason being that, in this way, the representation labels characterise the quantum area of the triangle the representation is associated to\footnote{ To understand this, let us consider a triangle $t$ with assigned representation $\rho_t$. The Lie algebra element associated to the bivector $B^{IJ}(t)$ would then be $*J(\rho_t)$. The area of $t$ can be expressed in terms of bivecotrs as $A^2=B(t)\cdot B(t)$, which gets translated into $A^2=J^{IJ}(\rho_t)\cdot J_{IJ}(\rho_t)$. If $\rho_t$ is irreducible and unitary we have, for the Riemannian and Lorentzian case, respectively, $A^2=2j(j+1)$ and $A^2=n^2-1$ or $A^2=-p^2-1$.}.\\
In order to assign the correct Hilbert space to each triangle, we first need to translate the simplicity constraint for the bivectors to constraints/requirements on the Lie algebra elements associated to such bivectors. In particular, for a given assignment of representations $\rho_t$ to triangles $t$, the condition $B(t)*B(t)=0$ translates to the condition that the second Casimir of the group vanishes in that representation. In the Riemannian case, such condition implies that the dual and the antiself dual part of the representation are the same, i.e. $\rho_f=(j, j)$. Instead, for the Lorentzian case, since irreducible unitary representations in the principal series are characterised by a pair $(n, p)$, where $n$ is a natural number, while $p$ a real number, the simplicity constraints translate to the condition that the representations are of the form $\rho_f=(0,p)$ or $\rho_f=(n,0)$. \\
By considering all possible representations, the Hilbert space associated to a single triangle is
\be
\mh_t=\oplus_j\mh^{(j,j)}\hspace{.5in}\mh_t=\oplus_n\mh^{(n,0)}\oplus_p\mh^{(0,p)}
\ee
for the Riemannian case and Lorentzian case, respectively.\\

Now that we have associated Hilbert spaces to each triangle, we can define the Hilbert space associated to a tetrahedron by tensoring the Hilbert spaces of the 4 triangles comprising the tetrahedron, which we call the tensor product Hilbert space.

\begin{figure}[htb]
 \begin{center}
 \psfrag{a}{$J_1$}\psfrag{c}{$J_2$}\psfrag{e}{$J_3$}\psfrag{f}{$J_4$}\includegraphics[scale=0.5]{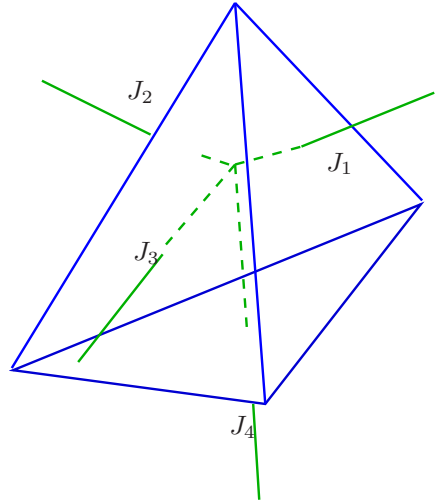}
\caption{The dual vertex associated to a tetrahedron in 4-dimensions. The links are labelled by representations.\label{fig:BCint} }
\end{center}
  \end{figure}

In this context the quantum space associated to a tetrahedron is an element of the tensor product of Hilbert space. However, there are certain constraints on the tensor product Hilbert space coming from both the simplicity constraint and the closure constraint. In particular, the simplicity constraints that refer to triangles sharing a common edge imply that the tensor product representation decomposes only into simple representations. \\
On the other hand, the closure constraint imposes the condition that the tensor product Hilbert space be the space of invariant tensors. Therefore, the Hilbert space of a tetrahedron is
\be
\mh=Inv(\mh_1\otimes\mh_2\otimes\mh_3\otimes\mh_4)
\ee 
where the individual $\mh_i$ are the Hilbert spaces associated to the four triangles comprising the tetrahedron. Each quantum state associated to a tetrahedron will, then, be an intertwiner of the four simple representations associated to the four triangles comprising the tetrahedron, i.e. $\phi: \mh_1\otimes\mh_2\otimes\mh_3\otimes\mh_4\rightarrow \Cl$. \\
Such intertwiners are called the Barrett-Crane intertwiners \cite{27,d111,d118}. A graphical characterisation of such an intertwiner is given in figure \ref{fig:BCint}.\\
The quantum states associated to each individual tetrahedron represent the building blocks for a general quantum state. Such states will be elements of the Hilbert spaces defined as the tensor product of the Hilbert spaces associated to a collection of tetrahedrons, i.e. 
\be
\mh_{\text{general}}:=\otimes_i Inv(\mh_{1i}\otimes\mh_{2i}\otimes\mh_{3i}\otimes\mh_{4i})
\ee
In particular $\mh_{\text{general}}$ will be a product of intertwiners for each tetrahedron $i$ with a sum over the labels coming from common triangles, shared by two tetrahedrons. It is straightforward to recognise a state living in $\mh_{\text{general}}$ as a spin network functions with edges (dual to triangles) labelled by representations and vertices (dual to tetrahedrons) labelled by intertwiners.\\
In this context, a single 4-simplex $\Delta$ will be the basic amplitude between quantum states referred to single tetrahedrons, i.e.
\be
F_{\Delta}:=\otimes_i Inv(\mh_{1i}\otimes\mh_{2i}\otimes\mh_{3i}\otimes\mh_{4i})\rightarrow \Cl
\ee
In terms of the intertwiners, $F_{\Delta}$ can be written as 
\be
F_{\Delta}=I^{\rho_{t_1}\rho_{t_2}\rho_{t_3}\rho_{t_4}}I^{\rho_{t_4}\rho_{t_5}\rho_{t_6}\rho_{t_7}}I^{\rho_{t_7}\rho_{t_8}\rho_{t_9}\rho_{t_{10}}}
\ee
 where the $\rho_{t_i}$ are the representations associated to the triangles $t_i$ (or dual faces $f^*(t_i)$). $F_{\Delta}$ represents the 10j-symbol.
The amplitude for a general quantum state is then given by a product of individual amplitude for each single 4-simplex, each glued along common tetrahedron.\\ As we can see, through geometrical quantisation, it is possible to view single 4-simplices as the main building blocks for defining transition amplitudes between spin networks. \\
The general form of the resulting amplitude, for non-fixed triangulations $\tau$ can be written as follows:
\be
Z=\sum_{\tau}Z(\mm, \tau)=\sum_{\tau}\sum_{\rho_{t}}\prod_{t\in C_2(K)}A_t\prod_{\Gamma\in C_3(K)}A_{\Gamma}\prod_{\Delta\in C_4(K)}A_{\Delta}
\ee
or in terms of the dual triangulation $\tau^*$
\be\label{equ:partfunc2}
Z=\sum_{\tau^*}\sum_{\rho_{f^*(t)}}\prod_{f^*(t)\in C_2(K^*)}A_{f^*(t)}\prod_{e^*\in C_1(K^*)}A_{e^*}\prod_{v^*\in C_0(K^*)}A_{v^*}
\ee 
which is of the form of \ref{equ:partfunc}.
\\ Therefore, through quantum simplicial geometry we derive an amplitude, whose form is analogous to \ref{equ:partfunc}. \\

In order to render \ref{ali:first} in the form of \ref{equ:partfunc} or \ref{equ:partfunc2} where the amplitude of each individual 4-simplex is considered independently, one introduces a refinement of the dual triangulation of the manifold in terms of the so called \emph{wedges}. \\
Essentially, a \emph{wedge} is the portion of the dual face which lies inside a 4-simplex. In particular, we know that dual edges $e^*\in K^*$ connect the barycentre $v=\hat{\sigma}^{(4)}$ of the 4-simplex with the barycentre $v^{'}=\hat{\sigma}^{(4)'}$ of a neighbouring 4-simplex through the barycentre $b_i=\hat{\sigma}^{(3)}$ of their common tetrahedron. Therefore, each edge $e^*\in K^*$ can be seen as composed of two edges $e^*=[v,v^{'}]=[v,b_i]\circ[b_i,v^{'}]=e^{*v}\circ (e^{*v^{'}})^{-1}$.

From the geometry of the dual triangulation $K^*$, it follows that each barycentre $v$ of a 4-simplex has five half (dual) edges incident at it $e_i^{*v}=[v, b_i]$, $i=0,\cdots,4$ labels the 5 boundary tetrahedrons $\sigma^{(3)}_i$ for each 4-simplex. If we then consider for $i<j$ the boundary triangle $\sigma^{(2)}_{ij}=\sigma^{(3)}_i\cap \sigma^{(3)}_j$ with barycentre $b_{ij}:=\hat{\sigma}{(2)}_{ij}$, a \emph{wedge} is defined to be the 2 dimensional polyhedron composed of the triangle $[v,b_i,b_{ij}]\cup [v, b_{ij}, b_j] $, which belongs to the baryonic refinement $B(K)$ of $K$ and  bounded by the loop $[v,b_i]\circ[b_i, b_{ij}]\circ[b_{ij}, b_j]\circ[b_j,v] $. \\
The collection of all wedges based at the barycentre $v$ of each 4-simplex is called a \emph{fundamental atom}. 

\begin{figure}[htb]
 \begin{center}
 \psfrag{E}{$J_{w}$}\psfrag{A}{$v$}\psfrag{D}{$b_i$}\psfrag{C}{$b_{ij}$}\psfrag{B}{$b_j$}\psfrag{a}{$g_{e_1}$}\psfrag{b}{$g_{e_2}$}
 \psfrag{c}{$g_{e_3}$}\psfrag{d}{$g_{e_4}$}
  \includegraphics[scale=0.5]{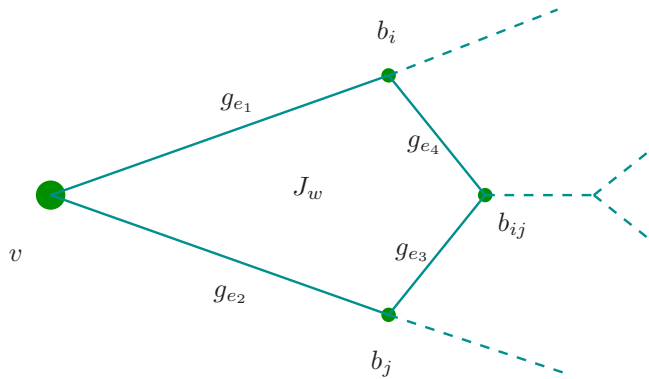}
\caption{A wedge formed by the half edges joining the barycentre $v $ of the 4-simlplex to the barycentre $b_i$ and $b_j$ of two boundary tetrahedrons through the barycentre $b_{ij}$ of their common triangle. $J_{w}$ is the representation associated to the wedge w\label{fig:wedge} }
\end{center}
  \end{figure}

It is straightforward to see that each dual face $f^*$ is composed out of those wedges which have the barycentre $b_{ij}$ in common. A graphical representation of a wedge is given in \ref{fig:wedge}.\\
The aim is now to express the boundary $\partial f^*$ of the dual face in terms of wedges. To this end, let us suppose that $\partial f^*=e^*_1\circ\cdots\circ e^*_n$, where each $e^*_k=[v^*_k, v^*_{k+1}]$, $k=1,\cdots, n$ with $v_{n+1}^*=v_{1}^*$. $b_k$ and $b_{k+1}$ are the barycentres of the tetrahedron shared by the 4-simplices dual to $v^*_k$ and $v_{k+1}^*$, respectively, while $b_{f^*}$ is the barycentre of the face $f^*$. We can then write $\partial f^*=\partial w_n\circ\partial w_{n-1}\circ\cdots\circ\partial w_1$ where each wedge $w_k=[b_f,b_k]\circ [b_k,v_k]\circ [v_k,b_{k-1}]\circ[b_{k-1}, b_f]$ for $k=1,\cdots, n$ with $b_0=b_n$. \\
It can then be shown that \cite{1b}
\be
\delta(U\partial(f^*))=\int\Big[\prod_{e^*\in E_{f^*}}^nd\mu_{H}(U([b_f,b_k]))\Big]\Big[\prod_{e^*\in E_{f^*}}\delta(U(\partial w_k)\Big]
\ee
Since a wedge is given by $w=[v^*,b_i]\circ[b_i, b_{ij}]\circ[b_{ij}, b_j]\circ[b_j,v^*] $, it follows that there exists a 1:2:1 correspondence between dual vertices $v^*$ and wedges, or alternatively, a 1:2:1 correspondence between \emph{wedges} and the faces dual to triangles with barycentre $b_{ij}$. This correspondence allows for a regrouping of the partition functions in terms of dual faces, dual edges and dual vertices. Therefore, omitting for the time being the simplicity constraint, we obtain the following expression:
\be
Z_P=\int\Big[\prod_{f^*\in C_2(K^*)}\prod_{e^*\in E_{f^*}}d\mu_H(U([b_f, b_k]))\Big]\prod_{v^*\in C_0(K^*)}\Big[\int\prod_{i=0}^4d\mu_H(U(e_i^v))[\prod_{w_k\text{ incident at } v^*}\delta(U(\partial w_k))]\Big]
\ee
 Now we have to impose the simplicity constraint as expressed for individual wedges. \\
To this end it is useful to note that 
\be
\prod_{w_k\text{incident at }v^*}\delta(U(\partial w_k))=\int\prod_{w_k\text{incident at }v^*}d^6[\int_{w_k}B]e^{iTr(\int_{w_k}BU(\partial w_k)}
\ee
is the delta distribution one obtains if the Plebanski action had been discretised directly on one 4-simplex only, and summed over all possible 4-simplices. In this context, the simplicity constraint is only imposed on triangles of each 4-simplices individually. \\
By applying the same approximations as done above we obtain
\ba
& &\int\prod_{w_k\text{incident at }v^*}d^6[\int_{w_k}B]\prod_{\alpha_v}\delta(C_{\alpha_v}(\{\int_{w_k}B\}))e^{iTr(\int_{w_k}BU(\partial w_k)}\nonumber\\
&\cong&\int\prod_{w_k\text{incident at }v^*}d^6[\int_{w_k}B]\prod_{\alpha_v}\delta(C_{\alpha_v}(\{\int_{w_k}X\}))e^{iTr(\int_{w_k}BU(\partial w_k)}\nonumber\\
&\cong&\prod_{\alpha_v}\delta(C_{\alpha_v}(\{\int_{w_k}X\}))\prod_{w_k\text{incident at }v^*}\delta(U(\partial w_k))
\ea
The constraint partition function thus becomes
\ba\label{ali:partfunc3}
Z_P&=&\int\Big[\prod_{f^*\in C_2(K^*)}\prod_{e^*\in E_{f^*}}d\mu_H(U([b_f, b_k]))\Big]\prod_{v^*\in C_0(K^*)}\prod_{\alpha_v}\delta(C_{\alpha_v}(\{\int_{w_k}X\}))\nonumber\\
& & \times \Big\{\int\prod_{i=0}^4d\mu_H(U(e_i^v)) [\prod_{w_k\text{ incident at } v^*}\delta(U(\partial w_k))]\Big\}
\ea
where the term in curly brackets represents the vertex amplitude. By rearranging the various terms and performing all the integrals inside the curly brackets, it is possible to show that \ref{ali:partfunc3} is exactly of the form of \ref{equ:partfunc}, \cite{1b}. This partition function, however, is no longer triangulation independent. The resulting spin foam model is called the \textbf{Barret-Crane} model.\\

Although the partition function \ref{ali:partfunc3} has the desired form of a partition function for a general BF-theory, the derivation of it is far from rigorous. We will now list the main conceptual and mathematical issues present in the Barrett-Crane model.  
\begin{itemize}
\item[1)] The Barrett-Crane model does not take into account the second class constraints present in both the Plebanski and Palatini action. As it was shown in \cite{48c}, if such constraints are taken into account, the measure present in the partition function should be augmented with a Jacobian coming from the Dirac brackets of the second class constraints.
\item [2)] There is no mathematical reason to consider only one solution of the simplicity constraint, ignoring the remaining four. In particular, if each solution was weighted with equal probability, even if the path integral was dominated by the classical configuration we would still not obtain the Palatini action. Thus, the correspondence between the Plebanski action and the Palatini action in the Barrett-Crane model is unnatural.
\item [3)] The simplicity constraints are inserted in by hand, rather than derived from integrating over the Lagrangian multiplier.
\item[4)] The $B$ field is substituted with the vector fields $X$ on the group before the integration over the $B$ field. This is done because one assumes, \emph{a priori}, the flatness of the connection. Such assumption is not justified at this stage.
\item [5)] The term $\prod_{f\in C(K^*)}\delta(U(\partial(f^*)))$ has support also over configuration with non flat connection. Such configurations are ignored.
\item[6)] The interaction terms are neglected. The constraints are only applied to individual 4-simplices separately.
\item[7)] Gauge invariance is lost when discretising the $B$ filed over the triangles of the simplicial complex $\tau$ and $F$ over the dual faces $f^*(t)$. This issue will be explained in more detail in Section \ref{s5.1.1}. 
\end{itemize}

It should be noted that, as for the 3-dimensional case, a strategy to solve the triangulation dependence of the partition function is through group field theory. A description of group field theory and its applications to spin foam models is given in Chapter \ref{chap:gft}. \\
In Section \ref{shsfm} we will describe a proposal made by the author and collaborators of an alternative model of spin foam, \cite{30a}.
\section{N-Point Functions}
\label{s5.2}

In the SFM literature, the first task that one addresses is the computation 
of the partition function. However, the partition function itself has 
no obvious physical meaning even if one imposes boundary conditions on 
the paths
(spin foams) to be integrated (summed) over. The hope is that  
SFM provide a formula for the physical inner product
of the underlying constrained canonical theory which starts from some 
kinematical Hilbert space $\cal H$. The purpose of this section is to 
sketch the connection between path integrals and n -- point functions 
for a general constrained theory. We will use reduced phase space 
quantisation as our starting point.

Although the quantisation process of a classical system with constraints was already described in 
Section \ref{s outline}, nonetheless, for pedagogical reasons, we will briefly summarise it in the following. 
\\
We assume that we are given a classical theory with first class 
constraints $\{F\}$ and possibly 
second class 
constraints $\{S\}$. We turn the system into a purely second class 
system by supplementing $\{F\}$ with suitable gauge fixing conditions 
$\{G\}$. The canonical Hamiltonian $H_c$ is a linear combination of the
primary constraints plus a piece $H'_0$ non -- 
vanishing on the constraint surface of the primary constraints (it 
could be identically zero). 
It can also be written as a first class piece $H_0$ and (some of) the 
first class constraints $F$. The 
gauge fixing conditions fix the Lagrange multipliers involved in the 
canonical Hamiltonian. One may split the complete set of canonical 
pairs $(q,p)$ on the full phase space into two sets $(\phi,\pi),\;(Q,P)$,
such that one can solve the system $S=F=G=0$, which 
defines the constraint surface for 
$(\phi,\pi)=f(Q,P)$ in terms of $Q,P$. The $Q,P$ are coordinates
on the reduced phase space which is equipped with the pull -- back 
symplectic structure\footnote{This symplectic structure coincides with the 
pull -- back of the degenerate symplectic structure on the 
full phase space corresponding to the 
Dirac bracket induced by the system $\{S,F,G\}$ \cite{420}.} 
induced by the embedding of the constraint surface specified by $f$. 

The gauge fixing conditions also induce a reduced Hamiltonian $H_r$ which 
only depends on $Q,P$ and which arises by computing the equations of 
motion 
for $Q,P$ with respect to $H_c$ and, then, restricting them to the gauge 
fixed values of the Lagrange multipliers and to the constraint surface.
Then $H_r$ is defined as the function of $Q,P$ only\footnote{For 
simplicity, we are 
assuming a gauge fixing which leads to a conservative reduced 
Hamiltonian.}, which generates these same equations of motion.
We are now in the situation of an ordinary Hamiltonian system equipped 
with a true Hamiltonian $H_r$. We quantise a suitable subalgebra of the
reduced Poisson algebra as a $\ast-$algebra $\mathfrak{A}$ and represent 
it on a Hilbert space $\cal H$. This Hilbert space is to be identified 
with the physical Hilbert space arising from reduced phase space 
quantisation. Let 
$t\mapsto U(t)$ be the unitary 
evolution induced by $H_r$, then the object of interest is the 
transition 
amplitude or n-point function
\be \label{5.14}
<\psi_f,U(t_f-t_n) a_n U(t_n - t_{n-1}) a_{n-1} .. U(t_2-t_1) a_1 
U(t_1-t_i)\psi_i>
\ee
between initial and final states $\psi_i,\psi_f$ at initial 
and final times $t_i,t_f$, respectively, with 
intermediate measurements of the operators $a_1,..,a_n\in \mathfrak{A}$ at 
$t_1<t_2<..<t_n$. 

Preferably one would like to be in a situation in which there is a
cyclic vector $\Omega$ for $\mathfrak{A}$ which is also a ground state 
for $H_r$. The existence of a cyclic vector is 
no restriction because representations of $\mathfrak{A}$ are always 
direct 
sums of cyclic representations. In this case $\mathfrak{A}$ is dense 
in $\cal H$ and we may, therefore, restrict attention to 
$\psi_i=\psi_f=\Omega$ by choosing appropriate $a_1,..,a_n$ in 
(\ref{5.14}). The existence of a vacuum state for $H_r$ means that 
zero is in the point spectrum of $H_r$. For 
simplicity, let us make this assumption. 

Let us abbreviate the 
Heisenberg time evolution as
$a_k(t):=U(t)^{-1}a_k U(t)$. 
In principle it  
would be sufficient to restrict the $a_k$ to 
be configuration operators $Q$ because their time evolution
contains sufficient information about $P$ as well. However, we will
stick to the more general case for reasons that will become clear later.
This gives us the n-point function
\be \label{5.15}
S(t_1,..,t_n):=\frac{<\Omega,U(t_f) a_n(t_n).. a_1(t_1) U(-t_i)\Omega>} 
{<\Omega,U(t_f-t_i)\Omega>} 
\ee
where we have properly normalised so to attain, for the 0 -- point function, the 
value unity. This has the advantage that certain infinities, that 
would otherwise arise, can be absorbed. Notice that since $\Omega$ is a 
ground state, the $U(t_f)$ and $U(t_i)$, as well as the 
denominator, could be dropped in (\ref{5.15}).

Now a combination of well known heuristic arguments \cite{420}, \cite{424},
reviewed 
in 
\cite{49b}, reveals the following:\\
consider any initial and final configuration $q_i,q_f$ on the full
phase space and denote by ${\cal P}((t_i,q_i),(t_f,q_f))$ the set 
of paths\footnote{This should be a suitable measurable space but 
we leave it unspecified.} in full configuration space between 
$q_i,q_f$ at times $t_i,t_f$, respectively. Consider
\be \label{5.25}
Z[j;q_i,q_f]=\lim_{-t_i,t_f\to \infty}
\int_{{\cal P}((t_i,q_i),(t_f,q_f))} \; 
[Dq\;Dp\;D\lambda\;D\mu]\;\delta[G]\;|\det[\{F,G\}]|\;\rho\;
e^{\frac{i}{\hbar} 
S[q,p,\lambda,\mu]}
\;e^{i\int_{t_i}^{t_f}\; dt\; j(t)\cdot q(t)}
\ee
Here $j$ is a current in the fibre bundle dual to that of $q$, 
$S[q,p,\lambda,\mu]$ is the canonical action, after performing the 
singular Legendre transform from the Lagrangian to the Hamiltonian 
formulation\footnote{The Lagrange multipliers $\lambda,\mu$ of the 
primary, 
first and 
second class constraints, respectively, play the role of the velocities 
which can not be solved in terms of the momenta in the process of 
the Legendre transform.}, and $\rho$ is a local function of $q,p$, which
is usually related to the Dirac bracket determinant $\det[\{S,S\}]$ 
\cite{424}.

Now, the primary constraints are always of the
form $\pi=f(Q,P,\phi)$ where we have split again the canonical pairs into
two groups. Thus, $S[q,p,\lambda,\mu]$
is linear in those momenta $\pi$ and we can integrate them out yielding
$\delta$ distributions of the form
$\delta[\lambda - (.)]\; \delta[\mu - (.)]$, which can be solved by
integrating over $\lambda,\mu$. If we assume that the dependence
of the remaining action on $P$ is only quadratic and that $G$ and
$|\det[\{F,G\}]|$ are independent of $P$, then we can integrate also over
$P$ which yields in general a Jacobian $I$ coming from the Legendre
transform. We can then write (\ref{5.25}) as
\be \label{5.25a}
Z[j;q_i,q_f]=\lim_{-t_i,t_f\to \infty}
\int_{{\cal P}((t_i,c_i),(t_f,c_f))} \;
[Dq]\;\delta[G]\;|\det[\{F,G\}]|\;\rho\;I\;
e^{\frac{i}{\hbar}
S[q]}
\;e^{i\int_{t_i}^{t_f}\; dt\; j(t)\cdot q(t)}
\ee
where proper substitutions of $\pi$, derived from solving the primary constraints
and of $P$ derived from the Legendre transformation, are understood. Here $S[q]$
is the original (covariant) Lagrangian action.

Defining $\chi[j]:=\frac{Z[j;\;q_i,q_f]}{Z[0;\;q_i,q_f]}$, the 
covariant or path integral n -- point 
functions 
\be \label{5.26}
S(t_1,..,t_n):=[\frac{\delta^n \chi[j]}{i^n \delta j(t_1) .. \delta 
j(t_n)}]_{j=0}
\ee
have the canonical or physical interpretation of 
\be \label{5.27}
<\Omega, T(a_1(t_1)..a_n(t_n)) \Omega>
\ee
where $T$ is the time ordering symbol, $\Omega$ is the aforementioned 
cyclic 
vacuum vector 
defined by the physical (or reduced) Hamiltonian $H_r$ induced by the 
gauge fixing $G$, $a_k(t)$ is the Heisenberg operator at time $t$
(evolved with respect to $H_r$)
corresponding to $a_k$ and $a_k$ classically corresponds to 
a component of $q$ 
evaluated on the constraint surface $S=F=G=0$. \\
The scalar product 
corresponds to a quantisation on the reduced phase space defined 
by $G$. Notice how the gauge 
fixing condition $G$ (or choice of clocks) prominently finds its way 
both into the 
canonical theory and into the path integral formula (\ref{5.25a}). 
In particular, notice that the seemingly similar expression
\be \label{5.28}
Z'[j;q_i,q_f]=\lim_{-t_i,t_f\to \infty}
\int_{{\cal P}((t_i,q_i),(t_f,q_f))} \; 
[Dq]\;
e^{\frac{i}{\hbar} 
S[q]}
\;e^{i\int_{t_i}^{t_f}\; dt\; j(t)\cdot q(t)}
\ee
does not have any obvious physical interpretation and, in addition, lacks 
the important measure factors $\rho,\;I$. \\
\\
Remarks:
\begin{itemize}
\item[1.]
One may be puzzled by the following: from ordinary gauge theories on  
background spacetimes such as Yang -- Mills theory on Minkowski space 
the path integral, or more precisely, the generating functional of the 
Schwinger functions (in the Euclidian formulation) does not require 
any gauge fixing in order to give the path integral a physical 
interpretation. One needs it only in order to divide out the gauge 
volume in a systematic way (Fadeev -- Popov identity), while the 
generating functional  
is independent of the gauge fixing. The gauge fixing also does not 
enter the construction of gauge invariant functions (such as Wilson 
loops). In our case, however, the gauge 
fixing condition is actually needed in order to formulate the physical 
time evolution and the preferred choice of gauge invariant functions on 
phase space. 

The difference between Yang -- Mills  
theory and, generally covariant systems, such as General Relativity,
that we are interested in here is that in GR the canonical 
Hamiltonian is in fact the generator of gauge 
transformations (spacetime diffeomorphisms) rather than physical time 
evolution and it is constrained to vanish. In contrast, in Yang -- 
Mills theory there is a preferred and gauge invariant Hamiltonian 
which is not constrained to vanish.  Thus, in order to equip the theory at hand 
with a notion of time, we have 
used the relational framework discovered in \cite{425}, which consists in 
choosing fields as clocks and rods with respect to which other fields 
evolve. Mathematically this is equivalent to a choice of gauge fixing.
Hence, in our case the gauge fixing plays a dual role: i) it
renders the generating functional less singular and ii) it defines physical time evolution. 
\item[2.]
The appearance of the $\delta$ distributions and functional 
(Fadeev -- Popov) determinants 
in (\ref{5.25}) indicates that we are not dealing with an ordinary 
Hamiltonian system, but rather with a constrained system. One can, in fact, 
get rid of the gauge fixing condition involved if one pays a price. The 
price is 
that if one considers instead of $q$ its gauge invariant extension
$\tilde{q}$ off the surface $G=0$ \cite{420,422}, then, since we 
consider
the quotient $Z[j]/Z[0]$ which leads to {\it connected n -- point 
functions} by the usual Fadeev -- Popov identity that exploits 
gauge invariance, we may
replace \cite{49b} (\ref{5.25}) by
\be \label{5.24}
\tilde{Z}[j,q_i,q_f]=
\int_{{\cal P}((t_i,q_i),(t_f,q_f))} \; [Dq\;Dp\;D\lambda\;D\mu]\;\rho 
e^{\frac{i}{\hbar} 
S[q,p,\lambda,\mu]}
\;e^{i\int_{t_i}^{t_f}\; dt\; j(t)\cdot \tilde{q}(t)}
\ee
However, (\ref{5.24}) is not very useful unless $\tilde{q}(q,p)$ is easy 
to calculate, which is typically not the case. Hence, we will refrain 
from doing so.  Nevertheless, no matter whether one deals with 
(\ref{5.25}) or (\ref{5.24}), the correlation functions depend on the 
gauge fixing $G$ or, in other words, on the choice of the clocks 
\cite{422,423} with respect to which one defines a physical reference 
system.
\item[3.] The correspondence between (\ref{5.26}) and 
(\ref{5.27}) also allows to reconstruct the physical inner product
from the n -- point functions: given arbitrary states $\psi,\psi'\in 
{\cal H}$ 
we find $a,a'\in \mathfrak{A}$ such that 
$||a\Omega-\psi||,\;||a'\Omega-\psi'||$ are arbitrarily small. Now pick 
any $t_i<t_0<t_f$, then
\be \label{5.27a}
<a\Omega,a'\Omega>=<\Omega,a^\dagger\;a'\Omega>
\ee
By assumption, the operator $a^\dagger a'$ can be written as a finite 
linear combination of monomials of homogeneous degree in the 
components of the operator $q$ which we write, suppressing indices for 
the components, as $q^n$. Then
\be \label{5.27b}
<\Omega, q^n \Omega>=\lim_{t_1,..,t_n\to t_0; 
t_n>..>t_1}\; <\Omega,q(t_n)..q(t_1)\Omega>
\ee
which can be expressed via (\ref{5.26}).
The existence of this coincidence limit of n -- point functions is 
often problematic in background dependent Wightman QFT, \cite{Haag} but 
their existence is actually the starting point of canonical quantisation
of background independent non -- Wightman QFT, as one can see from the 
identity (\ref{5.27b}).

\end{itemize}

\section{The Holst Spin Foam Model Via Cubulations}\label{shsfm}
In the previous Section we have shown that there are various issues in the Barrett-Crane model that need to be addressed. In this Section we will show how some of these issues can be solved if a slight departure from the model is taken. In particular, differently form the Barrett-Crane model, our starting point will be the Holst action \cite{7}. The advantage of starting from this action is that the simplicity constraints are explicitly solved, since one works entirely with tetrads from the beginning. 

More 
precisely, the Holst 
action uses a specific quadratic expression in the tetrads for the B 
field of BF-theory, which also depends on the Immirzi parameter 
\cite{8}. Hence, the Holst action depends on a specific, non degenerate 
linear 
combination of the four non degenerate solutions of the simplicity 
constraints and it is, thus, at the same time, 
more general and more restricted because the Holst path integral will 
not sum over the aforementioned five sectors of Plebanski's theory. 
It is debated how, the fact that one actually takes a sum over all histories 
with a mixture of positive and negative Palatini and topological actions, 
affects the semiclassical properties of the Plebanski path integral.

As observed in \cite{47a}, since the Holst action is quadratic in the 
tetrads one can, in principle, integrate out the tetrad in the resulting 
Gaussian integral. This has been 
sketched in \cite{47a}, however, the expressions given there are far from 
rigorous. In \cite{30a} we gave a rigorous expression where 
the correct measure factor \cite{48a}, resulting from the second class 
constraints 
involved in the Holst action, was included. This inclusion made sure that the path integral 
qualified as a reduced phase space quantisation of the theory, as it has 
been stressed in \cite{48b}. A similar analysis has been carried out 
for the Plebanski theory in \cite{48c}, however, the resulting measure 
factor is widely ignored in the SFM literature. The result of the 
Gaussian integral is an interesting determinant that displays the full
non linearity of Einstein's theory. When translating the remaining 
integral over the connection in the partition function into SFM 
language, that is, sums over 
vertex, edge and face representations, one sees that our model (\cite{30a}) differs 
drastically from all current SFM. 

The main observations, which led us to depart 
from the usual SFM approach where one works with simplicial cell 
complexes and define the cubulated SFM, are:
\begin{enumerate}
\item [1.]
In  \cite{30b, 46ah} it was demonstrated that 
current semiclassical states used in LQG do not assign good classical 
behaviour to the volume operator \cite{v7, 37} of LQG, unless the underlying 
graph has 
cubic topology. The fact that the volume operator plays a pivotal role for 
LQG because it defines triad operators and hence the dynamics, motivates the choice 
of  cubic 
triangulations (also called ``cubulations'') of the four manifold. 
Notice that any four manifold can be cubulated and that within each 
chart of an atlas the cubulation can be chosen to be regular (see e.g. 
\cite{v7, 37} and references therein). 
\item [2.]
The original motivation for considering 
simplicial 
cell complexes in current SFM comes from their closeness to BF-theory.
BF-theory is a topological QFT and, therefore, one would like to keep the
triangulation independence of the BF-SFM amplitude. That this is 
actually true is a celebrated result in BF-theory. In 
particular, in order to keep the triangulation independence, it is 
necessary to integrate the $B$ field over the triangles $t$ of the 
triangulation and the $F$ field over the faces $f$ bounding the loops in a 
dual graph
\cite{49}. However, GR is not a TQFT and, therefore, the requirement to 
have triangulation independence is somewhat obscure. Of course it is 
natural if one wants to exploit the properties of BF-theory but not if 
one takes a different route as we did in \cite{30a}. Hence, if we drop
that requirement, then it is much more natural to refrain from 
considering the dual graph in addition to the triangulation.

\item [3.]
The gauge group $SO(p,q)$ acts on the B field of BF-theory 
by the 
adjoint 
action and, on the connection $A$ underlying $F$, in the usual way. The 
question is where the gauge transformation acts on the discretised 
variables $B(t),\;A(\partial f)$ (flux and holonomy). It would be 
natural to have the gauge group act at the barycentres of $t$ and at 
the starting point of the loop $\partial f$, which will be a vertex of 
the dual graph. However, notice that the vertices of the dual graph and 
the triangles are disjoint from each other, since the edges of the graph are 
dual to the tetrahedra of the cell complex. Hence, at the level of 
the action,
{\it local} gauge invariance in discretised BF-theory 
is not manifest and, even less, in Plebanski theory. In fact, gauge 
invariance is related to the closure constraint in SFM which, as we will see, is a 
subtle issue. If one 
works just with 
a triangulation and drops the dual graph, then gauge invariance issues 
are 
easy to take care of. 
Hence, it is desirable to work with a triangulation that maximally 
simplifies the Gaussian integral. As we will show, this again leads to 
cubulations. This also nicely fits with the framework of Algebraic 
Quantum Gravity \cite{45}, \cite{418}, \cite{420} which, in its minimal version, is also 
formulated in terms of algebraic graphs of cubic topology only.
\end{enumerate}

It is also appropriate to mention further constraints in SFM, namely:
\\
SFM rely on a simplicial 
triangulation $\tau$ of the differential 4-manifold, as well as a dual 
graph $\tau^\ast$.
However, as shown in \cite{46a}, if one freely specifies the geometrical 
data (areas or fluxes)
on the faces of $\tau$, then inconsistencies in the values of 
the lengths of the edges of $\tau$ occur, unless so called Regge constraints, in 
addition to the 
simplicity constraints, are imposed. The underlying reason for these 
constraints is that Regge calculus is formulated directly in terms
of edge lengths, while in SFM one rather works with electrical fluxes 
or areas. However, 
a typical simplicial triangulation has far more faces than edges in 
$\tau$, so that assigning a length to an edge from given area values 
maybe ambiguous and/or inconsistent.\\
The imposition of such constraints is important for two reasons: i) if one wants to relate SFM to the 
established 
theory of Regge calculus \cite{46aa}; and ii) to capture the correct 
semiclassical limit. 

In fact, we recall that the underlying reason for these 
constraints is that Regge calculus is formulated directly in terms
of edge lengths, while in SFM, one rather works with electrical fluxes 
or areas. However, 
a typical simplicial triangulation has far more faces than edges in 
$\tau$, so that assigning a length to an edge from a given area value 
might be ambiguous and/or inconsistent. However, in the cubulated spin foam model developed in \cite{30a}, there is no necessity 
to relate it to the Regge action since the path 
integral is explicitly based on the Holst action.

\subsection{Cubulations}
\label{s5.1}

We will now describe the alternative spin foam model via the cubulation of the Holst action developed in \cite{30a}. As a first step we will analyse, in more detail, the reasons for adopting cubulations rather than simplicial
triangulations.

\subsubsection{Gauge invariance}
\label{s5.1.1}

Let us look more closely at the issue of gauge 
invariance for BF-theory. Here 
gauge invariance is not 
preserved 
locally (i.e. triangle wise) in the 
formula
$\int Tr(B\wedge F)=\sum_t Tr(B(t)F(f(t))$ if both $B$ and $F$ transform 
locally in the adjoint representation. In order to make the gauge 
transformations more local, one could discretise them. 
To see how this can be achieved, recall 
that by 
definition of a cell dual to a simplex\footnote{Recall that an 
n-simplex is denoted by $[p_0,..,p_n]$ where the points $p_i$ denote its
corners \cite{49a}, \cite{29}.}  
in a 
simplicial complex $\tau$, 
the face 
$f(t)$ is a union of triangles $[\hat{t},\hat{T},\hat{\sigma}]$ 
subject to the condition $t\subset \partial T,\;T\subset \partial \sigma$.
Here 
$\hat{(.)}$ denotes the barycentre (\cite{49a}, \cite{29}) of a simplex and 
$T,\;\sigma$ denote the
tetrahedra and four simplices in $\tau$, respectively. Both 
$t$ and $f(t)$ contain the barycentre $\hat{t}$ in their intersection, therefore we could {\it define} a disjoint action of the gauge group on
both $B(t),\;F(f(t))$ at $\hat{t}$. However, this is no longer 
possible when  
using the approximation
$\sum_t Tr(B(t) A(\partial f(t)))$ because now the only natural action 
of the gauge group on the loop holonomy is by adjoint action at a 
starting point on $\partial f(t)$. Now $\partial f(t)$ is a composition of
the half edges 
$[\hat{T},\hat{\sigma}]$ where $ T\subset \partial \sigma$, $(t\subset \partial T)$, but the fundamental degrees of freedom are the holonomies along 
the edges $e=[\hat{\sigma},\hat{\sigma'}]$ for $\sigma\cap 
\sigma'=T,\;(t\subset \partial T)$. 

Obviously, the only natural starting 
point of the loops is then at the vertices $\hat{\sigma}$ which are 
disjoint from the triangles $t$. But the triangles are also disjoint 
from the half edges, as a simple calculation reveals. Hence, in order 
to maintain gauge invariance one has to invent an unnatural 
discretised action 
of the gauge group. We do not know if such a consistent prescription 
can be found at all.

However, these complications that come from the 
fact that one is dealing simultaneously with a (simplicial) complex and 
its dual cell complex, are an ulterior motivation to work just with 
the triangulation.      

\subsubsection{Cubulations versus simplicial triangulations}
\label{s5.1.2}

The previous considerations do not specify the type of triangulations to 
be considered. As already said, the first motivation to 
use cubulations rather than simplicial triangulations is because the 
boundary graphs must contain cubical ones, in order to make sure 
that the corresponding boundary Hilbert space contains enough 
semiclassical states \cite{30b, 46ah}. However, there is an additional, 
more 
practical motivation for doing so which we are about to discuss.

Recall that the Holst action is given by 
\be \label{5.1}
S=-\frac{1}{\kappa}\int_M\; {\rm Tr}(G[A]\wedge e\wedge e)
=\frac{1}{\kappa}\int_M\; G_{IJ}[A]\wedge e^I\wedge e^J
\ee
Here $\kappa$ denotes Newton's constant
\be \label{5.2}
G[A]=2(\ast F[A]+\frac{1}{\gamma} F[A])
\ee
where 
$F_{IJ}=dA_{IJ}+A_{IK}\wedge A^K\;_{J}$ denotes the curvature of 
the connection $A$, $\gamma$ is the Immirzi parameter, and $\ast$ denotes 
the internal Hodge dual, that is, 
\be \label{5.3}
(\ast T)_{IJ}:=\frac{1}{2} \epsilon_{IJKL} \eta^{KM} \eta^{LN} T_{MN}
\ee
where $I,J,K,..=0,..,3$ and $\eta$ is the Minkowski or Euclidian metric
for structure group $G=SO(1,3)$ or $G=SO(4)$, respectively. As we have previously motivated, 
we plan to keep the co -- tetrad 1-forms $e^I$ rather 
than introducing a B field and thus 
the simplicity constraints are manifestly solved. 

In order to give 
meaning to a path integral formulation we consider a UV cutoff in terms 
of a triangulation $\tau$ of $M$ which we choose to be finite, thereby 
introducing an IR regulator as well.

Let us denote the two -- 
dimensional faces of $\tau$ by $f$ and the one dimensional edges of 
$\tau$ by $l$. We want to discretise (\ref{5.1}) in a manifestly (and 
locally) gauge invariant way, just using edges and faces. To do so we 
equip all edges with an orientation. Given an edge
$l$ consider
\be \label{5.4}
e^I_l:=\int_l \; [A(l(x))]^I\;_J e^J(x)
\ee
Here $l(x)$ for $x\in l$ denotes the segment of $l$ that starts at the 
starting point of $l$ and ends at $x$ and $[A(p)]^I\;J$ denotes the 
G valued holonomy
of $A$ along a path $p$. Under 
local gauge transformations
$g:\;M\to G$, 
(\ref{5.4}) transforms as $e^I_l\mapsto g^I\;_J(b(l))\; e^J_l$ where 
$b(l)$ denotes the beginning point of $l$. 

To avoid confusion, here $g\in G$ means the following: given the matrices $g^I\;_J$, set 
$\tilde{g}_{IJ}:=\eta_{IK}\; g^K\;_J$. Then $g\in G$ iff  
$\tilde{g}_{IK} \tilde{g}_{JL}\eta^{KL}=\eta_{IJ}$. This is 
equivalent with $(g^{-1})^I\;_{J}=\eta^{IL} g^K\;_L \eta_{KJ}$. In 
other words
\be \label{5.5}
\widetilde{(g^{-1})}=(\tilde{g})^T
\ee
If $g^I\;_J=[\exp(F)]^I\;_J$ 
for 
some generator $F^I\;_J$ then (\ref{5.5}) means that 
$\tilde{F}_{IJ}+\tilde{F}_{JI}=0$. With an abuse of notation one 
usually uses the same symbols $g,F$ and $\tilde{g},\tilde{F}$,  
respectively, but unless we are in the Euclidian regime we should pay 
attention to the index position.

Clearly, the curvature $F$ 
must be discretised in terms of the holonomy of $A$ along the closed 
loops $\partial f$ where we have also equipped the faces $f$ with a definite 
orientation. We have  
\ba \label{5.6}
F_{IJ}(f) 
&:=& \frac{1}{2}([\widetilde{A(\partial f)}]_{IJ}-
[\widetilde{(A(\partial 
f))^{-1}}]_{IJ})
\nonumber\\
&=& \frac{1}{2}([\widetilde{A(\partial f)}]_{IJ}-
[(\widetilde{(A(\partial 
f)})^T]_{IJ})
\nonumber\\
&=& \widetilde{A(\partial f)}_{[IJ]}
\nonumber\\
&\approx& \int_f \; F_{IJ}(x)
\ea
where we have used the non Abelian Stokes theorem for ``small'' 
loops, that is 
\be \label{5.7}
A(\partial f)\approx \exp(\int_f F)
\ee
and we have written $\tilde{F}_{IJ}(x):=F_{IJ}(x)$.
We may now define the antisymmetric matrix
\be \label{5.8}
G_{IJ}(f)=(\ast F(f))_{IJ}+\frac{1}{\gamma} F_{IJ}(f) 
\ee

If we imagine to use a simplicial triangulation, $M$ would be 
a disjoint (up to common tetrahedra) union of four simplices 
$\sigma=[p_0(\sigma),..,p_4(\sigma)]$. In this setting, for each $p_j(\sigma)$ we label the 
four boundary edges of $\sigma$ starting at $p_j(\sigma)$ by 
$l_\mu^j(\sigma)$ and the face (triangle) of $\sigma$ spanned by  
$l_\mu^j(\sigma)$ and $l_\nu^j(\sigma)$ are labelled by 
$f_{\mu\nu}^j(\sigma)$ with the convention 
$f_{\mu\nu}^j(\sigma)=-f_{\nu\mu}^j(\sigma)$.

The orientation of $l_\mu^j(\sigma)$ either coincides with the given 
orientation of the corresponding edge in $\sigma$ or it does not. 
In the former case we define $e_\mu^{I j}(\sigma):=e^I_{l_\mu^j(\sigma)}$ 
while in the latter we define  
$e_\mu^{I j}(\sigma):=[A(l_\mu^j(\sigma))^{-1} e_{l_\mu^j(\sigma)}]^I$.
Then we obtain 
\ba \label{5.9}
\kappa S &=& -\sum_{\sigma\in \tau}  \int_\sigma {\rm Tr}(G\wedge 
e\wedge e)
\nonumber\\
&\approx& 
\frac{1}{5}\sum_{\sigma\in \tau}\sum_{j=0}^4
\epsilon^{\mu\nu\rho\lambda} G_{IJ}(f_{\mu\nu}^j(\sigma)) 
e_\rho^{I j}(\sigma) \; e_\lambda^{J j}(\sigma)
\nonumber\\
&=:& \sum_{l,l'} G_{IJ}\;^{l,l'}\; e^I_l\; e^J_{l'}
\ea
where we have averaged over the corners of a 4 -- simplex.
For any simplicial triangulation the matrix 
$G_{IJ}^{ll'}$ (symmetric in the compound 
index $(I,l)$) is 
difficult
to write down explicitly due to bookkeeping problems, even in the case that we don't
average over the five corners of a 4 -- simplex. Moreover, since we 
intend to perform a Gaussian integral over the $e^I_l$, we need the 
determinant of that 
matrix. This is impossible to compute explicitly unless it is  
block diagonal in some sense.

The latter observation points to a possible solution. First of all
any manifold admits a cubulation, that is a triangulation by embedded 
hypercubes\footnote{An easy proof uses the fact that every manifold can be 
triangulated by simplices. Given a D -- simplex, consider the barycentre 
of each of its ${D+1 \choose p+1}$ sub -- $p$ -- simplices for $p=0,..,D$.
Connect the barycentre of any $p+1$ -- simplex with the barycentres of 
the $p$ -- simplices in its boundary. It is not difficult to see that
this defines a cubulation of the D -- simplex and that all p -- cubes,
thus defined, are the same ones in common q -- simplices of the original 
simplicial complex. In other words, every simplicial complex has a 
cubulated refinement.}  \cite{48e}. 
We now assume that $M$ has a countable cover 
by open sets $O_\alpha$. Consider a stratification 
by 4D regions $S_\alpha$ subordinate to it. Then $S_\alpha$ admits a 
regular cubulation, that is, the 1 -- skeleton of the cubulation of $M$ 
restricted to 
$S_\alpha$ can be chosen to be a regular cubic lattice. Non trivial 
departures from the regular cubulation only appear at the boundaries of 
the $S_\alpha$. We restrict attention to those $M$ admitting a 
cubulation, such that in every compact submanifold, the ratio of the 
number of cubes 
involved in the 
non -- regular regions divided by the number of cubes involved in the  
regular regions converges to zero when take the cubulation to the 
continuum. For those $M$, up to corrections which vanish in the
continuum limit, we can treat $M$ as if it would admit a global, regular 
cubulation. 

Given a regular cubulation $\tau$, consider its set of vertices. 
In 4D, each vertex $v$ is eight valent and there are four pairs of edges, 
such that the members of each pair are analytic continuations of each 
other while the tangents at $v$ of four members, from mutually different 
pairs, 
are linearly independent of each other. It is therefore possible to 
assign to each edge a direction $\mu=0,1,2,3$ and an orientation such 
that adjacent 
edges, in the same direction, have a common analytic continuation and 
agree in their orientation. We label the edges starting at $v$ in the $\mu$
direction by $l_\mu(v)$. Notice that this labelling exhausts all 
possible edges and unambiguously assigns an orientation to all of them.
The discretised co -- tetrad is then given by
\be \label{5.10}
e^I_\mu(v):=e^I_{l_\mu(v)}
\ee
Notice that the hypercubic lattice that results solves all our bookkeeping 
problems since we now may label each vertex by a point in $\mathbb{Z}^4$.

Next, given a vertex $v$ we denote by $v\pm \hat{\mu}$ the next 
neighbour vertex in the $\mu$ direction. We define the plaquette loop
in the $\mu,\nu$ plane at $v$ by
\be \label{5.11}
\partial f_{\mu\nu}(v):=l_\mu(v) \circ l_\nu(v+\hat{\mu}) \circ 
l_\mu(v+\hat{\nu})^{-1} \circ l_\nu(v)^{-1}
\ee
so that $\partial f_{\nu\mu}(v)=[\partial f_{\mu\nu}(v)]^{-1}$. Notice 
that, again, this labelling exhausts all minimal loops (definition \ref{def:minimalloop}) in the one skeleton 
of $\tau$.
The discretised ``curvature'' is therefore 
\be \label{5.12a}
G_{IJ}^{\mu\nu}(v):=\epsilon^{\mu\nu\rho\sigma} G_{IJ}(f_{\rho\sigma}(v))
\ee

We denote each 4D hypercubes in $\tau$ by $\sigma$. 
There is, then, a one to one correspondence between the vertices $v$ in the 0 -- 
skeleton of $\tau$ and the hypercubes given by assigning to $\sigma$
one of its corners $v=(z_0,..,z_3)$ which has the smallest values of all
$z_0,..,z_3\in \mathbb{Z}$. We then find 
\ba \label{5.12}
\kappa S &=& \sum_{\sigma} \int_M \; G_{IJ} \wedge e^I\wedge e^J
\nonumber\\
&\approx & \sum_v \sum_{I,J,\mu,\nu} \; G_{IJ}^{\mu\nu}(v)\; 
e^I_\mu(v)\; e^J_\nu(v)
\ea
The crucial observation is now the following: if we assemble pairs of indices 
into a joint index $A=(I,\mu),\;B=(J,\nu)$ etc. and let 
$e^A(v):=e^I_\mu(v),\;G_{AB}(v):=G^{\mu\nu}_{IJ}(v)$ etc. 
(Notice that by construction $G_{AB}(v)=G_{BA}(v)$ for all $v$), then 
(\ref{5.12}) can be written as 
\be \label{5.13}
\kappa S \approx \sum_v \; e^T(v)\; G(v)\; e(v)
\ee
This means that using (regular) cubulations the matrix
$G^{ll'}_{IJ}$ becomes block diagonal, where each block is labelled by a 
vertex and corresponds to the symmetric 16 x 16 matrix $G(v)$.
This is what makes the computation of the determinant of the huge matrix
with entries $G^{ll'}_{IJ}$ practically possible. As we will see,
the matrices $G(v)$ have a lot of intriguing symmetries which make the 
computation of their determinant an interesting task.\\ 
\\
Questions that arise in algebraic topology and still need to be addressed are:
\begin{itemize}
\item[1.] Given any D -- cubulation, does there exist a cubulated 
refinement such that one can consistently assign to every D cube 
$\sigma$ a vertex $v$ and, to all edges, an orientation such that there are 
precisely D edges outgoing from $v$? We call cubulations, for which 
this is possible {\it regular}. If that would be the case, we could 
generalise our discretisation from regular hypercubic lattices to 
arbitrary cubic ones 
and, thus, we should not make any error at the boundaries of the 
stratified regions mentioned above. 
\item[2.] If the answer to [1.] is negative, can one choose maximally 
regular cubulations as to minimise the error in our assumption of globally 
regular cubulations? In 3D some results on that issue seem to exist 
\cite{48e}.
\item[3.] Given maximally regular cubulations, can one make an error 
estimate resulting from the neglection of the non -- trivial topology?
\end{itemize}

\subsection{The Generating Functional of Tetrad N -- Point Functions}
\label{s5.3}

We now want to apply the general framework of section \ref{s5.2} to 
General Relativity in the Holst formulation. Classically, it is clear
that, without fermions all the geometry is encoded in the co-tetrad 
fields 
$e^I_\mu$ because, then, the spacetime connection is just the spin 
connection defined by the co-tetrad (on shell). If fermions are coupled, 
the same 
is still true in the second order formulation so that there is no 
torsion. But even in the first order formulation with torsion one 
can attribute the torsion to the fermionic degrees of freedom. Hence, 
we want to consider as a complete list of configuration fields the co -- 
tetrad. 

We will now make two assumptions about the choice of gauge fixing and 
the matter content of our system.
\begin{itemize}
\item[I.] The local measure factors $\rho,\;I$ ( in equation \ref{5.25a}) depend on the co -- tetrad
only analytically. This is actually true for the Holst action
\cite{48a}. See also \cite{48c}. 
\item[II.] The gauge fixing condition $G$ is independent of the 
co -- tetrad and the Fadeev -- Popov 
determinant $\det(\{F,G\})$ depends only analytically on
the co -- tetrad. 
With respect to the first class Hamiltonian and spatial diffeomorphism 
constraint, this 
can always be achieved by choosing suitable matter as a reference 
system, see e.g. \cite{26}, \cite{415}, \cite{416}, \cite{413}. However, in addition there is the Gauss -- 
law first class constraint. Here, it is customary to impose the time 
gauge condition \cite{47}, which asks that certain components of the 
tetrad vanish. This will also enable one to make the connection with 
canonical LQG, where one works in the time gauge in order to arrive at an
SU(2) rather than G connection. 

Fortunately, in this case it is 
possible to explicitly construct a complete set of G -- invariant 
functions of the tetrad, namely the four metric\footnote{In the presence 
of fermions there are additional gauge invariant functions also 
involving the fermions.}  $g_{\mu\nu}=e^I_\mu 
e^J_\nu \eta_{IJ}$ and if we only consider correlators of those, then we 
can get rid of the time gauge condition as indicated in 
section \ref{s5.2} (Fadeev -- Popov identity). In section \ref{s5.5} we will come 
back to this issue when trying to make the connection of the 
SFM, obtained with canonical LQG for which the time gauge is unavoidable.
We will then sketch how to possibly relax the assumptions made under 
[II.].
\end{itemize}
Under the assumptions made ([I.], [II.]) we consider the generating functional
$\chi(j,J):=Z[j,J]/Z[0,0]$ where 
\ba \label{5.29}
Z[j,J] &:=& \int [D\phi\;\;DA]\; \delta[G[A,\phi]]\; 
e^{i\int_M\; {\rm Tr}(J\wedge \phi)}\;
\times\\
&& \int\; [De]\; \rho[e,A,\phi]\; I[e,A,\phi]\; 
|\det[\{F,G\}]|[e,A,\phi]
\; e^{\frac{i}{\hbar} 
(S_g[e,A]+S_m[e,A,\phi])}
e^{i\int_M\; {\rm Tr}(j\wedge e)}\nonumber
\ea
Here $\phi$ denotes the matter configuration variable.
We have split the 
total action into the geometry 
(Holst) part $S_g$ and a matter part $S_m$, which typically depends non 
trivially, but analytically on $e$. Also the total current has been split into 
pieces $J,j$, each taking values in the bundles dual to those of $\phi,e$, 
respectively. 

A confusing and peculiar feature of first order actions,
such as the Holst or Palatini action, is that from a Lagrangian point of 
view both fields $e$ and $A$ must be considered as configuration variables.
In performing the Legendre transform \cite{48a}
one discovers that there are 
primary 
constraints, which relate certain combinations of $e$ to the momenta 
conjugate to $A$. One can solve these constraints and then $(A,e)$ appear 
as momentum and configuration coordinates of this partly reduced phase 
space. This is the reason why we
consider only correlations with respect to $e$.   

As done in path integral theory, we set 
\be \label{5.30}
\sigma[e,A,\phi]:=\rho[e,A,\phi]\; I[e,A,\phi]\; 
|\det[\{F,G\}]|[e,A,\phi]
\; e^{\frac{i}{\hbar} S_m[e,A,\phi]}
\ee
and write (\ref{5.29}) as
\be \label{5.31}
Z[j,J]:=\int [D\phi\;DA]\; \delta[G[A,\phi]]\; 
e^{i\int_M\; {\rm Tr}(J\wedge \phi)}\; 
\sigma[\frac{\delta}{i\delta j},A,\phi]\;
\{
\int\; [De]\;  e^{\frac{i}{\hbar} 
S_g[e,A]}
e^{i\int_M\; {\rm Tr}(j\wedge e)}\}
\ee
Of course $e^{iS_m}$ must be power expanded in a perturbation series in 
order to carry out the functional derivations with respect to $j$. 
Indeed, if 
we consider just the functional integration with respect to $e$ and 
think of $A$ and $\phi$ as external fields, then $S_g$, being 
quadratic in $e$, is analogous to the {\it free part}, while $S_m$, being only 
analytic in 
$e$, is like an {\it interaction part} of the action as far as the 
co-tetrad is 
concerned. Of course, in the computation of the physical tetrad n -- 
point functions all the functional derivatives involved in (\ref{5.31}) 
are eventually evaluated at $j=0$.   

It follows that the object of ultimate interest is the {\it Gaussian 
integral}
\be \label{5.32}
z[j;A]:=\int\; [De]\;  e^{-\frac{i}{\ell_P^2} \;\int_M\; {\rm 
Tr}(G\wedge 
e\wedge e+\ell_P^2 j\wedge e)}
\ee
which is computable exactly. However, it is not 
a standard Gaussian since i) the exponent is purely imaginary; and ii) the ``metric'' $G^{\mu\nu}_{IJ}(A)$ is indefinite so 
that
$z[j;A]$ would  be ill defined if the exponent was real\footnote{As 
usual
this prevents a ``Euclidian'' version of GR. Here Euclidian stands 
for Euclidian field theory with an analytic continuation to the 
imaginary axis of the real time variable involved (Wick rotation), which 
leads to a real exponent. 
This has nothing to do with Lorentzian or Euclidian signature GR.
In fact, most metrics do not have an analytic section so that Wick 
rotation is ill defined and, thus, the connection between the real and the 
Euclidian theory is veiled.}. 
In order to carry out this integral we must make the 
technical assumption that configurations $A$ for which $G$ is singular
have measure zero with respect to $DA$. 

This is the point where we have to regularise the path integral in order 
to perform the Gaussian integration\footnote{Actually we can formally solve 
the Gaussian integral without specifying
the triangulation, i.e. we can compute it in the continuum. However, 
one then has to regularise the resulting determinant which amounts to the 
same problem.} and we write the discretised version 
on a cubulation of $M$ as motivated in section \ref{s5.1}, i.e. we 
replace (\ref{5.32}) by the discretised version
\be \label{5.33a}
z[j;A]:=\int\; \prod_{v,I,\mu}\; de^I_\mu(v)\;  e^{\frac{i}{\ell_P^2} 
\;\sum_v\;[G^{\mu\nu}_{IJ}(v) e^I_\mu(v) e^J_\nu(v)+\ell_P^2 j^I_\mu(v)
e^I_\mu(v)]}
\ee
The results of appendices A and B in \cite{30a} now reveal that 
\be \label{5.33}
z[j;A]:=[\prod_{v}\; \frac{e^{\frac{i\pi}{4} {\rm 
ind}(G(v))}}{\sqrt{|\det(G(v))|}}] \; e^{-i\frac{\ell_P^2}{4}\sum_v 
[G^{-1}(v)]_{\mu\nu}^{IJ} j^\mu_I(v) j^\nu_J(v)}
\ee
where we have dropped a factor $\sqrt{\pi}^{16N}$ for a cubulation with
$N$ vertices because it is cancelled by the same factor coming from the 
denominator in $\chi(j,J)$, see (\ref{5.29}).

\subsection{Wick Structure, Graviton Propagator and SFM Vertex 
Structure}
\label{s5.4}

\subsubsection{Wick structure}
\label{s5.4.1}

Formula (\ref{5.33}) explicitly displays the main lesson of our 
investigation: The full $j$ dependence of the generating functional 
written as (\ref{5.31}) rests in (\ref{5.33}). We are interested 
in the n-th functional derivatives of (\ref{5.33}) at $j=0$. Now, similar 
as 
in free field theories, the corresponding n -- point functions 
vanish for $n$ odd. However, in contrast to free field theories,
for $n$ even, the $n-$point functions cannot be written in terms of 
polynomials of the 2-point function. The reason is that the 
``covariance'' $G^{-1}[A]$ of the Gaussian is not a background structure
but rather depends on the quantum field $A$ one has to 
integrate over. 
This renders the co -- tetrad theory to be non -- quasi -- free, that 
is, an
interacting theory. Nevertheless it is true that all Wick identities 
that have been derived for free field theories still hold also for 
the $n-$point tetrad functions {\it albeit in the sense of
expectation values or means with respect to $A$}. 

\subsubsection{Graviton Propagator}
\label{s5.4.2}

To illustrate the derivation of the graviton propagator, let us consider a fictive theory in which 
$\sigma(e,A,\phi),\;G(A,\phi)$ are both independent of 
$A$ and $e$. This is not a very physical assumption but it serves
to make some observations of general validity in the simplified context obtained by dropping the $\phi$ dependence. This simplification can be carried out since, due to the above assumptions, the 
generating functional factorises.
Thus, in our fictive theory we are looking at the generating functional
$\chi[j]=z[j]/z[0]$ where 
\be \label{5.34}
z[j]=\int\; [DA]\; z[j;A]=\int\; \prod_{v,\mu} \; d\mu_H(A(l_\mu(v))\; 
[\prod_v \frac{e^{\frac{i\pi}{4} {\rm 
ind}(G(v))}}{\sqrt{|\det(G(v))|}}]\;
e^{-\frac{i\ell_P^2}{4}\sum_v \; j^\mu_I(v)\; j^\nu_J(v)\; 
[G^{-1}(v)]^{IJ}_{\mu\nu}]}
\ee
and $\mu_H$ is the\footnote{In case of non -- compact $G$ the Haar 
measure is unique up to a normalisation constant which drops out in 
$\chi(j)$. The choice of the Haar measure instead of the Lebesgue measure 
is valid in the continuum limit of infinitely ``short'' edges as usual.} 
Haar measure 
on $G$. Now let 
\be \label{5.35}
<e_{\mu_1}^{I_1}(v_1)\; ..\; e_{\mu_n}^{I_n}(v_n)>
:=[\frac{\delta^n \chi[j]}{i^n 
\delta j^{\mu_1}_{I_1}(v_1)\; .. \; \delta j^{\mu_1}_{I_1}(v_1)}]_{j=0}
\ee
It is immediately clear that 
\be \label{5.36}
<e_{\mu_1}^{I_1}(v_1)\; e_{\mu_2}^{I_2}(v_2)>=0
\ee
unless $v_1=v_2$. This is reassuring since, as mentioned above, 
physically it makes only sense to consider correlators of $G-$invariant 
objects, such as the metric. The simplest $n-$point function of interest 
is, therefore, the 4-point function
\be \label{5.37}
< g_{\mu_1 \nu_1}(v_1) \; g_{\mu_2 \nu_2}(v_2)> 
=  
<e_{\mu_1}^{I_1}(v_1)\; e_{\nu_1}^{J_1}(v_1)
e_{\mu_2}^{I_2}(v_2)\; e_{\nu_2}^{J_2}(v_2)>\; \eta_{I_1 J_2}\; 
\eta_{I_2 J_2}
\ee
If we are interested in something like a graviton propagator we are 
interested in $v_1\not=v_2$ and obtain 
\be \label{5.38}
< g_{\mu_1 \nu_1}(v_1) \; g_{\mu_2 \nu_2}(v_2)> 
=[\frac{\ell_P^2}{2}]^4 
<[G(v_1)^{-1}]^{I_1 J_1}_{\mu_1 \nu_1}\; 
[G(v_2)^{-1}]^{I_2 J_2}_{\mu_2 \nu_2}>'
\ee
where for $F=F[A]$
\be \label{5.39}
<(F)>':=
\frac{
\int\; \prod_{v,\mu} \; d\mu_H(A(l_\mu(v))\; 
[\prod_v \frac{e^{\frac{i\pi}{4} {\rm 
ind}(G(v))}}{\sqrt{|\det(G(v))|}}]\;F[A]}
{\int\; \prod_{v,\mu} \; d\mu_H(A(l_\mu(v))\; 
[\prod_v \frac{e^{\frac{i\pi}{4} {\rm 
ind}(G(v))}}{\sqrt{|\det(G(v))|}}]
}
\ee
Notice that $G(v)^{-1}$ does not share the symmetries of $G(v)$, so 
$[G^{-1}(v)]^{(IJ)}_{\mu\nu}$ does not vanish automatically.\\
We are interested in correlators of the inverse 
matrix $G(v)^{-1}$ with respect to the joint Haar measure.  
Whether these have the correct behaviour in a situation where,  
instead of vacuum boundary states one chooses coherent states peaked 
on a classical background metric as suggested in \cite{46g,15,49a}, is 
currently under investigation.

\subsubsection{SFM Vertex Structure}
\label{s5.4.3}

Finally, in order to translate (\ref{5.39}) into spin foam language,
we should perform harmonic analysis on $G$ and write the integrand of 
the Haar measure in terms of irreducible representations of $G$. 
In particular, the vertex structure of a SFM is encoded in $z[0]$,
so that we are interested in the harmonic analysis of the function 
\be \label{5.40}
F(v):=\frac{e^{\frac{i\pi}{4} {\rm ind}(G(v))}}{\sqrt{\det(G(v))|}}
\ee
To derive its graph theoretical structure it is enough to find out 
which $F(v)$ depend on a given holonomy $A(l)$ and how. Recall that $F(v)$ 
is 
a function cylindrical over the graph 
$\gamma(v)=\cup_{\mu<\nu} \partial f_{\mu\nu}(v)$, which is the union of 
its respective plaquette loops . Consider a fixed 
edge $l=l_\mu(v)$. This is contained in $\gamma(v')$ if and only if 
it is contained in one of the plaquette loops $\partial f_{\mu \nu}(v')$ 
or $\partial f_{\nu \mu}(v')$ with $\mu<\nu$ or $\nu<\mu$, respectively.
In both cases it must coincide either with $l_\mu(v')$ or with 
$l_\mu(v'+\hat{\nu})$. Thus in either case we 
must have either $v'=v$ or $v'=v-\hat{\nu},\;\nu\not=\mu$. 

To better understand let us consider, for simplicity, that $G$ 
is compact (the non compact case has the same SFM vertex structure but 
the harmonic analysis is a bit more complicated). Then, each function 
$F(v)$ can be formally expanded into $SO(4)$ (or rather the 
universal cover $SU(2)\times 
SU(2)$) irreducible 
representations\footnote{This expansion would be rigorous if we 
knew that $F(v)$ is an $L_2$ function which is currently under 
investigation. We assume here that in any case we  may use the 
Peter \& 
Weyl theorem in a distributional sense.} with respect to the 
six plaquette holonomies $A(\partial f_{\mu\nu}(v)),\;\mu<\nu$. 
These representations $\pi$ are labelled by pairs of half integral spin 
quantum numbers, however, we will not need this for what follows.
Thus $F(v)$ admits an expansion of the form
\be \label{5.41}
F(v)=\sum_{\{\pi_{\mu\nu}\}}\; \iota'_{\{\pi_{\mu\nu}\}}\; \cdot
\;[\otimes_{\mu<\nu} \; \pi_{\mu\nu}(A(\partial f_{\mu\nu}(v)))]
\ee
where $\iota'_{\{\pi_{\mu\nu}\}}$ is a gauge invariant intertwiner
for the six -- tuple of irreducible representations 
$\{\pi_{\mu\nu}\}_{\mu<\nu}$. $\iota'_{\{\pi_{\mu\nu}\}}$ is independent of $v$, the only $v$ 
dependence rests in the holonomies. The expansion \ref{5.41} depends on the specific algebraic 
form of $F(v)$ which, itself, derives from the Holst action.

Let us define $\pi_{\nu\mu}:=\pi_{\mu\nu}$ for $\mu<\nu$. By writing 
the six plaquette holonomies in terms of four edge holonomies it is not 
difficult to 
see that $F(v)$ can also be written in the form  
\be \label{5.42}
F(v)=\sum_{\{\pi_{\mu\nu}\}}\; \iota_{\{\pi_{\mu\nu}\}}\;\cdot\;
[\otimes_{\mu,\mu\not=\nu} \; \pi_{\mu\nu}(A(l_\mu(v)))\;\otimes \;
\pi_{\mu\nu}A(l_\mu(v+\hat{\nu}))]
\ee
which displays explicitly the 16 
variables $A(l_\mu(v)),\;A(l_\mu(v+\hat{\nu}),\;\nu\not=\mu$ involved 
and 
it consists of 24=6 x 4 tensor product factors. In order to arrive at 
(\ref{5.42}) we had to rearrange the contraction indices which 
induced the change from $\iota'$ to $\iota$ and we also
made use of $\pi(A(l)^{-1})=\pi^T(A(l))$ for $G=SO(4)$. 

We may now carry out explicitly the integrals over edge holonomies in 
$z[0]$ by inserting the expansion (\ref{5.42}). We write 
symbolically\footnote{We rearrange the tensor products as if they were 
scalars but this can be corrected by performing corresponding 
rearrangements in the contraction structure of the intertwiners. We
assume this to be done without explicitly keeping track of it because 
it does not change the vertex structure.}
\ba \label{5.43}
z[0] &=& \int \; \prod_{v,\mu} \; d\mu_H(A(l_\mu(v)))\;\prod_{v'}\; 
F(v')
\nonumber\\
&=& \sum_{\{\pi^v_{\mu\nu}\}}\; [\prod_v \; 
\iota_{\{\pi^v_{\mu\nu}\}} \cdot]
\int \; \prod_{v,\mu} \; d\mu_H(A(l_\mu(v)))\;
[\otimes_{v',\mu,\mu\not=\nu} \; 
\pi^{v'}_{\mu\nu}(A(l_\mu(v')))\;\otimes 
\;
\pi^{v'}_{\mu\nu}(A(l_\mu(v'+\hat{\nu}))]
\nonumber\\
&=& \sum_{\{\pi^v_{\mu\nu}\}}\; [\prod_v \; 
\iota_{\{\pi^v_{\mu\nu}\}} \cdot]
\; \int \; \prod_{v,\mu} \; d\mu_H(A(l_\mu(v)))\;
[\otimes_{v',\mu,\mu\not=\nu} \; 
\pi^{v'}_{\mu\nu}(A(l_\mu(v')))\;\otimes 
\;
\pi^{v'-\hat{\nu}}_{\mu\nu}(A(l_\mu(v'))]
\nonumber\\
&=& \sum_{\{\pi^v_{\mu\nu}\}}\; [\prod_v \; 
\iota_{\{\pi^v_{\mu\nu}\}} \cdot]
\; \otimes_{v,\mu} [\int_G\; d\mu_H(g)\;
[\otimes_{\mu\not=\nu} \; 
\pi^{v}_{\mu\nu}(g)\;\otimes 
\;
\pi^{v-\hat{\nu}}_{\mu\nu}(g)]
\ea
Here in the second step we have shifted the vertex label in one of the 
tensor product factors in order to bring out the dependence on the 
$A(l_\mu(v))$. It follows that the end result of the integration is that, 
for each edge $l=l_\mu(v)$, there is a gauge invariant 
intertwiner.
\be \label{5.44}
\rho_{\{\pi^v_{\mu\nu},\pi^{v-\hat{\nu}}_{\mu\nu}\}_{\nu\not=\mu}}
:=
[\int_G\; d\mu_H(g)\;
[\otimes_{\mu\not=\nu} \; 
\pi^{v}_{\mu\nu}(g)\;\otimes 
\;
\pi^{v-\hat{\nu}}_{\mu\nu}(g)]
\ee
which intertwines {\it six} representations rather than four as in 
(constrained) BF-theory on simplicial triangulations. The origin of this 
discrepancy is of course that we are using cubulations rather than 
simplicial triangulations. These six representations involved for 
edge $l_\mu(v)$ correspond precisely to the six plaquette loops 
$\partial f_{\mu\nu}(v),\;\partial 
f_{\mu\nu}(v-\hat{\nu}),\;\nu\not=\mu$
of which $l_\mu(v)$ is a segment. Therefore, if we associate to each 
face $f=f_{\mu\nu}(v)$ an irreducible representation 
$\pi_f=\pi^v_{\mu\nu}$ and denote by $\{\pi\}$ the collection of all the 
$\pi_f$, 
then the  
basic building block
(\ref{5.44}) can be written in the more compact form 
\be \label{5.45}
\rho_l[\{\pi\}]=\int_G\; d\mu_H(g)\; \otimes_{l\subset \partial f}\;
\pi_f(g)
\ee
Likewise, if we denote $\iota_v[\{\pi\}]:=\iota_{\{\pi^v_{\mu\nu}\}}$,
then
\be \label{5.46}
z[0]=\sum_{\{\pi\}}\; [\prod_v\; \iota_v[\{\pi\}]\cdot]\;
\otimes_l \; \rho_l[\{\pi\}]
\ee
which of course hides the precise tensor product and contraction 
structure but it is still sufficient for our purposes. 

Formula (\ref{5.46}) is precisely the general structure of a SFM. 
Moreover, the intertwiner (\ref{5.45}) is the direct analogue of the 
intertwiner in BF-theory which defines the pentagon diagramme
\cite{45}. If we would try to draw a corresponding picture for our 
model, then for each vertex $v$ we would draw eight points, one for each 
edge $l$ incident at $v$. These edges are labelled by the 
intertwiner $\rho_l$. Given two points corresponding to edges 
$l,l'$ consider the unique face $f$ that has $l,l'$ in its 
boundary. Draw a line between each such points and label it by $\pi_f$,
the result is the {\it octagon diagramme}, see figure \ref{fig1}.

Consider the edges 
adjacent to $v$ which are $l_\mu(v),\;l_\mu(v-\hat{\mu}),\;\mu=0,1,2,3$. For 
$\mu\not=\nu$  we obtain four faces: a)
the face $f_{\mu\nu}(v)$ spanned by $l_\mu(v),\;l_\nu(v)$,\hspace{.1in} b)
the face $f_{\mu\nu}(v-\hat{\nu})$ spanned by $l_\mu(v),\;l_\nu(v-\hat{\nu})$,\hspace{.1in}
c) the face $f_{\mu\nu}(v-\hat{\mu})$ spanned by $l_\mu(v-\hat{\mu}),\;l_\nu(v)$ 
and d)
the face $f_{\mu\nu}(v-\hat{\mu}-\hat{\nu})$ spanned by $l_\mu(v-\hat{\mu}),\;l_\nu(v-\hat{\nu})$. 
The corresponding label on the 
lines is thus $\pi_{\mu\nu}^v,\; \pi_{\mu\nu}^{v-\hat{\nu}},\; 
\pi_{\mu\nu}^{v-\hat{\mu}},\; 
\pi_{\mu\nu}^{v-\hat{\mu}-\hat{\nu}}$, respectively. Therefore the octagon 
diagramme has eight points and 6 x 4 = 24 lines (each line connects two 
points). These correspond to the 
24 plaquettes that have a corner in $v$ which, for each 
$\mu<\nu$ are $f_{\mu\nu}(v),\; f_{\mu\nu}(v-\hat{\mu}),\;   
f_{\mu\nu}(v-\hat{\nu}),\;f_{\mu\nu}(v-\hat{\mu}-\hat{\mu})$.\\
In the 
case of $G=SO(4)$ each irreducible representation is labelled
by two spin quantum numbers. 

The intertwiner freedom is 
labelled by three irreducible representations of $SO(4)$ and there is 
one irreducible representation corresponding to a face. Thus the octagon 
diagramme depends on 3 x 8 + 24=48 irreducible representations of $SO(4)$ or 96 spin quantum 
numbers. Since each intertwiner (\ref{5.45}) factorises into {\it two} 
intertwiners \cite{1b} (one for the starting point and one for the 
beginning 
point of the edge but both depend on the same representations) we 
may actually collect those eight intertwiners associated to the same 
vertex. The collection of those eight factors is 
actually the analytic expression corresponding to the octagon diagramme
which, therefore, maybe called the {\it 96 j 
-- symbol}.
\begin{figure}[hbt]
\begin{center}
\psfrag{0+}{$l^+_0$}
\psfrag{1+}{$l^+_1$}
\psfrag{2+}{$l^+_2$}
\psfrag{3+}{$l^+_3$}
\psfrag{0-}{$l^-_0$}
\psfrag{1-}{$l^-_1$}
\psfrag{2-}{$l^-_2$}
\psfrag{3-}{$l^-_3$}
\includegraphics[scale=0.7]{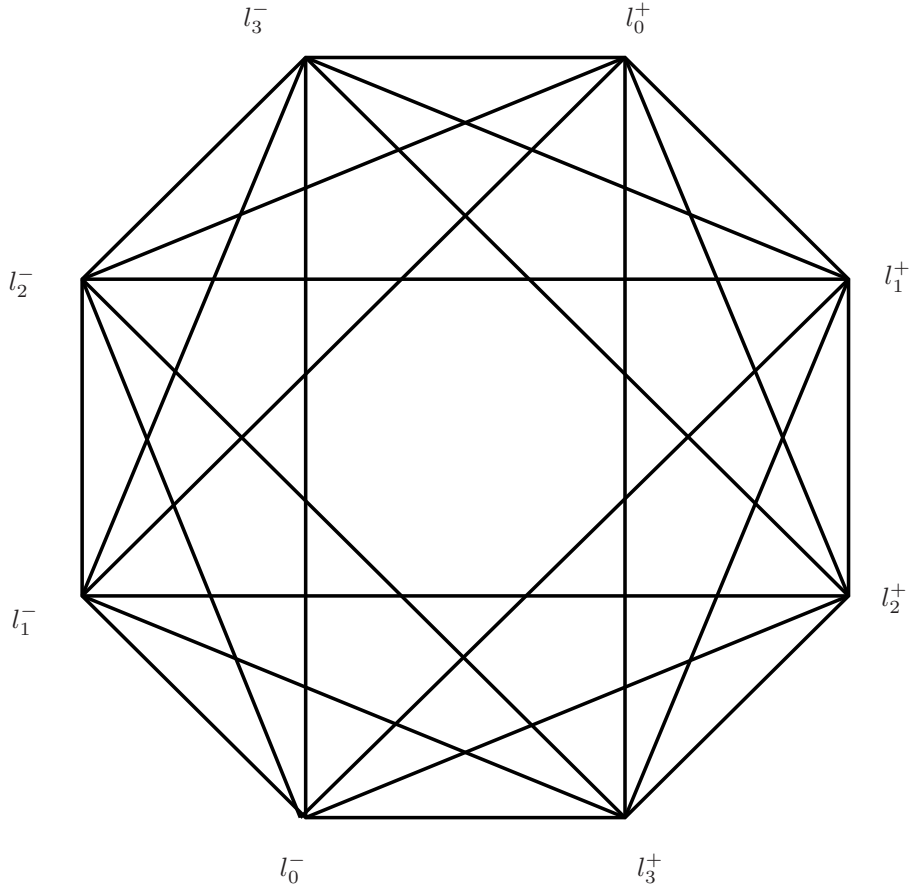}
\caption{The octagon diagramme associated to vertex $v$. The eight 
corners correspond to the eight edges 
$l=l_\mu^\sigma(v)=l_\mu(v+\frac{\sigma-1}{2}\hat{\mu}),\;\sigma=\pm$
adjacent to $v$. The line between
corners labelled by $l^\sigma_\mu(v),\;l^{\sigma'}_\nu(v)$ 
for $\mu\not=\nu$ corresponds 
to the face $f=f_{\mu\nu}^{\sigma\sigma'}(v)=
f_{\mu\nu}(v+\frac{\sigma-1}{2}\hat{\mu}+\frac{\sigma'-1}{2}\hat{\nu})$.
We should colour corners by intertwiners $\rho_l$ and lines by 
representations $\pi_f$ but refrain from doing so in order not to
clutter the diagramme. Altogether 48 irreducible 
representations of $Spin(4)$ (or 96 of $SU(2)$) are involved.}
\label{fig1}
\end{center}
\end{figure}
\\The decisive difference between (constrained) BF-theory and our model is 
however that in (constrained) BF-theory the analogue of the function 
$F(v)$ is a product of $\delta$ distributions, one for each face 
holonomy. The simplicity constraints just impose restrictions on the
representations and intertwiners, but this cannot change the fact that
there is factorisation in the face dependence. In our model, the face 
dependence does not factorise, hence, in this sense it is less local or 
more interacting.

\subsection{Relation between covariant and canonical connection}
\label{s5.5}

Another striking feature of the model presented above is the following: constrained 
BF-theory, that is, Plebanski theory, should be a candidate for quantum 
gravity. The Holst model should be equivalent to that theory, at least 
semiclassically, since the only difference 
between them lies in the technical implementation of the simplicity 
constraints. Now 
one of the most important property of the implementation of the 
simplicity constraints in usual SFM is that the irreducible Spin(4) 
representations that one sums over are the simple ones\footnote{If we 
label an irreducible representation of Spin(4) by a pair $(j_+,j_-)$ then a simple irreducible representation is 
one for which $j_+=j_-$ \cite{45}. There is a similar restriction if one 
works with arbitrary Immirzi parameter \cite{413}.}. In the cubulated SFM there is no such restriction. 
This is an important issue because the 
restriction to simple representations means that the underlying 
gauge theory is roughly SU(2), rather than Spin(4). This is 
correct if the SFM is to arise from canonical LQG which indeed is a 
SU(2) gauge theory. Thus, in usual SFM the simplicity constraints 
seem to already imply the gauge fixing of the ``boost'' part of the 
Spin(4) 
Gauss constraint that, at the classical level, is needed 
to pass from the Holst connection to the Ashtekar -- Barbero -- Immirzi
connection \cite{47}\footnote{Strictly speaking, that has not been established 
yet, as pointed out in \cite{426a}, where it is shown that the connection 
used in SFM is actually the spin connection and not the Holst 
connection.}. However, in the cubulated SFM no restrictions on the type of group 
representations are present.

However, what we have done in the previous section is incomplete, in fact, 
in order to properly define the n -- point functions we must 
gauge fix the generating functional with respect to the G Gauss 
constraints. Formally, this is not necessary if we only consider 
correlators of G invariant functions, such as the metric due to the fact that the 
infinite gauge group volume formally cancels out in the fraction 
$z[j]/z[0]$. However, in the case at hand it would seem that the formal arguments cannot be substantiated by hard proofs. Specifically, if we consider $G=SO(1,3)$, there is no measure  
known for gauge theories for non compact groups (see \cite{426b} for the 
occurring complications) and, 
thus, 
we are forced to gauge fix at least the boost part of the Gauss 
constraint. This is the same reason for which one uses the time gauge
in the canonical theory. We expect 
that implementing the time gauge fixing \cite{47} in a way similar to the 
implementation of the simplicity constraints in usual BF-theory will, 
effectively, reduce the gauge group to $SU(2)$. 

The idea to carry this out is, roughly speaking, as follows:\\
the time gauge is a set of constraints
$C[e]$ on the co -- tetrad $e$. By the usual manipulations we can 
pull the corresponding $\delta$ distribution out of the cotetrad 
fuctional integral and, formally, we obtain  
\be \label{5.47}
\chi_{{\rm ABI}}[j]=[\delta[C[\delta/\delta j]]\; \chi_{{\rm Holst}}[j]
\ee
where $\chi_{{\rm Holst}}$ is the generating functional of the previous 
section and $\chi_{{\rm ABI}}$ stands for the Ashtekar -- Barbero -- 
Immirzi path integral. 

Whether this really works in a rigorous fashion remains to be seen. 
However, we find it puzzling that the simplicity constraints in usual
SFM, which classically have nothing to do with the time gauge, should 
automatically yield the correct boundary Hilbert space. It seems 
intuitively clear that the time gauge must be imposed in the quantum 
theory in addition to the simplicity constraints, just like in the 
classical theory, as we suggest. Without imposing it, we do not see 
any sign of a restriction from $G$ to $SU(2)$ in the cubulated SFM where the simplicity constraints are solved differently.\\
This observation indicates that the usual SFM and the cubulated SFM are rather different from each other.

\chapter{Group Field Theory}
\label{chap:gft}
Group field theory (GFT) was originally born as a higher dimensional generalisation of the matrix model for 2-dimensions quantum gravity. However, an in depth study of the subject revealed its possible use as a candidate for a discretization independent formulation of spin foam models. In fact, previous discussions revealed that the partition function for BF-theory with constraints is dependent on the triangulation chosen.

This is not a desirable feature if we want to construct a background independent theory for quantum gravity. The very close similarities between GFT and spin foam suggested that the latter was a specification of the former but, in such a way, it would be triangulation independent. Moreover, GFT shows similarities with other approaches to quantum gravity, as for example dynamical triangulation, simplicial quantum gravity/Regge calculus and causal sets.

This suggests a deeper role played by GFT, namely as a structure which underlines any attempt to define a theory of quantum gravity in a background-independent way.\\
We will now briefly explain what GFT is.
\section{GFT Formalism}
Essentially GFT is a QFT on superspace\footnote{Roughfly speaking a QFT on superspace describes the evolution process of 3-geometries in terms of a perturbative expansion of sums of different topologies corresponding to Feynman diagrams and possible interaction processes of the 3-geometries itself. Thus, in this picture, the different spacetime topologies are represented by Feynman diagrams with boundaries and the amplitudes for such Feynman diagrams are given in terms of a sum over histories quantisation of gravity.} (space of 3 geometries), which is defined utilising simplicial description of spacetime, thus rendering the theory local\footnote{Alternatively, one can define GFT as a field theory over a group manifold, in which the field represents quantised (D-1)-simplex and in which no reference to spacetime is made. The states (which in momentum space are spin networks) are interpreted as triangulations of the (D-1) pseudo manifolds, topologically dual to the Feynman diagrams. Here we have called them pseudo manifolds rather than manifolds, since the data in the GFT diagrams do not restrict the simplices of dimensions equal or lower than (D-3) to have a particular characteristic, thus including also those which are not topologically equivalent to a sphere. }. In fact, a D dimensional simplicial space is identified with a D dimensional simplicial complex. \\
Such complexes can be constructed by gluing together certain D-dimensional ``atomic" elements, which have the topology of a D-dimensional ball along their D-1 boundaries. Thus, the fundamental building blocks of D-dimensional simplicial complexes can be considered to be, precisely, these D-1 dimensional boundary terms.

The realisation of a QFT of superspace in terms of these building blocks is what renders GFT local. In particular, one considers only the wave function on one D-1 dimensional simplicial complex, which is identified as a functional of the geometry and, then, quantises it. \\
In this scheme, the D-dimensional simplicial complex is identified with the interaction and evolution of the D-1 simplices on its boundary. Let us analyse the construction of GFT in more detail. \\Given a (D-1)-dimensional simplex, whose boundaries are (D-2)-dimensional simplicial complexes, the field utilised in GFT is denfined in terms of the following complex function:
\begin{equation}
\phi(g_1,g_2\cdots g_D):G^{\otimes D}\rightarrow\Cl
\end{equation}  
where G is any group and each group element $g_i$ is associated to one of the $D$, (D-2)-dimensional boundaries of the (D-1)-dimensional complex. There are two symmetries under which the field is invariant. These are i) an even permutation of the arguments of the field and ii) an invariance under a global action of the Lorenz SU(2) group.\\
The permutation invariance is a consequence of the fact that the order of the arguments in the field corresponds to the orientation of the (D-1)-dimensional simplex, to whose boundaries the group elements are assigned. Since even permutation of the group elements would correspond to similar orientations, one requires the field $\phi$ to be invariant under such permutation,i.e. 
\begin{equation}
\phi(g_1,g_2\cdots g_D)=\phi(g_{\pi(1)},g_{\pi(2)}\cdots g_{\pi(D)})
\end{equation}
where $\pi$ represents even permutations.\\
The Lorenz invariance, instead, is imposed through a projection operator as follows:
\begin{equation}
P_g\phi(g_1,g_2\cdots g_D)=\int_G dg\phi(g_1g,g_2g\cdots g_Dg)
\end{equation}
What this invariance exemplifies is that the (D-2)-dimensional simplices, to which the group elements are associated, are indeed the boundaries of a (D-1)-dimensional simplex. 
It is also possible to represent this field in configuration space using harmonic analysis on a group, thus obtaining
\begin{equation}\label{equ:configspace}
\phi(g_i)=\sum_{J_i \Lambda, k_i}\phi^{J_i\Lambda}_{k_i}\prod_iD^{J_i}_{k_il_i}(g_i)C^{J_1\cdots J_4\Lambda}_{l_1\cdots l_4}
\end{equation}
where the $J_i$ denote the representations of the group G, the $k_i$ the vector indices in the representation space, $C$ are the intertwiners and $\Lambda$ are some extra labels which will depend on the group in consideration.

It is precisely this definition of the fields in configuration space\footnote{It is worth noting at this point that the variables utilised in configuration space are the group elements of G, while the variables utilised in momentum space are the representations of the group G. This is precisely what happens in spin foam models when one labels the original simplex in terms of group elements $e$, and the dual simplex in terms of representations of the group.} which provides the link between GFT and both spin foam and loop quantum gravity. In fact, the states of GFT in momentum space are precisely the \emph{spin network states} of LQG and the boundary states in spin foams models. This is a consequence of the fact that GFT, as well as LQG, makes use of a description of gravity in terms of tetrads and connections instead of metric fields. \\
In this setting the group elements represent parallel transport of a connection along a path dual to the (D-2)-face, while the representations represent the volume of the same (D-2)-face.\\
Given the field $\phi$, the second quantisation is obtained by promoting the spin network functions to operators, choosing a field action and defining a partition function, which is defined, perturbatively, in terms of Feynman diagrams. This procedure presupposes a Fock space structure with creation and annihilation operators of (D-1)-simplices.

In fact, in GFT, the evolution of each quanta of (D-1)-dimension simplicial space is described through a scattering process, in which an initial state gets transformed to a final state through creation and annihilation of other quanta of (D-1)-simplicial space. Therefore, interaction and evolution is described in terms of D-simplices. \\
The fundamental interaction processes, for which a certain number of (D-1)-simplices gets annihilated and another number of such simplices gets created, correspond to the so called Pachner moves in D-dimensions. A sequence of such moves transforms any (D-1)-triangulation to another (D-1)-triangulation. This evolution picture uncovers the relation between GFT and spin foams. \\
In fact, we have previously stated that in momentum space the states in GFT are spin networks, thus the evolution of such states is given precisely by a 2-complex labelled by representations (spin foam) dual to a D-dimension simplex.\\
In the context of GFT, any D-simplex which represents a specific interaction process is described by a Feynman graph. Let us now analyse, in more detail, how this is done.\\
The classical field action is given by 
\begin{equation}
S_D(\phi,\lambda)=\frac{1}{2}\prod_{i=1}^{D}\int dg_id\tilde{g}_i\phi(g_i)K(g_i\tilde{g}_i^{-1})\phi(\tilde{g}_i)+\frac{\lambda}{D+1}\prod_{i\neq j=1}^{D+1}\int dg_{ij}\phi(g_{1j})\cdots\phi(g_{D+1j})V(g_{ij}g_{ji}^{-1})
\end{equation}
where $K$ is the kinetic term while $V$ is the interaction/vertex term. $K$ describes how the information and degrees of freedom  get transported between two (D-1) simplexes as seen from two different D-simplices, while $V $ describes the interaction of D+1 (D-1)-simplexes to form a D-simplex by gluing the common (D-2)-faces, that are pairwise linked at the interaction vertex (note that each $\phi$ contains in its argument a $g$ which is shared by another $\phi$.). \\
The preturbative expansion of the partition function is obtained through an expansion in terms of Feynman diagrams as follows:
\begin{equation}\label{equ:feyexpantion}
Z=\int D\phi e^{-S[\phi]}=\sum_{\Gamma}\frac{\lambda^N}{Sym\Gamma}Z(\Gamma)
\end{equation}
where $\Gamma$ represents a Feynman graph whose partition function is given by $Z(\Gamma)$. $N $ is the number of vertices and $Sym(\Gamma)$ is a symmetry factor, i.e. number of automorphisms of the Feynman diagram.

The edges of each Feynman graph are composed of various strands, each of which carries a representation $j_g$ on it. Each strand gets re-directed when it crosses an interaction vertex, it follows some path and then, eventually, ends up where it started, thus forming a closed surface. The collection of all these surfaces, together with the edges and vertices, form a two complex that, because of the chosen combinatorics of the arguments in the field, is topologically dual to a D-simplex. In this way each Feynman graph in the expansion can be associated to a D-simplex, which represents a particular scattering process. \\
Since in momentum space strands in each edge of the Feynman diagrams are labelled by representation, it is possible to identify such Feynman graphs with spin foams and the amplitudes for Feynman graphs with spin foam models.  
\begin{equation}\label{equ:amp}
Z(\Gamma)=\sum_{J_f}\prod_fA(J_f)\prod_eA_e(J_{f|e})\prod_vA_v( J_{f|v})
\end{equation}
Since each variable associated to a subsimplex carries a geometrical interpretation (ex length, area, volume), the amplitude in \ref{equ:amp} can be interpreted as a sum over histories for discrete quantum gravity on the specific dual triangulation of the Feynman graph in question. Interestingly, the converse is also true, namely, given a GFT it is always possible to obtain a spin foam model as a perturbative expansion.\\ The sum over Feynman graphs in \ref{equ:amp} then corresponds to a sum over spin foams and, equivalently, a sum over triangulations, which includes a sum over algebraic data (group elements/representations). The perturbative expansion given above allows for a computation of expectation values for GFT observables. Specifically, we get
\begin{equation}\label{equ:sum}
\langle\Psi_1|\Psi_2\rangle:=\sum_{\Gamma/\partial\Gamma=\gamma_{\Psi_1}\cup\gamma_{\Psi_2}}\frac{\lambda^N}{Sym\Gamma}Z(\Gamma)
\end{equation}
where, now, the sum over Feynman diagrams is restricted solely to two complex, whose boundary are spin networks. However, the topology corresponding to any such diagram is not necessarily trivial, since it can be any topology (you do not restrict the sum in the above equation). Instead, if we would like to make connection with LQG, it is conjectures that we would have to restrict the sum in \ref{equ:sum} to Feynman diagrams, whose associated topology is trivial. These are the so called tree diagrams\footnote{In these diagrams one neglects all quantum corrections and incodes only classical information, thus giving a definition of the 2-point function}. We thus obtain
\begin{equation}
\langle\Psi_1|\Psi_2\rangle:=\sum_{\Gamma_{\text{Tree}}/\partial\Gamma=\gamma_{\Psi_1}\cup\gamma_{\Psi_2}}\frac{\lambda^N}{Sym\Gamma}Z(\Gamma)
\end{equation}
The above would be a definition of the canonical inner product for a simplicial version of LQG. This implies that the utilisation of GFT might enable to solve one of the long standing problems of LQG, namely computing the solutions for the Hamiltonian constraint.\\
It is worth mentioning, at this point, the resemblances of GFT and, in particular, of the partition function of GFT to other approaches to quantum gravity. We have already seen the connection among GFT, LQG and spin foams. However, GFT also holds similarities with Regge calculus, Dynamical triangulation and causal sets.

In fact, as in Regge calculus, in GFT  one has a simplicial description of spacetime and a sum over geometrical data. As in dynamical triangulation in GFT, one performs a sum over triangulations dual to 2-complexes, while, by assuming an orientation of the 2-complexes (Feynman diagrams) it is possible to obtain an ordering of events (Feynman vertices), which is similar to causal sets. These similarities would suggest that GFT represents a fundamental structure necessary for any approach to a quantum theory of gravity.\\
Interestingly, it is also possible to couple matter to gravity in GFT. In this definition of GFT it is then possible to define both quanta of matter and gravity in the same way, such that in the perturbative expansion one obtains both Feynman diagrams of gravity and Feynman diagrams of any matter field theory. \\
The degrees of  matter fields and those of gravity should be correctly coupled so to reproduce the correct dynamical interaction between the two; this was done for spin foams in 3d in \cite{night}. \\
In this context the fields present in GFT are now two, the usual one $\phi$ associated to gravity, which represents a (D-1)-simplex (always working in D dimensions) with no particle on it, plus a field associated to matter:
\begin{equation}
\psi_s(g_1,g_2\cdots g_D,u):SU(2)^{\otimes D}\rightarrow\Cl
\end{equation}
Such a field, instead, represents a D-simplex with a particle of spin $s$ associated to a vertex, whose degrees of freedom are encoded in the variable $u$. The field $\psi_s$ has a global SU(2) symmetry obtained by a simultaneous right shift of all its arguments (i.e. $\psi_s(g_1,g_2\cdots g_D,u)=\psi_s(g_1g,g_2g\cdots g_Dg,ug)$ ).

If we consider a simple example in 3 dimensions we would obtain, in momentum space, that $\phi$ would represent a 3-valent spin network vertex (dual to a triangle), that gives closed spin network states when contracted to other such vertices. The field $\phi_s$, instead, would represent 4-valent spin network vertices (dual to tetrahedrons), which, when combined to other such 4-valent vertices, would give open spin networks. These latter spin networks represent both quantum gravity states and multi particle states. \\
The task of GFT is, then, to describe the dynamical evolution with creation and annihilation of the two above mentioned structures, in terms of spacetime Feynman diagrams and of matter Feynman diagrams, corresponding to a particle with spin $s$ embedded in the former.\\ The mass of such particle appears as a dynamical quantity in the interaction with gravity, i.e. it appears as a geometrical degree of freedom.
\section{GFT In 3-Dimensions Spin Foam Models}
\begin{figure}[htb]
 \begin{center}
 \includegraphics[scale=0.5]{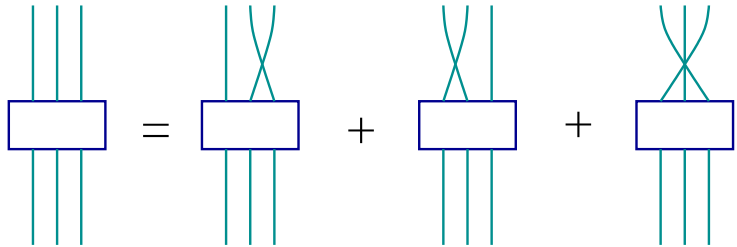}
\caption{Propagator in 3-dimensions where odd permutations are shown explicitly. Each line represents a delta function, while gauge invariance is represented by a box.\label{fig:kinetic} }
\end{center}
  \end{figure}

\begin{figure}[htb]
 \begin{center}
 \includegraphics[scale=0.5]{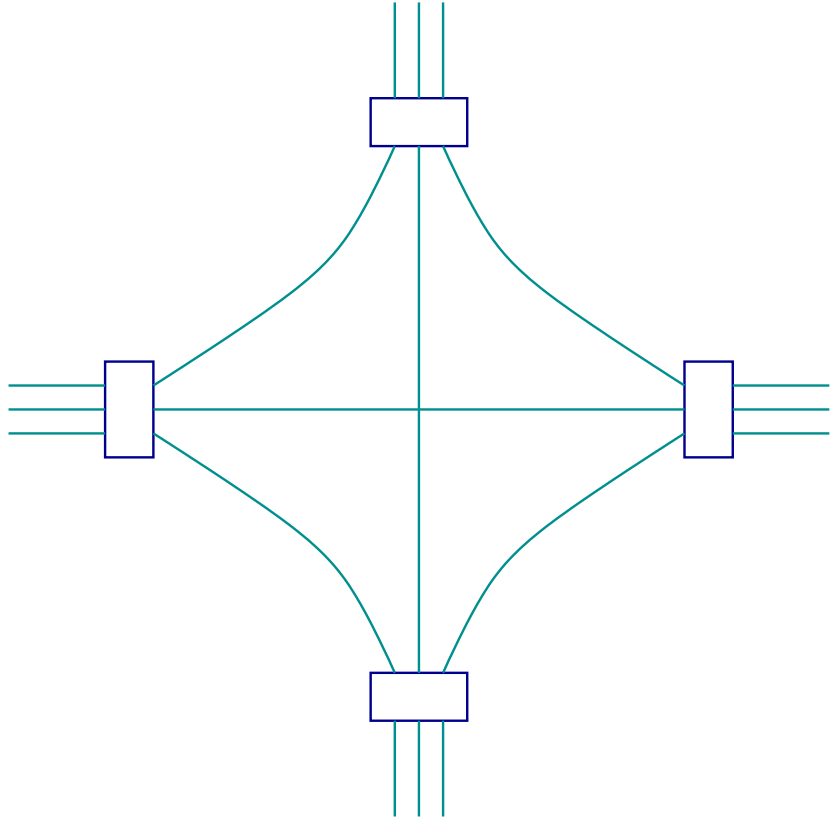}
\caption{The vertex in 3-dimensions. The permutations have been omitted. The vertex has the structure of a tetrahedron, where each set of incoming lines represents a triangle.\label{fig:potential} }
\end{center}
  \end{figure}

In Section \ref{ssfm3} we have seen how the partition function for a spin foam model in 3-dimensions is constructed (see equations  \ref{equ:transamp}). However, in order to make such a model a theory of gravity one has to consider the spacetime manifold as a dynamical quantity, therefore varying. In this respect, the partition function between two spin networks has to be implemented as sum over all possible 2-complexes interpolating the given spin networks. This sum over 2-complexes, which can alternatively be seen as a sum over triangulations, can be achieved through GFT.\\
In particular, in the 3-dimensional case at hand, the field will be a real function of three SU(2) elements $\phi(g_1, g_2, g_3)$, which undergoes the following symmetries:
\be
\phi(g_1, g_2, g_3)=(g_{\pi(1)}, g_{\pi(2)}, g_{\pi(3)})\hspace{.5in}\phi(g_1, g_2, g_3)=\phi(g_1g, g_2g, g_3g)
\ee
The action is then
\ba
S[\phi]&=&\frac{1}{2}\int dg_1dg_2dg_3\big[P_g(\phi(g_1,g_2,g_3))\big]^2\\
&-&\frac{\lambda}{4!}\int dg_1\cdots dg_6\big[P_{h_1}(\phi(g_1,g_2,g_3))\big]\big[P_{h_2}(\phi(g_3,g_5,g_4))\big]\big[P_{h_3}(\phi(g_4,g_2,g_6))\big]
\big[P_{h_4}(\phi(g_6,g_5,g_1))\big]\nonumber
\ea
where $P_g$ imposes gauge invariance $P_g\phi(g_1,g_2,g_3)=\int_{SU(2)}dg\phi(g_1,g_2,g_3)$
or, in a more compact form,
\be
S[\phi]=\frac{1}{2}\int dg_id\tilde{g}_jK(g_i,\tilde{g}_j)\phi(\tilde{g}_j)-\frac{\lambda}{4!}\int dg_{ij}V(g_{ij})\phi(g_{1j}\phi(g_{2j})\phi(g_{3j})\phi(g_{4j})
\ee
where $\phi(g_i)=\phi(g_1,g_2,g_3)$ and $\phi(g_{1j})=\phi(g_{12},\cdots,\phi g_{14})$. \\
The kinetic and the potential terms are defined, respectively, as follows:
\be
K(g_i,g_j)=\sum_{\pi}\int dg\prod_{i=1}^3\delta(g_ig\tilde{g}^{-1}_{\pi(j)})
\ee
\be
V=\int dh_i\prod_{i<j}\delta(h_ig_{ji}^{-1}g_{ij}h^{-1}_i)
\ee
Since the field $\phi$ is associated to a triangle and its arguments to the edges, the kinetic term represents the gluing of two triangles, while the potential $V$ represents the interaction of four triangles building up a tetrahedron. The graphic interpretation of these two terms is given in pictures \ref{fig:kinetic} and \ref{fig:potential}, respectively.\\
By gluing together vertices (interaction terms) along propagators (kinetic\footnote{The propagator is normally given by the inverse of the kinetic term which, in this case, coincides with the term itself.} terms) one obtains Feynman diagrams. It is now possible to define a perturbative expansion of the partition function in terms of such Feynman diagrams as follows:
\be\label{equ:gft3}
Z=\int d\phi e^{S[\phi]}=\sum_{\Gamma}\frac{\lambda}{sym[\Gamma]}Z({\Gamma})
\ee
as done in equation \ref{equ:feyexpantion}. \\
From the discussion of the previous Section we know that the sum over Feynman diagrams in the above equation corresponds to a sum over oriented 2-complexes dual to 3-dimensional triangulations\footnote{In this case the potential term which corresponds to a vertex in the 2-complex will be dual to the tetrahedron, the propagator will be dual to a triangle, while the surfaces formed by following around each strands is dual to the edges in the triangulation. }. By expressing the field $\phi$ in configuration space, as done in \ref{equ:configspace}, we obtain
\be
P_g\phi(g_1,g_2,g_3)=\sum_{j_1j_2j_3}\sqrt{\Delta_{j_1}\Delta_{j_2}\Delta_{j_3}}\Phi^{j_1j_2j_3}_{m_1m_2m_3}D_{m_1n_1}^{j_1}(g_1)D_{m_2n_2}^{j_2}(g_2)D_{m_3n_3}^{j_3}(g_3)
\ee
where $\Delta_i$ represents the dimension of the representation $j_i$ and $\Phi$ are the Fourier components of the field. The action thus becomes
\be
S[\phi]=\frac{1}{2}\sum_{j_1,j_2,j_3}|\Phi_{m_1m_2m_3}^{j_1j_2j_3}|^2-\frac{\lambda}{4!}\sum_{j_1\cdots j_6}\Phi_{m_1m_2m_3}^{j_1j_2j_3}\Phi_{m_3m_5m_4}^{j_3j_5j_4}\Phi_{m_4m_2m_6}^{j_4j_2j_6}\Phi_{m_6m_5m_1}^{j_6j_5j_1}\left[ \begin{array}{ccc}
j_1 & j_2 & j_3 \\
j_4 & j_5 & j_6 \end{array} \right]
\ee
As it can be seen from the above equation, the kinetic term results in a product of delta functions for the representations $j$ and the projections $m$, which indicate the gluing of triangles, while the potential term is the 6j-symbol and the delta terms, which represent the gluing of triangles to form a tetrahedron. In particular, we have
\ba K &=&\delta_{j_1\tilde{j}_1}\delta_{m_1\tilde{m}_1}\delta_{j_2\tilde{j}_2}\delta_{m_2\tilde{m}_2}\delta_{j_3\tilde{j}_3}\delta_{m_3\tilde{m}_3}\nonumber\\
V& =& \delta_{j_1\tilde{j}_1}\delta_{m_1\tilde{m}_1}\delta_{j_2\tilde{j}_2}\delta_{m_2\tilde{m}_2}\delta_{j_3\tilde{j}_3}\delta_{m_3\tilde{m}_3}\delta_{j_4\tilde{j}_4}\delta_{m_4\tilde{m}_4}\delta_{j_5\tilde{j}_5}\delta_{m_5\tilde{m}_5}\delta_{j_6\tilde{j}_6}\delta_{m_6\tilde{m}_6} \left[ \begin{array}{ccc}
j_1 & j_2 & j_3 \\
j_4 & j_5 & j_6 \end{array} \right]
\ea
The partition function for one Feynman diagram then becomes
\be
Z[\Gamma]=\Big(\prod_f\sum_{j_f}\Big)\prod_f\Delta_{j_f}\prod_v\left[ \begin{array}{ccc}
j_1 & j_2 & j_3 \\
j_4 & j_5 & j_6 \end{array} \right]_v
\ee
Inserting this result in \ref{equ:gft3}, we obtain
\be\label{equ:sumdiv}
Z=\sum_{\Gamma}\frac{\lambda^N}{sym[\Gamma]}\Big(\prod_f\sum_{j_f}\Big)\prod_f\Delta_{j_f}\prod_v\left[ \begin{array}{ccc}
j_1 & j_2 & j_3 \\
j_4 & j_5 & j_6 \end{array} \right]_v
\ee
This expression coincides with the expression for the partition function of a spin foam model in 3-dimensions, but augmented by a sum over all triangulations or, alternatively, over 2-complexes of both different and equal topology.\\
However the sum over topologies in \ref{equ:sumdiv} is bound to diverge. In \cite{d138} is was shown that, by adding an extra interaction term to the sum, this can be solved perturbatively.  The extra term should be of the form
\be
[P_{h_1}\phi(g_1, g_2, g_3)][P_{h_2}\phi(g_3, g_5, g_4)][P_{h_3}\phi(g_4 , g_5 , g_6)][P_{h_4}\phi(g_6 , g_2 , g_1)]
\ee
which represents a set of 4 triangles glued
together in such a way that two pairs of them share a single edge each, while two
other pairs share two edges. For this reason this term is called a ``pillow" in the literature.
\section{GFT In 4-Dimensions Spin Foam Models}
In this Section we will show how it is possible to derive the spin foam model of Section \ref{ssfm4} through GFT techniques, in such a way that a sum over triangulations is introduced in the definition of transition amplitudes.

We recall from Section \ref{sbft}, that the building blocks for constructing a general state in quantum 4-simplicial geometry are tetrahedrons. These tetrahedrons can be represented by a function of four group variables, each of which is associated to the four triangles comprising the tetrahedron. The group to be taken in consideration will differ if we are considering the Riemannian case ($spin(4)$) or the Lorentzian case ($sl(3,1)$). \\
In this setting the field will be the scalar function $\phi(g_1,g_2,g_3,g_4)=\phi(g_i)$. As for the 3-dimensional case, gauge invariance is given in terms of a projection operator $P_g(\phi(g_1,g_2,g_3,g_4))=\int_G dg\phi(g_1g,g_2g,g_3g,g_4g)$, while invariance under permutation is given by $\phi(g_1,g_2,g_3,g_4)=\phi(g_{\pi(1)},g_{\pi(2)},g_{\pi(3)},g_{\pi(4)})$. \\
However, differently from the 3-dimensional case, $\pi$ can identify different types of permutations, i.e. \emph{even}, \emph{mixed}, etc. As we will see, only \emph{even} permutations will allow to define a connection between Feynman diagrams and 2-complexes.

\begin{figure}[htb]
 \begin{center}
 \psfrag{a}{$j_1$}\psfrag{b}{$j_2$}\psfrag{c}{$j_3$}\psfrag{d}{$j_4$}
 \includegraphics[scale=0.5]{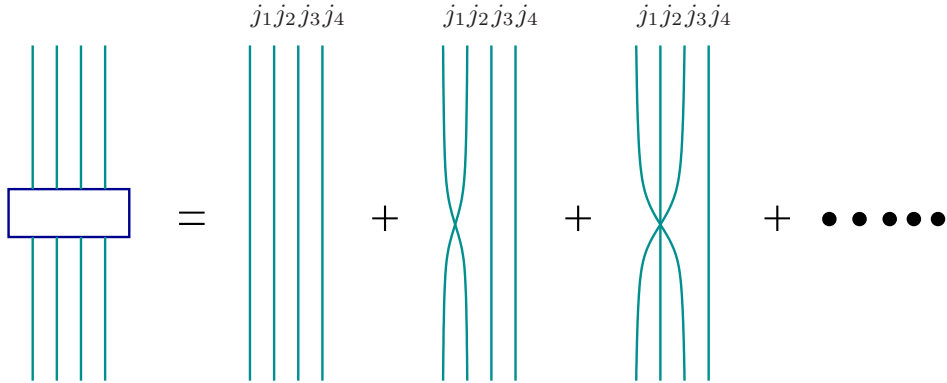}
\caption{The propagator term in 4-dimensions. Each strand carries a representation of the group. In this case, the box represents sums over given permutations of the ordering of the arguments.\label{fig:k4} }
\end{center}
  \end{figure}

\begin{figure}[htb]
 \begin{center}
  \includegraphics[scale=0.5]{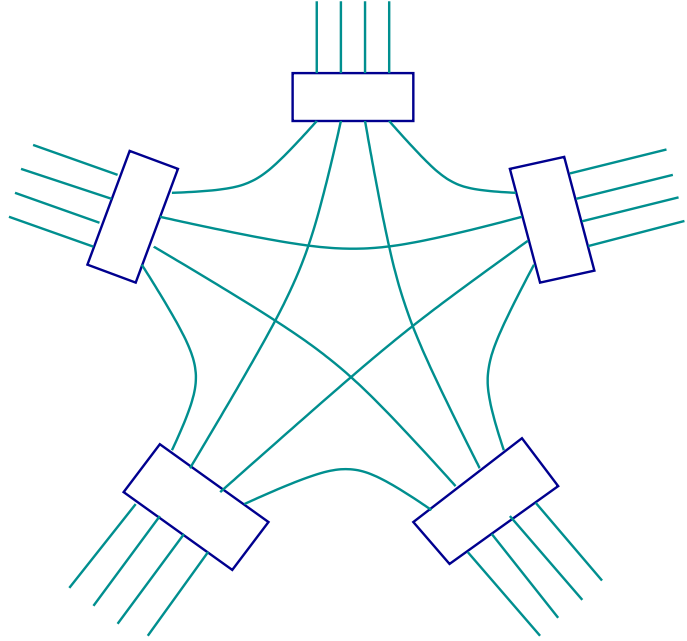}
\caption{The vertex in 4-dimensions. Each strand carries a representation of the group and the boxes have the same meaning as in the propagator. The combinatorial structure of the vertex operator is that of a 4-simplex, with five vertices which represent the five tetrahedrons comprising a 4-simplex. Each vertex (tetrahedron) has four lines coming out of it, which represent the four triangles in a tetrahedron.\label{fig:v4} }
\end{center}
  \end{figure}

If we were merely interested in the GFT representation of a general BF-theory, then the above elements would suffice to give us the desired action of the theory, as a $\phi^5$ action, i.e. $S=\int\phi^5+\lambda\phi^4$. However, to attain a theory of gravity we need to impose the analogue of the simplicity constraints which, in this case, are defined through a projection operator $P_h$ for $h\in H\subset G$: $P_h(\phi(g_1,g_2,g_3,g_4))=\Big(\prod_{i=1,\cdots4}\int_H dh_i\Big)\phi(g_1h,g_2h,g_3h,g_4h)$. Such constraint imposes that the representation has be be simple with respect to the subgroup $H\subset G$. \\
Different ways of imposing these extra constraints will lead to different versions of the GFT formulation of the Barrett-Crane model.\\
The form of the action is then 
\be
S[\phi]=\frac{1}{2}\int dg_i d\tilde{g}_i\phi(g_i)K(g_i,\tilde{g}_i)\phi(\tilde{g}_i)-\frac{\lambda}{5}\int dg_{ij}V(g_{ij})\phi(g_{1i})\phi(g_{2i})\phi(g_{3i})\phi(g_{4i})\phi(g_{5i})
\ee
where $\phi(g_{1i})=\phi(g_{12},g_{13},g_{14},g_{15})$, $K$ is the kinematic operator, whose inverse represents the propagator and $V$ is the potential term (vertex operator). See figures \ref{fig:k4} and \ref{fig:v4}, respectively. \\
Following the same procedure, as carried out at the beginning of chapter \ref{chap:gft}, we arrive at the definition of the partition function in terms of Feynman diagrams as follows:
\be\label{equ:feyspin}
Z=\sum_{\Gamma}\frac{\lambda^{v[\Gamma]}}{5!^v v! sym[\Gamma]}Z[\Gamma]
\ee
where $v[\Gamma]$ is the number of vertices in the Feynman diagram $\Gamma$ and $sym[\Gamma]$ is the symmetry factor.

\begin{figure}[htb]
 \begin{center}
\psfrag{A}{$A$} \psfrag{a}{$V_1$}\psfrag{b}{$V_2$}\psfrag{c}{$V_2$}\psfrag{d}{$V_4$}
  \includegraphics[scale=0.5]{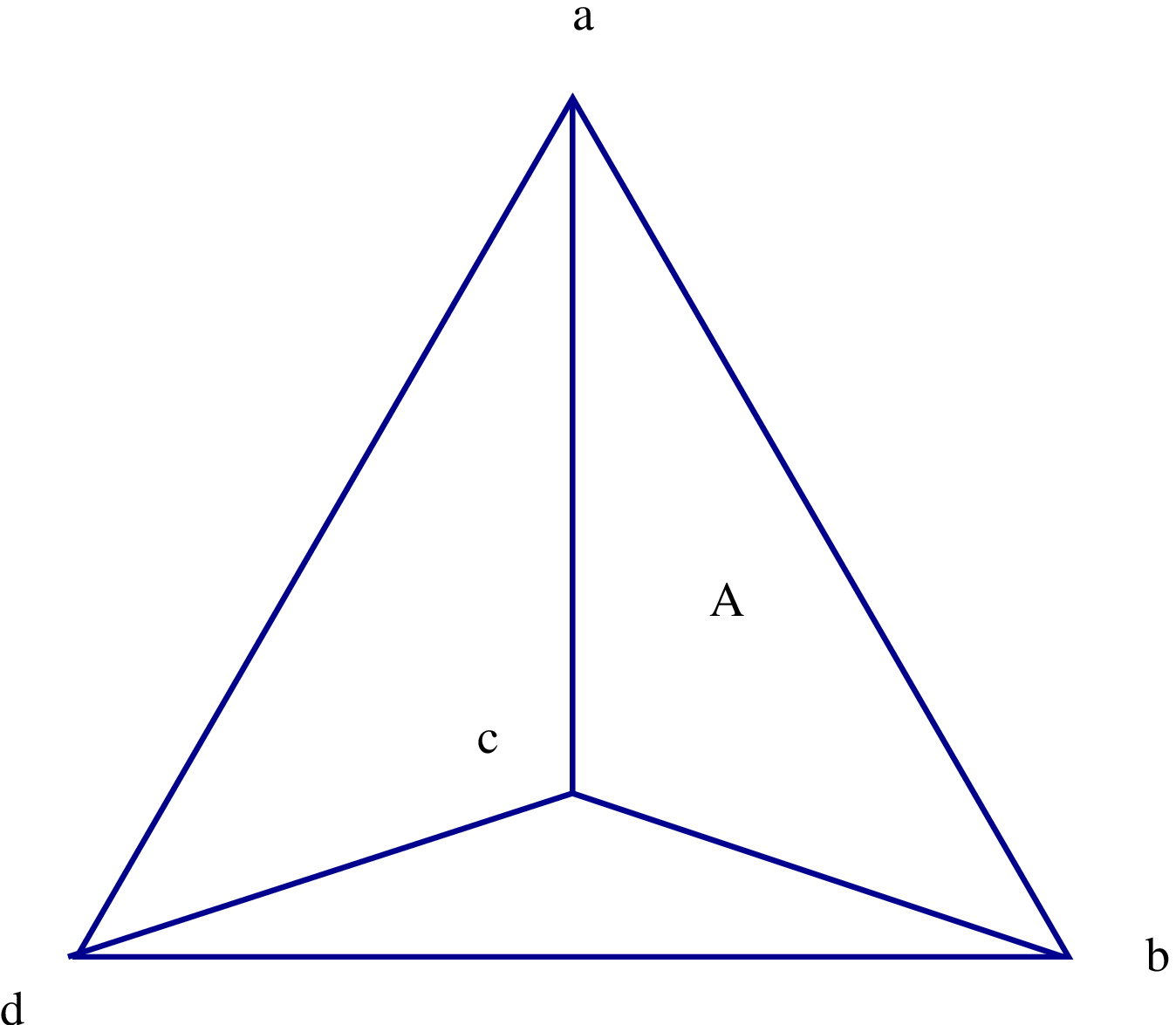}
\caption{3-simplex with an ordering of its boundary vertices.\label{fig:orientation} }
\end{center}
  \end{figure}

\begin{figure}[htb]
 \begin{center}
\psfrag{A}{$A$} \psfrag{a}{$V_1$}\psfrag{b}{$V_3$}\psfrag{c}{$V_2$}
  \includegraphics[scale=0.5]{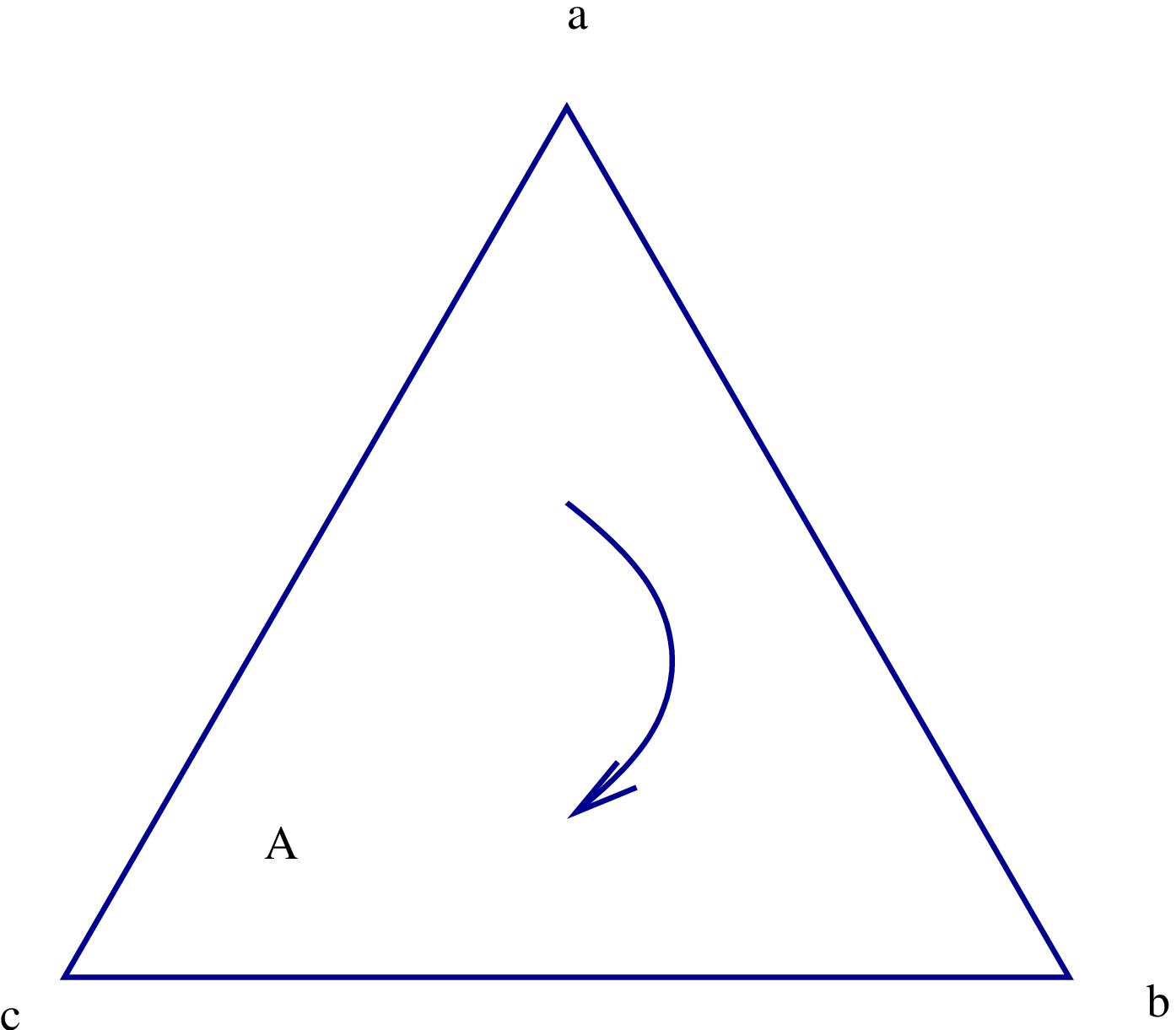}
\caption{Orientation of a triangle on the boundary of a tetrahedron induced by the ordering of the vertices of the tetrahedron .\label{fig:orientation triangle} }
\end{center}
  \end{figure}

As done for the 3-dimensional case, it is possible to associate to each Feynman diagram a 2-complex. Specifically, each of the four strands of a propagator goes through several vertices and propagators until, eventally, goes back to the starting point, thus forming a closed surface. \\
Moreover, since each strand in momentum space is labelled by a representation, these closed surfaces acquire the representation label of the strand that incloses them. The collection of all such faces together with the edges and the vertices forms a labelled 2-complex, i.e. a spin foam. Therefore, equation \ref{equ:feyspin} represents a spin foam model where $Z[\Gamma]$ is the amplitude for each spin foam $\Gamma$. \\
However, in order to obtain oriented 2-complexes in the above expansion, in \cite{d133}, \cite{tesid} it was shown  that only \emph{even} permutations of the field have to be taken into consideration. To understand why this is the case, we need to make a little digression on how orientations of simplices are defined. In particular, given an n-simplex $T$ an orientation of $T$ consists in a choice of ordering, up to \emph{even} permutations, of the (n + 1) 0-simplices (vertices) on its boundary. The (n+1) (n-1)-simplex on the boundary of $T$ are bounded by $n$ 0-simplices (vertices), i.e. the same vertices of the n-simplex but with one missing.  An orientation of these (n-1)-simplices can be obtained by considering an \emph{even} ordering of all the boundary points of the n-simplex, in which the missing point appears at the first place. \\
This induces an outgoing orientation of the (n-1)-simplex with respect to the n-simplex. To understand this, let us consider a simple example in 3-dimensions. In this case, an n-simplex would be a tetrahedron. We then define an ordering of its vertices as shown in figure \ref{fig:orientation}. Now, consider the triangle $A$, its boundary vertices are obtained from those of the tetrahedron with the exclusion of $V_4$. The orientation of this triangle is obtained by considering the following \emph{even} ordering of the vertices $V_4 V_1 V_3 V_2 $ and, then, dropping the missing one, thus, obtaining $ V_1 V_3 V_2 $. This orientation of the triangle $A$ is shown in figure \ref{fig:orientation triangle}.\\

Coming back to the general case, two n-simplices sharing an (n-1)-simplex have consistent orientation if
the shared (n-1)-simplex inherits opposite orientations from the two n-simplices. If all the n-simplices of a triangulation admit a consistent orientation, then we say that the triangulation is
orientable.\\
In the case of 2-complexes with 5-valent vertices and 4-valent edges, each vertex is given an orientation by the ordering of its adjacent edges up to \emph{even} permutation. This is a consequence of the fact that there is a 1:2:1 correspondence between the orientation of the (n-1)-simplices at the boundary of an n-simplex and the ordering of the boundary vertices. Specifically, each (n-1)-simplex can be paired with the vertex that does not belong to it, therefore an ordering of the points corresponds to an assignment of orientation to the (n-1)-simplices. \\
Therefore, in a 2-complex, an ordering of the vertices induces an orientation on the adjacent edges which, from the discussion above, corresponds to an ordering of the faces.\\
Similarly, as for a triangulation, we say that two
vertices joined by an edge have consistent orientation if the edge is given opposite orientation by the two vertices. A
2-complex is orientable if all its vertices can be consistently oriented.\\
If we now require the field $\phi$ to be invariant only under \emph{even} permutations, then the form of the action, writing down all the terms explicitly, would be
\ba
S[\phi]&=&\int \prod_{i-1}^4dg_i\phi(g_1 , g_2 , g_3 , g_4 )\phi(g_3 , g_2 , g_1 , g_4 )\\
& & +\frac{\lambda}{5!}\int\prod_{i=1}^{10}dg_i\phi(g_1 , g_2 , g_3 , g_4 ) \phi(g_4 , g_5 , g_6 , g_7 ) \phi(g_7 , g_3 , g_8 , g_9 ) \phi(g_9 , g_6 , g_2 , g_{10} ) \phi(g_{10} , g_8 , g_5 , g_1 )\nonumber
\ea
In this case the propagator only contains \emph{odd} permutations, while the vertex only allows for a pairing of the strands which causes \emph{odd} permutations only. Since the strands of the edges of the Feynman diagrams go through an equal number of vertices and propagators when forming a closed loop, they undergo an \emph{even} number of \emph{odd} permutation. Therefore the 2-complex is orientable. \\

It turns out that the 2-complexes defined above can be seen as dual to the triangulation obtained by the gluing of faces of co-dimension 1 of simplices. 
\\
Moreover, the sum over 2-complexes (or equivalent triangulations) is really a sum over all possible triangulations for a given topology, but also a sum of all triangulations of different topologies. In fact, the former is obtained from the different permutation within each propagator but all with the same pairing while, the latter, is obtained by all possible pairings. Therefore the use of GFT enables us not only to obtain a sum over triangulation, but also a sum over topologies\footnote{It should be noted that not every Feynman diagram is equivalent to a (oriented) 2-complex that triangulates a topological manifold. }. \\

As we mentioned at the beginning of this section, there are various versions of the GFT version of the Barrett-Crane model, which derive on how the projectors $P_g$ and $P_h$ are applied. In particular, considering as our basic fields the gauge invariant ones, i.e. $P_g\phi$ there are two possibilities. \\
For a detailed derivation, discussion and comparison the reader is referred to \cite{tesid}, \cite{d134}.
\begin{itemize}
\item It is possible to impose the combinations of projectors $P_gP_h$  as acting only on the interaction term, thus obtaining as an action
\ba
S[\phi]&=&\frac{1}{2}\int dg_1\cdot dg_4 [P_g\phi(g_1,g_2,g_3,g_4)]^2+\frac{1}{5}\int dg_1\cdot dg_{10}[P_gP_hP_g\phi(g_1,g_2,g_3,g_4)]\nonumber\\
& & \times[P_gP_hP_g\phi(g_4,g_5,g_6,g_7)] [P_gP_hP_g\phi(g_7,g_3,g_8,g_9)][P_gP_hP_g\phi(g_9,g_6,g_2,g_{10})]\nonumber\\
& &\times [P_gP_hP_g\phi(g_{10},g_8,g_5,g_1)]
\ea
The kinetic term is then
\be
K(g_i,\tilde{g}_j)=\sum_{\pi_e}\int dg\prod_i \delta(g_ig\tilde{g}^{-1}_{\pi_e(i)})
\ee
where $\pi_e$ indicates only \emph{even} permutations.\\
The vertex operator is
\be
V(g_{ij})=\frac{1}{5!}\int d\beta_id\tilde{\beta}_idh_{ij}\prod_{i<j}\delta(g_{ji}^{-1}\tilde{\beta}_ih_{ij}\beta^{-1}_i\beta_jh_{ji}\tilde{\beta}^{-1}_jg_{ij})
\ee
where $\beta, \beta^{'}\in SO(4)$ and $h_{ij}\in SO(3)$ \\
The amplitude for a single Feynman graph $\Gamma$ is derived to be
\be
Z(\Gamma)=\sum_{\rho}\prod_f\Delta_{\rho_f}\prod_e\frac{\Delta_{1234}}{\Delta_{\rho_{e_1}}\Delta_{\rho_{e_2}}\Delta_{\rho_{e_3}}
\Delta_{\rho_{e_4}}}\prod_v \mb^{BC}_v
\ee
where $\Delta_{\rho_{e_i}}$ represents the dimension of the representation $\rho_{e_i}$, $\Delta_{1234}$ is the number of possible 
intertwiner between the representations $\rho_{e_{i}}$, $i=1,\cdots 4$ and $\mb^{BC}_v$ is the vertex amplitude for the Barrett-Crane model.\\
This model is called the \emph{Perez-Rovelli GFT version of the BC model} for the Riemannian case.
\item The second possibility is to impose only the projection $P_h$ to both the kinetic and interaction terms, thus obtaining the following form of the action
\ba
S[\phi]&=&\frac{1}{2}\int dg_1\cdots dg_4 [P_hP_g\phi(g_1,g_2,g_3,g_4)]^2+\frac{1}{5}\int dg_1\cdots dg_{10}[P_hP_g\phi(g_1,g_2,g_3,g_4)]\nonumber\\
& & \times[P_hP_g\phi(g_4,g_5,g_6,g_7)][P_hP_g\phi(g_7,g_3,g_8,g_9)][P_hP_g\phi(g_9,g_6,g_2,g_{10})]\nonumber\\
& & \times [P_hP_g\phi(g_{10},g_8,g_5,g_1)]
\ea
The kinetic term is 
\be
K(g_i,\tilde{g}_j)=\sum_{\pi_e}\int dh_idgd\tilde{g}\prod_i \delta(g_igh_i\tilde{g}^{-1}_{\pi_e(i)})
\ee
while the potential term is
\be
V(g_{ij})=\frac{1}{5!}\int d\beta_id\tilde{\beta}_idh_{ij}\prod_{i<j}\delta(g_{ji}^{-1}\tilde{\beta}_ih_{ij}\beta^{-1}_i\beta_jh_{ji}\tilde{\beta}^{-1}_jg_{ij})
\ee
The resulting amplitude for a single Feynman graph is 
\be
Z(\Gamma)\sum_{\rho}=\prod_f\Delta_{\rho_f}\prod_{e^*}\Delta_{1234}^{-1}\prod_v \mb^{BC}_v
\ee
This model is called the \emph{DePietri-Freidel-Rovelli GFT version of the BC model} for the Riemannian case.
\end{itemize}
The above two examples represent a derivation of the GFT representation of the Barrett-Crane model with the advantage that now the sum is taken over all possible Feynman diagrams, as shown in equation \ref{equ:feyspin}. \\

Interestingly, the convergence behaviour for the two above models for a fixed triangulation has shown to be very different. In fact, on the one hand the DePietri-Freidel-Krasnov-Rovelli model diverges very rapidly even for simple triangulations. This divergence problem is caused by the rapid increase of the face amplitude, therefore for a triangulation in which very many 4-simplex share a common triangle (degenerate triangulations) this problem might be absent. On the other hand the Perez-Rovelli model is convergent for non-degenerate triangulations \cite{d187}. The convergence is determined by the term representing the gluing of 4-simplex along common tetrahedrons, obtained by integration over the group elements assigned to the common edges that are being glued. This implies that the most general configuration are the ones in which most of the faces are labelled by zero spin, while only few isolated ones are labelled by higher spins. The physical significance of this is not clear.

\chapter*{}
\vspace*{\fill}
\begingroup
\centering
\begin{center}
{\Huge{\textbf{PART II}}}
\end{center}
\endgroup
\vspace*{\fill}
\newpage
\chapter{Topos Theory In Physics}
\label{che:topostheory}
``We can't solve problems by using the same kind of thinking we used when we created them."\\
(Einstein)\\

The great revolution of the nineteenth century started with the theory of special and general relativity and culminated in quantum theory. However, up to date, there are still some fundamental issues with quantum theory that are yet to be solved. Nonetheless a great deal of effort in fundamental physics is spent on an elusive theory of quantum gravity which is an attempt to combine the two above mentioned theories which seem, as they have been formulated, to be incompatible. In the last five decades, various attempt to formulate such a theory of quantum gravity have been made, but none have fully succeeded in becoming \emph{the} quantum theory of gravity. One possibility of the failure for reaching an a agreement on a theory of quantum gravity might be presence of unresolved fundamental issues already present in quantum theory. Most approaches to quantum gravity adopt standard quantum theory as there starting point, with the hope that the unresolved issues of the theory will get solved along the way. However, it might be the case that these fundamental issues should be solved \emph{before} attempting to define a quantum theory of gravity. 

If one adopts this point of view, the questions that come next are: i) which are the main conceptual issues in quantum theory ii) How can these issues be solved within a new theoretical frame work of quantum theory.\\
Chris Isham, Andreas D\"oring, Jeremy Butterfield and others have proposed that the main issues in the standard quantum
formalism are: (A) the use of critical mathematical ingredients which seem to assume certain properties of space and/or
time which are not entirely justified. In particular it could be the case that such a priori assumptions of space and time are not compatible with a theory of quantum gravity. (B) The instrumental interpretation of quantum theory that denies the possibility of talking about systems without reference to an external observer. A consequence of this issue is the problematic notion of a closed system in quantum cosmology. \\

A possible way to overcome the above mentioned issues is through a reformulation of quantum theory in terms of a different mathematical framework called topos theory (see Appendix for a detailed definition). The reason for choosing topos theory is that it `looks like' sets and is equipped with an internal logic. As we will explain in detail in the following section, both these features are desirable, because they will allow for a reformulation of quantum theory which is more realist (thus solving issue (B)) and which does not rest on a priori assumptions about the nature of space and time.\\ 
The hope is that such a new formulation of quantum theory will shed some light on how a quantum theory of gravity should look like.

In the next section we will describe in detail the reformulation of quantum theory in terms of topos theory

\section{Topos formulation of Quantum Theory}
\label{s.topos}
In this section we will describe the
topos formulation of normal
quantum theory put forward by Chris Isham and Andreas D\"oring in
\cite{andreas1}, \cite{andreas2}, \cite{andreas3}, \cite{andreas4}
and \cite{andreas5} and by Chris Isham, Jeremy Butterfield, and
collaborators \cite{isham1}, \cite{isham2}, \cite{isham3},
\cite{isham4}, \cite{isham5}.

The main idea put forward by the authors in the above-mentioned
papers is that using topos theory to redefine the mathematical
structure of quantum theory leads to a reformulation of quantum
theory in such a way that it is made to `look like' classical
physics. Furthermore, this reformulation of quantum theory has the
key advantages that (i)  no fundamental role is played by the
continuum; and (ii) propositions can be given truth values without
needing to invoke the concepts of `measurement' or `observer`.
Before going into the detail of how this topos-based
reformulation of quantum theory  is carried out, let us first
analyse the reasons why such a reformulation is needed  in the
first place. These concern  quantum theory general and quantum
cosmology in particular.
\begin{itemize}
\item As it stands quantum theory is non-realist. From a mathematical perspective this is reflected in the Kocken-Specher theorem \footnote{
\textbf{Kochen-Specker Theorem}: if the dimension of $\Hi$ is
greater than 2, then there does not exist any valuation function
$V_{\vec{\Psi}}:\mathcal{O}\rightarrow\Rl$ from the set
$\mathcal{O}$ of all bounded self-adjoint operators $\hat{A}$ of
$\Hi$ to the reals $\Rl$ such that  for all
$\hat{A}\in\mathcal{O}$ and all $f:\Rl\rightarrow\Rl$, the following holds $V_{\vec{\Psi}}(f(\hat{A}))=f(V_{\vec{\Psi}}(\hat{A}))$.}. This theorem implies that any statement
regarding state of affairs, formulated within the theory, acquires
meaning contractually, i.e., after measurement. This implies that
it is hard to avoid the Copenhagen interpretation of quantum
theory, which is intrinsically non-realist.
\item Notions of `measurement' and `external observer' pose problems when dealing with cosmology. In fact, in this case there can be no external observer since we are dealing with a closed system. But this then implies that the concept of `measurement' plays no fundamental role, which in turn implies that the standard definition of probabilities in terms of relative frequency of measurements breaks down.
\item The existence of the Planck scale suggests that there is no \emph{a priori} justification for the adoption of the notion of a continuum in the quantum theory used in formulating quantum gravity.
\end{itemize}

These considerations led Isham and D\"oring to search for a
reformulation of quantum theory that is more realist\footnote{By a
`realist' theory we mean one in which the following conditions are
satisfied: (i) propositions form a Boolean algebra; and (ii)
propositions can always be assessed to be either true or false. As
will be delineated in the following, in the topos approach to
quantum theory both of these conditions are relaxed, leading to
what Isham and D\"oring called  a \emph{neo-realist} theory.} than
the existing one. It turns out that this can be achieved through
the adoption of topos theory as the mathematical framework with
which to reformulate Quantum theory.

One approach to reformulating quantum theory in a more realist way
is to re-express it in such a way that it `looks like' classical
physics, which is the paradigmatic example of a realist theory.
This is precisely the strategy adopted by the authors in
\cite{andreas1}, \cite{andreas2}, \cite{andreas3}, \cite{andreas4}
and \cite{andreas5}. Thus the first question is what is the
underlining structure which makes classical physics a realist
theory?

The authors identified this structure with the following elements:
\begin{enumerate}
\item The existence of a state space $S$.

\item  Physical quantities are represented by functions from the state space to the reals. Thus each physical quantity, $A$, is represented by a function
\begin{equation}
f_A:S\rightarrow \Rl
\end{equation}
\item Any propositions of the form ``$A\in \Delta$'' (``The value of the quantity A lies in the subset $\Delta\in\Rl$'') is represented by a subset of the state space $S$: namely, that subspace for which the proposition is true. This is just
\begin{equation}
f_A^{-1}(\Delta)=\{s\in S| f_A(s)\in\Delta\}
\end{equation}
The collection of all such subsets forms  a Boolean algebra,
denoted ${\rm Sub}(S)$.

\item States $\psi$ are identified with Boolean-algebra homomorphisms
\begin{equation}\label{equ:state}
\psi:{\rm Sub}(S)\rightarrow \{0,1\}
\end{equation}
from the Boolean algebra ${\rm Sub}(S)$ to the two-element
$\{0,1\}$. Here, $0$ and $1$ can be identified as `false' and
`true' respectively.

The identification of states with such maps follows from
identifying propositions with subsets of $S$. Indeed, to each
subset $f_A^{-1}(\{\Delta\})$, there is associated a
characteristic function
$\chi_{A\in\Delta}:S\rightarrow\{0,1\}\subset\Rl$ defined by
\begin{equation}
\chi_{A\in\Delta}(s)=\begin{cases}1& {\rm if}\hspace{.1in}f_A(s)\in\Delta;\\
0& \text{otherwise}  \label{eq:mam}.
\end{cases}
\end{equation}
Thus each state $s$ either lies in $f_A^{-1}(\{\Delta\})$ or it
does not. Equivalently, given a state $s$ every proposition about
the values of physical quantities in that state is either true or
false. Thus \ref{equ:state} follows
\end{enumerate}

The first issue in finding quantum analogues of 1,2,3, and 4 is to
consider the appropriate mathematical framework in which to
reformulate the theory. As previously mentioned the choice fell on
topos theory. There were many reasons for this, but a paramount
one is that in any topos (which is a special type of category)
distributive logic arise in a natural way: i.e., a topos has an
internal logical structure that is similar in many ways to the way
in which Boolean algebras arise in set theory. This feature is
highly desirable since requirement 3 implies that the subobjects
of our state space (yet to be defined) should form some sort of
logical algebra.

The second issue is to identify which  topos is the right one to
use. Isham et al achieved this by noticing that the possibility of
obtaining a `neo-realist' reformulation of quantum theory lied in
the idea of a \emph{context}. Specifically, because of the
Kocken-Specher theorem, the only way of obtaining quantum
analogues of requirements 1,2,3 and 4 is by defining them with
respect to commutative subalgebras (the `contexts') of the
non-commuting algebra, $\mathcal{B(H)}$, of all bounded operators
on the quantum theory's Hilbert space.

The set of all such commuting algebras (chosen to be von Neumann
algebras) forms a category, $\V(\Hi)$, called the \emph{context
category}. These contexts will represent classical `snapshots' of
reality, or `world-views'. From a mathematical perspective, the
reason for choosing
 commutative subalgebras as contexts is because, via
the Gel'fand transform\footnote{Given a commutative von Neumann
algebra V, the Gel'fand transform is a map \begin{align}
&V\rightarrow C(\Sig_V)\\
&\hat{A}\mapsto \bar{A}:\Sig_V\rightarrow\Cl
\end{align}
where $\Sig_V$ is the Gel'fand spectrum; $\bar{A}$ is such that
$\forall\lambda\in\Sig_V$ $\bar{A}(\lambda):=\lambda(\hat{A})$.},
it is possible to write the self-adjoint operators in such an
algebra as continuous functions from the Gel'fand
spectrum\footnote{ Given an algebra V, the \emph{Gel'fand
spectrum}, $\Sig_V$, is the set of all multiplicative, linear
functionals, $\lambda:V\rightarrow \Cl$, of norm 1.} to the
complex numbers. This is similar to how physical quantities are
represented in classical physics, namely as maps from the state
space to the real numbers.

The fact that the set of all \emph{contexts} forms a category is
very important. The objects in this category, $\mathcal{V(H)}$,
are defined to be the commutative von Neumann subalgebras of
$\mathcal{B(H)}$, and we say there is an arrow
$i_{V_2,V_1}:V_1\rightarrow V_2$ if $V_1\subseteq V_2$. The
existence of these arrows implies that relations between different
contexts can be formed.
 Then, given this category,
$\V(\Hi)$, of commutative von Neumann subalgebras, the topos  for
formulating quantum theory chosen by Isham et al is the topos of
presheaves over $\V(\Hi)$, i.e. $\Sets^{\V(\Hi)^{op}}$. Within
this topos they define the analogue of 1,2,3, and 4 to be the
following.
\begin{enumerate}
\item The state space is represented by the spectral presheaf $\Sig$.
\begin{definition}
The spectral presheaf, $\Sig$, is the covariant functor from the
category $\V(\Hi)^{op}$ to $\Sets$ (equivalently, the
contravariant functor from $\mathcal{V(H)}$ to $\Sets$) defined
by:
\begin{itemize}
\item \textbf{Objects}: Given an object $V$ in $\V(\Hi)^{op}$, the associated set $\Sig(V)$ is defined to be the Gel'fand spectrum of the (unital) commutative von Neumann sub-algebra $V$; i.e., the set of all multiplicative linear functionals $\lambda:V\rightarrow \Cl$ such that $\lambda(\hat{1})=1$
\item\textbf{Morphisms}: Given a morphism $i_{V^{'}V}:V^{'}\rightarrow V$ ($V^{'}\subseteq V$) in $\V(\Hi)^{op}$, the associated function $\Sig(i_{V^{'}V}):\Sig(V)\rightarrow
\Sig(V^{'})$ is defined for all $\lambda\in\Sig(V)$ to be the
restriction of the functional $\lambda:V\rightarrow\Cl$ to the
subalgebra $V^{'}\subseteq V$, i.e.
$\Sig(i_{V^{'}V})(\lambda):=\lambda_{|V^{'}}$
\end{itemize}
\end{definition}
\item Propositions, represented by projection operators in quantum theory, are identified with clopen subobjects of the spectral presheaf.
 A \emph{clopen} subobject
$\ps{S}\subseteq\Sig$ is an object such that for each context
$V\in \mathcal{V(H)}^{\op}$ the set $\ps{S}(V)$ is a clopen (both
closed and open) subset of $\Sig(V)$ where the latter is equipped
with the usual, compact and Hausdorff, spectral topology. Since
this a crucial step for the concepts to be developed in this thesis
we will briefly outline how it was derived. For a detailed
analysis the reader is referred to \cite{andreas1},
\cite{andreas2}, \cite{andreas3}, \cite{andreas4} and
\cite{andreas5}.

As a first step, we have to introduce the concept of
`daseinization'. Roughly speaking, what daseinization does is to
approximate operators so as to `fit' into any given context $V$.
In fact, because the formalism defined by Isham et al is
contextual, any  proposition  one wants to consider, has to be
studied within (with respect to ) each context $V\in\V(\Hi)$.

To see  how this works,  consider the case in which we would like
to analyse the projection operator $\hat{P}$ corresponding via the
spectral theorem to, say, the proposition ``$A\in\Delta$''. In
particular, let us take a context $V$ such that $\hat{P}\notin
P(V)$ (the projection lattice of $V$). We somehow need to define a
projection operator which does belong to $V$ and which is related
in some way to our original projection operator $\hat{P}$. This
was achieved in \cite{andreas1}, \cite{andreas2}, \cite{andreas3},
\cite{andreas4} and \cite{andreas5} by approximating $\hat{P}$
from above in $V$ with the `smallest' projection operator in $V$
greater than or equal to $\hat{P}$. More precisely, the
\emph{outer daseinization},
 $\delta^o(\hat P)$, of $\hat P$ is defined at each context $V$ by
 \begin{equation}
\delta^o(\hat{P})_V:=\bigwedge\{\hat{R}\in
P(V)|\hat{R}\geq\hat{P}\}
\end{equation}

This process of outer daseinization takes place for all contexts,
and hence gives, for each projection operator $\hat{P}$, a
collection of daseinized projection operators, one for each
context V, i.e.,
\begin{align}
\hat{P}\mapsto\{\delta^o(\hat{P})_V|V\in\V(\Hi)\}
\end{align}
Because of the Gel'fand transform, to each operator $\hat{P}\in
P(V)$ there is associated the map $\bar{P}:\Sig_V\rightarrow\Cl$
which takes values in $\{0,1\}\subset\Rl\subset\Cl$ since
$\hat{P}$ is a projection operator. Thus $\bar{P}$ is a
characteristic function of the subset
$S_{\hat{P}}\subseteq\Sig(V)$ defined by
\begin{equation}
S_{\hat{P}}:=\{\lambda\in\Sig(V)|\bar{P}(\lambda):=\lambda(\hat{P})=1\}
\end{equation}
Since $\bar{P}$ is continuous with respect to the spectral
topology on $\underline\Sigma(V)$, then
$\bar{P}^{-1}(1)=S_{\hat{P}}$ is a clopen subset of
$\underline\Sigma(V)$ since both $\{0\}$ and $\{1\}$ are both closed and open
subsets of the Hausdorff space $\Cl$.

Through the Gel'fand transform it is then possible to define a
bijective map from projection operators, $\delta(\hat{P})_V\in
P(V)$, and clopen subsets of the spectral presheaf $\Sig_V$ where,
for each \emph{context} V,
\begin{equation}\label{equ:smap}
S_{\delta^o(\hat{P})_V}:=\{\lambda\in\Sig_V|\lambda
(\delta^o(\hat{P})_V)=1\}
\end{equation}

This correspondence between projection operators and clopen
subsets of the spectral presheaf $\underline\Sigma$, implies the
existence of a lattice isomorphisms, for each $V$,
\begin{equation}
\mathfrak{S}:P(V)\rightarrow \Sub_{cl}(\Sig)_V\hspace{.2in}
\end{equation}
such that
\begin{equation}
\delta^o(\hat{P})_V\mapsto
\mathfrak{S}(\delta^o(\hat{P})_V):=S_{\delta^o(\hat{P})_V}
\end{equation}

It was shown in \cite{andreas1}, \cite{andreas2}, \cite{andreas3},
\cite{andreas4} and \cite{andreas5} that the collection of subsets
$S_{\delta(\hat{P})_V}$, $V\in\mathcal{V(H)}$, forms a subobject
of $\Sig$. This enables us to define the (outer) daseinization as
a mapping from the projection operators to the subobject of the
spectral presheaf given by
\begin{align}
\delta:&P(\Hi)\rightarrow \Sub_{cl}(\Sig)\\
&\hat{P}\mapsto(\mathfrak{S}(\delta^o(\hat{P})_V))_{V\in\V(\Hi)}=:\ps{\delta(\hat{P})}
\end{align}
We will sometimes denote $\mathfrak{S}(\delta^o(\hat{P})_V)$ as
$\ps{\delta(\hat{P})}_V$

Since the subobjects of the spectral presheaf form a Heyting
algebra, the above map associates propositions to a distributive
lattice. Actually, it is first necessary to show that the
collection of \emph{clopen} subobjects of $\underline\Sigma$ is a
Heyting algebra, but this was done by D\"oring and Isham.

Two particular properties of the daseinization map that are worth
mentioning are
\begin{enumerate}
\item $\delta(A\vee B)=\delta(A)\vee\delta(B)$ i.e. it preserves the ``or" operation
\item $\delta(A\wedge B)\leq\delta(A)\wedge\delta(B)$, i.e. it does not preserve the ``and" operation
\end{enumerate}
\item In classical physics a pure state, $s$, is a point in the state
space.  It is the smallest subset of the state space which has
measure one with respect to the Dirac measure $\delta_s$. This is
a consequence of the one-to-one correspondence which subsists
between pure states and Dirac measure. In particular, for each
pure state $s$ there corresponds a unique Dirac measure
$\delta_s$. Moreover, propositions which are true in a pure state
$s$ are given by subsets of the state space which have measure one
with respect to the Dirac $\delta_s$, i.e., those subsets which
contain s. The smallest such subset is the one-element set
$\{s\}$. Thus a pure state can be identified with a single point
in the state space.

In classical physics, more general states are represented by more
general probability measures on the state space. This is the
mathematical framework that underpins classical statistical
physics.

However, the spectral presheaf $\Sig$ has \emph{no}
points\footnote{In a topos $\tau$, a `point' (or `global element';
or just `element') of an object $O$ is defined to be a morphism
from the terminal object, $1_\tau$, to $O$.}: indeed, this is
equivalent to the Kochen-Specker theorem! Thus the analogue of a
pure state must  be identified with some other construction. There
are two (ultimately equivalent)\ possibilities:  a `state' can be
identified with (i) an element of $P(P(\Sig))$; or (ii) an element
of $P(\Sig)$. The first choice is called the \emph{truth-object}
option; the second is  the \emph{pseudo-state} option. In what
follows we will concentrate on the second option.

Specifically, given a pure quantum state $\psi\in\Hi$ we define
the presheaf
\begin{equation}
\ps{\w}^{\ket\psi}:= \ps{\delta(\ket\psi\langle\psi|)}
\end{equation}
such that for each stage V we have
\begin{equation}
\ps{\delta(\ket\psi\langle\psi|)}_V:=
\mathfrak{S}(\bigwedge\{\hat{\alpha}\in
P(V)|\ket\psi\langle\psi|\leq\hat{\alpha}\}) \subseteq\Sig(V)
\end{equation}
Where the map $\mathfrak{S}$ was defined in equation
(\ref{equ:smap}).

It was shown in \cite{andreas1}, \cite{andreas2}, \cite{andreas3},
\cite{andreas4} and \cite{andreas5} that the map
\begin{equation}
\ket\psi\rightarrow \ps{\w}^{\ket\psi}
\end{equation}
is injective. Thus for each state $\ket\psi$ there is associated a
topos pseudo-state, $\ps{\w}^{\ket\psi}$, which is defined as a
subobject of the spectral presheaf $\Sig$.

This presheaf $\ps{\w}^{\ket\psi}$ is interpreted as the smallest
clopen subobject of $\Sig$ which represents the proposition which
is totally true in the state $\psi$. Roughly speaking,  it is the
closest one can get to defining a point in $\Sig$.

\item For the sake of completeness we will also mention how a physical quantity is represented in this formalism. For a detailed definition and derivation of the terms the reader is referred to \cite{andreas1}, \cite{andreas2}, \cite{andreas3}, \cite{andreas4} and \cite{andreas5}

Given an operator $\hat{A}$, the physical quantity associated to
it is represented by a certain  arrow
\begin{equation}
\Sig\rightarrow \ps{\Rl}^{\leftrightarrow}
\end{equation}
where the presheaf $\ps{\Rl}^{\leftrightarrow}$ is the
`quantity-value object' in this theory; i.e. it is the object in
which physical quantities `take there values'. We note that, in
this quantum case, the quantity-value object is \emph{not necessarily} a
real-number object.
\end{enumerate}

Thus, by using a topos other than the topos of sets it is possible
to reproduce the main structural elements which would render any
theory as being `classical'.
\section{Single-Time Truth Values in the Language of Topos Theory}
\label{ssingletruth}
We are now ready to turn to the question of how truth values are assigned to
propositions, which in this case are represented by daseinized
operators $\delta(\hat{P})$. For this purpose it is worth thinking
again about classical physics. There, we know that a proposition
$\hat{A}\in\Delta$ is true for a given state $s$ if $s\in
f_{\hat{A}}^{-1}(\Delta)$, i.e., if $s$ belongs to those subsets
$f_{\hat{A}}^{-1}(\Delta)$ of the state space for which the
proposition $\hat{A}\in\Delta$ is true. Therefore, given a state
$s$, all true propositions of $s$ are represented by those
measurable subsets which contain $s$, i.e., those subsets which
have measure $1$ with respect to the measure $\delta_s$.

In the quantum case, a proposition of the form ``$A\in\Delta$'' is
represented by the presheaf $\ps{\delta(\hat{E}[A\in\Delta])}$
where $\hat E[A\in\Delta]$ is the spectral projector for the
self-adjoint operator $\hat A$ onto the subset $\Delta$ of the
spectrum of $\hat A$. On the other hand, states are represented by
the presheaves $\ps{\w}^{\ket\psi}$. As described above, these
identifications are obtained using the maps
$\mathfrak{S}:P(V)\rightarrow \Sub_{{\rm cl}}(\Sig_V)$,
$V\in\mathcal{V(H)}$, and the daseinization map
$\delta:P(\Hi)\rightarrow \Sub_{\rm {cl}}(\Sig)$, with the
properties that
\begin{eqnarray}
\{{\mathfrak{S}}(\delta(\hat{P})_V)\mid{V\in\V(\Hi)}\}
&:=&\ps{\delta(\hat{P})}\subseteq \Sig\nonumber\\
\{ {\mathfrak{S}}(\w^{\ket\psi})_V) \mid V\in\V(\Hi)\} &:=&
\ps{\w}^{\ket\psi}\subseteq \Sig
\end{eqnarray}
As a consequence, within the structure of formal, typed languages,
both presheaves $\ps{\w}^{\ket\psi}$ and $\ps{\delta(\hat{P})}$
are terms of type $P\Sig$ \cite{bell}.

We now want to define the condition by which, for each context
$V$, the proposition $(\ps{\delta(\hat{P})})_V$ is true given
$\ps{\w}^{\ket\psi}_V$. To this end we recall that, for each
context $V$, the projection operator $\w^{\ket\psi}_V$ can be
written as follows
\begin{align}
\w^{\ket\psi}_V&=\bigwedge\{\hat{\alpha}\in
P(V)|\ket\psi\langle\psi|\leq\hat{\alpha}\}\nonumber\\
&=\bigwedge\{\hat{\alpha}\in
P(V)|\langle\psi|\hat{\alpha}\ket\psi=1\}\nonumber\\
&=\delta^o(\ket\psi\langle\psi|)_V
\end{align}
This represents the smallest projection in P(V) which has
expectation value equal to one with respect to the state
$\ket\psi$. The associated subset of the Gel'fand spectrum is
defined as
$\ps{\w}^{\ket\psi}_V=\mathfrak{S}(\bigwedge\{\hat{\alpha}\in
P(V)|\langle\psi|\hat{\alpha}\ket\psi=1\})$. It follows that
$\ps{\w}^{\ket\psi}:= \{\ps{\w}^{\ket\psi}_V \mid{V\in\V(\Hi)}\}$
is the subobject of the spectral presheaf $\Sig$ such that at each
context $V\in\V(\Hi)$ it identifies those subsets of the Gel'fand
spectrum which correspond (through the map $\mathfrak{S}$) to the
smallest projections of that context which have expectation value
equal to one with respect to the state $\ket\psi$; i.e., which are
true in $\ket\psi$.

On the other hand, at a given context $V$, the operator
$\delta(\hat{P})_V$ is defined as
\begin{equation}
\delta^o(\hat{P})_V:=\bigwedge\{\hat{\alpha}\in
P(V)|\hat{P}\leq\hat{\alpha}\}
\end{equation}
Thus the sub-presheaf $\ps{\delta(\hat{P})}$ is defined as the
subobject of $\Sig$ such that at each context $V$ it defines the
subset $\ps{\delta(\hat{P})}_V$ of the Gel'fand spectrum $\Sig(V)$
which represents (through the map $\mathfrak{S}$) the projection
operator $\delta(\hat{P})_V$.

We are interested in defining the condition by which the
proposition represented by the subobject $\ps{\delta(\hat{P})}$ is
true given the state $\ps{\w}^{\ket\psi}$. Let us analyse this
condition for each context V. In this case, we need to define the
condition by which the projection operator $\delta(\hat{P})_V$
associated to the proposition $\ps{\delta(\hat{P})}$ is true given
the pseudo state $\ps{\w}^{\ket\psi}$. Since at each context $V$
the pseudo-state defines the smallest projection in that context
which is true with probability one: i.e., $(\w^{\ket\psi})_V$. For
any other projection to be true given this pseudo-state, this
projection must be a coarse-graining of $(\w^{\ket\psi})_V$, i.e.,
it must be implied by $(\w^{\ket\psi})_V$. Thus if
$(\w^{\ket\psi})_V$ is the smallest projection in $P(V)$ which is
true with probability one, then the projector $\delta(\hat{P})_V$
will be true if and only if $\delta(\hat{P})_V\geq
(\w^{\ket\psi})_V$. This condition is a consequence of the fact
that if $\langle\psi|\hat{\alpha}\ket\psi=1$ then for all
$\hat{\beta}\geq\hat{\alpha}$ it follows that
$\langle\psi|\hat{\beta}\ket\psi=1$.

So far we have defined a `truthfulness' relation at the level of
projection operators. Through the map $\mathfrak{S}$ it is
possible to shift this relation to the level of subobjects of the
Gel'fand spectrum:
\begin{align}
\mathfrak{S}((\w^{\ket\psi})_V)&\subseteq
\mathfrak{S}(\delta(\hat{P})_V)\label{equ:truthvalue}\\
\ps{\w^{\ket\psi}}_V&\subseteq
\ps{\delta(\hat{P})}_V\nonumber\\
\{\lambda\in\Sig(V)|\lambda
((\delta^o(\ket\psi\langle\psi|)_V)=1\}&\subseteq
\{\lambda\in\Sig(V)|\lambda((\delta^o(\hat{P}))_V)=1\}
\end{align}
What the above equation reveals is that, at the level of
subobjects of the Gel'fand spectrum, for each context $V$, a
`proposition' can be said to be (totally) true for given a
pseudo-state if, and only if, the subobjects of the Gel'fand
spectrum associated to the pseudo-state are subsets of the
corresponding subsets of the Gel'fand spectrum associated to the
proposition. It is straightforward to see that if
$\delta(\hat{P})_V\geq (\w^{\ket\psi})_V$ then
$\mathfrak{S}((\w^{\ket\psi})_V)\subseteq
\mathfrak{S}(\delta(\hat{P})_V)$ since for projection operators
the map $\lambda$ takes the values 0,1 only.

We still need a further abstraction in order to work directly with
the presheaves $\ps{\w}^{\ket\psi}$ and $\ps{\delta(\hat{P})}$.
Thus we want the analogue of equation (\ref{equ:truthvalue}) at
the level of subobjects of the spectral presheaf, $\Sig$. This
relation is easily derived to be
\begin{equation}\label{equ:tpre}
\ps{\w}^{\ket\psi}\subseteq\ps{\delta(\hat{P})}
\end{equation}

Equation (\ref{equ:tpre})  shows that whether or not a proposition
$\ps{\delta(\hat{P})}$ is `totally true' given a pseudo state
$\ps{\w}^{\ket\psi}$ is determined by whether or not the
pseudo-state is a sub-presheaf of the presheaf
$\ps{\delta(\hat{P})}$. With  motivation, we can now define the
\emph{generalised truth value} of the proposition ``$A\in\Delta$''
at stage $V$, given the state $\ps{\w}^{\ket\psi}$, as:
\begin{align}\label{ali:true}
v(A\in\Delta;\ket\psi)_V&= v(\ps{\w}^{\ket\psi}
\subseteq\ps{\delta(\hat{E}[A\in\Delta])})_V             \\
&:=\{V^{'}\subseteq V|(\ps{\w}^{\ket\psi})_V\subseteq
\ps{\delta(\hat{E}[A\in\Delta]))}_V\}\\ \nonumber
&=\{V^{'}\subseteq
V|\langle\psi|\delta(\hat{E}[A\in\Delta])\ket\psi=1\}
\end{align}
The last equality is derived by the fact that
$(\ps{\w}^{\ket\psi})_V\subseteq \ps{\delta(\hat{P})}_V $ is a
consequence of the fact that at the level of projection operator
$\delta^o(\hat{P})_V\geq (\w^{\ket\psi})_V$. But since
$(\w^{\ket\psi})_V$ is the smallest projection operator such that
$\langle\psi|(\w^{\ket\psi})_V\ket\psi=1$ then
$\delta^o(\hat{P})_V\geq (\w^{\ket\psi})_V$ implies that
$\langle\psi|\delta^o(\hat{P})\ket\psi=1$.

The right hand side of equation (\ref{ali:true}) means that the
truth value, defined at $V$, of the proposition ``$A\in\Delta$''
given the state $\ps{\w}^{\ket\psi}$ is given in terms of all
those sub-contexts $V^{'}\subseteq V$ for which the projection
operator $\delta(\hat{E}[A\in\Delta]))_V$ has expectation value
equal to one with respect to the state $\ket\psi$. In other words,
this \emph{partial} truth value is defined to be the set of all
those sub-contexts for which the proposition is totally true.

The reason all this works is that generalised truth values defined
in this way form  a \emph{sieve} on $V$; and the set of all of
these is a Heyting algebra. Specifically:
$v(\ps{\w}^{\ket\psi}\subseteq \ps{\delta(\hat{P})})_V$ is a
global element, defined at stage V, of the subobject classifier
$\Om:=(\Om_V)_{V\in\V(\Hi)}$ where $\Om_V$ represents the set of
all sieves defined at stage V. The rigorous definitions of both
sieves and subobject classifier are given below. For a detailed
analysis see \cite{topos7}, \cite{topos8}, \cite{andreas1},
\cite{andreas2}, \cite{andreas3}, \cite{andreas4} and
\cite{andreas5}
\begin{definition}
A \textbf{sieve} on an object $A$ in a topos, $\tau$, is a
collection, $S$, of morphisms in $\tau$ whose co-domain is A and
such that, if $f:B\rightarrow A\in S$ then, given any morphisms
$g:C\rightarrow B$ we have $f o g\in S$.
\end{definition}
An important property of sieves is the following. If
$f:B\rightarrow A$ belongs to a sieve $S$ on $A$, then the
pullback of S by f determines a principal sieve on B, i.e.
\begin{equation}
f^*(S):=\{h:C\rightarrow B|f o h \in S\}=\{h:C\rightarrow B\}=:
\;\downarrow\!\! B   \label{eq:principal}
\end{equation}
The \emph{principal sieve} of an object $A$, denoted
$\downarrow\!\! A$, is the sieve that contains the identity
morphism of $A$; therefore it is the biggest sieve on $A$.

For the particular case in which we are interested, namely sieves
defined on the poset $\V(\Hi)$, the definition of a sieve can be
simplified as follows:
\begin{definition}
For all $V\in\V(\Hi)$, a sieve $S$ on $V$ is a collection of
subalgebras $(V^{'}\subseteq V)$ such that, if $V^{'}\in S$ and
$(V^{''}\subseteq V^{'})$, then $V^{''}\in S$. Thus $S$ is a
downward closed set.
\end{definition}
In this case a maximal sieve on $V$ is
\begin{equation}
\downarrow\! V:=\{V^{'}\in\V(\Hi)|V^{'}\subseteq V\}
\end{equation}
The set of all sieves for each \emph{context} $V$ can be fitted
together so as to give the presheaf $\Om$ which is defined as
follows:
\begin{definition}
The presheaf $\Om\in \Sets^{\V(\Hi)^{op}}$ is defined as follows:
\begin{enumerate}
\item For any $V\in\mathcal{V(H)}$, the set $\Om(V)$ is defined as the set of all sieves on $V$.

\item Given a morphism $i_{V^{'}V}:V^{'}\rightarrow V$ $(V^{'}\subseteq V)$, the associated function in $\underline\Omega$ is
\begin{align}
\Om(i_{V^{'}V}):
&\Om(V)\rightarrow \Om(V^{'})\\
&S \mapsto \Om((i_{V^{'}V}))(S):=\{V^{''}\subseteq V^{'}|V^{''}\in
S\}
\end{align}
\end{enumerate}
\end{definition}

In order for the above definition to be correct we need to show
that indeed $\Om((i_{V^{'}V}))(S):=\{V^{''}\subseteq
V^{'}|V^{''}\in S\}$ defines a sieve on $V^{'}$. To this end we
need to show that $\Om((i_{V^{'}V}))(S):=\{V^{''}\subseteq
V^{'}|V^{''}\in S\}$ is a downward closed set with respect to
$V^{'}$. It is straightforward to see this.

As previously stated, truth values are identified with global
section of the presheaf $\Om$. The global section that consists
entirely of principal sieves is interpreted as representing
`totally true': in classical, Boolean logic, this is just  `true'.
Similarly, the global section that consists of empty sieves  is
interpreted as `totally false': in classical Boolean logic, this
is just `false'.

In the context of the topos formulation of quantum theory, truth
values for propositions are defined by equation (\ref{ali:true}).
However, it is important to emphasise that the truth values refer
to proposition at a \emph{given} time. It is straightforward to
introduce time dependence in natural way. For example, we could
use the curve $t\mapsto\ps{\w}^{\ket\psi_t}$ where $\ket\psi_t$
satisfies the usual time-dependent Schr\"odinger equation.

However, our intention is to follow a quite different path and to
extend the topos formalism to temporally-ordered collections of
propositions. Our goal is to construct  a quantum history
formalism in the language of topos theory. In particular, we want
to be able to assign generalised truth values to temporal
propositions. An important question  is the extent to which such
truth values can be derived from the  truth values of the
constituent propositions.

\section{The Temporal Logic of Heyting Algebras of Subobjects}
\subsection{Introducing the tensor product}
In this Section we begin to consider  sequences of propositions at
different times; these are commonly called `homogeneous
histories'. The goal is to assign truth value to such propositions
using a temporal extension of the topos formalism discussed in the
previous Sections.

As previously stated, in the consistent-history program, a central
goal is to get rid of the idea of state-vector reductions induced
by measurements. The absence of the state-vector reduction process
implies that given a  state $\psi(t_{0})$ at time $t_0$, the truth
value (if there is one) of a proposition ``$A_0\in\Delta_0$'' with
respect to $\psi(t_{0})$ should not influence the truth value of a
proposition ``$A_1\in\Delta_1$'' with respect to
$\psi(t_{1})=\hat{U}(t_1,t_0)\psi(t_0)$, the evolved state at time
$t_1$. This suggests that, if it existed, the truth value of  a
homogeneous history should be computable from the truth values of
the constituent single-time propositions.

Of course, such truth values do not exist in standard quantum
theory. However, as we have discussed in the previous Sections,
they \emph{do} in the topos approach to quantum theory.
Furthermore, since there is no explicit state reduction in that
scheme, it seems  reasonable to try to assign a generalised truth
value to a homogeneous history by employing the topos truth values
that can be assigned to the constituent single-time propositions
at each of the time points  in the temporal support of the
proposition.

With this in mind let us consider the (homogeneous) history
proposition $\hat{\alpha}=$ ``the quantity $A_1$ has a value in
$\Delta_1$ at time $t_1$, and then the quantity $A_2$ has value in
$\Delta_2$ at time $t_2$, and then $\ldots$ and then the quantity
$A_n$ has value in $\Delta_n$ at time $t_n$'' which is a
time-ordered sequence of different propositions at different given
times (We are assuming that $t_1<t_2<\cdots<t_n$). Thus $\alpha$
represents a homogeneous history. Symbolically, we can write
$\alpha$ as
\begin{equation}
\alpha=(A_1\in\Delta_1)_{t_1}\sqcap(A_2\in\Delta_2)_{t_2}
\sqcap\ldots\sqcap (A_n\in\Delta_n)_{t_n}
\end{equation}
where the symbol `$\sqcap$' is the temporal connective `and then'.

The first thing we need to understand is how to ascribe some sort
of `temporal structure' to the Heyting algebras of subobjects of
the spectral presheaves at the relevant times. What we are working
towards here is the notion of the `tensor product' of Heyting
algebras. As a  first step towards motivating the definition, let
us reconsider the history theory of classical physics in this
light.

For classical history theory, the topos under consideration is
$\Sets$. In this case the state spaces $\Sigma_i$ for each time
$t_i$, are topological spaces and we can focus on  their Heyting
algebras of open sets. For simplicity we will concentrate on
two-time histories, but the arguments generalise at once to any
histories whose temporal support is a finite set.

Thus, consider propositions $\alpha_1$, $\beta_1$ at time $t_1$
and $\alpha_2$, $\beta_2$ at time $t_2$, and let\footnote{We will
denote the set of open subsets of a topological space, $X$, by
$\Sub_\op(X)$.} $S_1, S^{'}_1\in \Sub_\op(\Sigma_1)$ and
$S_2,S^{'}_2\in \Sub_\op(\Sigma_2)$ be the open
subsets\footnote{Arguably, it is more appropriate to represent
propositions in classical physics with Borel subsets, not just
open ones. However, will not go into this subtlety here.} that
represent them. Now  consider the homogeneous history propositions
$\alpha_1\sqcap\alpha_2$ and $\beta_1\sqcap\beta_2$, and the
inhomogeneous proposition
$\alpha_1\sqcap\alpha_2\vee\beta_1\sqcap\beta_2$. Heuristically,
this proposition is true (or the history is \emph{realised}) if
either history $\alpha_1\sqcap\alpha_2$ is realised, or history
$\beta_1\sqcap\beta_2$ is realised.   In the classical history
theory, $\alpha_1\sqcap\alpha_2$ and $\beta_1\sqcap\beta_2$ are
represented by the subsets (of $\Si_1\times\Si_2$) $S_1\times S_2$
and $S_1'\times S_2'$ respectively.  However, it is clearly not
possible to represent the inhomogeneous proposition
$(\alpha_1\sqcap\alpha_2)\vee(\beta_1\sqcap\beta_2)$ by any subset
of $\Si_1\times\Si_2$ which is itself of the product form
$O_1\times O_2$.

What if instead we consider the proposition
$(\alpha_1\vee\beta_1)\sqcap(\alpha_2\vee\beta_2)$, which is
represented by the subobject $S_1\cup S^{'}_1\times S_2\cup
S^{'}_2$: symbolically, we write
\begin{equation}
(\alpha_1\vee\beta_1)\sqcap(\alpha_2\vee\beta_2)\mapsto S_1\cup
S^{'}_1\times S_2\cup\label{rep:alorbtandalorb} S^{'}_2
\end{equation}
This history has a different meaning from
$(\alpha_1\sqcap\alpha_2)\vee(\beta_1\sqcap\beta_2)$, since it
indicates that at time $t_1$ either proposition $\alpha_1 $ or
$\beta_1$ is realised, and subsequently, at time $t_2$, either
$\alpha_2$ or $\beta_2$ is realised. It is clear intuitively that
we then have the equation
\begin{equation}\label{equ:v2}
(\alpha_1\vee\beta_1)\sqcap(\alpha_2\vee\beta_2):=
(\alpha_1\sqcap\beta_2)\vee(\alpha_1\sqcap\alpha_2)
\vee(\beta_1\sqcap\alpha_2)\vee(\beta_1\sqcap\beta_2)
\end{equation}
The question that arises now is how to represent these
inhomogeneous histories in such a way that equation (\ref{equ:v2})
is somehow satisfied  when using the representation of
$(\alpha_1\vee\beta_1)\sqcap(\alpha_2\vee\beta_2)$ in equation
(\ref{rep:alorbtandalorb}).

The point is that if we take just the product
$\Sub_\op(\Sigma_1)\times \Sub_\op(\Sigma_2)$ then we cannot
represent inhomogeneous histories, and therefore cannot find a
realisation of the right hand side of equation (\ref{equ:v2}).
However, in the case at hand the answer is obvious since we know that
$\Sub_\op(\Sigma_1)\times \Sub_\op(\Sigma_2)$ does not exhaust the
open sets in the topological space $\Si_1\times\Si_2$. By itself,
$\Sub_\op(\Sigma_1)\times \Sub_\op(\Sigma_2)$ is the collection of
open sets in the \emph{disjoint union} of $\Si_1$ and $\Si_2$, not
the Cartesian product.

In fact, as we know, the subsets of $\Si_1\times\Si_2$  in
$\Sub_\op(\Sigma_1)\times \Sub_\op(\Sigma_2)$ actually form a
\emph{basis} for the topology on  $\Si_1\times\Si_2$: i.e., an
arbitrary open set can be written as a \emph{union} of elements of
$\Sub_\op(\Sigma_1)\times \Sub_\op(\Sigma_2)$. It is then clear
that the representation of the inhomogeneous history
$(\alpha_1\sqcap\alpha_2)\vee(\beta_1\sqcap\beta_2)$ is
\begin{equation}
(\alpha_1\sqcap\alpha_2)\vee(\beta_1\sqcap\beta_2)  \mapsto
S_1\times S_1'\cup S_2\times S_2'
\end{equation}
It is easy to check that equation (\ref{equ:v2}) is satisfied in
this representation.

It is not being too fanciful to imagine that we have here made the
transition from the product Heyting algebra
$\Sub_\op(\Sigma_1)\times \Sub_\op(\Sigma_2)$ to a \emph{tensor}
product; i.e., we can tentatively postulate the relation
\begin{equation}
   \Sub_\op(\Sigma_1)\otimes \Sub_\op(\Sigma_2)\simeq
   \Sub_\op(\Si_1\times\Si_2)     \label{subsubcross}
\end{equation}

The task now is to see if some meaning can be given in general to
the tensor product of Heyting algebras and, if so, if it is
compatible with equation (\ref{subsubcross}). Fortunately this is
indeed possible although it is easier to do this in the language
of \emph{frames} rather than Heyting algebras. Frames are easier
to handle is so far as the negation operation is not directly
present. However, each frame gives rise to a unique Heyting
algebra, and vice versa (see below). So nothing is lost this way.

All this is described in detail in the book by Vickers \cite{sv}.
In particular, we have the following definition.
\begin{definition}
A frame A is a poset such that the following are satisfied
\begin{enumerate}
\item Every subset has a join
\item Every finite subset has a meet
\item \emph{Frame distributivity}: $x\wedge\bigvee Y=\bigvee\{x\wedge y:y\in Y\}$

i.e., binary meets distribute over joins. Here $\bigvee Y$
represents the join of the subset $Y\subseteq A$
\end{enumerate}

\end{definition}

We now come to something that is of fundamental importance in our
discussion of topos temporal logic: namely, the definition of the
tensor product of two frames:
\begin{definition}\label{def:tensor}\cite{sv}
Given two frames A and B, the tensor product $A\otimes B$ is
defined to be the frame represented by the following presentation
\begin{align}
\mathcal{T}&\langle a\otimes b,a\in A\text{ and }b\in B|\nonumber\\
 &\bigwedge_i(a_i\otimes b_i)=\big(\bigwedge_ia_i\big)\otimes
 \big(\bigwedge_ib_i\big) \label{Def:TP1}\\
&\bigvee_i(a_i\otimes b)=\big(\bigvee_ia_i\big)\otimes b \label{Def:TP2}\\
&\bigvee_i(a\otimes b_i)=a\otimes\big(\bigvee_ib_i\big)
\label{Def:TP3}
\end{align}
\end{definition}
In other words, we form the formal products, $a\otimes b$, of
elements $a\in A$, $b\in B$ and subject them to the relations in
equations (\ref{Def:TP1})--(\ref{Def:TP3}). Our intention is to
use the tensor product as the temporal connective, $\sqcap$,
meaning `and then'. It is straight forward to show that equations
(\ref{Def:TP1})--(\ref{Def:TP3}) are indeed satisfied with this
interpretation when `$\lor$' and `$\land$' are interpreted as `or'
and `and' respectively.

We note that there are injective maps
\begin{eqnarray}
                i:A&\rightarrow& A\otimes B\nonumber\\
                  a&\mapsto& a\otimes{\rm true}
\end{eqnarray}
and
\begin{eqnarray}
                j:B&\rightarrow& A\otimes B\nonumber\\
                  b&\mapsto& {\rm true}\otimes b
\end{eqnarray}

These frame constructions are easily  translated into the setting
of Heyting algebras with the aid of the following theorem
\cite{sv}
\begin{theorem}\label{the:fh}
Every frame A defines a complete Heyting algebra (cHa) in such a
way that the operations $\wedge$ and $\vee$ are preserved, and the
implication relation $\rightarrow$ is defined as follows
\begin{equation}
a\rightarrow b=\bigvee\{c:c\wedge a\leq b\}
\end{equation}
\end{theorem}
Frame distributivity implies that $(a\rightarrow b)\wedge a\leq
b$, from which it follows
\begin{equation}
c\leq a\rightarrow b\hspace{.5in}\text{iff}\hspace{.5in}c\wedge
a\leq b
\end{equation}
This is  the definition of the pseudo-complement in the Heyting
algebra.

Now that we have the definition of the tensor product of frames,
and hence the definition of the tensor product of Heyting
algebras, we are ready to analyse quantum history propositions in
terms of topos theory.

Within a topos framework, propositions are identified with
subobjects of the spectral presheaf. Thus for example, given two
systems $S_1$ and $S_2$, whose Hilbert spaces are $\Hi_1$ and
$\Hi_2$ respectively, the propositions concerning each system are
identified with elements of $\Subone$ and $\Subtwo$ respectively
via the process of `daseinization'. We will return later to the
daseinization of history propositions, but for the time being we
will often, with a slight abuse of language, talk about elements
of $\Sub(\Sig)$ as `being' propositions rather than as
`representing propositions via the process of daseinization'.

With this in mind, since both $\Subone$ and $\Subtwo$ are Heyting
algebras, it is possible to use definition (\ref{def:tensor}) to
define the tensor product $\Subone\otimes\Subtwo$ which is itself
a Heyting algebra. We propose to use such tensor products to
represent the temporal logic of history propositions.

Because of the existence of a one-to-one correspondence between
Heyting algebras and frames, in the following we will first
develop a temporal logic for frames in quantum theory and then
generalise to a temporal logic for Heyting algebras by utilising
Theorem \ref{the:fh}. Thus we will consider $\Subone$, $\Subtwo$
and $\Subone\otimes\Subtwo$ as frames rather than Heyting
algebras, thereby not taking into account the logical connectives
of implication  and negation. These will then be reintroduced by
applying Theorem \ref{the:fh}.
\begin{definition}
$\Sub(\Sig^{\Hi_1})\otimes \Sub(\Sig^{\Hi_2})$ is the frame whose
generators are of the form $\ps{S_1}\otimes\ps{S_2}$ for
$\ps{S_1}\in \Sub(\Sig^{\Hi_1})$ and $\ps{S_2}\in
\Sub(\Sig^{\Hi_2})$, and such that the following relations are
satisfied
\begin{align}\label{ali:rel}
\bigwedge_{i\in I}(\ps{S_{1}^i}\otimes\ps{S_{2}^i})&=
\big(\bigwedge_{i\in I}
\ps{S_{1}^i} \big)\otimes\big(\bigwedge_{j\in I}\ps{S_{2}^j}\big)\\
\bigvee_{i\in I}\big(\ps{S_{1}^i}\otimes\ps{S_{2}}\big)&=
(\bigvee_{i\in I}\ps{S_{1}^i})\otimes\ps{S_{2}}\\
\bigvee_{i\in I}(\ps{S_{1}}\otimes\ps{S_{2}^i})&=
\ps{S_{1}}\otimes\big(\bigvee_{i\in I}\ps{S_{2}^i}\big)
\end{align}
\end{definition}
for an arbitrary index set $I$. From the above definition it
follows that a general element of $\Sub(\Sig^{\Hi_1})\otimes
\Sub(\Sig^{\Hi_2})$ will be of the form $\bigvee_{i\in
I}\big(\ps{S_{1}^i}\otimes\ps{S_{2}^i}\big)$.

\subsection{Realising the tensor product in a topos}
We propose to use, via daseinization, the Heyting algebra
$\Subone\otimes \Subtwo$ to represent the temporal logical
structure with which to handle (two-time) history propositions in
the setting of topos theory. A homogeneous history
$\alpha_1\sqcap\alpha_2$ will be represented by the daseinized
quantity
$\ps{\delta(\hat\alpha_1)}\otimes\ps{\delta(\hat\alpha_2)}$ and
the inhomogeneous history
$(\alpha_1\sqcap\alpha_2)\vee(\beta_1\sqcap\beta_2)$ by
$\ps{\delta(\hat\alpha_1)}\otimes\ps{\delta(\hat\alpha_2)}\lor
\ps{\delta(\hat\beta_1)}\otimes\ps{\delta(\hat\beta_2)}$, i.e. we
denote
\begin{equation}
(\alpha_1\sqcap\alpha_2)\vee(\beta_1\sqcap\beta_2)\mapsto
\ps{\delta(\hat\alpha_1)}\otimes\ps{\delta(\hat\alpha_2)}\lor
\ps{\delta(\hat\beta_1)}\otimes\ps{\delta(\hat\beta_2)}
\end{equation}
Here, the `$\lor$' refers to the `or' operation in the Heyting
algebra $\Subone\otimes \Subtwo$.

Our task now is to relate this, purely-algebraic representation,
with one that involves subobjects of some object in some topos. We
suspect that there should be some  connection with
$\Sub(\Sig^{\Hi_1\otimes\Hi_2})$, but at this stage it is not
clear what this can be. What we need is a topos in which there is
some object whose Heyting algebra of sub-objects is isomorphic to
$\Sub(\Sig^{\Hi_1})\otimes \Sub(\Sig^{\Hi_2})$: the connection
with $\Sub(\Sig^{\Hi_1\otimes\Hi_2})$ will then hopefully become
clear.

Of course, in classical physics the analogue of
$\Sig^{\Hi_1\otimes\Hi_2}$ is just the Cartesian product
$\Si_1\times\Si_2$, and then, as we have indicated above, we have
the relation $\Sub_\op(\Sigma_1)\otimes \Sub_\op(\Sigma_2)\simeq
\Sub_\op(\Si_1\times\Si_2)$. This suggests that, in the quantum
case, we should start by looking at the `product'
$\Sig^{\Hi_1}\times\Sig^{\Hi_2}$. However, here we immediately
encounter the problem that $\Sig^{\Hi_1}$ and $\Sig^{\Hi_2}$ are
objects in \emph{different} topoi\footnote{Of course, in the case
of temporal logic, the Hilbert spaces $\Hi_1$ and $\Hi_2$ are
isomorphic, and hence so are the associated topoi. However, their
structural roles in the temporal logic are clearly different. In fact,
in
the closely related situation of composite systems it will
generally be the case that $\Hi_1$ and $\Hi_2$ are \emph{not}
isomorphic. Therefore, in the following, we will not exploit this
particular  isomorphism. }, and so we cannot just take their `product' in the
normal categorial way.

To get around this  let us consider heuristically what defining something like
`$\Sig^{\Hi_1}\times\Sig^{\Hi_2}$' entails. The fact that $\Sets^{\Hi_1}$
and $\Sets^{\Hi_2}$ are independent topoi strongly suggests  that
we will need something in which the contexts are \emph{pairs}
$\langle V_1,V_2\rangle$ where $V_1\in\Ob(\V(\Hi_1)$ and
$V_2\in\Ob(\V(\Hi_2))$. In other words, the base category for our
new presheaf topos will be the product category
$\V(\Hi_1)\times\V(\Hi_2)$, defined as follows:
\begin{definition}
The category $\V(\Hi_1)\times\V(\Hi_2)$ is such that
\begin{itemize}
\item\textbf{Objects}: The objects are pairs of abelian von Neumann subalgebras
$\langle V_1,V_2\rangle$ with $V_1\in \V(\Hi_1)$ and $\V(\Hi_2)$
\item\textbf{Morphisms}: Given two such pair,  $\langle V_1,V_2\rangle$ and $\langle V_1^{'},V^{'}_2\rangle$, there exist an arrow $l:\langle V_1^{'},V_2^{'}\rangle\rightarrow \langle V_1,V_2\rangle$ if and only if  $V_1^{'}\subseteq V_1$ and $V_2^{'}\subseteq V_2$; i.e., if and only if  there exists a morphism $i_1:V_1^{'}\rightarrow V_1$ in $\V(\Hi_1)$ and a morphism $i_2:V_2^{'}\rightarrow V_2$ in $\V(\Hi_2)$.
\end{itemize}
\end{definition}

This product category $\V(\Hi_1)\times\V(\Hi_2)$ is related to the
constituent categories, $\V(\Hi_1)$ and $\V(\Hi_2)$ by the
existence of the functors\begin{align} p_1&:\V(\Hi_1)\times
\V(\Hi_2)\rightarrow \V(\Hi_1)
\label{p1}\\
p_2&:\V(\Hi_1)\times\V(\Hi_2)\rightarrow \V(\Hi_2)\label{p2}
\end{align}
which are defined in the obvious way. For us, the topos
significance  of these functors lies in the following fundamental
definition and theorem.

\begin{definition} \cite{topos7}, \cite{sv}
A \emph{geometric morphism} $\phi:\tau_1\rightarrow \tau_2$
between topoi $\tau_1$ and $ \tau_2$ is defined to be a pair of
functors $\phi_*:\tau_1\rightarrow \tau_2$ and
$\phi^*:\tau_2\rightarrow \tau_1$, called respectively the
\emph{inverse image} and the \emph{direct image} part of the
geometric morphism, such that
\begin{enumerate}
\item $\phi^*\dashv  \phi_*$ i.e., $\phi^*$ is the left adjoint of $\phi_*$
\item $\phi^*$ is left exact, i.e., it preserves all finite limits.
\end{enumerate}
\end{definition}
In the case of presheaf topoi, an important  source of such
geometric morphisms arises from functors between the base
categories,  according to the following theorem.
\begin{theorem} \label{GeomMo} \cite{topos7}, \cite{sv}
A functor $\phi:A\rightarrow B$ between two categories $A$ and
$B$, induces a geometric morphism (also denoted $\phi$)
\begin{equation}
\theta:\Sets^{A^{op}}\rightarrow \Sets^{B^{op}}
\end{equation}
of which the inverse image part
$\theta^*:\Sets^{B^{op}}\rightarrow \Sets^{A^{op}}$ is such that
\begin{equation}
F\mapsto \theta^*(F):=F\circ \theta
\end{equation}

\end{theorem}

Applying these results to the functors in equations
(\ref{p1})--(\ref{p2}) gives the geometric morphisms between the
topoi\footnote{We are here exploiting the trivial fact that, for
any pair of categories ${\cal C}_1,{\cal C}_2$, we have $({\cal
C}_1\times{\cal C}_2)^\op\simeq{{\cal C}_1}^\op\times{{\cal
C}_2}^\op$.} $\Sets^{\V(\Hi_1)^\op}$, $\Sets^{\V(\Hi_2)^\op}$ and
$\Sets^{(\V(\Hi_1)\times\V(\Hi_2))^\op}$
\begin{align}
p_1&:\Sets^{(\V(\Hi_1)\times\V(\Hi_2))^\op}
\rightarrow \Sets^{\V(\Hi_1)^\op}\\
p_2&:\Sets^{(\V(\Hi_1)\times \V(\Hi_2))^\op}\rightarrow
\Sets^{\V(\Hi_2)^\op}
\end{align}
with associated left-exact functors
\begin{align}
p^*_1&:\Sets^{\V(\Hi_1)^\op}\rightarrow
\Sets^{(\V(\Hi_1)\times\V(\Hi_2))^\op}\\
p^*_2&:\Sets^{\V(\Hi_2)^\op}\rightarrow
\Sets^{(\V(\Hi_1)\times\V(\Hi_2))^\op}
\end{align}

This enables us to give a meaningful definition of the `product'
of $\Sig^{\Hi_1}$ and $\Sig^{\Hi_2}$ as
\begin{equation}
\Sig^{\Hi_1}\times\Sig^{\Hi_2}:= p_1^*(\Sig^{\Hi_1})\times
p_2^*(\Sig^{\Hi_2}) \label{Def:Sig1timesSig2}
\end{equation}
where the `$\times$' on the right hand side of equation
(\ref{Def:Sig1timesSig2}) is the standard categorial product in
the topos $\intt$.

We will frequently write the product, $p_1^*(\Sig^{\Hi_1})\times
p_2^*(\Sig^{\Hi_2})$, in the simpler-looking form
`$\Sig^{\Hi_1}\times\Sig^{\Hi_2}$' but it must always be born in
mind that what is really meant is the more complex form on the
right hand side of (\ref{Def:Sig1timesSig2}). The topos $\intt$
will play an important role in what follows. We will call it the
`intermediate topos' for reasons that will appear shortly.

We have argued that (two-time) history propositions, both
homogeneous and inhomogeneous, should be represented in the
Heyting algebra $\Subone\otimes\Subtwo$ and we now want to assert
that the topos that underlies such a possibility is precisely the intermediate
topos $\intt$.

The first thing to notice is that the constituent
single-time propositions can be represented in the pull-backs
$p_1^*(\Sig^{\Hi_1})$ and $p_2^*(\Sig^{\Hi_2})$ to the topos
$\intt$, since we have that, for example, for the functor $p_1$,
\begin{equation}
        p_1^*(\Sig^{\Hi_1})_{\langle V_1,V_2\rangle}:=\Sig^{\Hi_1}_{V_1}
\end{equation}
for all stages $\langle V_1,V_2\rangle$. Further more
\begin{equation}
p_1^*(\Sig^{\Hi_1})\times p_2^*(\Sig^{\Hi_2})_{\langle
V_1,V_2\rangle}:= \Sig^{\Hi_1}_{V_1}\times\Sig^{\Hi_2}_{V_2}
\label{p1p2times}
\end{equation}
so that it is clear that we can represent two-time homogeneous
histories in this intermediate topos.

However, at this point everything looks similar to the
corresponding classical case. In particular we have
\begin{equation}
\Sub(p_1^*(\Sig^{\Hi_1}))\times \Sub(p_2^*(\Sig^{\Hi_2})) \subset
\Sub(p_1^*(\Sig^{\Hi_1})\times p_2^*(\Sig^{\Hi_2}))
\end{equation}
which is a proper subset relation because, as is clear from
equation (\ref{p1p2times}) the general subobject of
$\Sig^{\Hi_1}\times\Sig^{\Hi_2}:=  p_1^*(\Sig^{\Hi_1}) \times
p_2^*(\Sig^{\Hi_2})$ will be a `$\lor$' of product sub-objects in the
Heyting algebra $\Sub(\Sig^{\Hi_1})\times\Sub(\Sig^{\Hi_2})$. In fact,
we have the following theorem:
\begin{theorem}\label{the:conj}
There is an isomorphism of  Heyting algebras
\begin{equation}
   \Subone\otimes\Subtwo\simeq     \Sub(\Sig^{\Hi_1}\times\Sig^{\Hi_2})
\end{equation}

\end{theorem}
In order to show there is an isomorphism between the algebras we
will first construct an isomorphism between the associated frames,
the application of theorem \ref{the:fh} will then lead to the
desired isomorphisms between Heyting algebras. Because of the fact
that the tensor product is given in terms of  relations on
product elements, it suffices to define $h$ on products
$\ps{S_1}\otimes\ps{S_2}$ and show that the function thus defined
preserves these relations

The actual definition of $h$\ is the obvious one:
\begin{eqnarray}\label{ali:h1}
h:\Sub(\Sig^{\V(\Hi_1)})\otimes
\Sub(\Sig^{\V(\Hi_2)})&\rightarrow&
\Sub(\Sig^{\Hi_1}\times\Sig^{\Hi_2})\nonumber\\
\ps{S}_1\otimes \ps{S}_2&\mapsto& \ps{S}_1\times \ps{S}_2
(:=p_1^*\ps{S_1}\times p_2^*\ps{S_2})
\end{eqnarray}
and the main thing is to show that equations (\ref{ali:rel}) are
preserved by $h$.

To this end consider the following
\begin{equation}
h\big(\bigvee_i\ps{S}^i_1\otimes
\ps{S}_2\big):=(\bigvee_i\ps{S}^i_1)\times \ps{S}_2
\end{equation}
For a given context $\langle V_1\ V_2\rangle $ we have
\begin{align}\label{ali:v}
h\big(\bigvee_i\ps{S}^i_1\otimes\ps{S}_2\big)_{\langle V_1\
V_2\rangle}&=((\bigvee_i\ps{S}^i_1)\times
\ps{S}_2)_{\langle V_1\ V_2\rangle}\nonumber\\
&=(\bigcup_i\ps{S}^i_1)_{V_1}\times
(\ps{S}_2)_{V_2}\nonumber\\
&=\bigcup_i(\ps{S}^i_1)_{V_1}\times
(\ps{S}_2)_{V_2})\nonumber\\
&=\bigcup_i(\ps{S}^i_1\times
\ps{S}_2)_{\langle V_1\ V_2\rangle}\nonumber\\
&=\bigcup_i\big(h(\ps{S}_1^i\otimes \ps{S}_2)\big)_{\langle
V_1\ V_2\rangle}\nonumber\\
&=\bigg(\bigvee_i h(\ps{S}_1^i\otimes \ps{S}_2) \bigg)_{\langle
V_1\ V_2\rangle}
\end{align}
where the third equality follows from the general property of
products $(A\cup B)\times C=A\times C\cup B\times C$. It follows
that
\begin{equation}
h\big(\bigvee_i\ps{S}^i_1\otimes\ps{S}_2\big)=
\bigvee_ih\big(\ps{S}^i_1\otimes\ps{S}_2)
\end{equation}
There is a very similar proof of
\begin{equation}
h\big(\bigvee_i\ps{S}_1\otimes\ps{S}^i_2\big)=
\bigvee_ih(\ps{S}_1\otimes\ps{S}^i_2)
\end{equation}

Moreover \begin{align} h\big(\bigwedge_{i\in I}\ps{S}^i_1\otimes
\ps{S}_2^i\big)_{\langle V_1,V_2\rangle}&= h\bigg(\bigwedge_{i\in
I}\ps{S}_1^i\otimes \bigwedge_{j\in I}\ps{S}_2^j\bigg)_{\langle
V_1,V_2\rangle}= \big(\bigwedge_{i\in
I}\ps{S}_1^i\big)_{V_1}\times
\big(\bigwedge_{j\in I}\ps{S}_2^j\big)\nonumber\\
&=\bigcap_{i\in I}\ps{S}_{V_1}^i\times\bigcap_{j\in
I}\ps{S}_{V_2}^j
=\bigcap_{i\in I}\big(\ps{S}^i_{V_1}\times\ps{S}^i_{V_2}\big)\nonumber\\
&=\bigwedge_{i\in I}(\ps{S^i_1}\times\ps{S^i_2})_{\langle
V_1,V_2\rangle}= \bigwedge_{i\in
I}h(\ps{S^i}_1\otimes\ps{S^i}_2)_{\langle V_1,V_2\rangle}
\end{align}
from which it follows that
\begin{equation}
h\big(\bigwedge_{i\in I}\ps{S}^i_1\otimes\ps{S}_2^i\big)=
\bigwedge_{i\in I}h(\ps{S^i}_1\otimes\ps{S^i}_2)
\end{equation}
as required.

The injectivity of $h$ is obvious. The surjectivity follows from
the fact than any element$, \ps{R}$, of
$\Sub(\Sig^{\Hi_1}\times\Sig^{\Hi_2})$ can be written as
$\ps{R}=\vee_{i\in I}(\ps{S^i_1}\times\ps{S^i_2})= \vee_{i\in
I}h(\ps{S^i_1}\otimes\ps{S^i_2}) =h\big(\vee_{i\in
I}\ps{S^i_1}\otimes\ps{S^i_2}\big)$ (because $h$ is a homomorphism
of frames)

Thus the frames $\Sub(\Sig^{\Hi_1})\otimes \Sub(\Sig^{\Hi_2})$ and
$\Sub(\Sig^{\Hi_1}\times\Sig^{\Hi_2})$ are isomorphic. The
isomorphisms of the associated Heyting algebras then follows from
Theorem \ref{the:fh}.

\subsection{Entangled stages}
The discussion above reinforces the idea that homogeneous history
propositions can be represented by subobjects of products of
pullbacks of single-time spectral presheaves.

However, in this setting there can be no notion of entanglement of
contexts since the contexts are just pairs $\langle
V_1,V_2\rangle$; i.e., objects in the product category
$\V(\Hi_1)\times\V(\Hi_2)$. To recover `context entanglement' one
needs to use the context category $\V(\Hi_1\otimes\Hi_2)$, some of
whose objects are simple tensor products $V_1\otimes V_2$ (which,
presumably, relates in some way to the pair $\langle
V_1,V_2\rangle$) but others are `entangled' algebras of the form
$W=V_1\otimes V_2+ V_3\otimes V_4$. Evidently, the discussion
above does not apply to contexts of this more general type.

To explore this further  consider the following functor
\begin{align}
\theta:\V(\Hi_1)\times \V(\Hi_2)&\rightarrow \V(\Hi_1\otimes\Hi_2)\\
\langle V_1,V_2\rangle&\mapsto V_1\otimes V_2\label{V1V2->V1xV2}
\end{align}
where equation (\ref{V1V2->V1xV2}) refers to the action on the
objects in the category $\V(\Hi_1)\times\V(\Hi_2)$; the action on
the arrows is obvious.

According to Theorem 5.2 this gives rise to a geometric morphism,
$\theta$, between topoi, and an associated left-exact functor,
$\theta^*$:
\begin{align}
&\theta:\Sets^{\V(\Hi_1)}\times \Sets^{\V(\Hi_2)}\rightarrow
\Sets^{\V(\Hi_1\otimes\Hi_2)}\\
&\theta^*:\Sets^{\V(\Hi_1\otimes\Hi_2)}\rightarrow
\Sets^{\V(\Hi_1)}\times \Sets^{\V(\Hi_2)}
\end{align}

In particular, we can consider the pull-back
$\theta^*(\Sig^{\Hi_1\otimes\Hi_2})$ which,  on pairs of contexts,
is:
\begin{equation}
(\theta^*\Sig^{\Hi_1\otimes\Hi_2})_{\langle V_1,
V_2\rangle}:=(\Sig^{\Hi_1\otimes\Hi_2})_{\theta\langle V_1,
V_2\rangle}=\Sig^{\Hi_1\otimes\Hi_2}_{V_1\otimes
V_2}\end{equation} Thus the pull-back,
$\theta^*(\Sig^{\Hi_1\otimes\Hi_2})$ of the spectral presheaf of
$\Hi_1\otimes\Hi_2$ to the intermediate topos $\intt$ completely reproduces
$\Sig^{\Hi_1\otimes\Hi_2}$ at contexts of the
tensor-product form $V_1\otimes V_2$.

However, it is clear that, for all contexts $V_1, V_2$ we have
\begin{equation}
\Sig^{\Hi_1\otimes\Hi_2}_{V_1\otimes V_2}\cong
\Sig^{\Hi_1}_{V_1}\times\Sig^{\Hi_2}_{V_2}
\end{equation}
since we can define an isomorphic function
\begin{equation}
\mu:\Sig^{\Hi_1}_{V_1}\times \Sig^{\Hi_2}_{V_2}\rightarrow
\Sig^{\Hi_1\otimes\Hi_2}_{V_1\otimes V_2}
\end{equation}
where, for all $\hat{A}\otimes \hat{B}\in V_1\otimes V_2$, we have
\begin{equation}
\mu(\langle\lambda_1,\lambda_2\rangle)(\hat{A}\otimes
\hat{B}):=\lambda_1(\hat{A})\lambda_2(\hat{B})
\end{equation}

The fact that, for all contexts of the form  $V_1\otimes V_2$, we
have $\Sig^{\Hi_1\otimes\Hi_2}_{V_1\otimes
V_2}\cong\Sig^{\Hi_1}_{V_1}\times\Sig^{\Hi_2}_{V_2}$, means that,
\begin{equation}
        \theta^*(\Sig^{\Hi_1\otimes\Hi_2})\simeq
        \Sig^{\Hi_1}\times\Sig^{\Hi_2}
\end{equation}
in the intermediate topos $\intt$. Thus, in the topos $\intt$, the
product $\Sig^{\Hi_1}\times\Sig^{\Hi_2}$ is essentially the
spectral presheaf $\Sig^{\Hi_1\otimes\Hi_2}$ but restricted to
contexts of the form $V_1\otimes V_2$. Thus $\intt$ is an
`intermediate' stage in the progression from the pair of topoi
$\Sets^{\V(\Hi_1)^\op}$, $\Sets^{\V(\Hi_2)^\op}$ to the topos
$\Sets^{\V(\Hi_1\otimes\Hi_2)^\op}$ associated with the full
tensor-product Hilbert space $\Hi_1\otimes\Hi_2$. This explains
why we called $\intt$ the `intermediate' topos.

The choice of $\intt$ as the appropriate topos to use in the
setting of quantum temporal logic reflects the fact that, although
the full topos for quantum history theory is
$\Sets^{\V(\Hi_1\otimes\Hi_2)^\op}$, never-the-less, to account
for both homogeneous and inhomogeneous history propositions it
suffices to use the intermediate topos. However, if we do use  the
full topos $\Sets^{\V(\Hi_1\otimes\Hi_2)^{\op}}$ a third type of
history proposition arises. These `entangled, inhomogeneous
propositions' cannot be reached/defined by single-time
propositions connected through temporal logic.

The existence of such propositions is a consequence of the fact
that in the topos $\Sets^{\V(\Hi_1\otimes\Hi_2)^\op}$, the
\emph{context} category $\V(\Hi_1\otimes\Hi_2)$ contains
`entangled' abelian Von Neumann subalgebras $W$: i.e., subalgebras
of the form $V_1\otimes V_2+V_3\otimes V_4$ which cannot be
reduced to a pure tensor product $W_1\otimes W_2$. For such
contexts it is not possible to define a clear relation between a
history proposition and individual single-time propositions.

To clarify what is going on let us return for a moment to the HPO
formalism of consistent history theory. There, a time-ordered
sequence of individual time propositions (i.e., a homogeneous
history) is identified with the tensor product of projection
operators
$\hat{P}_1\otimes\hat{P}_2\otimes\cdots\otimes\hat{P}_n$. We get a
form of `entanglement' when we consider inhomogeneous propositions
$\hat{P}_1\otimes\hat{P}_2\vee \hat{P}_3\otimes\hat{P}_4$ that
cannot be written as $\hat Q_1\otimes\hat Q_2$. However, this type
of entanglement, which comes from logic, is not exactly the
same as the usual entanglement of quantum mechanics (although
there are close connections).

To understand this further  consider a simple example in ordinary
quantum theory of an entangled pair of spin-up spin-down
particles. A typical entangled state is
\begin{equation}
|\uparrow\rangle|\downarrow\rangle-|\downarrow\rangle|\uparrow\rangle
\end{equation}
and the projector operator associated with this state is
\begin{equation}
\hat{P}_{\text{entangled}}=
(|\uparrow\rangle|\downarrow\rangle-|\downarrow\rangle|
\uparrow\rangle)(\langle\uparrow|\langle\downarrow|-
\langle\downarrow|\langle\uparrow|)
\end{equation}
However, the projection operator $\hat{P}_{\text{entangled}}$ is
not the same as the projection operator
$\hat{P}_{ud}\vee\hat{P}_{du}$ where
$\hat{P}_{ud}:=(|\uparrow\rangle|\downarrow\rangle)(\langle\downarrow|
\langle\uparrow|)$ and $\hat{P}_{du}:=(|\downarrow\rangle|
\uparrow\rangle)(\langle\uparrow|\langle\downarrow|)$. This
implies that
$\hat{P}_{\text{entangled}}\neq\hat{P}_{ud}\vee\hat{P}_{du}$.

When translated to the history situation, this implies that a
projection operator onto an entangled state in
$\Hi_1\otimes\Hi_2$, cannot be viewed as being an inhomogeneous
history proposition: it is something different. The precise
temporal-logic meaning, if any, of these entangled projectors
remains to be seen.

\chapter{Histories Approach to Quantum Theory}
\label{cha:hist}
\section{Consistent Histories}
Consistent histories theory was born as an attempt to describe
closed systems in quantum mechanics, partly in light of a desire
to construct quantum theories of cosmology. In fact, the
Copenhagen interpretation of quantum mechanics cannot be applied
to closed systems, since it rests on the notion of probabilities
defined in terms of a sequence of repeated measurements by an
external observer. Thus it enforces a cosmologically
inappropriate division between system and observer. The
consistent-history formulation avoids this division, since it
assigns probabilities without making use of the measurements and
the associated state vector reductions.

In the standard Copenhagen interpretation of quantum theory,
probability assignments to sequences of measurements are computed
using the von Neumann reduction postulate which, roughly speaking,
determines a measurement-induced change in the density matrix that
represents the state.

Specifically, let us consider a density matrix $\rho(t_0)$ defined
at time $t_0$, which in the Schrodinger picture evolves to
$\rho(t_1)$ at time $t_1$ through the time evolution operator
$\hat{U}(t_1,t_0)=e^{-i(t_1-t_0)\hat H}$; i.e.,
\begin{equation}
\rho(t_1)=\hat{U}(t_1,t_0)\rho(t_0)\hat{U}(t_1,t_0)^{-1}
\end{equation}
Suppose at time $t_1$ we measure a property represented by the
projection operator $\hat{P}$. If the result of such a measurement
is retained then, according to the Von Neumann reduction
postulate the density matrix gets transformed to
\begin{equation}
\rho_{\text{red}}(t_1):=\frac{\hat{P}(t_1)\rho(t_0)\hat{P}(t_1)}{tr(\hat{P}(t_1)\rho(t_0))}
 \end{equation}
Here, $tr(\hat{P}(t_1)\rho(t_0))$ represents the probability of
finding the property represented by the projection operator
$\hat{P}(t_1)$, namely
\begin{equation}
Prob(\hat{P}=1;
\rho(t_1))=tr(\hat{P}(t_0)\overbrace{\hat{U}(t_1,t_0)\rho(t_0)\hat{U}(t_1,t_0)^{\dagger}}^{\text{evolution
of }
\rho})=tr(\overbrace{\hat{U}(t_1,t_0)^{\dagger}\hat{P}(t_0)\hat{U}(t_1,t_0)}^{\hat{P}(t_1)}\rho(t_0))=tr(\hat{P}(t_1)\rho(t_0))
 \end{equation}
where $\hat
P(t_1):=\hat{U}(t_1,t_0)^{\dagger}\hat{P}\hat{U}(t_1,t_0)$ is the
Heisenberg-picture evolution of $\hat{P}$.

If we then want to perform a subsequent measurement at time
$t_2>t_1$, say, of the property represented by an operator
$\hat{Q}$ then, the \emph{conditional probability} of finding this
property at time $t_2$, given that we found the property
represented by $\hat{P}$ at time $t_1$ (which corresponds to the
eigenvalue $1$ of $\hat P$) is
\begin{equation}
tr(\hat{Q}(t_2)\rho_{\text{red}}(t_1))=\frac{tr(\hat{Q}(t_2)\hat{P}(t_1)\rho(t_0)\hat{P}(t_1))}{tr(\hat{P}(t_1)\rho(t_0))}
\end{equation}
Here, $\hat
Q(t_2):=\hat{U}(t_2,t_1)^{\dagger}\hat{Q}\hat{U}(t_2,t_1)$.

If we now consider the \emph{joint probability}  of obtaining
$\hat{P}=1$ at time $t_1$ and $\hat{Q}=1$ at time $t_2$, given the
initial sate $\rho(t_0)$, we get the following expression:
\begin{equation}
tr(\hat{Q}(t_2)\hat{P}(t_1)\rho(t_0)\hat{P}(t_1)\hat{Q}(t_2))
\end{equation}
Then, generalising to $n$ measurements at $n$ linearly-ordered
time points, the joint probability is
\begin{align}\label{ali:pro}
&Prob(\hat{P_1}=1\text{ at time } t_1\text{ and }\hat{P_2}=
1\text{ at time } t_2\text{ and }\cdots\hat{P_n}=
1\text{ at time } t_n\text{ and };\rho(t_0))=\\
&tr(\hat{P}_n(t_n)\cdots\hat{P}_1(t_1)\rho(t_0)\hat{P}_1(t_1)\hat{P}_n(t_n))
\end{align}
It is clear that, in this Copenhagen interpretation, equation
(\ref{ali:pro}) makes fundamental use of the notion of
measurement-induced, state-vector reduction.

The consistent history formalism was developed in order to make
sense of equation (\ref{ali:pro}) but without invoking the notion
of measurement. This requires introducing the decoherence
functional, $d$, which is a map from the space of all histories to
the complex numbers. Specifically, given two histories (sequences
of projection operators)
$\alpha=(\hat\alpha_{t_1},\hat\alpha_{t_2},\cdots,\hat\alpha_{t_n})$
and
$\beta=(\hat\beta_{t_1},\hat\beta_{t_2},\cdots,\hat\beta_{t_n})$
the decoherence functional is defined as
\begin{equation}
d_{\rho,\hat H}(\alpha,\beta)=tr(\tilde{C}^{\dagger}_{\alpha}\rho
\tilde{C}_{\beta})=tr(\hat{C}^{\dagger}_{\alpha}\rho
\hat{C}_{\beta})
\end{equation}
where $\rho$ is the initial density matrix, $\hat H$ is the
Hamiltonian, and $\tilde{C}_{\alpha}$ represents the `class
operator' which is defined in terms of the Schrodinger-picture
projection operator $\alpha_{t_i}$ as
\begin{equation}
\tilde{C}_{\alpha}:=\hat{U}(t_0,t_1)\alpha_{t_1}\hat{U}(t_1,t_2)\alpha_{t_2}\cdots\hat{U}(t_{n-1},t_n)\alpha_{t_n}\hat{U}(t_{n},t_0)
\end{equation}
Thus $\tilde{C}_{\alpha}$ represents the history proposition
``$\alpha_{t_1}$ is true at time $t_1$, and then $\alpha_{t_2}$ is
true at time $t_2$, $\cdots$, and then $\alpha_{t_n}$ is true at
time $t_n$''. It is worth noting that the class operator can be
written as the product of Heisenberg-picture projection operators
in the form
$\hat{C}_\alpha=\hat{\alpha}_{t_{n}}(t_n)\hat{\alpha}_{t_{n-1}}(t_{n-1})\cdots
\hat{\alpha}_{t_1}(t_1)$. Generally speaking this is not itself a
projection operator.

A more axiomatic definition of a decoherence functional is as
follows:
\begin{definition}
A decoherence functional is a complex-valued function
$d:\mathcal{UP}\times\mathcal{UP}\rightarrow \Cl$ defined on pairs
of histories $\alpha=(\hat{\alpha}_{t_1}, \hat{\alpha}_{t_2}\cdots
\hat{\alpha}_{t_n})$ and
$\beta=(\hat{\beta}_{t^{'}_1},\hat{\beta}_{t^{'}_2}\cdots
\hat{\beta}_{t^{'}_n})$ (the temporal supports\footnote{The
\emph{temporal support} of a history $(\hat{\alpha}_{t_1},
\hat{\alpha}_{t_2}\cdots \hat{\alpha}_{t_n})$ is the set
$\{t_1,t_2,\ldots,t_n\}$. Here it is assumed that these time
points satisfy $t_1<t_2<\cdots<t_n$.} need not be the same) such
that the following properties hold:
\begin{enumerate}
\item \emph{Hermiticity}: $d(\alpha,\beta)=d^*(\beta,\alpha)$
\item \emph{Positivity}: $d(\alpha,\alpha)\geq 0$ for all $\alpha\in\mathcal{UP}$
\item \emph{Normalization}: $\sum_{i}d(\alpha_i,\alpha_i)=1$ for all collections $\alpha_1,\alpha_2,\ldots$ whose elements are pairwise disjoint and whose sum is the unit history.
\item \emph{Null triviality}: $d(0,\alpha) = 0$ for all $\alpha\in\mathcal{UP}$.
\item \emph{Additivity}: Given two disjoint\footnote{The meanings of `disjoint' and the $\lor$-operation are given below.} histories $\alpha$ and $\beta$ then, for all $\gamma\in\mathcal{UP}$, $d(\alpha\vee\beta,\gamma)=d(\alpha,\gamma)+ d(\beta,\gamma)$
\end{enumerate}
\end{definition}

 The physical meaning associated to the quantity $d(\alpha,\alpha)$ is that it is the probability of the history $\alpha$ being realized. However, this interpretation can only be ascribed in a non-contradictory way if the history $\alpha$ belongs to a special set of histories, namely a \emph{consistent set}. In order to rigorously define what a consistent set is we will first give the axiomatic definition of the consistent-histories approach to quantum mechanics put forward by Gell-Mann and Hartle. For an in-depth analysis of the axioms and definition of consistent-history theory the reader is referred to \cite{consistent1}, \cite{consistent3}, \cite{consistent4} and references therein.

\paragraph{The main ideas of the consistent-history formalism}
\begin{enumerate}
\item The main ingredients in the consistent history formalisms are a space $\mathcal{D}$ of decoherence functionals and a space $\mathcal{UP}$ of histories which contains both \emph{homogeneous histories} and
\emph{inhomogeneous histories}

\item A \emph{homogeneous history} is any sequentially-ordered sequence of projection operators $\hat{\alpha}_1,\hat{\alpha}_2,\cdots\hat{\alpha}_n$.

\item An important notion is that of `coarse graining'. This notion can be defined for histories with the same time support and  for histories in which the time support of one  is a proper subset of the time support, of the other. Specifically, a homogeneous history $\alpha=(\hat{\alpha}_{t_1},\hat{\alpha}_{t_2}\cdots,\hat{\alpha}_{t_n})$ is said to be \emph{finer} than a history
$\beta=(\hat{\beta}_{t^{'}_1},
\hat{\beta}_{t^{'}_2}\cdots,\hat{\beta}_{t^{'}_m})$, denoted
$\alpha\leq\beta$, if (i) the temporal support of $\beta$ is equal
to, or a proper subset of, the temporal support of $\alpha$; and
(ii) such that for every $t_i$ in the temporal support of $\beta$,
we have $\hat{\alpha}_{t_i}\leq\hat{ \beta}_{t_i}$. Here $\leq$
denotes the usual partial ordering of projection operators.

\item The set of all homogeneous histories can be equipped with a partial ordering, $\leq$, in which $\alpha\leq\beta$ means that $\beta$ is coarser than $\alpha$; or, equivalently, $\alpha$ is finer than $\beta$.

\item Two homogeneous histories, $\alpha=(\hat{\alpha}_{t_1},\hat{\alpha}_{t_2}\cdots,\hat{\alpha}_{t_n})$  and $\beta=(\hat{\beta}_{t^{'}_1},\hat{\beta}_{t^{'}_2}\cdots,\hat{\beta}_{t^{'}_m})$, are said to be \emph{disjoint}, or \emph{orthogonal}, (denoted
$\alpha\perp\beta$) if (i) their temporal supports have at least
one point in common; and (ii)  for each such point $t_i$,
$\hat{\beta}_{t_i}$ is disjoint from $\hat{\alpha}_{t_i}$, i.e.,
these operators project onto orthogonal subspaces of $\Hi$ with
$\hat{\beta}_{t_i}\hat{\alpha}_{t_i}=0=\hat{\alpha}_{t_i}\hat{\beta}_{t_i}$.
It follows that if two histories are orthogonal to each other, the
realization of one history excludes the realization of the other.

\item There exists a \emph{unit} history, $1$, (a history which is always realized) and a \emph{null} history, $0$, (a history which is never realized). Given any history $\alpha$ then $0\leq\alpha\leq 1$.

\item A history $\alpha$ is said to be \emph{fine-grained} if the only history which if finer than $\alpha$ is the null history or $\alpha$ itself. Such histories are represented by time-ordered sequence of projection operators whose ranges are one-dimensional subspaces of the Hilbert space.

\item A set of histories $\{\alpha^1, \alpha^2,\ldots, \alpha^n\}$ is said to be \emph{exclusive} if $\alpha^i\perp\alpha^j $
for all $i,j=1,2,\cdots N$.

\item A set of histories,  $\{\alpha^1, \alpha^2,\ldots, \alpha^n\}$, is said to be \emph{exhaustive} (or \emph{complete}) if it is \emph{exclusive} and $\alpha^1\vee\alpha^2\vee\cdots\vee\alpha^n=1$ (see below for a discussion of $\vee$).

\item
\begin{definition}
A set $\mathcal{C}$ of histories $\{\alpha^1,\alpha^2,\ldots,
\alpha^n\}$ is said to be \emph{consistent} with respect to a
given decoherence functional, $d$, if all of the following
conditions are satisfied:
\begin{enumerate}
\item $\mathcal{C}$ is \emph{exclusive};
\item $\mathcal{C}$ is \emph{exhaustive} (\emph{complete});
\item $d(\alpha,\beta)=0$ for all $\alpha,\beta\in \mathcal{C}$ such that $\alpha\neq\beta$
\end{enumerate}
\end{definition}
Only within a consistent set does the axiomatic definition of
consistent histories have any physical meaning. In fact, it is
only within a given consistent set that the probability
assignments as defined in equation (\ref{ali:pro}), are consistent.
Each decoherence functional defines a consistent set(s) such that
the assignments in equation (\ref{ali:pro}) are possible.

\item The definition of the join $\vee$ is straightforward when the two histories have the same time support and differ in their values only at one point $t_i$. In this case $\alpha\vee\beta:=(\alpha_{t_1},\alpha_{t_2},\cdots, \alpha_{t_i}\vee\beta_{t_i}, \cdots \alpha_{t_n})=
(\beta_{t_1},\beta_{t_2},\cdots, \beta_{t_i}\vee\alpha_{t_i},
\cdots \beta_{t_n})$ is a homogeneous history and satisfies the
relation
$\hat{C}_{\alpha\vee\beta}=\hat{C}_{\alpha}\vee\hat{C}_{\beta}$.

The problem arises when the time supports are different, in
particular when the two histories $\alpha$ and $\beta$ are
disjoint. The join of such histories would take us outside the
class of homogeneous histories. Similarly the negation of a
homogeneous history would not itself be a homogeneous history.

\item An \emph{inhomogeneous history} arises when  two disjoint homogeneous histories are joined using the logical connective ``or''($\vee$) or when taking the negation ($\neg$) of a history proposition. Specifically, given two disjoint homogeneous histories $\alpha$ and $\beta$ we can meaningfully talk about the inhomogeneous histories $\alpha\vee\beta$ and $\neg \alpha$. Such histories are generally not a just a sequence of projection operators, but when computing the decoherence functional they are represented by the operator $\hat{C}_{\alpha\vee\beta}:=\hat{C}_{\alpha}\vee\hat{C}_{\beta}$ and $\hat{C}_{\neg\alpha}:=\hat{1}-\hat{C}_{\alpha}$
\end{enumerate}

Gell Mann and Hartle tried to solve the problem of representing
inhomogeneous histories using  path integrals on the configuration
space, $Q$, of the system. The representation of the decoherence
functional using a path integral from initial time $t_0$ to final
time $t_1$ is
\begin{equation}
d(\alpha,\beta):=\int_{q\in\alpha,q^{'}\in\beta}DqDq^{'}e^{-i(S[q]-S[q^{'})\hbar}\delta(q(t_1)-q^{'}(t_1)\rho(q(t_0)-q^{'}(t_0))
\end{equation}
In this formalism the histories $\alpha$ and $\beta$ are seen as
subsets of the paths of Q. Then a pair of histories is said to be
disjoint if they are disjoint subsets of the path space Q. Seen as
path integrals, the additivity property of the decoherence
functional is easily satisfied, namely
\begin{equation}\label{equ:ad}
d(\alpha\vee\beta, \gamma)=d(\alpha,\gamma)+ d(\beta,\gamma)
\end{equation}
where $\gamma$ is any subset of the path space Q.

Similarly, the negation of a history proposition $\neg\alpha$ is
represented by the complement of the subset $\alpha$ of Q,
therefore
\begin{equation}\label{equ:neg}
d(\neg\alpha,\gamma)=d(1,\gamma)-d(\alpha,\gamma)
\end{equation}
where 1 is the unit history.

The above properties in (\ref{equ:ad}) and \ref{equ:neg} are well
defined in the context of path integrals. But what happens when
defining the decoherence functional on a string of projection
operators? Gell Mann and Hartle solved this problem by postulating
the following definitions for the class operators when computing
decoherence functionals:
\begin{align}\label{equ:clas1}
\tilde{C}_{\alpha\vee\beta}&:=
\tilde{C}_{\alpha}+\tilde{C}_{\beta}\nonumber\\
\tilde{C}_{\neg\alpha}&:=1-\tilde{C}_{\alpha}
\end{align}
if $\alpha$ and $\beta$ are disjoint histories. The right hand
side of these equations are indeed operators that represent
$\alpha\vee\beta$ and $\neg\alpha$ when computing the decoherence
functional but as objects in the consistent-history formalism, it
is not really clear what $\alpha\vee\beta$ and $\neg\alpha$ are.

In fact, as defined above, a homogeneous history is a time-ordered
sequence of projection operators, but there is no analogous
definition of $\alpha\vee\beta$ or $\neg\alpha$. One might try to
define the inhomogeneous histories $\neg\alpha$ and
$\alpha\vee\beta$ component-wise so that, for a simple two-time
history $\alpha=(\hat\alpha_{t_1},\hat\alpha_{t_2})$, we would
have
\begin{equation}\label{equ:wrong}
\neg\alpha=\neg(\hat{\alpha}_{t_1},\hat{\alpha}_{t_2}):=
(\neg\hat{\alpha}_{t_1},\neg\hat{\alpha}_{t_2}).
\end{equation}
However, this definition of the negation operation is wrong. For
$\alpha$ is the temporal proposition ``$\alpha_1$ is true at time
$t_1$, and then $\alpha_2$ is true at time $t_2$'', which we shall
write as $\hat{\alpha}_{t_1}\sqcap\hat{\alpha}_{t_2}$. It is then
intuitively clear that the negation
 of this proposition should be
\begin{equation}\label{equ:right}
\neg(\hat{\alpha}_{t_1}\sqcap\hat{\alpha}_{t_2})=
(\neg\hat{\alpha}_{t_1}\sqcap\hat{\alpha}_{t_2})
\vee(\hat{\alpha}_{t_1}\sqcap\neg\hat{\alpha}_{t_2})
\vee(\neg\hat{\alpha}_{t_1})\sqcap\neg(\hat{\alpha}_{t_2})
\end{equation} which is  not in any obvious sense the same as (\ref{equ:wrong}).

A similar problem arises with the  ``or'' ($\vee$) operation:
given two homogenous histories $(\alpha_1,\alpha_2)$ and
$(\beta_1,\beta_2)$, the ''or'' operation defined component-wise
is
\begin{equation}
(\alpha_1,\alpha_2)\vee(\beta_1,\beta_2):=(\alpha_1\vee\beta_1,
\alpha_2\vee\beta_2)
\end{equation}
This history would be true (realized) if both
$(\alpha_1\vee\beta_1)$ and $(\alpha_2\vee\beta_2)$ are true,
which implies that either an element in each of the pairs
$(\alpha_1,\alpha_2)$ and $(\beta_1,\beta_2)$ is true, or both
elements in either of the pairs $(\alpha_1,\alpha_2)$ and
$(\beta_1,\beta_2)$ are true. But this contradicts with the actual
meaning of the proposition
$(\alpha_1,\alpha_2)\vee(\beta_1,\beta_2)$, which states that
either history $(\alpha_1,\alpha_2)$ is realized or history
$(\beta_1,\beta_2)$ is realized. In fact the `or' in the
proposition $(\alpha_1,\alpha_2)\vee(\beta_1,\beta_2)$ should
really be as follows:
\begin{equation}\label{equ:tor}
({\alpha}_1\sqcap{\alpha}_2)\vee({\beta}_1\sqcap{\beta}_2)=(\neg(
{\alpha}_1\sqcap{\alpha}_2)\wedge({\beta}_1\sqcap{\beta}_2))\vee
(({\alpha}_1\sqcap{\alpha}_2)\wedge\neg({\beta}_1\sqcap{\beta}_2))
\end{equation}
Thus for the proposition
$({\alpha}_1\sqcap{\alpha}_2)\vee({\beta}_1\sqcap{\beta}_2)$ to be
true, both elements, in either of the pairs
$(\alpha_1\sqcap\alpha_2)$ and $(\beta_1\sqcap\beta_2)$ have to be
true, but not all four elements at the same time. If instead we
had the history proposition from equation (16),
$({\alpha}_1\vee{\beta_{1}})\sqcap({\alpha}_2\vee{\beta}_2)$,
 this would be equivalent to
\begin{equation}\label{equ:tor2}
({\alpha}_1\vee{\beta_{1}})\sqcap({\alpha}_2\vee{\beta}_2):=
({\alpha}_1\sqcap{\alpha}_2)\vee({\alpha}_1\sqcap{\beta}_2)\vee
({\beta}_1\sqcap{\beta}_2)\vee({\beta}_1\sqcap
{\alpha}_2)\geq({\alpha}_1\sqcap{\alpha}_2)\vee({\beta}_1
\sqcap{\beta}_2)
\end{equation}
This shows that it is not possible to define inhomogeneous
histories component-wise. Moreover, the appeal to path integrals
when defining $\tilde{C}_{\alpha\vee\beta}$ is
realization-dependent and does not uncover what
$\tilde{C}_{\alpha\vee\beta}$ actually is.

However, the right hand side of  equations (\ref{equ:clas1}) have
a striking similarity to the single-time propositions in quantum
logic. In fact, given two single-time propositions P and Q, which
are disjoint, the proposition $P\vee Q$  is simply represented by
the projection operator $\hat{P}+\hat{Q}$; similarly, the negation
$\neq P$ is represented by the operator $\hat{1}-\hat{P}$.

This similarity of the single-time propositions with the right
hand side of the equations (\ref{equ:clas1}) suggests that, somehow,
it should be possible to identify history propositions with
projection operators. 

Obviously these projection operators cannot
be the class operators since,  generally, these are not projection
operators. The claim that a logic for consistent histories can be
defined, such that each history proposition is represented by a
projection operator on some Hilbert space, is also motivated by the
fact that the statement that a certain history is ''realized" is
itself a proposition. Therefore, the set of all such histories
could possess a lattice structure similar to the lattice of
single-time propositions in standard quantum logic.

These considerations led Isham to construct the so-called HPO
formalism. In this new formalism of consistent histories it is
possible to identify the entire set $\mathcal{UP}$ with the
projection lattice of some `new' Hilbert space.  In the following
Section we will describe this formalism in more detail.

\section{The HPO Formulation of Consistent Histories}
As shown in the previous Section, the identification of a
homogeneous history $\alpha$ as a projection operator on the
direct sum $\oplus_{t\in\{t_1,t_2\cdots t_n\}}\Hi_t$ of $n$ copies
of the Hilbert space $\Hi$ does not lead to a satisfactory
definition of a quantum logic for  histories.

A solution to this problem was put forward by Isham in
\cite{temporall}. In this paper he introduces an alternative
formulation of consistent histories, namely the HPO (History
Projection Operator) formulation. The key idea is to identify
homogeneous histories with tensor products of projection
operators: i.e.,
$\alpha=\hat{\alpha}_{t_1}\otimes\hat{\alpha}_{t_2}\otimes\cdots
\otimes\hat{\alpha}_{t_n}$. This definition was motivated by the
fact that, unlike a normal product, a \emph{tensor} product of
projection operators is itself a projection operators since
\begin{eqnarray}
(\hat{\alpha}_{t_1}\otimes\hat{\alpha}_{t_2})^2=
(\hat{\alpha}_{t_1}\otimes\hat{\alpha}_{t_2})
(\hat{\alpha}_{t_1}\otimes\hat{\alpha}_{t_2})&:=&
\hat{\alpha}_{t_1}\hat{\alpha}_{t_1}\otimes\hat{\alpha}_{t_2}
                \hat{\alpha}_{t_2}\nonumber\\
&=&\hat{\alpha}^2_{t_1}\otimes\hat{\alpha}^2_{t_2}\\
&=&\hat{\alpha}_{t_1}\otimes\hat{\alpha}_{t_2}
\end{eqnarray}
and
\begin{align}
(\hat{\alpha}_{t_1}\otimes\hat{\alpha}_{t_2})^{\dagger}&:=
\hat{\alpha}_{t_1}^{\dagger}\otimes\hat{\alpha}_{t_2}^{\dagger}\\
&=\hat{\alpha}_{t_1}\otimes\hat{\alpha}_{t_2}
\end{align}

For this alternative definition of a homogeneous history, the
negation operation coincides with equation (\ref{equ:right}):
\begin{align}
\neg(\hat{\alpha}_{t_1}\otimes\hat{\alpha}_{t_2})
=\hat{1}\otimes\hat{1}-\hat{\alpha}_{t_1}\otimes\hat{\alpha}_{t_2}&=
(\hat{1}-\hat{\alpha}_{t_1}) \otimes\hat{\alpha}_{t_2}+
\hat{\alpha}_{t_1} \otimes
(\hat{1}-\hat{\alpha}_{t_2})+(1-\hat{\alpha}_{t_1} )\otimes (1 -
\hat{\alpha}_{t_2})\\ \nonumber &= \neg
\hat{\alpha}_{t_1}\otimes\hat{\alpha}_{t_2} + \hat{\alpha}_{t_1}
\otimes \neg\hat{\alpha}_{t_2} + \neg \hat{\alpha}_{t_1} \otimes
\neg\hat{\alpha}_{t_2}
\end{align}
Moreover, given two disjoint homogeneous histories
$\alpha=(\hat{\alpha}_{t_1},\hat{\alpha}_{t_2})$ and
$\beta=(\hat{\beta}_{t_1},\hat{\beta}_{t_2})$ then, since
$\hat{\alpha}_{t_1}\hat{\beta}_{t_1}=0$ and/or
$\hat{\alpha}_{t_2}\hat{\beta}_{t_2}=0$ it follows that the
projection operators that represent the two propositions are
themselves disjoint ,i.e.,
$(\hat{\alpha}_{t_1}\otimes\hat{\alpha}_{t_2})(\hat{\beta}_{t_1}\otimes\hat{\beta}_{t_2})=0$.
It is now possible to define $\alpha\vee\beta$ as
\begin{equation}
(\hat{\alpha}_{t_1}\otimes\hat{\alpha}_{t_2})\vee
(\hat{\beta}_{t_1}\otimes\hat{\beta}_{t_2}):=(\hat{\alpha}_{t_1}\otimes\hat{\alpha}_{t_2})+(\hat{\beta}_{t_1}\otimes\hat{\beta}_{t_2})
\end{equation}

In the HPO formalism, homogeneous histories are represented by
`homogeneous' projection operators in the lattice
$P(\otimes_{t\in\{t_1,t_2\cdots t_n\}}\Hi_{t})$, while
inhomogeneous histories are represented by inhomogeneous
operators. Thus, for example,
$\hat{P}_1\otimes\hat{P}_2\vee\hat{R}_1\otimes\hat{R}_2=\hat{P}_1\otimes\hat{P}_2+\hat{R}_2\otimes\hat{R}_2$
would be the join of the two elements $\hat{P}_1\otimes\hat{P}_2$
and  $\hat{R}_2\otimes\hat{R}_2$ as defined in the lattice
$P(\otimes_{t\in\{t_1,t_2\}}\Hi_{t})$.

Mathematically, the introduction of the tensor product is quite
natural. In fact , as shown in the previous section, in the
general history formalism a homogenous history is an element of
$\oplus_{t\in\{t_1,t_2\cdots t_n\}}
P(\Hi_t)\subset\oplus_{t\in\{t_1,t_2\cdots t_n\}}B(\Hi_t)$ which
is a vector space. The vector space structure of
$\oplus_{t\in\{t_1,t_2\cdots t_n\}}B(\Hi_t)$ is utilised when
defining the decoherence functional, since the map
$(\hat{\alpha}_{t_1},\hat{\alpha}_{t_2},\cdots
\hat{\alpha}_{t_n})\rightarrow
tr(\hat{\alpha}_{t_1}(t_1)\hat{\alpha}_{t_2}(t_2)\cdots
\hat{\alpha}_{t_n}(t_n))$ is multi-linear.

However, tensor products are defined through the universal
factorization property, namely: \\
given a finite collection of vector
spaces $V_1$, $V_2$, $\cdots$, $V_n$, any multi-linear map
$\mu:V_1\times V_2\times\cdots\times V_n\rightarrow W$ uniquely
factorizes through a tensor product, i.e. the diagram
\[\xymatrix{
V_1\otimes V_2\cdots\otimes V_n\ar[rr]^{\mu^{'}}&&W\\
&&\\
V_1\times V_2\cdots \times V_n\ar[uu]^{\phi}\ar[rruu]^{\mu}&&\\
}\] commutes. Thus the map
$\phi:(\hat{\alpha}_{t_1},\hat{\alpha}_{t_2},\cdots
\hat{\alpha}_{t_n})\mapsto
\hat{\alpha}_{t_1}\otimes\hat{\alpha}_{t_2}\otimes\cdots
\hat{\alpha}_{t_n}$ arises naturally.

At the level of algebras, the map $\phi$ is defined in the obvious
way as
\begin{equation}
\phi:\oplus_{t\in\{t_1,t_2\cdots t_n\}}B(\Hi_t)\rightarrow
\otimes_{t\in\{t_1,t_2\cdots t_n\}}B(\Hi_t)
\end{equation}
This map is many-to-one, since $(\lambda A)\otimes
(\lambda^{-1}B)=A\otimes B$. However, if we restrict only to
$\oplus_{t\in\{t_1,t_2\cdots
t_n\}}P(\Hi_t)\subseteq\oplus_{t\in\{t_1,t_2\cdots t_n\}}B(\Hi_t)$,
then the map becomes one-to-one, since for all projection operators\\
$\hat{P}\in\oplus_{t\in\{t_1,t_2\cdots t_n\}}P(\Hi_t)$ ,
$\lambda\hat{P}$ ($\lambda\neq 0$, $\hat{P}\neg 0$) is a
projection operator if and only if $\lambda=1$.

In this scheme, the decoherence functional is computed using the
map
\begin{align}
D:&\otimes_{t\in\{t_1.t_2\cdots t_n\}}
B(\Hi)\rightarrow B(\Hi)\\
&(\hat{A}_1\otimes\hat{A}_2\cdots\otimes\hat{A}_n)\mapsto
(\hat{A}_n(t_n)\hat{A}_{n-1}(t_{n-1})\cdots\hat{A}_1(t_1))
\end{align}
Since this map is linear, it can be extended to include
inhomogeneous histories. Furthermore,  the class operators
$\hat{C}$ can be defined as a map from the projectors on the
Hilbert space $\otimes_{t\in\{t_1,t_2\cdots t_n\}}\Hi$, seen as a
subset of all linear operators on $\otimes_{t\in\{t_1,t_2\cdots
t_n\}}\Hi$ to the operators on $\Hi$
\begin{equation}
\hat{C}_{\alpha}:=D(\phi(\alpha))
\end{equation}
and again extended to inhomogeneous histories by linearity .

This map satisfies the relations
$\tilde{C}_{\alpha\vee\beta}=\tilde{C}_{\alpha}\vee\tilde{C}_{\beta}$
and $\tilde{C}_{\neg\alpha}=1-\tilde{C}_{\alpha}$, and hence their
justification by path integrals is no longer necessary.

The HPO formalism can be extended to non-finite temporal supports
by using an infinite (continuous if necessary) tensor product of
copies of $B(\Hi)$. The interested reader is referred to \cite{consistent3}.

\chapter{Topos Formulation Of The HPO Formalism}\label{cha:toposhistory}
\section{Direct product of truth values}We are now interested in defining truth values for history
propositions. In single-time topos quantum theory, truth values
are assigned through the evaluation map, which is a
state-dependent map from the algebra of history propositions to
the Heyting algebra of truth values. In the history case, for this
map to be well-defined it has to map the temporal  structure of
the Heyting algebras of subobjects to some temporal structure of
the algebras of truth values. In the following Section we will
analyse how this mapping takes place.

Let us consider a homogeneous history proposition $\hat{\alpha}=$
``the quantity $A_1$ has a value in $\Delta_1$ at time $t_1$, and
then the quantity $A_2$ has a value in $\Delta_2$ at time $t_1=2$,
and then $\ldots$ and then the quantity $A_n$ has a value in
$\Delta_n$ at time $t_n$'. Symbolically, we can write $\alpha$ as
\begin{equation}
\alpha=(A_1\in\Delta_1)_{t_1}\sqcap(A_2\in\Delta_2)_{t_2}
\sqcap\ldots\sqcap (A_n\in\Delta_n)_{t_n}
\end{equation}
where the symbol `$\sqcap$' is the temporal connective `and then'.

In the HPO formalism, $\alpha$ is represented by a tensor product
of the spectral projection operators, $\hat E[A_k\in\Delta_k]$
associated with each single-time proposition ``$A_k\in\Delta_k$'',
$k=1,2,\ldots, n$:
\begin{equation}
\hat{\alpha}=\hat{E}[A_1\in\Delta_1]_{t_1}\otimes
\hat{E}[A_2\in\Delta_2]_{t_2}\otimes\cdots\otimes
\hat{E}[A_n\in\Delta_n]_{t_n}
\end{equation}
We will return later to the role of this HPO
representation of histories in topos theory.

In order to ascribe a topos truth value to the homogeneous history
$\alpha$, we will first consider the truth values of the
individual, single-time propositions
``$(A_1\in\Delta_{1})_{t_1}$'', ``$(A_2\in\Delta_{2})_{t_2}$'',
\ldots, ``$(A_n\in\Delta_{n})_{t_n}$''. These truth values are
elements of $\Gamma\Om^{\Hi_{t_k}}$, $k=1,2,\ldots, n$:,  i.e.
global sections of the subobject classifier in the appropriate
topos, $\Sets^{\V(\Hi_{t_k})^\op}$. We will analyse how these
truth values can be combined to obtain a truth value for the
entire history proposition $\alpha$. For the sake of simplicity we
will restrict ourselves to two-time propositions, but the
extension to $n$-time slots is trivial.

Since there is no state-vector reduction, one can hope to define
the truth value of the entire history $\alpha:=
(A_1\in\Delta_1)_{t_1}\sqcap(A_2\in\Delta_2)_{t_2}$  in terms of
the truth values of the individual propositions at times $t_1$ and
$t_2$. In particular, since we are conjecturing that the truth
values at the two times are independent of each other, we expect
an equation something like that\footnote{Since there is no
state-vector reduction the existence of an operation $\sqcap$ between truth values ,
that satisfies equation (\ref{TVtwo-time}) is plausible.  In fact,
unlike the normal logical connective `$\land$', the meaning of the
temporal connective `$\sqcap$' implies that the propositions it
connects do not `interfere' with each other, since they are
asserted at different times: it is thus a sensible first guess to
assume that their truth values are independent.

The distinction between the temporal connective `$\sqcap$' and the
logical connective `$\wedge$' is discussed in detail in various papers
by Stachow and Mittelstaedt \cite{s.logicalfoundations}
,\cite{s.modeltheoretic}, \cite{m.timedep},
\cite{m.quantumlanddecoherence}. In these papers they analyse
quantum logic using the ideas  of game theory. In particular they
define logical connectives in terms of sequences of subsequent
moves of possible attacks and defenses. They also introduce the
concept of `commensurability property' which essentially defines
the possibility of quantities being measured at the same time or
not.\\
 The definition of \emph{logical connectives} involves  both
possible attacks and defenses, as well as the satisfaction of the
commensurability property, since logical connective relate
propositions which refer to the same time. On the other hand, the
definition of \emph{sequential connectives} does not need the
introduction of the commensurability properties, since sequential
connectives refer to propositions defined at different times, and
thus can always be evaluated together. The commensurability
property introduced by Stachow and Mittelstaedt can be seen as the
game theory analogue of the commutation relation between operators
in quantum theory. We note that, the same type of analysis can be
applied as a justification of Isham's choice of the tensor product,
as temporal connective in the HPO theory.}
\begin{equation}
v\big((A_1\in\Delta_1)_{t_{1}}\sqcap (A_2\in\Delta_2)_{t_2};
\ket\psi_{t_{1}}\big) = v\big(A_1\in\Delta_1;
\ket\psi_{t_1}\big)\sqcap
v\big(A_2\in\Delta_2;\ket\psi_{t_{2}}\big)\label{TVtwo-time}
\end{equation}
where $\ket\psi{_{t_2}}$ is the unitary evolution of
$\ket\psi{_{t_2}}$. The `$\sqcap$' ,on the right hand side, remains
to be defined as some sort of temporal connective on the Heyting
algebras $\Sets^{\V(\Hi_{t_1})^\op}$ and
$\Sets^{\V(\Hi_{t_1})^\op}$.

However, at this point we  hit the problem that
$v\big(A_1\in\Delta_1; \ket\psi_{t_1}\big)$ and
$v\big(A_2\in\Delta_2;\ket\psi_{t_{2}}\big)$ are global elements
of the subobject classifiers $\Om^{\Hi_{t_1}}$ and
$\Om^{\Hi_{t_2}}$ in  the topoi $\Sets^{\V(\Hi_{t_1})^\op}$ and
$\Sets^{\V(\Hi_{t_2})^\op}$, respectively. Since these topoi are
different from each other, it is not obvious how the the
`$\sqcap$' operation on the right hand side of equation
(\ref{TVtwo-time}) is to be defined.

On the other hand, since $\Ga\Om^{\Hi_{t_1}}$ and
$\Ga\Om^{\Hi_{t_2}}$ are Heyting algebras, we can take their
tensor product $\Ga\Om^{\Hi_{t_1}}\otimes \Ga\Om^{\Hi_{t_2}}$. By
analogy with what we did earlier with the Heyting algebras of
subobjects of the spectral presheaves, it is natural to interpret
the `$\sqcap$' on the right hand side of equation
(\ref{TVtwo-time}) as this tensor product, so that we end up with
the plausible looking equation
\begin{equation}
v\big((A_1\in\Delta_1)_{t_{1}}\sqcap (A_2\in\Delta_2)_{t_2};
\ket\psi_{t_{1}}\big) = v\big(A_1\in\Delta_1;
\ket\psi_{t_1}\big)\otimes
v\big(A_2\in\Delta_2;\ket\psi_{t_{2}}\big)\label{TVtwo-timeT}
\end{equation}

The problem now is to find a topos for which the Heyting algebra
$\Ga\Om^{\Hi_{t_1}}\otimes \Ga\Om^{\Hi_{t_2}}$ is well defined. This is
reminiscent of the problem we encountered earlier when trying to
represent inhomogeneous histories in a topos, and the answer is
the same: pull everything back to the intermediate topos
$\inttonetwo$. Specifically, let us define
\begin{equation}
        \Om^{\Hi_{t_1}}\times\Om^{\Hi_{t_2}}:=
        p_1^*(\Om^{\Hi_{t_1}})\times p_2^*(\Om^{\Hi_{t_2}})
\end{equation}
which is an object in $\inttonetwo$. In fact, it is easy to check
that it is the \emph{subobject classifier} in the intermediate
topos, and it is defined at stage $\langle
V_1,V_2\rangle\in\Ob(\V(\Hi_{t_1})\times\V(\Hi_{t_2}))$ by
\begin{equation}
(\Om^{\Hi_{t_1}}\times\Om^{\Hi_{t_2}})_{\la V_1,V_2\ra}:=
        \Om^{\Hi_{t_1}}_{V_1} \times \Om^{\Hi_{t_2}}_{V_2}
\end{equation}
and we have the important  result that there is an
isomorphism\begin{equation}
j:\Ga\Om^{\Hi_{t_1}}\otimes\Ga\Om^{\Hi_{t_2}}
\rightarrow\Ga(\Om^{\Hi_{t_1}}\times\Om^{\Hi_{t_2}}):=
\Ga\big(p_1^*(\Om^{\Hi_{t_1}})\times p_2^*(\Om^{\Hi_{t_2}})\big)
\simeq\Ga\big(p_1^*(\Om^{\Hi_{t_1}})\big)\times
\Ga\big(p_2^*(\Om^{\Hi_{t_2}})\big)
\end{equation}
given by
\begin{equation}
        j(\omega_1\otimes\omega_2)(\la V_1,V_2\ra):=
        \la\omega_1(V_1),\omega_2(V_2)\ra\label{Def:j}
\end{equation}The proof of this result is similar to that of Theorem 5.3 and
will not be written out here.

For us, the significant implication of this result is that the
truth value $v\big((A_1\in\Delta_1)_{t_{1}}\sqcap
(A_2\in\Delta_2)_{t_2}; \ket\psi_{t_{1}}\big)$ of the history
proposition $(A_1\in\Delta_1)_{t_{1}}\sqcap
(A_2\in\Delta_2)_{t_2}$ can be regarded as an element of the
Heyting algebra $\Ga(\Om^{\Hi_{t_1}}\times\Om^{\Hi_{t_2}})$, whose
`home' is the intermediate topos $\inttonetwo$. Thus a more
accurate way of writing equation (\ref{TVtwo-timeT}) is
\begin{equation}
v\big((A_1\in\Delta_1)_{t_{1}}\sqcap (A_2\in\Delta_2)_{t_2};
\ket\psi_{t_{1}}\big) = j\Big(v\big(A_1\in\Delta_1;
\ket\psi_{t_1}\big)\otimes
v\big(A_2\in\Delta_2;\ket\psi_{t_{2}}\big)\Big)\label{TVtwo-timeT2}
\end{equation}

\subsection{The representation of HPO histories}
In this Section we will pull together what has been said above in order to obtain
a topos analogue of the HPO formalism of  quantum history
theory.

First we recall that in the HPO formalism, a history proposition
$\alpha=\alpha_1\sqcap\alpha_2$ is identified with the tensor
product of the projection operators $\ha_1$ and $\ha_2$
representing the single-time propositions $\alpha_1$ and
$\alpha_2$, respectively, i.e. $\ha=\ha_1\otimes\ha_2$. One main
motivation  for introducing the tensor product has been a desire to
make sense of the negation operation of homogeneous history
propositions, as given intuitively by equation (\ref{equ:right}).

In fact, in the original approaches to consistent-histories theory
the temporal connective `and then' was simply associated to the
operator product, thus the proposition
$\alpha=\alpha_1\sqcap{\alpha}_2$ was represented by
$\hat\alpha=\ha_1\ha_2$. But this identification loses any logical
meaning, since, given projection operators $\hat{P}$ and $\hat{Q}$
the product $\hat{P}\hat{Q}$ is generally not itself a projection
operator.

However,, if one defines the sequential connective $\sqcap$ in
terms of the tensor product, such that $\alpha={\alpha}_1\sqcap
{\alpha}_2$ is represented by $\ha=\ha_1\otimes \ha_2$, then $\ha$
\emph{is} a projection operator. Furthermore, one obtains the
right definition for the negation operation, specifically
\begin{equation}
\neg(\ha_1\otimes\ha_2)=(\neg \ha_1 \otimes
\ha_2)+(\ha_1\otimes\neg \ha_2)+(\neg \ha_1 \otimes\neg \ha_2)
\end{equation} where we identify $+$ with $\vee$
\footnote{This is correct since the projectors which appear on the
right hand side of the equation are pair-wise orthogonal, thus the
`or', $\vee$, can be replaced by the summation operation $+$ of
projector operators.}.

We will now proceed by considering history propositions, as
defined by the HPO formalism, as individual entities and, then, apply
the machinery defined in  \cite{andreas1}, \cite{andreas2},
\cite{andreas3}, \cite{andreas4}, \cite{andreas5} and
\cite{andreas6} to derive a topos version of the history
formalism. Thus (i) the `and then', $\sqcap$, on the right hand
side of equation (\ref{TVtwo-time}) is represented by the tensor
products of the Heyting algebras $\Ga\Om^{\Hi_{t_1}}$ and
$\Ga\Om^{\Hi_{t_2}}$ (as in equation (\ref{TVtwo-timeT})); and
(ii) the `and then' on the left hand side of  equation
(\ref{TVtwo-time}) will be represented, initially, by the tensor
product of the associated spectral projectors (i.e. using the HPO
formalism) and, then, `daseinized' to become the tensor product
between the Heyting algebras $\Sub(\Sig^{\Hi_{t_1}})$ and
$\Sub(\Sig^{\Hi_{t_2}})$

We have argued in the previous Sections that (two-time)
inhomogeneous history propositions can be represented as
subobjects of the spectral presheaf in the intermediate topos
$\intt$. In particular, the homogeneous history
 $\alpha_1\sqcap\alpha_2$
is represented by the presheaf
$\ps{\delta(\ha_1)}\otimes\ps{\delta(\ha_2)}\subseteq\Sig^{\Hi_{t_1}}\times
\Sig^{\Hi_{t_2}}\simeq\theta^*\big(\Sig^{\Hi_{t_1}}\otimes\Sig^{\Hi_{t_2}}\big)$.
On the other hand, the HPO-representative, $\ha_1\otimes\ha_2$,
belongs to $\Hi_{t_1}\otimes\Hi_{t_2}$ and, hence, its
daseinization, $\ps{\delta(\ha_1\otimes\ha_2)}$, is a subobject of
the spectral presheaf $\Sig^{\Hione\otimes\Hitwo}$, which is an
object in the topos $\Sets^{\V(\Hione\otimes\Hitwo)^\op}$. As
such, $\ps{\delta(\ha_1\otimes\ha_2)}$ is defined at every stage
in $\V(\Hione\otimes\Hitwo)$, including entangled ones of the form
$W=V_1\otimes V_2+V_3\otimes V_4$. However, since by its very
nature, the tensor product
$\ps{\delta(\ha_1)}\otimes\ps{\delta(\ha_2)}$ is defined only in
the intermediate topos $\inttonetwo$, in order to compare it with
$\ps{\delta(\ha_1\otimes\ha_2)}$ it is necessary to first
pull-back the latter to the intermediate topos using the geometric
morphism $\theta^*$. However, having done that, it is easy to
prove that
\begin{equation}\label{equ:p}
\theta^*\big(\ps{\delta(\ha_1\otimes \ha_2)}\big)_{\la V_1,
V_2\ra}=\ps{\delta(\ha_1)}_{V_1}\otimes\ps{\delta(\ha_2)}_{V_2}
\end{equation}
for all $\la V_1,V_2\ra\in\V(\Hione)\times\V(\Hitwo)$. A
marginally less accurate way of writing this equation is
\begin{equation}
\ps{\delta(\ha_1\otimes \ha_2})_{V_1\otimes V_2}=
\ps{\delta(\ha_1)}_{V_1}\otimes\ps{\delta(\ha_2)}_{V_2}\label{dasa1ota2}
\end{equation}

We  need to be able to daseinize inhomogeneous histories as well
as homogeneous ones but, fortunately, here we can exploit  one of
the important features of daseinization, namely, that it preserves
the `$\lor$'-operation, i.e. at any stage $V$ we have
$\delta(\hat Q_1\lor\hat Q_2)_V=\delta(\hat Q_1)_V\lor\delta(\hat
Q_2)_V$. Thus, for an inhomogeneous history of the form
$\alpha:=(\alpha_1\sqcap\alpha_2)\vee(\beta_1\sqcap\beta_2)$ we
have the topos representation
\begin{eqnarray}
\ps{\delta(\hat\alpha)}&=&\ps{\delta(\ha_1\otimes\ha_2\lor
\hat\beta_1\otimes\hat\beta_2)}\nonumber\\
&=&\ps{\delta(\ha_1\otimes\ha_2)}\,\cup\,\ps{\delta(\hat\beta_1\otimes\hat\beta_2)}
\end{eqnarray}
which, using equation (\ref{dasa1ota2}), can be rewritten as
\begin{equation}
\ps{\delta(\ha)}_{V_1\otimes V_2}=\ps{\delta(\ha_1)}_{V_1}\otimes
\ps{\delta(\ha_2)}_{V_2}\cup\;
\ps{\delta(\hat{\beta}_1)}_{V_1}\otimes
\ps{\delta(\hat{\beta}_2)}_{V_2} \label{toposRepalpha}
\end{equation}
This is an important result for us.

Let us now consider a specific two-time history $\alpha:=
(A_1\in\Delta_1)_{t_1}\sqcap(A_2\in\Delta_2)_{t_2}$ and try to
determine its truth value in terms of the truth values of the
single-time propositions of which it is composed. Let the initial
state be $\ket\psi_{t_1}\in\Hi_{t_1}$ and let us first construct
the truth value of the proposition ``$(A_1\in\Delta_1)_{t_1}$''
(with associated spectral projector $\hat E[A_1\in\Delta_1]$) in
the state $\ket\psi_{t_1}$. To do this we must  construct the
pseudo-state associated with $\ket\psi_{t_1}$. This is defined at
each context $V\in\Ob(\V({\cal H}_{t_1}))$ as
\begin{equation*}
\ps{\w}_V^{\ket\psi_{t_{1}}}:=
\ps{\delta\big(\ket\psi_{t_1}\,{}_{t_1}\!\bra\psi\big)}_V
\end{equation*}
which form the components of the presheaf
$\ps{\w}^{\ket\psi_{t_{1}}}\subseteq \Sig^{\Hi_1}$. The truth
value of the proposition ``$(A_1\in\Delta_1)_{t_1}$'' at stage
$V_1$, given the pseudo-state $\ps{\w}^{\ket\psi_{t_{1}}}$, is
then the global element of $\Om^{\Hi_{t_1}}$ given by
\begin{align}\label{ali:t1}
v(A_1\in\Delta_1; \ket\psi_{t_1})(V_1)&= \{V'\subseteq V_1\mid
\ps{\w}^{\ket\psi_{t_{1}}}_{V'}\subseteq
\ps{\delta(\hat{E}[A_1\in\Delta_1])}_{V'}\}\\
&=\{V'\subseteq
V_1\mid{}_{t_1}\!\bra\psi\delta\big(\hat{E}[A_1\in\Delta_1]
\big)_{V'}\ket\psi_{t_1}=1\}
\end{align}
for all $V_1\in\Ob(\V(\Hi_{t_1}))$.

As there is no state-vector reduction in the topos quantum theory,
the next step is to evolve the state $\ket\psi_{t_1}$ to time
$t_2$ using the usual, unitary time-evolution operator
$\hat{U}(t_1,t_2)$, thus
$\ket\psi_{t_2}=\hat{U}(t_1,t_2)\ket\psi_{t_1}$. Of course, this
vector still lies in $\Hi_{t_1}$. However, in the spirit of the
HPO formalism, we will take its isomorphic copy (but still denoted
$\ket\psi_{t_2}$) in the  Hilbert space
$\Hi_{t_2}\simeq\Hi_{t_1}$.

Now we consider the truth value of the proposition
``$(A_2\in\Delta_2)_{t_2}$'' in this evolved state
$\ket\psi_{t_2}$. To do so we employ the pseudo-state
\begin{equation}\label{equ:pseudo}
\ps{\w}_{V_2}^{\ket\psi_{t_{2}}}=
\ps{\w}_{V_2}^{\hat{U}(t_2,t_1)\ket\psi_{t_{1}}} =
\ps{\delta(|\psi\rangle_{t_{2}}\,{}_{t_{2}}\!\bra\psi)}_{V_2}
=\ps{\delta\left(\hat{U}(t_2,t_1)|\psi\rangle_{t_{1}}\,
{}_{t_{1}}\!\bra\psi\hat{U}(t_2,t_1)^{-1}\right)}_{V_2}
\end{equation}
at all stages $V_2\in\Ob{(\V(\Hi_2))}.$ Then  the truth value of
the proposition ``$(A_2\in\Delta_2)_{t_2}$'' (with associated
spectral projector $\hat{E}[A_2\in\Delta_2]$) at stage
$V_2\in\Ob(\V(\Hi_2))$ is
\begin{align}\label{ali:t2}
v\big(A_2\in\Delta_2;\ket\psi_{t_{2}}\big)(V_2) &=\{V'\subseteq
V_2\mid \ps{\w}^{\ket\psi_{t_{2}}}_{V'}\subseteq
\ps{\delta\big(\hat{E}[A_2\in\Delta_2]\big)}_{V'}\}\\
&=\{V'\subseteq V_2\mid
{}_{t_2}\!\bra\psi\delta\big(\hat{E}[A_2\in\Delta_2]\big)_{V'}
\ket\psi_{t_2} =1\}\nonumber
\end{align}

We would now like to define truth values of daseinized history
propositions of the form $\ps{\delta(\ha_1\otimes \ha_2)}$. To do
so we need to construct the appropriate pseudo states. A state in
the tensor product Hilbert space $\Hi_{t_1}\otimes\Hi_{t_2}$ is
represented by $\ket\psi_{t_1}\otimes\ket\psi_{t_2}$ where, for
reasons explained above,
$\ket\psi_{t_2}=\hat{U}(t_2,t_1)\ket\psi_{t_1}$. To each such
tensor product of states, we can associate the tensor product
pseudo-state:
\begin{equation}
\ps{\w}^{\ket\psi_{t_{1}}\otimes
\ket\psi_{t_{2}}}:=\ps{\delta\big(|\psi_{t_{1}}\otimes
\psi_{t_{2}}\rangle\langle\psi_{t_{2}}\otimes\psi_{t_{1}}|\big)}=
\ps{\delta\big(\ket\psi_{t_1}\,{}_{t_{1}}\!\bra\psi\otimes
\ket\psi_{t_2}\,{}_{t_{2}}\!\bra\psi\big)}
\end{equation}

On the other hand, for contexts $V_1\otimes V_2\in
\Ob(\V(\Hi_1\otimes\Hi_2))$ we have
\begin{align}\label{ali:w}
\ps{\w}^{\ket\psi_{t_{1}}}_{V_1}\otimes
\ps{\w}^{\ket\psi_{t_{2}}}_{V_2}&=
\ps{\delta\big(\ket\psi_{t_1}\,{}_{t_{1}}\!\bra\psi\big)}_{V_1}\otimes
\ps{\delta\big(\ket\psi_{t_2}\,{}_{t_{2}}\!\bra\psi\big)}_{V_2}\\
&=\ps{\delta\big(\ket\psi_{t_1}\,{}_{t_{1}}\!\bra\psi\otimes
\ket\psi_{t_2}\,{}_{t_{2}}\!\bra\psi\big)}_{V_1\otimes V_2}
\end{align}
so that
\begin{equation}
\ps{\w}^{\ket\psi_{t_{1}}}_{V_1}\otimes
\ps{\w}^{\ket\psi_{t_{2}}}_{V_2}= \ps{\w}^{\ket\psi_{t_{1}}\otimes
\ket\psi_{t_{2}}}_{V_1\otimes V_2}
\end{equation}
or, slightly more precisely
\begin{equation}
\ps{\w}^{\ket\psi_{t_{1}}}_{V_1}\otimes
\ps{\w}^{\ket\psi_{t_{2}}}_{V_2}=
\theta^*\big(\ps{\w}^{\ket\psi_{t_{1}}\otimes
\ket\psi_{t_{2}}}\big)_{\la V_1, V_2\ra}
\end{equation}

Given the pseudo-state $\ps{\w}^{\ket\psi_{t_{1}}}\otimes
\ps{\w}^{\ket\psi_{t_{2}}}\in \Sub_{\cl}(\Sig^{\Hi_{t_1}})\otimes
\Sub_{\cl}(\Sig^{\Hi_{t_2}})$ we want to consider the truth value
of the subobjects of the form $\ps{S}_1\otimes \ps{S}_2$ (more
precisely, of the homogeneous history proposition represented by
this subobject) as a global element of
$\Om^{\Hi_{t_1}}\times\Om^{\Hi_{t_2}}$. This is given by
\begin{align}\label{ali:truthvalue1}
v\big(\ps{\w}^{\ket\psi_{t_{1}}}\otimes
\ps{\w}^{\ket\psi_{t_{2}}}&\subseteq\ps{S}_1\otimes
\ps{S}_2\big)\big( \la V_1, V_2\ra\big)\nonumber\\
:= &\{\la V'_1, V'_2\ra\subseteq \la V_1, V_2\ra\mid
\big(p_1^*(\ps{\w}^{\ket\psi_{t_{1}}})\times
p_2^*(\ps{\w}^{\ket\psi_{t_{2}}})\big)_{\la V'_1,
V'_2\ra}\subseteq(\ps{S}_1\times \ps{S}_2)_{
\la V'_1, V'_2\ra}\}\nonumber\\
\simeq\,&\{V^{'}_1\subseteq V_1\mid \ps{\w}^{\ket\psi_{t_{1}}}_{
V^{'}_1}\subseteq(\ps{S}_1)_{ V^{'}_1}\}\times \{V^{'}_2\subseteq
V_2\mid \ps{\w}^{\ket\psi_{t_{1}}}_{ V^{'}_2}\subseteq(\ps{S}_1)_{
V^{'}_2}\}\\\nonumber =\,&\big\la
v\big(\ps{\w}^{\ket\psi_{t_{1}}}\subseteq\ps{S}_1\big)(V_1),
v\big(\ps{\w}^{\ket\psi_{t_{2}}}\subseteq\ps{S}_2\big)(V_2)\big\ra\\
=&j\big(\,v(\ps{\w}^{\ket\psi_{t_{1}}}\subseteq\ps{S}_1)\otimes
v(\ps{\w}^{\ket\psi_{t_{2}}}\subseteq\ps{S}_2)\big)(\la
V_1,V_2\ra)
\end{align}
where $j:\Ga\Om^{\Hi_{t_1}}\otimes\Ga\Om^{\Hi_{t_2}}
\rightarrow\Ga(\Om^{\Hi_{t_1}}\times\Om^{\Hi_{t_2}})$ is discussed
in equation (\ref{Def:j}). Thus we have
\begin{equation}
v\big(\ps{\w}^{\ket\psi_{t_{1}}}\otimes
\ps{\w}^{\ket\psi_{t_{2}}}\subseteq\ps{S}_1\otimes \ps{S}_2\big)=
j\big(\,v(\ps{\w}^{\ket\psi_{t_{1}}} \subseteq\ps{S}_1)\otimes
v(\ps{\w}^{\ket\psi_{t_{2}}}\subseteq\ps{S}_2)\big)
\end{equation}
where the link with equation (\ref{TVtwo-time}) is clear.  In
particular, for the homogenous history $\alpha:=
(A_1\in\Delta_1)_{t_1}\sqcap(A_2\in\Delta_2)_{t_2}$ we have the
generalised truth value
\begin{eqnarray}
v\big((A_1\in\Delta_1)_{t_{1}}\sqcap (A_2\in\Delta_2)_{t_2};
\ket\psi_{t_{1}}\big)&=&v\big(\ps{\w}^{\ket\psi_{t_{1}}}\otimes
\ps{\w}^{\ket\psi_{t_{2}}}\subseteq \ps{\delta\big(\hat
E[A_1\in\Delta_1]\big)}\otimes
\ps{\delta\big(\hat E[A_2\in\Delta_2]\big)}\\
&=&j\Big(v\big(\ps{\w}^{\ket\psi_{t_{1}}}
\subseteq\ps{\delta\big(\hat E[A_1\in\Delta_1]\big)}\otimes
v\big(\ps{\w}^{\ket\psi_{t_{2}}}\subseteq \ps{\delta\big(\hat
E[A_2\in\Delta_2]\big)}\Big) \nonumber\hspace{1cm}
\end{eqnarray}
This can be extended to inhomogeneous histories with the aid of
equation (\ref{toposRepalpha}).

The discussion above shows that D\"oring-Isham topos scheme for
quantum theory can be extended to include propositions about the
history of the system in time. A rather striking feature of the
scheme is the way that the tensor product of projectors used in
the HPO history formalism is `reflected' in the existence of a
tensor product between the Heyting algebras of sub-objects of the
relevant presheaves. Or, to put it another way, a type of
`temporal logic' of Heyting algebras can be constructed using the
definition of the Heyting-algebra tensor product.

As we have seen,  the topos to use for all this  is the
`intermediate topos' $\inttonetwo$ of presheaves over the category
$\V(\Hi_{t_1})\times\V(\Hi_{t_2})$. The all-important spectral
presheaf in this topos is essentially the presheaf
$\Sig^{\Hi_{t_1}\otimes\Hi_{t_2}}$ in the topos
$\Sets^{\V(\Hi_{t_1}\otimes\Hi_{t_2})^\op}$, but restricted to
`product' stages $V_1\otimes V_2$ for $V_1\in\Ob(\V(\Hi_{t_1}))$
and $V_2\in\Ob(\V(\Hi_{t_2}))$. This restricted presheaf can be
understood as a `product' $\Sig^{\Hi_{t_1}}\times
\Sig^{\Hi_{t_2}}$. A key result in this context is our proof in
Theorem 5.3 of the existence of a Heyting algebra isomorphism
$h:\Sub(\Sig^{\Hi_{t_1}})\otimes \Sub(\Sig^{\Hi_{t_2}})\rightarrow
\Sub(\Sig^{\Hi_{t_1}}\times\Sig^{\Hi_{t_2}})$.

Moreover, as we have shown,  the evaluation map of history propositions
maps the temporal structure of history propositions to the
temporal structure of truth values, in such a way that the
temporal-logic properties are preserved.

A fundamental feature of the topos analogue of the HPO formalism
developed above is that the notion of consistent sets, and thus of
the decoherence functional, plays no  role. In fact, as was shown
above, truth values can be ascribed to \emph{any} history
proposition independently of whether it belongs to a consistent
set or not. Ultimately, this is because the topos formulation of
quantum theory makes no fundamental use of the notion of
probabilities, which are such a central notion in the
(instrumentalist) Copenhagen interpretation of quantum theory.
Instead, the topos approach deals with `generalised' truth values
in the Heyting algebra of global elements of the subobject
classifier.  This is the sense in which the theory is
`neo-realist'.

Reiterating, the standard consistent histories approach makes use
of the Copenhagen concept of probabilities which must satisfy the
classical summation rules and, thus, can only be applied to
``classical'' sets of histories, i.e. consistent sets of histories
defined using the decoherence functional. The topos formulation of
the HPO formalism abandons the concept of probabilities and
replaces them with truth values defined at particular stages,
i.e. abelian Von Neumann subalgebras. These stages are
interpreted as the classical snapshots of the theory. In this
framework there is no need for the notion of consistent set and,
consequently, of decoherence functional. Thus the topos
formulation of consistent histories avoids the issue of having
many incompatible, consistent sets of proposition, and can assign
truth values to any history proposition.

It is interesting to note that, in the consistent history
formulation of \emph{classical} physics, we do not have the notion
of decoherence functional since, in this case, no history
interferes with any other. Since, as previously stated, one of the
aims of re-expressing quantum theory in terms of topos theory  was
to make it ``look like'' classical physics, it would seem that, at
least as far as the notion of decoherence functional is involved,
the resemblance has been successfully demonstrated.
\section{Summary and discussion}
The consistent histories interpretation of quantum theory was born in the light of making sense of
quantum theory as applied to a closed system. A central ingredient in the consistent-histories approach
is the notion of the decoherence functional which defines consistent sets of propositions, i.e. propositions
which do not interfere with each other. Only within these consistent sets can the Copenhagen notion of
probabilities be applied. Thus, only within a given consistent set is it possible to use quantum theory
to analyse a closed system. 

Unfortunately there are many incompatible consistent sets of propositions,
which can not be grouped together to form a larger set. This feature causes several problems in the
consistent histories approach, since it is not clear how to interpret this plethora of consistent sets or how
to select a specific one, if needed.
In standard quantum theory the problem is overcome by the existence of an external observer who
selects what observable to measure. This is not possible when dealing with a closed system since, in this
case, there is no notion of external observer.
As mentioned in previous Sections, attempts have been made to interpret this plethora of
consistent sets, including one by Isham \cite{consistent2} that used topos theory albeit in a very different way from
what we have described in this thesis.

Rather, we derive a formalism for analysing history propositions, which
does not require the notion of consistent sets, thus avoiding the problem of incompatible sets from the
outset. In particular we adopt the topos formulation of quantum theory put forward by Isham and D¨oring
in \cite{andreas1}, \cite{andreas2}, \cite{andreas3}, \cite{andreas4}, \cite{andreas5} and \cite{andreas} and apply it to situations in which the propositions, to be evaluated, are
temporally-ordered propositions, i.e. history propositions. In the above mentioned papers, the authors
only define truth values for single time propositions, but in this thesis we have extended their scheme to sequences
of propositions defined at different times. In particular we have shown how to define truth values of
homogeneous history propositions in terms of the truth values of their individual components. 

In order
to achieve this we exploit the fact that, in the histories approach, there is no state-vector reduction
induced by measurement, since we are in the context of a closed system. We take the absence of state-vector
reduction to imply that truth values of propositions, at different times, do not ‘interfere’ with each
other, so that it is reasonable to try to define truth values of the composite proposition in terms of the
truth values of the individual, single-time propositions.

In the setting of topos theory, propositions are identified with subobject of the spectral presheaf. We have
shown that for (the example of two-time) history propositions the correct topos to utilise is the `intermediate topos' $\Sets^{\V(\Hi_{t_1})\times\V(\Hi_{t_2} ))^\op}
\cong\theta^*\Sets^{\V(\Hi_1\otimes \Hi_2)^\op}$ whose category of contexts only contains
pure tensor products of Abelian von Neumann subalgebras.
The reason why this topos was chosen instead of the full topos $\Sets\V(\Hi_1\otimes \Hi_2)$ is because of its
relation to the tensor product, $Sub(\Hi_{t_1})\otimes Sub(\Hi_{t_2})$, of Heyting algebras $Sub(\Hi_{t_1})$ and $Sub(\Hi_{t_2})$
However, the full topos is interesting as there are entangled contexts, i.e. contexts which are not pure
tensor products. For such contexts it is impossible to define a history proposition as a temporally ordered
proposition, or a logical `or' of such. Moreover, in our formalism, because of the absence of
state-vector reduction, the truth value of a proposition at a given time does not influence the truth value
of a proposition at a later time as long as the states, in terms of which such truth values are defined, are
the evolution (through the evolution operator) of the same states at different times. These means that
the pseudo-states at different times are related in a causal way.
To analyse in detail the dependence between history propositions and individual time components,
the notion of temporal logic in the context of Heyting algebras is introduced. Specifically 
the temporal structure of the Heyting algebra of propositions $\theta^*\Big(Sub(\Hi_{t_1}\otimes\Hi_{t_2})\Big)$ was identified with the tensor
product of Heyting algebras of single-time propositions $Sub(\Hi_{t_1})\otimes Sub(\Hi_{t_2})$, i.e. the
two algebras are isomorphic. 

It is then possible to define an evaluation map within the intermediate
topos $\Sets^{(\V(\Hi_{t_1})\times\V(\Hi_{t_2}))^\op}$ and show that such a map correctly preserves the temporal structure of the
history propositions it evaluates.
There are still a number of open questions that need to be addressed. In particular it would be very
important to analyse the precise temporal-logical meaning, if there were one, of entangled inhomogeneous
propositions and, thus, extend the topos formalism of history theory to the full topos $\Sets^{\V(\Hi_{t_1}\otimes\Hi_{t_2})^\op}$ .
Such an extension would be useful since it would shed light on composite systems in general in the
context of topos theory, something that is still missing.

The topos-centred history formalism described in this thesis, does not require the notion of
consistent sets. However, in standard consistent-history theory, the importance of consistent sets lies in
the fact that, given such a set, the formalism can be interpreted as saying that it is `as if' the quantum
state had undergone a state-vector reduction. This phenomenon allows for predictions of events in a
closed system, i.e. the assignment of probabilities to the possible outcomes.
Given the importance of such consistent sets, their absence in the topos formulation of the history
formalism is striking. 
Since the decoherence functional assigns probabilities to histories, a related issue is that of defining
the notion of a probability within the topos formulation of history theory. The introduction of such
probabilities would allow us to assign truth values to `second-level propositions', i.e. propositions of the
form ``the probability of the history $\alpha$ being true is p". This type of proposition is precisely of the form
dealt with in \cite{consistent2}.

Another interesting topic for further investigation would be the connection, if any, with the path
integral formulation of history theory. In fact, in a recent work by A. D¨oering, \cite{andreas} it was shown that
it is possible to define a measure within a topos. A very interesting new research programme would be
to analyse whether such a measure can be used in the context of the topos formulation of consistent
histories developed in the present paper to recover the path-integral formulation of standard quantum
theory. This analysis would require the definition of probabilities different from \emph{one} discussed above,
since the path integral was introduced precisely to define the decoherence functional between histories.
\chapter{Conclusion}
The topic of this thesis is the discussion and the development of two approaches to quantum theory:\\
loop quantum gravity (LQG) and the topos approach to quantum gravity. \\
We have started by discussing the general framework of LQG, analysing, in detail, the semiclassical properties of the volume operator. \\
Such analysis was carried out with respect to both \emph{area coherent states} and \emph{flux coherent states}. The result of our analysis has shown that the \emph{area coherent states} should be abandoned as tools for analysing semicalssical properties of the volume operator, for the following reasons:
\begin{enumerate}
\item artificial rescaling of the coherent state label is required. 
\item Particular embeddings of the 4-valent and 6-valent graphs are required. However, it has been shown that the combinations of Euler angles, for which such embeddings are attained, have measure zero in SO(3), and are, therefore, negligible. 
\item Impossibility to eliminate the embedding dependence (the ‘staircase problem).
\end{enumerate}
On the other hand, the \emph{flux coherent states} can be utilised for performing semiclassical analysis, as long as the graph we take in consideration has valence \emph{six}.\\
This result has heavy repercussions on \emph{spin foam models} (SFM), which provide the dynamical aspects of LQG. In fact, the current SFM are all based on boundary spin networks of valence four. \\
Since the volume operator plays a pivotal role for LQG, as it defines triad operators and
hence the dynamics, the impossibility of obtaining six valent boundary spin networks in spin foams is of particular importance.\\

Only 4-valent spin networks emerge in the current spin foam models because the manifold is discretised in terms of 4-simplices, and spin network arises as dual simplices of the boundary tetrahedrons (of the 4-simplices). \\
However, if the manifold is discretised in terms of hypercubes, whose boundaries are 3 dimensional cube, then the resulting spin networks (dual simplices) would be six valent. This observation has motivated the development and analysis of a possible SFM defined in terms of cubic triangulations of the four manifold, also called ``cubulations". This model has only been constructed at a heuristic level, but it already exhibits the following advantages over the current SFM:
\begin{enumerate}
\item It avoids simplicity constraints since the starting point is the Holst action, rather than the Plebanski action.
\item The $B$ field of BF-theory transforms by the adjoint action of the gauge group, while the
connection $A$ underlying $F$ transforms in the usual way. This implies that, in the current SFM, local gauge invariance of the Plebanski action is not manifest. \\
However, if one does not consider dual graphs, but only works with the triangulation,  gauge invariance issues can be solved.
\item It overcomes the difficulty to relate SFM to the Regge calculus since, differently from current SFM, the starting point is the Holst action.
\end{enumerate}
However, the cubulated SFM has still many open issues such as:
\begin{enumerate}
\item \emph{continuum limit}:  cubulations suggest a naive but natural notion of continuum
limit, which consists in studying the behaviour of the correlation functions under barycentric refinement of
the hypercubes at fixed IR regulator (boundary surface). Of course, in the spirit of the AQG framework
[20] one could also say that the continuum limit has already been taken, provided that one works with
infinite cubulations. Moreover, one works directly with infinite IR. 
\item Even though we can work at finite UV and IR regulators, it is still hard to compute the determinant of the covariance matrix of the co-tetrad Gaussian and to determine its index. Since these covariances are highly correlated, the
practical computation of the n--point functions, at least in the macroscopic regime, will be possible only if
the corresponding non trivial measure has some kind of cluster property [54].
\end{enumerate}
This ends the discussion concerning LQG and its dynamical aspect defined in terms of SFM.\\

The second part of this thesis is concerned with a possible topos approach to quantum gravity. Such an approach needs a reformulation, in terms of topos theory, of the theories involved in. \\
In Section \ref{s.topos} we have discussed the topos reformulation of quantum theory, which suggests a more realist interpretation of the theory. \\ Such an interpretation is preferable since it overcomes the conceptual difficulties related to the notion of closed system and the Kochen-Specker no-go theorem inherent in the standard Copenhagen interpretation of the theory. \\
However, a radical new way of thinking about what a theory of physics \emph{is}, emerges. Consequently, a different interpretation of the concepts of space, time and matter is required.

In order to make connection with a possible theory of quantum gravity, in Chapter \ref{cha:toposhistory} we have explained how a formulation of history quantum theory can be carried out in terms of topos theory. This reformulation is very important, since it allows the possibilities of defining any quantum statements about four-metrics. \\
In particular, in this new topos approach of history theory it has been shown that Heyting-algebra valued truth values can be assigned to any history proposition, i.e. it is no longer necessary to consider just `consistent' sets of propositions. This is an advantage over the older consistent history formalism, in which the process of choosing which consistent set of history propositions to employ, when defining quantum statements, is really problematic.\\
Therefore, the topos formulation of history theory sets the stage for a framework in which truth values can be assigned to \emph{any} proposition about spacetime.

Both the topos version of quantum theory and the history theory are only the first steps towards a theory of quantum gravity in terms of topos theory. A lot of work is still needed. However, the prescription of how a theory of quantum gravity should be derived, is the same as the one used for reformulating quantum theory and history quantum theory in the language of topos theory. \\
In particular, these theories are the result of an interplay between four main ingredients:
\begin{enumerate}
\item[1.] The physical system under consideration. 
\item[2.] The type of theory one is set out to analyse (classical or quantum)
\item[3.] The corresponding correct topos with which to express such a theory. The choice of such a topos will depend on the theory type and on the system under consideration.
\item[4.] The formal language or underlying logic associated to the system.
\end{enumerate}
A theory of physics is then identified with finding a representation, in a certain topos, of the formal language that is attached to the system.\\
This strategy revealed itself successful, for both quantum theory and history theory, with advantages and enrichment over the standard formulations of the theories in both cases.  The hope is that this same strategy can reveal itself fruitful for defining a possible quantum theory of gravity.
\chapter{Appendix}
\section{Category Theory}
``Category theory allows you to work on structures without the need first to pulverise them into set theoretic dust" (Corfiel).
The above quote explains, in a rather pictorial way, what category theory, and in particular Topos theory, are really about.
In fact, Category theory, and in particular Topos theory, allows one to abstract from the specification of points (elements of a set) and 
functions between these points to a universe of discourse in which the basic elements are arrows, and any property is given 
in terms of compositions of arrows.\\
Let us analyse, in a more rigorous way, what a Category is.  
 \begin{definition} \label{def:cat} \cite{topos3} \cite{topos7} \cite{topos8}
A \textbf{category} consists of two things:
\begin{enumerate}
\item a collection of objects
\item a collection of morphisms between these objects such that the following conditions hold:
\end{enumerate}
\begin{itemize}
\item \textbf{composition condition}: given two morphisms $f:a\rightarrow b$ and $g:b\rightarrow c$ 
with dom g=cod f then there exists the composite map $gof:a\rightarrow c$
\item \textbf{associative law}: given $a\xrightarrow{f}b\xrightarrow{g}c$ then $(h \circ(g \circ f))=((h \circ g)\circ f)$,
  i.e. the following diagram commutes: 
   \[\xymatrix{
\ar[dd]_{(h o g)o f}&a\ar[rr]^f\ar@{-->}[rrdd]^{g o f}\ar[dd]_{h o (g o f)}&&b\ar[dd]^g\ar[lldd]^{h o g}\\
&&&\\
&d\ar[rr]^h&&c\\
}\]

\item \textbf{identity law}: for any object b in the category there exists a morphism $1_b:b\rightarrow b$ 
called identity arrow such that, given
any other two morphisms $f:a\rightarrow b$ and $g:b\rightarrow c$, we then have  $1_b o f=f$   
and $g o 1_b=g$, i.e. the following diagram commutes:
\end{itemize}

\[\xymatrix{
a\ar[rr]^f\ar[rrdd]^f&&b\ar[rrdd]^g\ar[dd]^{1_b}&&\\
&&&&\\
&&b\ar[rr]^g&&c\\
}\]

\end{definition}

\subsection{Examples of Categories}
In this Section we will analyse some example of categories. For more detail see \cite{topos3} \cite{topos7}
\begin{enumerate}
\item\textbf{Simple example}\\
A two element category:
\[\xymatrix{
0\ar@(ul,ur)[]^{i_0}\ar[rr]^{f_{01}}&&1\ar@(ul,ur)^{i_1}
}\]
This category has 3 arrows:
\begin{itemize}
\item $i_0:0\rightarrow 0$ identity on 0
\item $i_1:1\rightarrow 1$ identity on 1
\item $f_{01}:0\rightarrow 1$
\end{itemize}
It is easy to see that the composition arrow are:
$i_0 \circ i_0=i_0$ ,$i_1 \circ i_1=i_1$ ,$i_1 \circ f_{01}=f_{01}$ and $f_{01} \circ i_1=f_{01}$. 
\item \textbf{More complex example}: 
\emph{Comma Category}\\
This category has as objects arrows with fixed domain or codomain.
For example consider the comma category $C\downarrow\mathds{R}$ where:
\begin{itemize}
\item Objects: given $A,B\in C$, the objects in $C\downarrow\mathds{R}$ are arrows whose codomain is $\mathds{R}$, i.e. $f:A\rightarrow\mathds{R}$ and $g:B\rightarrow\mathds{R}$, also written as: (A,f) and (B,g)
\item An arrow between objects $f$ and $g$ is a function $k:A\rightarrow B$ such that

\[\xymatrix{
A\ar[rr]^k\ar[ddr]_f&&B\ar[ddl]^g\\
&&\\
&\mathds{R}& \\
}\]

commutes in $C\downarrow\mathds{R}$\\
The above definition of arrows in $C\downarrow\mathds{R}$ implies the following:
\begin{itemize}
\item \emph{Composition}

\[\xymatrix{
A\ar[rrrr]^{j \circ i}\ar[drr]^j\ar[ddrr]_f&&&&C\ar[ddll]^h\\
&&B\ar[rru]^j\ar[d]^g&&\\
&&\mathds{R}&&\\
}\]
\item \emph{Identity}\\
The identity arrow on $f:A\rightarrow\mathds{R}$ is: $id_A:(A,f)\rightarrow(A,f)$
\end{itemize}

It should be noted that a comma category is equivalent to the category of bundles over $\mathds{R}$ 
iff C is not concrete, whereby a 
concrete category is a category in which, roughly speaking, all objects are sets possibly 
carrying some additional structure, 
all morphisms are functions between those sets, and the composition of morphisms is the 
composition of functions. The 
prototypical concrete category is Set, the category of sets and functions.
\end{itemize}
\item\textbf{Complex example}: 
\emph{Category $Sets^{\mathscr{C}^{op}}$}\\
Given a contravariant (see section \ref{sfunc}) between a Category $\mathscr{C}$ and Sets then we can form a 
category $Sets^{\mathscr{C}^{op}}$\footnote{It should be noted that $\mathscr{C}^{op}$ represents the 
opposite of the category $\mathscr{C}$. Objects in $\mathscr{C}^{op}$ are the same as the objects in 
$\mathscr{C}$, while the morphisms are the inverse of the morphisms in $\mathscr{C}$,
i.e. $\exists$ a $\mathscr{C}^{op}$-morphisms $f:A\rightarrow B$ iff $\exists$ a $\mathscr{C}$-
morphisms $f:B\rightarrow A$.} such that we have the following:
\begin{itemize}
\item Objects:\\
all contravariant functors $P:\mathscr{C}\rightarrow Sets$ 
\[\xymatrix{
&&1&&\\
A\ar[rru]^h&&&&B\ar[llu]_g\\
&&o\ar[rru]_f\ar[llu]^k&&\\
&&\Downarrow^P&&\\
&&P(1)\ar[rrd]^{P(g)}\ar[lld]_{P(h)}&&\\
P(A)\ar[rrd]_{P(k)}&&&&P(B)\ar[lld]^{P(f)}\\
&&P(0)&&\\
}\]
\item Arrows: \\
all natural transformation $N:P\rightarrow P^`$ between contravariant functors such that given a function 
$f:D\rightarrow C$ the following diagram commutes
\[\xymatrix{
PC\ar[rr]^{Pf}\ar[dd]_{N_C}&& PD\ar[dd]^{N_D}\\
&&\\
P^`C\ar[rr]_{P^`f}&& P^`D\\
}\]

where a $\emph{natural transformation}$ is defined as follows:
\begin{definition} 
\label{def:nat}
A \textbf{natural transformation} from $Y:\mathscr{C}\rightarrow set$ to 
$X:\mathscr{C}\rightarrow set$ 
is an  assignment of an arrow 
$N:Y\rightarrow X$ that associates to each object A in $\mathscr{C}$ an arrow 
$N_A:Y(A)\rightarrow X(A)$ in Set such that, for any 
$\mathscr{C}$-arrow $f:A\rightarrow B$ the following diagram commutes
\[\xymatrix{
A\ar[dd]^f&&Y(B)\ar[rr]^{N_B}\ar[dd]_{Y(f)}&&X(B)\ar[dd]^{X(f)}\\
&&&&\\
B&&Y(A)\ar[rr]^{N_A}&&X(A)\\
}\]
i.e.
\begin{equation*}
N_A \circ Y(f)=X(f) \circ N_B
\end{equation*}
\end{definition}
where $N_A:Y(A)\rightarrow X(A)$ are the components on N while N is the \emph{natural transformation}.\\
From this diagram it is clear that the two arrows $N_A$ and $N_B$ turn the Y-picture of $f:A\rightarrow B$
into the respective X-picture.\\  
We can now define the following:

\begin{itemize}
\item \emph{Identity maps} for objects X in $\mathcal{S}^{\mathscr{C}^{op}}$ are identified with maps 
$i_X$ 
whose components $i_{X_A}$ are the identity maps of X(A) in $\mathcal{S}$  
\item \emph{Composition maps} in $\mathcal{S}^{\mathscr{C}^{op}}$ :
consider X,Y and Z that belong to $\mathcal{S}^{\mathscr{C}^{op}}$, such that there exist maps 
$X\xrightarrow{N}Y$
and $Y\xrightarrow{M}Z$ between them. We can then form a new map $X\xrightarrow{M \circ N}Y$, whose 
components would be $(M \circ N)_A=M_A o\circ N_A$, i.e. graphically we would have 
\[\xymatrix{
X(A)\ar[rrrr]^{X(f)}\ar[dd]_{N_A}&&&&X(B)\ar[dd]^{N_B}\\
&&&&\\
Y(A)\ar[rrrr]^{Y(f)}\ar[dd]_{M_A}&&&&Y(B)\ar[dd]^{M_B}\\
&&&&\\
Z(A)\ar[rrrr]^{Z(f)}&&&&Z(B)\\
}\]

\end{itemize}

\end{itemize}

$Sets^{\mathscr{C}^{op}}$ is called the category of presheaves.
The Category $Sets^{\mathscr{C}^{op}}$ is very important since, as it will be shown later on, $Sets^{\mathscr{C}^{op}}$ is 
actually a Topos. From now on we will refer to $Sets^{\mathscr{C}^{op}}$ as the Topos of Presheaves.

\end{enumerate}

\subsection{Elements and arrows in a category}
n Category theory it is convenient to define categorical concepts externally, i.e. 
by reference to connections with other 
categories. This connections is established by functions, therefore we will describe categorical concepts by functions.
\begin{itemize}
\item \textbf{Monic arrow}\\
 Monic arrow is the "arrow-analogue" of an injective function.
\begin{definition}
An arrow $f:a\rightarrow b$ in a Category C is monic in C if for any parallel pair 
$g:c\rightarrow a,\hspace{.2in}h:c\rightarrow a$ of arrows, the equality $f \circ g = f \circ h$ implies that $h=g$, i.e $f$ is left 
cancellable.
Monic arrows are denoted as:

\[\xymatrix{
*++{a}\ar@{>->}[rr]&&b\\
}\]

\end{definition}
We now want to show how it is possible to derive a monic function from an injective one and vice versa.
\begin{proof}
Consider an injective function $f:a\rightarrow b$ (i.e. if f(x)=f(y) then x=y) and a pair of parallel functions 
$g:c\rightarrow a,\hspace{.2in}h:c\rightarrow a$ such that 
\[\xymatrix{
c\ar[rr]^{g}\ar[dd]_{h}&&a\ar[dd]^{f}\\
&&\\
a\ar[rr]_f&&b\\
}\]
commutes, then f o g = f o h.\\ 
Now if \begin{align*}\hspace{.2in}x\in C\Longrightarrow \hspace{.2in}&f \circ g(x)= f \circ h(x)\\
&f(g(x))=f(h(x))
\end{align*}
Since $f$ is injective it follows that $g(x)=h(x)$, i.e f is left cancellable.
Vice versa, let f be left cancellable, consider the following diagram
\[\begin{xy}0;/r15mm/:
,(0,0)="x",*@{},*+!U{0}
,(2,0)="z",*@{},*+!{}
,{\ellipse(0.5,1.6){}}
,(2,0.5)="t",*@{},*+!U{x}
,(2,-0.5)="f",*@{},*+!U{y}
,{\ar"x";"t"}
,{\ar"x";"f"}
,(5,0)="p",*@{},*+!U{f(x)=f(y)}
,{\ellipse(1,1){}}
,{\ar"t";"p"}
,{\ar"f";"p"}
,(1,0.5),*@{},*+!{g}
,(1,-0.5),*@{},*+!{h}
,(3.5,0),*@{},*+!{f}
 \end{xy}\]

then $f \circ g=f \circ h$ since $f(x)=f(y)$, where $x=g(0)$ and $y=h(0)$.
Since f is left cancellable by assumption we get: $g=h$, therefore $x=y$ for $f(x)=f(y)$, i.e. $f$ is injective.
\end{proof}
\item\textbf{Epic arrow}\\
Epic arrow is the "arrow-analog" of a surjective function.
\begin{definition}
 An arrow $f:a\rightarrow b$ in a Category C is epic in C if for any parallel pair $g:b\rightarrow c,\hspace{.2in}h:b\rightarrow c$ of arrows, the equality $g \circ f =h \circ f$ implies that $h=g$, i.e $f$ is right cancellable.
Monic arrows are denoted as:\[\xymatrix{
*++{a}\ar@{->>}[rr]&&b\\
}\]
\end{definition}
An epic is a dual\footnote{If A is a statement in the language of categories, then the dual $A^{op}$ of A is the statement 
obtained by replacing domain by codomain (and vice versa) and $h=g \circ f$ by $h=f \circ g$, therefore arrow and composites in A 
are reversed in $A^{op}$. A theorem which is true in A will automatically be true in $A^{op}$}  of a monic
\item\textbf{Iso arrow}\\
An iso arrow is the "arrow-analogue" of a bijective function.
\begin{definition}
A C-arrow $f:a\rightarrow b$ is iso, or invertible in C if there is a C-arrow $g:b\rightarrow a$ such that $g \circ f=1_a$ and $f \circ g=1_b$, therefore $g$ is the inverse of $f$, i.e. $g=f^{-1}$.
\end{definition}
\begin{theorem}
g is unique.
\end{theorem}
\begin{proof}
Consider $g^` \circ f=1_a$ and $f \circ g^`=1_b$, then we have\\
$g^`=1_a \circ g^`= (g \circ f) \circ g^` = g \circ (f \circ g^`) = g \circ 1_b = g$
\end{proof}
An iso arrow has the following properties:
\begin{enumerate}
\item \emph{An iso arrow is always monic}
\begin{proof}
consider an iso $f$, such that $f \circ g = f \circ h$ ($f:a\rightarrow b$ and $g,h:c\rightarrow a$)
then $g = 1_a \circ g = (f^{-1} \circ f) \circ g = f^{-1} \circ (f \circ g) = f^{-1} \circ (f \circ h) = (f^{-1} \circ f) \circ h = h $, therefore f is left cancellable
\end{proof}
\item \emph{An iso arrow is always epic}
\begin{proof}
consider an iso $f$ such that $g \circ f = h \circ f$  ($f:a\rightarrow b$ and $g,h:b\rightarrow c$)
$g= g \circ 1_b = g \circ (f \circ f^{-1}) = (g \circ f) \circ f^{-1} = (h \circ f) \circ f^{-1} = h \circ (f \circ f^{-1}) = h$,
therefore f is right cancellable
\end{proof}
\end{enumerate}
It should be noted not all arrows which are monic and epic are iso, for example: inclusion map is both monic and epic but it is not iso, otherwise it would have an inverse and as a set function it would have to be a bijection, but it is not.
In poset even though all functions are monic and epic, only iso is the identity map. In fact consider a function $f:p\rightarrow q$ this implies that $p\leq q$ if $f$ is an iso it implies that $f^{-1}:q\rightarrow p$ exists, therefore $g\leq p$, but from the antisymmetry property $p\leq q$ and $g\leq p$ imply that $p=q$, therefore $f=1_p$ is a unique arrow.
\item \textbf{Subobjects}
\begin{definition}
A subobject of a C-object d is an equivalence class of C-arrow which are monics with codomain d
i.e. of the form \[\xymatrix{
*++{a}\ar@{>->}[rr]&&d\\
}\]
\end{definition}
This definition implies that the inclusion relation between subobjects of d is defined as follows:
given\[\xymatrix{
*++{f:a}\ar@{>->}[rr]d&&*++{g:b}\ar@{>->}[rr]&&d\\
}\] 
$f\subseteq g$ iff $\exists$ a C-arrow \[\xymatrix{
*++{h:a}\ar@{>->}[rr]&&b\\
}\] 
such that the following diagram commutes\\
\[\xymatrix{
*++{b}\ar@{>->}[rrd]^g&&\\
&&*++{d}\\
*++{a}\ar@{>->}[uu]^h\ar@{>->}[rru]_f&&\\
}\]
i.e $f= g \circ h$. Since $f$ and $g$ are monic it follows that $h$ is monic, therefore $h$ is a subobject of $d$.
We have then showed that $f\subseteq g$ iff $f$ factors through $g$.
It follows that the collection $Sub(d)$ forms a partial ordered set where $[f]\leq[g]$ iff f=gh.
\item \textbf{Elements}
\begin{definition}
Given a category $\mathscr{C}$, with terminal object 1, then an element of a $\mathscr{C}$-object b is a C-arrow $x:1\rightarrow b$
\end{definition}
\begin{example}
In $\mathcal{Set}$, an element $x\in A$, can be identified with the singleton subset $\{*\}$,therefore with an arrow 
$\{*\}\rightarrow A$ from the terminal object to A (see definition of terminal object)
\end{example}
\item \textbf{Products}
\begin{definition}
A \textit{product} of two objects A and B in a category $\mathscr{C}$ is a third $\mathscr{C}$-object
$A\times B$ together with a pair of $\mathscr{C}$-projection arrows:\\
$pr_A:A\times B\rightarrow A$ and $pr_B:A\times B\rightarrow B$\\
such that, given any other pair of  $\mathscr{C}$-arrows $f:C\rightarrow A$ and $g:C\rightarrow B$,
there exists a unique arrow $\langle f,g\rangle:C\rightarrow A\times B$
such that the following diagram commutes
\[\xymatrix{
&&C\ar[rrdd]^g\ar@{-->}[dd]^{\langle f,g\rangle}\ar[lldd]_f&&\\
&&&&\\
A&&A\times B\ar[rr]_{pr_B}\ar[ll]^{pr_A}&&B\\
}\]

i.e.
\begin{equation*}
pr_A o\langle f,g\rangle=f\hspace{.2in}and\hspace{.2in}pr_b o\langle f,g\rangle=g
\end{equation*}
\end{definition}
\item \textbf{Co-products}
\begin{definition}
A \textit{co-product} of two objects A and B in a category $\mathscr{C}$ is a third $\mathscr{C}$-object
$A + B$ together with a pair of $\mathscr{C}$-arrows:\\
$i_A:A\rightarrow A + B$ and $i_B:B\rightarrow A + B$\\
such that, given any other pair of  $\mathscr{C}$-arrows $f:A\rightarrow C$ and $g:B\rightarrow C$,
there exists a unique arrow $[f,g]:A + B\rightarrow C$
which makes the following diagram commute
\[\xymatrix{
A\ar[rr]^{i_A}\ar[rrdd]_f&&A+B\ar[dd]^{[f,g]}&&B\ar[ll]_{i_B}\ar[ddll]^g\\
&&&&\\
&&C&&\\
}\]
i.e. the co-product is the dual of the product
\end{definition}
\end{itemize}
\section{Example of Categories in Quantum Mechanics and General Relativity}
In this Section we will delineate three different categories that arise in Quantum Mechanics, 
namely the category $\mathcal{O}$ \cite{isham1} \cite{isham2} of self-adjoint operators, the category
$\mathcal{W}$ \cite{isham1} \cite{isham2} of Boolean subalgebras of the lattice $P(\mathcal{H})$ and the category Hilb \cite{isham12}
 of Hilbert spaces.
We will then analyse the category nCob \cite{isham12} which arise in General relativity and also the relation between Hilb and nCob
\subsection{Categories in Quantum Mechanics}
\subsubsection{The Category $\mathcal{O}$ of bounded self-adjoint operators}
\begin{definition} 
the \textbf{Set $\mathcal{O}$} of bounded self-adjoint operators is a \textbf{category}, such that
\begin{itemize}
\item the objects of $\mathcal{O}$ are the self-adjoint operators   
\item given a function $f:\sigma(\hat{A})\rightarrow\mathds{R}$ (from the spectrum 
of $\hat{A}$ to the Reals), such that $\hat{B}=f(\hat{A})$,
then there exists a morphism $f_{\mathcal{O}}:\hat{B}\rightarrow\hat{A}$ in $\mathcal{O}$ between operators 
$\hat{B}$ and $\hat{A}$ 
\end{itemize}
\end{definition}
To show that the category $\mathcal{O}$, so defined, is a category (see Definition \ref{def:cat}), 
we need to show that it 
satisfies the identity
law and composition law. 
This can be shown in the following way:
\begin{itemize}
\item \textit{Identity Law}:
given any $\mathcal{O}$-object $\hat{A}$ the identity arrow is defined as the arrow 
$id_{\mathcal{O}_A}:\hat{A}\rightarrow\hat{A}$ that corresponds to the arrow $id:\mathds{R}\rightarrow\mathds{R}$
in $\mathds{R}$.
\item \textit{Composition Condition}:
given two $\mathcal{O}$-arrows $f_{\mathcal{O}}:\hat{B}\rightarrow\hat{A}$ and 
$g_{\mathcal{O}}:\hat{C}\rightarrow\hat{B}$ such that $\hat{B}= f(\hat{A})$ and 
$\hat{C}= g(\hat{B})$, then the composite function $f_{\mathcal{O}}\circ g_{\mathcal{O}}$ in 
$\mathcal{O}$ corresponds to the composite function $f \circ g:\mathds{R}\rightarrow\mathds{R}$
in $\mathds{R}$.
\end{itemize}
The category  $\mathcal{O}$, as defined above, represents a pre-ordered set\footnote{A \emph{pre-ordered} set is a set with the property that, between 
any two objects
there is at most one arrow. This entails that there exists a 
binary relation R between the objects of the pre-ordered set such that
the following holds:
\begin{enumerate}
\item aRa (reflexivity)
\item if aRb and bRc then aRc (transitivity) 
\end{enumerate} 
}.
In fact, the function $f:\sigma(\hat{A})\rightarrow\mathds{R}$ is unique up to isomorphism, therefore 
it follows that for any two objects in 
$\mathcal{O}$ there exists, 
at most, one morphism between them, i.e. $\mathcal{O}$ is a pre-ordered set.
However, $\mathcal{O}$ fails to be a poset\footnote{A \emph{poset} is a pre-ordered set with the extra property 
of being 
antisymmetric: $pRq\hspace{.2in}and\hspace{.2in}qRp\hspace{.2in}\Rightarrow p=q$} since it lacks the 
antisymmetry property .
In fact it can be the case that two operators $\hat{B}$ and $\hat{A}$ in $\mathcal{O}$ are 
such that $\hat{A}\neq\hat{B}$ but they are related by $\mathcal{O}$-arrows
$f_{\mathcal{O}}:\hat{B}\rightarrow\hat{A}$ and 
$g_{\mathcal{O}}:\hat{A}\rightarrow\hat{B}$ in such a way that:
\begin{equation}
g_{\mathcal{O}}\circ f_{\mathcal{O}}=id_B\hspace{.1in}and\hspace{.1in}f_{\mathcal{O}}\circ g_{\mathcal{O}}
=id_A      \label{eq:eq}
\end{equation}
(It should be noted that if $\hat{B}$ and $\hat{A}$ are related in such a way, then $W_A=W_B$ since $\hat{B}=f(\hat{A})\hspace{.1in}\Longrightarrow\hspace{.1in} W_B\subseteq W_A$ and
$\hat{A}=f(\hat{B})\hspace{.1in}\Longrightarrow\hspace{.1in} W_A\subseteq W_B$ )
It is possible to transform the set of self-adjoint operators into a poset by defining a new 
category $[\mathcal{O}]$ in
which the objects are taken to be equivalence classes of operators, whereby two operators are
considered to be equivalent if the $\mathcal{O}$-morphisms  relating them satisfies equation
 \ref{eq:eq}.
\subsubsection{Category $\mathcal{W}$ of Boolean subalgebras}
\begin{definition} \label{def:W}
The \textbf{category $\mathcal{W}$} of Boolean subalgebras of the lattice $P(\mathcal{H})$ has:
\begin{itemize}
\item as objects, the individual Boolean subalgebras, i.e.elements $W\in\mathcal{W}$ which represent 
spectral algebras associated with different operators. 
\item as morphisms, the arrows between objects of $\mathcal{W}$, such that a morphism 
$i_{W_1W_2}:W_1\rightarrow W_2$ exists iff $W_1\subseteq W_2$.
\end{itemize}
\end{definition}
From the definition of morphisms it follows that there is, at 
most, one morphisms between any two elements of $\mathcal{W}$, therefore $\mathcal{W}$
forms a poset under subalgebras inclusion $W_1\subseteq W_2$.
To show that $\mathcal{W}$, as defined above is indeed a category, we need to define the identity
arrow and the composite arrow.

The identity arrow in $\mathcal{W}$ is defined as $id_W:W\rightarrow W$, which corresponds 
to $W\subseteq W$ whereas, given two $\mathcal{W}$-arrows $i_{W_1W_2}:W_1\rightarrow W_2$ 
$(W_1\subseteq W_2)$ and 
$i_{W_2W_3}:W_2\rightarrow W_3$ $(W_2\subseteq W_3)$ the composite 
$i_{W_2W_3}\circ i_{W_1W_2}$ corresponds to $W_1\subseteq W_3$. 
\begin{example}
An example of the category $\mathcal{W}$ can be formed in the following way:
consider a category formed by four objects $\hat{A}$,$\hat{B}$,$\hat{C}$,$\hat{1}$,
such that the spectral decomposition is the following:
\begin{align*}
\hat{A}&=a_1\hat{P}_1+a_2\hat{P}_2+a_3\hat{P}_3\\
\hat{B}&=b_1(\hat{P}_1\vee\hat{P}_2)+b_2\hat{P}_3\\
\hat{C}&=C_1(\hat{P}_1\vee\hat{P}_3)+c_2\hat{P}_2
\end{align*}
then the spectral algebras are the following:
\begin{align*}
W_A&=\{\hat{0},\hat{P}_1,\hat{P}_2,\hat{P}_3,\hat{P}_1\vee\hat{P}_3,\hat{P}_1\vee\hat{P}_2,\hat{P}_3\vee\hat{P}_2,\hat{1}\}\\ 
W_B&=\{\hat{0},\hat{P}_3,\hat{P}_1\vee\hat{P}_2,\hat{1}\}\\
W_C&=\{\hat{0},\hat{P}_2,\hat{P}_1\vee\hat{P}_3\hat{1}\}\\
W_1&=\{\hat{1}\}
\end{align*}
The relation between the spectral algebras is given by the following diagram:
\[\xymatrix{
&&W_B\ar[rrd]&&\\
W_1\ar[rru]\ar[rrd]&&&&W_A\\
&&W_C\ar[rru]&&
}\]
where the arrows are subset inclusions.
\end{example}
\textbf{Relation between categories}\\
The categories, as defined above, can be related to another through the spectral algebra functor. 
\begin{definition}
The \textbf{spectral algebra functor} is a contravariant functor $W:\mathcal{O}\rightarrow\mathcal{W}$,
such that:
\begin{itemize} 
\item each object $\hat{A}\in\mathcal{O}$ is mapped to the object $W_A\in\mathcal{W}$ where $W_A$ is 
the spectral algebra of $\hat{A}$ 
\item given an $\mathcal{O}$-arrow $f_{\mathcal{O}}:\hat{B}\rightarrow\hat{A}$ then the 
corresponding $\mathcal{W}$-arrow is $i_{W_AW_B}:W_A\rightarrow W_B$ which is defined as subset 
inclusion.
\end{itemize}
\end{definition} 
The above definition of morphisms in W as subset inclusions is motivated by the following reasoning:
let us consider an object $\hat{A}\in\mathcal{O}$ whose spectral algebra is $W_A\in\mathcal{W}$.
If there exists a map $f_{\mathcal{O}}:\hat{B}\rightarrow\hat{A}$, such that
$\hat{B}=f(\hat{A})$, then from the Spectral Theorem it follows that 
the spectral algebra $W_B$ 
of $\hat{B}$ is a subalgebra of $W_A$ i.e. $W_B\subseteq W_A$. Therefore, to each map 
$f_{\mathcal{O}}:\hat{B}\rightarrow\hat{A}$, there corresponds a  
unique map $i_{W_BW_A}:W_B\rightarrow W_A$ which represents subset inclusion.
\subsubsection{Category Hilb (Hilbert spaces)}
Given the collection of all possible Hilbert spaces, it is possible to transform this collection into a Category in its own right
by defining the following:
\begin{itemize}
\item Objects of Hilb are defined as (arbitrary) Hilbert spaces
\item Morphisms in Hilb are identified as bounded linear operators between the various Hilbert spaces.
\end{itemize}
In order to rigorously prove that Hilb, as defined above is a category, we need to prove the following:
\begin{enumerate}
\item \textbf{composition condition}
\item \textbf{associative law}
\item \textbf{identity law}
\end{enumerate}
1) and 3) are straitforward to prove: 1) given $T:H\rightarrow H^1$ and $G:H^1\rightarrow H^2$ we then get 
$G\hspace{.01in}\circ\hspace{.01in} T:H\rightarrow H^2$.\\
3) $1_H:H\rightarrow H$. Condition 2) follows. 
It is possible to show that Hilb is a *-Category and a Monoidal category. This is a desirable feature since the category nCob 
(defined below) shares the same properties (definition \ref{ssmcat} \ref{sscat}).
Why are these extra definitions needed? The answer lies in the existence of the inner product and tensor product in the 
Hilbert space. In fact, bounded 
linear operators do not preserve the inner product which is irrelevant in transforming the collections of Hilbert spaces in a 
category (from a mathematical point of view), but it is relevant for using the Hilbert space in the context of Quantum Mechanics.

Moreover in any "normal category" the tensor product would be equivalent to the Cartesian product, condition that does not 
agree in a Quantum Mechanical setting.
Therefore the extra properties of Hilb being a *-Category and a Monoidal category account for the inner product and tensor 
product, respectively.

We will not go into the detail of how these two categories are implemented in Quantum Mechanics, 
the exact detail can be found in \cite{isham12}. What is important, at this stage, is that it has been proved possible to 
describe Quantum mechanics in terms of a category, which is very similar to the category nCob (defined below) through which
General Relativity is described. This, then, creates the platform for applying an equivalent topos theory to both 
General Relativity and Quantum Gravity. This would seem a desirable aim since it might shed new light on a possible way of
uniting the above two theories.   
\subsection{Category nCob in General Relativity}
It is possible to describe General Relativity in terms of the category nCob in which we have the following:
\begin{itemize}
\item Objects are identified with (arbitrary) (n-1)-dimensional manifolds which represent space at a given time.
\item Morphisms are identified with n-dimensional manifolds which represent spacetime (also called cobordism). The conditions 
on this cobordism are such that given two (n-1)-manifolds S and $S_1$, then M is a cobordism between S and $S_1$ iff the 
boundary of M is the union of S and $S_1$.
It is useful to think of M as a process which changes the Topological structure of space, i.e. process of time passing such that 
its effects (time) are identified with Topological changes in space.
\end{itemize}
Within this framework we identify the following:
\begin{enumerate}
 \item \emph{Composition}: given $M:S\rightarrow S_1$ and $M^1:S_1\rightarrow S_2$ the composite is $M^1M:S\rightarrow S_2$ 
 such that associativity is satisfied : $(M^2M^1)M=M^2(M^1M)$ 
 \item\emph{Identity}: $1_S:S\rightarrow S$ such that $1_S \circ M=M$ and $M \circ 1_S=M$
 \end{enumerate}
 It can be shown that nCob is both a *-Category and a Monoidal Category (see \cite{isham12} for detail) 
 \subsubsection{Relation between nCob and Hilb}
 Given the category nCob and Hilb, it is possible to create a covariant functor $Z:nCob\rightarrow Hilb$ such that for any 
 (n-1)-manifold S it assigns a Hilbert space of states Z(S) and, given a cobordism $M:S\rightarrow S_1$ we obtain the corresponding 
 function $Z(M):Z(S)\rightarrow Z(S_1)$. \\Z(M) is such that the following conditions are satisfied:
 \begin{itemize}
 \item given $M:S\rightarrow S_1$ and $M^1:S_1\rightarrow S_2$ then\\
 $Z(M^1 \circ M)= Z(M^1) \circ Z(M)$
 \item $Z(1_S)=1_{Z(S)}$ where S=(n-1)-dimensional manifold.
 \end{itemize}
 J.C. Baez identified this functor as  a representation of a Topological Field Theory (for detail see \cite{isham12})   
\subsection{Monoidal Category}
\label{ssmcat}
A Monoidal category $\mathcal{M}$ is a one object category equipped with a binary operation on that object and a unit 
element.In the situation in which the object in $\mathcal{M}$ is a category, then $\mathcal{M}$ is defined as follows
\begin{definition} 
A monoidal category $\mathcal{M}$ is a triplet (M, *, i), such that
\begin{itemize}
\item M is a category
\item * is a functor $M\times M\rightarrow M$
\item $i\in M$ such that $\forall\hspace{.02in}x\in M$ i * x = x * i = x
\end{itemize}
\end{definition}
The * functor can be identified with the tensor product, direct sum or direct product according to which category 
M one is taking into consideration.
\subsection{*-Category}
\label{sscat}
\begin{definition} 
A *-category is a category in which for each morphisms $f:a\rightarrow b$ there is associated a morphism $f^*:b\rightarrow a$ 
such that the following are satisfied
\begin{itemize}
\item $1^*_a=1_a$
\item $(fg)^*=g^*f^*$
\item $f^{**}=f$
\end{itemize}
\end{definition} 
 
\section{Topos Theory}
In this Section we will describe what a Topos is \cite{topos3} \cite{topos7} and we will illustrate this 
definition with some examples. Since the Topos we will be most concerned with 
is the Topos of Presheaves, we will pay particular attention to examples given within that Topos.

A Topos, as previously stated, is a category in which a number of basic constructions of a category are always possible. 
A number of known categories are, in fact, Topoi.

\begin{definition} \label{def:topos}
A \textbf{Topos} is a category T with the following extra properties:
\begin{itemize}
\item T has an initial (0) and a terminal (1) object
\item T has pullbacks
\item T has pushouts
\item T has exponentiation, i.e. T is such that for every pair of objects X and Y in T exists 
the map $Y^X$
\item T has a subobject classifier 
\end{itemize}
\end{definition}
Let us analyse each property individually.
\subsection{Initial and Terminal objects}
\subsubsection{Initial Object}
\begin{definition}  \label{def:ini}
An \emph{initial object} in a category $\mathscr{C}$ is a $\mathscr{C}$-object 0 such that, 
for every other $\mathscr{C}$-object A, there exists one and only one $\mathscr{C}$-arrow from 0 
to A.
\end{definition}
\textbf{Examples}
\begin{enumerate}
\item 
In $C\downarrow\mathds{R}$ the initial object is $f:\emptyset\rightarrow\mathds{R}$, 
such that the following diagram commutes:
\[\xymatrix{
\emptyset\ar[rr]^k\ar[ddr]_f&&A\ar[ddl]^g\\
&&\\
&\mathds{R}& \\
}\]
\item In Set the initial object is the 0 element.
\item In the Topos of Presheaves $\mathcal{S}^{\mathscr{C}^{op}}$ we have the following 
definition for an initial object:
\begin{definition}
A \emph{initial object} in $\mathcal{S}^{\mathscr{C}^{op}}$ is the constant functor 
$0:\mathscr{C}\rightarrow\mathcal{S}$
that maps every $\mathscr{C}$-object to the empty Set $\emptyset$ and every 
$\mathscr{C}$-arrow to 
the identity arrow on $\emptyset$.
\end{definition}
An initial object is the dual of a terminal object. 
\end{enumerate}
\subsubsection{Terminal Object}
\begin{definition} \label{def:term}
A \emph{terminal object} in a category $\mathscr{C}$ is a $\mathscr{C}$-object 1 such that, 
given any other $\mathscr{C}$-object A, there exists one and only one $\mathscr{C}$-arrow from A 
to 1. 
\end{definition}
\textbf{Examples} 
\begin{enumerate}
\item
in $C\downarrow\mathds{R}$ the terminal object is ($\mathds{R}$, $id_{\mathds{R}}$),
\[\xymatrix{
A\ar[rr]^k\ar[ddr]_f&&\mathds{R}\ar[ddl]^{id_{\mathds{R}}}\\
&&\\
&\mathds{R}& \\
}\]
commutes ($\therefore$ k=f)
\item For example in set (S) a terminal object is a singleton $\{*\}$, since given any other element $A\in S$ there exist 1 and only 1 arrow $A\rightarrow\{*\}$.

\item 
A terminal object in the Topos of presheaves $\mathcal{S}^{\mathscr{C}^{op}}$ 
is defined as follows:
\begin{definition}
A \textbf{terminal object} in $\mathcal{S}^{\mathscr{C}^{op}}$ is the constant functor 
$1:\mathscr{C}\rightarrow\mathcal{S}$
that maps every $\mathscr{C}$-object to the one element Set $\{0\}$ and every 
$\mathscr{C}$-arrow to 
the identity arrow on $\{0\}$.
\end{definition}
\end{enumerate}

\subsection{Pullback}
\begin{definition} \label{def:pull}
A \textit{pullback} or \textit{fibered product} of a pair of functions $f:A\rightarrow B$ and $g:B\rightarrow C$ in a category 
$\mathscr{C}$ is a pair of $\mathscr{C}$-arrows $h:D\rightarrow A$ and $k:D\rightarrow B$, such that 
the following conditions are satisfied:
\begin{enumerate}
\item $f \circ h=g \circ k$ i.e the following diagram commutes

\[\xymatrix{
D\ar[rr]^{k}\ar[dd]^{h}&&B\ar[dd]^{g}\\
&&\\
A\ar[rr]_f&&C\\
}\]
  
One usually writes $D=A\times_C B$ 
\item Given two functions $i:E\rightarrow A$ and $j:E\rightarrow B$, where $f \circ i=g \circ j$,
then there exists a unique $\mathscr{C}$-arrow l from E to D such that the outer rectangle of 
the following diagram commutes
\[\xymatrix{
E\ar@/^/[rrrrd]^j\ar@{-->}[rrd]^l\ar@/_/[rrddd]_j&&&&\\
&&D\ar[rr]^k\ar[dd]^h&&B\ar[dd]^g\\
&&&&\\
&&A\ar[rr]_f&&C\\
}\]

i.e.
\begin{equation*}
i=h \circ l\hspace{.2in}j=k \circ l
\end{equation*} 
We then say that $f$ (respectively $g$) has been pulled back along $g$ (respectively $f$) 
\end{enumerate}
\end{definition}
\textbf{Examples}
\begin{enumerate}
\item
If A, C, D and B where sets then 
 $D=A \times_C B=\{(a,b)\in A \times B | f(a)=g(b)\}\subseteq A \times B$ 
\item
Pullbacks exist in any (functor category) topos of presheaves $Sets^{C^{op}}$.
In fact, if $X,Y,B\in Sets^{C^{op}}$, then $P\in Sets^{C^{op}}$ is a pullback in $Sets^{C^{op}}$ iff
\[\xymatrix{
P(C)\ar[rr]^{k}\ar[dd]_{h}&&Y(C)\ar[dd]^{g}\\
&&\\
X(C)\ar[rr]_l&&B(C)\\
}\]
is a pullback in set.
This implies that $P=(X \times_B Y)_C\cong X(C)\times_{B(C)} Y(C)$.
Specifically, the above diagram implies that $P:\mathscr{C}\rightarrow Set$ assigns to each object $C\in\mathscr{C}$ an object P(C), thus obtaining 
in $\mathcal{S}^{\mathscr{C}^{op}}$ the following pullback cube:
\[\xymatrix{
P(C)\ar[rrr]^{k}\ar[rd]^{P(f)}\ar[dd]^{h}&&&Y(C)\ar[dd]^{g}\ar[rd]^{Y(f)}&\\
&P(A)\ar[dd]\ar[rrr]&&&Y(A)\ar[dd]\\
X(C)\ar[rrr]^l\ar[rd]^{X(f)}&&&B(C)\ar[rd]{B(f)}&\\
&X(A)\ar[rrr]&&&B(A)\\
}\]
such that for each $f:A\rightarrow C$ in $\mathscr{C}$we obtain the unique arrow $P(f):P(C)\rightarrow P(A)$ in $\mathcal{S}^{\mathscr{C}^{op}}$
\end{enumerate}

\subsection{Pushouts}
A pushout is essentially the dual of a pullback, therefore it has co-products where the pullback has products and the 
direction of all arrows has to be reversed.
By the duality principle all categories that have a pullback must also have a pushout. Therefore, for the sake of brevity, we will omit
any further elaboration.

\subsection{Exponentiation}
\begin{definition} \label{def:exp}
An \emph{exponentiation} from a $\mathscr{C}$-object A to a $\mathscr{C}$-object 
B is a map $f:A\rightarrow B$ denoted $B^A$ together with an evaluation map 
$ev:B^A\times A\rightarrow B$ with the property that, given any other $\mathscr{C}$-object C and 
$\mathscr{C}$-arrow $g:C\times A\rightarrow B$, there exists a unique arrow 
$\hat{g}:C\rightarrow B^A$, such that the following diagram commutes
\[\xymatrix{
B^A\times A\ar[rr]^{ev}&&B\\
&&\\
C\times A\ar@{-->}[uu]^{\hat{g}\times 1_A}\ar[rruu]_g&&\\
}\]
\end{definition} 
The definition of exponentiation implies the following:
\begin{definition} \label{def:one}
objects of $B^A$ are in one-to-one correspondence with maps of the form $A\rightarrow B$.
To see this, let us consider the following commuting diagram
\[\xymatrix{
B^A\times A\ar[rr]^{ev}&&B\\
&&\\
1\times A\ar@{-->}[uu]^{\hat{f}\times 1_A}\ar[rruu]_f&&\\
}\]
where $f:1\times A\rightarrow B$ is unique but $1\times A\equiv A$, therefore to each element of $B^A$ there corresponds a unique function $A\rightarrow B$.
\end{definition}
\textbf{Examples}
\begin{itemize}
\item In Set: given two objects $A$ and $B$, the exponential $B^A$ is defined as follows
\be
B^A=set=\{f|f\text{is a function from A to B}\}\ee in this case the evaluation map 
would be the
following: $ev(\langle f,x\rangle)=f(x)$ with $x\in A$
\item In $Sets^{\mathscr{C}^{op}}$ the exponentiation can be defined as follows:\\
consider $F\in Sets^{\mathscr{C}^{op}}$, such that given an object $a\in\mathscr{C}$ F defines a functor 
$F_a:\mathscr{C}\downarrow a\rightarrow Set$
such that to each object $f:b\rightarrow a\in\mathscr{C}\downarrow a$ it assigns an object F(b), and to each arrow
$h:f\rightarrow g$ such that the diagram
\[\xymatrix{
b\ar[rr]^h\ar[ddr]_f&&c\ar[ddl]^g\\
&&\\
&a& \\
}\] commutes,
it assigns the arrow $F(h):F(c)\rightarrow F(b)$.
Given this context, we define the exponential $G^F:\mathscr{C}\rightarrow Set$ between the contravariant functors F and G,
as follows:\\
$G^F(a)=Nat[F_a,G_a]$, i.e. the elements of $G^F(a)$ are the collection of all natural transformations from 
$F_a$ to $G_a$. The arrows in $G^F(a)$ are, instead, defined in the following way: given a function $k:a\rightarrow d$ we get:
$G^F(k):Nat[F_d,G_d]\rightarrow Nat[F_a,G_a]$. \\
To better understand this definition let us consider 
the function $\alpha\in Nat[F_d,G_d]$ and $\theta\in Nat[F_a,G_a]$, such that the action of $G^F(k)$ can be illustrated as 
follows:
\[\xymatrix{
F_d\ar[dd]_{\alpha}&&\\
&&\\
G_d&&
}
\xymatrix{
&&\\
&\ar@{=>}[r]^{G^F(k)}&}
\xymatrix{&&F_a\ar[dd]^{\theta}\\
&&\\
&&G_a\\
}\]
i.e an arrow in $G^F(k)$ assigns to each natural transformation from 
$F_d$ to $G_d$, a natural transformation  from $F_a$ to $G_a$ iff there exist a function $F(k):F(d)\rightarrow F(a)$, 
and a function $G(h):G(d)\rightarrow G(c)$ such that $h=k \circ f$ for some $f:c\rightarrow a$ and 
(from definition of F(k) and G(h)) the following diagram commutes
\[\xymatrix{
a\ar[rr]^k&&d\\
&&\\
&&c\ar[lluu]^f\ar[uu]_{kof=h}\\
}\]
therefore $\alpha$ and $\theta$ have components $\theta_f=\alpha_{kof}$.
In this formulation the evaluation function would be the following: $ev:G^F\times F\rightarrow G$ in $Sets^{\mathscr{C}^{op}}$. 
This map has components  $ev_a:G^F(a)\times F(a)\rightarrow G(a)$ where $ev_a(\langle\theta,x\rangle)=\theta_{1_a}(x)=\alpha_{ko1_a}(x)$, 
$\theta\in Nat[F_a,G_a]$ and $x\in F(a)$
\end{itemize}

\subsection{Subobject Classifier}
\subsubsection{Subobjects}
In order to define what a subobject classifier is we first need to understand what a subobject 
(categorical version of a subset) is, and what it means for an element to belong or not to 
a certain subobject.\\
For this purpose let us consider a specific example in Set, which is a type of Category.
Given a subset A of S i.e $A\subseteq S$, the notion of being a subset can be expressed mathematically
using the so called characteristic function: $\chi_A:S\rightarrow\{0,1\}$, which is 
defined as follows:
\begin{equation}
\chi_A(x)=\begin{cases}0& if\hspace{.1in}x\notin A\\
1& if\hspace{.1in}x\in A  \label{eq:mam}
\end{cases}
\end{equation}
(here we interpret 1=true and 0=false). The role of the characteristic function is to determine what 
elements belong to a certain subset.\\
Remembering that in any category subobjects are identified as monic arrows, we define the value true as follows:
\ba
true:1=\{0\}&\rightarrow& 2=\{0,1\}\nonumber\\
 0&\mapsto& 1\ea
It can be easily seen that $A=\chi_A^{-1}(1)$, which is equivalent to saying that the diagram
\begin{diagram} \label{dia:11}
 \[\xymatrix{
*++{S}\ar@{^{(}->}[rr]\ar[dd]&&A\ar[dd]^{\chi_A}\\
&&\\
1\ar[rr]_T&&2\\
}\]
\end{diagram}
is a pullback.\\
\textbf{Example}\\
Consider the the Topos of presheaves $Sets^{\mathscr{C}^{op}}$, a subobject of a presheaf is defined as follows: 
\begin{definition} \label{def:sub}
Y is a \textbf{subobject of a presheaf} X if there exists a natural transformation 
$i:Y\rightarrow X$ 
which is defined 
componentwise as $i_a:Y(A)\rightarrow X(A)$ and where $i_a$ defines a subset embedding,
i.e. $Y(A)\subseteq X(A)$.
\end{definition}
Since Y is itself a presheaf, the maps between the objects of Y are the restrictions of the 
corresponding maps between the objects of X.
This can be easily seen with the aid of the following diagram:

\[\begin{xy} 0;/r15mm/:
,(0,0)="x",*@{},*+!U{Y(A)}
,{\ellipse(0.5,1.6){}}
,{\ellipse(0.3,0.8){}}
,(1.8,-0.3),*@{},*+!{Y(f)}
,(0.6,1.5),*@{},*+!{X(A)}
,(3.6,1)="z",*@{},*+!{}
,(0.4,1)="y",*@{},*+!{}
,{\ar"y";"z"}
,(4.6,1.5),,*@{},*+!{X(B)}
,(2,1.1),,*@{},*+!{X(f)}
,(4,0)="p"*@{},*+!U{Y(B)}
,{\ellipse(0.5,1.6){}}
,{\ellipse(0.3,0.8){}}
,{\ar"x";"p"}
 \end{xy}\]

An alternative way of expressing this condition is through the following commutative diagram:

\[\xymatrix{
Y(A)\ar[rr]^{Y(f)}\ar[dd]_{i_A}&&Y(B)\ar[dd]^{i_B}\\
&&\\
X(A)\ar[rr]_{X(f)}&&X(B)\\
}\]

\subsection{Subobject Classifier}
Motivated by the definition of a subobject in Sets, we construct the following definition for a subobject classifier in 
a general 
category.
\begin{definition} \label{def:clas}
Given a Category with a terminal object 1, a subobject classifier is an object $\Omega$, 
together with a monic arrow $\mathcal{T}:1\rightarrow\Omega$ such that, given a monic $\mathscr{C}$-arrow 
$f:a\rightarrow b$, there exists one and only one $\chi_f$ arrow, such 
that the following is a pullback  
\[\xymatrix{
*++{a}\ar@{>->}[rr]\ar[dd]&&b\ar[dd]^{\chi_f}\\
&&\\
1\ar[rr]_T&&\Omega\\
}\]


\end{definition}
\vspace{.5in}
\begin{axiom} \label{axi:1}
Given a category $\mathscr{C}$, then there exists an Isomorphisms
\be y:Sub_C(X)\cong Hom_C(X,\Omega)\hspace{.1in}\forall X\in C\ee
\end{axiom}
In order to prove the above axiom we need to show that y is a) injective and b) surjective. Since the prove of the above 
theorem in topos is quite complicated and needs definitions, not yet given, we will use an analogous proof in Sets, which 
essentially has the 
same strategy as the proof in topos, but it is much more intuitive.
In Sets we can write the above axiom as follows:
\begin{axiom}
The collection of all subsets of S denoted by $\mathcal{P}(S)$, and 
the collection of all maps from S to the set $\{0,1\}=2$ denoted by $2^S$ are isomorphic, i.e. the function $y:\mathcal{P}(S)\rightarrow2^S$ which, in terms of single elements of $\mathcal{P}(S)$ is $A\rightarrow\chi_A$, is a bijection.
\end{axiom}

\begin{proof}
Let us consider the diagram \ref{dia:11}\\
a) \emph{y is injective (1:2:1)}:\\
 consider the case in which $\chi_A=\chi_B$ where
 \begin{equation*}
 \chi_B(x)=\begin{cases}1& iff\hspace{.1in}x\in B\\
 0& iff\hspace{.1in}x\notin B
 \end{cases}
 \end{equation*}
 It follows that since the two functions are the same, to the codomain 1 they both associate the 
 same domain, therefore A=B\\
 b) \emph{y is surjective (onto)}:
 given any function $f\in2^S$
 then there must exist a subset A of S, such that $A_f=\{x:x\in D\hspace{.1in}and\hspace{.1in} f(x)=1\}$, 
 i.e. $A_f=f^{-1}(\{1\})$ 
 therefore $f=\chi_{A_f}$
\end{proof}
\subsubsection{Elements of the subobject classifier}
In the simple Set case $\Omega\cong\{0,1\}$, therefore the elements of $\Omega$ are simply 0 and 1. This is not the case for 
a general Topos. In fact in what follows we will prove that the elements of a subobject classifier in  Topos are sieves.
Since the notion of sieves is quite complicated we will describe it in detail in the next subsection, and then prove that 
sieves so described correspond to elements of a subobject classifier.
\subsubsection{Sieve}
In order to define elements of a subobject we first need to be familiar with the notion of sieve.
\begin{definition}
A \textbf{sieve} on an object $A\in\mathscr{C}$ is a collection S of morphisms in $\mathscr{C}$ 
whose codomain is A and such that, if $f:B\rightarrow A\in S$ then, given any morphisms 
$g:C\rightarrow B$ we have
$f o g\in S$, i.e. S is closed under left composition:
\[\xymatrix{
B\ar[rr]^{f}&&A\\
&&\\
C\ar[uu]^{g}\ar[rruu]_{fog}&&\\
}\]
\end{definition} 
For example in a poset a sieve is an upper set. Specifically, given a poset C, a sieve on 
$p\in C$ is any subset S of C, such that if $r\in S$ the 1) $p\leq r$ 2) $r^`\in S\hspace{.1in}\forall r\leq r^`$.
\\
A map $\Omega_{qp}:\Omega_p\rightarrow\Omega_q$ between sieves exists iff $p\leq q$ then, given 
$S\in\Omega_q$, $\Omega_{qp}$ is defined as follows: 
\begin{equation*}
\Omega_{qp}(S):=\uparrow p\cap S
\end{equation*}
where $\uparrow p:=\{r\in C|p\leq r\}$\\
 
An important property of sieves is the following: if $f:B\rightarrow A$ belongs to $S$ which is a 
sieve on $A$, then the pullback of $S$ by $f$ determines a principal sieve on $B$, i.e.
\begin{equation*}
f^*(S):=\{h:C\rightarrow B|f o h \in S\}=\{h:C\rightarrow B\}=\downarrow B   
\end{equation*}
\[\xymatrix{
&&C\ar[dl]\\
&B\ar[dl]&\\
A&&D\ar[ll]\\
}\hspace{.2in}
\xymatrix{
&&\\
\ar@{=>}[rr]^{f^*}&&}\hspace{.2in}
\xymatrix{
&&\\
B\ar@(ul,ur)[]^{i_B}&&C\ar[ll]
}\]
The principal sieve of an object A, denoted by $\downarrow A$, is the sieve that contains the 
identity morphism of A therefore it is the biggest sieve on A.\\
An important property of sieves is that the set of sieves defined on an object
 forms an Heyting algebra (definition \ref{shey}), with partial ordering given by subset inclusion.\\

\subsection{Elements as Sieves}
The elements $\Omega$ in a Topos are derived from the following theorem:

\begin{theorem} \label{the:2}
Subpresheaves can be identified with sieves 
\end{theorem}
In order to prove the above theorem we need the following lemma:
\begin{lemma}
\textbf{Yoneda Lemma}: Given an arbitrary presheaf P on a category C and a functor y from C to the set of contravariant
functors on C, i.e $y:C\rightarrow Sets^{\mathscr{C}^{op}}$ elementwise $A\rightarrow Hom_C(-,A)$;
there exists a bijective correspondence between natural transformations
$y(A)\rightarrow P$ and elements of the set P(A):\be
\theta:Hom_C(y(A),P)\rightarrow^{\sim} P(A)\ee
defined for $\alpha:y(A)\rightarrow P$ by $x(\alpha)=\alpha_A(1_A)$
\end{lemma}
\begin{example}
for each element A on a category $\mathscr{C}$ we define a presheaf y(A) such that:
\begin{itemize}
\item 
Given an object D of $\mathscr{C}$ we have \\
$y(A)D=Hom_\mathscr{C}(D,A)$
\item 
Given a morphism $\alpha B\rightarrow D$ and $\theta:D\rightarrow A$ we obtain:
$y(A)(\alpha):Hom_\mathscr{C}(D,A)\rightarrow Hom_\mathscr{C}(B,A)$\\
$y(A)(\alpha)(\theta)=\theta o \alpha$
\end{itemize}
Given any morphism on C of the form $f:A\rightarrow A_1$ then there exists a natural transformation $y(A)\rightarrow y(A_1)$
therefore, y is actually a functor from the category C to the set of presheaves defined on $\mathscr{C}$, i.e 
$y:\mathscr{C}\rightarrow Sets^{\mathscr{C}^{op}}$, 
such that to each object of y (which is defined as a contravariant functor which assigns to an object in $\mathscr{C}$ 
a presheaf on that
object) there corresponds an element of a Presheaf on $\mathscr{C}$, precisely an element of the presheaf which is the codomain of y.   
\end{example}
We can now prove theorem \ref{the:2}
\begin{proof}
Let us consider $\Omega$ to be a subobject classifier of $\hat{C}=Sets^{C^{op}}$. Given a presheaf 
$y(C)=Hom_{\hat{C}}(-,C): C^{op}\rightarrow Sets$,
we know from \ref{axi:1} that $Sub_{\hat{C}}(Hom_\mathscr{C}(-,C))\cong Hom_{\hat{C}}(Hom_\mathscr{C}(-,C),\Omega)$, therefore,
form Yonedas lemma it follows that 
$Hom_{\hat{C}}(Hom_{\mathscr{C}}(-,C),\Omega)=\Omega(A)$.
Thus the subobject classifier $\Omega$ must be a presheaf $\Omega:\mathscr{C}\rightarrow Set$ such that 
\begin{align*}
\Omega(A)=& Sub_{\hat{C}}(Hom_{\mathscr{C}}(-,C))\\
& =\{S|S\hspace{.1in}a\hspace{.1in}subfunctor\hspace{.1in}of\hspace{.1in}Hom_{\mathscr{C}}(-,C)\}
\end{align*}
Now if $Q\subset Hom_{\mathscr{C}}(-,C)$ is a subfunctor then the set 
$S=\{f|\hspace{.1in}for\hspace{.1in}some\hspace{.1in}object\hspace{.1in}A,\hspace{.1in}f:A\rightarrow C
\hspace{.1in}and\hspace{.1in}f\in Q(A)\}$ 
is a sieve on C. \\ 
Conversely given a sieve S on C we define 
$Q(A)=\{f|f:A\rightarrow C\hspace{.1in}and\hspace{.1in}f\in S\}\subseteq Hom_{\mathscr{C}}(A,C)$ 
which produces a presheaf $Q:\mathscr{C}\rightarrow Set$ which is a subfunctor of $Hom_{\mathscr{C}}(-,C)$.
Since the transformation function from Q to S is a bijection (as can be seen from above definition) we can conclude that a
Sieve on $A$ is equivalent to a subfunctor of $Hom_{\mathscr{C}}(-,C)$ 
\end{proof}
Given the above proof we can now define a subobject classifier in the topos of presheaves in a more rigorous way.
\subsection{Subobject Classifier In The Topos Of Presheaves}
\begin{definition}
A \textbf{Subobject Classifier} $\Omega$ is a presheaf 
$\Omega:\mathscr{C}\rightarrow\mathcal{S}^{\mathscr{C}^{op}}$ such that to each object $A\in\mathscr{C}$
there corresponds an object $\Omega(A)\in\mathcal{S}^{\mathscr{C}^{op}}$ which represents the set 
of all sieves
on A, and to each 
$\mathscr{C}$-arrow $f:B\rightarrow A$ there corresponds an 
$\mathcal{S}^{\mathscr{C}^{op}}$-arrow $\Omega(f):\Omega(A)\rightarrow\Omega(B)$ such that
$\Omega(f)(S):=\{h:C\rightarrow B|f o h\in S\}$ is a sieve on B, where $\Omega(f)(S)\equiv f^*(S)$ 
\end{definition}
We now want to show that this definition of subobject classifier is in agreement with definition \ref{def:clas}.
In order to do that we need to define the analogue of arrow 
\textit{true} (T) and the \textit{character function} in Topos.
\begin{definition}
$\textbf{T}:1\rightarrow\Omega$ is the natural transformation that has components $T_A:\{0\}\rightarrow\Omega(A)$ 
given by $T_A(0)=\downarrow A$ = principal sieve on A (see appendix)
\end{definition}
To understand how T works, let us consider a monic arrow  
$f:F\rightarrow X$ in $\mathcal{S}^{\mathscr{C}^{op}}$
which is 
defined componentwise as $f_A:F(A)\rightarrow X(A)$ and represents subset inclusion.
Now we define the character $\chi^f:X\rightarrow\Omega$ of $f$ which is a natural transformation in 
the
topos of presheaves, such that the components $\chi_A^f$ represent functions from 
X(A) to $\Omega(A)$, as shown in the following diagram:\begin{diagram} \label{dia:ni}
\[\xymatrix{
*++{F(A)}\ar@{^{(}->}[rr]^{f_A}\ar[dd]&&X(A)\ar[dd]^{\chi^f_A}\\
&&\\
\{0\}\ar[rr]^T&&\Omega(A)\\
}\]
\end{diagram}
where $\{0\}\equiv 1$.
From the above diagram we can see that $\chi^f_A$ assigns to each element x of X(A)
a sieve $\Omega(A)$ on A. For a function to belong to the sieve $\Omega(A)$ on A we require that
the following diagram commutes:
\begin{diagram}                                          
  \[\xymatrix{
*++{F(A)}\ar@{^{(}->}[rr]\ar[dd]_{F(f)}&&X(A)\ar[dd]^{X(f)}\\
&&\\
*++{F(B)}\ar@{^{(}->}[rr]&&X(B)\\
}\]            
\end{diagram}                             
therefore
\begin{equation}
\chi_A^F(x):=\{f:B\rightarrow A|X(f)(x)\in F(B)\}    \label{eq:char}
\end{equation}
What equation \ref{eq:char} means is that we require F(f) to be the restriction of X(f) to F(A).
This condition is expressed by the following diagram:
 \begin{diagram}                                           
  \[\begin{xy} 0;/r15mm/:
(0,0)="x",*@{},*+!U{}
,(0,0.2),*@{},*+!U{X(f)}
,(1,1.7),*@{},*+!U{F(A)}
,(2.8,1.7),*@{},*+!U{X(A)}
,(1.2,-1.6),*@{},*+!U{X(B)}
,(-0.7,-1.35),*@{},*+!U{F(B)}
,(0.1,1.4)="f"*@{*},*+!LD{x}
,(1,1)="o"*@{},*+!LD{}
,{\ellipse(1.6,0.8){}}
,{\ellipse(0.9,0.3){}}
,(-0.6,-1)="p"*@{},*+!R{}
,{\ellipse(1.6,0.8){}}
,{\ellipse(0.9,0.3){}}
,{\ar"f";"p"}
\end{xy}\]
\end{diagram}
i.e. $f$ belongs to $\Omega(A)$ iff X(f) maps x into F(B). 
$\chi_{A}^F(x)$ as defined by equation \ref{eq:char}
represents a sieve on A.
\begin{proof}
Consider the following commuting diagram which represents subobjects F of the presheaf X:
\begin{diagram} \label{dia:ci}
 \[\xymatrix{
F(A)\ar[rr]^{F(f)}\ar@{^{(}->}[dd]&&F(B)\ar[rr]^{F(g)}\ar@{^{(}->}[dd]&&F(C)\ar@{^{(}->}[dd]\\
&&&&\\
X(A)\ar[rr]^{X(f)}&&X(B)\ar[rr]^{X(g)}&&X(C)\\
}\]
\end{diagram}
If $f:B\rightarrow{A}$ belongs to $\chi_{A}^F(x)$
then, given $g:C\rightarrow B$ it follows that $f \circ g$ belongs to $\chi_{A}^F(x)$,
since from diagram \ref{dia:ci} it can be deduced that $X(f o g)(x)\in F(C)$. 
This is precisely the definition of a sieve so we have proved that
$\chi_{A}^F(x):=\{f:B\rightarrow A|X(f)(x)\in F(B)\}$  is a sieve.  
\end{proof}
As a consequence of \ref{axi:1} the condition of being a subobject classifier can be 
restated in the following way:
\begin{definition}
$\Omega$  is a 
\textbf{subobject classifier} iff there is a ``one to one" correspondence between subobject of X and morphisms
from X to $\Omega$.
\end{definition}
Given this alternative definition of a subobject classifier, it is easy to prove that $\Omega$ 
is a subobject classifier. In fact, from equation \ref{eq:char}, 
we can see that indeed there is a 1:2:1 correspondence between subobject of X and characteristic 
morphism (character) $\chi$.\\
Moreover for each morphism $\chi:X\rightarrow\Omega$ we have
\begin{align*}
F^{\chi}(A):&=\chi^{-1}_{A}\{1_{\Omega(A)}\}\\
&=\{x\in X(A)|\chi_{A}(x)=\downarrow A\}\\
&=\text{subobject of X} 
\end{align*} 
\subsection{Global And Local Sections}
Other important features of topos theory are the local and global sections.
\begin{definition} \label{def:glo}
A \textbf{global section} or \textbf{global element} of a presheaf X in $\mathcal{S}^{\mathscr{C}^{op}}$ is a map 
$k:1\rightarrow X$ from the terminal object 1 to the presheaf X.
\end{definition} 
What $k$ does is to assign to 
each object $A$ in $\mathscr{C}$ an element $k_A\in X(A)$ in the corresponding object of the presheaf $X$.  
The assignment is such that, given a function $B\rightarrow A$ the following relation holds
\begin{equation}
X(f)(k_A)=k_B     \label{eq:k}
\end{equation}
What \ref{eq:k} uncovers, is that the elements of $X(A)$, assigned by the global section k, are mapped 
into each other by the morphisms in $X$.
Presheaves with a local or partial section can exist even if they do not have a global section.
\begin{definition} \label{def:lo}
A \textbf{local} or \textbf{partial section} of a presheaf X in $\mathcal{S}^{\mathscr{C}^{op}}$ is a map 
$\rho:U\rightarrow X$ where $U$ is a subobject of the terminal object $1$.
\end{definition}
In a presheaf, a subobject $U$ of $1$ can 
either be the empty set $\emptyset$, or a singleton $\{*\}$.
From the above definition it is clear that a local section is an assignment of an element of an 
object of $X$ to the corresponding subobject $U$ of $1$ in $\mathscr{C}$. This assignment is said to be 
``closed
downwards", i.e. given a subobject $U(A)=\{*\}$ of $1$ and a $\mathscr{C}$-morphisms $f:B\rightarrow A$
then we have $U(B)=\{*\}$. To illustrate let us  
consider a category with 4 elements $\{A,B,C,D\}$, such that the following relations hold between 
the elements:
\[\xymatrix{
 A\ar[rr]^f\ar[dd]_i&&B\ar[dd]^g\\
 &&\\
 D\ar[rr]_p&&C\\
 }\] 
Given a subobject $U$ of $1$ we then have the following relations
\[\xymatrix{
 U(A)\ar[rr]^{U(f)}\ar[dd]_{U(i)}&&U(B)\ar[dd]^{U(g)}\\
 &&\\
 U(D)\ar[rr]_{U(p)}&&U(C)\\
 }\]
If $U(A)=\emptyset$ then $U(f)$ is either the unique function $\emptyset\rightarrow\{*\}$ iff $U(B)=\{*\}$
or $\emptyset\rightarrow\emptyset$ iff $U(B)=\emptyset$.
If instead $U(A)=\{*\}$ then the only possibility is that $U(B)=\{*\}$ since there does not exist a
function 
$\{*\}\rightarrow\emptyset$.
Therefore $\rho$ assigns to particular subsets of objects $A\in X$, elements
$\rho_A$. These objects A are called the domain of $\rho$ $(dom\hspace{.05in}\rho)$ and are such 
that the following conditions are satisfied:
\begin{itemize}
\item The domain is closed downwards, i.e. if $A\in dom\hspace{.05in}\rho$ and if there exists a map 
$f:B\rightarrow A$ then $B\in dom\hspace{.05in}\rho$
\item If $A\in dom\rho$ and if there exists a map $f:B\rightarrow A$, then the following condition 
is satisfied:
\begin{equation*}
X(f)(\rho_A)=\rho_B 
\end{equation*}
\end{itemize}  
\section{Heyting algebra}
\begin{definition} \label{shey}
 A \textbf{Heyting Algebra H} is a \textbf{relative pseudo complemented distributive lattice}.
 \end{definition}
 The property of being \emph{distributive} means that the following equations are satisfied for any 
$S_i\in\textbf{H}$ 
\begin{align*}
S_1\wedge(S_2\vee S_3)&=(S_1\wedge S_2)\vee(S_1\wedge S_3)\\
S_1\vee(S_2\wedge S_3)&=(S_1\vee S_2)\wedge(S_1\vee S_2)
\end{align*}
 The property of being \textbf{relative pseudo complemented lattice} means that for any two elements 
$S_1,S_2\in\textbf{H}$ there exist a third element $S_3\in\textbf{H}$, such that:
\begin{enumerate}
\item $S_1\cap S_3\subseteq S_2$
\item $\forall S\in\textbf{H}\hspace{.2in}S\subseteq S_3\hspace{.2in}iff\hspace{.2in}S_1\cap S\subseteq S_2$
\end{enumerate}
where $S_3$ is defined as the \textit{pseudo complement} of $S_1$ relative to $S_2$, i.e. the 
greatest element of the set $\{S:S_1\cap S\subseteq S_2\}$, and it is denoted 
as $S_1\Rightarrow S_2$.\\
 A particular feature of the Heyting algebra is the negation operation.
The negation of an element S is defined to be the pseudo-complement of $S$ i.e. $\neg S:=S\Rightarrow 0$,
therefore we can 
write 
\begin{equation*}
\neg S:=\{f:B\rightarrow A|\forall g:C\rightarrow B, f o g \notin S\}
\end{equation*}
The above equation entails that $\neg S$ is the least upper bound of the set $\{x:S\cap x=0\}$, i.e.
the biggest set that does not contain any element of S. 
From the above definition of negation operation it follows that the Heyting algebra does not satisfy 
the law of excluded middle, i.e. given any element S of an Heyting algebra we have the following 
relation: $S\vee\neg S\leq1$.
\begin{proof}
Let us consider $S\vee\neg S=S\cup\neg S$, this represents the least upper bound of S and $\neg S$ 
therefore, given any other element $S_1$ in the Heyting algebra such that $S\leq S_1$ and 
$\neg S\leq\ S_1$, then $S\vee\neg S\leq S_1$. But since for any S we have $S\leq1$ and $\neg S\leq1$
it follows that $S\vee\neg S\leq1$.
\end{proof}

\section{Sets}
\label{sset}
\begin{definition}  a \emph{pre-ordered} set is a set with the property that, between 
any two objects
there is, at most, one arrow. This entails that there exists a 
binary relation R between the objects of the pre-ordered set such that
the following holds:
\begin{enumerate}
\item aRa (reflexivity)
\item if aRb and bRc then aRc (transitivity) 
\end{enumerate} 
\end{definition}

\begin{definition} \label{def:poset} a \emph{poset} is a pre-ordered set with the extra property 
of being 
antisymmetric: $(pRq, qRp)\Rightarrow p=q$
\end{definition}
\section{Functors}
\label{sfunc}
We will now briefly explain the concept of a functor.\\
Generally speaking a functor is a transformation from one category $\mathscr{C}$
to another category
$\mathscr{D}$, such that
the categorical structure of the domain $\mathscr{C}$ is preserved, i.e. gets mapped onto 
$\mathscr{D}$.\\
There are two types of functors:
\begin{enumerate}
\item \textbf{Covariant Functor}
\item \textbf{Contravariant Functor}
\end{enumerate}
\begin{enumerate}
\item \begin{definition}:
A \textbf{covariant functor} from a category $\mathscr{C}$ to a category $\mathscr{D}$ is a map 
$F:\mathscr{C}\rightarrow\mathscr{D}$ that assigns to each $\mathscr{C}$-object a 
$\mathscr{D}$-object F(a) and to each $\mathscr{C}$-arrow $f:a\rightarrow b$ a $\mathscr{D}$-arrow
$F(f):F(a)\rightarrow F(b)$, such that the following are satisfied:
\begin{enumerate}
\item $F(1_a)=1_{F(a)}$
\item $F(f o g)=F(f) o F(g)$ for any $g:c\rightarrow a$ 
\end{enumerate}
\end{definition}
It is clear, from the above, that a covariant functor is a transformation that preserves both:
\begin{itemize}
\item the domain's and the codomain's identities;
\item the composites of functions i.e. it preserves the direction of the arrows.
\end{itemize}
This can be easily seen with the aid of the following diagram;
 \begin{diagram}
 \[\xymatrix{
a\ar[rr]^f\ar[rrdd]_h&&b\ar[dd]^g\\
&&\\
&&c\\
}
\xymatrix{\ar@{=>}[rr]^F&&}
\xymatrix{
F(a)\ar[rr]^{F(f)}\ar[rrdd]_{F(h)}&&F(b)\ar[dd]^{F(g)}\\
&&\\
&&F(c)\\
}\]
\end{diagram}

\item \begin{definition}

A \textbf{contravariant functor} from a category $\mathscr{C}$ to a category $\mathscr{D}$ is a 
map $X:\mathscr{C}\rightarrow\mathscr{D}$ that assigns to each $\mathscr{C}$-object a $\mathscr{D}$-object X(a) and to each $\mathscr{C}$-arrow $f:a\rightarrow b$ a $\mathscr{D}$-arrow
$X(f):X(b)\rightarrow X(a)$, such that the following are satisfied
\begin{enumerate}
\item $X(1_a)=1_{X(a)}$
\item $X(f o g)=X(g) o X(f)$ for any $g:c\rightarrow a$ 
\end{enumerate}
\end{definition}
A diagrammatic representation of a contravariant functor is the following:
\begin{diagram}
\[\xymatrix{
a\ar@{>}[rr]^f\ar[rrdd]_h&&b\ar[dd]^g\\
&&\\
&&c\\
}
\xymatrix{\ar@{=>}[rr]^X&&}
\xymatrix{
F(a)&&X(b)\ar[ll]^{X(f)}\\
&&\\
&&X(c)\ar[lluu]^{X(h)}\ar[uu]^{X(g)}\\
}\]
\end{diagram}

As we can see from the above diagram, a contravariant functor in mapping arrows from one category 
to the next which reverses
the directions of the arrows by mapping domains to codomains and vice versa.
\end{enumerate}


\newpage
\chapter*{}
{\LARGE\bf{Acknowledgments}}\\[40pt]
I would like to express my gratitude to my advisor Professor Thomas Thiemann for all
the support, advice, time, stimulating discussions and help he has gaven me through out these three years of my PhD. \\
I would also like to thank the referees of the dissertation, Professor Christopher J. Isham, Professor Jan Plefka and Professor Thomas Thiemann for their time. \\
In particular Professor Christopher J. Isham and Doctor Andreas D\"oring for the very useful and inspiring discussions during my PhD.\\
A special thank goes to my parents and my granny for constant support throughout.\\
Many thanks to Nicolas Behr for making my office life much more enjoyable and to Johannes Tambornino for helping me with my German.\\
Finally I would like to thank all my friends and colleagues for having been always there for me. 
\\
Thank you all
\newpage
\chapter*{}
{\LARGE\bf {Selbst\"andigkeitserkl\"arung}}\\[40pt]

Hiermit erkl\"re ich, Cecilia Flori, dass ich diese Arbeit
selbstst\"andig verfasst und dabei auf keine aanderen Hilfsmittel als jene
im Text angegebenen zurueckgegriffen habe\\[50pt]
Cecilia Flori
\end{document}